\documentclass[natbib,smallextended]{svjour3}       % onecolumn (second format)

\usepackage{color}
\definecolor{myorange}{RGB}{240,50,40}
\definecolor{mydarkblue}{RGB}{20,50,200}
\definecolor{mydarkgreen}{RGB}{50,180,50}
\definecolor{lightgrey}{RGB}{180,180,180}

\usepackage[colorlinks=true,linkcolor=mydarkblue,citecolor=mydarkblue,urlcolor=mydarkblue,breaklinks]{hyperref}
\usepackage{geometry}                % See geometry.pdf to learn the layout options. There are lots.
\usepackage{graphicx}
\usepackage{amssymb}
\usepackage{amsmath}
\usepackage{microtype}
\usepackage{aas_macros}
\usepackage{xfrac}
\usepackage[normalem]{ulem}
\usepackage{makecell}
\usepackage[T1]{fontenc}
\usepackage{lmodern}
\usepackage{mathrsfs}

\newcommand{\vecb}[1]{\ensuremath\boldsymbol{#1}}

\newcommand{\ii}{\ensuremath{\textrm{i}}}
\newcommand{\bnabla}{\ensuremath{\boldsymbol{\nabla}}}

\newcommand{\Msun}{ h^{-1}{\rm M_{ \odot}}}
\newcommand{\hMpc}{ h^{-1}{\rm Mpc}}
\newcommand{\hGpc}{ h^{-1}{\rm Gpc}}
\newcommand{\hkpc}{ h^{-1}{\rm kpc}}
\newcommand{\ihMpc}{ h\,{\rm Mpc}^{-1}}

\newcommand{\tnsr}[1]{\ensuremath{\boldsymbol{\mathsf{#1}}}}

\journalname{Living Reviews in Computational Astrophysics}

\begin{document}

\title{Large-scale dark matter simulations}
\author{Raul E. Angulo and Oliver Hahn}

\authorrunning{Raul E. Angulo and Oliver Hahn} % if too long for running head

\institute{Raul E. Angulo \at
	     Donostia International Physics Center (DIPC), Donostia-San Sebastian, Spain \\
IKERBASQUE, Basque Foundation for Science, 48013, Bilbao, Spain. \\
              ORCID : 0000-0003-2953-3970 \\
              \email{reangulo@dipc.org}           \\
           \and
           Oliver Hahn \at
            Department of Astrophysics, University of Vienna, T\"urkenschanzstra{\ss}e 17, 1180 Vienna, Austria\\
	   Department of Mathematics, University of Vienna, Oskar-Morgenstern-Platz 1, 1090 Vienna, Austria \\ 
	   Laboratoire Lagrange, Universit\'e C\^ote d'Azur, CNRS, Bvd. de l'Observatoire, 06304 Nice, France\\
	   ORCID : 0000-0001-9440-1152 \\
	   \email{oliver.hahn@univie.ac.at}
}

\date{Received: date / Accepted: date}

\maketitle

%%%%%%%%%%%%%%%%%%%%%%%%%%%%%%%%%%%%%%%%%%%%%%%%%%%%
%%%%%%%%%%%%%%%%%%%%%%%%%%%%%%%%%%%%%%%%%%%%%%%%%%%%

\begin{abstract}
We review the field of collisionless numerical simulations for the large-scale structure of the Universe. We start by providing the main set of equations solved by these simulations and their connection with General Relativity. We then recap the relevant numerical approaches: discretization of the phase-space distribution (focusing on $N$-body but including alternatives, e.g., Lagrangian submanifold and Schr\"odinger--Poisson) and the respective techniques for their time evolution and force calculation (Direct summation, mesh techniques, and hierarchical tree methods). We pay attention to the creation of initial conditions and the connection with Lagrangian Perturbation Theory. We then discuss the possible alternatives in terms of the micro-physical properties of dark matter (e.g., neutralinos, warm dark matter, QCD axions, Bose-Einstein condensates, and primordial black holes), and extensions to account for  multiple fluids (baryons and neutrinos), primordial non-Gaussianity and modified gravity. We continue by discussing challenges involved in achieving highly accurate predictions. A key aspect of cosmological simulations is the connection to cosmological observables, we discuss various techniques in this regard: structure finding, galaxy formation and baryonic modelling, the creation of emulators and light-cones, and the role of machine learning. We finalise with a recount of state-of-the-art large-scale simulations and conclude with an outlook for the next decade.

%% this? https://jnep.sumdu.edu.ua/download/PACS_index.pdf
\end{abstract}

%%%%%%%%%%%%%%%%%%%%%%%%%%%%%%%%%%%%%%%%%%%%%%%%%%%%
%%%%%%%%%%%%%%%%%%%%%%%%%%%%%%%%%%%%%%%%%%%%%%%%%%%%

\newpage
\setcounter{tocdepth}{3}
\tableofcontents

%% Introduction
%% \newpage
\section{Introduction}
The current standard cosmological model describes a Universe accelerated by a cosmological constant ($\Lambda$) and dominated by cold dark matter (CDM), where structure arose from minute initial perturbations -- seeded in the primordial quantum Universe -- which collapsed on ever larger scales over cosmic time \citep{Planck:2020,Alam:2017,Betoule:2014}. The nonlinear interplay of these ingredients have formed the cosmic web and the intricate network of haloes harbouring galaxies and quasars.

Over the last decades, numerical simulations have played a decisive role in establishing and testing this $\Lambda$CDM paradigm. Following pioneering work in the 1980s, numerical simulations steadily grew in realism and precision  thanks to major advances in algorithms, computational power, and the work of hundreds of scientists. As a result, various competing hypotheses and theories could be compared with observations, guiding further development along the years.  Ultimately, $\Lambda$CDM was shown to be \textit{quantitatively} compatible with virtually all observations of the large-scale structure of the Universe, even for those that involve nonlinear physics and that are inaccessible to any method other than computer simulations \citep[see e.g.][]{Springel:2006,Vogelsberger:2020}. 

Nowadays, simulations have become the go-to tool in cosmology for a number of tasks: i) the interpretation of observations in terms of the underlying physics and cosmological parameters; ii) the testing and aiding of the development of perturbative approaches and analytic models for structure formation; iii) the production of reliable input (training) data for data-driven approaches and emulators; iv) the creation of mock universes for current and future large-scale extragalactic cosmological surveys, from which we can quantify statistical and systematic errors; v) the study of the importance of various aspects of the cosmological model and physical processes, and determining their observability.  

Despite the remarkable agreement between simulations and the observed Universe, there are indications that $\Lambda$CDM might not be ultimately the correct theory \citep{Planck:2020,Riess:2019,Asgari:2021,Abbott:2021}. Future cosmological observations will provide enough data to test competing explanations by probing exceedingly large sub-volumes of the Universe in virtually all electromagnetic wavelengths, and including increasingly fainter objects and smaller structures \citep[e.g.][]{Euclid,JPAS,DESI,LSST,SO,eROSITA}. These observations will be able to test the physics of our Universe beyond the standard model: from neutrino masses, over the nature of dark matter and dark energy, to the inflationary mechanism. Since these observations are intimately connected to the nonlinear regime of structure formation, any optimal exploitation of the cosmological information will increasingly rely on numerical simulations. This will put numerical simulations in the spotlight of modern cosmology: they can either confirm or break the standard $\Lambda$CDM paradigm, and therefore will play a key role in the potential discovery of new physics. 

The required accuracy and precision to make predictions of the necessary quality poses many important challenges and requires a careful assessment of all the underlying assumptions. This is the main topic of this review; we cover the ample field of cosmological simulations, starting from the fundamental equations to their connection with astrophysical observables, highlighting places where current research is conducted.

\subsection{Large-scale simulations}

\begin{figure}
\begin{centering}
\includegraphics[width=\textwidth]{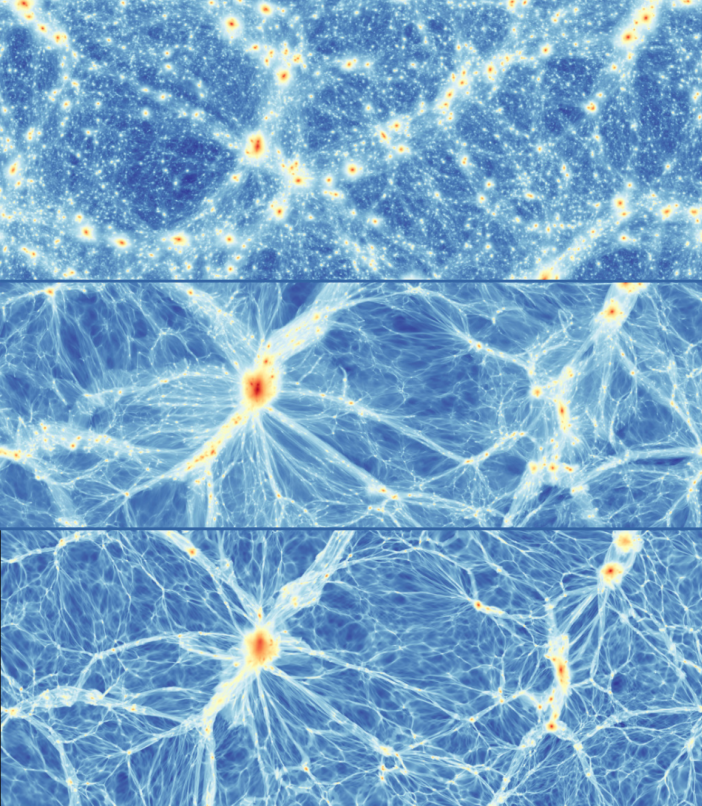} 
\caption{The distribution of dark matter in thin slices as predicted by a cosmological $N$-body simulation. Each panel shows a region $40\hMpc$ wide with different levels of thickness -- $40$, $2$, and $0.1\hMpc$, from top to bottom -- which highlight different aspects of the simulated density field, from the distribution of dark matter halos in the top panel, to the filamentary nature of the nonlinear structure in the bottom panel. Figure adapted from \cite{Stucker:2018}. \label{fig:cosmic_web}}
\end{centering}
\end{figure}

The spatial distribution and dynamics of the large-scale structure give us access to fundamental aspects of the Universe: its composition, the physical laws, and its initial conditions. For instance, the overall shape of the galaxy power spectrum is sensitive to the abundance of baryons and dark matter; the anisotropy induced by redshift-space distortions can constrain the cosmic density-velocity relation which in turn is set by the gravity law; and high-order cumulants of the galaxy field can encode non-Gaussian signatures inherited from an early inflationary period. 

To extract this information from observations of the large-scale distribution of galaxies, quasars or other tracers, we must rely on a model that predicts their observed distribution as a function of cosmic time for a given cosmological model. That is, we require a prediction for the distribution of the mass density and the velocity field together with the properties and abundance of collapsed objects. Furthermore, we need to predict how galaxies or other astronomical objects will populate these fields. This is a well-posed but very challenging problem, due to the complexity and nonlinear nature of the underlying physics. 

On large scales and/or at early times, density fluctuations are small and the problem can be tackled analytically by expanding perturbatively the relevant evolution equations. However, a rigorous perturbation theory can only be carried out on scales unaffected by shell-crossing effects. On intermediate scales, instead, so far only effective models with several free parameters exist (that are themselves either fitted to simulations or tested with simulations). On smaller very nonlinear scales, the only way to accurately solve the problem is numerically.

We illustrate the complicated nature of the nonlinear universe in Fig.~\ref{fig:cosmic_web}, which shows the simulated matter field in a region $40\hMpc$ wide. In the top panel we can appreciate the distribution of dark matter halos and their variety in terms of sizes, masses, and shapes. In the middle and bottom panels we show the same region but on much thinner slices which emphasise the small-scale anisotropies and ubiquitous filamentary structure.

When one is concerned with the large-scale structure of the Universe, the dynamics is dominated by gravity, and baryons and dark matter can be considered as a single pressureless (cold) fluid. This poses an ideal situation for computer simulations: the initial conditions as well as the relevant physical laws are known and can be solved numerically considering only gravitational interactions. We review in detail the foundations of such  numerical simulations in Sects.~\ref{sec:gr}, \ref{sec:discretizations}, \ref{sec:time_integration}, \ref{sec:gravity}, and \ref{sec:ICs}. Specifically, in Sect.~\ref{sec:gr} we describe the derivation of the  relevant equations solved by numerical simulations, in Sect.~\ref{sec:discretizations} how they can be discretised by either an $N$-body approach or by alternative methods. As we will discuss later, considering different discretisations will be crucial to test the robustness of the predictions from $N$-body simulations. In Sect.~\ref{sec:time_integration} we discuss how to evolve the discretised equations in time, and pay attention to common approaches for computational optimization, whereas in Sect.~\ref{sec:gravity} we discuss various numerical techniques to compute gravitational interactions. Finishing our recap of the main numerical techniques underlying large-scale simulations, in Sect.~\ref{sec:ICs} we discuss several aspects of how to set their initial conditions.

The importance of numerical solutions for structure formation is that they provide currently the only way to predict the highly nonlinear universe, and potentially extract cosmological information from observations at all scales. In contrast, if one is restricted to scales where perturbative approaches are valid and shell-crossing effects are negligible (i.e., have only a sub-per cent impact on summary statistics), then a huge amount of cosmological information is potentially lost. 

The primary role of numerical simulations for cosmological constraints has already been demonstrated for several probes focusing mostly on small scales, and in setting constraints on the properties of the dark matter particle, as exemplified by constraints from the Ly-$\alpha$ forest, the abundance and properties of Milky-Way satellites, and strong lensing modelling. They all rely on a direct comparison of observations with numerical simulations, or with models calibrated and/or validated using numerical simulations. 
%Specifically, the strongest constraint on the mass of the hypothetical dark matter particle arises from the number of Milky Way satellites contrasted with the predictions of $N$-body simulations augmented with a galaxy formation modelling. 
In Sect.~\ref{sec:dark_matter} we discuss several ways in which the distinctive properties of various potential dark matter candidates can be modelled numerically, including neutralinos, warm dark matter, axions, wave dark matter, and decaying and self-interacting dark matter.  In the future, this will also be the case for large-scale clustering, and we discuss the current and potential challenges to be address in Sect.~\ref{sec:numerics}. 

\subsection{Upcoming challenges}

Given that simulations are increasingly used for the inference of cosmological parameters, a question of growing importance is therefore how one can demonstrate the correctness of a given numerical solution. For instance, any simulation-based evidence -- of massive neutrinos or a departure from GR -- would necessarily need to prove its accuracy in modelling the relevant physics in the nonlinear regime. A proper understanding is of paramount importance: simulators need to demonstrate that a potential discovery relying on simulations is not simply a numerical inaccuracy or could be explained by uncertain/ill understood model parameters (i.e. the uncertainties due to \emph{all} ``free'' parameters must be quantified). We devote Sect.~\ref{sec:dark_matter} to this topic. 

Unfortunately still only a limited set of exact solutions is known for which strong convergence can be demonstrated. For useful predictions, the correctness of the solution has to be established in a very different, much more non-linear regime. However, far from these relatively simplistic known solutions. There has been significant progress in this direction over the last decade: from large code comparisons, a better understanding of the effect of numerical noise and discreteness errors, the impact of code parameters that control the accuracy of the solution, to the quality of the initial conditions used to set-up the simulations. These tests, however, presuppose that the $N$-body method itself converges to the right solution for a relatively small number of particles. Clear examples where the $N$-body approach fails have emerged: most notably near the free-streaming scale in warm dark matter simulations, and in the relative growth of multi-fluid simulations. Very recently, methods that do not rely on the $N$-body approach have become possible which have allowed to test for such errors. Although so far no significant biases have been measured for the statistical properties of CDM simulations, many more careful and systematic comparisons will be needed in the future.  

In parallel, there has been significant progress in the improved modelling of the multi-physics aspect of modern cosmological simulations. For instance, accurate two-fluid simulations are now possible capturing the distinct evolution of baryons and cold dark matter, as are simulations that quantify the non-linear impact of massive neutrinos with sophisticated methods. Further, the use of Newtonian dynamics has been tested against simulations solving general relativistic equations in the weak field limit, and schemes to include relativistic corrections have been devised. We discuss all these extensions, which seek to improve the realism of numerical simulations, in Sect.~\ref{sec:dark_matter}. 

An important aspect for confirming or ruling out the $\Lambda$CDM model will be the design of new cosmic probes that are particularly sensitive to departures from $\Lambda$CDM. For this it is important to understand the role of non-standard ingredients on various observables and in structure formation in general. This is an area where rapid progress has been seen with advances in the variety and sophistication of models, for instance regarding the actual nature of dark matter and simulations assuming neutralinos, axions, primordial black holes, or wave dark matter. Likewise, modifications of general relativity as the gravity law, and simulations with primordial non-Gaussianity have also reached maturity.

To achieve the required accuracy and precision that is necessary to make optimal use of upcoming observational data, it is blatantly clear that, ultimately, \emph{all} systematic errors in simulations must be understood and quantified, and the impact of \emph{all} the approximations made must be under control. One of them is that on nonlinear scales baryonic effects start to become important since cold collisionless and collisional matter start to separate, and baryons begin being affected by astrophysical processes. Hence, it becomes important to enhance the numerical simulation with models for the baryonic components, either for the formation of galaxies or for the effects of gas pressure and feedback from supernovae or supermassive black holes. We discuss different approaches to this problem in Sect.~\ref{sec:analysis}.

In parallel to increasing the accuracy of simulations, the community is focusing on improving their ``precision''. Cosmological simulations typically push international supercomputer centers and are among the largest calculations. New technologies and algorithmic advances are an important part of the field of cosmological simulations, and we review this in Sect.~\ref{sec:state-of-the-art}. We have seen important advances in terms of adoption of GPUs and hardware accelerators and new algorithms for force calculations with improved precision, computational efficiency, and parallelism. Thank to these, state-of-the-art simulations follow regions of dozens of Gigaparsecs with trillions of particles, constantly approaching the level required by upcoming extragalactic surveys. 

The field of cosmological simulations have been also reviewed in the last decade by other authors, who have focused on different aspects of the field and from different perspectives. For more details, we refer the reader to the excellent reviews by \cite{Kuhlen:2012}, \cite{Frenk:2012}, \cite{Dehnen:2011}, and by \cite{Vogelsberger:2020}.

%%%%

\subsection{Outline}

In the following we briefly outline the contents of each subsection of this review.

\noindent {\bf Section~\ref{sec:gr}:} This section provides the basic set of equations solved by cosmological dark matter simulations. We emphasise the approximations usually adopted by most simulations regarding the weak-field limit of General Relativity and the properties of dark matter. This section sets the stage for various kinds of simulations we discuss afterwards.

\noindent {\bf Section~\ref{sec:discretizations}:} This section discusses the possible numerical approaches for solving the equations presented in Section 2. Explicitly, we discuss the $N$-body approach and alternative methods such as the Lagrangian submanifold, full phase-space techniques, and Schr\"odinger-Poisson.

\noindent {\bf Section~\ref{sec:time_integration}:} This section derives the time integration of the relevant equations of motion. We discuss the symplectic integration of the dynamics at second order. We also review individual and adaptive timestepping, and the integration of quantum Hamiltonians. 

\noindent {\bf Section~\ref{sec:gravity}:} We review various methods for computing the gravitational forces exerted by the simulated mass distribution. Explicitly, we discuss the Particle-Mesh method solved by Fourier and mesh-relaxation algorithms, Trees in combination with traditional multipole expansions and Fast multipole methods, and their combination.  

\noindent {\bf Section~\ref{sec:ICs}:} In this section we outline the method for setting the initial conditions for the various types of numerical simulations considered. Explicitly, we review the numerical algorithms employed (Zel'dovich approximation and higher-order formulations) as formulated in Fourier or configuration space. 

\noindent {\bf Section~\ref{sec:dark_matter}:} This section is focused on simulations relaxing the assumption that all mass in the Universe is made out of a single cold collisionless fluid. That is, we discuss simulations including both baryons and dark matter; including neutrinos; assuming dark matter is warm; self-interacting; made out of primordial black holes; and cases where its quantum pressure cannot be neglected on macroscopic scales. We also discuss simulations easing the restrictions that the gravitational law is given by the Newtonian limit of General Relativity, and that the primordial fluctuations were Gaussian.

\noindent {\bf Section~\ref{sec:numerics}} This section discusses the current challenge for high accuracy in cosmological simulations. We consider the role of the softening length, cosmic variance, mass resolution, among others numerical parameters. We also review comparisons of $N$-body codes and discuss the validity of the $N$-body discretization itself. 

\noindent {\bf Section~\ref{sec:analysis}:} This section covers the connection between simulation predictions  and cosmological observations. We discuss halo finder algorithms, the building of merger trees, and the construction of ligthcones. We also briefly review halo-occupation distribution, subhalo-abundance-matching, and semi-analytic galaxy formation methods.

\noindent {\bf Section~\ref{sec:state-of-the-art}:} This section provides a list of state-of-the-art large-scale numerical simulations. We put emphasis in the computational challenge they face in connection with current and future computational trends and observational efforts.

\noindent In the final section, we conclude with an outlook for the future of cosmological dark matter simulations.

%% Dark Matter Models
%% \newpage

\section{Gravity and dynamics of matter in an expanding Universe}
\label{sec:gr}

Large-scale dark matter simulations are (mostly) carried out in the weak-field, non-relativistic, and collisionless limit of a more fundamental Einstein--Boltzmann theory. Additionally, since these simulations neglect any microscopic interactions in the collisionless limit right from the start (we will not consider them until we discuss self-interacting dark matter), one operates in the Vlasov-Einstein limit. This is  essentially a continuum description of the geodesic motion of microscopic particles since only (self-)gravity is allowed to affect their trajectories. Despite these simplifications, this approach keeps the essential non-linearities of the system, which gives rise to their phenomenological complexity.

In this section, we first derive the relevant relativistic equations of motion in a Hamiltonian formalism. We then take the non-relativistic weak-field limit by considering perturbations around the homogeneous and isotropic FLRW (Friedmann-Lema\^itre--Robertson--Walker) metric. This Vlasov--Poisson limit yields ordinary non-relativistic equations of motion, with the twist of a non-standard time-dependence due to the expansion of space in a general FLRW space time. Due to the preservation of symplectic invariants, the expansion of space leads to an intimately related contraction of momentum space to preserve overall phase space volume. 

With the general equations of motion at hand, we consider the cold limit (owed to a property of \emph{cold} dark matter (CDM) that is well constrained by observations) which naturally arises since the expansion of space (or rather the compression of momentum space) reduces any intrinsic momentum dispersion of the particle distribution over time. In the cold limit, the distribution function of dark matter takes a particularly simple form of a low-dimensional submanifold of phase space. These discussions are aimed to provide a formal foundation for the equations of motion as well as a motivation for many of the techniques and approximations discussed and reviewed in later sections.

\subsection{Equations of motion and the Vlasov equation}
Since we are, due to the very weak interactions of dark matter, interested mostly in the collisionless limit, we are essentially looking at freely moving particles in a curved space-time. To describe the motion of these particles, it is much easier to work with Lagrangian coordinates in phase space, i.e. the positions and momenta of particles. In general relativity, we have in full generality an eight dimensional relativistic extended phase space of positions $x_\mu$ and their conjugate momenta\footnote{We will use greek letters for indices that run over both time (0) and space (1-3), and Latin letters when only space is implied (1-3). The summation over repeated indices is implied unless otherwise indicated.} $p^\mu$. Kinetic theory in curved space time is discussed in many introductory texts on general relativity in great detail (e.g. \citealt{Ehlers:1971,Misner:1973,Straumann:2013,Choquet:2015} for the curious reader), but mostly without connection to a Hamiltonian structure. For our purposes, we can eliminate one degree of freedom by considering massive particles and neglecting processes that alter the mass of the particles. In that case the mass-shell condition $p^\mu p_\mu = -m^2 = {\rm const}$ holds\footnote{Note that for the speed of light we set $c=1$ in this section (it does not appear explicitly in the non-relativistic end results), but we explicitly keep the gravitational constant $G$.} and allows us to reduce dynamics to 3+3 dimensional phase space with one parameter (e.g. time) that can be used to parameterise the motion. Note that we employ throughout Einstein's summation convention, where repeated indices are implicitly summed over, unless explicitly stated otherwise.

\paragraph{Geodesic motion of massive particles.} In the presence of only gravitational interactions, the motion of particles in general relativity is purely geodesic by definition. Let us therefore begin by considering the geodesic motion of a particle moving between two points $A$ and $B$. The action for the motion along a trajectory\footnote{We will, unless indicated otherwise, use majuscule letters when describing the (time-parameterised) position and momentum of individual particles, while we will use minuscule letters to indicate Eulerian or general coordinates.}  $(X(t),P(t))$, parametrised by coordinate time $t$, between the spacetime points $A$ and $B$ is 
\begin{equation}
S(A,B) = \int_{A}^{B}  P_\mu {\rm d}X^\mu   = \int_A^B \left[ P_0 {\rm d}t + P_i {\rm d}X^i \right] = \int_A^B\left[ P_i \frac{{\rm d}X^i}{{\rm d}t} + P_0  \right] {\rm d}t. \label{eq:geodesic_action}
\end{equation}
From Eq.~(\ref{eq:geodesic_action}), one can immediately read off that the Lagrangian $\mathscr{L}$ and Hamiltonian $\mathscr{H}$ of geodesic motion are given by the usual Legendre transform pair \citep[e.g.][]{Goldstein:2002}
\begin{equation}
\mathscr{L}:=P_i \frac{{\rm d}X^i}{{\rm d}t} - \mathscr{H},\qquad\textrm{with}\qquad \mathscr{H}:=-P_0,\quad {\rm d}t := {\rm d}X^0
\end{equation}
respectively, meaning that $P_0$ represents the Hamiltonian itself (as one finds also generally in extended phase space, cf. \citealt{Lanczos:1986}). It is easy to show in a few lines of calculation that the coordinate-time canonical equations of motion in curved spacetime are then given by two dynamical equations\footnote{We denote partial derivatives in index notation, when it is obvious w.r.t. which coordinate one differentiates, using a comma, i.e. $f_{,i} = \partial f/\partial x^i$.} \citep[e.g.][]{Choquet:2015}
\begin{equation}
\frac{{\rm d}X^\mu}{{\rm d}t} = \frac{P^\mu}{P^0} \qquad\textrm{and}\qquad
\frac{{\rm d}P_\mu}{{\rm d}t} = F_\mu\qquad\textrm{with}\quad F_\mu := -g_{\alpha\beta,\mu}(X)\, \frac{P^\alpha P^\beta}{2P^0}.\label{eq:equations_of_motion}
\end{equation}
The Christoffel symbols of the metric simplify here to this simple partial derivative due to the mass-shell condition, but otherwise these two equations are equivalent to the geodesic equation. Note the formal similarity of these equations compared to the non-relativistic equations, with the `gravitational interaction' absorbed into the derivative of the metric. Eqs.~(\ref{eq:equations_of_motion}) determine the particle motion given the metric. The metric in turn is determined by the collection of all particles in the Universe through Einstein's field equations, which we will address in the next section.

\paragraph{Statistical Mechanics.} When considering a large number of particles, it is necessary to transition to a statistical description and consider the phase-space distribution (or density) function of particles in phase-space over time, i.e. on $(\vecb{x},\vecb{p},t) \in \mathbb{R}^{3+3+1}$, rather than individual microscopic particle trajectories $(\vecb{X}(t),\vecb{P}(t))$. The on-shell phase space distribution function $f_{\rm m}(\vecb{x},\vecb{p},t)$ can be defined e.g. through the particle number, which is a Lorentz scalar, per unit phase space volume. This phase-space density is then related to the energy-momentum tensor as
\begin{equation}
T^{\mu\nu} :=  \frac{1}{\sqrt{|g|}} \int_{\mathbb{R}^3} {\rm d}^3p\;f_{\rm m}(\vecb{x},\vecb{p},t) \,\frac{p^\mu p^\nu}{p^0} , \label{eq:gr_energy_mom_tensor}
\end{equation}
where $g$ is the determinant of the metric. For purely collisionless dynamics, the evolution of $f_{\rm m}$ is therefore determined by the on-shell Einstein-Vlasov equation \citep[e.g.][]{Choquet-Bruhat:1971,Ehlers:1971}
\begin{equation}
\label{eq:vlassov}
\widehat{L}_{\rm m} \,f_{\rm m} = 0,\qquad\textrm{with}\qquad \widehat{L}_{\rm m} := \frac{\partial}{\partial t} + \frac{p^i}{p^0}\frac{\partial}{\partial x^i} + \frac{F_i}{p^0} \frac{\partial}{\partial p_i}
\end{equation}
where $\widehat{L}_{\rm m}$ is the on-shell Liouville operator in coordinate time. This equation relates Hamiltonian dynamics and incompressibility in phase space: the Vlasov equation is simply the continuum limit of Hamiltonian mechanics with only long-range gravitational interactions (i.e., geodesic motion). This can be seen by observing that particle trajectories $\left(X^i(t),P_i(t)\right)$following Eqs.~(\ref{eq:equations_of_motion}) solve the Einstein-Vlasov equation as characteristic curves, i.e. ${\rm d}f_{\rm m}\left(X^i(t),P_i(t),t\right)/{\rm dt}=0$.

\subsection{Scalar metric fluctuations and the weak field limit}% -- or what most $N$-body simulations solve}
\paragraph{Metric perturbations.} The final step needed to close the equations is made through Einstein's field equations $G_{\mu\nu} = 8\pi G T_{\mu\nu}$.  The field equations connect the evolution of the phase-space density $f_{\rm m}$, which determines the stress-energy tensor $T^{\mu\nu}$, with the force 1-form $F_i$, which is determined by the metric. The results presented above are valid non-perturbatively for an arbitrary metric. Here, we shall make a first approximation by considering only scalar fluctuations using two scalar potentials $\phi$ and $\psi$\footnote{In full generality, initially purely scalar fluctuations in a dust fluid will source vector and after shell-crossing also tensor fluctuations. They remain however small since velocities never become relativistic. We are glancing here over various details of a rigorous Newtonian limit and refer the interested reader to \cite{Kopp:2014,Milillo:2015} for a rigorous treatment.}. This \emph{approximation} is valid if velocities are non-relativistic, i.e. $\left|P_i/P^0\right|\ll 1$. In this case, the only dynamically relevant component of $T^{\mu\nu}$ is the time-time component. Let us thus consider the metric (which corresponds to the ``Newtonian gauge'' with conformal time), following largely the notation of \cite{Bartolo:2007},
\begin{equation}
{\rm d}s^2=a^2(\tau)\left[ -\exp(2\psi)\,{\rm d}\tau^2 + \exp(-2\phi)\,{\rm d}x^i {\rm d}x_i \right] \label{eq:perturbed_metric}
\end{equation}
where $x$ are co-moving coordinates. The metric determinant is given by $\sqrt{|g|}=a^4\exp\left(\psi-3\phi\right)$.

The kinetic equation in GR is simply a geodesic transport equation and will thus only depend on the gravitational ``force'' 1-form $F_i$, which can be readily computed for this metric to be
\begin{equation}
F_i = a^2 p^0 \exp\left(-2\psi\right) \left(\psi_{,i}+\phi_{,i}\right)   - \frac{m^2}{p^0} \phi_{,i}
\end{equation}
If the vector and tensor components are non-relativistic (see e.g. \citealt{Kopp:2014,Milillo:2015} for a rigorous derivation of the Newtonian limit), we are left only with a constraint equation from the time-time component of the field equations. The time-time component of the Einstein tensor $G_{\mu\nu}$ is found to be
\begin{equation}
G^0_{\phantom{0}0} = \frac{\exp(2\phi)}{a^2} \left( (\nabla\phi)^2 -2\nabla^2\phi\right) - 3 \frac{\exp(-2\psi)}{a^2} \left(\frac{a'}{a}-\phi'\right)^2,
\end{equation}
where a prime indicates a derivative w.r.t. $\tau$. Inserting this in the respective field equation and performing the weak field limit (i.e. keeping only terms up to linear order in the potentials) one finds the following constraint equation 
\begin{equation}
%\frac{3}{2}\left(\phi' -2\frac{a'}{a}\psi\right)+ \nabla^2 \phi -\frac{3}{2}\frac{a'}{a} = 4\pi G a^2 \rho.
-3\mathcal{H}\left(\phi'+\mathcal{H}\psi\right) + \nabla^2 \phi + \frac{3}{2}\mathcal{H}^2 = 4\pi G a^2 \rho, 
\end{equation}
where $\rho:=T^0_{\phantom{0}0}$, $\mathcal{H}:= a'/a$, and $G$ is Newton's gravitational constant. Note that this equation alone does not close the system, since we have no evolution equation for $a(\tau)$ yet. 

\paragraph{Separation of background and perturbations.} The usual assumption is that backreaction can be neglected, i.e. the homogeneous and isotropic FLRW case is recovered with $\phi,\psi\to 0$ and density $\rho\to \overline{\rho}(\tau)$. For a discussion of the averaging problem in cosmology, see e.g. \cite{Paranjape:2009}. In this case, $a(\tau)$ is given by the solution of this equation in the absence of perturbations which becomes the usual Friedmann equation
\begin{equation}
\mathcal{H}^2 = \frac{8\pi G}{3} a^2 \overline{\rho}\qquad\textrm{with }\qquad\overline{\rho} =: \left(\Omega_{\rm r} a^{-4}+\Omega_\nu(a)+\Omega_{\rm m} a^{-3} + \Omega_{\rm k} a^{-2} + \Omega_\Lambda \right)\rho_{{\rm c},0},
\end{equation}
where $\rho_{{\rm c},0} := \frac{3H_0^2}{8 \pi G}$ is the critical density of the Universe \emph{today}, $H_0$ is the Hubble constant, and the $\Omega_{X\in\{{\rm r},\nu,{\rm m},{\rm k},\Lambda\}}$ are the respective density parameters of the various species in units of this critical density (at $a=1$). Note that massive neutrinos $\Omega_\nu(a)$ have a non-trivial scaling with $a$ (see Sect.~\ref{sec:HDM} for details). In the inhomogeneous case one can subtract out this FLRW evolution -- neglecting by doing so any non-linear coupling, or `backreaction', between the evolution of $a$ and the inhomogeneities -- and finds finally
\begin{equation}
-3\mathcal{H}\left(\phi'+\mathcal{H}\psi\right) + \nabla^2 \phi = 4\pi G a^2 (\rho-\overline{\rho}), \label{eq:relativist_poissoneq}
\end{equation}
This is an inhomogeneous diffusion equation (cf.\ e.g., \citealt{Chisari:2011,Hahn:2016}) reflecting the fact that the gravitational potential does not propagate instantaneously in an expanding Universe so that super-horizon scales, where density evolution is gauge-dependent, are screened.  

\subsection{Newtonian cosmological simulations}

\paragraph{Newtonian gravity.} In the absence of anisotropic stress the two scalar potentials in the metric (\ref{eq:perturbed_metric}) coincide and one has $\psi=\phi$. One can further show that on sub-horizon scales (where $\rho$ must be gauge independent, see e.g. Appendix~A of \citealt{Hahn:2016}) one then recovers from Eq.~(\ref{eq:relativist_poissoneq}) the non-relativistic Poisson equation,
\begin{eqnarray}
\nabla^2 \phi = 4\pi G a^2 (\rho-\overline{\rho}).
\end{eqnarray}
Note that this Poisson equation is however a priori invalid on super-horizon scales. Formally, when carrying out the transformation that removed the extra terms from Eq.~(\ref{eq:relativist_poissoneq}), the Poisson source $\rho$ has been gauge transformed to the synchronous co-moving gauge. If a simulation is initialised with density perturbations in the synchronous gauge and other quantities are interpreted in the Newtonian gauge, then the Poisson equation consistently links the two.  In addition, we have in the non-relativistic weak field limit that $p^0=a^{-1} m$ so that we also recover the Newtonian force law
\begin{equation}
F_i \to -am\phi_{,i}\,. \label{eq:NewtonForce}
\end{equation}
Note that such gauge mixing can be avoided and horizon-scale effects can be rigorously accounted for by choosing a more sophisticated gauge \citep{Fidler:2016,Fidler:2017} in which the force law is required to take the form of Eq.~(\ref{eq:NewtonForce}) and coordinates and momenta are interpreted self-consistently in this `Newtonian motion' gauge to account for leading order relativistic effects. A posteriori gauge transformations exist to relate gauge-dependent quantities, but remember that observables can never be gauge dependent.

\paragraph{Non-relativistic moments.} For completeness and reference, we also give the components of the energy-momentum tensor (\ref{eq:gr_energy_mom_tensor}) as moments of the distribution function $f_{\rm m}$ in the non-relativistic limit
\begin{subequations}
\begin{eqnarray}
\rho := T^0_{\phantom{0}0} &=& \frac{m}{a^{3}} \int_{\mathbb{R}^3} {\rm d}^3p\,f_{\rm m}(\vecb{x},\vecb{p},t)\, \label{eq:moment_0}\\
\pi_i := T^0_{\phantom{0}i} &=& \frac{1}{a^4}\int_{\mathbb{R}^3} {\rm d}^3p\,f_{\rm m}(\vecb{x},\vecb{p},t) \,p_i\, \label{eq:moment_1}\\
\Pi_{ij} := T_{ij} &=& \frac{1}{ma^5} \int_{\mathbb{R}^3} {\rm d}^3p\,f_{\rm m}(\vecb{x},\vecb{p},t) \,p_i \,p_j\,,\label{eq:moment_2}
\end{eqnarray}
\end{subequations}
defining the mass density $\rho$, momentum density $\vecb{\pi}$ and second moment $\Pi_{ij}$, which is related to the stress tensor as $\Pi_{ij} - \pi_i \pi_j / \rho$.

\paragraph{The equations solved by standard $N$-body codes.} Finally, the equations of motion in \emph{cosmic time} ${\rm d}t = a\, {\rm d}\tau$, assuming the weak-field non-relativistic limit, are 
\begin{equation}
\frac{{\rm d}{X}^i}{{\rm d}t} = \frac{P^i}{m}=\frac{{P}_i}{ma^2}\qquad\textrm{and}\qquad\frac{{\rm d}{P}_i}{{\rm d}t} = -m\frac{\partial\phi}{\partial X^i} \label{eq:canonical_eom_cosmo}
\end{equation}
with the associated Vlasov--Poisson system
\begin{subequations}
\begin{eqnarray}
\frac{\partial f_{\rm m}}{\partial t} + \frac{{p}_i}{ma^2} \,\frac{\partial f_{\rm m}}{\partial {x}^i} - m\,\frac{\partial\phi}{\partial x^i} \,\frac{\partial f_{\rm m}}{\partial p_i} =0 \label{eq:cosmo_vlasov_poisson}\\
\nabla^2\phi = 4\pi G a^2 (\rho - \overline{\rho}) \label{eq:cosmo_vlasov_poisson2} \\
\rho = \frac{m}{a^{3}} \int_{\mathbb{R}^3} {\rm d}^3p\,\,f_{\rm m}(\vecb{x},\vecb{p},t) 
\end{eqnarray}
\end{subequations}
where $\overline{\rho}(t)$ is the spatial mean of $\rho$ that is used also in the Friedmann equation $\mathcal{H}^2 = \frac{8\pi G}{3} a^2 \overline{\rho}$ which determines the evolution of $a(t)$. It is convenient to change to a co-moving matter density $a^{-3}\rho$, eliminating several factors of $a$ in these equations. In particular, the Poisson equation can be  written as
\begin{equation}
\nabla^2 \phi = \frac{3}{2}H_0^2 \Omega_{\rm m} \delta / a, \label{eq:Poisson_equation}
\end{equation}
if gravity is sourced by matter perturbations alone so that $\rho(\vecb{x},t) = (1+\delta(\vecb{x},t))\,\Omega_{\rm m} \rho_{{\rm c},0} a^{-3}$. Note that we have also introduced here the fractional overdensity $\delta := \rho/\overline{\rho}-1$.

\subsection{Post-Newtonian simulations}
\label{sec:postnewton_simulations}

While traditionally all cosmological simulations were carried out in the non-relativistic weak-field limit, neglecting any back-reaction effects on the metric, the validity and limits of this approach have been questioned \citep{Buchert:2005,Heinesen:2020}. In addition, with upcoming surveys reaching horizon scales, relativistic effects need to be quantified and accounted for correctly. Since such effects are only relevant on very large scales, where perturbations are assumed to be small, various frameworks have been devised to interpret the outcome of Newtonian simulations in relativistic context \citep{Chisari:2011,Hahn:2016}, which suggested in particular that some care is necessary in the choice of gauge when setting up initial conditions. Going even further, it turned out to be possible to define specific fine-tuned gauges, in which the gauge-freedom is used to absorb relativistic corrections, so that the equations of motion are strictly of the Newtonian form \citep{Fidler:2015,Adamek:2017,Fidler:2017b}. This approach requires only a modification of the initial conditions and a re-interpretation of the simulation outcome. Alternatively, relativistic corrections can also be included at the linear level by adding additional large-scale contributions computed using linear perturbation theory to the gravitational force computed in non-linear simulations \citep{Brandbyge:2017}.

Going beyond such linear corrections, recently the first Post-Newtonian cosmological simulations have been carried out \citep{Adamek:2013,Adamek:2016} which indicated however that back-reaction effects are small and likely irrelevant for the next generation of surveys. Most recently, full GR simulations are now becoming possible \citep{Giblin:2016,East:2018,Macpherson:2019,Daverio:2019} and seem to confirm the smallness of relativistic effects. The main advantage of relativistic simulations is that relativistic species, such as neutrinos, can be included self-consistently. In all cases investigated so-far, non-linear relativistic effects have however appeared to be negligibly small on cosmological scales. However, such simulations will be very important in the future to verify the robustness of standard simulations regarding relativistic effects on LSS observables (e.g. gravitational lensing, gravitational redshifts, e.g. \citealt{Cai:2017}, or the clustering of galaxies on the past lightcone, e.g. \citealt{Breton:2019,Guandalin:2021,Lepori:2021,Coates:2021}, all of which have been proposed as tests of gravitational physics on large scales).

\subsection{Cold limit: the phase space sheet and perturbation theory}
\label{sec:coldlimit_pt}
\paragraph{The cold limit.} All observational evidence points to the colder flavours of dark matter (but see Sect.~\ref{sec:dark_matter} for an overview over various dark matter models). A key limiting case for cosmological structure formation is therefore that of an initially perfectly cold scalar fluid (i.e. vanishing stress and vorticity).  In this case, the dark matter fluid is at early enough times fully described by its density and (mean) velocity field, which is of potential nature. The higher order moments (\ref{eq:moment_2}) are then fully determined by the lower order moments (\ref{eq:moment_0}-\ref{eq:moment_1}) and the momentum distribution function at any given spatial point is a Dirac distribution so that $f_{\rm m}$ is fully specified by only two scalar degrees of freedom, a density $n(\vecb{x})$, and a velocity potential, $S(\vecb{x})$, at some initial time, i.e.
\begin{equation}
f_{\rm m}(\vecb{x},\vecb{p},t) = n(\vecb{x},t)\;\delta_D\left(\vecb{p} - m \nabla_x S(\vecb{x},t) \right). \label{eq:Eulerian_phasesheet}
\end{equation}
Since $S$ is differentiable, it endows phase space with a manifold structure and the three-dimensional hypersurface of six-dimensional phase space on which $f$ is non-zero is called the `Lagrangian submanifold'. In fact, if at any time one can write $\vecb{p}=m\bnabla S$, then Hamiltonian mechanics guarantees that the Lagrangian submanifold preserves its manifold structure, i.e. it never tears or self-intersects. It can however fold up, i.e. lead to a multi-valued field $S(\vecb{x},t)$, invalidating the functional form~(\ref{eq:Eulerian_phasesheet}). Prior to such shell-crossing events (as is the case at the starting time of numerical simulations) this form is, however, perfectly meaningful and by taking moments of the Vlasov equation for this distribution function, one obtains a Bernoulli-Poisson system which truncates the infinite Boltzmann hierarchy already at the first moment, leaving only two equations \citep{Peebles:1980} in terms of the density contrast $\delta= n/\overline{n}-1$ and the velocity potential $S$,
\begin{equation}
\frac{\partial \delta}{\partial t} + a^{-2} \bnabla\cdot \left((1+\delta) \bnabla S\right) =0 \qquad\textrm{and}\qquad \frac{\partial S}{\partial t} + \frac{1}{2a^2} \left(\bnabla S\right)^2 + \phi = 0, \label{eq:BernoulliPoisson}
\end{equation}
supplemented with Poisson's equation (Eq.~\ref{eq:Poisson_equation}). Note that this form brings out also the connection to Hamilton-Jacobi theory. After shell-crossing, this description breaks down, and all moments in the Boltzmann hierarchy become important. 

\paragraph{Eulerian perturbation theory.} For small density perturbations $|\delta|\ll 1$, it is possible to linearise the set of equations (\ref{eq:BernoulliPoisson}). One then obtains the ODE governing the linear instability of density fluctuations
\begin{equation}
\delta^{\prime\prime} + \mathcal{H} \delta^\prime - \frac{3}{2}H_0^2 \Omega_{\rm m} a^{-1} \delta = 0. \label{eq:linear_density_pt}
\end{equation}
The solutions can be written as $\delta(\vecb{x},\tau) = D_+(\tau) \delta_+(\vecb{x}) + D_-(\tau) \delta_-(\vecb{x})$ and in $\Lambda$CDM cosmologies given in closed form \citep{Chernin:2003,Demianski:2005} as 
\begin{equation}
D_+(a) = a \, \phantom{}_{2}F_1\left( \sfrac{1}{3},\, 1,\, \sfrac{11}{6};\,-f_\Lambda(a)\right)\qquad\textrm{and}\qquad D_-(a)=\sqrt{1+f_\Lambda(a)} \,a^{-\sfrac{3}{2}}, \label{eq:LCDM_growthfactor}
\end{equation}
where $ f_\Lambda := \Omega_\Lambda / (\Omega_{\rm m} a^{-3})$, and $\phantom{}_2F_1$ is Gauss' hypergeometric function. In general cases, especially in the presence of trans-relativistic species such as neutrinos, Eq.~(\ref{eq:linear_density_pt}) needs to be integrated numerically however. Moving beyond linear order, recursion relations to all orders in perturbations of Eqs.~(\ref{eq:BernoulliPoisson}) have been obtained in the 1980s \citep{Goroff:1986} and provide the foundation of standard Eulerian cosmological perturbation theory (SPT; cf. \citealt{Bernardeau:2002} for a review).

\begin{figure}
\begin{centering}
\includegraphics[width=\textwidth]{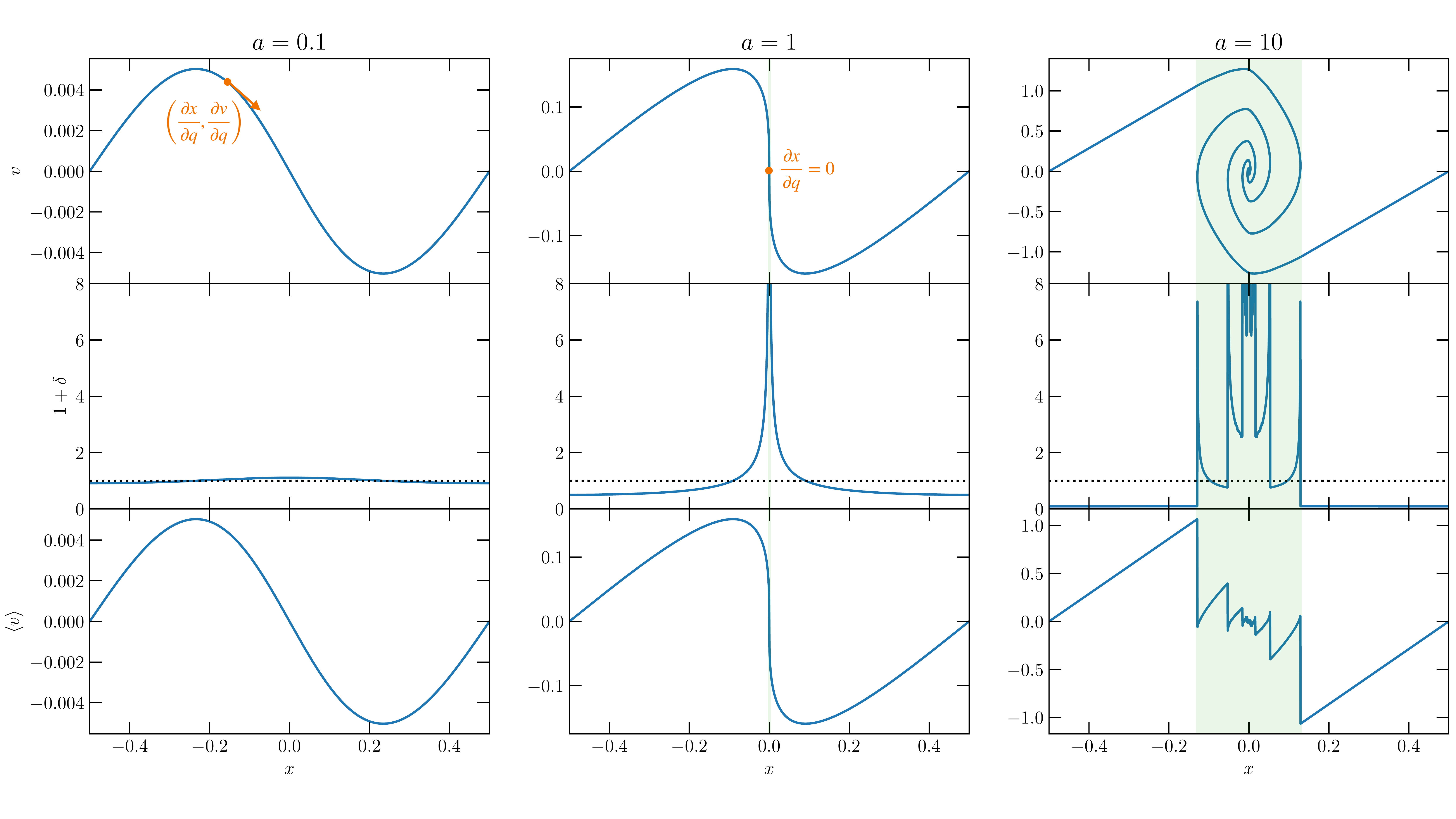}
\caption{\label{fig:plane_wave}Evolution of a single-mode perturbation from early times ($a=0.1$, left panels), through shell-crossing (at $a=1$, middle panels), to late times ($a=10$, right panels). The top row shows the phase space, where the cold distribution function occupies only a one-dimensional line. Density is shown in the middle row, with singularities of formally infinite density appearing at and after the first shell-crossing. The bottom panels show the mean fluid velocity ($\langle \vecb{v}\rangle = \vecb{\pi}/\rho$, Eq.~\ref{eq:moment_1}), which is identical to the phase-space diagram up to the first shell-crossing, but develops a complicated structure with discontinuities in the multi-stream region (indicated by green shading). Since the distribution function has a manifold structure, its tangent space (indicated in orange) can be evolved in a ``geodesic deviation'' equation, or it can be approximated by tessellations. Caustics appear where $\partial x/\partial q = 0$. $N$-body simulation do not track this manifold structure, and sample only the distribution function.}
\end{centering}
\end{figure}

\paragraph{Lagrangian perturbation theory.} Alternatively to considering the Eulerian fields, the dynamics can be described also through the Lagrangian map, i.e. by considering trajectories $\vecb{x}(\vecb{q},t) = \vecb{q} + \vecb{\Psi}(\vecb{q},t)$ starting from Lagrangian coordinates $\vecb{q} = \vecb{x}(\vecb{q},t=0)$. It becomes then more convenient to write the distribution function~(\ref{eq:Eulerian_phasesheet}) in terms of the Lagrangian map, i.e. 
\begin{equation}
f_{\rm m}(\vecb{x},\vecb{p},\tau) = \delta_D\left(\vecb{x} - \vecb{q}-\vecb{\Psi}(\vecb{q},\tau)\right)\;\delta_D\left(\vecb{p} - m a \vecb{\Psi}^\prime(\vecb{q},t) \right). \label{eq:Lagrangian_phasesheet}
\end{equation}
Mass conservation then implies that the density is given by the Jacobian ${\rm J} := \det J_{ij} := \det \partial x_i/\partial q_j$ as
\begin{equation}
1+\delta(\vecb{q},\tau) = \left|{\rm J}\right|^{-1} := \left| \det \bnabla_q \otimes \vecb{x} \right|^{-1} = \left| \det \delta_{ij} + \partial \Psi_i / \partial q_j \right|^{-1}, \label{eq:lagrange_density}
\end{equation}
which is singular if any eigenvalue of $\bnabla_q\otimes\vecb{x}$ vanishes. This is precisely the case when shell crossing occurs. The canonical equations of motion (\ref{eq:canonical_eom_cosmo}) can be combined into a single second order equation, which in conformal time reads
\begin{equation}
\vecb{x}^{\prime\prime} +\mathcal{H}\vecb{x}^\prime + \bnabla_x \phi(\vecb{x}) = 0, \label{eq:CDM_EOM}
\end{equation}
where we now consider trajectories not for single particles but for the Lagrangian map, i.e. $\vecb{x}=\vecb{x}(\vecb{q},t)$. By taking its divergence, this can be rewritten as an equation including only derivatives w.r.t. Lagrangian coordinates
\begin{equation}
{\rm J}\left(\delta_{ij}+\Psi_{i,j}\right)^{-1} \left( \Psi_{i,j}^{\prime\prime}+\mathcal{H}\Psi_{i,j}^\prime \right) = \frac{3}{2}\mathcal{H}^2 \Omega_{\rm m}\left( {\rm J} - 1 \right). \label{eq:main_lpt_equation}
\end{equation}
In Lagrangian perturbation theory (LPT), this equation is then solved perturbatively using a truncated time-Taylor expansion of the form $\vecb{\Psi}(\vecb{q},\tau) = \sum_{n=1}^\infty D(\tau)^n \vecb{\Psi}^{(n)}(\vecb{q})$ \citep{Buchert:1989,Buchert:1994,Bouchet:1995,Catelan:1995}. At first order $n=1$, restricting to only the growing mode, one finds the famous Zel'dovich approximation \citep{Zeldovich:1970}
\begin{equation}
\vecb{x}(\vecb{q},\tau) = \vecb{q} + D_+(\tau) \bnabla_q \nabla_q^{-2} \delta_+(\vecb{q}),
\end{equation}
where $\delta_+(\vecb{q})$ is, as above, the growing mode spatial fluctuation part of SPT. All-order recursion relations have also been obtained for LPT \citep{Rampf:2012a,Zheligovsky:2014,Matsubara:2015}. LPT solutions are of particular importance for setting up initial conditions for simulations. Both SPT and LPT are valid only prior to the first shell-crossing since the (pressureless) Euler-Poisson limit of Vlasov--Poisson ceases to be valid after. This can be easily seen by considering the evolution of a single-mode perturbation in the cold self-gravitating case, as shown in Fig.~\ref{fig:plane_wave}. Prior to shell-crossing, the mean-field velocity $\langle \vecb{v}\rangle(\vecb{x},\,t)$ coincides with the full phase-space description $\vecb{p}(\vecb{x}(\vecb{q};\,t);\,t)/m$. Then the DF from Eq.~(\ref{eq:Eulerian_phasesheet}) guarantees that the (Euler/Bernoulli-) truncated mean field fluid equations describe the full evolution of the system. This is no longer valid after shell-crossing, accompanied by infinite densities where $\det \partial x_i/\partial q_j=0$, when the velocity becomes multi-valued. Nevertheless, LPT provides an accurate bridge between the early Universe and that at the starting redshift of cosmological simulations, which can then be evolved further deep into the nonlinear regime. This procedure will be discussed in detail in Sect.~\ref{sec:ICs}.

\paragraph{Heuristic models based on PT.} Additionally, perturbation theory has been used as the backbone for approximate, but computationally extremely fast, descriptions of the nonlinear structure (see \citealt{Monaco:2016} or a review and \citealt{Chuang:2015b} for a comparison of approaches). These methods overcome the shell-crossing limitation of perturbation theory in different ways. The introduction of a viscuous term in the \textit{adhesion model} \citep{Gurbatov:1985,Kofman:1988,Kofman:1992} prevents the crossing of particle trajectories. In an alternative approach, \cite{KitauraHess:2013} replaced the LPT displacement field on small-scales by values motivated from spherical collapse \citep{Bernardeau:1994,Mohayaee:2006}. A similar idea is implemented in {\sc Muscle} \citep{Neyrinck:2016,Tosone:2021}. Numerous models have been developed based on an extension of the predictions of LPT via various small-scale models that aim to capture the collapse into halos or implement empirical corrections: {\sc Patchy} \citep{Kitaura:2014}, {\sc PTHalos} \citep{Scoccimarro:2002,Manera:2013}, {\sc EZHalos} \citep{Chuang:2015}, {\sc HaloGen} \citep{Avila:2015}. Finally, {\sc Pinocchio} \citep{Monaco:2002,Monaco:2013} and {\sc WebSky} \citep{Stein:2020} both combine LPT displacements with an ellipsoidal collapse model and excursion set theory to predict the abundance, mass accretion history, and spatial distribution of halos. The low computational cost of these approaches makes them useful for the creation of large ensembles of ``simulations'' designed at constructing covariance matrices for large-scale structure observations, or for a direct modelling of the position and redshift of galaxies. However, due to their heuristic character, their predictions need to be constantly quantified and validated with full $N$-body simulations.

\subsection{Deformation and evolution in phase space}
\label{sec:phasespace_deformation}
The canonical equations of motion describe the motion of points in phase space over time. Moving further, it is also interesting to consider the evolution of an infinitesimal phase space volume, spanned by $({\rm d}\vecb{x},{\rm d}\vecb{p})$, which is captured by the ``geodesic deviation equation''.

As we have just discussed, in the cold case, the continuum limit leads to all mass occupying a thin phase sheet in phase space and one can think of the evolution of the system as the mapping $\vecb{q}\mapsto (\vecb{x},\vecb{p})$ between Lagrangian and Eulerian space (c.f. Fig.~\autoref{fig:plane_wave}). One can take the analysis of deformations of pieces of phase space one level further beyond the cold case by considering a general mapping of phase space onto itself, i.e. $(\vecb{q},\vecb{w})\mapsto(\vecb{x},\vecb{p})$ (but note that this definition is formally not valid as $a\to0$ since in the canonical cosmological case the momentum space blows up in this limit). The associated phase-space Jacobian matrix $\tnsr{D}$, which reflects the effect in Eulerian space of infinitesimal changes to the Lagrangian coordinates, is 
\begin{equation}
\tnsr{D} := \frac{\partial (\vecb{x},\vecb{p})}{\partial (\vecb{q},\vecb{w})} = 
\begin{bmatrix}
\frac{\partial \vecb{x}}{\partial \vecb{q}} & \frac{\partial \vecb{x}}{\partial \vecb{w}} \\
\frac{\partial \vecb{p}}{\partial \vecb{q}} & \frac{\partial \vecb{p}}{\partial \vecb{w}}
\end{bmatrix}
=:
\begin{bmatrix}
\tnsr{D}_{\rm xq} & \tnsr{D}_{\rm xw} \\
\tnsr{D}_{\rm pq} & \tnsr{D}_{\rm pw} 
\end{bmatrix},
\end{equation}
where in the last equality, we have split the 6D tensor into four blocks. Its dynamics are fully determined by the canonical equations of motion, which we can obtain after a few steps as  \citep{Habib:1995,Vogelsberger:2008}.
\begin{equation}
\dot{\tnsr{D}} =  \frac{\partial (\dot{\vecb{x}},\dot{\vecb{p}})}{\partial (\vecb{q},\vecb{w})}
=
\left(\begin{bmatrix}
 \bnabla_x\otimes\bnabla_p  &  \bnabla_p\otimes\bnabla_p   \\ 
- \bnabla_x\otimes\bnabla_x &  \bnabla_p\otimes\bnabla_x
\end{bmatrix} 
\mathscr{H} \right)
\cdot \tnsr{D} =: \tnsr{H}\cdot\tnsr{D}. \label{eq:GDE_EOM}
\end{equation}
This is equation is called the ``geodesic deviation equation'' (GDE) in the literature and it quantifies the relative motion in phase space along the Hamiltonian flow. For separable Hamiltonians $\mathscr{H}=T(\boldsymbol{p},t)+V(\boldsymbol{x},t)$, the coupling matrix $\tnsr{H}$ becomes
\begin{equation}
\tnsr{H} = \begin{bmatrix}
\tnsr{0}& (\bnabla_p\otimes\bnabla_p)T  \\
-(\bnabla_x\otimes\bnabla_x)V & \tnsr{0}
\end{bmatrix},
\quad\textrm{and in cosmic time:}\quad
\tnsr{H} = \begin{bmatrix}
\tnsr{0} & \frac{1}{ma^{2}} \delta_{ij} \\
-m\phi_{,ij} & \tnsr{0}
\end{bmatrix},
\end{equation}
and shows a coupling to the gravitational tidal tensor \citep{Vogelsberger:2008,Vogelsberger:2011}. The evolution of $\tnsr{D}$ can be used to track the evolution of an infinitesimal environment in phase space around a trajectory $(\vecb{x}(\vecb{q},\vecb{w};\,t),\,\vecb{p}(\vecb{q},\vecb{w};\,t))$. In particular, from Eq.~(\ref{eq:lagrange_density}) follows that zero-crossings of the determinant of the $\tnsr{D}_{\rm xq}$ block correspond to infinite-density caustics, so that it can be used to estimate the local (single) stream density, and count the number of caustic crossings. Infinite density caustics would cause singular behaviour in the evolution of $\tnsr{D}$, so that its numerical evolution \emph{has} to be carried out with sufficient softening \citep{Vogelsberger:2011,Stuecker:2019}. Since it is sensitive to caustic crossings, the GDE can be used to quantify the distinct components of the cosmic web (\citealt{Stuecker:2019}, see also Sect.~\ref{sec:cosmic_web}). The GDE has an intimate connection also to studies of the emergence of chaos in gravitationally collapsed structures since it quantifies the divergence of orbits in phase space and has an intimate connection to Lyapunov exponents \citep{Habib:1995}. An open problem is how rapidly a collapsed system achieves efficient phase space mixing since discreteness noise in N-body simulations could be dominant in driving phase space diffusion if not properly controlled \citep{Stuecker:2019,Colombi:2020}.

%% Numerical techniques
%% \newpage
\section{Discretization techniques for Vlasov--Poisson systems}
\label{sec:discretizations}

\begin{figure}
\begin{centering}
\includegraphics[width=\textwidth]{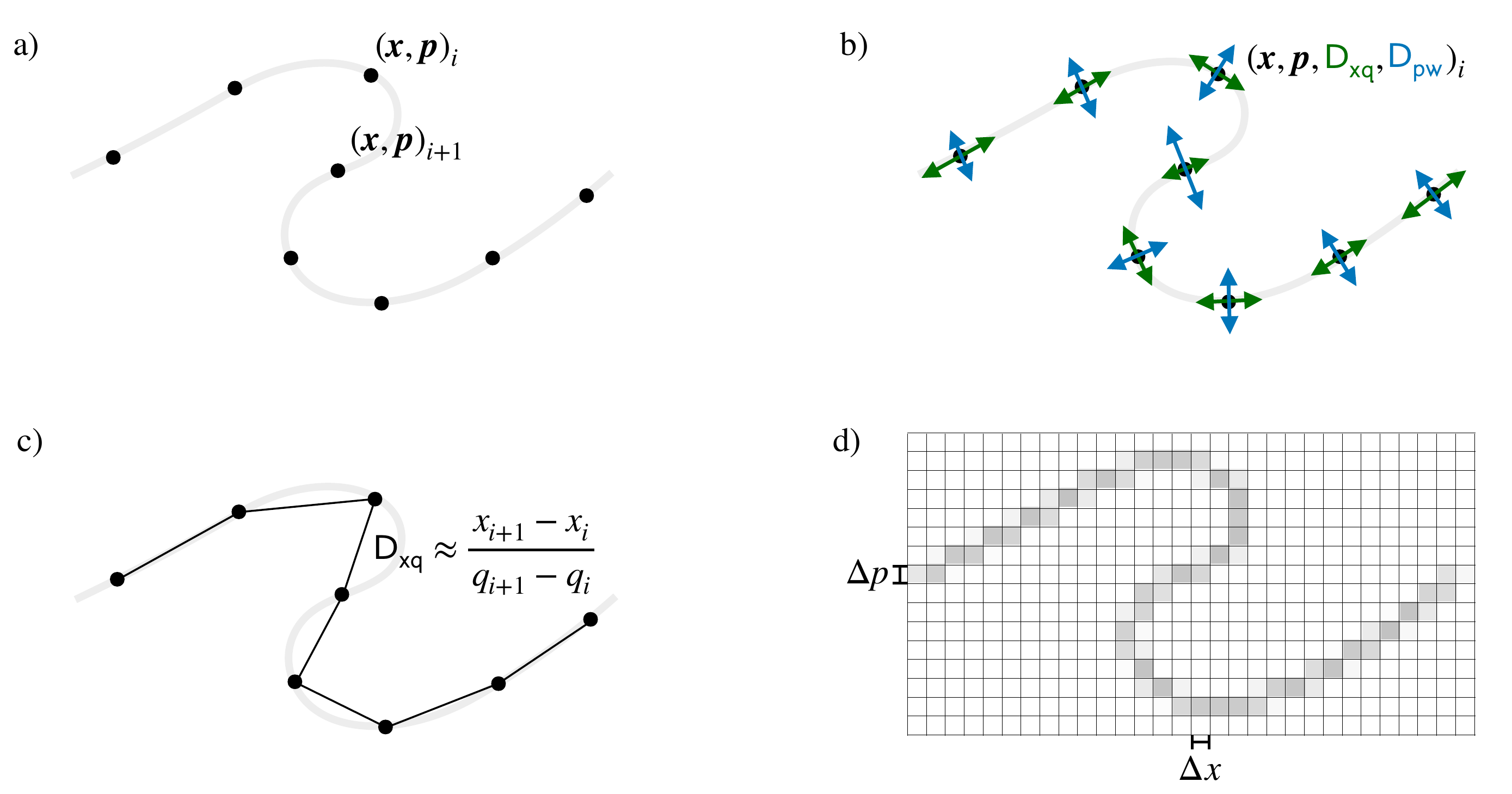} 
\caption{\label{fig:discretisations} Discretisations used in the numerical solution of Vlasov--Poisson: a) the $N$-body method which samples the fine grained distribution function (light gray line) at discrete locations, b) the `GDE' method that can evolve the local manifold structure along with the particles (the green eigenvectors of $\mathsf{D_{xq}}$ are tangential to the Lagrangian submanifold, c) the sheet tessellation method, which uses interpolation (here linear) between particles to approximate the Lagrangian submanifold with a tessellation, d) a finite volume discretisation of the full phase space with uniform resolution $\Delta x$ in configuration space and $\Delta p$ in momentum space. }
\end{centering}
\end{figure}

The macroscopic collisionless evolution of non-relativistic self-gravitating classical matter in an expanding universe is governed by the cosmological Vlasov--Poisson (VP) set of Eqs.~(\ref{eq:cosmo_vlasov_poisson}--\ref{eq:cosmo_vlasov_poisson2}) derived above. VP describes the evolution of the density $f(\vecb{x},\vecb{p},t)$ in six-dimensional phase space over time. Due to the non-linear character of the equations and the attractive (focusing) nature of the gravitational interaction, intricate small-scale structures (filamentation) emerge in phase space already in 1+1 dimensions as shown in Fig.~\ref{fig:plane_wave}, and chaotic dynamics can arise in higher dimensional phase space. Various numerical methods to solve VP dynamics have been devised, with intimate connections also to related techniques in plasma physics. The $N$-body approach is clearly the most prominent and important technique to-day, however, other techniques have been developed to overcome its shortcomings in certain regimes and to test the validity of results. A visual representation of the various approaches to discretise either phase space or the distribution function is shown in Fig.~\ref{fig:discretisations}. 

\subsection{The \texorpdfstring{$N$}{N}-body technique}
The most commonly used discretisation technique for dark matter simulations is the $N$-body method, which has been used since the 1960s as a numerical tool to study the Hamiltonian dynamics of gravitationally bound systems such as star and galaxy clusters \citep{Hoerner:1960,Aarseth:1963,Henon:1964} by Monte-Carlo sampling the phase space of the system. They started being used to study cosmological structure formation beginning in the early 1970s \citep{Peebles:1971,Press:1974,Miyoshi:1975}, followed by an explosion of the field in the first half of the 1980s \citep{Doroshkevich:1980,Klypin:1983,White:1983,Centrella:1983,Shapiro:1983,Miller:1983}. These works demonstrated the web-like structure of the distribution of matter in the Universe and established that cold dark matter (rather than massive neutrinos) likely provides the ``missing'' (dark) matter. By the late 1990s, the resolution and dynamic range had increased sufficiently so that it became possible to study the inner structure of dark matter haloes, leading to the discovery of universal halo profiles \citep{Navarro:1997}, the large abundance of substructure in CDM subhaloes \citep{Moore:1999,Klypin:1999}, and predictions of the mass function of collapsed structures in the Universe over a large range of masses and cosmic time \citep{Jenkins:2001}. The $N$-body method is now being used in virtually all large state-of-the-art cosmological simulations as the method of choice to simulate the gravitational collapse of cold collisionless matter (cf.\ Sect.~\ref{sec:state-of-the-art} for a review of state-of-the-art simulations). In ``total matter'' (often somewhat falsely called ``dark matter only'') simulations, the $N$-body mass distribution serves as a proxy of the potential landscape in which galaxy formation takes place. And also in multi-physics simulations that simulate the distinct evolution of dark matter and baryons (see also Sect.~\ref{sec:multi_species_amplitudes}), collisionless dark matter is solved via the $N$-body method, while a larger diversity of methods are employed to evolve the collisional baryonic component \citep[see e.g.][for a review]{Vogelsberger:2020}.

\paragraph{The $N$-body discretisation. } Underlying all these simulations is the fundamental $N$-body idea: the Vlasov equation is the continuum version of the Hamiltonian equations of motion, which implies that phase-space density is conserved along Hamiltonian trajectories. The non-linear coupling in Hamilton's equations arises through the coupling of particles with gravity via Poisson's equation, which is only sourced by the density field. Therefore, as long as a finite number of $N$ particles is able to fairly sample the density field, the evolution of the system can be approximated by these discrete trajectories. 

A practical complication is that the (formally) infinitely extended mass distribution has to be taken into account. Most commonly, this complication is solved by restricting the simulation to a finite cubical volume $V=L^3$ of co-moving linear extent $L$ with periodic boundary conditions. This can formally be written as considering infinite copies of this fundamental cubical box. The effective $N$-body distribution function is then given by a set of discrete \emph{macroscopic} particle locations and momenta $\left \{ \left(\vecb{X}_i(t),\vecb{P}_i(t)\right),\,i=1\dots N\right\}$ along with the infinite set of periodic copies, so that 
\begin{eqnarray}
f_N(\vecb{x}, \vecb{p}, t) = \sum_{\vecb{n}\in\mathbb{Z}^3} \sum_{i=1}^N \frac{M_i}{m}\,\delta_D(\vecb{x}-\vecb{X}_i(t)-\vecb{n} L )\,\delta_D(\vecb{p}-\vecb{P}_i(t)), \label{eq:n_body_df}
\end{eqnarray}
is an unbiased sampling of the true distribution function. Here $M_i$ is the effective particle mass assigned to an $N$-body particle, $m$ the actual microscopic particle mass, and $\vecb{X}_i(t)$ and $\vecb{P}_i(t)$ are the position and momentum of particle $i$ at time $t$. The most widespread  choice of discretisation is one in which all particles are assumed to have equal mass, $M_i = \overline{M} = \Omega_m \rho_{\rm c,0} V / N$. Note, however, that using different masses is also possible and sometimes desirable (e.g., for multi-resolution simulations, see Sect.~\ref{sec:zoom_simulations}).

\paragraph{Initial conditions. } Since the particles are to sample the full six-dimensional distribution function, a key question is how the initial positions $\vecb{X}_i(t_0)$ and momenta $\vecb{P}_i(t_0)$ should be chosen. For cold systems, we have derived a consistent approach above and it is given by Eq.~(\ref{eq:Eulerian_phasesheet}) which can be readily evaluated from a discrete sampling of the Lagrangian manifold alone, i.e. by choosing an (ideally homogeneous and isotropic) uniform sampling in terms of Lagrangian coordinates $\vecb{Q}_i$ for each $N$-body particle (discussed in more detail in Sect.~\ref{sec:pre_ics_discreteness}) and then obtaining the Eulerian position $\vecb{X}_i$ and momentum $\vecb{P}_i$ at some initial time $t_0$ from the Lagrangian map $\vecb{\Psi}(\vecb{Q}_i,t_0)$ (see Sect.~\ref{sec:ICs_from_LPT} for more details). This means that, in the case of a cold fluid, the particles sample the mean flow velocity exactly. The situation is considerably more involved if the system has a finite temperature which requires to sample not only the three-dimensional Lagrangian submanifold but the full six-dimensional phase space density. This implies that in some sense each particle in the cold case needs to be sub-divided into many particles that sample the momentum part of the distribution function. Particularly for hot distribution functions, where the momentum spread is large compared to mean momenta arising from gravitational instability (such as in the case of neutrinos), this has caused a formidable challenge due to the large associated sampling noise. To circumvent these problems various solutions have been proposed, e.g. using a careful sampling of momentum space using shells in momentum modulus and an angle sampling based on the healpix sphere decomposition \citep{Banerjee:2018}, or reduced variance sampling based on the control variates method \citep{Elbers:2020}. Such avenues are discussed in more detail in the context of massive neutrino simulations in Sect.~\ref{sec:HDM} below.

\paragraph{Equations of motion. } Once the initial particle sampling has been determined, the subsequent evolution is fully governed by VP dynamics. Moving along characteristics of the VP system, the canonical equations of motion for particle $i=1\dots N$ in cosmic time are obtained from ${\rm d}f(\vecb{X}_i(t),\,\vecb{P}_i(t),\,t)/{\rm d}t=0$ as
\begin{equation}
\dot{\vecb{X}}_i = \frac{{\vecb P}_i}{M_i a^2}\qquad\textrm{and}\qquad
\dot{\vecb{P}}_i = -M_i \left.\bnabla_{x} \phi \right|_{\vecb{X}_i}. \label{eq:nbody_eom}
\end{equation}
These are consistent with a cosmic-time Hamiltonian system with a pair interaction potential $I(\vecb{x},\vecb{x}')$ of the form
\begin{subequations}
\begin{align}
%H = \sum_{i=1}^N \frac{p_i^2}{2m_i a^2} + V(\vecb{x}_1,\dots,\vecb{x}_N),\quad\textrm{with}\quad  \nabla^2 V = \frac{3H_0^2\Omega_m}{2a} \left( \frac{\rho}{\overline{\rho}}-1\right).
\mathscr{H} &= \sum_{i=1}^N \left[ \frac{P_i^2}{2M_i a^2} + \frac{1}{2}  M_i \sum_{j\neq i} I(\vecb{X}_i,\vecb{X}_j) \right]\\
\textrm{where}& \quad \sum_{j\neq i} I(\vecb{X}_i,\vecb{X}_j) = \phi(\vecb{X}_i) \quad\textrm{with}\quad \nabla_x^2\phi   = \frac{3H_0^2\Omega_{\rm m}}{2a} \left( \frac{\rho}{\overline{\rho}}-1\right). \label{eq:nbody_poisson}
\end{align}
\end{subequations}
The resulting acceleration term is given by 
\begin{equation}
\vecb{g}(\vecb{x}) := -\left. \bnabla_x \phi \right|_{\vecb{x}} = \frac{G}{a} \int_{\mathbb{R}^3} {\rm d}^3x' \, \rho(\vecb{x'},t) \frac{\vecb{x}'-\vecb{x}}{\left\| \vecb{x}'-\vecb{x}\right\|^3}, \label{eq:nbody_acc}
\end{equation}
which has no contribution from the background $\overline{\rho}$ for symmetry reasons \citep{Peebles:1980}. The co-moving configuration space density $\rho$ that provides the Poisson source arises from Eq.~(\ref{eq:n_body_df}) and is given by %gravitational acceleration is computed from the discrete density distribution
\begin{equation}
\rho(\vecb{x},t) = \int_{\mathbb{R}^3} {\rm d}^3p\,\int_{\mathbb{R}^3} {\rm d}^3x'\,m\,f_N(\vecb{x}',\vecb{p},t) \, W(\vecb{x}'-\vecb{x}) = \sum_{\vecb{n}\in\mathbb{Z}^3} \sum_{i=1}^NM_i\,\,W\left(\vecb{x}-\vecb{X}_i(t)-\vecb{n} L \right). \label{eq:n_body_density}
\end{equation}
Here, we additionally allowed for a regularisation kernel $W(\vecb{r})$ that is convolved with the discrete $N$-body density in order to improve the regularity of the density field and speed up convergence to the collisionless limit (or so one hopes). It represents a softening kernel (also called `assignment function' depending on context) that regularises gravity and accounts for the fact that each particle is not a point-mass (like a star or black hole), but corresponds to an extended piece of phase space so that two-body scattering between the effective particles is always artificial and must be suppressed. We discuss how the acceleration obtained from this infinite sum is solved in practice in Sect.~\ref{sec:gravity}.

\paragraph{Discreteness effects. } The quality of the force calculation rests on how good an approximation the force associated with the density from Eq.~(\ref{eq:n_body_density}) is. The hope is that by appropriate choice of $W$ and an as large as possible number $N$ of particles, the evolution remains close to the true collisionless dynamics and microscopic collisions remain subdominant. Due to the discrete nature of the particles, problems of the $N$-body approach are known to arise when force and mass resolution are not matched in which case the evolution of the discrete system can deviate strongly from that of the continuous limit \citep{Centrella:1983,Centrella:1988,Peebles:1989,Melott:1989,Diemand:2004,Melott:1997,Splinter:1998,WangWhite:2007,Melott:2007,MelottFragmentation:2007,Marcos:2008,Bagla:2009}. A slow convergence to the correct physical solution can, however, usually be achieved by keeping the softening so large that individual particles are never resolved. At the same time, if the force resolution in CDM simulations is not high enough at late times, then sub-haloes are comparatively loosely bound and prone to premature tidal disruption, leading to the `overmerging' effect and the resulting orphaned galaxies (i.e., if the subhalo hosted a galaxy, it would still be a distinct system rather than having merged with the host), e.g. \cite{Klypin:1999b,DiemandVel:2004,vandenBosch:2018a}. In this case, one would want to choose the softening as small as possible. We discuss this in more detail in Sect.~\ref{sec:softening}.

More sophisticated choices of $W$ beyond a global (possibly time-dependent) softening scale are possible, for instance, the scale can depend on properties of particles, such as the local density (leading to what is called ``adaptive softening''). We discuss the aspect of force regularisation by softening in more detail in Sect.~\ref{sec:softening}. The gravitational acceleration that follows from Eq.~(\ref{eq:n_body_density}) naturally has to take into account the cosmological Poisson equation, i.e., include the subtraction of the mean density and assume periodic boundary conditions. All aspects related to the time integration of cosmological Hamiltonians will be discussed in Sect.~\ref{sec:time_integration}, those related to computing and evaluating gravitational interactions efficiently in Sect.~\ref{sec:gravity} below. 

\subsection{Phase space deformation tracking and Lagrangian submanifold approximations}
\label{sec:tessellation_methods}

For large enough $N$, the $N$-body method is expected to converge to the collisionless limit. Nonetheless, an obvious limitation of this approach is that the underlying manifold structure is entirely lost as the particles retain only knowledge of positions and momenta and all other quantities (e.g. density, as well as other mean field properties) can only be recovered by coarse-graining a larger number of particles. Two different classes of methods, that we shall discuss next, have been developed over recent years that overcome this key limitation in various ways. The first class is based on promoting particles (which are essentially vectors) to tensors and re-write the canonical equations of motion to evolve them accordingly, resulting in equations of motion reminiscent of the geodesic deviation equation (GDE) in general relativity. The second class retains the particles but promotes them to vertices of a tessellation whose cells provide a discretisation of the manifold.   

\subsubsection{Tracking deformation in phase space -- the GDE approach}
We already discussed in Sect.~\ref{sec:phasespace_deformation} how infinitesimal volume elements of phase space evolve under a Hamiltonian flow. In particular, Eq.~(\ref{eq:GDE_EOM}) is the canonical equation of motion for the phase-space Jacobian matrix. In the `GDE' approach, instead of evolving only the vector $(\vecb{X}_i,\vecb{P}_i)$ for each $N$-body particle, one evolves in addition the tensor $\tnsr{D}_i$ for each particle (cf.\ \citealt{Vogelsberger:2008,White:2009}, but see also \citealt{Habib:1995} who derive a method to compute Lyapunov exponents based on the same equations). Of particular interest is the $\tnsr{D}_{\rm xq}$ sub-block of the Jacobian matrix since it directly tracks the local (stream) density associated with each $N$-body particle through $\delta_i+1 = \left(\det \tnsr{D}_{{\rm xq},i}\right)^{-1}$. The equations of motion for the relevant tensors associated to particle $i$ are 
\begin{subequations}
\begin{align}
\dot{\tnsr{D}}_{{\rm xq},i} &= \frac{1}{m a^{2}}\,\tnsr{D}_{{\rm pq},i}\label{eq:gde1}\\
\dot{\tnsr{D}}_{{\rm pq},i} &=- m\,\tnsr{D}_{{\rm xq},i} \cdot \left.\tnsr{T}\right|_{\vecb{X}_i}\quad\textrm{where}\quad\tnsr{T} := \nabla_x\otimes\nabla_x \phi,
\end{align}
\end{subequations}
and are solved alongside the $N$-body equations of motion (\ref{eq:nbody_eom}) by computing the tidal tensor \tnsr{T}. One caveat with the GDE approach is that the evolution of $\tnsr{D}_{\rm xq}$ is determined not by the force but by the tidal field -- which contains one higher spatial derivative of the potential than the force -- and therefore is significantly less regular than the force field (see the detailed discussion and analysis in \cite{Stuecker:2020} who have also studied the stream density evolution in virialised halos, based on a novel low-noise force calculation). This approach thus requires larger softening to achieve converged answers than a usual $N$-body simulation, and possibly cannot be shown to converge in the limit of infinite density caustics.

Evolving $\tnsr{D}_{\rm xq}$ provides additional information about cosmic structure that is not accessible by standard $N$-body simulations. For instance, solving for the GDE enabled \cite{Vogelsberger:2008,Vogelsberger:2011} to estimate the number of caustics in dark matter haloes, which might add a boost to the self-annihilation rate of CDM particles, or the amount of chaos and mixing in haloes. A key result of \cite{Vogelsberger:2011} was that despite the large over-densities reached in collapsed structures, each particle is nonetheless inside of a stream with a density not too different from the cosmic mean density. This is possible since haloes are built like p\^ate feuillet\'ee as a layered structure of many stretched and folded streams, as can be seen in panel a) of Fig.~\ref{fig:phasespace_evolution}.

\begin{figure}
\begin{centering}
\includegraphics[width=\textwidth]{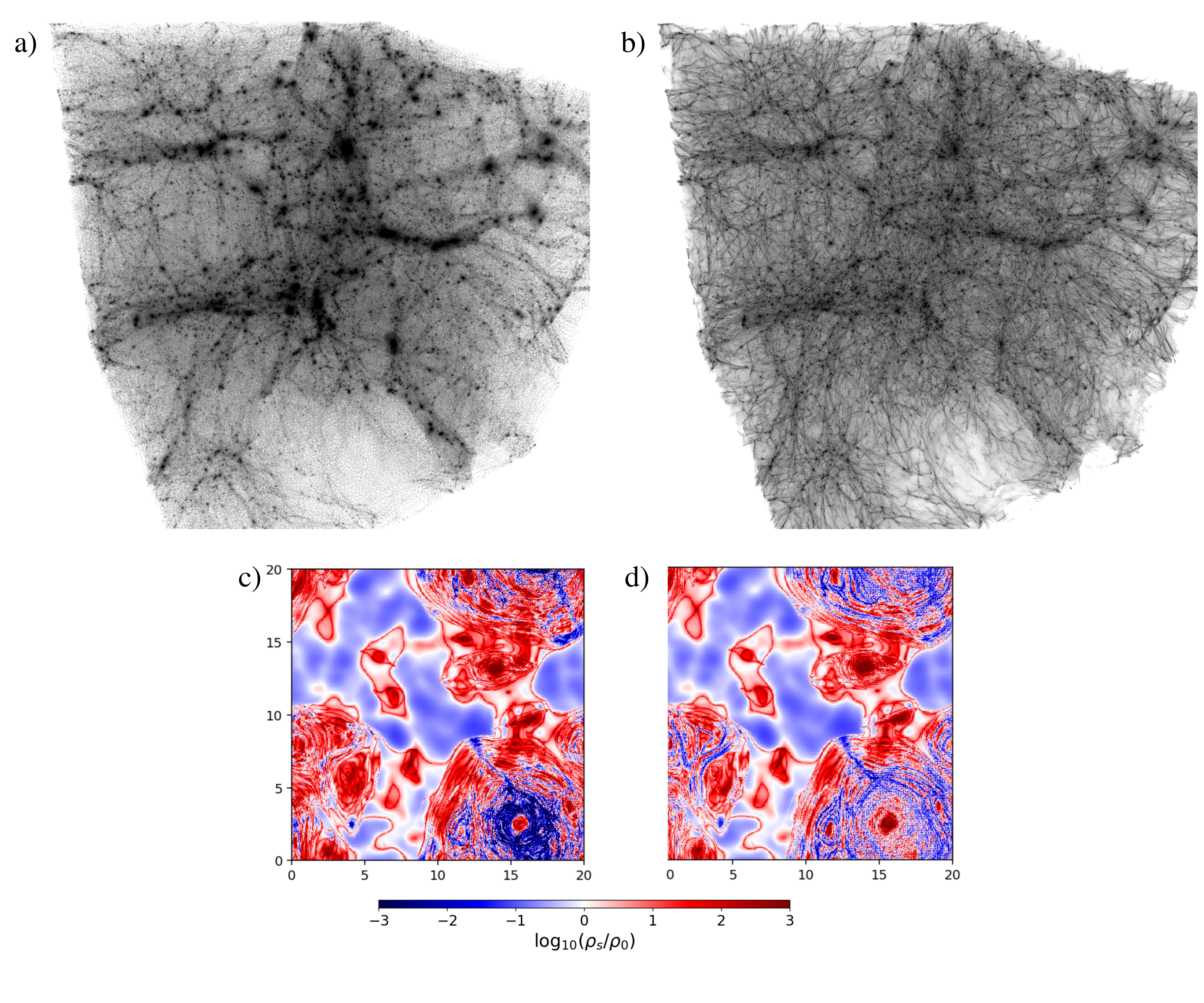} 
\caption{\label{fig:darkmattersheet} Panels a-b: The density field obtained from the same set of $N$-body particles as a simple particle $N$-body density in a) and in terms of a phase space sheet interpolation in terms of tetrahedral cells as described in \cite{Abel:2012} in b). Panels c-d: The GDE method and the sheet tessellation method provide direct access to the stream density, which is shown in Lagrangian $\vecb{q}$-space for c) the GDE approach and d) the sheet tessellation approach. [Panels a) and b) courtesy \cite{Abel:2012}, c) and d) courtesy \cite{Stuecker:2020}]}
\end{centering}
\end{figure}

\subsubsection{The dark matter sheet and phase space interpolation}
\label{sec:dmsheetsims}
A different idea to reconstruct the Lagrangian submanifold from existing $N$-body simulations was proposed by \cite{Abel:2012} and \cite{Shandarin:2012} who noted that a tessellation of Lagrangian space, constructed by using the initial positions of the $N$-body particles at very early times as vertices (i.e., the particles generate a simplicial complex on the Lagrangian submanifold), is topologically preserved in Hamiltonian evolution. This means that initially neighbouring particles can be connected up as non-overlapping tetrahedra (in the case of a 3D submanifold of 6D phase space). Their deformation and change of position and volume reflect the evolution of the phase-space distribution (and thus changes in the density field). A visual impression of the difference between an $N$-body density and this tessellation based density is given in Fig.~\ref{fig:darkmattersheet}. No holes can appear through Hamiltonian dynamics, but since the divergence of initially neighbouring points depends on the specific dynamics (notably the Lyapunov exponents that described the divergence of such trajectories), the edge connecting two vertices can become a tangled curve due to the complex dynamics in bound systems, e.g., \cite{Laskar:1993,Habib:1995}. As long as the tetrahedra edges still approximate well the submanifold, the simplicial complex provides access to a vast amount of information about the distribution of matter in phase space in an evolved system that is difficult or even impossible to reconstruct from $N$-body simulations. Most notably, it yields an estimate of density that is local but defined everywhere in space, shot-noise free, and produces sharply delineated caustics of dark matter after shell-crossing \citep{Abel:2012}, leading also to new rendering techniques for 3D visualisation of the cosmic density field \citep{Kaehler:2012,Igouchkine:2016,Kaehler:2018}, and very accurate estimators of the cosmic velocity field \citep{Hahn:2015,Buehlmann:2019}.

Since the density is well defined everywhere in space just from the vertices, and reflects well the anisotropic motions in gravitational collapse, \cite{Hahn:2013} have proposed that this density field can be used as the source density field when solving Poisson's equation as part of the dynamical evolution of the system. The resulting method, where few $N$-body particles define the simplicial complex that together determine the density field, solves the artificial fragmentation problem of the $N$-body method for WDM initial conditions \citep{Hahn:2013}. The complex dynamics in late stages of collapse, however, limits the applicability of a method with a fixed number of vertices. This problem was later solved by allowing for higher order reconstructions of the Lagrangian manifold from $N$-body particles -- corresponding in some sense to non-constant-metric finite elements in Lagrangian space --, and dynamical refinement \citep{HahnAngulo:2016,Sousbie:2016}. For systems that exhibit strong mixing (phase or even chaotic, such as dark matter haloes), following the increasingly complex dynamics by inserting new vertices becomes quickly prohibitive (e.g., \citealt{Sousbie:2016,Colombi:2020} report an extremely rapid growth of vertices over time in a cosmological simulation with only moderate force resolution). \cite{Stuecker:2019} have carried out a comparison of the density estimated from phase space interpolation and that obtained from the GDE and found excellent agreement between the two except in the center of halos. The comparison between the two density estimates, shown in Lagrangian space, is reproduced in the bottom panels of Fig.~\ref{fig:darkmattersheet}.

\begin{figure}
\begin{centering}
\includegraphics[width=\textwidth]{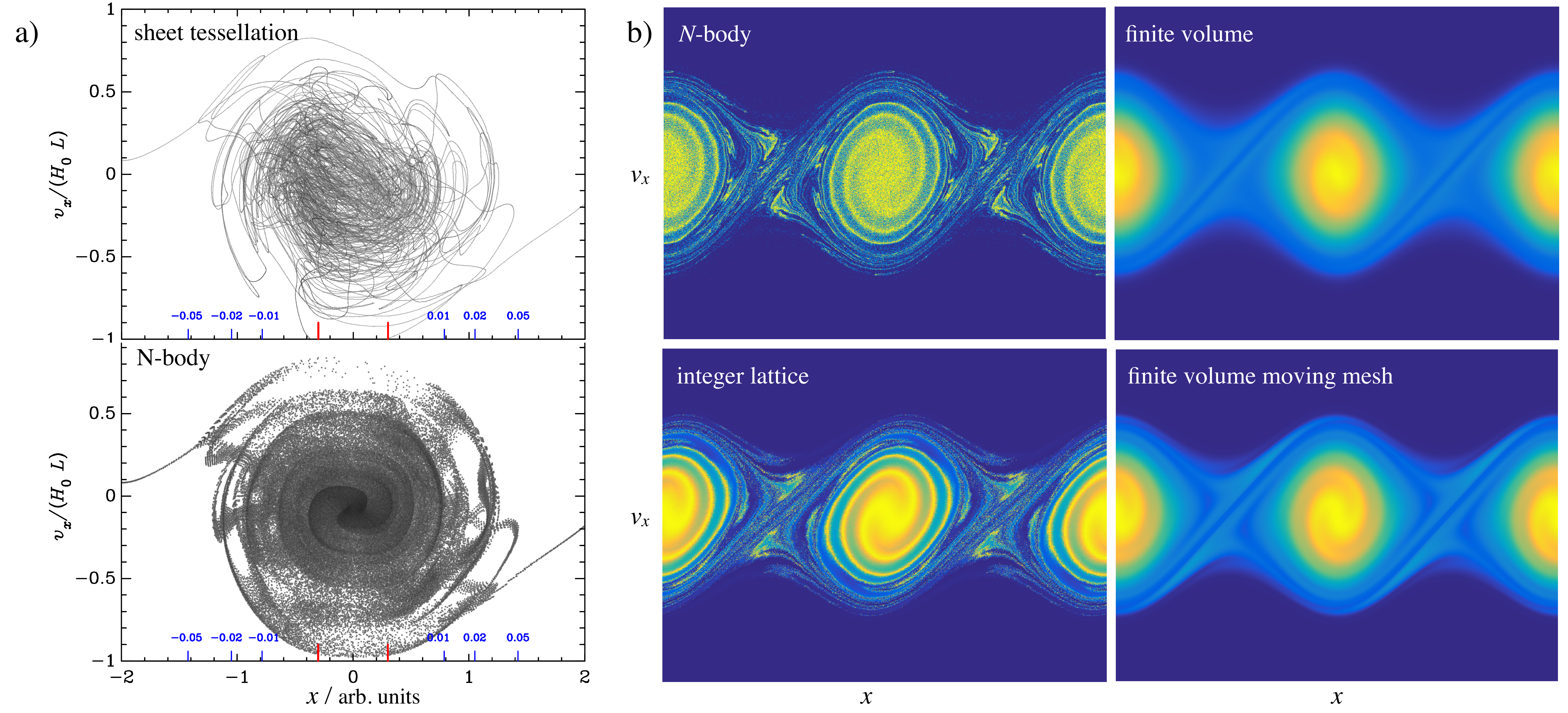} 
\caption{\label{fig:phasespace_evolution} Comparison of evolved structures from $N$-body simulations with other discretisation approaches. Panels a: Comparison of a 1+1 dimensional phase space cut from simulations of a three-dimensional collapse of a CDM halo using a sheet tessellation with refinement (top, cf.\ \citealt{Sousbie:2016}), and a reference $N$-body particle mesh simulation (bottom). The panels show an infinitely thin slice in the sheet case, and a finitely thin projection in the $N$-body case. Panels b: Simulations of collapse in 1+1D phase space with a particle mesh $N$-body method, the integer lattice method proposed by \cite{MoczLattice:2017}, as well as two finite volume approaches, one where slabs in velocity space are allowed to continuously move against each other (`moving mesh'). Panels a) are adapted from \cite{Colombi:2020}, panels b) from \cite{Mocz:2017}.}
\end{centering}
\end{figure}

The path forward in this direction lies likely in the use of hybrid $N$-body/sheet methods that exploit the best of both worlds as proposed by \cite{Stuecker:2019}.  Panel a) of Fig.~\ref{fig:phasespace_evolution} shows a 1+1D cut through 3+3D phase space for the case of a CDM halo comparing the result of a sheet-based simulation, where the cut results in a finite number continuous lines (top) and the equivalent results for a thin slice from an $N$-body simulation. The general impact of spurious phase space diffusion driven by the $N$-body method is still not very well understood with detailed comparison between various solvers under way (e.g., \citealt{Halle:2019,Stuecker:2019,Colombi:2020}). 

For hot distribution functions, such as e.g. neutrinos, the phase space distribution of matter is not fully described by the Lagrangian submanifold. While a 6D tessellation is feasible in principle, it has undesirable properties due to the inherent shearing along the momentum dimensions. However, Lagrangian submanifolds can still be singled out to provide a foliation of general six-dimensional phase space by selecting multiple Lagrangian submanifolds that are offset from each other initially by constant momentum vectors as proposed by \cite{Dupuy:2014,KatesHarbeck:2016}.  

Another version of phase space interpolation has been discussed in the context of collisionless dynamics by \cite{Colombi:2008,Colombi:2014}, the so-called `waterbag' method, which allows for general non-cold initial data but is restricted to 1+1 dimensional phase space. In this approach one exploits that the value of the distribution function is conserved along characteristics. If one now traces out isodensity countours of $f$ in phase space, one finds a sequence of $n$ (closed) lines defining the level set $\left\{ (x,p)\;|\;f(x,p,t_0)=f_i\right\}$ of 1+1D phase space with $i=1\dots n$ at the initial time $t_0$. In 1+1D, these are closed curves. The curves can then be approximated using a number of vertices and interpolation between them. Moving the vertices along characteristics then guarantees that they remain part of the level set at all times as phase space density is conserved along characteristics. The number of vertices can be adaptively increased in order to maintain a high quality representation of the set contour interpolation at all times. The acceleration of the vertices can be conveniently defined in terms of integrals over the contours (cf.\ \citealt{Colombi:2014}).  

\subsection{Full phase space techniques}
\label{sec:full_phasespace}

Almost as old as the $N$-body approach to solve gravitational Vlasov--Poisson dynamics are approaches to directly solve the continuous problem for an incompressible fluid in phase space (cf. \citealt{Fujiwara:1981} for 1+1 dimensions). By discretising phase space into cells of finite size in configuration and momentum space ($\Delta x$ and $\Delta p$ respectively) standard finite volume, finite difference methods or semi-Lagrangian techniques for incompressible flow can be employed. The main disadvantage of this approach is that memory requirements can be prohibitive since, without adaptive techniques or additional sophistications, the memory needed to evolve a three-dimensional system scales as $\mathcal{O}(N_x^3\times N_p^3)$ to achieve a linear resolution of $N_x$ cells per configuration space dimension and $N_p$ cells per momentum space dimension. Only rather recently this has become possible at all as demonstrated for gravitational interactions by \cite{Yoshikawa:2013} and \cite{Tanaka:2017}. The limited resolution that can be afforded in 3+3 dimensions leads to non-negligible diffusion errors even with high order methods, so that this direct approach is arguably best suited for hot mildly non-linear systems such as e.g. neutrinos \citep{Yoshikawa:2020,Yoshikawa:2021}, as the resolution required for colder systems is prohibitive. As a way to reduce such errors, \cite{Colombi:2017} proposed a semi-Lagrangian `metric' method that uses a generalisation of the `GDE' deformation discussed above to improve the interpolation step and reduce the diffusion error in such schemes.

As another way to overcome the diffusion problem, integer lattice techniques have been discussed (cf.\ \citealt{Earn:1992}), which exploit that if the time step is matched to the phase space discretisation, i.e., $\Delta t = m (\Delta x / \Delta p)$, then the configuration space advection is exact and a reversible Hamiltonian system can be obtained for the lattice model discretisation. While this approach does not overcome the $\mathcal{O}(N^6)$ memory scaling problem of a full phase space discretisation technique, recently \cite{Mocz:2017} have proposed important optimisations that might allow $\approx \mathcal{O}(N^4)$ scaling by overcomputing, but that, to our knowledge, have not been demonstrated yet in $3+3$ dimensional simulations. Results obained by \cite{Mocz:2017} comparing the various techniques are shown in Fig.~\ref{fig:phasespace_evolution}.

\subsection{Schr\"odinger--Poisson as a discretisation of Vlasov--Poisson}
An entirely different approach to discretise the Vlasov--Poisson system by exploiting the quantum-classical correspondence has been proposed by \cite{Widrow:1993} in the 1990s. Hereby one exploits that full information about the system, such as density, velocity, etc., can be recovered from the (complex) wave function, and phase space is discretised by a (here tuneable, not physical) quantisation scale $\hbar=2\Delta x\Delta p$. Since the Schr\"odinger--Poisson system converges in the limit of $\hbar\to0$ to Vlasov--Poisson \citep{Zhang:2002}, it can be used as a UV modified analogue model also for classical dynamics if one restricts attention (i.e. smoothes) on scales larger than $\hbar$. It is important to note that the phase of the wave function is intimately related to the Lagrangian submanifold -- both are given by a single scalar degree of freedom. For this reason, the Schr\"odinger--Poisson analogue has the advantage that it provides a full phase space theory with only a three-dimensional field (the wave function) that needs to be evolved. Following the first implementation by \cite{Widrow:1993}, this model has found renewed interest recently \citep{Uhlemann:2014,Schaller:2014,Kopp:2017,Eberhardt:2020,Garny:2020}. It is important to note that the underlying equations are identical to those of `fuzzy dark matter' (FDM) models of ultralight axion-like particles, which we discuss in more detail in Sect.~\ref{sec:fuzzy_dm}, in the absence of a self-interaction term. In the case of FDM, the quantum scale $\hbar/m_{\rm FDM}$ is set by the mass $m_{\rm FDM}$ of the microscopic particle and is (likely) not a numerical discretisation scale dictated by finite memory.

%\subsection*{Concluding remarks}
%In this section we have reviewed several possible discretizations of the matter distribution function in the Universe. We discussed the traditional $N$-body approach, but also several alternative techniques. While such methods cannot rival the $N$-body discretization at present time, we would like to emphasise that they are of crucial importance in the field of computational cosmology. First of all, as we have highlighted, there is no formal proof of the correctness of N-body simulations in the sense of convergence to the continuum limit, or known analytic solutions: the possible systematic errors are simply quantified by a comparison of simulations at different resolution (together with basic checks such as a global energy and momentum conservation). In this context, alternative discretizations appear essential in validating traditional simulations employed in interpreting observations of the Universe, in the precision era. Furthermore, in the future we might see further developments in combining different discretizations, either for different dynamical stages of structure formation, or for different components (e.g. CDM vs. neutrinos), which might build on the strengths and complementarities of different approaches.

%% Time Integration
%% \newpage

\section{Time evolution}
\label{sec:time_integration}
As we have shown above, large-scale dark matter simulations have an underlying Hamiltonian structure, usually with a time-dependent Hamiltonian. Mathematically, such Hamiltonian systems have a very rigid underlying structure, where the phase-space area spanned by canonically conjugate coordinates and momenta is conserved over time. Consequentially, specific techniques for the integration of Hamiltonian dynamical systems exist that preserve such underlying structure even in a numerical setting. For this reason, this section focuses almost exclusively on integration techniques for Hamiltonian systems as they arise in the context of cosmological simulations.

\subsection{Symplectic integration of cosmological Hamiltonian dynamics}
\label{sec:symplectic_integration}
In the cosmological $N$-body problem, Hamiltonians arising in the Newtonian limit are typically of the non-autonomous but separable type, i.e. can be written
\begin{equation}
\mathscr{H} = \alpha(t) \, T(\vecb{P}_1,\dots,\vecb{P}_N) + \beta(t)\, V(\vecb{X}_1,\dots,\vecb{X}_N),
\end{equation}
where $\vecb{X}_i$ and $\vecb{P}_i$ are canonically conjugate, and $\alpha(t)$ and $\beta(t)$ are time-dependent functions that absorb all explicit time dependence (i.e. all factors of `$a$' are pulled out of the Poisson equation for $V$). In cosmic time $t$ one has $\alpha=a(t)^{-2}$ and $\beta=a(t)^{-1}$, which is not a clever choice of time coordinate since then the time dependence appears in both terms which complicates higher order symplectic integration schemes as we discuss below. The unique best choice is to not forget the relativistic origin of this Hamiltonian and consider time as a coordinate in extended phase space (cf.~\citealt{Lanczos:1986}), using a parametric time $\tilde{t}$ with ${\rm d}\tilde{t} = a^{-2} {\rm d}t$ so that $\alpha=1$ and $\beta=a(t)$. This coincides with the ``super-conformal time'' first introduced by \cite{Doroshkevich:1973} and extensively discussed by \cite{Martel:1998} under the name ``super-comoving'' coordinates. 

Grouping coordinates and momenta together as $\vecb{\xi}_j:=(\vecb{X},\vecb{P})_j$ and remembering that the equations of motion can be written in terms of Poisson brackets\footnote{ The Poisson bracket $\left\{\cdot,\cdot\right\}$ is defined for canonically conjugate coordinates $(\vecb{X}_i,\vecb{P}_i)_{i=1,\dots,N}$ and any two (possibly time-dependent) functions $A$ and $B$ of these coordinates as $\left\{A,B\right\} = \sum_{i=1}^{N}\left(\frac{\partial A}{\partial \vecb{X}_i}\cdot\frac{\partial B}{\partial \vecb{P}_i}-\frac{\partial A}{\partial \vecb{P}_i}\cdot\frac{\partial B}{\partial \vecb{X}_i}\right)$. In particular, one finds the properties $\left\{A,B\right\}=-\left\{B,A\right\}$ and $\frac{{\rm d}A}{{\rm d}t}=\frac{\partial A}{\partial t}+\left\{A,\mathscr{H}\right\}$ \citep[e.g.][]{Goldstein:2002}.} as $\dot{\vecb{P}}_j = \left\{ \vecb{P}_j,\,\mathscr{H} \right\}$ and $\dot{\vecb{X}}_j = \left\{ \vecb{X}_j,\,\mathscr{H} \right\}$, one can write the canonical equations as a first order operator equation
\begin{equation}
\dot{\vecb{\xi}}_j = \hat{\mathscr{H}}(t)\, \vecb{\xi}_j\qquad\textrm{with}\qquad \hat{\mathscr{H}}(t):= \left\{ \cdot,\, \mathscr{H}(t)\right\} = \left\{ \cdot,\, \alpha T\right\} + \left\{ \cdot,\, \beta V\right\} =: \hat{D}(t) + \hat{K}(t), \label{eq:operator_hamiltonian}
\end{equation}
which defines the drift and kick operators $\hat{D}$ and $\hat{K}$, respectively. This first order operator equation has the formal solution
\begin{equation}
\vecb{\xi}_j(t) = \mathcal{T}  \exp\left[ \int_0^t {\rm d}t' \hat{\mathscr{H}}(t')\right] \vecb{\xi}_j(0), \label{eq:hamilton_formalsolution}
\end{equation}
where Dyson's time-ordering operator $\mathcal{T}$ is needed because the operator $\hat{\mathscr{H}}$ is time-dependent. Upon noticing that the kick acts only on the momenta and it depends only on $V$ (and therefore on the positions), and that the drift acts only on the positions and depends only on the momenta, one can seek for time-explicit operator factorisations that split the coordinate and momentum updates in the form
\begin{equation}
 \mathcal{T}  \exp\left[ \int_t^{t+\epsilon} {\rm d}t' \hat{\mathscr{H}}(t')\right]  \simeq \exp\left[ \epsilon_n \hat{K}\right]\cdots \exp\left[ \epsilon_3 \hat{D} \right] \,\exp\left[ \epsilon_2 \hat{K} \right] \,\exp\left[ \epsilon_1 \hat{D} \right] + \mathcal{O}(\epsilon^m). \label{eq:operator_factorisation}
\end{equation} 
with appropriately chosen coefficients $\epsilon_j$ that in general depend on (multiple) time integrals of $\alpha$ and $\beta$ \citep{Magnus:1954,Oteo:1991,Blanes:2009}. This is a higher-order generalisation of the Baker--Campbell--Hausdorff (BCH) expansion in the case that $\alpha$ and $\beta$ are constants \citep{Yoshida:1990}. The cancellation of commutators in the BCH expansion by tuning of the coefficients $\epsilon_j$ determines the order of the error exponent $m$ on the right hand side of Eq.~(\ref{eq:operator_factorisation}). It is important to note that if both $\alpha$ and $\beta$ are time-dependent then the generalised BCH expansion contains unequal-time commutators and the error is typically at best $\mathcal{O}(\epsilon^3)$. It is therefore much simpler to consider only the integration in extended phase space in super-conformal time, in which no unequal-time commutators appear and standard higher order BCH expansion formulae can be used. While some $N$-body codes (e.g., {\sc Ramses}, \citealt{Teyssier:2002}) use super-conformal time, one finds numerous other choices of integration time for second order accurate integrators in the literature (e.g., \citealt{Quinn:1997,Springel:2005b}). In order to allow for generalisations to higher orders, we discuss here how to construct an extended phase-space integrator. Consider the set of coordinates $(\vecb{X}_j,\,a)$, $j=1\dots N$, including the cosmic expansion factor, with conjugate momenta $(\vecb{P}_j,\,p_a)$ along with the new extended phase-space Hamiltonian in super-conformal time
\begin{equation}
\tilde{\mathscr{H}} := \sum_j \frac{{P}_j^2}{2M} + a V(\vecb{X}_{1\dots N}) + a^2\mathcal{H}(a) p_a,
\end{equation}
where $\mathcal{H}(a)$ was the conformal-time Hubble function. Then the second order accurate ``leap-frog'' integrator is found when $\epsilon_1=\epsilon_3=\epsilon/2$ and $\epsilon_2=\epsilon$ in Eq.~(\ref{eq:operator_factorisation}) (all higher orders $\epsilon_{3\dots n}=0$) after expanding the operator exponentials to first order into their generators $\exp[\epsilon \hat{D}]\simeq I + \epsilon \hat{D}$. The final integrator takes the form
\begin{equation}
\vecb{\xi}_j(\tilde{t}+\epsilon) = \left(I+\frac{\epsilon}{2}\hat{D}\right)\left(I+\epsilon\hat{K}\right)\left(I+\frac{\epsilon}{2}\hat{D}\right) \vecb{\xi}_j(\tilde{t}) 
\end{equation}
or explicitly as it could be implemented in code
\begin{subequations}
\begin{align}
\vecb{X}_j(\tilde{t}+\epsilon/2)& = \vecb{X}_j(\tilde{t}) + \frac{\epsilon}{2M} \; \vecb{P}_j(\tilde{t}) \label{eq:leapfrog_drift1}\\
a(\tilde{t}+\epsilon/2) & =  a(\tilde{t}) + \frac{\epsilon}{2} \; a(\tilde{t})^2 \, \mathcal{H}(a(\tilde{t}))\\
\vecb{P}_j(\tilde{t}+\epsilon) &= \vecb{P}_j(\tilde{t}) - \epsilon\,a(\tilde{t}+\epsilon/2)\; \bnabla_{\vecb{X}_j} V\left(\vecb{X}_{1\dots N}(\tilde{t}+\epsilon/2)\right) \\
\vecb{X}_j(\tilde{t}+\epsilon)& = \vecb{X}_j(\tilde{t}+\epsilon/2) + \frac{\epsilon}{2M} \; \vecb{P}_j(\tilde{t}+\epsilon) \\\label{eq:leapfrog_drift2}
a(\tilde{t}+\epsilon) & =  a(\tilde{t}+\epsilon/2) + \frac{\epsilon}{2}\; a(\tilde{t}+\epsilon/2)^2 \, \mathcal{H}(a(\tilde{t}+\epsilon/2))\;.
\end{align}
\end{subequations}
Note that the supplementary equation ${\rm d} a/{\rm d}\tilde{t} = \partial \tilde{\mathscr{H}}/\partial p_a = a^2\mathcal{H}(a)$, can in principle also be integrated inexpensively to arbitrarily high precision in general cases, for EdS one has $\tilde{t} = -2/(H_0 \sqrt{a})$ and in $\Lambda$CDM 
\begin{equation}
\tilde{t} = -\frac{2}{H_0 \sqrt{\Omega_m a}} \phantom{X}_2F_1\left(-\sfrac{1}{6},\sfrac{1}{2},\sfrac{5}{6};-f_\Lambda(a)\right),
\end{equation}
where $f_\Lambda := \Omega_\Lambda / (\Omega_{\rm m} a^{-3})$ as in Eq.~(\ref{eq:LCDM_growthfactor}), which has to be inverted numerically to yield $a(\tilde{t})$. Since this is a symplectic integration scheme, it will conserve the energy associated with the Hamiltonian $\tilde{\mathscr{H}}$.

Eqs.~(\ref{eq:leapfrog_drift1})--(\ref{eq:leapfrog_drift2}) represent the drift-kick-drift (DKD) form of a second order integrator. It is trivial to derive also the respective kick-drift-kick (KDK) form. Based on this, it is possible to construct also higher order integrators, see e.g., \cite{Yoshida:1990} for a derivation of operators up to 8th order that involve, however, positive and negative time coefficients. Also, alternative symplectic formulations with purely positive coefficients are possible, see \cite{Chin:2001} for a 4th order method. An exhaustive discussion of symplectic and other geometric integrators and their properties can be found in \cite{Hairer:2006} and \cite{Blanes:2016}. In cosmological simulations, the second order leap frog is, however, the most commonly used integrator to date, arguably due to its robustness, simplicity, slim memory footprint, and easy integration with hierarchical time-stepping schemes (see below). We are not aware of production implementations of higher-order symplectic integrators used in cosmological simulations. 

A long-time evolution operator can be constructed by many successive applications of the KDK or DKD propagators. Writing out the product, it is easy to see that the last and first half-step operators from two successive steps can often be combined into a single operator (if $\hat{A}$ below is time independent, otherwise usually at second order). Then, in the long product of operators, combining
\[
\dots \exp(\epsilon_B \hat{B})\exp(\frac{\epsilon_A}{2} \hat{A})\exp(\frac{\epsilon_A}{2} \hat{A})\exp(\epsilon_B \hat{B})\dots=\dots\exp(\epsilon_B \hat{B})\exp(\epsilon_A \hat{A})\exp(\epsilon_B \hat{B})\dots,
\]
implies that in continued stepping only two interleaved steps have to be made per time step, not three, and that the splitting into three sub-steps just serves to symmetrise the scheme and interleave the steps. Half-steps are only made at the very beginning and end of the stepping -- or whenever one needs synchronous dynamical variables (e.g. for output or analysis).

\subsection{Multi-stepping, adaptive steps and separation of time-scales}
\subsubsection{Time step criteria}

A challenge in cosmological simulations is the large dynamic range from quasilinear flow on large-scales to very short dynamical times in the centres of dark matter halos: we seek to
simultaneously simulate large underdense regions of the universe -- where particles move on long timescales together with massive clusters -- where density contrasts can reach $10^4$--$10^5$ times the average density and with very short timescales. In the absence of an adaptive or hierarchical time-stepping scheme, the criteria discussed below yield a global timestep $\Delta t = \min_i \Delta t_i$ dictated by the $N$-body particle with the smallest time step.

A simple condition for choosing a time step is the Courant--Friedrichs--Lewy (CFL) criterion, which requires that particles travel less than a fraction of one force resolution element, $\Delta x$, over the time step. Specifically
\begin{equation}
\label{eq:CFL}
\Delta t_i = C \frac{\Delta x}{ \| \vecb{P}_i / M_i \|} \,,
\end{equation}
where $0<C<1$ is a free parameter, usually $C \sim 0.25$. While commonly used in the case of Vlasov--Poisson, we are not aware of explicit derivations of this value from stability criteria as in the case of hyperbolic conservation laws. A closely related criterion is $\Delta t_i = C \sqrt{\Delta x/\|\vecb{A}_i\|}$, where $\vecb{A}_i := -\bnabla \phi|_{\vecb{X}_i}$ is the acceleration. This condition sets a global timestep that is commonly used in simulations where forces are computed via the PM algorithm (e.g., \citealt{Merz:2005}). Other criteria are also possible, for instance the {\sc ABACUS} code \citep{Garrison:2016}, in addition to Eq.~\ref{eq:CFL}, uses a heuristic condition based on the global maximum value of the RMS velocity over the maximum acceleration of particles in a volume element. 

In the case of tree or direct summation based methods, the scale played by the mesh grid resolution is taken over by the softening length. Therefore, a simple criterion is to estimate for each particle a time-scale
\begin{eqnarray}
\Delta t_i \simeq \eta \sqrt{\frac{\varepsilon}{\|\boldsymbol{A}_i\|}} \,, \label{eq:std_timestep_crit}
\end{eqnarray}
where $\varepsilon$ is the gravitational force softening scale,  and $\eta$ is a dimensionless accuracy parameter. This is the most common time-stepping criterion adopted in large-scale simulations and it is used, for instance, in {\sc PKDGRAV-3} (with $\eta = 0.2$) and {\sc GADGET}.

Several authors have argued that a more optimal timestepping criterion should be based on the tidal field rather than the acceleration \citep{Dehnen:2011,Stuecker:2019,Grudic:2019}. This is also motivated by the fact that a constant (global) acceleration does not change the local dynamics of a system, see e.g. \cite{Stucker:2021a}. Additionally, using tides avoids invoking the non-physical scale $\varepsilon$, and one has
\begin{equation}
\Delta t_i \simeq \frac{\eta}{ \sqrt{\left\| \; \tnsr{T}(\vecb{X}_i) \; \right\|}} \,,
\end{equation}
where $\tnsr{T}=\bnabla\otimes\bnabla \phi$ is the tidal field tensor, and $\| \cdot \|$ is e.g. the Frobenius matrix norm. This tidal criterion typically yields shorter timesteps in the innermost parts and longer timesteps in the outer parts of haloes compared to the standard criterion (\ref{eq:std_timestep_crit}). A caveat is that it is not trivial to get a robust estimate of the tidal field, since it is one spatial order less smooth than the acceleration field entering (\ref{eq:std_timestep_crit}) and so in principle time-step fluctuations due to noise might amplify integration errors.

An additional global timestep criterion, regardless of the dynamics of $N$-body particles, is sometimes used in large-scale cosmological simulation when high resolution is not required. These criteria are usually tied to the scale-factor evolution and e.g. of the form
\begin{equation}
\label{eq:delta_loga}
\Delta \log(a) < B \,,
\end{equation}
where $B \sim 0.01$. These kind of criteria are also usually employed in PM codes and {\sc COLA} (which we will discuss in Sect.~\ref{sec:cola_fastpm}), which typically adopt timesteps equally spaced in the expansion factor or in its logarithm, or more complicated functions (e.g., $\Delta a/a = (a_1^{-2} + a_2^{-2})^{-0.5}$ with $a_1$ and $a_2$ being free parameters in {\sc Fast-PM}). Different authors have advocated different options and number of steps justified simply by convergence rates of a set of desired summary statistics \citep{White:2014,Feng:2016,Tassev:2013,Izard:2016}. The criterion in Eq.~\ref{eq:delta_loga} is also commonly used together with other conditions even in high-resolution simulations, since it appears to be necessary for a precise agreement with linear theory predictions at high redshift. 

We note that different timestep criteria (and combinations) are adopted by different codes, usually heuristically motivated. The optimal choice seems to depend on details of the simulation (redshift, force accuracy, etc), which suggest there could be better strategies to choose the timestep. This could be very important as it has a significant impact in the overall computational cost of a simulation. For instance, by adjusting the timesteps, 	\cite{Sunayama:2016} finds a factor of 4 reduction in the CPU time in $N$-body simulations while still accurately recovering the power spectrum and halo mass function (after a correction of masses). As far as we know, no systematic study of the optimal general time-stepping strategy has been published for large-scale cosmological simulations taking into account the target accuracy needed for upcoming observations. 

For some applications, a global timestep is sufficient to obtain accurate results on the mildly nonlinear regime, as it has been adopted in some large-scale simulations codes. However, as the resolution of a simulation increases, the minimum value of $\Delta t_i$ quickly decreases, usually as the result of a small number of particles in short orbit inside dark matter halos. To avoid that the shortest time-scale dictates an intractably small global time step, it is desirable to have individually adaptive time-steps. Some care needs to be taken to consistently allow for this in a time integrator, as we discuss next.

\subsubsection{Hierarchical / block time stepping}
\label{sec:block_timestep}

For systems of particles with widely different time-scales, a division into `fast' and `slow' particles is advantageous. Given a second order accurate integrator and a splitting of the Hamiltonian into $\mathscr{H} = T + V_{\rm slow} + V_{\rm fast}$, then the following $n$-fold sub-cycling scheme is also second order accurate \citep{Hairer:2006}
\begin{equation}
\left(I+\frac{\epsilon}{2}\hat{K}_{\rm slow}\right) \left[ \left(I+\frac{\epsilon}{2n}\hat{K}_{\rm fast}\right) \left(I+\frac{\epsilon}{n}\hat{D}\right) \left(I+\frac{\epsilon}{2n}\hat{K}_{\rm fast}\right) \right]^n \left(I+\frac{\epsilon}{2}\hat{K}_{\rm slow}\right). \label{eq:timestep_subcycle}
\end{equation}
The gain is that in one such KDK timestep, while the fast particles are kicked $2n$ times, the slow particles are kicked only twice. Since the force computation is the algorithmically slowest part of the update, this leads to a computational speed up.

This sub-cycling idea can be generalised to the block time step scheme (BTS, sometimes also called `hierarchical time stepping') to update particles on their relevant time scales \citep{Hayli:1967,Hayli:1974,McMillan:1986} using a time-quantised recursive version of the sub-cycling scheme above. By fixing $n=2$, but applying the formula recursively, i.e., splitting the `fast' part itself into a `slow' and `fast' part and so on, one achieves a hierarchical time integration scheme with time steps $\epsilon_\ell = 2^{-\ell} \epsilon_0$ on recursion level $\ell$. The power-two quantisation means that while a particle on level $\ell$ makes one timestep, a particle on level $\ell+1$ makes two, and one on level $\ell+2$ makes four in the same time interval. Since the scheme is self-similar, after every sub step with Eq.~(\ref{eq:timestep_subcycle}) on level $\ell$, all particles on levels $\ell'>\ell$ have carried out complete time steps, and can be re-assigned a new time bin. In this way, each particle can be assigned to a close to optimal time step ``bin'' which is consistent with its local timestep criterion. This multi-stepping approach is adopted in {\sc PKDGRAV-3} and {\sc GADGET}. 

This scheme can be further optimized. As written above, the kick $\hat{K}$ on level $\ell$ involves the interaction with particles on all other levels, i.e. the `fast' particles interact with the `slow' particles on the `fast' timescale, while it is likely sufficient that they do so on the `slow' time scale. For this reason, a variation, e.g. implemented in {\sc GADGET4} \citep{Springel:2020}, is that the kick of particles on level $\ell$ is computed using the gravitational force of particles {\it only} on all levels $\ell' \ge \ell$. Such hierarchical integration schemes have recently been extended to higher order integrators in the non-cosmological case \citep{Rantala:2021}. In principle, secondary trees need to be built in every level of the hierarchy of timesteps, however, this would require a significant amount of computing time. Therefore, a optimization strategy, e.g. adopted in {\sc PKDGRAV-3}, is to build a new secondary tree only if a timestep level contains a small fraction of particles compared to the previous level where the tree was built.

%%%%%%%%%%%%%%%%%%%%%%%%%%%%%%%%%%%%%%%%%%%%%%%%%%%%%

\subsection{Symplectic integration of quantum Hamiltonians}
\label{sec:quantum_integration}
The integration of a classical Hamiltonian in operator notation, Eq.~(\ref{eq:operator_hamiltonian}) is basically identical to that of the Schr\"odinger-Poisson system which is an effective description of non-relativistic scalar field dark matter, Eq.~(\ref{eq:Gross-Pitaevskii}). One can therefore use the entire machinery of operator splitting developed above, with differences only in the form of the kick and drift operators, as they now act on the wave function, rather than the set of conjugate coordinates and momenta. As above, the best choice is again superconformal time so that one has a quantum Hamiltonian acting on wave functions $\psi$ with associated Poisson equation:
\begin{equation}
i\hbar \frac{\partial\psi}{\partial \tilde{t}} = \hat{\mathscr{H}}\psi \qquad\textrm{with}\qquad  \hat{\mathscr{H}} = \frac{\hat{p}^2}{2m}+a(\tilde{t})\,\hat{V}(\hat{q})\qquad\textrm{and}\qquad  \nabla^2 \hat{V} = \frac{3}{2}H_0^2 m \Omega_X ( \left| \psi \right|^2-1). 
\end{equation}
With the formal solution identical to that in Eq.~(\ref{eq:hamilton_formalsolution}) apart from a factor ${\rm i}/\hbar$ in the exponent, note also that the mass $m$ is the actual microscopic particle mass and not a coarse-grained effective mass. The main difference and advantage compared to the classical case is that drift and kick operators $\hat{D}$ and $\hat{K}$ need not be represented through their infinitesimal generators, but can be directly taken to be the operator exponentials 
\begin{equation}
\hat{D}(\tilde{t},\tilde{t}+\epsilon) = \exp\left( -\epsilon\,\frac{i}{\hbar} \frac{\hat{p}^2}{2m} \right),\qquad\textrm{and}\qquad \hat{K}(\tilde{t},\tilde{t}+\epsilon) = \exp\left( -\epsilon a(\tilde{t}+\epsilon/2)\frac{i}{\hbar} \hat{V} \right).
\end{equation}
The kick operator is purely algebraic, i.e. it is simply a scalar function multiplying the wave function. The same is true for the drift operator in Fourier space, where

\begin{equation}
\hat{\tilde{D}}(\tilde{t},\tilde{t}+\epsilon) := \mathcal{F}\,\hat{D}(\tilde{t},\tilde{t}+\epsilon) = \exp\left( -\epsilon\, \frac{{\rm i}\hbar}{2m}k^2 \right) \label{eq:qm_drift}
\end{equation}
is simply a scalar function; $\mathcal{F}$ is the Fourier transform operator with inverse $\mathcal{F}^{-1}$, $\vecb{k}$ the Fourier-conjugate wave number (or momentum) to coordinate $\vecb{x}$. One can thus formulate a split-step spectral integration scheme (e.g., \citealt{Taha:1984,Woo:2009}), where drift operators are simply applied in Fourier space, i.e. before a drift, the wave function is transformed, and after it is transformed back. The time coefficients again have to be matched to cancel out commutators in the BCH relation, so that a second order DKD time step is e.g. given by
\begin{equation}
\psi(\vecb{x},t+\epsilon) = \mathcal{F}^{-1} \hat{\tilde{D}}(\tilde{t}+\epsilon/2,\tilde{t}+\epsilon) \mathcal{F} \hat{K}(\tilde{t},\tilde{t}+\epsilon) \mathcal{F}^{-1} \hat{\tilde{D}}(\tilde{t},\tilde{t}+\epsilon/2) \mathcal{F}\,\psi(\vecb{x},t).
\end{equation}
The chaining of multiple steps eliminates two of the Fourier transforms for the steps that are not at the beginning or end of the time integration, leaving one forward and one backward transform per time step. The use of Fourier transforms and the spectrally accurate drift operator (\ref{eq:qm_drift}) which contains the exponentiated Laplacian $\hat{p}^2$ to all orders significantly increases the convergence rate of the scheme even for large time steps. 

The limiting factor of a spectral approach is that the spatial discretisation is necessarily uniform for Fourier based methods. It is clear however that the use of AMR techniques to locally increase the resolution inhibits (to our knowledge) the simple use of spectral approaches, so that smaller stencils and finite-difference expansions become necessary. Since gravity and the expanding Universe contract gravitationally bound structures to ever smaller scales, methods with higher dynamic range are needed to probe the interior of halos. To our knowledge, no spatially adaptive spectral method has been developed yet. Instead, methods with higher dynamic range resort to finite difference methods with AMR, which has been successfully used to model the interior dynamics of scalar field dark matter haloes \citep{Schive:2014,Mina:2020,Schwabe:2020}. By discretising the kick operator, one has to resort to the generator formulation again, i.e. $\hat{D}_\epsilon =  1+\epsilon \frac{{\rm i}\hbar}{m}\nabla^2 + \mathcal{O}(\epsilon^2)$, where then the Laplacian can be approximated with finite differences. This imposes strong CFL-like time-stepping constraints. 

%%%%%%%%%%%%%%%%%%%%%%%%%%%%%%%%%%%%%%%%%%%%%%%%%%%%%%%%%

\subsection{Acceleration methods: {\sc COLA} and {\sc FastPM}}
\label{sec:cola_fastpm}

One of the main short-comings of symplectic integration schemes is that they have to evolve a time-dependent Hamiltonian, which can require many time steps even during the quasi-linear early phase of structure formation. In contrast, perturbation theory is very accurate during this regime. For instance, first-order LPT (a.k.a. the Zel'dovich approximation) yields exact results for one-dimensional problems up to shell-crossing and, therefore, the solution could be obtained in a single time-step. This implies that (mostly in low resolution large-scale simulations) a considerable amount of computing time can be spent to accurately evolve the particles during the quasi-linear phase, since too large timesteps during this phase can lead to significant large-scale errors. This has motivated several methods aimed at incorporating results from perturbation theory in the time-integration of an $N$-body scheme. These approaches have been widely-adopted to more efficiently create large ensembles of simulations that achieve high accuracy on mildly non-linear scales. We review the main ideas and implementations next.

\subsubsection*{COLA: COmoving Lagrangian Acceleration}

In the {\sc COLA} approach \citep{Tassev:2013}, the idea is to consider motion relative to a pre-computed LPT solution (specifically 1 and 2LPT in all implementations we are aware of), by considering the equations of motion of CDM, Eq.~(\ref{eq:CDM_EOM}), relative to the motion in $n$LPT, as quantified by the order $n$ truncated Lagrangian map $\vecb{x}_j = \vecb{q}_j+\vecb{\Psi}_{{\rm LPT},j}(t)$ for particle $j$. In all existing work on {\sc COLA} that we are aware of, an ad-hoc modification of the leapfrog drift and kick operators is made to reflect this change in the equations of motion. One can however write such a transformation rigorously as a canonical transformation to new coordinates $(\vecb{X}_j,\vecb{P}_j)$ with a generating function $\mathscr{F}_3(\vecb{p}_j,\vecb{X}_j,\tilde{t}) = ( M\dot{\vecb{\Psi}}_{{\rm LPT},j}-\vecb{p}_j)\cdot\vecb{X}_j$ so that $\vecb{X}_j=\vecb{x}_j$ and $\vecb{P}_j=\vecb{p}_j - M\dot{\vecb{\Psi}}_{{\rm LPT},j}$. The {\sc COLA} Hamiltonian then becomes (where the dot now indicates a derivative w.r.t. superconformal time)
\begin{equation}
\mathscr{H}_{\rm COLA} = \sum_j\frac{(\vecb{P}_j+M\dot{\vecb{\Psi}}_{{\rm LPT},j})^2}{2 M} + a V(\vecb{X}_{1\dots N})  + M \sum_j \ddot{\vecb{\Psi}}_{{\rm LPT},j}\cdot \vecb{X}_j.
\end{equation}
It is immediately obvious that this Hamiltonian has explicit time-dependence in both the kinetic and the potential part which will complicate the development of symplectic splitting schemes. The equations of motion now reflect the motion relative to the LPT solution of the form
\begin{equation}
\dot{\vecb{X}}_j = \frac{\vecb{P}_j}{M} + \dot{\vecb{\Psi}}_{{\rm LPT},j}\qquad\textrm{and}\qquad \dot{\vecb{P}}_j = -a\,\bnabla_{\vecb{X}_j} V - M\ddot{\vecb{\Psi}}_{{\rm LPT},j}.
\end{equation}
Existing {\sc COLA} implementations ignore the Dyson time-ordering and simply modify the drift and kick operators to become
\begin{subequations}
\begin{eqnarray}
\hat{D}_{\rm COLA}(\tilde{t},\tilde{t}+\epsilon) \vecb{X}_j &=& \hat{D}(\tilde{t},\tilde{t}+\epsilon) \vecb{X}_j + \vecb{\Psi}_{{\rm LPT},j}(\tilde{t}+\epsilon) -  \vecb{\Psi}_{{\rm LPT},j}(\tilde{t}), \\
\hat{K}_{\rm COLA}(\tilde{t},\tilde{t}+\epsilon) \vecb{P}_j &=& \hat{K}(\tilde{t},\tilde{t}+\epsilon) \vecb{P}_j + M\dot{\vecb{\Psi}}_{{\rm LPT},j}(\tilde{t}+\epsilon) -  M\dot{\vecb{\Psi}}_{{\rm LPT},j}(\tilde{t}),
\end{eqnarray}
\end{subequations}
which is accurate at first order since none of the unequal time commutators can be expected to vanish (\citealt{Tassev:2013} discuss an ad-hoc improvement to reduce errors in their Appendix~A.3.2). Despite the low order, this method allows for a very rapid approximate evolution in the quasi-linear regime, at the expense of having to store the fields needed to compute the $n$LPT trajectory of each fluid element. The above modifications are widely used in, for instance, the generation of a large number of mock catalogues for estimates of clustering covariance matrices. In such cases, ensembles of thousands of low-mass resolution, each of them with typically 10 time steps, are performed. An MPI-parallel implementation of the {\sc COLA} algorithm is publicly available for the implementation of \cite{Koda:2016}\footnote{Available from \url{https://github.com/junkoda/cola_halo/}} and in the form of the {\sc L-PICOLA}\footnote{Available from \url{https://cullanhowlett.github.io/l-picola/}} code \citep{Howlett:2015}, and a modified sCOLA algorithm allowing a more efficient spatial decomposition and ``zoom simulations'' has been recently proposed \citep{Tassev:2015} and is implemented in the {\sc Simbelmyne} code \citep{Leclercq:2020}\footnote{Available from \url{https://bitbucket.org/florent-leclercq/simbelmyne}.}. Therefore, it might be worthwhile exploring non-symplectic integration schemes in COLA, which could be rigorously higher order (given that there is no obvious benefit from symplectic integration anyway) and improve the performance of the method.

\subsubsection*{FastPM}
An alternative that does not require to compute or store the LPT displacement fields, thus saving computer memory, while relying on PT input to speed up the evolution was proposed as the {\sc FastPM} method by \cite{Feng:2016}\footnote{Available from \url{https://github.com/fastpm/fastpm}.}. In this approach, the prefactors for drift and kick operators receive an ad-hoc modification such that they contain the expected contribution of non-constant acceleration and velocities computed in the Zeldovich approximation. Note, however, that it has been argued that no such modifications are needed to obtain an accurate time integration, as long as the time stepping is chosen appropriately (\citealt{Klypin:2018}; see also \citealt{Sunayama:2016}). The performance is similar to that of {\sc COLA}, allowing approximate simulations with very few time steps that can be accurate on large scales. As for {\sc COLA}, the order of convergence has, to our knowledge, not been discussed in the literature. The {\sc FastPM} approach has been recently extended to include the modelling of massive neutrinos \citep{Bayer:2020}, and also ported to {\sc TensorFlow} \citep{ModiTensorFlow:2020}.

%% Gravity Calculation
%%%%%%%%%%%%%%%%%%%%%%%%%%%%%%%%%%%%%%%%%%%%%%%%%%%%%%%%%
%% \newpage
\section{Gravity calculation}
\label{sec:gravity}

After having discussed the discretization in time and space (i.e. mass, for Lagrangian schemes) of the evolution equations, we now turn to the problem of computing the gravitational interactions of the simulated mass distribution. This step is usually the most time-consuming aspect of a modern $N$-body simulation and thus also where most numerical approximations are made and where various parallelization strategies have the largest impact. Depending on the problem at hand, the targeted numerical accuracy, and the computer architecture employed, several different methods exists that are in different senses 'optimal'. Modern state-of-the-art codes typically exploit all these existing techniques. Optimal algorithmic complexity, $\mathcal{O}(N)$ in the number $N$ of particles, is achieved e.g. by the Fast Multipole Method (FMM), which is very promising for very large particle count simulations and is used e.g., in the {\sc PKDGRAV3} or {\sc Gadget-4} codes, and the geometric multigrid method, used e.g., in the {\sc RAMSES} code. The newest codes also readily utilise thousands of GPUs to generate simulated Universes for the upcoming generations of cosmological observations.

In the following, we provide a brief overview of the main methods and the main ideas behind them.  Regarding the general topic of gravity calculations in $N$-body, we also refer to other reviews for further details on these methods \citep{Dehnen:2011}.

%%%%%%%%%%%%%%%%%%%%%%%%%%%%%%%%%%%%%%%%%%%%%%%%%%%%%%%%%
\subsection{Mesh-based methods}
A robust and fast method to solve for the gravitational interactions of a periodic system is provided by the particle-mesh (PM) method \citep{Doroshkevich:1980,Hockney:1981}. Derived from the particle-in-cell (PIC) technique developed in plasma physics, they are among the oldest numerical methods employed to study cosmological structure formation. The techniques  described here can be employed not only for the $N$-body discretisations, but are readily applicable also e.g. for full phase space or integer lattice methods (cf. Sects.~\ref{sec:tessellation_methods} and \ref{sec:full_phasespace}), see also \cite{Miller:1968}, and even in the case of Schr\"odinger-Poisson systems \citep{Woo:2009}.

\subsubsection{Force and potential determination -- spectral calculation}

Considering a periodic domain of side length $L$, we want to solve the cosmological Poisson equation~(Eq.~\ref{eq:nbody_poisson}). Assume that both density $\rho$ and potential $\phi$ are periodic in $[-L/2,L/2)$ and can be expanded in a Fourier series, i.e. 
\begin{equation}
\rho(\vecb{x})=\sum_{\vecb{n}\in\mathbb{Z}^3} \tilde{\rho}_{\vecb{n}}\exp\left({\rm i} k_0\, \vecb{x}\cdot\vecb{n}\right),\quad\textrm{with}\quad k_0:=\frac{2\pi}{L}
\end{equation} 
and identically for $\phi(\vecb{x})$ with coefficients $\tilde{\phi}_{\vecb{n}}$. It then follows from Poisson's equation~(Eq.~\ref{eq:Poisson_equation})  that their Fourier coefficients obey the algebraic relation
\begin{equation}
-k_0^2\left\|\vecb{n}\right\|^2\,\tilde{\phi}_{\vecb{n}} = 4\pi G a^{-1} \left(  \tilde{\rho}_{\vecb{n}} - \overline{\rho}\,\delta_D(\vecb{n}) \right)\quad\textrm{for all}\quad\vecb{n}\in\mathbb{Z}^3.
\end{equation}
This equation imposes the consistency condition $\tilde{\rho}_{\vecb{n}=\vecb{0}}=\overline{\rho}$, i.e. the mean Poisson source {\em must} vanish. In practice, this is achieved in PM codes by explicitly setting to zero the $\vecb{n}=0$ mode (a.k.a. the ``DC mode'', in analogy to AC/DC electric currents). For the acceleration field $\vecb{g} = -\nabla\phi$, one finds $\tilde{\vecb{g}}_{\vecb{n}} = -{\rm i}k_0 \vecb{n} \tilde{\phi}_{\vecb{n}}$. The solution for potential and acceleration can thus be conveniently computed using the Discrete Fourier transform (DFT) as

\begin{equation}
\tilde{\phi}_{\vecb{n}} = \left\{ 
\begin{array}{cl}
-\frac{4\pi G }{a k_0^2} \frac{\tilde{\rho}_{\vecb{n}}}{\|\vecb{n}\|^2}& \quad\textrm{if}\quad\vecb{n}\neq \vecb{0} \\
0 & \quad\textrm{otherwise }
\end{array}
\quad,
\right.
\qquad
\tilde{\vecb{g}}_{\vecb{n}} = \left\{
\begin{array}{cl}
\frac{4\pi G}{a k_0} \frac{{\rm i}\,\vecb{n}\tilde{\rho}_{\vecb{n}}}{\|\vecb{n}\|^2}& \quad\textrm{if}\quad\vecb{n}\neq \vecb{0} \\
0 & \quad\textrm{otherwise }
\end{array}
\right.
.
\end{equation}
If one considers a uniform spatial discretisation of both potential $\phi_{\vecb{m}}:=\phi_{i,j,k} := \phi(\vecb{m} h)$ and density $\rho_{\vecb{m}}$, with $i,j,k\in[0\dots N_g-1]$, mesh index $\vecb{m}:=(i,j,k)^T$, and grid spacing $h:=L/N_g$, then the solution can be directly computed using the Fast-Fourier-Transform (FFT) algorithm at $\mathcal{O}(M\log M)$ for $M=N_g^3$ grid points. Many implementations exist, the {\sc FFTW} library\footnote{Available from \url{https://www.fftw.org}} \citep{FFTW05} is one of the most commonly used with support for multi-threading and MPI. In the case of the DFT, the Fourier sum is truncated at the Nyquist wave number, so that $\vecb{n} \in (-N_g/2,N_g/2]^3$.

Note that instead of the exact Fourier-space Laplacian, $-k_0^2 \| \vecb{n} \|^2$, which is implicitly truncated at the Nyquist wave numbers, sometimes a finite difference version is used in PM codes such as {\sc Fast-PM} \citep{Feng:2016} (c.f. \ref{sec:cola_fastpm}). Inverting the second order accurate finite difference Laplacian in Fourier space yields\footnote{At second order one has $\partial_x^2 f = (\Delta x)^{-2}\left[ \delta_D(x+\Delta x)-2\delta_D(x)+\delta_D(x-\Delta x)\right] \ast f +\mathcal{O}(\Delta x^3)$, and one obtains Eq.~(\ref{eq:discrete_laplace}) by Fourier transformation of the stencil operator. This exercise can be repeated also for higher order stencils, respective expressions are given e.g. in \cite{Hahn:2011}.}
\begin{equation} 
\tilde{\phi}_{\vecb{n}}^{\rm FD2} = \left\{
\begin{array}{cl}
 -\frac{\pi G \Delta x^2 }{a} \;\tilde{\rho}_{\vecb{n}}\;\left( \sin^2\left[  \frac{\pi n_x}{N_g} \right] + \sin^2\left[  \frac{\pi n_y}{N_g} \right] + \sin^2\left[  \frac{\pi n_z}{N_g} \right]\right)^{-1}& \quad\textrm{if}\quad\vecb{n}\neq \vecb{0} \\
0 & \quad\textrm{otherwise. }
\end{array}
\right. \label{eq:discrete_laplace}
\end{equation}

This kernel has substantially suppressed power on small scales compared to the Fourier space Laplacian, which reduces aliasing (see the discussion in the next section). It also reduces the effect of anisotropies due to the mesh on grid scales. 

Solving Poisson's equation in Fourier space with FFTs becomes less efficient if boundary conditions are not periodic, or if spatial adaptivity is necessary. For isolated boundary conditions, the domain has to be zero padded to twice its size per linear dimension, which is an increase in memory by a factor of eight in three dimensions. This is a problem on modern architectures since memory is expensive and slow, while floating-point operations per second (FLOP) are much cheaper to have in comparison. A further problem of FFT methods is their parallelization: a multidimensional FFT requires a global transpose of the array. This leads to a very non-local communication pattern and the need to transfer all of the data multiple times between computer nodes per force calculation.

Additionally, if high resolution is required, as is often the case in cosmology due to the nature of gravity as an attractive force, the size of the grid can quickly become the computational bottleneck. One possibility is to introduce additional higher resolution meshes \citep{Jessop:1994,Suisalu:1995,Pearce:1997,Kravtsov:1997,Teyssier:2002}, deposit particles onto them and then solve using an adaptive ``relaxation method'' such as the adaptive multigrid method (see below), or by employing the periodic FFT solution as a boundary condition. Adaptive algorithms are typically more complex` due to the more complicated data structures involved. 

It is also possible to employ another (or many more) Fourier mesh extended over a particular region of interest in a so-called "zoom simulation", c.f. Sect.~\ref{sec:zoom_simulations},  if higher force resolution is required in a few isolated subregions of the simulation volume.  A problem related of this method is that, for a finite grid resolution, Fourier modes shorter than the Nyquist frequency will be incorrectly aliased to those supported by the Fourier grid \citep{Hockney:1981}, which causes a biased solution to the Poisson equation. The magnitude of this aliasing effect depends on the mass assignment scheme and can be reduced when PM codes are complemented with other force calculation methods, as discussed below in Sect.~\ref{sec:ewald_p3m}, since then the PM force is usually UV truncated. 

Instead of adding a fine mesh on a single region of interest, it is possible to add it {\it everywhere} in space. This approach is known as two-level PM or PMPM, and has been used for carrying out {\sc Cosmo-$\pi$}, the largest $N$-body simulation to date (c.f. Sect.~\ref{sec:state-of-the-art}). This approach has the advantage that, for a cubical domain decomposition, all the operations related to the fine grid can be performed locally, i.e. without communication among nodes in distributed-memory systems, which might result in significant advantages specially when employing hundreds of thousands of computer nodes.

For full phase-space techniques, the PM approach also is preferable if a regular mesh already exists in configuration space onto which the mass distribution can then be easily projected. The Fourier space spectral solution of the Poisson equation can also be readily employed in the case of Schr\"odinger-Poisson discretisations on a regular grid. In this case, the Poisson source is computed from the wave function which is known on the grid, so that $\rho_{\vecb{m}} = \psi_{\vecb{m}} \psi_{\vecb{m}}^\ast$.

%%%%%%%%%%%%%%%%%%%%%%%%%%%%

\subsubsection{Mass assignment schemes}
\label{sec:mass-assignment-schemes}
Grid-based methods always rely on a charge assignment scheme \citep{Hockney:1981} that deposits the mass $m_i$ associated with a particle $i$ at location $\vecb{X}_i$ by interpolating the particle masses in a conservative way to grid point locations $\vecb{x}_{\vecb{n}}$ (where $\vecb{n}\in \mathbb{N}^3$ is a discrete index, such that e.g. $\vecb{x}_{\vecb{n}} = \mathbf{n}\,\Delta x$ in the simplest case of a regular (cubic) grid of spacing $\Delta x$). This gives a charge assignment of the form
\begin{figure}
\begin{center}
\includegraphics[width=\textwidth]{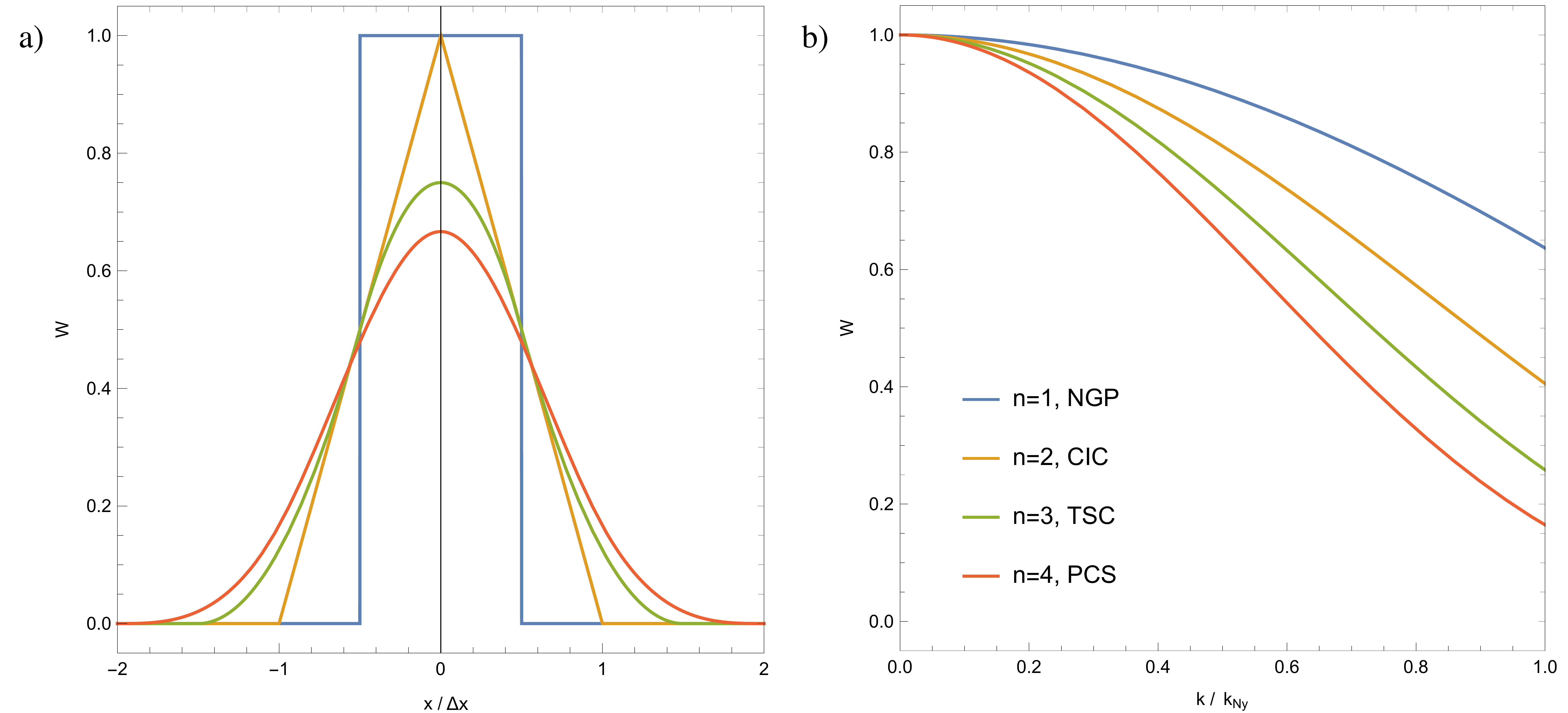} 
\end{center}
\caption{\label{fig:assignment_kernels}Common particle-mesh mass assignment kernels in real space (panels a), and Fourier space (panels b) of increasing order: $n=1$ NGP, $n=2$ CIC, $n=3$ TSC, $n=4$ PCS. Note that the NGP kernel is not continuous, CIC is continuous but not differentiable, TSC is continuously differentiable, and PCS is twice differentiable. The support of the assignment functions is $n\Delta x$ per dimension, and they converge to a normal distribution for $n\to\infty$. Due to their increasing smoothness, they also act as increasingly stronger low-pass filters.}
\end{figure}
\begin{equation}
\rho_{\vecb{n}} = \int_{\mathbb{R}^3} {\rm d}^3x'\,\hat{\rho}(\vecb{x}') \,W_{3D}(\vecb{n}\,\Delta x-\vecb{x}')\quad\textrm{with}\quad\hat{\rho}(\vecb{x}):=\sum_{i=1}^N M_i \delta_D(\vecb{x}-\vecb{X}_i),
\end{equation}
where the periodic copies in the density were dropped since periodic boundary conditions are assumed in the Poisson solver. Charge assignment to a regular mesh is equivalent to a single convolution if $M_i=M$ is identical for all particles. The most common particle-grid interpolation functions (cf.\ \citealt{Hockney:1981}) of increasing order are given for each spatial dimension by\footnote{The abbreviations commonly used to describe the different orders are owed to history: NGP = nearest grid point, CIC = cloud in cell, TSC = triangular shaped cloud, PCS = piecewise cubic spline. It would be more useful to simply call them ``$n$th order assignment kernels''.}
\begin{subequations}
\begin{eqnarray}
W_{\rm NGP}(x) &=& \frac{1}{h}\left\{ \begin{array}{ll} 
1 & \quad \textrm{for }\left| x \right| \le \frac{\Delta x}{2}\\
0 & \quad \textrm{otherwise}
\end{array}\right. \\
W_{\rm CIC}(x) &=& \frac{1}{h}\left\{ \begin{array}{ll} 
1-\frac{\left| x\right|}{\Delta x} &\quad  \textrm{for }\left|x\right| < \Delta x \\
0 &\quad  \textrm{otherwise}
\end{array}\right. \label{eq:cic_kernel}\\
W_{\rm TSC}(x) &=& \frac{1}{\Delta x}\left\{ \begin{array}{ll} 
\frac{3}{4} - \left( \frac{x}{\Delta x}\right)^2 & \quad \textrm{for }\left|x\right|\le \frac{\Delta x}{2}\\
\frac{1}{2}\left( \frac{3}{2} - \frac{\left|x\right|}{\Delta x}\right)^2 & \quad \textrm{for }\frac{\Delta x}{2}\le \left| x \right| < \frac{3\Delta x}{2}\\
0 & \quad \textrm{otherwise}
\end{array}\right. \\
W_{\rm PCS}(x) &=& \frac{1}{\Delta x}\left\{ \begin{array}{ll} 
\frac{1}{6} \left[ 4 - 6\left(\frac{x}{\Delta x}\right)^2  + 3 \left(\frac{|x|}{\Delta x}\right)^3 \right] & \quad \textrm{for }\left|x\right|\le \Delta x\\
\frac{1}{6}\left( 2 - \frac{\left|x\right|}{\Delta x}\right)^3 & \quad \textrm{for } \Delta x \le |x| < 2 \Delta x\\
0 & \quad \textrm{otherwise}
\end{array}\right.
\end{eqnarray}
\end{subequations}
The three-dimensional assignment function is then just the product $W_{3D}(\vecb{x})=W(x)\,W(y)\,W(z)$, where $\vecb{x}=(x,y,z)^T$. It can be easily shown that interpolating using the inverse of these operators from a field increases the regularity of the particle density field, and thus also has a smoothing effect on the resulting effective gravitational force. This can also be seen directly from the Fourier transform of the assignment functions which have the form (per dimension)
\begin{equation}
%\widetilde{W}(k) = h \, \left[  \frac{\sin\left( k/k_0\right)}{\left( k/k_0\right)} \right]^n,
\tilde{W}_{n}(k) = \left[{\rm sinc }\frac{\pi}{2}\frac{k}{k_{\rm Ny}}\right]^n\quad\textrm{with}\quad{\rm sinc}\,x = \frac{\sin x}{x}.
\end{equation}
where $n=1$ for NGP, $n=2$ for CIC, $n=3$ for TSC, $n=4$ for PCS interpolation, and $k_{\rm Ny}:=\pi/\Delta x$ is the Nyquist wave number. NGP leads to a piecewise constant, CIC to a piece-wise linear, TSC to a piecewise quadratically (i.e., continuous value and first derivative), and PCS to piecewise cubically changing acceleration as a particle moves between grid points. The real space and Fourier space shape of the kernels is shown in Fig.~\ref{fig:assignment_kernels}. Note that the support is always $n \Delta x$, i.e. $n$ cells, per dimension and thus increases with the order, and by the central limit theorem $\tilde{W}_n$ converges to a normal distribution as $n\to\infty$. Hence, going to higher order can impact memory locality and communication ghost zones negatively. Since an a priori unknown number of particles might deposit to the same grid cell, special care needs to be taken to make the particle projection thread safe in shared-memory parallelism \citep{Ferrell:1994}.

Alternatively to these mass assignment kernels for particles, it is possible to project phase-space tessellated particle distributions (cf.\ Sect.~\ref{sec:tessellation_methods}) exactly onto the force grid \citep{Powell:2015,Sousbie:2016}. In practice, when using such sheet tessellation methods, for a given set of flow tracers, the phase-space interpolation can be constructed and sampled with $M$ "mass carrying" particles which can then be deposited into the grid. Since the creation of mass carriers is a local operation, $M$ can be arbitrarily large and thus the noise associated to $N$-body discreteness can be reduced systematically. This approach has been adopted by \cite{Hahn:2013} and \cite{Angulo:2013b} to simulate warm dark matter while suppressing the formation of {\it artificial fragmentation}, as we will discuss in greater detail in Sect.~\ref{sec:wdm}

The same mass assignment schemes can be used to reversely interpolate values of a discrete field back to the particle positions $\left\{\vecb{X}_i\right\}$ by inverting the projection kernel. It has to be ensured that the same order is used for both mass deposit and interpolation of the force to the particle positions, i.e., that deposit and interpolation are mutually inverse. This is an important consistency since, otherwise, (1) exact momentum conservation is not guaranteed, and (2) self-forces can occur allowing particles to accelerate themselves (cf.\ \citealt{Hockney:1981}). It is important to note that due to the grid discretisation, particle separations that are unresolved by the discrete grid are aliased to the wrong wave numbers, which e.g. can cause certain Fourier modes to grow at the wrong rate. Aliasing can be ameliorated by filtering out scales close to the Nyquist frequency, or by using interlacing techniques where by combination of multiple shifted deposits individual aliasing contributions can be cancelled at leading order \citep{Chen:1974,Hockney:1981}. Such techniques are important also when estimating Fourier-space statistics (i.e., poly-spectra) from density fields obtained using above deposit techniques (see Sect.~\ref{sec:analysis} for a discussion).

\subsubsection{Relaxation methods and multi-scale}
\label{sec:multigrid}
In order to overcome the limitations of Fourier space solvers (in particular, the large cost of the global transpose on all data necessary along with the lack of spatial adaptivity), a range of other methods have been developed. The requirement is that the Poisson source is known on a grid, which can  also be an adaptively refined `AMR' grid structure. On the grid, a finite difference version of the Poisson equation is then solved, e.g., for a second-order approximation in three dimension the solution is given by the finite difference equation:
\begin{equation}
\phi_{i-1,j,k}+\phi_{i+1,j,k}+\phi_{i,j-1,k}+\phi_{i,j+1,k}+\phi_{i,j,k-1}+\phi_{i,j,k+1}-6\phi_{i,j,k} = \Delta x^2\,f_{i,j,k} \,,
\end{equation}
where indices refer to grid point locations as above, $\Delta x$ is the grid spacing, and $f_{i,j,k} := 4\pi G (\rho_{i,j,k}-\overline{\rho})/a$ is the Poisson source. This can effectively be written as a matrix inversion problem $\tnsr{A} \phi = f$ where the finite difference stencil gives rise to a sparse matrix $\tnsr{A}$ and the solution sought is $\phi=\tnsr{A}^{-1}f$. Efficient methods exist to solve such equations. A particularly powerful one, that can directly operate even on an AMR structure, is the adaptive multigrid method \citep{Brandt:1977,Trottenberg:2001}, which is used e.g., by the {\sc RAMSES} code \citep{Teyssier:2002}. It combines simple point relaxation (e.g., Jacobi or Gauss--Seidel iterations) with a hierarchical coarsening procedure which spreads the residual correction exponentially fast across the domain. Some additional care is required at the boundaries of adaptively refined regions. Here the resolution of the mesh changes, typically by a linear factor of two, and interpolation from the coarser grid to the ghost zones of the fine grid is required. In the one-way interface type of solvers, the coarse solution is obtained independently of the finer grid, and then interpolated to the finer grid ghost zones to serve as the boundary condition for the fine solution \citep{Guillet:2011}, but no update of the coarse solution is made based on the fine solution. This approach is particularly convenient for block-stepping schemes (cf.\ Sect.~\ref{sec:block_timestep}) where each level of the grid hierarchy has its own time step by solving e.g. twice on the fine level while solving only once on the coarse. A limitation of AMR grids is however that the force resolution can only change discontinuously by the refinement factor, both in time -- if one wants to achieve a resolution that is constant in physical coordinates -- and in space -- as a particle moves across coarse-fine boundaries. On the other hand, AMR grids contain self-consistently an adaptive force softening (see Sect.~\ref{sec:softening}), if the refinement strategy is tied to the local density or other estimators \citep{Hobbs:2016}.

Depending on the fragmentation of the finer levels due to the dynamic adaptivity, other solvers can be more efficient than multigrid, such as  direct relaxation solvers \citep{Kravtsov:1997} or conjugate gradient methods. However, it is in principle more accurate to account for the two-way interface and allow for a correction of the coarse potential from the fine grid as well, as discussed e.g. by \cite{Johansen:1998,Miniati:2007}. Note that, once a deep grid hierarchy has developed, global Poisson solves in each fine time step are usually prohibitive for numerical algorithms. For this reason optimizations are often employed to solve for the gravitational acceleration of only a subset of particles in multi-stepping schemes. In the case of AMR, some care is necessary to interpolate boundary conditions also in time to avoid possible spurious self-interactions of particles.

%%%%%%%%%%%%%%%%%%%%%%%%%%%%%%%%%%%%%%%%%%%%%%%%%%%%%%%%%
\subsection{Direct P2P Summation}
\label{sec:direct_summation}

As discussed above, mesh-based methods bring along an additional discretisation of space. This can be avoided by computing interactions directly at the particle level from Eqs.~(\ref{eq:nbody_poisson}-\ref{eq:nbody_acc}). In this case, the gravitational potential at particle $i$'s location, $\vecb{X_i}$, is given by the sum over the contribution of all the other particles in the system along with all periodic replicas of the finite box, i.e.
\begin{equation}
%\label{eq:force}
\label{eq:potential_periodic}
\phi(\vecb{x}_i) = - a^{-1} \sum_{\vecb{n}\in\mathbb{Z}^3} \left[ {\sum_{\substack{j=1\\ {i\neq j}}}^N}\frac{G M_j}{\|\vecb{X}_i-\vecb{X}_j-\vecb{n}L \|} + \varphi_{{\rm box},L}(\vecb{X}_i-\vecb{n}L)\right]. 
\end{equation}
Note that we neglected force softening for the moment, i.e. we set $W(\vecb{x})=\delta_D(\vecb{x})$. Here $\varphi_{{\rm box},L}$ is the potential due to a box $[0,L)^3$ of uniform background density $\overline{\rho}=\Omega_m\rho_c$ that guarantees that the density $\rho-\overline{\rho}$ sourcing $\phi$ vanishes when integrated over the box.

This double sum is slowly convergent with respect to $\vecb{n}$, and in general there can be spurious forces arising from a finite truncation (but note that the sum is unconditionally convergent if the box has no dipole, e.g., \citealt{Ballenegger:2014}). A fast and exact way to compute this expression is provided by means of an \textit{Ewald summation} \citep{Ewald:1921}, in which the sum is replaced by two independent sums, one in Fourier space for the periodic long-range contribution, and one in real space, for the non-periodic local contribution, which both converge rapidly. It is then possible to rewrite Eq.~(\ref{eq:potential_periodic}) employing the position of the nearest replica, which results into pairwise interactions with a modified gravitational potential. This potential needs to be computed numerically, thus, in {\sc GADGET3}, for instance, it is tabulated and then interpolated at runtime, whereas in {\sc GADGET4}, the code relies on a look-up table of a Taylor expansion with analytic derivatives of the Ewald potential. We summarise in more detail how this is achieved in Sect.~\ref{sec:ewald_p3m}, where we discuss in particular how the FFT can be efficiently used to execute the Fourier summation.

This direct summation of individual particle-particle forces is $\mathcal{O}(N^2)$, that is, quadratic in the number of particles and thus becomes quickly computationally prohibitive. In addition, since it is a highly non-local operation, it would require a considerable amount of inter-process communication. In practice, this method is sometimes used to compute short-range interactions, where the operation is local and can exploit the large computational power provided by GPUs. This is, for instance, the approach followed by the {\sc HACC} code \citep{Habib:2016}, when running one of the largest simulations to date with 3.6 trillion particles; and also by the {\sc ABACUS} code \citep{Garrison:2018}. Direct summation enabled by GPUs has also been adopted by \cite{Racz:2019} for {\it compactified} simulations, where there is an additional advantage that only a small subset of the volume has to be followed down to $z=0$ (cf.\ Sect.~\ref{sec:supersample}).

%%%%%%%%%%%%%%%%%%%%%%%%%%%%%%%%%%%%%%%%%%%%%%%%%%%%%%%%%
\subsection{Particle mesh Ewald summation, force splitting and the \texorpdfstring{P$^3$M}{P3M} method}
\label{sec:ewald_p3m}
Beyond the poor $\mathcal{O}(N^2)$ scaling of the direct P2P summation (for which we discuss the solutions below), another important limitation of the na\"ive infinite direct summation is the infinite periodic contribution in Eq.~(\ref{eq:potential_periodic}). At the root of the solution is the Ewald summation \citep{Ewald:1921}, first used for cosmological simulations by \cite{Bouchet:1988}, in which the total potential or acceleration is split into a short and a long range contribution, and where the short range contribution is summed in real space, while the long range contribution is summed in Fourier space where it converges due to its periodic character much faster. One thus introduces a `splitting kernel' $\mathcal{S}$ so that

\begin{equation}
\phi(\vecb{x}) = \phi_{\rm lr}(\vecb{x})+ \phi_{\rm sr}(\vecb{x}) := \mathcal{S}\ast \phi + (1-\mathcal{S})\ast\phi.
\end{equation}

The long-range contribution $\phi_{\rm lr}$ can be computed using the PM method on a relatively coarse mesh. On the other hand, the short-range contribution $\phi_{\rm sr}$, can be computed from the direct force between particles only in their immediate vicinity -- since the particles further away contribute through the PM part. Instead of the direct force, which gives then rise to the P$^3$M method, modern codes often use a tree-method (see next section) for the short range force (this is e.g., what is implemented in the {\sc GADGET2} code by \cite{Springel:2005b}, see also \citealt{WangFMM:2020}).

The splitting kernel effectively spreads the mass over a finite scale $r_s$ for the long range interaction, and corrects for the residual with the short range interaction on scales $\lesssim r_s$. Many choices are a priori possible, \cite{Hockney:1981}, e.g., propose a sphere of uniformly decreasing density, or a Gaussian cloud. The latter is, e.g., used in the {\sc GADGET} codes.

In terms of the Green's function of the Laplacian $\mathcal{G}(\vecb{r}) = -1/(4\pi\|\vecb{r}\|)$, the formal solution for the cosmological Poisson equation reads $\phi = \frac{4\pi G}{a} \left(\rho-\overline{\rho}\right) \ast \mathcal{G}$. For a Gaussian cloud of scale $r_s$, one has in real and Fourier space
\begin{equation}
\mathcal{S}(r; r_s) = (2\pi r_s^2)^{-3/2} \exp\left( -\frac{r^2}{2r_s^2} \right),\qquad \tilde{\mathcal{S}}(k; r_s) = \exp\left[-\frac{1}{2}k^2 r_s^2\right].
\end{equation}
The `dressed' Green's functions $\mathcal{G}_{\rm lr} := \mathcal{G}\ast S$ and $\mathcal{G}_{\rm sr} := \mathcal{G}\ast(1-\mathcal{S})$ then become explicitly in real and Fourier space  % in Fourier and real space obtained by multiplication or convolution respectively, i.e.
\begin{subequations}
\begin{alignat}{3}
  \mathcal{G}_{\rm lr}(r; r_s) &= - \frac{1 }{4\pi \,r} \,{\rm erf\left[ \frac{r}{\sqrt{2}r_s} \right]}, \qquad& \tilde{\mathcal{G}}_{\rm lr}(k; r_s) &= -\frac{1}{k^2}\exp\left[-\frac{1}{2}k^2r_s^2 \right], \\
  \mathcal{G}_{\rm sr}(r; r_s) &= - \frac{1 }{4\pi \,r} \,{\rm erfc\left[ \frac{r}{\sqrt{2}r_s} \right]},&  \tilde{\mathcal{G}}_{\rm sr}(k; r_s) &= -\frac{1}{k^2}\left(1-\exp\left[-\frac{1}{2}k^2r_s^2 \right]\right).
\end{alignat}
\end{subequations}
Instead of the normal Green's functions, one thus simply uses these truncated functions and obtains a hybrid solver. In order to use this approach, one chooses a transition scale of order the grid scale, $r_s\sim \Delta x$, and then replaces the PM Green's function with $\mathcal{G}_{\rm lr}$. Instead of the particle-particle interaction in the direct summation or tree force (see below), one uses $\mathcal{G}_{\rm sr}$ for the potential, and $\boldsymbol{\nabla} \mathcal{G}_{\rm sr}$ for the force.

While the long range interaction already includes the periodic Ewald summation component if solved with Fourier space methods, when evaluating the periodic replica summation for the short-range interaction,  the evaluation can be restricted to the nearest replica in practice due to the rapid convergence with the regulated interaction. In addition, since PM forces are exponentially suppressed on scales comparable to $r_s$ which is chosen to be close to the grid spacing $\Delta x$, aliasing of Fourier modes is suppressed.

Note that another more aggressive near-far field combination is adopted by the {\sc ABACUS} code. In this approach, the computational domain is first split into a uniform grid with $K^3$ cells. Interactions of particles separated by less than approximately $2 L/K$ are computed using direct summation (neglecting Ewald corrections); otherwise are computed using a high-order multipole ($p=8$) representation of the force field in the $K-$grid. Since two particles only interact via either the near- or far-field forces, and the tree structure is fixed to the $K$-grid, this allows for several optimizations and out-of-core computations. The price is discontinuous force errors with a non-trivial spatial dependence, as well as reduced accuracy due to the lack of Ewald corrections. This, however, might be acceptable for some applications and, as we will see in Sect.~\ref{sec:code_comparison}, {\sc ABACUS} performs well when compared to other state-of-the-art codes.

%%%%%%%%%%%%%%%%%%%%%%%%%%%%%%%%%%%%%%%%%%%%%%%%%%%%%%%%%

\subsection{Hierarchical tree methods}
\label{sec:tree}

Assuming that it is acceptable to compute gravitational forces with a given specified accuracy, there are ways to circumvent the $\mathcal{O}(N^2)$ and non-locality problem of direct summation. A common approach is to employ a hierarchical tree structure to partition the mass distribution in space and compute the gravitational potential jointly exerted by groups of particles, whose potential is expanded to a given multipole order \citep{BarnesHut:1986}. Thus, instead of particle-particle interactions, particle-node interactions are evaluated. Since the depth of such a tree is typically $\mathcal{O}(\log N)$, the complexity of the evaluation of all interactions can be reduced to $\mathcal{O}(N\log N)$. This can be further reduced to an ideal $\mathcal{O}(N)$ complexity with the fast multipole method (FMM, see below).

There are several alternatives for constructing tree structures. The most common choice is a regular octree in which each tree level is subdivided into 8 sub-cells of equal volume, this is for instance used by {\sc GADGET}. Another alternative, used for instance in old versions of {\sc PKDGRAV} are binary trees in which a node is split into only two daughter cells. This in principle has the advantage to adapt more easily to anisotropic domains, and a smoother transition among levels, at the expense of a higher cost in walking the tree or the need to go to higher order multipole expansions at fixed force error. The tree subdivision continues until a maximum number $M$ of particles per node is reached ($M=1$ in {\sc GADGET2-3} but higher in {\sc GADGET4} and {\sc PKDGRAV}).

The main advantage brought by tree methods is that the pairwise interaction can be expanded perturbatively and grouped among particles at similar locations, thus reducing dramatically the number of calculations that needs to be carried out. The key philosophical difference with respect to direct summation is that one seeks to obtain the result at a desired accuracy, rather than the exact result to machine precision. This difference allows a dramatic improvement in algorithmic complexity. Another key aspect is that hierarchical trees are well suited for hierarchical (adaptive) timesteps. 

Tree methods have for a long time been extraordinarily popular for evaluating the short range interactions also in hybrid tree-PM methods, as pioneered by \cite{Bagla:2002,Bagla:2003}, or more recent FMM-PM \citep{Gnedin:2019,WangFMM:2020,Springel:2020} approaches, thus supplementing an efficient method for periodic long-range interactions with an efficient method which is not limited to the uniform coarse resolution of FFT-based approaches (or also discrete jumps in resolution of AMR approaches). We discuss some technical aspects of these methods next.

\subsubsection{Hierarchical multipole expansion} 

In the `Barnes~\&~Hut tree' algorithm \citep{Appel:1985,BarnesHut:1986}, particle-node interactions are evaluated instead of particle-particle interactions. Let us consider a hierarchical octree decomposition of the simulation box volume $\mathcal{V}:=[0,L_{\rm box}]^3$ at level $\ell$ into cubical subvolumes, dubbed `nodes', $\mathcal{S}^\ell_{i=1\dots N_\ell}$ of side length $L_{\rm box}/2^\ell$, where $N_\ell=2^{3\ell}$, so that $\bigcup_i \mathcal{S}^\ell_i = \mathcal{V}$ and $\mathcal{S}^\ell_i\cap \mathcal{S}^\ell_{j\neq i} = \emptyset$ on each level gives a space partitioning. Let us consider the gravitational potential due to all particles contained in a node $\vecb{X}_j\in S^\ell_i$. The partitioning is halted when only one (but typically a few) particle is left in a node.  We shall assume isolated boundary conditions for clarity, i.e. we neglect the periodic sum in Eq.~(\ref{eq:potential_periodic}). Thanks to the partitioning, the gravitational interaction can be effectively localised with respect to the `tree node' pivot at location $\vecb{\lambda}\in \mathcal{S}^\ell_i$, so that the distance $\| \vecb{X}_j - \vecb{\lambda} \| \le \sqrt{3} L_{\rm box}/2^\ell =: r_\ell$ is by definition bounded by the `node size` $r_\ell$ and can serve as an expansion parameter. To this end, one re-writes the potential due to the particles in the node subvolume $\mathcal{S}^\ell_i$
\begin{equation}
\phi^\ell_i(\vecb{x}) \propto \sum_{\vecb{X}_j\in\mathcal{S}_i^\ell} \frac{M_j}{\|\vecb{x}-\vecb{X}_j\|} =  \sum_{\vecb{X}_j\in\mathcal{S}_i^\ell} \frac{M_j}{\|(\vecb{x}-\vecb{\lambda})-(\vecb{X}_j-\vecb{\lambda})\|} = \sum_{\vecb{X}_j\in\mathcal{S}_i^\ell} \frac{M_j}{\|\vecb{d}+\vecb{\lambda}-\vecb{X}_j\|}  \label{eq:node_expansion}
\end{equation}
where $\vecb{d}:=\vecb{x}-\vecb{\lambda}$. This can be Taylor expanded to yield the `P2M' (particle-to-multipole) kernels

\begin{align}
  \begin{split}
\frac{1}{\|\vecb{d}+\vecb{\lambda}-\vecb{X}_j\|} = &\underbrace{\frac{1}{\|\vecb{d}\|}}_{\textrm{monopole}} + \underbrace{\frac{d_k}{\|\vecb{d}\|^3} \left(X_{j,k}-\lambda_k\right)}_{\textrm{dipole}\; \mathcal{O}(r_\ell/d^2)} + \\
&+ \underbrace{\frac{1}{2}\frac{d_kd_l}{\| \vecb{d} \|^5} \left( 3(X_{j,k}-\lambda_k)(X_{j,l}-\lambda_l) -\delta_{kl} \|\vecb{X}_j-\vecb{\lambda} \|^2  \right)}_{\textrm{quadrupole}\;\mathcal{O}(r_\ell^2/d^3)} 
%\underbrace{\frac{1}{2} \vecb{d}^T \frac{3(\vecb{X}_j-\vecb{\lambda})(\vecb{X}_j-\vecb{\lambda})^T - \tnsr{I}(\vecb{X}_j-\vecb{\lambda})^2}{\|\vecb{d}\|^5}\vecb{d}}_{\textrm{quadrupole}}
+\dots,
  \end{split}\label{eq:multipole_expansion}
\end{align}
which converges quickly if $\|\vecb{d}\| \gg r_\ell$. The multipole moments depend only on the vectors $(\vecb{X}_j-\vecb{\lambda})$ and can be pre-computed up to a desired maximum order $p$ during the tree construction and stored with each node. In doing this, one can exploit that multipole moments are best constructed bottom-up, as they can be translated in an upward-sweep to the parent pivot and then co-added -- this yields an `upwards M2M' (multipole-to-multipole) sweep. Note that if one sets $\vecb{\lambda}$ to be the centre of mass of each tree node, then the dipole moment vanishes. The complexity of such a tree construction is $\mathcal{O}(N\log N)$ for $N$ particles.

When evaluating the potential $\phi(\vecb{x})$ one now proceeds top-down from the root node at $\ell=0$ in a `tree walk' and evaluates M2P (multipole-to-particle) interactions between the given particle and the node. Since one knows that the error in $\phi^\ell_i(\vecb{x})$ is $\mathcal{O}\left( (r_\ell/d)^p \right)$, one defines a maximum  `opening angle' $\theta_{\rm c}$ and requires in order for the multipole expansion $\phi^\ell_i(\vecb{x})$ to be an acceptable approximation for the potential due to the mass distribution in $\mathcal{S}^\ell_i$ that the respective opening angle obeys
\begin{equation}
 \frac{r_\ell}{\|\vecb{d}\|} <\theta_{\rm c}.
 \label{eq:openning}
\end{equation}
Otherwise the procedure is recursively repeated with each of the eight child nodes. Since the depth of a (balanced)  octree built from a distribution of $N$ particles is typically $\mathcal{O}(\log N)$, a full potential or force calculation has an algorithmic complexity of $\mathcal{O}(N\log N)$ instead of the $\mathcal{O}(N^2)$ of the direct summation. The resulting relative error in a node-particle interaction is \citep{Dehnen:2002}
\begin{equation}
\delta \phi \leq \frac{\theta_c^{p+1}}{1-\theta_c} \frac{M_{\rm node}}{\|\mathbf{d}\|}, 
\end{equation}
where $M_{\rm node}$ is the node mass (i.e. the sum of the masses of all particles in $\mathcal{S}^\ell_i$), and $p$ is the order of the multipole expansion. Eq.~\ref{eq:openning} error estimate is a purely geometric criterion, independent of the magnitude of $M_{\rm node}$ and the multipole moments, as well as the actual value of the gravitational acceleration. It is also independent of the magnitude of the interaction, i.e. neglecting that far nodes contribute more than nearby ones to the total interaction. 

An alternative method, proposed by \cite{Springel:2001b}, is to use a dynamical criterion by comparing the expected acceleration with the force error induced by a given node interaction. Specifically, when evaluating the particle-node interactions for particle $j$ one sets
\begin{equation}
\theta_{{\rm c},j} = \left(\alpha \|\vecb{A}_j\| \frac{\|\vecb{d}\|^2}{G M_{\rm node}}\right)^{1/p},
\end{equation}
where $\|\vecb{A}_j\|$ is the modulus of the gravitational acceleration (which could be estimated from the force calculation performed in a previous step), and $\alpha$ is a dimensionless parameter that controls the desired accuracy. Note, however, that for relatively isotropic mass distributions, the uncertainty of a given interaction might not be representative of the uncertainty in the total acceleration.

We highlight that the expressions (\ref{eq:node_expansion})-(\ref{eq:multipole_expansion}) are valid for the non-periodic particle-node interactions, but for periodic boundary conditions additional terms arise owing to the modified Green's function as seen in Eq.~(\ref{eq:potential_periodic}). The Green's function is also modified in the case when tree interactions are combined with other methods such as PM in a tree-PM method (see Sect.~\ref{sec:ewald_p3m}). This implies in principle also modified error criteria (or opening angles), however, this is often neglected.

So far, performing the multipole expansion only to monopole order (with nodes centered at the center of mass) has been a popular choice for $N$-body codes. The reason behind this is that a second-order accurate expression is obtained with very low memory requirements (one needs to simply store the centre of mass of tree nodes instead of the geometric centre, which is enough when moderate accuracy is sought. However, in search of higher accuracy, a larger number of codes have started to also consider quadrupole and octupole terms, which requires more memory and computation but allows a less aggressive opening criteria. This has been advocated as the optimal combination that provides the most accurate estimate at a fixed computational cost \citep{Dehnen:2011,Potter:2016}, although the precise optimal order depends on the required accuracy \citep{Springel:2020}. In the future, further gains from higher order terms can be obtained as computer architectures evolve towards higher FLOP/byte ratios.

A problem for tree codes used for cosmological simulations is that on large scales and/or at high redshift the mass distribution is very homogeneous. This is a problem since the net acceleration of a particle is then the sum of many terms of similar magnitude but opposite sign that mostly cancel. Thus, obtaining accurate forces requires a low tolerance error which increases the computational cost of a simulation. For instance, the {\sc Euclid Flagship} simulation \citep{Potter:2016}, which  employed a pure Tree algorithm (c.f. Sect.~\ref{sec:state-of-the-art}), spends a considerable amount of time on the gravitational evolution at high redshift. Naturally, this problem is exacerbated the larger the simulation and the higher the starting redshift. 

A method to address this problem, that was proposed by \cite{Warren:2013} and implemented in the {\sc 2HOT} code, is known as `background subtraction''. The main idea is to add the multipole expansion of a local uniform negative density to each interaction, which can be computed analytically for each cubic cell in a particle-node interaction. Although this adds computational cost to the force calculation, it results in an important overall reduction of the cost of a simulation since many more interactions can be represented by multipole approximations at high redshift. As far as we know, this has not been widely adopted by other codes.

A further optimization that is usually worth carrying out on modern architectures is to prevent tree refinement down to single particles (for which anyway all multipoles beyond the monopole vanish). Since the most local interactions end up being effectively direct summation anyway, one can get rid of the tree overhead and retain a `bucket' of $10^{2-3}$ particles in each leaf node rather than a single individual particle. All interactions within the node, as well as those which would open child nodes are carried out in direct summation. While algorithmically more complex, such a direct summation is memory-local and can be highly optimized and e.g. offloaded to GPUs, providing significant speed-up over the tree.

%%%%%%%%%%%%%%%%%%%%%%%%%%%%%%%%%%%%%%%%%%%%%%%%%%%%%%%%%

\subsubsection{Fast-Multipole Method}
\label{sec:FMM}

Despite the huge advantage with respect to direct summation, a single interaction of a particle with the tree is still computationally expensive as it has a $\mathcal{O}(\log N)$ complexity for a well-balanced tree. Furthermore, trees as described above have other disadvantages, for instance, gravitational interactions are not strictly symmetric. This leads to a violation of momentum conservation. A solution to these limitations is provided by fast multipole methods (FMM), originally proposed by \cite{Greengard:1987} and extended to Cartesian coordinates by \cite{Dehnen:2000,Dehnen:2002}. These algorithms take the idea of hierarchical expansions one step further by realising that significant parts of the particle-node interactions are redundantly executed for particles that are within the same node. In order to achieve a $\mathcal{O}(1)$ complexity per particle, the node-node interaction should be known and translated to the particle location. This is precisely what FMM achieves by symmetrising the interaction to node-node interactions between well-separated nodes, which are separately Taylor expanded inside of the two nodes. Up to recently, FMM methods have not been widespread in cosmology, presumably due to a combination of higher algorithmic and parallelization complexity. The advantages of FMM are becoming evident in modern $N$-body codes, which simulate extremely large numbers of particles and seek high accuracy, and thus FMM has been adopted in {\sc PKDGRAV}, {\sc GADGET-4}, and {\sc SWIFT}. We only briefly summarize the main steps of the FMM algorithm here, and refer the reader to the reviews by, e.g., \cite{Kurzak:2006,Dehnen:2011} for details on the method.

The FMM method builds on the same hierarchical space decomposition as the Barnes\&Hut tree above and shares some operators. For the FMM algorithm, three steps are missing from the tree algorithm outlined in the previous section: a `downward M2L' (multipole-to-local) sweep,  which propagates the interactions back down the tree after the upward M2M sweep, thereby computing a local field expansion in the node. This expansion is then shifted in `downward L2L' (local-to-local) steps to the centers of the child nodes, and to the particles in a final `L2P' (local-to-particle) translation. As one has to rely on the quality of the local expansion in each node, FMM requires significantly higher order multipole expansions compared to standard Barnes\&Hut trees to achieve low errors. Note that for a Cartesian expansion in monomials $x^ly^mz^n$ at a fixed order $p=l+m+n$, one has $(p+1)(p+2)/2$ multipole moments, i.e. $(p+1)(p+2)(p+3)/6$ for all orders up to incl. $p$, i.e. memory needed for each node scales as $\mathcal{O}(p^3)$, and a standard implementation evaluating multipole pair interactions scales as $\mathcal{O}(p^6)$. For expansions in spherical harmonics, one can achieve $\mathcal{O}(p^3)$ scaling \citep{Dehnen:2014}. Note that for higher order expansions one can rely on known recursion relations to obtain the kernel coefficients \citep{Visscher:2010} allowing arbitrary order implementations. Recently, it was demonstrated that a trace-free reformulation of the Cartesian expansion has a slimmer memory footprint \citep{Coles:2020} (better than 50\% for $p\ge8$). The same authors provide convenient {\sc Python} scripts to auto-generate code for optimized expressions of the FMM operators symbolically. It is important to note that the higher algorithmic complexity lends itself well to recent architectures which favour high FLOP-to-byte ratio algorithms \citep{Yokota:2012}.

While the FMM force is symmetric, \cite{Springel:2020} report however that force errors can be much less uniform in FMM than in a standard tree approach, so that it might be required to randomise the relative position of the expansion tree w.r.t. the particles between time steps in order to suppress the effect of correlated force errors on sensitive statistics for cosmology. In principle also isotropy could be further improved with random rotations. Note that errors might have a different spatial structure with different expansion bases however. 

The FMM method indeed has constant time force evaluation complexity for each $N$-body particle. This assumes that the tree has already been built, or that building the tree does not have $\mathcal{O}(N\log N)$ complexity (which is only true if it is not fully refined but truncated at a fixed scale). Note however that for FMM solvers, it is preferable to limit the tree depth to a minimum node size or at least use a larger number of particles in a leaf cell for which local interactions are computed by direct `P2P' (particle-to-particle) interactions. Also, tree construction has typically a much lower pre-factor than the `tree walk'. Note further that many codes use some degree of `tree updating' in order to avoid rebuilding the tree in every timestep.

In order to avoid explicit Ewald summation, some recent methods employ hybrid FFT-FMM methods, where essentially a PM method is used to evaluate the periodic long range interactions as in tree-PM and the FMM method is used to increase the resolution beyond the PM mesh  for short-range interactions \citep{Gnedin:2019,Springel:2020}.

%% Initial Conditions
%% \newpage 

\section{Initial conditions}
\label{sec:ICs}

In previous sections we have discussed how to discretise a cold collisionless fluid, compute its self-gravity, and evolve it in time. Closing our review of the main numerical techniques for cosmological simulations, in this section we present details on how to compute and set up their initial conditions.

The complicated non-linear structures that are produced at late times in cosmological simulations develop from minute fluctuations around homogeneity in the early Universe that are amplified by gravitational collapse. While the fluctuations remain small (i.e. at early times, or on large scales) they permit a perturbative treatment of the underlying non-linear coupled equations. Such perturbative techniques belong to the standard repertoire of analytic techniques for the study of the cosmic large-scale structure, see e.g., \cite{Bernardeau:2002} for a review, and Sect.~\ref{sec:coldlimit_pt} for a concise summary. At late times and on smaller scales, shell-crossing and deeply non-linear dynamics limit the applicability of perturbative techniques. While some attempts have been made to extend PT beyond shell-crossing \citep{Taruya:2017,Pietroni:2018,Rampf:2019}, or by controlling such effects in effective fluid approaches, e.g., \cite{Baumann:2012}, the evolved non-linear universe is still the domain of simulations. At the same time, PT of various flavours is used to set up the fields that provide the initial conditions for fully non-linear cosmological simulations.
%%%%%%%%%%%%%%%%%%%%%%%%%%%%%%%%%%%%%%%%%%%%%%%%%%%%%%%%%%

\subsection{Connecting simulations with perturbation theory}
\label{sec:sims_and_pt}
The physics governing the infant phase of the Universe, that is dominated by a hot plasma tightly coupled to radiation and linked through gravity to dark matter and neutrinos, is considerably more complex than the purely geodesic evolution of collisionless gravitationally interacting paricles outlined in Sect.~\ref{sec:gr}. Since density fluctuations are small, this phase can be treated accurately by perturbative techniques at leading order. State-of-the-art linear-order Einstein-Boltzmann (EB) codes that numerically integrate these coupled multi-physics systems are e.g., {\sc Camb} \citep{Lewis:1999bs} and {\sc Class} \citep{Lesgourgues:2011,Blas:2011}\footnote{The {\sc Camb} software package can be obtained from \url{https://camb.info}, and {\sc Class} is available from \url{https://lesgourg.github.io/class_public/class.html}.}. These codes usually evolve at least dark matter, baryons, photons and (massive) neutrinos and output Fourier-space transfer functions for density $\delta_X$ and velocity divergence $\theta_X$ for each of the species $X$ as well as the total matter density fluctuations at specifiable output times. Typically equations are integrated in synchronous gauge, in a frame comoving with dust. The use of the output of these Einstein-Boltzmann solver for non-linear simulations that (in the case of $N$-body simulations) model only Newtonian gravity and no relativistic species, let alone baryon-photon coupling, requires still some numerical considerations that we discuss next. The inclusion or non-inclusion of relativistic species makes a difference of several per cent in the background evolution and therefore the growth rate between redshifts $z=100$ and $z=0$ \citep{Fidler:2017}, implying that it is crucial to be aware of what physics is included in the calculations for the initial conditions and in the non-linear cosmological code. Usually, it is sufficiently accurate to combine output for  density perturbations  in synchronous gauge with Newtonian gauge velocities when working with Newtonian equations of motion, but also self-consistent gauge choices exist which allow the correct inclusion of relativistic effects even within Newtonian simulations.  

\begin{figure}
\begin{centering}
\includegraphics[width=\textwidth]{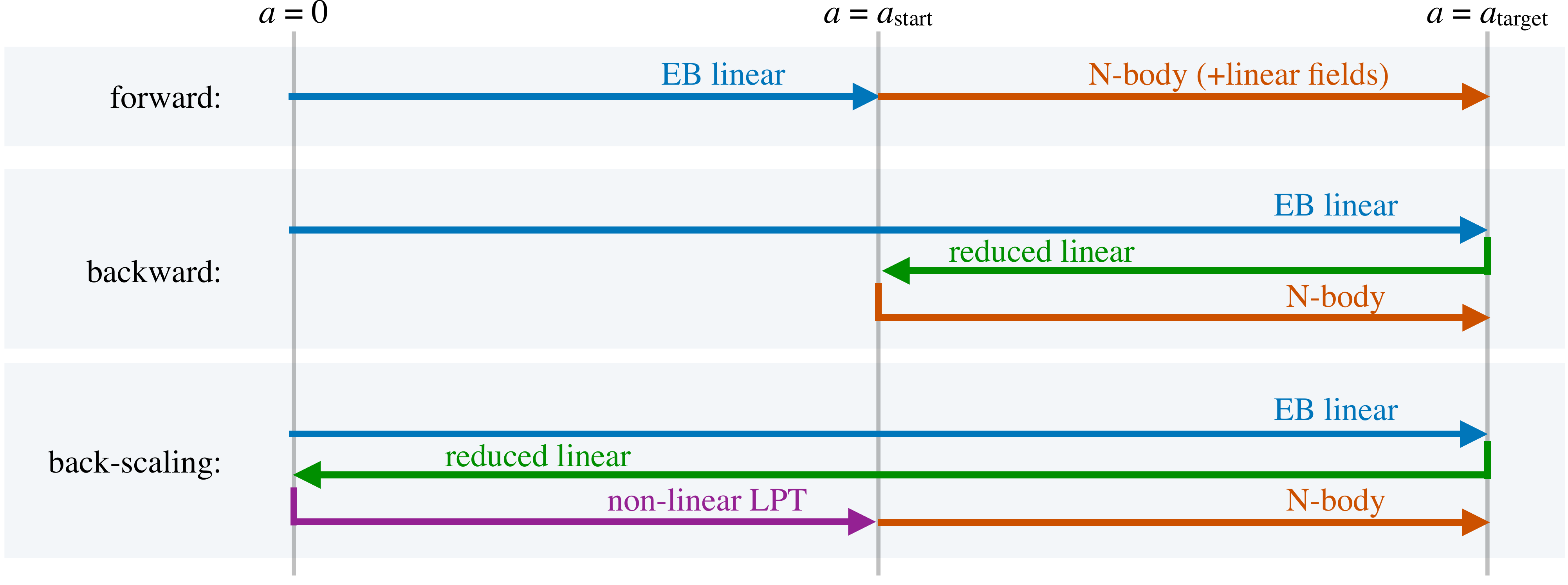} 
\caption{\label{fig:ic_setups} Different setups used to initialise $N$-body simulations with the output from a linear Einstein--Boltzmann (EB) solver such as {\sc Camb} or {\sc Class} to bridge the gap between missing physics in $N$-body codes on the one hand, and missing non-linearities in the EB codes on the other hand. `EB linear' represents the linear full-physics evolution through the EB code, `reduced linear' a reduced linear model (including physics captured in the $N$-body code), `non-linear LPT' the non-linear evolution using Lagrangian perturbation theory to some finite order valid prior to shell-crossing, and `N-body' the full non-linear evolution. In the `forward' approach additional fields (e.g., neutrinos, relativistic corrections) need to be added to match the EB solution at $a_{\rm target}$.}
\end{centering}
\end{figure}

The main approaches adopted when using the output from an EB solvers to set up initial conditions are illustrated in \autoref{fig:ic_setups} and are:

\paragraph*{Forward method }: In this approach, the output of the Einstein-Boltzmann code  at some early time $z_{\rm start} \gtrsim 100$ is used. To avoid errors of several per cent at low redshift on all scales, relativistic components {\em must} be included in the background evolution of the non-linear solution in this case. Also the significant evolution of the horizon between $z>100$ and $0$ means that for very large-scale simulations, relativistic corrections can become important and should be included as well for high precision. Since all corrections beyond the background are significant only on large scales, they remain perturbative. In the {\sc Cosira} approach \citep{Brandbyge:2017}, which is also used e.g., in the {\sc Euclid} flagship simulations \citep{Potter:2017}, they are added as corrections at the linear level (subtracting essentially the Newtonian linear gravity from the relativistic), convolving the resulting correction with the random phases of the initial conditions, and adding this realisation-specific correction to the gravity solver step of the $N$-body code. 

Care has to be taken that these corrections never significantly contribute to non-linear terms as the corresponding back-reaction on the linear field is neglected. While requiring a direct integration of the Einstein-Boltzmann solver with the $N$-body code, this approach can readily  include also a treatment of massive neutrinos at linear order (\citealt{Tram2018}, and see discussion in Sect.~\ref{sec:HDM}). If all relevant physics is included, this approach guarantees very good agreement between Einstein--Boltzmann and $N$-body on linear scales. Due to the very nature of this approach, it does not allow the use of high-order Lagrangian perturbation theory since only linear quantities are known at the starting time $z_{\rm start}$, which therefore has to be pushed to early times in order to not itself affect the simulation results (see discussion in Sect.~\ref{sec:accuracy_ic}). If fields $\delta_X$ and $\theta_X$ are known for a species $X$ at the starting time, then they can be converted into leading order consistent Lagrangian maps with respective velocities through the relations

\begin{equation}
\vecb{x}(\vecb{q};\,z_{\rm start}) = \vecb{q} - \bnabla \nabla^{-2} \delta^{\rm EB}_X(\vecb{q};\,z_{\rm start})\qquad\textrm{and}\qquad \vecb{v}(\vecb{q};\,z_{\rm start}) = \bnabla \nabla^{-2} \theta^{\rm EB}_X(\vecb{q};\,z_{\rm start}). \label{eq:IC_forward}
\end{equation}

\paragraph*{Backward method }: An alternative approach to the forward method that allows to couple a non-linear simulation with reduced physical models to the EB solutions is given by the `backward method' (not to be confused with the `backscaling' method below). Here, the linearised set of equations  solved by the non-linear code are used to integrate backwards in time, i.e. from $z_{\rm target}$ where the output of the EB code is known, to the starting time of the non-linear simulation $z_{\rm start}$. It is thus possible to reduce the multi-species universe e.g., to just two species, total matter and massive neutrinos \citep{Zennaro:2017}, at $z_{\rm target}$, evolve them backwards in time under the linearised physics of the $N$-body code, and then provide ICs using the prescription~(\ref{eq:IC_forward}). The leading order evolution of the `active' fluids takes into account scale-dependent growth and agrees reasonably well at high redshifts with the full EB solution. The limitation of this approach is that any decaying modes that are present at $z_{\rm target}$ must still be small at $z_{\rm start}$. This can be achieved well for neutrinos (with sub per cent errors for $z_{\rm start}\lesssim100$ with $\sum m_\nu \lesssim 0.3\,{\rm eV}$, see \citealt{Zennaro:2017}) due to their small contribution to the total energy budget, but is more limited e.g., for baryons. Again, this approach is then restricted to being used for first order accurate displacements and velocities as only linear fields are known.

\paragraph*{Backscaling method }: The arguably simplest approach, the back-scaling method, avoids the complications of differences of physics. At the same time, it is the one rigorously consistent with Lagrangian perturbation theory. It is also the traditionally used approach to set up $N$-body ICs. In this method, one uses the Einstein-Boltzmann code to evolve the linear multi-physics equations to a target redshift $z_{\rm target}$ and then re-scales the total matter perturbation $\delta^{\rm EB}(z_{\rm target})$ as output by the EB-code to arbitrary times using the linear theory growth factor defined by Eq.~(\ref{eq:linear_density_pt}) as
\begin{equation}
\tilde{\delta}_m(k;\, z_{\rm start}) = \frac{D_+(z_{\rm start})}{D_+(z_{\rm target})} \, \tilde{\delta}_m^{\rm EB}(k;\,z_{\rm target}). \label{eq:backscaling_IC}
\end{equation}
The main advantage of this approach is that by definition the correct linear theory is obtained in the vicinity of $z=z_{\rm target}$ including e.g., relativistic and neutrino effects, without having to include this physics into the $N$-body code. This comes at the price that at early times, it might not in general agree with the full EB solution since a plethora of modes captured by the higher-dimensional EB system are reduced to scale-independent growth under the linear growing mode only. However, this is not a problem if non-linear coupling is unimportant, which is an excellent assumption precisely at early times. Backscaling can also be rigorously extended to simulations including multiple fluids coupled through gravity \citep{Rampf:2020} by including additional isocurvature modes. It can in principle also account for scale-dependent evolution if it can be modelled at the $D_+(k;\,a)$ level. This method, w.r.t.\ the total matter field in $\Lambda$CDM agrees exactly with the `backward method' above if the decaying mode is neglected. Arguably the biggest advantage of the backscaling method is that it connects naturally with high order Lagrangian perturbation theory, as we will discuss next.

\subsection{Initial conditions from Lagrangian perturbation theory}
\label{sec:ICs_from_LPT}

\begin{figure}
\begin{centering}
  \includegraphics[width=\textwidth]{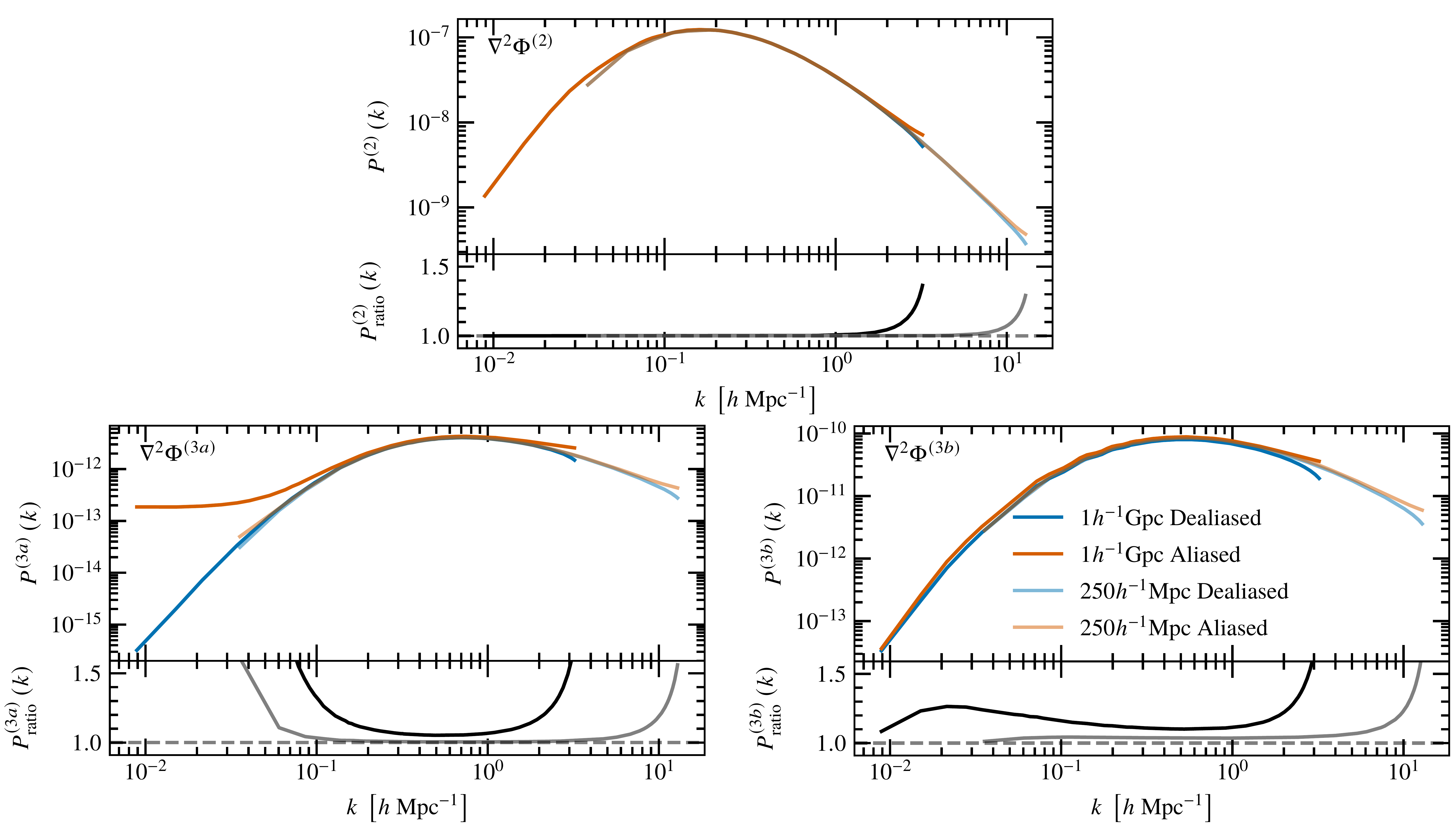}   
\caption{The power spectrum of various fields contributing to displacements in Lagrangian Perturbation Theory up to 3rd order. The top panel shows the power spectrum of $\nabla^2 \phi^{(2)}$ that contributes to the 2nd-order displacements (cf.\ Eq.~\ref{eq:2LPT}), whereas the bottom left and right panel show the spectrum of $\nabla^2 \phi^{(3a)}$ and $\nabla^2 \phi^{(3b)}$ which correspond to the first and second terms of the 3rd order LPT contribution $\phi^{(3)}$, given in Eq.~(\ref{eq:3LPT}). In each case we display the results measured in a $250\hMpc$ and $1000\hMpc$ boxes, with and without correct de-aliasing, as indicated by the figure. The ratio between the aliased and de-aliased solutions with respect to the analytic expectation is shown in the bottom panels. Figure adapted courtesy \cite{Michaux:2020}. \label{fig:lpt}}
\end{centering}
\end{figure}

With the increasing precision requirements of $N$-body simulations over the last 20 years, it quickly became clear that first order accurate initial conditions are insufficient. Those are ICs that follow Eq.~(\ref{eq:IC_forward}), which together with the back-scaled input spectrum from Eq.~(\ref{eq:backscaling_IC}) amounts to the Zel'dovich approximation (ZA; \citealt{Zeldovich:1970}). The reason is that linear ICs suffer from significant transients -- decaying truncation errors between the ICs and the true non-linear evolution \citep{Scoccimarro:1998,Crocce:2006} -- since higher order non-Gaussian moments of the density distribution are not accurately captured (see discussion in Sect.~\ref{sec:accuracy_ic}). Higher order is needed to follow correctly the higher order moments of the density distribution, i.e. first order captures only the variance, second in addition also the skewness and third even the kurtosis \citep{Munshi:1994}. These `transients' can only be suppressed by going to very early starting times, when the linear Gaussian approximation is accurate, at the cost of typically larger numerical errors \citep{Michaux:2020}. The alternative to early starts is to go to higher order perturbation theory beyond the ZA, where the displacement field $\vecb{\Psi}$ of the Lagrangian map, $\vecb{x}(\vecb{q},\,\tau)=\vecb{q}+\vecb{\Psi}(\vecb{q},\,\tau)$, is expanded in a Taylor series to order $n$ in the linear theory growth factor $D_+$ yielding the $n$LPT approximation
\begin{equation}
\vecb{\Psi}(\vecb{q},\tau) = \sum_{j=1}^n D_+(\tau)^j \; \vecb{\Psi}^{(j)}(\vecb{q}). \label{eq:1LPT}
\end{equation}
This includes only growing modes, but remains regular as $D_+\to0$, i.e. $a\to0$, with the key property that $\vecb{x}\to\vecb{q}$, i.e. the initial state is indeed a homogeneous unperturbed universe. Since the density is uniform in the $a\to0$ limit (where the limit is taken in the absence of radiation, otherwise one can use $a\sim10^{-2}$ during matter domination for simplicity), the growing mode perturbations are at leading order encapsulated in a single potential $\phi^{(1)}(\vecb{q})$ which can be connected to the back-scaling relation from above to the EB code output. This yields the famous `Zel'dovich approximation'
\begin{equation}
 \vecb{\Psi}^{(1)} = -\bnabla \phi^{(1)} \quad\textrm{with}\quad  \phi^{(1)} :=  -\nabla^{-2} \lim_{a\to 0} \frac{D_+(a)}{a}\,\delta_m(\vecb{q};\, a),
\end{equation}
that can be used to set up simulation initial conditions by displacing Lagrangian fluid elements (e.g., $N$-body particles) consistent with $\vecb{\Psi}$ and giving them velocities according to $\dot{\vecb{\Psi}}$. 
Inserting this ansatz order by order in Eq.~(\ref{eq:main_lpt_equation}) returns the well known order-truncated $n$-th order LPT forms. Specifically, the 2LPT contribution to the displacement field has the form
\begin{equation}
\label{eq:2LPT}
\vecb{\Psi}^{(2)} = \bnabla \phi^{(2)}\quad\textrm{with}\quad  \phi^{(2)} = -\frac{3}{14}\nabla^{-2} \left[ {\phi}^{(1)}_{,ii} {\phi}^{(1)}_{,jj} - {\phi}^{(1)}_{,ij}{\phi}^{(1)}_{,ij}\right],
\end{equation}
while at third order, i.e. 3LPT, the displacement field starts to have both longitudinal (i.e. irrotational) and transverse (i.e. solenoidal) components \citep{Catelan:1995,Rampf:2012b}

% \begin{subequations}
\label{eq:3LPT}
\begin{eqnarray}
\vecb{\Psi}^{(3)} = \bnabla \phi^{(3)}+\bnabla\times \vecb{A}^{(3)} \quad&\textrm{with}&\quad \phi^{(3)} = \frac{1}{3}\nabla^{-2} \left[ \det \phi^{(1)}_{,ij} \right] -\frac{5}{21}\nabla^{-2}\left[\phi^{(2)}_{,ii}\phi^{(1)}_{,jj}-\phi^{(2)}_{,ij} \phi^{(1)}_{,ji}\right] \nonumber\\
\quad&\textrm{and}&\quad \vecb{A}^{(3)} = \frac{1}{7}\nabla^{-2}\left[\boldsymbol{\nabla}\phi^{(2)}_{,i}\times\boldsymbol{\nabla}\phi^{(1)}_{,i}\right].
\end{eqnarray}
% \end{subequations}

The transverse part appears at 3LPT order and preserves the potential nature of the flow in Eulerian space. Newtonian gravity (i.e. Hamiltonian mechanics coupled to only a scalar potential) cannot produce vorticity (i.e. it exactly preserves any that might be present in the initial conditions, which however in cosmological context would appear as a decaying mode that blows up as $a\to0$). For the truncated $n$LPT series this is only true at the respective order, i.e. $\bnabla_x \times \vecb{v} = \bnabla_x \times \dot{\vecb{\Psi}} =\mathcal{O}(D_+^n)$ \citep{Uhlemann:2019}. For systems of more than one pressureless fluid, this approach can be readily generalised taking into account isocurvature modes to all, and decaying modes to first order currently \citep{Rampf:2020,Hahn:2020}. Note that the relations given above for $n$LPT have small corrections in $\Lambda$CDM \citep{Bouchet:1995,Rampf:2020}, which are however not important when initial conditions are generated at $z\gg1$, when it is safe to assume an Einstein-de~Sitter cosmology.

In Fig.~\ref{fig:lpt} we show the power spectrum of the source terms, $\nabla^2 \phi^{(n)}$, for 2LPT and 3LPT. As expected, we see that the higher-order fields have significantly smaller amplitude than the linear-order density power spectrum. However, these higher-order contributions are required to improve the faithfulness of the simulated field and are in fact important for correctly predicting certain cosmological statistics. In addition, note that when computing the high-order potentials $\phi^{(n)}$ and $\vecb{A}^{(n)}$ (with $n\ge2$) some care has to be taken to avoid aliasing \citep{Orszag:1971,Michaux:2020,Rampf:2020b}, which can be important even on large scales, as shown in the bottom panels of Fig.~\ref{fig:lpt}. %, i.e. it is not correct to simply multiply the fields in real space .

Note that recently, \cite{Rampf:2020b} were the first to numerically implement the full $n$LPT recursion relations so that fields to arbitrary order can be used for ICs in principle, only limited by computer memory. This is included in the publicly-available {\sc MonofonIC Music-2} code\footnote{Available from \url{https://bitbucket.org/ohahn/monofonic}.}.

%%%%%%%%%%%%%%%%%%%%%%%%%%%%%%%%%%%%%%%%%%%%%%%%%%%%%%%%%%

\subsection{Generating Gaussian realisations of the perturbation fields}

The previous results were generic, for numerical simulations, one has to work with a specific realisation of the Universe, which we will discuss next. The specific case of the realisation of {\em our} Universe is discussed as well.

\subsubsection{Unconstrained realisations} 
Many inflationary cosmological models predict that scalar metric fluctuations are very close to Gaussian \citep{Maldacena:2003,Acquaviva:2003,Creminelli:2003}. As we have shown above, Lagrangian perturbation theory, focusing on the fastest growing mode, is built up from a single initial scalar potential $\phi^{(1)}$ as defined in Eqs.~(\ref{eq:backscaling_IC}) and (\ref{eq:1LPT}), while the forward and backward approaches work with multiple fields $\delta_m$, $\theta_m$, and possibly others. These specify expectation values for a random realisation. Since these are all linear fields (i.e. we can assume that non-linear corrections can be neglected at $a=0$ for back-scaling, and at $a_{\rm start}$ for the forward and backward method) they will be statistically homogeneous and isotropic Gaussian random fields fully characterised by their two-point function. A general real-valued Gaussian homogeneous and isotropic random field $\phi(\vecb{x})$ can be written as the Fourier integral
\begin{equation}
\phi(\vecb{x}) = \frac{1}{(2\pi)^3}\int_{\mathbb{R}^3}{\rm d}^3k\, {\rm e}^{\ii \vecb{k}\cdot \vecb{x}} \tilde{\varphi}(k)\,\tilde{W}(\vecb{k}) , \label{eq:GaussianWhiteField}
\end{equation}
where $\tilde{W}(\vecb{k})$ is a complex-valued three-dimensional random field (also known as ``white noise" since its power spectrum is $\vecb{k}$ independent) with 
\begin{equation}
\tilde{W}(\vecb{k}) = \overline{\tilde{W}(-\vecb{k})},\qquad \langle\tilde{W}(\vecb{k})\rangle=0,\qquad\textrm{and}\qquad\langle\tilde{W}(\vecb{k})\,\overline{\tilde{W}(\vecb{k}^\prime)}\rangle=\delta_D(\vecb{k}-\vecb{k}^\prime), 
\end{equation}
and $\tilde{\varphi}(k)$ is the (isotropic) field amplitude in Fourier space as computed by the Einstein-Boltzmann code. This is often given in terms of a transfer function $\tilde{T}(k)$ which is related to the field amplitude as $\tilde{\varphi}(k)\propto k^{n_s/2}\tilde{T}(k)$, where $n_s$ is the spectral index of the primordial perturbations from inflation. If the power spectrum $P(k)$ is given, then setting $\tilde{\varphi}(k) = (2\pi)^{3/2}\sqrt{P(k)}$ yields the desired outcome
\begin{equation}
\langle\tilde{\phi}(\vecb{k})\,\overline{\tilde{\phi}(\vecb{k}^\prime)}\rangle=(2\pi)^3\,P(k)\,\delta_D(\vecb{k}-\vecb{k}^\prime).
\end{equation}
In order to implement these relations numerically, the usual approach when generating initial conditions is to replace the Fourier integral~(\ref{eq:GaussianWhiteField}) with a discrete Fourier sum that is cut off in the IR by a `box mode' $k_0=2\pi/L$ and in the UV by a `Nyquist mode' $k_{\rm Ny}:=k_0 N/2$ so that the integral can be conveniently computed by a DFT of size $N^3$. Naturally, fluctuations on scales larger than the box and their harmonics cannot be represented (but see below in Sect.~\ref{sec:supersample}). This is usually not a problem as long as $k_0\ll k_{\rm NL}$, where $k_{\rm NL}$ is the scale where non-linear effects become important, since then non-linearities can be assumed to not be strongly coupling to unresolved IR modes and are also not sourced dominantly by the very sparsely populated modes at the box scale (which would otherwise break isotropy in the non-linear structure). For simulations evolved to $z=0$ this implies typically box sizes of at least $300\,h^{-1}{\rm Mpc}$ comoving.

\subsubsection{Reduced variance sampling}
\label{sec:fixed-and-paired}

\begin{figure}
\begin{center}
\includegraphics[width=\textwidth]{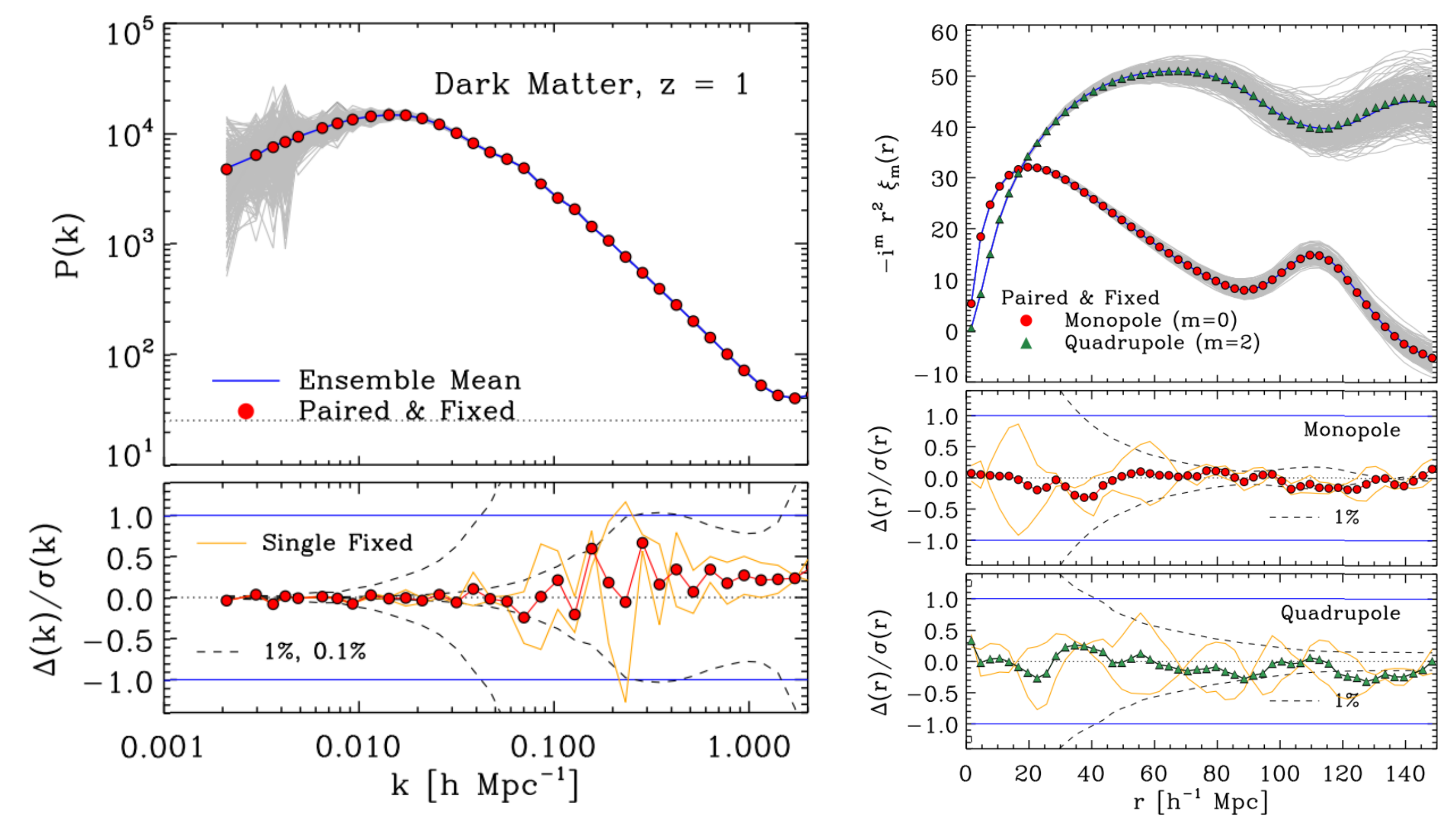} 
\end{center}
\caption{The nonlinear dark matter clustering at $z=1$ in simulations with different initial conditions. The left panel shows the real-space power spectrum whereas the right panel shows the monopole and quadrupole of the redshift space correlation function. Gray lines display the results for an ensemble of 300 simulations with random Gaussian initial  conditions, whereas the symbols show the results from a single Paired-\&-Fixed pair of simulations. In the bottom panels we show the difference between the ensemble mean and the Paired-\&-Fixed results in units of the standard deviation of the ensemble measurements. Note the drastic reduction of noise of this method, specially on large scales. Figure adapted courtesy of \cite{Angulo:2016} \label{fig:paired_fixed}}
\end{figure}

The noise field $\tilde{W}$ naturally has a polar decomposition $\tilde{W}(\vecb{k}) =: \tilde{A}(\vecb{k})\, {\rm e}^{\ii \theta(\vecb{k})}$, where $A$ obeys a Rayleigh distribution (i.e. a $\chi$-distribution with two degrees of freedom) and $\theta$ is uniform on $[0,2\pi)$. The power spectrum associated with $W$ transforms as 
\begin{equation}
\langle \tilde{W}(\vecb{k})\, \overline{\tilde{W}(\vecb{k}^\prime)} \rangle = \langle \tilde{A}(\vecb{k}) \,\tilde{A}(\vecb{k}^\prime)\rangle = \delta_D(\vecb{k}-\vecb{k}^\prime),
\end{equation}
i.e., it is independent of the phase $\theta$. In any one realization of $\tilde{W}(\vecb{k})$, one expects fluctuations (``cosmic variance'') in $\langle\tilde{W}(\vecb{k})\overline{\tilde{W}(\vecb{k}^\prime)}\rangle$ around the ensemble average of the order of the inverse square root of the number of discrete modes in some finite interval $k\dots k+dk$ available in the simulation volume. 

The amplitude of this cosmic variance can be dramatically reduced by simply fixing $A$ to its expectation value \citep{Angulo:2016}, i.e,. by  `sampling' $A$ from a Dirac distribution $A(\vecb{k})=\delta_D(\vecb{k})$. This technique is commonly referred to as ``Fixing''. Clearly, the resulting field has much reduced degrees of freedom and a very specific non-Gaussian character. In principle, this introduces a bias in the nonlinear evolution of such a field, e.g., the ensemble averaged power spectrum at $z=0$ differs from that obtained from an ensemble of Gaussian realizations. However, using perturbation theory, \cite{Angulo:2016} showed that the magnitude of this bias in the power spectrum and bispectrum is always smaller (by a factor equal to the number of Fourier modes) compared to mode-coupling terms of physical origin. In fact, it was found empirically that simulations initialised with such ICs produce highly accurate estimates of non-linear clustering (including power spectra, bispectra, halo mass function, and others) and that the level of spurious non-Gaussianity introduced is almost undetectable for large enough simulation boxes \citep{Angulo:2016,Villaescusa-Navarro:2018,Klypin:2020}.

The main advantage of ``Fixing'' is that it allows to avoid giant boxes or large ensembles of simulations in order to obtain measurements with low cosmic variance. A further reduction in cosmic variance can be achieved by considering pairs of simulations where the second simulation is initialised with $\overline{\tilde{W}(\vecb{k})}$ instead of $\tilde{W}(\vecb{k})$ \citep{Pontzen:2016,Angulo:2016}. This is equivalent to inverting the phase $\theta$, and averages out the leading order non-linear term contributing to cosmic variance \citep{Pontzen:2016}. This technique has been adopted in several state-of-the-art dark matter and hydro-dynamical simulations \citep{Angulo:2020,Chuang:2019,Knabenhans:2019,Anderson:2019}.

The performance of this approach in practice is shown in Fig.~\ref{fig:paired_fixed} which compares the $z=1$ power spectrum and multipoles of the redshift space correlation function. The mean of 300 $L=3000\hMpc$ simulations is displayed as a solid line whereas the prediction of a single pair of `fixed' simulations is shown as red circles. In the bottom panels we can appreciate the superb agreement between both measurements, with relative differences being almost always below one per cent.

\subsubsection{Numerical universes: spatially stable white noise fields}

A key problem in generating numerical realisations of random fields is that certain properties of the field should not depend on the exact size or resolution of the field in the numerical realisation. This means that it is desirable to have a method to generate white Gaussian random fields which guarantees that the large-scale modes remain identical, no matter what the resolution of the specific realisation. A further advantage is gained if this method can be parallelised, i.e. if the drawing does not necessarily have to be sequential in order to be reproducible. This problem has found several solutions, a selection of which we discuss below.

{\sc N-GenIC} \citep{Springel:2005a}\footnote{Available from \url{https://www.h-its.org/2014/11/05/ngenic-code/}} and derivates produce spatially stable white noise by drawing white noise in Fourier space in a reproducible manner. I.e. for two simulations with $k_{\rm Ny}$ and $k_{\rm Ny}^\prime>k_{\rm Ny}$, all modes that are representable in both, i.e. $-k_{\rm Ny}\le k_{x,y,z}\le k_{\rm Ny}$, are identical, and for the higher resolution simulation, new modes are added between $-k_{\rm Ny}^\prime \le k_{x,y,z} < -k_{\rm Ny}$ and $-k_{\rm Ny}< k_{x,y,z} \le k^\prime_{\rm Ny}$. The random number generation is parallelised by a stable 2+1 decomposition of Fourier space. A shortcoming of this approach of sampling modes is that drawing in Fourier space is inherently non-local so that it cannot be generalised to zoom simulations, where high-resolution is desired only in a small subvolume, without generating the full field first.

{\sc Panphasia} \citep{Jenkins:2013} \footnote{The stand-alone {\sc Fortran} library can be downloaded from \url{http://icc.dur.ac.uk/Panphasia.php}, but it is also built into the publicly available {\sc Music-2} IC-code.} has overcome this shortcoming by relying on a hierarchical decomposition of the white noise field in terms of cleverly chosen octree basis functions. Essentially instead of drawing Fourier modes, one draws the coefficients of this hierarchical basis thus allowing to add as much small-scale information as desired at any location in the three-dimensional volume. 

{\sc Music-1} \citep{Hahn:2011} subdivides the cubical simulation volume into subcubes for which the random numbers can be drawn independently and in parallel in real space, since each cube carries its own seed. Refined regions are combined with white noise from the overlapping volume from the coarser levels, just as in the {\sc N-GenIC} approach, thus enforcing the modes represented at the coarser level to be present on the finer level.

%%%%%%%%%%%%%%%%%%%%%%%%%%%%%%%%%%%%%%%%%%%%%%%%%%%%%%%%%%
\subsubsection{Zoom simulations}
\label{sec:zoom_simulations}

In cases in which the focus of a simulation is the assembly of single objects or particular regions of the universe, it is neither desirable nor affordable to achieve the necessary high resolution throughout the entire simulation volume. In such situations, `zoom simulations' are preferable, where the (mass) resolution is much higher in a small sub-volume of the entire box. This can be achieved in principle by generating high-resolution initial conditions, and then degrading the resolution outside the region of interest (as followed e.g. by the {\sc ZInCo} code \citealt{Garaldi:2016}). This approach is, however, limited by memory, and for deeper zooms, refinement methods are necessary. The basic idea is that in the region of interest nested grids are inserted on which the Gaussian noise is refined. Some special care must be taken when applying the transfer function convolution and resolution of the Poisson equation. Such approaches were pioneered by \cite{Katz:1994,Bertschinger:2001}, and then extended to higher accuracy and 2LPT perturbation theory using a tree-based approach by \cite{Jenkins:2010} which is implemented in a non-public code used by the {\sc Virgo} consortium, and using multi-grid techniques by \cite{Hahn:2011}  in the publicly available {\sc MUSIC-1} code\footnote{Available from \url{https://bitbucket.org/ohahn/music}}. A recent addition is the {\sc GENET-IC} code\footnote{Available from \url{https://github.com/pynbody/genetIC}} \citep{Stopyra:2020} that focuses on the application of constraints (cf.\ Sect.~\ref{sec:constrained_realisations}) to such zoom simulations but currently only supports first order LPT. An example of a particularly deep zoom simulation is the `voids in voids' simulation shown in Fig.~\ref{fig:vvv}.

%%%%%%%%%%%%%%%%%%%%%%%%%%%%%%%%%%%%%%%%%%%
%%%%%%%%%%%%%%%%%%%%%%%%%%%%%%%%%%%%%%%%%%%

\subsubsection{Super-sample covariance and ensembles}
\label{sec:supersample}

On the scale of the simulated box, the mean overdensity and its variance are expected to be non-zero for realistic power spectra. This is a priori in contradiction with periodic boundary conditions (but see Sect.~\ref{sec:separate_universe}) and so the mean overdensity of the  volume in the vast majority of cosmological simulations is enforced to be zero for consistency. Hence, when an ensemble of simulations is considered, the variance of modes $k<k_0$ is zero despite all of them having different initial white noise fields and providing fair ensemble averages for modes $k_0\lesssim k\lesssim k_{\rm Ny}$. This implies that the component of the covariance that is due to large-scale overdensities and tides is underestimated \citep{Akitsu:2019} which is sometimes referred to as \emph{super-sample covariance} -- in analogy to a similar effect present in galaxy surveys \citep{Li:2018}-- , and can be an important source of error in covariance matrices derived from ensembles of simulations, especially if the simulated boxes are of small size \citep{Klypin:2019}. Such effects can be circumvented in ``separate universe'' simulations that are discussed in Sect.~\ref{sec:separate_universe}.

Furthermore, it is important to note that for  Fourier summed realisations of $\phi$, the correspondence between power spectra and correlation functions is broken since the discrete (cyclic) convolution does not equal the continuous convolution, i.e.,
\begin{subequations}
\begin{align}
\phi(\vecb{x}) \neq \phi^K(\vecb{x}) &:= {\rm DFT}^{-1}\left[ \tilde{W}(\vecb{k})\, \tilde{\phi}(k) \right]  & \textrm{and}&\\
\phi(\vecb{x}) \neq \phi^R(\vecb{x}) &:= {\rm DFT}^{-1}\left[\tilde{W}\right]\circledast \phi(\|\vecb{x}\|) &\textrm{where} \;\phi(r) &:= \frac{1}{2\pi^2}\int_0^\infty \frac{\sin kr}{kr} \tilde{\phi}(k) k^2 {\rm d}k, \label{eq:real_space_sampling} 
\end{align}
\end{subequations}
where `$\circledast$' symbolises a discrete cyclic convolution. This implies that real space and Fourier space statistics on discrete finite numerical Universes coincide neither with one another nor with the continuous relation. It is always possible to consider such real space realisations $\phi^R$ instead of Fourier space realisations $\phi^K$ \citep{Pen:1997,Sirko:2005}.  In the absence of super-sample covariance, the correct statistics is always recovered on scales $k_0\ll k\ll k_{\rm Ny}$. Since the real-space kernel $ \phi(r)$ is effectively truncated at the box scale, it does not impose the box overdensity to vanish and therefore also samples density fluctuations on the scale of the box, which must be absorbed into the background evolution (cf.\ Sect.~\ref{sec:separate_universe}).

The real-space sampling, by definition, reproduces the two-point correlation function also in smaller boxes if one allows the mean density of the box to vary. In fact, \cite{Sirko:2005} argued that this approach yields better convergence on statistics that depend more sensitively on real-space quantities (such as the halo mass function that depends on the variance of the mass field on a given scale), and also an accurate description of the correlation function for scales $r \gtrsim L_{\rm box}/10$. However, \cite{Orban:2013} demonstrated that the correct correlation function can still be correctly recovered by accounting for the integral constraint in correlation function measurements, i.e. by realising that the $\phi^K$ sampling implicitly imposes $\int_0^{R}r^2 \xi(r){\rm d}r=0$ already over a scale $R$ related to the box size $L_{\rm box}$ instead of only in the limit $R\to\infty$. A better estimator can therefore be obtained by simply subtracting this expected bias.

Note that an alternative approach to account for non-periodic boundary conditions have been recently proposed \citep{Racz:2018,Racz:2019}. In such `compactified' simulations an infinite volume is mapped to the surface of a four-dimensional hypersphere. This compactified space can then be partitioned onto a regular grid with which it is possible to simulate a region of the universe without imposing periodic boundary conditions. This approach has the advantage for some applications that it naturally provides an adaptive mass resolution which increases near to the center of the simulation volume where a hypothetical observer is located. 

%%%%%%%%%%%%%%%%%%%%%%%%%%%%%%%%%%%%%%%%%%%%%%%%%%%%%%%%%%

\subsubsection{Separate universe simulations}

\begin{figure}
\includegraphics[width=\textwidth]{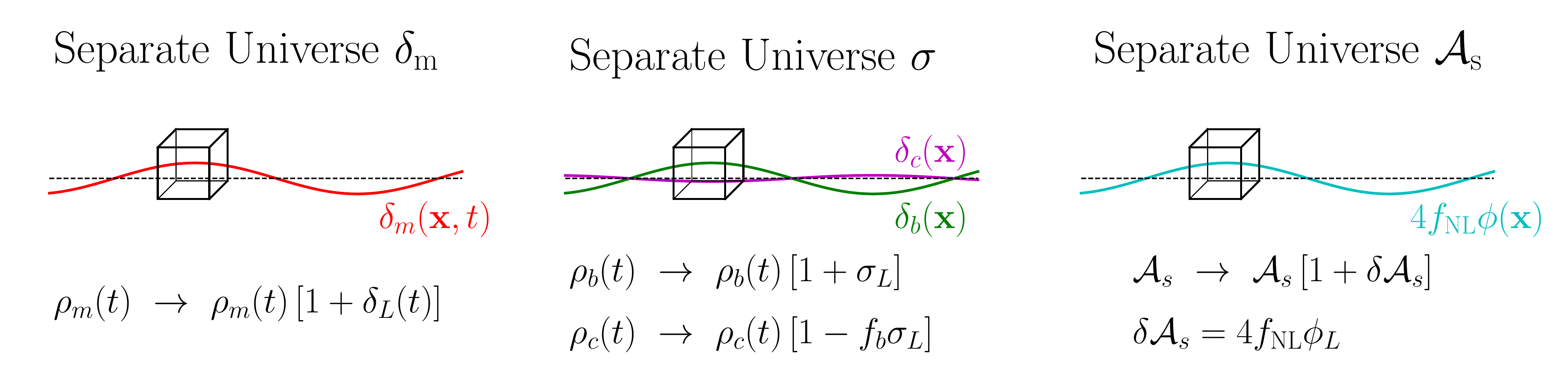} 
\caption{Schematic illustration of various kinds of \textit{Separate Universe} simulations. Structure formation embedded into a large-scale fluctuation $\delta_L$ is equivalent to that of a simulation with a modified background matter parameter, $\rho_m$; structure formation with inside large-scale compensated isocurvature fluctuation, $\sigma_L$ is equivalent to a modification in the background baryon and cold dark matter densities, $\rho_c$ and $\rho_m$; and finally structure formation inside a large potential fluctuation, as originated by primordial non-gaussianity, can be captured in the Separate Universe formalism by a change in the amplitude of fluctuations $A_s$. Figure courtesy of \cite{Voivodic:2020}.
\label{fig:sepu}} 
\end{figure}

\label{sec:separate_universe}

How small scales are affected by the presence of fluctuations on larger scales is not only important to understand finite-volume effects and ensembles of simulations, as we discussed in the previous section, but it also is a central question for structure formation in general. For instance, this response is an essential ingredient in models for biased tracers or in perturbation theory. These interactions can be quantified in standard $N$-body simulations, however, a method referred to as ``separate universe simulations'' provides a more controlled way to carry out experiments where perturbations larger than the simulated volume are systematically varied to yield accurate results even with modest simulation volumes. A key advantage of the separate universe technique is that it allows to quantify the dependence of small-scale structure formation on large-scale perturbations. For instance, changing the effective mean density, e.g., of two simulations to $+\delta_0$ and $-\delta_0$, one can compute the response of the power spectrum by taking a simple derivative $d P / d \delta_0 \simeq (P(k;+\delta_0) - P(k;-\delta_0))/2\delta_0$ (e.g., \citealt{Li:2014}), which can be extended also to higher orders.

The main idea behind the separate universe formalism \citep{Sirko:2005,Baldauf:2011,Sherwin:2012,Li:2014,Wagner:2015,Baldauf:2016} is that a long wavelength density fluctuation can be absorbed in the background cosmology. In other words, larger-than-box fluctuations simply lead to a non-zero overdensity $\delta_0={\rm const.}$ of the box that must be absorbed into the background density in order to be consistent with the periodic boundary conditions of the box, i.e. one matches
\begin{equation}
\rho(a) [1 + D_+(a) \delta_0] =: \breve{\rho}(a) ,
\end{equation}
and thus structure in a given region embedded inside a region of overdensity $\delta_0$ (today, i.e. at $a=1$) is equivalent to that of an isolated region of the universe evolving with a modified set of cosmological parameters indicated by `$\breve{\phantom{a}}$'. Specifically, the modified Hubble parameter, and matter, curvature, and dark energy density parameters become
\begin{equation}
\breve{H_0} := H_0 \Delta_H, \qquad
\breve{\Omega}_m := \Omega_{\rm m} \Delta_H^{-2}, \qquad
\breve{\Omega}_K := 1 - \Delta_H^{-2},  \qquad
\breve{\Omega}_{\rm \Lambda} := \Omega_{\rm \Lambda} \Delta_H^{-2} ,
\end{equation}
where for a simulation initalised at $a_{\rm ini}$
\begin{equation}
\Delta_H := \sqrt{1 - \frac{5\Omega_m}{3} \frac{D_+(a_{\rm ini})}{a_{\rm ini}} \delta_0}.
\end{equation}
Note that although these expressions are exact, solutions only exist for $\delta_0 <\frac{3}{5\Omega_m} \frac{a_{\rm ini}}{D_+(a_{\rm ini})}$. For larger background densities, the whole region is expected to collapse.
An important aspect is that the age of the universe should match between the separate universe box and that of the unperturbed universe, thus the scale factors are not identical but relate to each other as
\begin{equation}
\breve{a} := a \left(1 - \frac{1}{3} D_+(a) \delta_0 \right).
\end{equation}
Also, when initialising a simulation, the perturbation spectrum from which the ICs are sampled should be rescaled with the growth function $\breve{D}_+(\breve{a})$ based on the `separate universe' cosmological parameters. 

Separate universe simulations have been successfully applied to many problems, that can be roughly split into two groups. The first group includes measurements of the value and mass dependence of local and nonlocal bias parameters for voids and halos. The second group includes quantification of the response of the nonlinear power spectrum and/or bispectrum to global processes. For instance, \cite{Barreira:2019} studied the role of baryonic physics on the matter power spectrum. Other studies include measurements of the linear and quadratic bias parameters associated to the abundance of voids \citep{Chan:2020,Jamieson:2019b};  the correlation between large-scale quasar overdensities and small-scale Ly-$\alpha$ forest \citep{Chiang:2017}; halo assembly bias \citep{Baldauf:2016,Lazeyras:2017,Paranjape:2017}; and cosmic web anisotropy \citep{Ramakrishnan:2020}. 
 
Naturally, a simple change in the mean density only modifies the isotropic expansion. More realistically, a given volume will also be exposed to anisotropic deformation due to a large-scale tidal field. \cite{Schmidt:2018} (see also \citealt{Stuecker:2020,Masaki:2020,Akitsu2020}) have demonstrated that such a global anisotropy can be accounted for by a modification of the force calculation in numerical simulations. These simulations have been used to study the role of large-scale tidal fields in the abundance and shape of dark matter halos and the response of the anisotropic power spectrum, and will be very useful also for studies of coherent alignment effects of haloes and galaxies which are important to understand intrinsic alignments in weak gravitational lensing. We note that the separate Universe approach can also be generalised to study the impact of compensated isocurvature modes (where the relative flucutations of baryons and dark matter changes while leaving fixed the total matter fluctuations) \citep{BarreiraCabass:2020} or of modifications to the primordial gravitational potential \citep{BarreiraPNG:2020a}, as illustrated in Fig.~\ref{fig:sepu}.

Another limitation of the original separate universe formulation, is that the long wavemode is assumed to evolve only due to gravitational forces. This means that the scale on which $\delta_0$ is defined has to be much larger than any Jeans or free streaming scale. This condition might be violated if, e.g., neutrinos are considered since their evolution cannot be represented as a cold matter component. This limitation was avoided in the approach of \cite{Hu:2016}, who introduced additional degrees of freedom in terms of ``fake'' energy densities tuned to mimic the correct expansion history and thus growth of the large-scale overdensity. This approach has been applied to study inhomogeneous dark energy \citep{Chiang:2017,Jamieson:2019a} and massive neutrinos \citep{Chiang:2018}.

%%%%%%%%%%%%%%%%%%%%%%%%%%%%%%%%%%%%%%%%%%%%%%%%%%%%%%%%%%

\subsubsection{Real universes: constrained realisations}
\label{sec:constrained_realisations}

\begin{figure}
\begin{centering}
\includegraphics[width=0.5\textwidth]{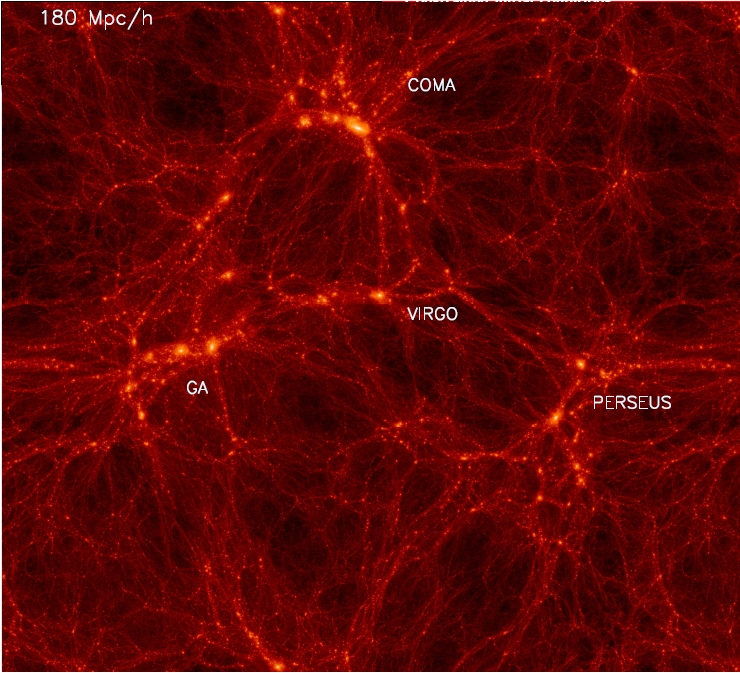}
\includegraphics[width=0.47\textwidth]{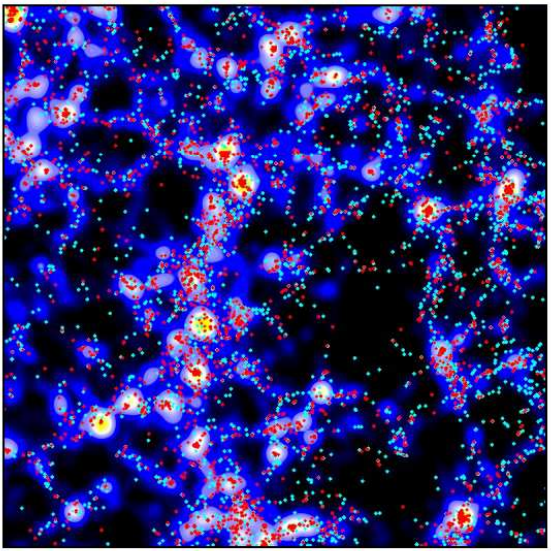}\\
\includegraphics[width=0.85\textwidth]{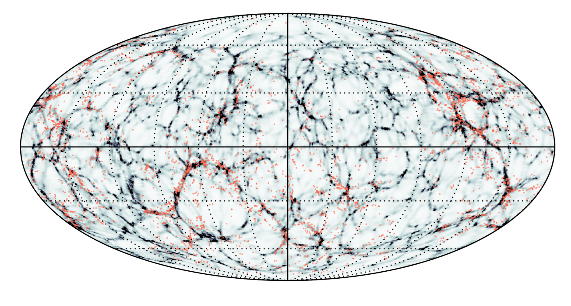} 
\caption{Various $N$-body simulations of initial fields constrained by local observations. The top left panel shows one of the simulations of local Universe carried out by the CLUES collaboration, where the initial conditions were constrained using the CosmicFlows observations \citep{Tully:2016}. In the top right panel we show a $200\hMpc$ wide slice of the density field in the \textit{ELUCID} simulation \citep{ElucidIII:2016}. The simulated dark matter is shown in a black-blue color scale whereas red/cyan symbols show the location of red/blue galaxies in the SDSS DR7 observations. Similarly, in the bottom panel we show a Mollweide projection of $\log(2+\delta)$ of a PM simulation whose initial conditions were set by requiring agreement with the observations of the 2M++ catalogue \citep{Lavaux:2011} shown as red dots. Composite figure courtesy of \cite{ElucidIII:2016,Jasche:2019} and the CLUES collaboration\label{fig:constrained_sims}.}
\end{centering}
\end{figure}

So far we have discussed how to generate random fields which produce numerical universes in which we have no control over where structures, such as galaxy clusters or voids, would form. In order to carry out cosmological simulations of a realisation that is consistent with the observed cosmic structure surrounding us, it is therefore desirable to impose additional constraints.This could, for instance, shed light on the specific formation history of the Milky Way and its satellites, inform about regions where a large hypothetical DM annihilation flux is expected, or quantify the role of cosmic variance in the observed galaxy properties (see \citealt{Yepes:2014} for a review.)

The simplest way to obtain such constrained initial conditions is by employing the so-called  Hoffman--Ribak (HR) algorithm \citep{Hoffman:1991}.  Given an unconstrained realisation of a field $\phi$, we seek to find a new constrained field $\check{\phi}$ that fulfils $M$ (linear) constraints. In general, this can be expressed in terms of kernels $H$ by requiring $(H_j\star\check{\phi})(\vecb{x}_j) = \check{c}_j$, where $\check{c}_j$ is the desired value of the $j$-th constraint centered on $\vecb{x}_j$. The Gaussian field $\check{\phi}$ obeying these constraints can be found using the Hoffman--Ribak algorithm \citep{Hoffman:1991} by computing
\begin{equation}
\tilde{\check{\phi}}(\vecb{k}) = \tilde{\phi}(\vecb{k}) + P(k) \, \tilde{H}_i(\vecb{k}) \,\xi_{ij}^{-1}(\check{c}_j-c_j),
\end{equation}
with
\begin{equation}
c_j = \frac{1}{(2\pi)^3}\int_{\mathbb{R}^3}{\rm d}^3k\, \overline{\tilde{H}_j(\vecb{k})}\,\tilde{\phi}(\vecb{k})\qquad\textrm{and}  \qquad \xi_{ij} = \frac{1}{(2\pi)^3} \int_{\mathbb{R}^3}{\rm d}^3k\, \overline{\tilde{H}_i(\vecb{k})}\,\tilde{H}_j(\vecb{k})\,P(k),
\end{equation}
where $c_j$ is the covariance between constraint $j$ and the unconstrained field, and $\xi_{ij}$ is the $M\times M$ constraint covariance matrix (\citealt{VanDeWeygaert:1996}; see also \citealt{Kravtsov:2002} and \citealt{Klypin:2003} for further implementation details). A possible constraint is e.g., a Gaussian peak on scale $R$ at position $\vecb{x}_i$ so that $\tilde{H}_i(\vecb{k})=\exp\left[-k^2R^2/2+\ii \vecb{k}\cdot\vecb{x}_i\right]$, constraints on differentials and integrals of the field can be easily taken into account in the Fourier-space kernel.

Using this algorithm, various simulations were able to reproduce the local distribution of galaxies and famous features of the local Universe such as the Local and Perseus-Pices Superclusters, and the Virgo and Coma clusters as well as the Great Attractor (e.g., \citealt{Sorce:2014}). This is illustrated in the top left panel of Fig.~\ref{fig:constrained_sims}, which displays a simulation from the CLUES collaboration\footnote{\url{https://www.clues-project.org/}} \citep{Gottloeber:2010,Carlesi:2016}. Traditionally, the observational constraints were mostly set by radial velocity data (which was assumed to be more linear than density, thus more directly applicable for constraining the primordial density field), for instance, from dataset such as the ``Cosmic Flows 2'' \citep{Tully:2016}.

A further technique related to constrained realisations has been termed as `genetic modification' \citep{Roth:2016,Stopyra:2020}. Its main idea is to impose constraints, not in order to be compatible with the Local Universe, but to perform controlled numerical experiments. For instance, the mass of a halo can be altered by imposing a Gaussian density peak at the halo location with the desired height and mass. In this way, it is possible to, e.g., isolate the role of specific features (e.g., formation time, major merger, etc) in the formation of a halo or of the putative galaxy it might host, seeking a better understanding of the underlying physics.

The HR approach has several limitations, for instance, it does not account for the Lagrangian--Eulerian displacements. A general problem of all reconstruction methods is that small scales are difficult to constrain since those scales have shell-crossed in the $z=0$ Universe, so that information from distinct locations has been mixed or even lost. Constrained simulations therefore resort to trial and error by running many realisations of the unconstrained scales until a desired outcome is achieved (e.g., a Milky Way Andromeda pair). To speed up this process, \cite{Sorce:2020} recently proposed an important speed up by using pairs of simulation carried out with the Paired and Fixed method (cf.\ Sect.~\ref{sec:fixed-and-paired}), and \cite{Sawala:2021} quantified the influence of unconstrained initial phases on the final Eulerian field. Another improvement in terms of a ``Reverse Zel'dovich Approximation'' has been proposed to estimate the Lagrangian position of local structure \citep{Doumler:2013a,Doumler:2013b}.

An alternative route to numerical universes in agreement with observations is followed in Bayesian inference frameworks \citep{Kitaura:2008,Jasche:2010,Ata:2015,Lavaux:2016}. In this approach, LPT models or relatively low-resolution $N$-body simulations \citep{ElucidI:2014} are carried out as a forward model mapping a random field to observables. The associated large parameter space of typically millions of dimensions of the IC field is then explored using Hamiltonian Monte Carlo until a realisation is found that compares favourably with the observations. The white noise field can then be stored and used for high resolution simulations that can give insights into the formation history of the local Universe. For instance, \cite{Hess:2013,Libeskind:2018,Sawala:2021b} simulated the initial density field derived from observations of the local Universe, and \cite{ElucidIII:2016,ElucidII:2017} created a constrained simulation compatible with the whole SDSS DR7 galaxy distribution, which is shown in the top right panel of Fig.~\ref{fig:constrained_sims}. 

%%%%%%%%%%%%%%%%%%%%%%%%%%%%%%%%%%%%%%%%%%%%%%%%%%%%%%%%%%

\subsection{Initial particle loads and discreteness}
\label{sec:pre_ics_discreteness}
\begin{figure}
\includegraphics[width=\textwidth]{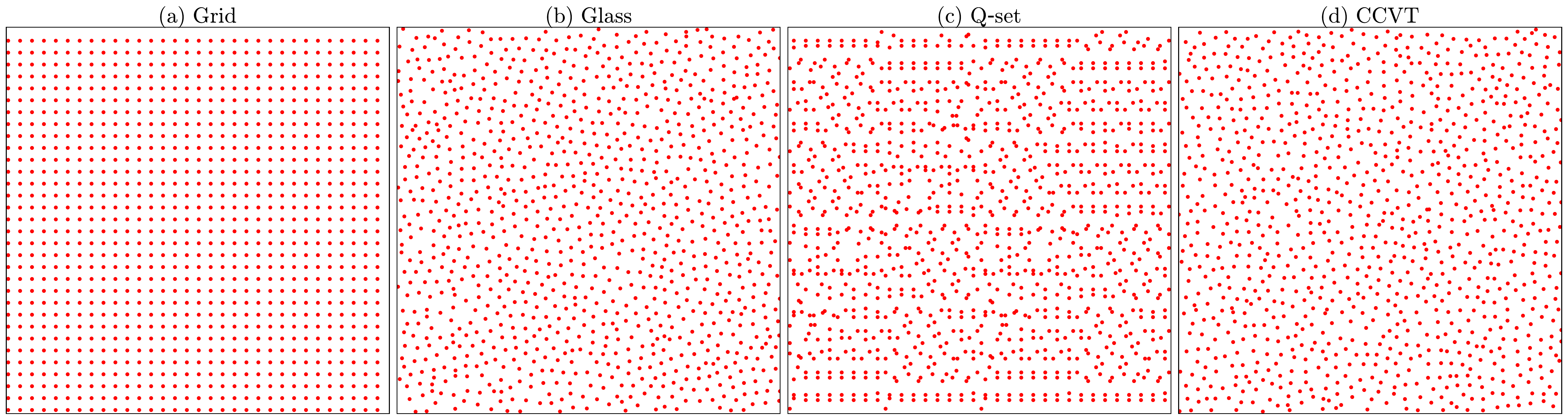} 
\caption{\label{fig:initial_particle_load} Illustration in two dimensions of various  particle distributions employed in the generation of initial conditions for cosmological numerical simulations. From left to right: a regular simple cubic grid lattice, a glass, and particle loads from quaquaversal tiling, and capacity constrained Voronoi tessellations. Figure adapted courtesy \cite{Liao:2018}.}
\end{figure}

Given displacement and velocity fields from Lagrangian perturbation theory and a random realisation of the underlying phases, one is left with imposing these displacement and velocity perturbations onto a set of particles, i.e., a finite set of initially unperturbed Lagrangian coordinates $\vecb{q}_{1\dots N}$. These correspond to $N$-body particles representing the homogeneous and isotropic initial state of the numerical universe. Drawing the $\vecb{q}_j$ from a Poisson process would be the naive choice, however, its intrinsic power spectrum is usually well above that of the matter fluctuations at the starting time, so that it introduces a large amount of stochastic noise and is gravitationally unstable even in the absence of perturbations. Other choices are regular Bravais lattices (a regular simple cubic grid being the most obvious choice), which are gravitationally stable, have no stochasticity, but are globally anisotropic. Higher-order Bravais lattices such as body-centered or face-centered lattices are more isotropic than the simple cubic lattice. A gravitationally stable arrangement with broken global symmetry can be obtained by evolving a Poisson field under a repulsive interaction \citep{White:1994} until it freezes into a glass-like distribution. The resulting particle distribution is more isotropic than a regular lattice, and has a white-noise power spectrum on scales smaller than the mean inter-particle separation, which decreases as $k^{4}$ on larger scales and therefore has more noise than a Bravais lattice. Also other alternative initial particle loads have been proposed, among them quaquaversal tilings \citep{Hansen:2007}, and capacity constrained Voronoi tessellations (CCVT, \citealt{Liao:2018}), both of which have a $k^4$ power spectrum. Example initial particle distribution for various cases are shown in Fig.~\ref{fig:initial_particle_load}. 

After the creation of the initial particle load, the displacement $\vecb{\Psi}(\vecb{q}_{1\dots N})$ and velocity fields $\dot{\vecb{\Psi}}(\vecb{q}_{1\dots N})$ are interpolated to the particle locations $\vecb{q}_{1\dots N}$, thereby defining the initial perturbed particle distribution with growing mode velocities at the simulation starting time. In the case of \textit{Bravais} lattices, the modes coincide directly with the modes of the DFT in the case of the simple cubic lattice, so that no interpolation is necessary, or can be obtained by Fourier interpolation. For the other pre-initial conditions typically CIC interpolation is used, c.f. Sect.~\ref{sec:mass-assignment-schemes}. Since CIC interpolation acts as a low-pass filter, the resulting suppression of power is usually corrected by de-convolving the displacement and velocity fields with the CIC interpolation kernel (Eq.~\ref{eq:cic_kernel}). 

Since the symmetry of the fluid is always broken at the particle discretisation scale, the specific choice of pre-initial condition impacts the growth rate and growth isotropy at the scale of a few particles. While in the Bravais cases, this deviation from the fluid limit is well understood \citep{Joyce:2005,Marcos:2006,Marcos:2008}, for glass, CCVT and quaquaversal tilings such analysis has not been performed. One expects that they are affected by a stochastic component in the growth rate with an amplitude comparable to that of the regular lattices. Such deviations of the discrete fluid from the continuous fluid accumulate during the pre-shell-crossing phase when the flow is perfectly cold so that over time the simulations, which are initialised with the fluid modes, relax to the growing modes of the discrete fluid.

%% dark matter candidates etc.
%% \newpage

\section{Beyond the cold collisionless limit: physical properties and models for dark matter candidates}
\label{sec:dark_matter}

So far, in this review article we have focused on the case where all the mass in the universe corresponds to a single cold and collisionless fluid. This is also the assumption of the vast majority of ``dark matter only'' or gravity-only simulations, and it is justified by the fact that gravity dominates long-range interactions, and that dark matter is the most abundant form of matter in the Universe ($\Omega_{\rm c}/\Omega_{\rm m}\approx 84\%$). 

In this section we discuss simulations where these assumptions are relaxed in several ways, either to improve the realism of the simulated system, or to explore the detectability and physics associated to the nature of the dark matter particle. In each case, we briefly discuss the physical motivation, its numerical implementation along with potential modifications necessary to the initial conditions, and summarise the main results.

In the first part of this section we will discuss simulations where dark matter is not assumed to be perfectly cold, but instead have a non-zero temperature as is the case when made out of WIMPs, QCD Axions, or generically Warm Dark Matter. We also discuss cases where dark matter is not a classical particle but made instead of ultra light axion-like particles, or of primordial black holes. We also consider cases where dark matter is not assumed to be perfectly collisionless, but instead display microscopic interactions, such in the case of self-interacting and decaying dark matter.

In the second part of this section we consider simulations that seek a more accurate representation of the Universe. Specifically, we discuss simulations where the mass density field is composed of two distinct fluids representing dark matter and baryons, and simulations that include massive neutrinos. For completeness, we also discuss simulations with non-Gaussian initial conditions and modified gravity.

\subsection{Weakly-interacting massive particles}
\label{sec:wimp}

\begin{figure}
\includegraphics[width=\textwidth]{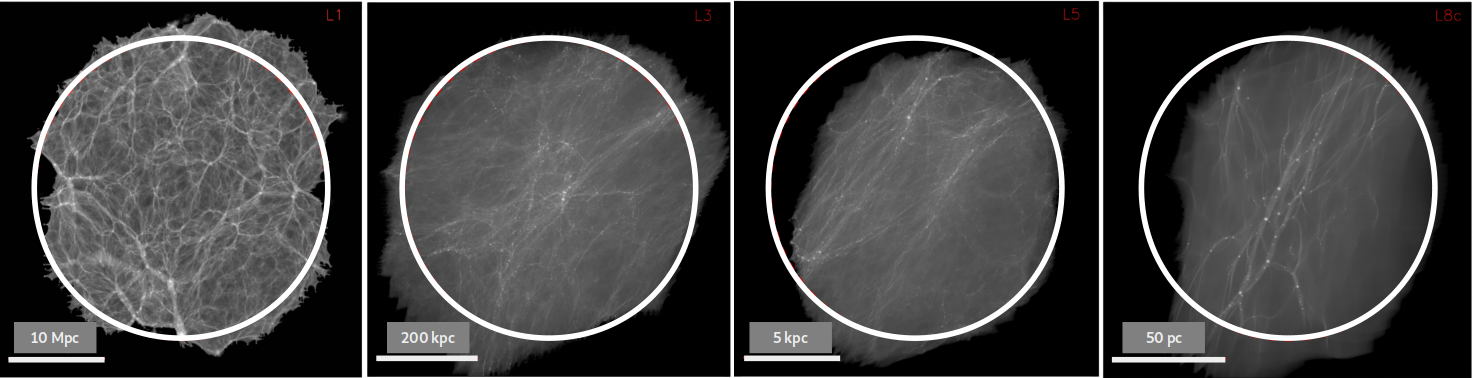} 
\caption{Progressive zooms onto smaller regions of a simulated nonlinear dark matter field at $z=0$. From left to right, each image shows a smaller region by factors of $5$, $40$, and $100$. Note that in the rightmost panel shows a region approximately $150$ pc wide where the smallest visible clumps corresponds to the smallest dark matter halos expected to form in an scenario where the dark matter particle is made out of $\gtrsim100$~GeV neutralinos. \label{fig:vvv} Figure adapted courtesy \cite{Wang:2020}.}
\end{figure}

Historically, the favoured candidate for dark matter has been a weakly-interacting massive particle (WIMP). WIMP is a generic name for a hypothetical particle with a very cold distribution function due to its early decoupling and high mass (in the GeV scale). For many decades, WIMPs were a strong contender to be the cosmic DM, motivated by the observation that if the cross-section was set by the electroweak scale, then it would result in a relic abundance matching the measured density of dark matter in the Universe. This coincidence has been termed as the ``WIMP miracle'' (see \citealt{Bertone:2004} for a classic review). However, the non-detection of supersymmetric particles at the LHC to-date \citep{Bertone:2010}, and the strong constraints from direct detection experiments begin to challenge the explanation of dark matter through thermally produced WIMP particles \citep{Roszkowski:2018}. Nevertheless, massive WIMPs remain compatible with all observational constraints and are one of the best motivated dark matter candidates. See \cite{Bertone:2018} for a recent review of various experimental searches.

A concrete example of a WIMP in supersymmetric extensions to the Standard Model is the lightest neutralino. These particles are stable and weakly interacting and should have a mass in excess of $\gtrsim 100~{\rm Gev}$. On astrophysical scales, neutralinos can be described as a perfectly cold collisionless fluid. However, the finite temperature at which these particles decouple implies that they have small but non-zero random microscopic velocities. As a consequence, neutralinos can stream freely out of perturbations of sizes $\sim0.7$pc which means that the formation of halos of masses $\lesssim10^{-8}\,M_{\odot}$ is strongly suppressed, and the typical mass of the first halos to collapse is about one Earth mass \citep{Hofmann:2001,Green:2004,Loeb:2005}.

For perfectly cold fluids, the distribution function reduces to a Dirac-$\delta$-distribution in momentum space, i.e., in any point in space there is a unique particle momentum. This corresponds to the ``single stream'' or ``monokinetic'' regime as usually referred to in fluid mechanics and plasma physics (see also the discussion in Sect.~\ref{sec:coldlimit_pt}). For a warmer fluid, such as that describing a neutralino field, the cold limit is also an excellent approximation for the distribution function. This is because thermal random velocities are small compared to the mean velocities arising from gravitational instability, specially at late times where the former are adiabatically cooled due to the expansion of the Universe whereas the latter keep increasing due to structure formation.

Consequently, numerical simulations of structure formation with neutralinos assume them to be perfectly cold, follow them with traditional $N$-body integration methods, and incorporate their free streaming effects only in the initial power spectrum suppressing the amplitude of small scale modes. Note however, that the very central parts of halos and caustics could be affected by intrinsic neutralino velocity dispersion, providing e.g., maximum bounds on the density.

One important challenge associated with these types of simulations is the huge dynamic range of the scales involved. For instance, to resolve the full hierarchy of possible structures, it would require about $10^{23}$ $N$-body particles. For this reason, neutralino simulations have focused on the formation of the first structures at high redshifts and over small volumes. Usually, these simulations involve zooming (cf.\ Sect.~\ref{sec:zoom_simulations}) into the formation of a small number of halos \citep{Diemand:2005,Ishiyama:2009,Anderhalen:2013,Ishiyama:2014,Angulo:2014}. Another alternative is to carry out a suite of nested zoom simulations \citep{Gao:2005b}, which has been recently extended to the free streaming mass by \cite{Wang:2020} by re-simulating low density regions embedded into larger underdensities and so on. A selection of projected density fields from this simulation suite are shown in Fig.~\ref{fig:vvv} which displays progressive zooms by factors of $\sim50$ in scale, from 10\,Mpc to 50\,pc.

There is a consensus from all these simulations that the first microhalos have a mass of about $10^{-6}\,M_{\odot}$ and start collapsing at a redshift of $z\sim300$. At those mass scales and resfhifts, structure formation is different than at galactic scales: the spectrum of CDM density fluctuations has a slope close to ${-3}$, i.e. $P(k) \propto k^{-3}$, which causes a large range of scales to collapse almost simultaneously. This also implies that the formation of the first microhalos is immediately followed by a rapid mass growth and many major mergers. Whether these microhalos can survive tidal forces inside Milky-Way halos is still an open question, which could have implications for the detectability of a potential dark matter self-annihilation signal.

Another, perhaps unexpected, outcome of these simulations is that the density profile of these microhalos is significantly different from that of halos on larger mass scales. Ever since standard CDM simulations (i.e., those without any free-streaming effects) reached adequate force and mass resolution, they revealed that the internal density profiles of collapsed structures were well described by a simple functional form \citep{Navarro:1997}, referred to as NFW profile, in terms of a dimensionless radial coordinate $x := r/r_s$ and density $\varrho := \rho(r) / \rho_0$ (where $r_s$ and $\rho_0$ are parameters that vary from halo to halo)
\begin{equation}
\label{eq:NFW}
\varrho(x) = \frac{1}{x (1 + x^2)}%\quad\textrm{, with concentration}\quad c:=r_s/R_{\rm vir},
\end{equation} 
regardless of the mass of the halo, cosmic epoch, cosmological parameters, etc. \citep{Ascasibar:2008,Neto:2007,Brown:2020}. Despite its importance and several proposed explanations that involve, for instance, a fundamental connection with the mass accretion history \citep{Ludlow:2014,Ludlow:2016}, maximum entropy or adiabatic invariance arguments \citep{Taylor:2001,Dalal:2010,Pontzen:2013,ElZant:2013,Juan:2014}, or even the role of numerical noise as the main driver \citep{Baushev:2015}, there is not yet a consensus on the physical explanation behind this result. This is even more puzzling when contrasted with analytic predictions, which suggest single power laws as the result of gravitational collapse (see e.g.,  the self-similar solution of secondary infall by \citealt{Bertschinger:1985}).

In contrast, most neutralino simulations find that the initial internal structure of microhalos is better described by a single power law profile $\sim r^{-1.5}$, as first pointed out by \cite{Diemand:2005} \citep[see also][]{Ishiyama:2009,Anderhalen:2013,Ishiyama:2014,Ogiya:2016,Angulo:2017,Delos:2019b}. This very steep profile would make microhalos very resilient to tidal disruption by the Milky Way or by binary stars and also would enhance their corresponding emission from dark matter self annihilation, making them potentially detectable by future experiments \citep{Diemand:2006,Ishiyama:2009,Ishiyama:2014,Delos:2019a}, although their abundance is affected by free streaming \citep{Ishiyama:2020}. This power-law profile, however, quickly mutates to a NFW-like profile for higher mass halos. Several authors have argued that subsequent mergers drive this transformation and determine the final density profile \citep{Ogiya:2016,Angulo:2017}. These results have been further supported by \cite{Colombi:2020} using the idealised collapse of three sine waves simulated with a Lagrangian tessellation algorithm. In contrast, \cite{Wang:2020} which is the only neutralino-like simulation that has reached $z\sim0$, finds density profiles consistent with an NFW shape. 

These inconsistencies among simulations leave the question about the `true' density profiles of the first halos forming in a WIMP dark matter model currently unanswered. The role of numerical artefacts vs. physical processes (such as the dynamical state of halos, their mass accretion histories, or environmental effects) still need to be better understood. The alternative discretization methods discussed in Sect.~\ref{sec:discretizations} might play an important role in this. 

\subsection{Axions}

\begin{figure}
\includegraphics[width=\textwidth]{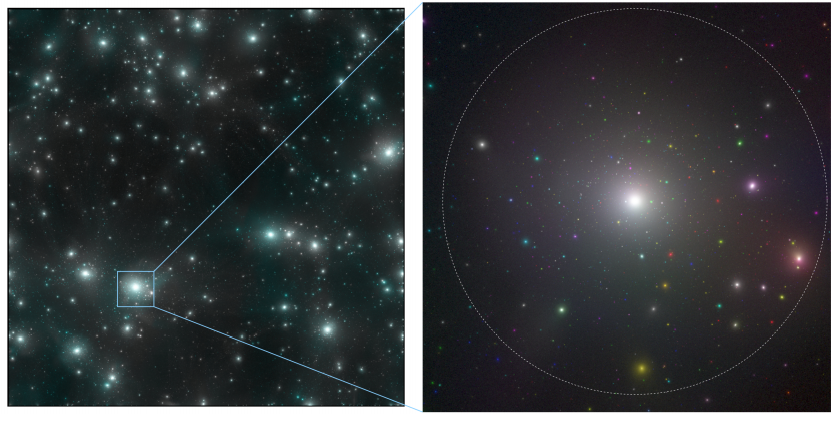} 
\caption{The projected mass field at $z=99$ in a cosmological simulation assuming the dark matter is composed of QCD axions. The left panel shows the whole simulation box of 0.86\,pc, whereas the right panel zooms into the largest axion minicluster of mass $\sim10^{-9}\,\Msun$ where the white circles denotes the radius enclosing a mean density 200 times the background value. Note that, unlike most cosmological simulations, there is a lack of filamentary structure owing to the scale-independent power spectrum of fluctuations expected in the primordial axion field on the scales displayed by the left panel. Figure courtesy of \cite{Eggemeier:2020}. \label{fig:axion}}
\end{figure}

The QCD (Peccei--Quinn) axion is another hypothetical particle, originally proposed to solve the strong charge-parity problem  \citep[see][for reviews]{Duffy:2009,Marsh:2016}, that has been considered as a possible dark matter candidate. Axions, despite being very light ($m \sim 10^{-5}$--$10^{-3}\,{\rm eV}$), are extremely cold due to non-thermal production. Therefore they gravitationally cluster on small scales and behave as cold dark matter on cosmological scales.

As for WIMPs, primordial fluctuations in the axion field exist down to very small scales up to an eventual smoothing produced by free streaming velocities. In the case of axions, this occurs on sub-parsec scales. The smallest structures are expected to be ``mini halos'' of masses $10^{-12}\,\Msun$ (set roughly by the mass inside the horizon when the axions become nonrelativistic) and $10^{12}\,{\rm cm}$ radii. These minihalos are even smaller than those in neutralino cosmologies and would form even earlier: their typical collapse is expected to occur during the radiation dominated epoch of the Universe at $z\sim1000$. If a large fraction of axion miniclusters survive tidal disruption during structure growth, they could be potentially detected in femto- and pico-lensing experiments \citep{Kolb:1996,Fairbairn:2018}, or alternatively, the resulting tidal streams could impact indirect and direct detection in cavity experiments seeking to detect the transformation of axions into photons \citep{Tinyakov:2016,OHare:2017,Knirck:2018}. Therefore, numerical simulations are required to explore structure formation and evolution of axion structures.

Analogously to WIMPs and neutralinos, cosmological simulations including axions assume them to be in the cold limit, i.e., to have zero velocity dispersion at the starting redshift. The subsequent gravitational evolution can be followed by standard $N$-body codes and algorithms, but because the first halo collapse is expected to occur on extremely small mass scales and even before the matter-radiation equality, simulations typically have sizes of $\sim1-10\hkpc$ and evolve from $z_{\rm start} \sim 10^6$ up to  $z \sim 100$ where the fluctuations on the scale of the box start to become nonlinear.

The initial conditions of axion simulations are very different to that of standard dark matter simulations. Whereas most dark matter fluctuations are given by adiabatic fluctuations on all scales, the initial distribution of axions (on the small scales targeted by axion simulations) is set by the formation and decay of topological defects and axion self interactions after the QCD phase transition at the end of inflation. Thus, the initial distribution is not predicted by e.g., Boltzmann codes but has to be followed by QCD lattice simulations \citep{Vaquero:2018,Buschmann:2019}. Because of these complications, only recently the first self-consistent cosmological simulation of the formation of structure in an axion dark matter field was carried out \citep{Eggemeier:2020}. On large scales, however, different patches of the universe would have been uncorrelated, and thus it is common to assume the statistics to be given by a white-noise power spectrum whose amplitude is set to match QCD axion simulations, and with a cut-off at modes that entered the horizon when the axions were relativistic \citep{Xiao:2021}. 

These simulations find that indeed axion miniclusters form at very high redshifts ($z>150'000$) with an abundance that continuously grows covering a broad range of masses, $10^{-15}$--$10^{-8}\,\Msun$ at $z\sim100$. The abundance of axion miniclusters can be reasonably well approximated by analytic arguments based on Press--Schechter and Excursion set theory with the appropriate spectrum of fluctuations. Such calculations allow an estimate of the abundance of these axion miniclusters down to redshifts beyond the end of simulations. These axion miniclusters have density profiles in reasonably good agreement with an NFW profile and extremely high concentrations $c \sim 100-400$, consistent with the classical picture where concentration reflects the density of the universe at collapse.

These simulations make interesting predictions for the abundance and internal properties of miniclusters at high redshifts. However, uncertainties exist that are related to the survival of these halos to redshift zero and in an environment such as that of the Milky Way in terms of the host halo tidal field and stellar encounters. While perhaps the recursive-zooming techniques discussed in the previous section could be extended to the axion scales, it would be a formidable computational challenge. Another alternative could be to explore the impact of the tidal field of the Milky Way halo using, for instance, idealised simulation techniques that have been employed to study the survival of CDM halos. Combining these tools it will be possible to perform robust prediction for upcoming observations sensitive to the axion structure.

\subsection{Warm Dark Matter (WDM)}
\label{sec:wdm}

For both WIMPs and QCD axions, their microscopic nature has essentially no implications for the properties of galaxies and for the observable large-scale scale structure of the Universe. However, dark matter could be a particle whose mass and free streaming scale is comparable to that of galaxies. Such particles are generically referred to as warm dark matter and a concrete examples are sterile neutrinos or gravitinos.

Additional motivation for considering dark matter alternatives, such as WDM, is that the physical properties of DM could offer a solution to the tension between the predictions of $\Lambda$CDM and observations of the abundance, spatial distribution, and internal properties of dwarf galaxies \citep{Klypin:1999,Moore:1999,Zavala:2009,Power:2013,Papastergis:2011,Papastergis:2015,Klypin:2015,Boylan-Kolchin:2011}. Although baryonic physics and the effects of galaxy formation are found to be able to resolve them even within CDM (or at least ameliorate the tensions) \citep{Navarro:1996,RomanoDiaz:2008,Pontzen:2012,Brooks:2013,Sawala:2016,Brook:2015,Chan:2015}, there is no consensus on the inevitability nor the magnitude of these effects \citep{Oman:2015}.

Constraining the mass of a hypothetical warm dark matter particle is an active area of research. Currently, the most competitive methods are: i) the abundance of Milky Way satellites, as measured by extragalactic surveys, ii) the small-scale properties of intergalactic gas as measured by the Ly-alpha forest, and iii) the abundance and properties of small halos and subhalos as measured by perturbations to strong lensing systems. All these methods have different statistical power and systematics, yet they agree on an upper limit for the dark matter free streaming scale, which currently translates to a mass of about $m_{\chi} \gtrsim 6~{\rm keV}$ for a thermal relic \citep{Enzi:2020}.

Since all these observations involve nonlinear structures, predictions from numerical simulations are crucial for the robustness of these constraints. However, as we will discuss below, simulations of WDM structure formation have encountered serious challenges which are only now starting to be overcome.

As for the case of WIMPs, numerical simulations describe warm DM as a perfectly cold fluid but with an initial power spectrum modified to account for the small-scale power suppression due to free streaming effects, erasing small-scale fluctuations below a characteristic free-streaming scale. These initial thermal velocities will decay adiabatically so that they become irrelevant compared to gravitationally induced velocities during structure formation. For instance, even for a relatively warm particle of $m_{\chi} \sim 250\,{\rm eV}$, the velocity dispersion is expected to be 0.28\,km/s, to be compared to about 100 to 1000~km/s of typical velocity dispersion in galactic-sized halos. (Nevertheless, there have been attempts to directly sample the WDM velocity in $N$-body simulations, however, this appears to introduce significant numerical noise, see e.g., \citealt{Colin:2008,Maccio:2012,Leo:2017} and our discussion regarding neutrinos in Sect.~\ref{sec:HDM}.) While small during the time of the collapse of structures, the free streaming velocities impose phase-space constraints \citep{Tremaine:1979}, preventing the formation of arbitrarily high densities in caustics and halo centers, in contrast to perfectly cold CDM.

The most important signature of free streaming velocities in structure formation is imprinted as a suppression in the transfer function relative to CDM. There have been several parameterisations for this effect. One of the most widely-used is \cite{Bode:2001} for a WDM particle of mass $m_\chi$ and number of degrees of freedom $g_\chi$
\begin{eqnarray}
 T_{\rm WDM}(k) &=& T_{\rm CDM}(k) \times \left[1 + (\alpha\,k)^{2\nu}\right]^{-5/\nu} \label{eq:WDM_transfer_Bode}\\
\textrm{where}\quad\alpha &=& 0.05 \left(\frac{\Omega_{\rm m}}{0.4}\right)^{0.15}\times \left(\frac{h}{0.65}\right)^{1.3}  
\left( \frac{m_{\chi}}{1\,{\rm keV}}  \right)^{-1.15}  \left(\frac{1.5}{g_\chi}\right)^{0.29}\,h^{-1}{
\rm Mpc}, \nonumber
\end{eqnarray}
which is based on a fit to a linear Einstein-Boltzmann calculation, and where $\nu\simeq1-1.2$, but  see also \cite{Schneider:2013,Viel:2005} for other popular alternatives. 

\begin{figure}
\includegraphics[width=\textwidth]{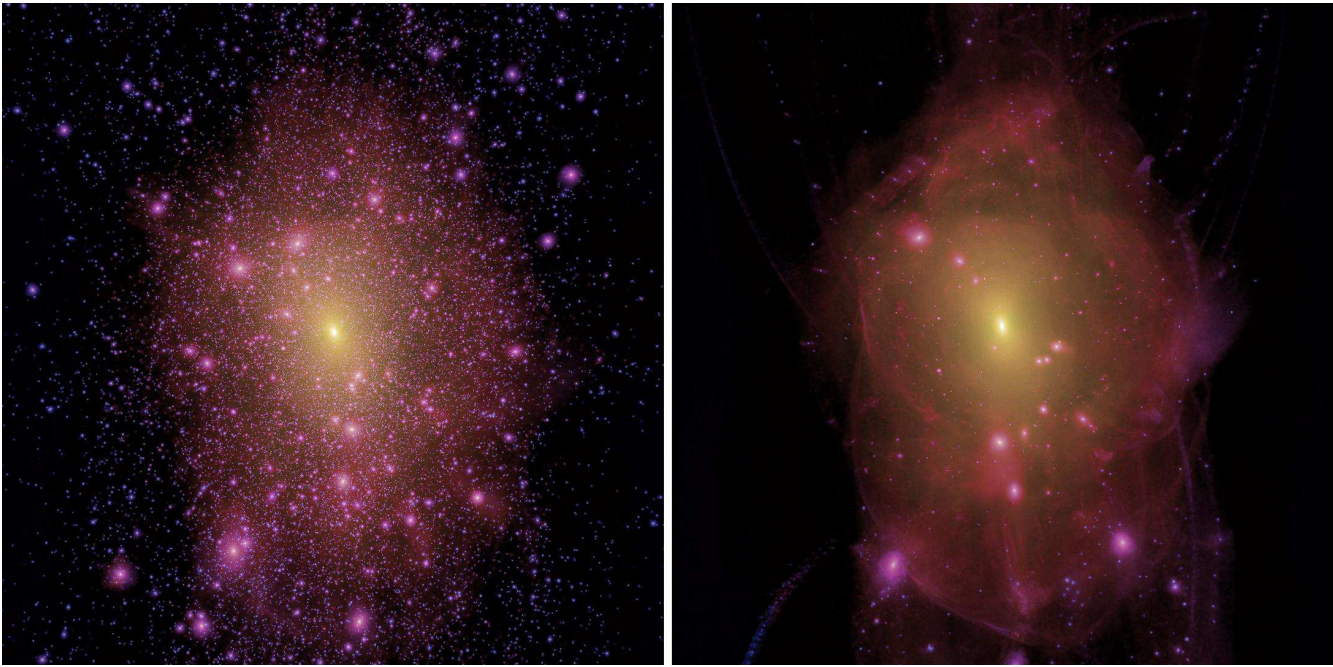} 
\caption{The squared density field of a simulated halo with mass $\sim1.2\times10^{12}\Msun$ assuming a cold dark matter model in the left panel; and a warm dark matter in the right panel. In this WDM model, fluctuations below $k = 4.5\ihMpc$ are suppressed, which consistent with dark matter being made of resonantly-produced sterile neutrinos of mass 2\,keV. Note that the suppression of small scale fluctuations induced by free streaming of warm dark matter results into a lack of low mass subhalos but caustics in the halo outskirts become more visible. Figure adapted courtesy \cite{Lovell:2012}. \label{fig:aquarius_wdm} }
\end{figure}

Warm DM simulations have shown that the abundance of halos decreases exponentially with respect to that in CDM for scales below the half-mode mass, defined as the scale where the transfer function decreases by a factor of 2 with respect to  CDM. The abundance of subhalos also is strongly suppressed, though in a smoother fashion owing to the mixture of mass scales caused by tidal stripping, i.e., a broad range of halo masses give origin to any given subhalo mass scale. The lack of small scale structure also leads to a later collapse of halos, which simulations have found is inherited as a reduction in the concentration of halos. These effects can be appreciated in Fig.~\ref{fig:aquarius_wdm} which compares a zoom simulation of a Milky-Way sized halo assuming a perfectly cold dark matter and a $\sim2$\,keV WDM particle. The lack of small scale structures and low mass subhalos in the WDM case, as discussed above, is plainly visible. Finally, another interesting question concerns the internal structure of WDM halos near the free streaming scale. As in the case of neutralino simulations, simulations have found indications that WDM halos have inner density profiles that are steeper than their CDM counterparts \citep{Colin:2008,Polisensky:2015,Ogiya:2018,Delos:2019b}.

\begin{figure}
\includegraphics[width=\textwidth]{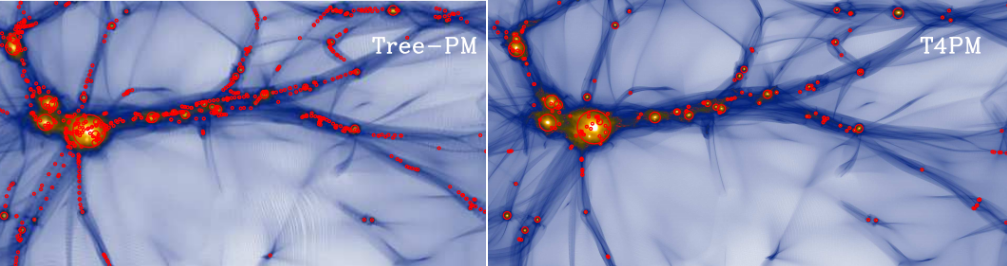} 
\caption{ The simulated density field of Warm Dark Matter simulations illustrating the ``artificial fragmentation'' problem. The left panel shows the result of a standard $N$-body simulation where dark matter halos are highlighted as red circles. Note that filaments are broken up into pieces, which are referred to as ``artificial fragmentation'', whose abundance  increases high with higher resolution. These fragments originate from discreteness errors in $N$-body simulations, which are considerably suppressed in calculations employing phase-space tessellations as is shown in the right hand side panel. \label{fig:fragments} Figure adapted courtesy \cite{Angulo:2012}.}
\end{figure}

Warm Dark Matter simulations are also interesting from a theoretical point of view, since the full hierarchy of nonlinear objects is within reach of a single numerical simulations. Thus, they provide an ideal test-bed for numerical convergence and the accuracy of simulations in general. This is in stark contrast to standard $N$-body CDM simulation where the smallest resolved scale is always set by numerical limitations rather than physical processes, i.e., the simulated transfer has an effective small-scale cut-off at the scale of the mean inter-particle separation.

Unfortunately, WDM simulations have revealed a serious shortcoming of the $N$-body method. Ever since the first WDM simulations, they showed the presence of a population of low-mass halos of purely numerical origin. These \emph{artificial fragments} dominate by factors of 100--1000 in number over those of physical origin \citep{Klypin:1993,Avila-Reese:2001,Bode:2001,WangWhite:2007,Angulo:2012}. This problem is ameliorated, but only slowly, with increasing mass resolution, roughly as $N_{\rm p}^{1/3}$ \citep{WangWhite:2007,Schneider:2013}. An illustration of these fragments is shown in Fig.~\ref{fig:fragments}, where we can see in the bottom left region of the image that a vertical filament has been split into small halos, highlighted by the red circles. These fragments likely originate from correlated and highly anisotropic force errors driven by particle discreteness in regions of anisotropic collapse \citep{Hahn:2013,Power:2016}.

There have been several attempts to solve the problem of artificial fragmentation. At a practical level, there are proposals based on identifying and removing artificial halos according to their Lagrangian properties or by comparing the results of simulations at different mass resolutions \citep{Lovell:2014,Schneider:2013}. At a more fundamental level, there are proposals to improve simulation techniques and directly prevent the formation of these spurious micro-halos. Although with limited success, these proposals include adaptive softening based on the local density \citep{Power:2016} and anisotropic softening based on the local moment of inertia \citep{Hobbs:2016}. Another interesting idea was proposed by \cite{Myers:2016} who showed that the key to accurate $N$-body simulations is a decorrelation in time of discretization errors. This was achieved by adding an artificial velocity dispersion to particles and periodically remapping the set of $N$-body particles. To our knowledge, however, this idea has only been investigated in up to 2D and in idealised test problems. Among all the proposed solutions, perhaps the most successful has come from the reformulation of the $N$-body method in terms of Lagrangian phase-space elements (see Sect.~\ref{sec:tessellation_methods}). 

In the context of Warm Dark Matter, \cite{Angulo:2013b} showed that by using these Lagrangian methods it was possible to properly quantify the mass function below the cut-off scale. More recently, \cite{Stucker:2021} extended these results as a function of the properties of the cut-off in the initial transfer function by using a hybrid Lagrangian-$N$-body approach \citep{Stuecker:2019}, where dynamically simple regions, such as filaments, voids, and sheets are evolved using a phase-space interpolation whereas halos are simulated using $N$-body discretization. The lack of artificial fragments in these simulations have also allowed to investigate in more detail the observability of warm dark matter in general and in lensing in particular \citep{Richardson:2021}. 

One might be concerned that artificial fragments could affect the properties of larger halos that have accreted them (e.g., by gravitationally heating their centers). These new generation of Lagrangian simulations are, however, demonstrating that these effects are small and that the $N$-body method is reliable for halos whose mass is above those dominated by artificial fragmentation. This is an important validation for current constraints on the WDM particle mass, which heavily rely on the correctness of $N$-body simulations.

%%%%%%%%%%%%%%%%%%%%%%%%%%%%%%%%%%%%%%%%%%%%%%%%%%%%%%%%%%%%

\subsection{Fuzzy Dark Matter: Quantum Hamiltonians and condensates}
\label{sec:fuzzy_dm}

A particle can no longer be treated as classical in cosmological context if it is so light that its de~Broglie wavelength corresponds to astrophysical scales. If the particle has a high enough number density, then occupation numbers will be so high that it can form a Bose--Einstein-Condensate. Specific examples are generic ultra-light bosons with a wave length of $\sim10$~kpc \citep{Press:1990,Frieman:1995} and ultralight axions with masses $m\gtrsim 10^{-23}{\rm eV}$ arising from string theory, e.g., \cite{Svrcek:2006}. All these candidates are generically referred to as `Fuzzy Dark Matter' (FDM) and have the distinctive property that they display genuine quantum effects on astrophysical scales (see \citealt{Marsh:2016,Hui:2021,Ferreira:2021} for recent reviews).

These quantum effects lead to a suppression of small scale structure (due to an effective ``quantum pressure''), and as a consequence claims have been made that FDM can alleviate various of the small-scale problems of CDM mentioned in Sect.~\ref{sec:wdm}, qualitatively similar to WDM (cf.\ \citealt{Hui:2017} for an overview). On the other hand, constraints from comparing hydrodynamical simulations with observations of the Ly-$\alpha$ forest \citep{Irsic:2017b,Armengaud:2017} currently rule out masses smaller than $10^{-21}$~eV, which is higher than what is needed to solve these problems \citep{Kobayashi:2017}. Note, however, that these simulations only model the effects of FDM on the initial transfer function and neglect the effect of quantum pressure in their dynamics, thus there is a debate whether this could affect these constraints \citep{ZhangQP:2018,Armengaud:2017,Nori:2019}. Nonetheless ultralight axions remain an interesting dark matter candidate with a distinct signature due to the associated large-scale quantum effects, and that could co-exist together with other forms of dark matter. As for other DM candidates, the observable signatures of FDM are located in the non-linear regime, thus accurate simulation is essential to rule out or eventually constrain the properties of FDM.

\begin{figure}
\begin{centering}
\includegraphics[width=0.59\textwidth]{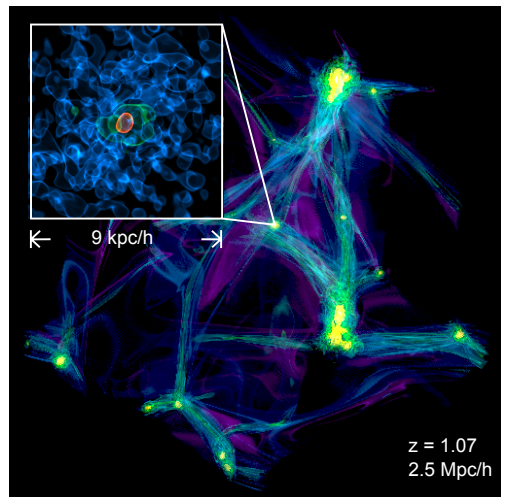} 
\includegraphics[width=0.4\textwidth]{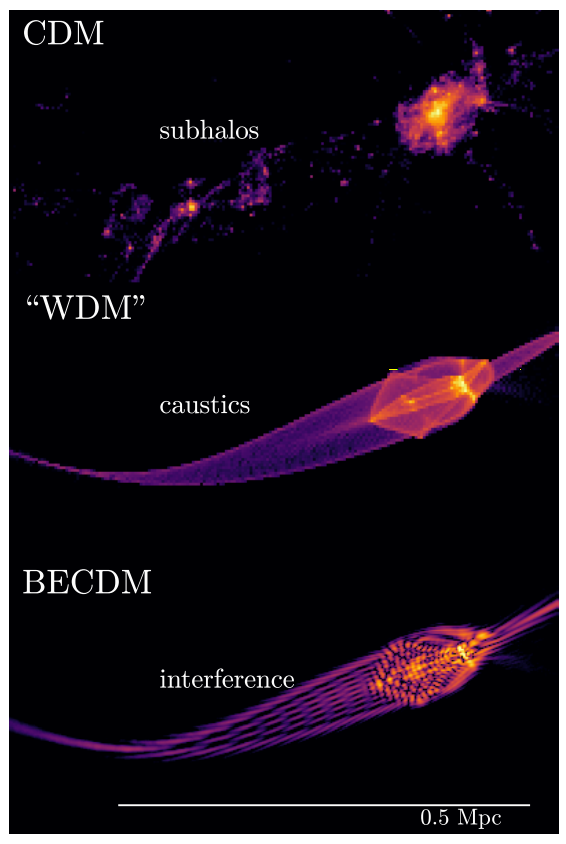} 
\caption{The simulated cosmic density field when dark matter is assumed to be made of \textit{Fuzzy Dark Matter} (FDM), where quantum effects are important on astrophysical scales. The left panel shows a simulation at $z\sim1$ from \cite{Veltmaat:2018}, whose inset zooms into a collapsed structure where a granular behaviour and a dense solitonic core are evident. The right panel shows a comparison of a filament at $z \sim 7$ simulated assuming CDM, WDM, or FDM (or BECDM), from top to bottom. Note that the filament is broken into small low-mass halos in CDM, whereas in WDM and FDM such small-scale structure is absent due to the smoothing effects of free-streaming velocities and quantum effects, respectively. In addition, in the FDM case, an interference pattern appears as a result of multi-streaming in collapsed regions. Figured adapted courtesy of \cite{Veltmaat:2018} and \cite{Mocz:2020}. \label{fig:fdm}}
\end{centering}
\end{figure}

Fuzzy dark matter is described by the Schr\"odinger-Poisson (SP; also known as Gross-Pitaevskii-Poisson in the context of condensates) system for a single complex wave function $\psi$. This is the non-relativistic limit of a Klein-Gordon field after averaging over rapid oscillations of the scalar field, leaving two scalar degrees of freedom (amplitude and phase). The resulting system is typically expressed through a Hamiltonian of the form
\begin{equation}
\hat{\mathscr{H}} = \frac{\hat{p}^2}{2m a^2} + \frac{m}{a}\hat{V} + \frac{\lambda m}{a^{3}} (1+\delta) \quad\textrm{with}\quad  \nabla^2 V = \frac{3}{2}H_0^2 \Omega_m \delta, \quad\textrm{and}\quad \delta =\left| \psi \right|^2-1, \label{eq:Gross-Pitaevskii}
\end{equation}
where $\lambda$ represents a possible self-coupling of the field, and the expansion of the Universe has already been subtracted out as for the co-moving Vlasov--Poisson system. Alternatively, it is possible to transform the SP equations into fluid equations which resemble those of a classical fluid with an additional `quantum pressure' term so that: 
\begin{equation}
\label{eq:quantum_pressure}
\frac{\partial\vecb{u}}{\partial t} +\frac{1}{a}(\vecb{u}\cdot\bnabla) \vecb{u}= -\frac{1}{a} \bnabla V - \frac{1}{a^3} \bnabla Q \quad\textrm{with}\quad Q := - \frac{\hbar}{2m} \frac{\nabla^2 \sqrt{\rho}}{\sqrt{\rho}}
\end{equation}
where $m$ is the axion mass, $\rho:=1+\delta=|\psi|^2$ the density, and $\vecb{u}:=\frac{\hbar}{m}\bnabla {\rm arg}(\psi)$ is the velocity. $Q$ can be regarded as a `quantum pressure' term (prominent in Bohmian quantum mechanics) and is in general singular wherever quantum interference effects are present (since phase jumps, ${\rm arg}(\psi)+2\pi$, are accompanied by $\rho\to0$). This equation together with the continuity equation for $\rho$ are also known as the  \textit{Madelung} formulation \citep{Madelung:1927} of quantum mechanics. 

The first aspect to consider in FDM numerical simulations is the effect on the initial conditions. At linear order, $Q\simeq-\frac{\hbar}{4m}\nabla^2\delta$ which introduces an effective Jeans length below which structure growth is suppressed:
\begin{equation}
k_{\rm J} = \left(\frac{16 \pi G \rho_{\rm b} m^2}{\hbar^2}\right)^{1/4} \underset{a = a_{\rm eq}}{\simeq} 9\, {\rm Mpc^{-1}}\; \left( \frac{m}{10^{-22} eV} \right)^{1/2},
\end{equation}
where $\rho_{\rm b}$ is the physical mater density, and $a_{\rm eq}$ is the expansion factor at matter-radiation equality. The exact effect on the initial transfer function can be computed directly including the relevant modifications to Boltzmann solvers (e.g., {\sc AxionCAMB}, \citealt{Hlozek:2015}), or approximated as
\begin{equation}
T_{\rm FDM}(k) \approx T_{\rm CDM}(k) \times \left(\frac{\cos x^3}{1+x^8} \right)
\end{equation}
where $x \simeq 1.61 (m/10^{-22})^{1/18} k/k_{\rm J}(a_{\rm eq})$ \citep{Hu:2000}.

There are essentially two approaches in the literature to follow the nonlinear evolution of the axion field. The first one is to directly solve the associated Schr\"odinger equation for the complex field $\psi$. Essentially, these codes adopt Eulerian discretisation schemes where the axion field is described on a (static or moving, fixed or adaptive) grid \citep{Schive:2014,Schwabe:2016,Mocz:2017,Edwards:2018,Veltmaat:2018,Mina:2020,May:2021}. We have discussed schemes for the numerical integration in Sect.~\ref{sec:quantum_integration}. The integrator receives  modifications if self-interactions of the field ($\lambda\neq0$) are included. Spectral Eulerian approaches are limited by their lack of spatial adaptivity, so that all recent Eulerian schemes resort to finite difference approximations with AMR or hybrid schemes.

An arguably simpler approach has been followed by other authors by resorting to the Madelung formulation and including the quantum pressure term using an SPH-based estimate \citep{Veltmaat:2016,Mocz:2015,Nori:2018,Zhang:2018,Hopkins:2019}. Due to the singular nature of the quantum pressure term, this approach will have difficulty to resolve quantum interference effects, which has put some doubts on the results (but see below). During phases of smooth evolution without topological changes \citep{Mocz:2015} have shown that SPH methods can capture the evolution accurately. This technique has been successfully employed to obtain the first FDM constraints from the Ly-$\alpha$ forest \citep{Irsic:2017,Nori:2019} (arguably since these constraints are more sensitive to the suppression of perturbations already in the initial conditions rather than details of the structure in collapsed regions).

The large-scale dynamics in an FDM universe is identical to CDM since Schr\"odinger--Poisson becomes Vlasov--Poisson in the $\hbar/m\to0$ limit (cf.\ \citealt{Widrow:1993,Zhang:2002}), and thus, FDM should resemble the filamentary structure of CDM. Numerical simulations have confirmed this and illustrations of the qualitative agreement can e.g., be found in \cite{Uhlemann:2014,Kopp:2017,Mocz:2018}. On smaller scales, structure formation is suppressed in a way that resembles a WDM cosmology with a free streaming length comparable to the FDM effective Jeans length. On even smaller scales, two effects appear. The first one is the presence of a central long-lived solitonic core in the centers of halos \citep{Schive:2014} whose properties are correlated with the mass of the host halo \citep{Chavanis:2011,Chavanis:2011b,Chen:2017,Bar:2018}. The existence of this central mass excess has been confirmed via Eulerian and Lagrangian simulations, and its presence or absence in observations of galaxies has been used to argue in favor or against FDM \citep{Desjacques:2019,deMartino:2020,Pozo:2020,Burkert:2020}.

Another key signature of FDM on small scales is the presence of a distinctive granular structure associated with interference patterns and variations in density. This, for instance, can be appreciated in Fig.~\ref{fig:fdm}. This feature appears to be a prediction in all Eulerian simulation codes, however, it is absent in Lagrangian ones. The reason behind this is that the Madelung formulation formally diverges when $\rho \rightarrow 0$, which introduces a singularity that limits the ability of the corresponding numerical methods to correctly capture the behavior of the system. It is thus remarkable that the solitonic cores can be reproduced also in Madelung simulations, which is likely a consequence of them arising from hydrostatic equilibrium in the thermodynamic $\rho\gg1$ regime, unaffected by errors arising in earlier, colder, stages of collapse.

Regardless of such numerical problems, there is consensus that interference patterns should appear in FDM and, in fact, they could provide a clear evidence of  its existence. For instance, the granularity and the rapid oscillations of the Klein--Gordon field could perturb strong lenses and/or affect the frequency of light from pulsars in a way that could be detected by Pulsar Timing Arrays \citep{Khmelnitsky:2014,Porayko:2014,deMartino:2017}, or from binary pulsars\citep{Blas:2017}. Current constraints are still weak ($m > 10^{-23}$ eV) \citep{Porayko:2018,Kato:2020}, but the next generation of PTAs and multiple other probes proposed should significantly improve upon these limits, which together with future advances and improvements in numerical simulations, could scrutinise this interesting DM candidate.

%%%%%%%%%%%%%%%%%%%%%%%%%%%%%%%%%%%%%%%%%%%%%%%%%%%%%%%%%%%%

\subsection{Primordial black holes}

\begin{figure}
\includegraphics[width=\textwidth]{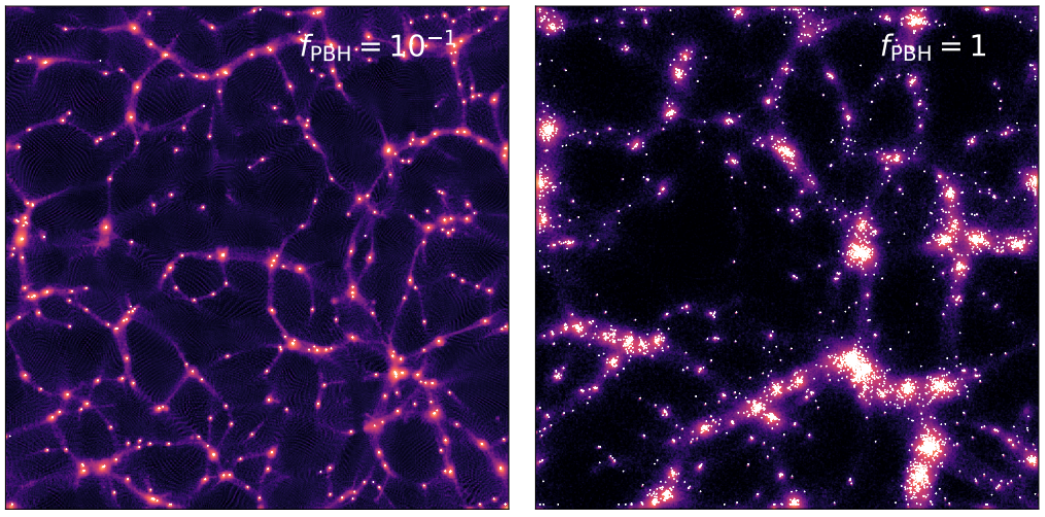} 
\caption{The projected matter distributionat $z=99$ as predicted by numerical simulations assuming primordial black holes (PBH) make up 10 or 100\% of the dark matter (left and right panels, respectively). The simulations correspond to box sizes of $30\hkpc$ and assume $20\Msun$ as the mass of the PBHs. Note that for larger PBH fractions, structures collapse earlier containing multiple PHB, whereas for low fractions, PBH are mostly founds in isolation but are surrounded by ``standard'' dark matter particles. Figure adapted courtesy \cite{Inman:2019}.  \label{fig:pbh}}
\end{figure}

Even though large-scale primordial fluctuations are found to follow a $k^{n_s\simeq0.96}$ nearly scale-invariant spectrum from the CMB and observations of the large-scale structure of the Universe, small scales are essentially unconstrained. It is therefore possible that large energy-density fluctuations on small scales were generated during inflation. These fluctuations could have collapsed to form a population of black holes \citep{Carr:2016,Carr:2020} as soon as they enter the horizon as early as at the matter-radiation equality. 

An interesting possibility is that these primordial black holes (PBH) could make up a significant fraction (or possibly all) of the dark matter without recourse to particles beyond the standard model. This possibility is even more interesting given the recent detection of many black hole mergers detected by LIGO, with measured black hole masses following an astrophysically unexpected distribution. Recent lattice QCD calculations \citep{Borsanyi:2016} now predict the equation of state during the QCD phase transition, which can lead to distinct features in the PBH mass spectrum \citep{Byrnes:2018,Carr:2021}.

Various astrophysical observations have put limits on the abundance and mass of these PBHs. The lower mass limit is determined by requiring the evaporation time due to Hawking radiation to be larger than the Hubble time, which puts the minimum mass to make up all of the dark matter at $m_{\rm PBH}\gtrsim10^{15}\,{\rm g}$. Up to stellar masses, PBH dark matter is severely constrained by astrophysical observations (microlensing, close binary disruption), at higher than stellar masses it is constrained by allowed spectral distortions of the CMB. Current constraints have ruled out PBH of virtually any mass as the total dark matter with a monochromatic PBH mass function, except for a window around $10^{-16}-10^{-10}\,{\rm M_{\odot}}$ \citep{Carr:2020c}, which narrows to a small region around $10^{-16}\,{\rm M_{\odot}}$ if constraints from white dwarf and neutron star disruptions are included. However, a broad spectrum of PBH masses would render them unconstrained by any single observation. Furthermore, since the physics behind PBH formation and that behind an hypothetical DM particle are in principle unrelated, it is possible that both types of dark matter coexist, but CDM clustering around massive PBHs does constrain that scenario \citep{Adamek:2019b,Carr:2020b}.

The evolution of PBHs is determined by collisional dynamics, thus it can be correctly captured by traditional $N$-body codes for small or zero softening length. Due to the strong accelerations possible in pair encounters and the need to resolve the formation and disruption of PBH binaries, integration techniques known from star cluster simulations need to be employed \citep{Aarseth:2009}, but traditional integrators are sufficient when one is interested in the large-scale distribution of PBHs.

Concerning initial conditions, on sufficiently large scales, PBH fluctuations follow that imprinted in the particle dark matter (CDM), while on small scales the PBHs' distribution can be assumed to be Poissonian. Recent numerical simulations of the particle DM and PBH have employed standard $N$-body codes \citep{Inman:2019,Tkachev:2020}, usually augmented with models for BH-BH mergers. In Fig.~\ref{fig:pbh}, we show the density field at $z=99$ of one of such simulations assuming that 10\% or 100\% of the DM is in the form of PBHs. For low DM fractions in the form of PBHs, \cite{Inman:2019} found that DM halos form around single a PBH which have steep power-law density profiles. As they consider higher PBH fractions, halos contain a larger number of PBHs and display broken power laws. In addition, the formation of small halos occurs earlier the larger the PBH fraction, which could have observable consequences (in e.g., the formation of first stars or reionisation). Currently, these simulations are limited by computational resources. In the future, one can expect that these simulations will be able to tackle questions such as the formation of PBH binaries already in the radiation-dominated epoch, potentially informing the fraction of PBH mergers detectable by future gravitational wave experiments.

Note that if the small-scale fluctuations produced by inflation are not large enough to collapse into PBHs, they could create a population of dense low-mass dark matter halos, usually referred to as ultra-compact minihalos \citep{Berezinsky:2003,Bringmann:2012}. These objects are found to have a steep inner density profile, $\rho\sim r^{-9/4}$ or $r^{-1.5}$, in simulations \citep{Gosenca:2017,Delos:2018a}, and they are expected leave their own distinctive observational signature that could be constrained by microlensing \citep{Ricotti:2009,LiUCMH:2012}, pulsar time delays \citep{Clark:2016}, or by (the lack of) a gamma-ray signal if DM self-annihilates \citep{Bringmann:2012,Gosenca:2017,Delos:2018b}.

%%%%%%%%%%%%%%%%%%%%%%%%%%%%%%%%%%%%%%%%%%%%%%%%%%%%%%%%%%%%

\subsection{Self-interacting dark matter (SIDM), decaying and dissipative dark matter}

For many purposes, dark matter can be considered to have no other interactions besides gravity. This 'collisionless' assumption is justified since after freeze-out (in the case of thermal production of the dark matter particles), non-gravitational particle-particle interactions must be very weak given current observational constraints. Nevertheless, any physically motivated particle DM model must have a non-zero interaction cross-section in order for those particles to be produced in the early Universe in the first place. 

There are various possible interaction mechanisms. For instance, dark matter can interact and annihilate with itself (e.g., if it were a Majorana particle) producing standard model particles and (gamma-ray) photons. This has fueled searches of annihilation products in galaxy clusters, Milky way satellites, the galactic centre, or as a diffuse extragalactic background (see \citealt{Leane:2020} for a recent review). This `indirect detection' of particle dark matter heavily relies on $N$-body simulations which can pinpoint the most likely places for a putative detection, as well as the expected emission (see e.g., \citealt{Springel:2008b,Zavala:2010} and \citealt{Kuhlen:2012,Fornasa:2015,Zavala:2019} for reviews). Although these searches have not been successful so far \citep{Ackermann:2012,Abdallah:2016}, future facilities such as the Cherenkov Telescope Array will offer new prospects for a detection \citep{Doro:2013}.

Other possible self-interaction considers the weakly collisional regime with elastic binary processes $\chi\chi\to\chi\chi$. This case is generically known as {\em self-interacting dark matter}, and the cross-sections can be large enough to be relevant for structure formation. For instance, weak collisionality was found to have an effect on the density profiles of isolated halos \citep{Burkert:2000,Kochanek:2000}, predominantly by isotropising the core velocity dispersion and reducing the density. In principle, such systems can undergo a gravothermal catastrophe leading to core collapse, but current constraints on the cross-section put the time-scale for this at $\gtrsim100~{\rm Gyr}$ \citep{Koda:2011}. The resulting observational signatures enable constraints of such `richer' dark sector physics (see e.g., \citealt{Tulin:2018} for a recent review of the topic), which as for other DM candidates, required a detailed simulation counterpart.

Self-interacting DM requires a modelling of microscopic short-range particle scattering in $N$-body simulations. These interactions alter the overall the dynamics which is no longer purely geodesic, and thus it requires upgrading from the Vlasov--Poisson (cf.\ Eq.~\ref{eq:vlassov}) to the Boltzmann--Poisson system of equations by adding a scattering balance term
\begin{equation}
\frac{\partial f}{\partial t} + \frac{\vecb{p}}{ma^2} \cdot \bnabla_x f - m \bnabla_x\phi \cdot \bnabla_p f = \Gamma_{\rm in} - \Gamma_{\rm out},
\end{equation}
where $\Gamma_{\rm in}$ is the instanteneous rate of change of $f(\vecb{x},\vecb{p},t)$ due to all scattering events that lead to particles ending up in an infinitesimal phase space volume ${\rm d}^3x\,{\rm d}^3p$ around the point $(\vecb{x},\vecb{p})$, and $\Gamma_{\rm out}$ the respective rate for scattering out of that volume.

Assuming a two-particle momentum `in state' $(\vecb{p},\, \vecb{p}_1)$ is scattered into the 'out state' $(\vecb{p}',\,\vecb{p}'_1)$, then the scattering rate balance, given a differential scattering cross section ${\rm d}\sigma_\chi / {\rm d}\Omega$, is
\begin{equation}
\Gamma_{\rm in}-\Gamma_{\rm out} =  \,\int_{\mathbb{R}^3} {\rm d}^3p_1 \int_{0}^{4\pi} {\rm d}\Omega\,\frac{{\rm d}\sigma_\chi}{{\rm d}\Omega} \frac{\left\| \vecb{p}-\vecb{p}_1\right\|}{ma^2} \,\left[ f(\vecb{x},\vecb{p}'; t) f(\vecb{x},\vecb{p}'_1;t)- f(\vecb{x},\vecb{p}; t) f(\vecb{x},\vecb{p}_1;t) \right],
\end{equation}
where the centre-of-mass scattering (solid) angle $\Omega$ contains the remaining two degrees of freedom in $(\vecb{p}',\,\vecb{p}'_1)$ allowed by symmetries for binary elastic collisions, i.e. after accounting for momentum and energy (and particle number) conservation. Unfortunately, if one inserts the $N$-body distribution function (\ref{eq:n_body_df}) in the scattering rate $\Gamma:=\Gamma_{\rm in}-\Gamma_{\rm out}$, one does not arrive at a practicable discretisation since $\Gamma$ is zero unless two $N$-body particles are located at the exact same location. Since each $N$-body particle anyway represents an element of phase space in a coarse-grained sense, practical approaches \citep{Vogelsberger:2012,Rocha:2013}  have to resort to estimating $\Gamma$ with smeared out particles, relying on an SPH-like approach where (\ref{eq:n_body_df}) is replaced with
\begin{equation}
f_N(\vecb{x}, \vecb{p}, t) = \sum_{\vecb{n}\in\mathbb{Z}^3} \sum_{i=1}^N \frac{M_i}{m} \,W\left(\left\|\vecb{x}-\vecb{X}_i(t)-\vecb{n} L \right\|;\,h_i\right)\,\delta_D(\vecb{p}-\vecb{P}_i(t)), \label{eq:smoothed_n_body_df}
\end{equation}
where we recall that $M_i$ and $\vecb{P}_i$ are the mass and momentum of $N$-body particles, and $W(r;\,h)$ is a smooth kernel function with finite support and $h_i$ a bandwidth (or smoothing scale) parameter chosen so that several other particles are found within a distance $h_i$ from particle $i$ (the summation over periodic copies is of course irrelevant for the self-interaction). This distribution function can then be used to express the scattering rate experienced by particle $i$ as (see \citealt{Rocha:2013} for details of the calculation)
\begin{equation}
  \begin{aligned}
\Gamma_i &= \sum_j \Gamma_{i\mid j} = \sum_j \frac{\sigma_\chi}{ma^2} \left\| \vecb{P}_i -\vecb{P}_j \right\| V_{ij},\\ &\quad\textrm{with}\quad V_{ij} = \int_{\mathbb{R}^3} {\rm d}^3x \;W(\|\vecb{x}-\vecb{X}_i\|; h_i)\,W(\|\vecb{x}-\vecb{X}_j\|;h_j)
  \end{aligned}
\end{equation}
representing the overlap between the particle kernels. Since numerically $\Gamma_{i\mid j} = \Gamma_{j\mid i}$ is not guaranteed, \cite{Rocha:2013} proposed to use a symmetrised scattering rate $\Gamma_{ij}:=(\Gamma_{i\mid j}+ \Gamma_{j\mid i})/2$. Once the scattering rate $\Gamma_{ij}$ is known, the scattering probability over a time step $\Delta t$ is $\mathcal{P}_{ij} = \Gamma_{ij}\Delta t$. Then, one can realise an elastic scattering event by sampling this probability over one timestep and randomising the direction of the relative velocity vector of particles $i$ and $j$ in the centre-of-mass frame. Such a Monte Carlo approach to the weakly collisional regime has been compared and validated with a non-ideal fluid model e.g., by \cite{Koda:2011}.

Dark matter self-interactions have been included in many cosmological simulations (e.g., \citealt{Yoshida:2000,Vogelsberger:2012,Rocha:2013}). An important focus of such simulations is on dwarf galaxies (e.g., \citealt{Zavala:2013}) and on simulations of galaxy clusters (e.g., \citealt{Brinckmann:2018,Banerjee:2020} for recent examples) since merging clusters provide some of the strongest constraints on the cross-section of dark matter self-interaction \citep{Harvey:2015,Kahlhoefer:2015,Robertson:2017}. Note that self-interactions are also included in the 'ETHOS' effective dark matter model \citep{Cyr-Racine:2016,Lovell:2018}, which we will discuss in the next subsection.

Another possible interaction is {\em decaying dark matter}, where the dark matter decays into massless species on cosmological time scales. Current constraints from the CMB require that, if all dark matter decays into photons (or dark radiation), then the decay rate is at most $\Gamma^{-1} \gtrsim 160 {\rm Gyr}$ \citep{Audren:2014}. However, a much smaller fraction of DM can decay and still leave a cosmological signature. In this case, the corresponding increase in energy density of relativistic species at late times needs to be accounted for in the background evolution of numerical simulations. Additional non-homogeneous relativistic corrections can, e.g., be absorbed in a similar way as for trans-relativistic massive neutrinos (see Sect.~\ref{sec:HDM}) using linear theory corrections to the gravitational potential \citep{Dakin:2019} or using a gauge approach \citep{Fidler:2017}. Beyond a background contribution, dark matter decay and annihilation could also contribute to a heating of the gas in low mass halos \citep[e.g.][]{Schoen:2015}, and the inter-galactic medium (IGM), which can be modelled e.g., in hydrodynamic cosmological simulations \citep{Iwanus:2017,List:2019a}, or using machine learned modifications to matter only simulations \citep{List:2019}, similar to those employed to include baryonic effects or modify cosmological parameters (see Sect.~\ref{sec:machine_learning}).

While Ockham's razor might push us to consider simple DM models, there is a priori no reason that the physics of the dark sector could not be significantly richer, with multiple dark species and internal degrees of freedom. An intermediate space between self-interacting dark matter and decaying dark matter is occupied by {\em dissipative dark matter}, which allows for up-scattering to an excited state $\chi\chi\to\chi'\chi'$ with a subsequent decay, e.g., $\chi'\to\chi+X$ under emission of a light or massless particle '$X$'. Such processes can efficiently remove energy from the centres of halos and lead to halo core collapse in less than a Hubble time \citep{Essig:2019,Huo:2020} leaving testable signatures in dwarf galaxies. They could also contribute to the formation of supermassive black holes already at high redshift \citep{Choquette:2019}. Note, however, that dissipative dark matter interactions are constrained by the non-detection of DM acoustic oscillations \citep{Cyr-Racine:2014}, and absence of a significant thin `dark disk' in the Milky Way \citep{Schutz:2018}.

%%%%%%%%%%%%%%%%%%%%%%%%%%%%%%%%%%%%%%%%%%%%%%%%%%%%%%%%%%%%

\subsection{Effective Descriptions}

In the previous subsections we considered in detail specific DM candidates. However, many more alternatives exist which span a broad range of masses, interactions, production mechanisms, etc. For instance, sterile neutrinos produced by resonant or non-resonant transitions, production in the decay of a parent particle which might result in thermal or non-thermal distributions, or mixed models in which DM is made out multiple particles with different properties. Furthermore, new candidates are being constantly proposed.

Since it is unpractical to carry out numerical simulations for every possible particle and covering their respective degrees of freedom, effective descriptions have been proposed. The basic idea is that a large fraction of physically-viable candidates can be mapped to a generic  model with a few free parameters. For instance, since nonlinear large-scale structure expected in a given DM model depends mostly on the initial transfer function, deviations from the CDM transfer function can be parameterised, and a given DM candidate can be mapped to a particular point in this parameter space. Specifically, \cite{Murgia:2017} have argued that a large class of DM models can be described using a generalisation of the WDM modification given in Eq.~(\ref{eq:WDM_transfer_Bode})
\begin{equation}
\label{eq:t_murgia}
T_\chi(k) = T_{\rm CDM}(k)\times\left[1 + \left(\alpha k\right)^{\beta} \right]^{\gamma}
\end{equation}
where $\alpha$, $\beta$, and $\gamma$ are the three free parameters of the model (but note a strong degeneracy between $\alpha$ and $\gamma$). This form could be extended with additional free parameters to describe a broader range of models, especially at high wavenumbers (such as `dark oscillations'). On the other hand, these scales are expected to have a very minor impact on structure formation. 

A different effective description has been proposed in terms of the particle physics Lagrangian in the so-called Effective Theory of Structure Formation (ETHOS) \citep{Cyr-Racine:2016}. The physical parameters of a given DM model (DM particle mass, coupling constants, number of degrees of freedom, mediator mass, etc.) map into effective parameters that determine the initial transfer function (e.g., $\alpha$, $\beta$, and $\gamma$ in Eq.~\ref{eq:t_murgia}) and an effective (velocity-dependent) cross section. 

An advantage of these effective approaches is that only a relatively small number of simulations need to be carried out for the free parameters of the effective description \citep{Vogelsberger:2016,Stucker:2021b}. With these, predictions for the nonlinear structure, their observable signatures, and comparison with observations, can be obtained for any DM model. This has been done for the first galaxies and reionization, the Ly-$\alpha$-forest, the abundance of dwarf galaxies, and gravitational lensing \citep{Murgia:2017,Lovell:2018,DiazRivero:2018,Lovell:2019,Bose:2019}.

This represents an example where $N$-body simulations are directly employed in constraining cosmological parameters and fundamental physics -- the properties of dark matter in this case. As we will argue later, this approach can also be applied to  observations of the large-scale structure but it is of paramount importance to demonstrate the accuracy and robustness of the corresponding numerical predictions.

%%%%%%%%%%%%%%%%%%%%%%%%%%%%%%%%%%%%%%%

\subsection{Multiple species with distinct initial perturbation amplitudes}

Due to the dominance of collisionless dark matter and the coldness of both dark matter and baryons during the structure formation epoch, large-scale simulations usually represent the total matter component with a single collissionless fluid. This fluid is commonly referred to as dark matter but in reality it represents dark matter and any other massive, non-relativistic component in the Universe. While this is a good approximation in many cases, the advent of more precise observations and, as a consequence, stricter accuracy requirements for numerical simulations demands the simulation of multiple fluids as the other known massive components in the universe beside dark matter -- baryons and neutrinos --  have different initial distributions and relative velocities which affects late-time structure.

In this section we review several efforts and approaches to simulating the gravitational co-evolution of multiple fluids, specifically baryons and neutrinos. This is a veritable challenge, that only recently is becoming possible to tackle thanks to multiple novel algorithms and development in numerical techniques.

\subsubsection{Baryons}
\label{sec:multi_species_amplitudes}

\begin{figure}
\includegraphics[width=\textwidth]{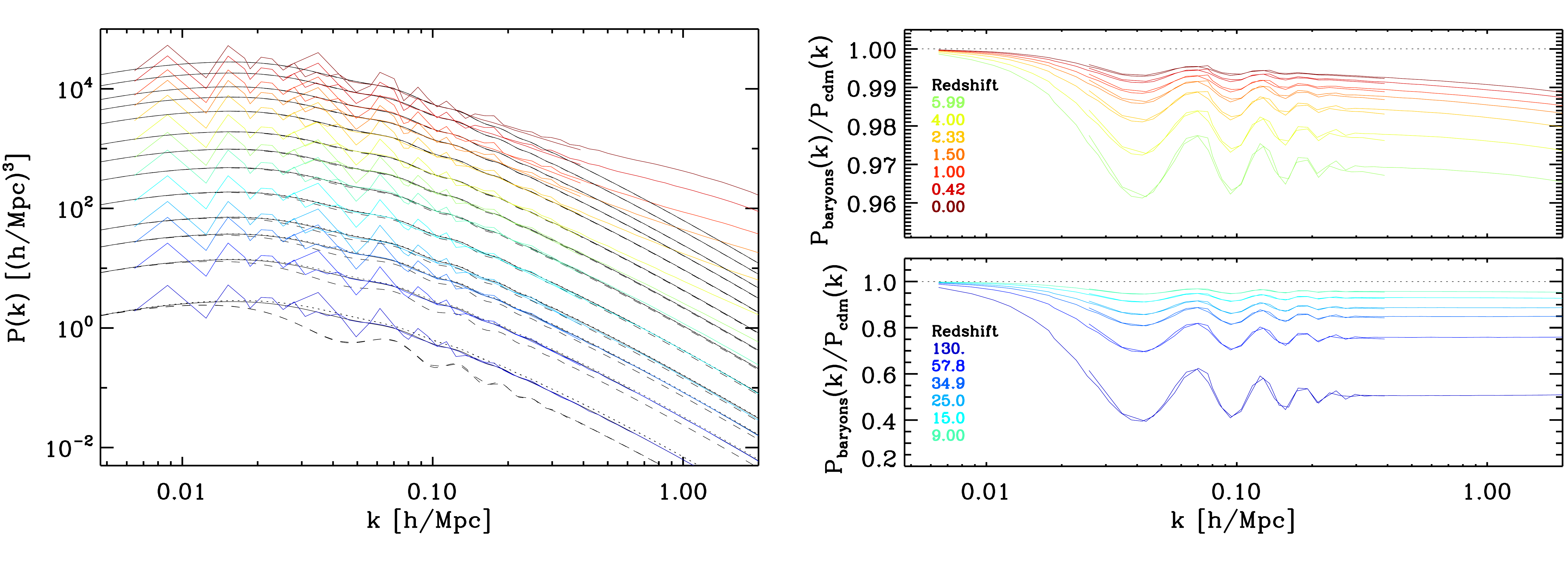} 
\caption{The time evolution of cold dark matter plus baryon fluctuations from $z=130$ up to $z=0$. The left panels show the linear-theory power spectrum for the total mass, cold dark matter, and baryons as solid, dotted, and dashed lines, respectively. Coloured lines show the mass power spectrum predicted by an $N$-body simulation. The right panels show the ratio of the power spectra of baryons to that of cold dark matter. The amplitude of fluctuations at high-$z$ is typically a factor of 2 smaller for baryons that for dark matter. These differences are progresively reduced at lower redshifts to about a few percent at late times. \label{fig:cdm_baryons} Figure adapted courtesy \cite{Angulo:2013a}. \label{fig:baryon-cdm}}
\end{figure}

If we consider only primordial adiabatic modes, at very early times and on super-horizon scales, baryons and dark matter have an identical spatial distribution as imprinted by the quantum fluctuations processed by inflation. However, the tight coupling of baryons with the radiation field via Compton scattering prior to recombination renders its subsequent evolution distinctly different from that of the dark matter.
Radiation pressure opposes the growth of baryonic overdensities creating oscillations in density and temperature which are damped on small scales because of an imperfect coupling and Jeans damping. On the other hand, dark matter is not expected to couple to radiation and mostly grows unimpeded. After recombination, baryons and photons effectively decouple and the same gravitational interactions dominate the growth of baryon and dark matter overdensities. However, the starting point for baryons and dark matter is already different. In addition to this effect caused by recombination on purely adiabatic perturbations, primordial isocurvature perturbations could add additional differences among baryons and dark matter.

The linear evolution of a system of multiple fluids coupled by gravity can be modelled and computed accurately by Einstein-Boltzmann solvers, which, in fact, show that the power spectrum of density fluctuations at $z \sim 100$ (the typical starting redshift of numerical simulations) are still significantly different. For instance, the baryonic acoustic oscillations are barely present in the dark matter, there is a relative velocity between baryons and dark matter, and the fluctuation amplitues in baryons are approximately half of those in dark matter, even on gigaparsec scales. This can be seen in Fig.~\ref{fig:baryon-cdm} which shows the time evolution of the power spectra for total mass, baryons, and dark matter as predicted by linear perturbation theory and by a $N$-body simulation.

Although one could think that simulating gravitationally baryons and dark matter is a straightforward extension of standard $N$-body codes, achieving an accurate evolution as well as initial conditions has proven to be rather challenging, and it is an example of a situation where certain discretizations of the underlying equations can result in a considerable numerical error.

A range of studies \citep{Yoshida:2003,OLeary:2012,Angulo:2013a} have shown that a na\"ive simulation of such a baryon-DM two-fluid system leads to the incorrect evolution of the relative baryon-dark matter perturbations. Spurious particle coupling dominates over the real differences on small scales which then quickly propagates to large scales. In early studies, these errors could only be suppressed if forces were smoothed on scales larger than the mean inter-particle separation, either with a large fixed softening or one adapting with the local density \citep{OLeary:2012,Angulo:2013a}. 

Other proposed solutions \citep{Yoshida:2003,Bird:2020} relate to the initial particle load (cf.\ Sect.~\ref{sec:pre_ics_discreteness}), of the two-fluid simulations -- adopting at least one glass distribution for one component reduces the spurious coupling even at high force resolution. However, \cite{Hahn:2020} demonstrated that the dominant error contribution comes from the constant mode (the density difference $\delta_b-\delta_c$ approaches a constant at late times) which can be absorbed at all orders of LPT into a simple variation of the relative masses of CDM and baryon particles.  This, therefore, enables an extension of Lagrangian Perturbation theory for multiple fluids  \citep{Rampf:2020} which allowed more accurate initial conditions and also lower starting redshifts. The inclusion of relative velocities beyond linear order in the ICs is however still an unsolved problem. In contrast to this back-scaling approach, missing physics in the non-linear solvers (such as the small residual coupling to radiation at $z\gtrsim100$) makes the `forward' approach to initial conditions for CDM+baryon simulations inaccurate at low $z$, so that in general `back-scaling' should be preferred (cf.\ Sect.~\ref{sec:sims_and_pt}). 

Simulations by \cite{Angulo:2013a} showed that the nonlinear total matter power spectrum is largely unaffected by the single/two fluid distinction with the $z=0$ results differing by less than 0.1\% at $k \sim 1 \ihMpc$. However, baryon-dark matter differences could be imprinted in halo formation, as expected on theoretical grounds. This question was indeed investigated by using simulations with adaptive softening \citep{Khoraminezhad:2020} and by using an extension of the separate universe approach \citep{BarreiraCabass:2020}, which does not suffer from the numerical inaccuracies described above. These authors showed that halo formation is actually sensitive to baryon-dark matter fluctuations, in agreement with analytic arguments \citep{Chen:2019,Schmidt:2016}. At a fixed matter overdensity, halos tend to form more efficiently in regions with smaller baryon-dark matter differences. While small, this additional dependence creates coherent fluctuations even on large (BAO) scales which could affect observational constraints on neutrino masses and from baryonic acoustic oscillations \citep{Chen:2019}. On much smaller scales, baryons stream pass dark matter collapsed structure \citep{Tseliakhovich:2010}, which delays their eventual accretion and thus is expected to affect structure formation in a correlated way. Numerical simulations of this effect have shown that neglecting this streaming leads to an overestimation of the abundance of high redshift low-mass halos \citep{Tseliakhovich:2010,Dalal:2010b,Park:2020}. Finally, we note that although current limits on isocurvature modes are tight, a subclass where the total matter fluctuations are left invariant is largely unconstrained. These, so-called compensated isocurvature modes, could be strongly constrained by the next generation of LSS surveys by using the observational signatures of distinct  large-scale spatial distribution of baryons and dark matter \citep{Hotinli:2019,BarreiraCabass:2020}. 

We note that in addition to the above, another (in principle unrelated) problem is that caused by baryonic physics itself (finite Jeans scale, gas cooling, UV heating, feedback, etc), which will be discussed below (cf.\ Sect.~\ref{sec:baryonic_effects}). 

\subsubsection{Massive neutrinos}
\label{sec:HDM}

Neutrinos are one of the fundamental particle families in the Standard Model of particle physics, where they are expected to be massless given the allowed symmetries. However, observations of flavor oscillations in solar and atmospheric neutrinos indicate that they do have mass, which could be a signature of physics beyond the standard model.     
The measurement of the absolute mass scale of neutrinos is very important, since, when combined with measurements of the neutrino-mass splitting coming from neutrino oscillation experiments, they could distinguish whether neutrinos are Majorana or Dirac particles. This in turn could indicate the kind of extension required to the standard model and thus answering one of the fundamental questions in physics.

Recent results from the KATRIN experiment find an electron neutrino mass upper bound of 0.8~eV (at 90\% CL) from the study of the electron endpoint energy in tritium decay experiments \citep{Aker:2021}. However, the large-scale structure of the Universe provides currently the most accurate method to constrain the total mass of neutrinos, but note that future experiments such as PTOLEMY \citep{Betti:2019} could be competitive with forecasted errors of $10^{-3}$eV. Current constraints on neutrino masses from large-scale structure and the CMB are $\sum m_{\nu} < 0.12\,{\rm eV}$ \citep{Palanque:2019,Planck:2020}, but the upcoming generation of large-scale surveys have the potential to achieve an accuracy of $\sigma[\sum m_{\nu}] = 0.01 - 0.03\,{\rm eV}$ when combined with CMB lensing \citep{Boyle:2020,Chen:2021}. Therefore, precise numerical simulations of cosmic structure formation in the presence of neutrinos, and particularly their interplay with cold dark matter has become increasingly important. 

Neutrinos affect both the expansion history and the clustering of matter in the Universe. The neutrino temperature is intimately connected to the photon temperature as $T_\nu = (4/11)^{1/3}T_{\rm CMB}\simeq1.95~{\rm K}$. As the Universe expands, neutrinos cool and become non-relativistic at $z \simeq 189 (\sum m_{\nu} / 0.1\,{\rm eV})$. While they are relativistic, they contribute to the total energy density of the universe as $\Omega_\nu(a) = N_{\rm eff}(7/8)(4/11)^{4/3}\Omega_\gamma a^{-4}$ (where for massive neutrinos we always explicitly indicate the time-dependence in the density parameter), while as a non-relativistic species they contribute as 
\begin{equation}
\Omega_{\nu}(a)  =  \frac{\sum m_{\nu}}{93.14\,{\rm eV}} \,h^{-2}\,a^{-3} .
\end{equation}
For intermediate cases, the neutrino mass fraction needs to be obtained numerically from the Fermi-Dirac distribution function
\begin{equation}
\Omega_{\nu}(a) = a^{-3} \sum_j \left(\frac{m_{\nu,j}}{5.32~{\rm meV}}\right)^4 \int_0^\infty {\rm d}y\,\frac{y^2\sqrt{1+y^2/a^2}}{\exp\left[\beta_j y\right]+1},
\end{equation}
where $\beta_j:= m_{\nu,j}c^2/(k_BT_\nu)$.

Since they are still relativistic when they decouple at $z\sim10^9$, they have a typical velocity of about a few hundreds km/sec today. Therefore at late times they contribute mostly as perturbations through Newtonian gravity rather than as a relativistic species through the background, so that for a CDM+baryon+neutrino simulation one would have to solve the Poisson equation
\begin{align}
\nabla^2 \phi &= \frac{3 H_0^2}{2 a} \Omega_{\rm m} \bigl(f_{\rm c} \delta_{\rm c} + f_{\rm b} \delta_{\rm b} + f_\nu(a)\,\delta_{\nu}\bigr)\\
&\textrm{where} \quad f_{\rm c}:=\frac{\Omega_{\rm c}}{\Omega_{\rm m}},\quad f_{\rm b}:=\frac{\Omega_{\rm b}}{\Omega_{\rm m}},\quad f_\nu(a):= \frac{\Omega_\nu(a)}{\Omega_{\rm m}a^{-3}}.\nonumber
\label{eq:poisson_eq1}
\end{align}
Massive neutrinos therefore, depending on their relative importance through their total mass fraction, need to be included in simulations of structure formation.

\begin{figure}
\includegraphics[width=\textwidth]{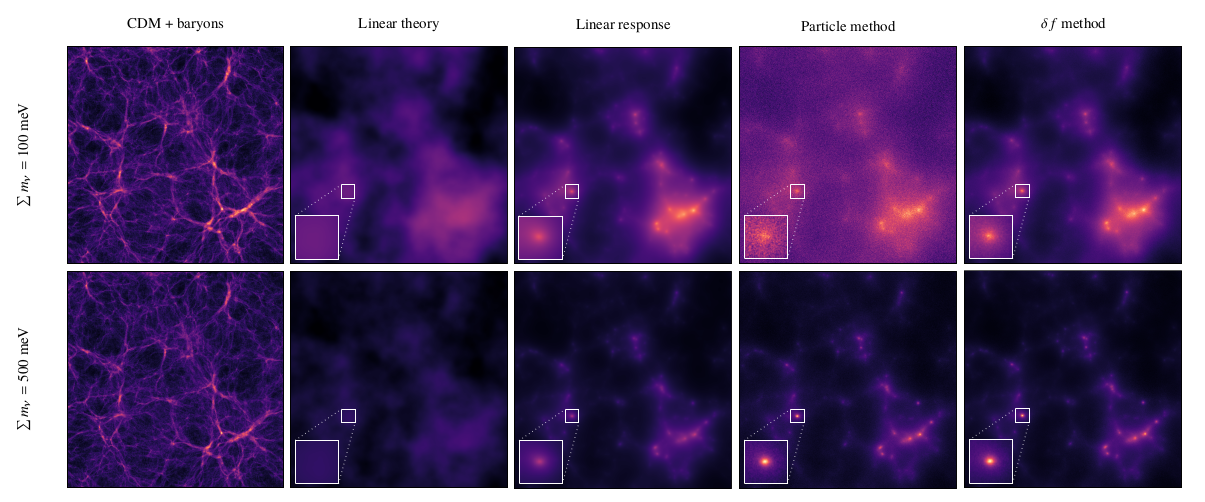} 
 \caption{\label{fig:neutrinos} Comparison of the simulated density field of massive neutrinos with total mass $\sum m_{\nu}=0.5eV$ and $0.1eV$ (top and bottom rows, respectively). Each panel shows the results of various computational methods; linear perturbation theory, linear response, particle method, and $\delta f$ method. For comparison, the leftmost column shows the cold mass (cold dark matter and baryons) over the same volume. Note the inset in each panel showing a zoom into a massive dark matter halo. Figure adapted courtesy \cite{Elbers:2020}.}
\end{figure}

Many different approximations and discretisations were proposed to include neutrinos into non-linear simulations of structure formation. These approaches adopt different degrees of simplification, but, as we will see below, they all agree to a large extent for the small neutrino masses compatible with observations.

In the simplest approach, massive neutrinos are simply included in the Einstein-Boltzmann solver, and total matter initial conditions for the $N$-body simulation are generated in a normal `back-scaling' approach \citep{Agarwal:2011,Upadhye:2014}. More recent approaches however try to capture also the time-evolution of the massive neutrino component. Since they are hot dark matter, the evolution of neutrinos obeys the Vlasov--Poisson equations with a hot distribution function once they are non-relativistic. In contrast to cold matter, neutrinos remain fairly linear so that it is in principle possible to solve the VP equation directly in phase space. This indeed has been recently achieved by \cite{Yoshikawa:2020,Yoshikawa:2021}, who followed CDM using traditional $N$-body techniques and solved the VP equation describing neutrinos as an incompressible phase space fluid using a 6D Cartesian grid.  This solution is however computationally very expensive. Another approach is to apply the $N$-body approach to sample the six-dimensional distribution function of neutrinos, i.e. at each location, multiple particles sample the possible momentum magnitude and directions given by a Fermi-Dirac distribution. This approach can capture the nonlinear evolution of the neutrino fluid and traditionally had been regarded as the gold standard in the field. Consequently, it has been adopted in some very large simulations from which most of the current knowledge about the role of neutrinos in structure formation stems \citep{Brandbyge:2008,Viel:2010,Castorina:2015}. This $N$-body approach has been extended to relativistic simulations by \cite{Adamek2017}, which accurately take into account the transition of neutrinos from the relativistic to the non-relatvitistc regime, along with the contribution of both their energy density and their anisotropic stress to metric perturbations in the weak field regime.

Unfortunately, a large number of neutrino particles are needed to reduce discreteness noise, which rapidly increases the computational cost of these simulations \citep{Emberson:2017}, since the error on the mean momentum (and thus essentially the growing mode) only decreases with the Poisson noise. Some alternatives have been proposed to minimise the impact of noise. One of them, \cite{Ma:1994}, used a pairing of neutrino particles sampling exactly opposite momenta to enforce local momentum conservation. This guarantees that at the very start of the simulation, the mean field is exactly recovered, but is washed out shortly afterwards. To circumvent this problem, \cite{Banerjee:2018} sampled the directional distribution in a more regular fashion (using the {\sc Healpix} decomposition of the sphere rather than uncorrelated random directions). More recently, \cite{Elbers:2020} proposed another method to reduce the Poisson noise by sampling only the deviations from the linear solution with particles, which shifts the sampling error from the expectation value of the mean to the expectation value of the deviation from the linear solution (referred to as the  `$\delta f$ method').

Alternatively, to avoid these numerical inaccuracies and reduce the computational cost, one can take advantage of the fact that neutrinos are never expected to cluster significantly, owing to their large peculiar velocities, and thus can be treated perturbatively. In the simplest approach, neutrinos are described on a grid which is evolved according to linear theory (i.e. the neutrino density obtained with a linear Einstein-Boltzmann solver is convolved with the random phases of the simulation), which is then co-added to the total large-scale gravitational potential field \citep{Brandbyge:2009}. An advantage of this approach is that the neutrino perturbations can be computed incorporating also general relativistic effects \citep{Tram2018}. A disadvantage is that it cannot correctly capture the nonlinear evolution of neutrino perturbations and that momentum and energy conservation are not guaranteed as there is no back-reaction of non-linear matter on the neutrinos (which is however negligible if the neutrinos are light).

A refinement can be obtained with the so-called linear-response approaches where, although neutrinos are treated perturbatively, the full nonlinear dark matter field is used as a source in the perturbative solution for the neutrinos \citep{AliHamoud:2013}. This approach appears to work extremely well for all but the slowest moving neutrinos which can be captured in dark matter halos and develop significant nonlinearities. A further refinement was proposed by \cite{Bird2018} where only the initially coldest neutrinos are sampled with $N$-body particles, and an even more efficient implementation was recently proposed by \cite{Chen:2020}.

A somewhat different approach is to solve the neutrino evolution using a fluid approach through a Boltzmann hierarchy expansion. One formulation has been proposed by \cite{Dakin2017} who considered the first three moments of the Boltzmann equation: solving for the continuity and Euler equations in Eulerian space (i.e. the lowest two moments of the Boltzmann equation) while describing the third moment, capturing the stress tensor, at linear order. These terms could also be estimated using $N$-body particles \citep{Banerjee:2016}, or by decomposing the neutrino phase-space into shells of equal speed and evolve them with hydrodynamic equations \citep{Inman:2020}. 

Neutrinos have also been incorporated in approximate $N$-body methods (cf.\ Sect.~\ref{sec:cola_fastpm} for a review of such approaches). Recently, \cite{Bayer:2020} proposed an extension to the {\sc FastPM} algorithm by modifying the kick and drift operator; and \cite{Wright:2017} extended the {\sc COLA} algorithm by incorporating the scale-dependent effect of neutrinos in the growth factors. On the other hand, an accurate prediction for the impact of neutrinos can also be achieved in post-processing by using cosmology-rescaling methods \citep{Zennaro:2019,Angulo:2010,Contreras:2020a}, or as a gauge transformation \citep{Partmann:2020}.

Finally, when neutrinos are treated in a Newtonian framework at low redshift, simulation initial conditions cannot be accurately used in a ``forward'' approach (c.f. Sect.~\ref{sec:sims_and_pt}), which would result in significant inaccuracies (see e.g., \citealt{Bastos:2021}). \cite{Zennaro:2017} proposed a back-scaling approach where for a given target redshift, the linear power spectrum predicted by linear Boltzmann solvers are integrated backwards in time accounting only with gravitational interactions. This is in principle unstable for long times since decaying modes are excited that blow up during the backwards evolution, but due to the low mass of neutrinos, this approach works well for starting times $z_{\rm start}\lesssim 100$. In this way, when setting up first order accurate initial conditions as in Eq.~(\ref{eq:IC_forward}), particles are set-up at the starting redshifts so that when evolved with gravitational interactions they retrieve the correct large-scale density fluctuations (see also the discussion on `backward' initial conditions in Sect.~\ref{sec:sims_and_pt}).

Despite the huge differences among the various methodologies to include neutrinos into non-linear simulations, all these approaches agree extremely well on their predictions for the impact of (light) massive neutrinos on the large-scale structure, especially for the allowed neutrino masses where differences are typically at the sub-percent level. Therefore, currently the cosmic evolution of neutrinos appears to be a robust prediction of numerical simulations. This is illustrated in Fig.~\ref{fig:neutrinos} which displays the projected density field of neutrinos in a box of $180\hMpc$ employing various numerical approaches discussed in this section. Here we can clearly see how linear theory predictions correctly capture the large-scale distribution of neutrinos but underestimate the high-density tail. Additionally, discreteness noise is evident in the particle method. Nevertheless, the agreement of all methods on large scales is remarkable, especially for lighter neutrinos. In the figure insets we see, however, how these numerical methods differ in the accuracy with which describe neutrinos in the nonlinear regime and the associated degree of noise.

In Fig.~\ref{fig:neutrinos} one can also see that neutrinos follow the same large-scale patterns as the dark matter. In contrast, on smaller scales, the large velocity dispersion of the neutrino field stops its growth and the field remains smooth. This causes a well-known scale dependent suppression in the total matter clustering. On intermediate scales below the neutrino free streaming, the suppression has an amplitude roughly proportional to $\Delta P/P \sim -10 f_{\nu}$, slightly larger than the linear theory prediction $\Delta P/P \sim-8 f_{\nu}$ \citep{Brandbyge:2008}. On even smaller scales, the differences decrease due to nonlinearities (see, e.g., \citealt{Hannestad:2020}).

Usually, numerical simulations consider a degenerate mass state for neutrinos, i.e. assume that all neutrino states have the same mass. This is incompatible with oscillation experiments, but justified as cosmic structure is mostly sensitive to the total neutrino mass. However, the linear power spectrum is slightly different between the normal and inverted hierarchy \citep{Lesgourgues:2004,Jimenez:2010}. Although the effect is small -- below 0.4--0.6\% for weak lensing and galaxy clustering observables \citep{Archidiacono:2020} -- numerical simulations have shown that this signature is imprinted and even enhanced during nonlinear clustering, which opens up the possibility of a marginal detection with future large-scale structure surveys \citep{Wagner:2012}. However, this will require significant advances in the modelling of baryonic effects, redshift space distortions, and galaxy bias.

The role of neutrinos has also been explored via the `separate universe' technique (cf.\ Sect.~\ref{sec:separate_universe}) \citep{Chiang:2018}. These simulations have shown that the linear bias of dark matter halos is scale-dependent in the presence of neutrinos. These findings were later confirmed by the same authors but using standard $N$-body simulations (where the contribution of neutrinos was artificially increased to enhance their effect) \citep{Chiang:2019}, who further showed that the scale dependence persists regardless of whether it is defined with respect to the total or cold mass (CDM plus baryons) power spectrum. 

%%%%%%%%%%%%%%%%%%%%%%%%%%%%%%%%%%%%%%%%%%%%%%%%%%%%%%%%%%%%

\subsection{Primordial non-Gaussianity and small-scale features from inflation}
\label{sec:nongaussianity}

So far, we have considered the case where primordial fluctuations in the Universe were Gaussian. This is motivated by the fact that original quantum fluctuations are known to be Gaussian, thus, if the subsequent physics and evolution were linear, it would result in the primordial seeds for structure formation being also Gaussian. This, however, does not need to be the case since general inflationary models predict various degrees of primordial non-Gaussianity (PNG). These originate from, for instance, non-linear dynamics due to self-interactions in single-field inflation, or from correlations of the inflaton field  with additional (light or heavy) fields (see \citealt{Bartolo:2004} for a review). 

The departures from Gaussianity are expected to be of the same order of the second-order corrections to linear-perturbation theory. Since the observed amplitude of primordial perturbations is $\mathcal{O}(10^{-5})$, the natural expectation of these corrections is thus $\mathcal{O}(10^{-10})$. Therefore, the deviations from Gaussianity are expected to be very small, and thus they can be generically (agnostic of specific models) prescribed by expanding the true primordial Bardeen potential, $\Phi$, around a Gaussian field $\phi^{(1)}$ (enforcing $\langle \phi^{(1)} \rangle=0$) \citep[cf.][]{Creminelli:2007}:
\begin{equation}
\begin{aligned}
\Phi(\vecb{x}) &=  \phi^{(1)}(\vecb{x}) + f_{\rm NL} \int_{\mathbb{R}^3} {\rm d}^3y \left[ \phi^{(1)}(\vecb{x}) \phi^{(1)}(\vecb{y}) W(\vecb{x},\vecb{y}) - \langle \phi^{(1)}(\vecb{x}) \phi^{(1)}(\vecb{y}) W(\vecb{x},\vecb{y}) \rangle \right] + \label{eq:nongaussianity}\\
& + g_{\rm NL} \times \left(\textrm{3-point combinations of }\phi^{(1)}\right) + \tau_{\rm NL} \times \left(\textrm{4-point combinations of }\phi^{(1)}\right) + \dots 
% g_{\rm NL} \mathcal{O}\left( \left[ \phi^{(1)} \right]^3\right)
% + \tau_{\rm NL} \mathcal{O}\left(\left[ \phi^{(1)} \right]^4\right)
\end{aligned} 
\end{equation}
where $f_{\rm NL},g_{\rm NL},\tau_{\rm NL},\dots $ are referred to as the non-Gaussianity parameters, which quantify the level and structure of PNG; and $W$ is a kernel defining the type of non-Gaussianity considered. The most common type in the literature is the so-called ``local type'' quadratic non-Gaussianity, which is defined by non-zero $f_{\rm NL}$ and $W = \delta_D(\vecb{x}-\vecb{y})$, and is a probe of multi-field inflationary models. Other kinds of configurations, e.g., equilateral and orthogonal types, are also (albeit less frequently) considered, which are expected to be sensitive to different aspects of inflation

From the above arguments, one expects $f_{\rm NL} \sim 1$, with the simplest inflationary models of single-field slow-roll predicting $f_{\rm NL} \lesssim 1$, but with several others (including non-inflationary cosmologies) predicting $f_{\rm NL} \gtrsim 1$. We note that if inflation occurs at high energies, then primordial gravitational waves could distinguish alternative models, but if it occurs at low energies, measuring the value of $f_{\rm NL}$ might be the only source of information about the early Universe.

Constraints from the analysis of the CMB fluctuations, as measured by the Planck satellite, are $f_{\rm NL}^{\rm local} = -0.9 \pm 5.1$; $f_{\rm NL}^{\rm equil} = -26 \pm 47$; and $f_{\rm NL}^{\rm ortho} = -38 \pm 24$ at the 68\% confidence level \citep{PlanckfNL:2016,PlanckfNL:2020}, which is in agreement with most inflationary models. Upcoming polarization and small-scale CMB measurements are expected to improve these constraints \citep{CMBS4}. An alternative for stronger constraints on $f_{\rm NL}$ relies on measurements of the late time large-scale structure of the Universe \citep{Alvarez:2014}. Although current LSS  constraints are significantly weaker than those from the CMB ($\sigma[f_{\rm NL}^{\rm local}] \sim 20-100$) \citep{Leistedt:2014,Giannantonio:2014,Ho:2015,Castorina:2019,Mueller:2021}, forecasts anticipate that future surveys could constrain $f_{\rm NL}$ at the level of $\sigma[f_{\rm NL}] \sim 0.5-3$ \citep{Giannantonio:2012,Yamauchi:2014,Camera:2015,Ferraro:2015}. However, as the interpretation of upcoming surveys relies on non-linear physics and biased tracers, numerical simulation of primordial non-Gaussianity is essential.

Numerical simulations of primordial non-Gaussianity can be carried out in the same fashion as Gaussian simulations, with the difference being only in the initial conditions. For a \emph{local} PNG type, $\phi$ can be trivially computed from any given value of $f_{\rm NL}$. The only numerical consideration is that, as for higher-order LPT implementations, Orszag's rule needs to be applied to avoid aliasing in the computation of field convolutions. For \emph{orthogonal} and \emph{equilateral} types, the generation of initial conditions is more complicated as it involves the generation of a random field subject to a constraint on its bispectrum, but \cite{Scoccimarro:2012} and \cite{Regan:2012} proposed a general algorithm to efficiently achieve such a task. Another subtlety regards the exact definition of $\phi^{(1)}$ and its value at the simulation starting redshift. As for the case of multi-fluids, one can define $\phi^{(1)}$ directly at that redshift or apply back-scaling, a freedom that affects the comparison among simulations for different values of $f_{\rm NL}$ (cf.\ \citealt{Pillepich:2010}). Alternatively, the question of the role of PNG in structure formation can be investigated using the separate universe approach (cf.\ Sect.~\ref{sec:separate_universe}). As explored by \cite{BarreiraPNG:2020a,BarreiraPNG:2020b}, a change in the initial power spectrum amplitude can mimic a primordial non-Gaussianity of the local type.

Simulations of primordial non-Gaussianity have found essentially three effects relevant for structure formation. The first one is that a local-type PNG caused a small modification in the nonlinear matter power spectrum \citep{Wagner:2010} which can be understood in perturbation theory as additional mode-coupling terms. The abundance of halos is also affected by the existence of primordial non-Gaussianity, with positive (negative) values of $f_{\rm NL}$ increasing(decreasing) the number of very massive halos \citep{Dalal:2008,Wagner:2012b}, and with qualitatively similar results for $g_{\rm NL}$ and $\tau_{\rm NL}$ \citep{LoVerde:2011}. However, these effects can become very degenerate with astrophysical processes and nonlinear evolution, which cautions on their use as cosmological probes. 

A clearer signature of local PNG arises on the very large-scale clustering of galaxies or quasars. PNG modifies the abundance of biased tracers on scales, $k<10^{-2}\ihMpc$, in a way that the clustering statistics receive an additional contribution proportional to the non-Gaussian term in Eq.~(\ref{eq:nongaussianity}) \citep{Dalal:2008,McDonald:2008,Baumann:2013,Assassi:2015}. This has been confirmed by a number of $N$-body simulations \citep{Dalal:2008,Grossi:2009,Pillepich:2010,Scoccimarro:2012,Wagner:2012b}. In practice, this contribution appears as a scale-dependent term proportional to $b_{\phi} f_{\rm NL} k^{-2}$ in e.g., the power spectrum, which can dominate on large scales over the contribution of the standard linear bias parameter, $b_1$, and thus can be used to place constraints on $f_{\rm NL}$ \citep{Slosar:2008}. Numerical simulations have shown that a similar scale-dependent effect exists for other kinds of non-Gaussianities \citep{Desjacques:2010,Scoccimarro:2012,Shandera:2011}.

The amplitude of the `non-Gaussianity bias', $b_{\phi}$, can be related to the value of the linear bias parameter, $b_1$, using analytic arguments: $b_{\phi} \propto (b_1 - 1)$ \citep{Slosar:2008,Matarrese:2008}, with which a much more predictive model can be constructed with the obvious benefits of increased constraining power. However, results from $N$-body simulations have revealed that the precise relation depends on the kind of halo considered (or the property used to select them) and the specific galaxy formation physics \citep{Slosar:2008,Scoccimarro:2012,Desjacques:2009,Reid:2010,BarreiraPNG:2020a,Barreira:2021}, since it depends on the details in which the formation of a given nonlinear object responds to a change in the large-scale potential. This might hinder the ability to robustly place constraints on PNG in future surveys \citep{BarreiraPNG:2020b}. However, as a potential detection of PNG is within reach of the upcoming generation of LSS surveys, there is certainly motivation to seek a better understanding of these effects in all LSS statistics. 

\subsection{Modified Gravity}
\label{sec:modified_gravity}
Throughout this review we have assumed General Relativity (GR) as the theory of gravity. And so far no experiments or astrophysical observations have been found to be inconsistent with GR. A particularly significant prediction of GR was the recently detected gravitational wave (GW) event with the co-incident detection of an electromagnetic counterpart, which has put stringent constraints on the class of modifications to GR allowed \citep{Creminelli:2017,Ezquiaga:2017}. GR is also consistent with a plethora of cosmological observations and specifically with tests based on redshift space distortions \citep{Mueller:2018,Barreira:2016,HernandezMonteagudo:2020}. However, departures from GR on cosmological scales are still possible and have received significant attention as a potentially important piece to understand the accelerated expansion of the Universe. In fact, testing gravity on cosmological scales is one of the primary science goals of the upcoming generation of large-scale structure observations and gravitational wave detectors \citep{Alam:2020,Belgacem:2019}. 

To properly interpret future observations it is thus crucial to understand structure formation in modified gravity, for which numerical simulations are indispensable (see \citealt{Llinares:2021,Baldi:2012,BaojiuLi:2018} for specialised reviews). A large number of modifications to gravity has been proposed (reviewed, e.g., in \citealt{Clifton:2012} and \citealt{Koyama:2016}). In the context of large-scale structure, the class that has received most attention are those with a `screening' mechanism. These models can become indistinguishable from GR in high density regions, e.g., via the gravitational potential \citep[chameleon][]{Khoury:2004a,Khoury:2004b}, its gradient \citep[k-Mouflage][]{Babichev:2009}, or its second-order derivatives (e.g. its Laplacian or the density) \citep[Vainshtein][]{Vainshtein:1972}. They are in agreement with many local gravity tests, but they may depart significantly from GR on large scale so that they could be probed by cosmic measurements.

The two most explored modifications to GR are the $f(R)$ and DGP (short for Dvali, Gabadadze and Porrati, \citealt{Dvali:2000}) models, which are regarded as representative of the kind of MG models currently available. These two gravity models feature an equal speed for photons and gravity, and naturally contain a screening mechanism. Note that although the initial motivation for such models was to explain the accelerated expansion of the universe, nowadays they are predominantly explored to study deviations from GR. 

For the case of $f(R)$, there is an additional term in the Einstein-Hilbert action that is a function of the Ricci curvature $R$, which leads to equations of motion with a gravitational potential modified as 
\begin{equation}
\nabla^2 \Phi = \nabla^2 \Phi_{\rm GR} - \frac{1}{2} \nabla^2  f_R
\quad\textrm{with}\quad \nabla^2 f_R =-\frac{a^2}{3}[\Delta R + 8 \pi G  \bar{\rho} \delta ],
\end{equation}
where $\Phi_{\rm GR}$ is the gravitational potential in GR, $\Delta R$ is the perturbation to the Ricci curvature, which could be written  in terms of a scalar field $f_R \equiv d f(R)/dR$ (e.g., in the Hu \& Sawicki model $f(R) := -M^2 \frac{c_1 (-R/M^2)^n}{c_2(-R/M^2)^n+1}$ with $M^2:=H_0 \Omega_{\rm m}$ and $n$, $c_1$ and $c_2$ being model parameters with $c_1/c_2^2 \propto f_{R0}$, the value of the scalar field today). Note that the theory is unscreened in low density regions where there is an additional `fifth-force', which has as an upper limit of 1/3 of the GR value. 

In the case of (the normal branch of) DGP, which is screened by means of a Vainshtein mechanism in regions where the Laplacian of the potential is large, the modified Poisson equation reads
\begin{equation}
\nabla^2 \Phi = \nabla^2 \Phi_{\rm GR} + \frac{1}{2} \nabla^2\phi
\quad\textrm{with}\quad \nabla^2 \phi + \frac{r_c^2}{3\beta a^2} \left[ (\nabla^2\phi)^2 - (\nabla_i \nabla_j \phi)^2 \right] = \frac{8\pi G a^2}{3\beta} \delta \bar{\rho}
\end{equation}
where $\beta = 1 + 2 H r_c \left( 1+\frac{\dot{H}}{3 H^2} \right)$ and $r_c$ is a free parameter below which gravity becomes 4-dimensional.

One can see that these gravity models generically modify the Poisson equation by adding an additional term, whose amplitude itself is dynamically determined by a non-linear equation. Numerical simulations thus need to solve for these fields in each timestep. This is usually achieved by representing these fields on an adaptive grid and solving for their values via relaxation and multi-grid methods (cf.\ Sect.~\ref{sec:multigrid}). Note that because of this overhead, traditionally, MG simulations were significantly slower than $\Lambda$CDM simulations, although recent advances (and suitable approximations) have reduced their computational cost and nowadays they have similar execution times (e.g., \citealt{Barreira:2015,Winther:2014,BoseLi:2017}). 

Note that in general, MG can produce differences with respect to $\Lambda$CDM already at high redshift. In such cases, the initial conditions of simulations need to be made with power spectra computed with Boltzmann codes incorporating such effects, such as e.g., {\sc MG-CAMB} \citep{Hojjati:2011} or {\sc hi-CLASS} \citep{Zumalacarregui:2017}. However, it is common to simulate models that depart from GR only at low redshifts, thus, the initial conditions follow that of standard simulations.

Several codes exist that simulate various kinds of modified gravity \citep{Oyaizu:2008,Schmidt:2009,Zhao:2011,Li:2012,Brax:2012,BLi:2013a, BLi:2013b,Llinares:2014,LlinaresMota:2014,Puchwein:2013,Arnold:2019,Ruan:2021,HAguayo:2021}, usually extending well-established GR codes such as RAMSES, Gadget, and AREPO. \cite{Winther:2015} carried out a comparison of various $N$-body codes for $f(R)$, DGP, and Symmetron models by simulating a $L=250\hMpc$ box with $N=512^3$ particles. These authors found a very good agreement among different codes and for multiple statistics (for instance $<1\%$ difference in the power spectrum up to $k \sim 5\ihMpc$), which has supported the validity of the MG observables predicted by numerical simulations. These MG simulations have revealed that the amplitude of the power spectrum is enhanced relative to $\Lambda$CDM due to the additional force. This enhancement is roughly scale-independent for DGP with a larger amplitude at low redshifts, about 15(3)\% for models with $H_0 r_{\rm c}=1(5)$ at $z=0$. In contrast, departures with respect to $\Lambda$CDM are scale dependent for $f(R)$, with a fractional increase of $\sim25(5)\%$ at $k\sim10\ihMpc$ for $f_{R0}=10^{-5}(10^{-6})$. The abundance of massive halos is also affected by MG, with an increase on all mass scales for DGP and a mass-dependent effect in $f(R)$ as a result of large haloes being effectively screened (see \citealt{Winther:2015} and references therein for more details).

Modified gravity has also been incorporated in approximate methods. Specifically, \cite{Winther:2017} and \cite{Valogiannis:2017} implemented modifications to {\sc COLA} (cf.\ Sect.~\ref{sec:cola_fastpm}) by computing 2LPT displacements in generic MG models (up to second order) and included various screening mechanisms. These authors found that the changes relative to $\Lambda$CDM in the power spectrum and halo mass function were accurately captured by COLA (within a few per cent up to $k \sim 3\ihMpc$). A similar accuracy was reached by \cite{MeadFR:2015} who extended cosmology-rescaling algorithms, so that the effects of MG in an $N$-body simulation could be incorporated in post-processing.

As for dark matter candidates, a large number of possible modifications to the equations of motion are allowed, each of which with their own free parameters. This is a difficulty when generic predictions are required, e.g., in data analysis, or to perform a systematic scanning of models with numerical simulations. For this reason, there has been significant work to formulate an effective parameterization whose parameters can then be constrained \citep{Lombriser:2016,Thomas:2020}.

Deviations from GR can be quantified in a general way \footnote{Sometimes this is referred to as \textit{Parameterised Post-Friedmannian approach}, in analogy to the Parameterised Post-Newtonian approach developed to measure deviations from Newtonian gravity in Solar System observations} by modifying two of the Einstein equations (in Fourier space) as
\begin{subequations}
\begin{eqnarray}
-k^2 \tilde{\Phi}(\vecb{k}) &=& 4\pi G a^2 \overline{\rho}(a)\, \mu(k,a) \, \tilde{\delta}(\vecb{k}) \\
\tilde{\Psi}(\vecb{k}) &=& \gamma(k,a)\, \tilde{\Phi}(\vecb{k})
\end{eqnarray}
\end{subequations}
where $\Psi$ and $\Phi$ are the two gravitational potentials (note that the lensing potential is given by $(\Psi+\Phi)/2$), $\gamma$ is the gravitational slip, and $\mu$ is a generic function that captures the modifications between the relation of densities and the gravitational potential (i.e. a scale and time-dependence of the gravitational `constant'). Therefore, $\mu = \gamma = 1$ yields the GR limit, but in general $\mu$ and $\gamma$ can be complicated functions of time and scale. In the linear regime, many theories (including $f(R)$ and DGP) can be exactly mapped to specific values and forms of $\gamma$ and $\mu$. However, in the nonlinear regime this is less straightforward and parameterisations have been proposed based on spherical collapse \citep{Lombriser:2016}, or a post-Friedmann formalism \citep{Thomas:2020}, which have been argued to be valid even down to very small scales.

Following this philosophy, \cite{Cui:2010} and \cite{Srinivasan:2021} used parameterised deviations from GR and carried out a suite of simulations with different parameter values. So far, both of these approaches have considered only a time dependence for $\mu$ (e.g., \citealt{Srinivasan:2021} considered piecewise constant values of $\mu$ in redshift), but it is likely that in the future extensions to scale dependence will be possible. Additionally, recently \cite{Hassani:2020} showed that simulations adopting the \cite{Lombriser:2016} parametrisation can result in percent-level agreement in the power spectrum with respect to direct $f(R)$ and DGP simulations. These results are very promising, although a more exhaustive exploration of other summary statistics will be required in the future.  Note that in this case there is no need to solve for the evolution of the scalar field, which makes this kind of simulations significantly easier to carry out. 

Another advantage of such parameterisations is that they could be readily applied to observations in the linear regime (e.g., \citealt{Blake:2020}). For instance, \cite{Mueller:2018} used a similar parameterisation to constrain deviations from GR using redshift space distortions in the BOSS survey. These constraints are expected to get significantly tighter with the newest generation of galaxy surveys and also by combining different probes (higher order $N$-point functions, marked statistics, redshift space distortion), for which we anticipate, the results of numerical simulations will be crucial. An illustration of this are the recent results of \cite{He:2016,He:2018}, who by comparing the observed small-scale galaxy clustering against $N$-body simulations, could rule out $f(R)$ models with $|f_{R0}| \gtrsim 10^{-6}$. Additionally, the advent of emulators for modified gravity (c.f. Sect. \ref{sec:emulators}) will make possible to perform those observational tests in a much more accurate manner while testing a broader set of departures from GR.

%\vspace{1cm}

\subsection*{Closing remark}
In this section we have reviewed several possible extensions to the physically simplest $\Lambda$CDM simulations. Some of these modifications are more speculative in nature, as e.g., modified gravity, whereas others are strongly motivated by physical experiments, e.g., massive neutrinos. For an adequate interpretation of future datasets it will be important to explore the interplay of such modifications and the possible degeneracies that arise among them when interpreting cosmological data. For instance, the power spectrum suppression that baryons display relative to CDM, could potentially be misinterpreted as a signature of massive neutrinos. In this direction, for instance, \cite{Kuo:2018} carried out simulations with decaying Warm Dark Matter, and \cite{Schwabe:2020} mixing fuzzy and particle Dark Matter. Additionally, \cite{Baldi:2014,Baldi:2018,Hashim:2018} have performed simulations with this focus, where they consider cosmological scenarios with both modified gravity and warm dark matter, with modified gravity and massive neutrinos, and with primordial non-Gaussianities and interacting dark energy. As the quest for new physics continues to ever smaller effects on the large-scale structure of the Universe, these kind of simulations will be increasingly important in the future, and they will be required to guarantee the robustness when a given $\Lambda$CDM extension becomes ruled out or favoured.

%% Analysis
%% \newpage 
\section{Numerical considerations and the challenge of high-accuracy simulations}
\label{sec:numerics}

Numerical simulations start to play an increasingly central role in the interpretation of observational data and in the quantitative inference of physical properties of the Universe. It thus becomes essential to ensure the high precision and accuracy of simulation results. In this section we discuss several key aspects in this regard.

\subsection{Box size and mass resolution}

\begin{figure}
\begin{center}
    \includegraphics[width=0.46\textwidth]{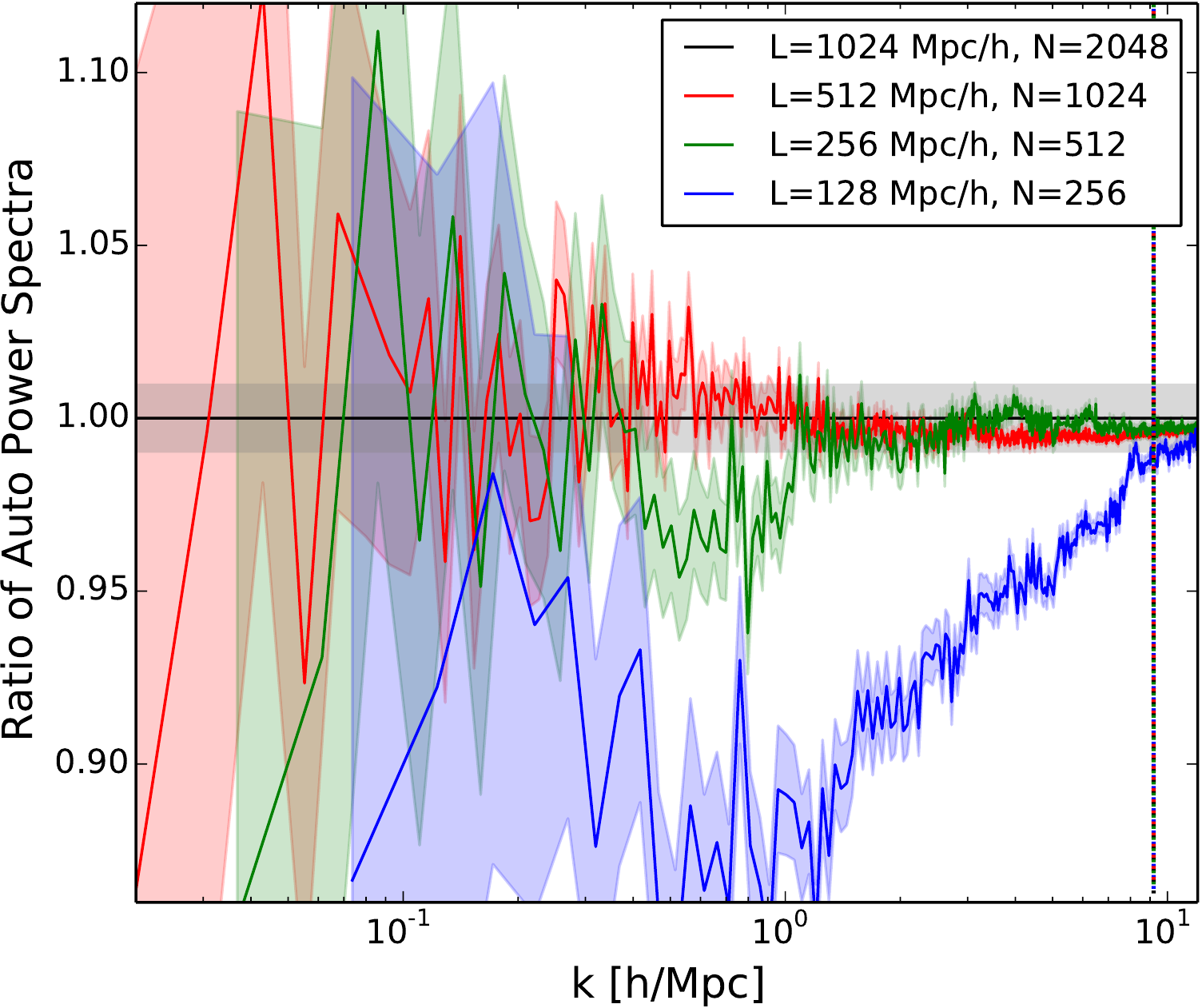}
    \hspace{8mm}
    \includegraphics[width=0.46\textwidth]{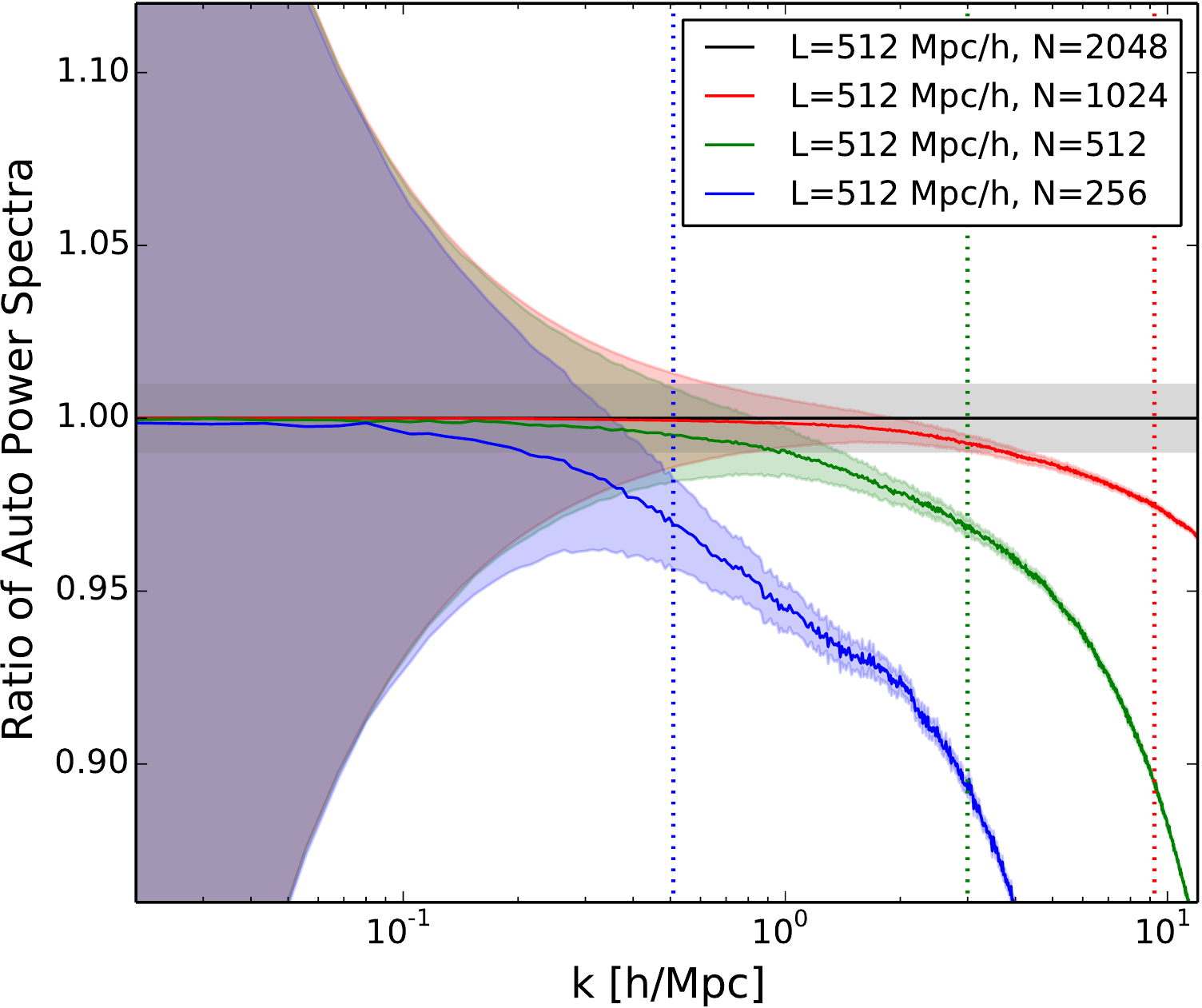}
\end{center}
\caption{\label{fig:pk_convergence}
Dependence of the nonlinear matter power spectrum at $z=0$ on the volume and mass resolution of a $N$-body simulation. The left panel shows the fractional change in the power spectrum for various box sizes $L \in [128, 256, 512, 1024]\hMpc$ at a fixed mass resolution. The right panel shows instead the change produced by increasing the mass resolution while keeping fixed the size of the simulation box. Note that for achieving percent convergence (marked by the shaded region) up to $k \sim 3\ihMpc$, boxes of at least $\sim 500\hMpc$ with particles less massive than $10^{9}\Msun$ are required. Figure adapted courtesy \cite{Schneider:2016}}
\end{figure}

Two basic properties of a cosmological simulation are the size of the simulated box and the mass of the particles employed, both of which affect the nonlinear structure formed.
For a simulation of a given side length, $L$, Fourier modes below the fundamental mode $k < k_0 = 2 \pi/L$ are effectively set to zero. Thus, structure grows as if embedded in a region at the cosmic mean density and devoid of tidal forces, which is a biased representation of actual finite regions in the universe. \cite{Power:2006} carried out a systematic study of the impact of these missing large modes finding that, while the internal properties of dark matter halos are unaffected, their abundance is strongly suppressed in small boxes, especially so at high masses.

Such a reduction in the number of dark matter halos also affects the nonlinear power spectrum, suppressing its amplitude on quasi-linear scales. This has been studied recently by several authors who all find that a boxsize of $\gtrsim 1000\,\hMpc$ is required at $z=0$ to obtain a converged measurement -- i.e. independent of further increases of box size --  of the density power spectrum and covariance matrices at the per cent level \citep{Schneider:2016,Mohammed:2014,Klypin:2019}. Specifically, the nonlinear evolution of baryonic acoustic oscillations depends on the correct modelling of large-scale flows which also requires similarly large boxes \citep{Crocce:2006b}. The result of a study investigating the effect of box size and mass resolution on the power spectrum by \cite{Schneider:2016} is reproduced in Fig.~\ref{fig:pk_convergence}, where in the left panel the box size is varied at fixed mass resolution and in the right panel the mass resolution is varied at fixed box size, each with significant impact on the matter density spectrum.

The effects of the boxsize can be ameliorated by changing the equations of motion such that they account for the fundamental (or DC) mode \citep{Sirko:2005,Gnedin:2011}, and sampling an ensemble of simulations representative of the variance at the box scale. Such an approach is particularly important when carrying out simulation ensembles to compute covariance matrices, as these are sensitive to overdensities on scales larger than the box (see Sect.~\ref{sec:supersample}). Additionally, by setting up initial conditions matching statistics in real space rather than Fourier space \citep{Sirko:2005}, it is possible to obtain e.g., halo mass functions that are less biased even for smaller boxes.  Also, recent developments for modelling tidal fields larger than the box size \citep{Schmidt:2018} might prove useful in the future to achieve better convergence. Further, compactified simulations might offer an alternative path as they naturally incorporate very large-scale modes \citep{Racz:2018}.

While the finite simulation volume modifies the formation of halos mostly at the high mass end, finite mass resolution determines the smallest halo resolvable in a given simulation box. Thus, if halos that contribute significantly to the nonlinear power spectrum are not resolved, then the spectrum will be biased low. \cite{Schneider:2016} argued that a mass resolution of at least $10^{9}\,\Msun$ is required to achieve per cent level convergence at $k \sim 1\,\ihMpc$ at $z=0$, as smaller halos contribute negligibly to the power spectrum. Higher wavenumbers will be set by the inner regions of halos, which  might be affected by two-body relaxation and other effects if not resolved with an adequate number of particles.

Furthermore, initial fluctuations resolved with a small number of particles will suffer from large numerical errors. Specifically, the abundance of halos resolved with approximately less than $100$ particles is overestimated as halo finders identify statistical upward fluctuations as real objects \citep{Warren:2006}. Note however that the actual rate of convergence depends on the halo finder algorithm (see Sect.~\ref{sec:halos}), as well as other numerical parameters such as force accuracy and the softening length.

%%%%%%%%%%%%%%%%%%%%%%%%%%%%%%%%%%%%%%%%%%%%%%%%%

\subsection{Close encounters and regularization in \texorpdfstring{$N$}{N}-body methods}
\label{sec:softening}

\begin{figure}
\begin{centering}
\includegraphics[width=0.6\textwidth]{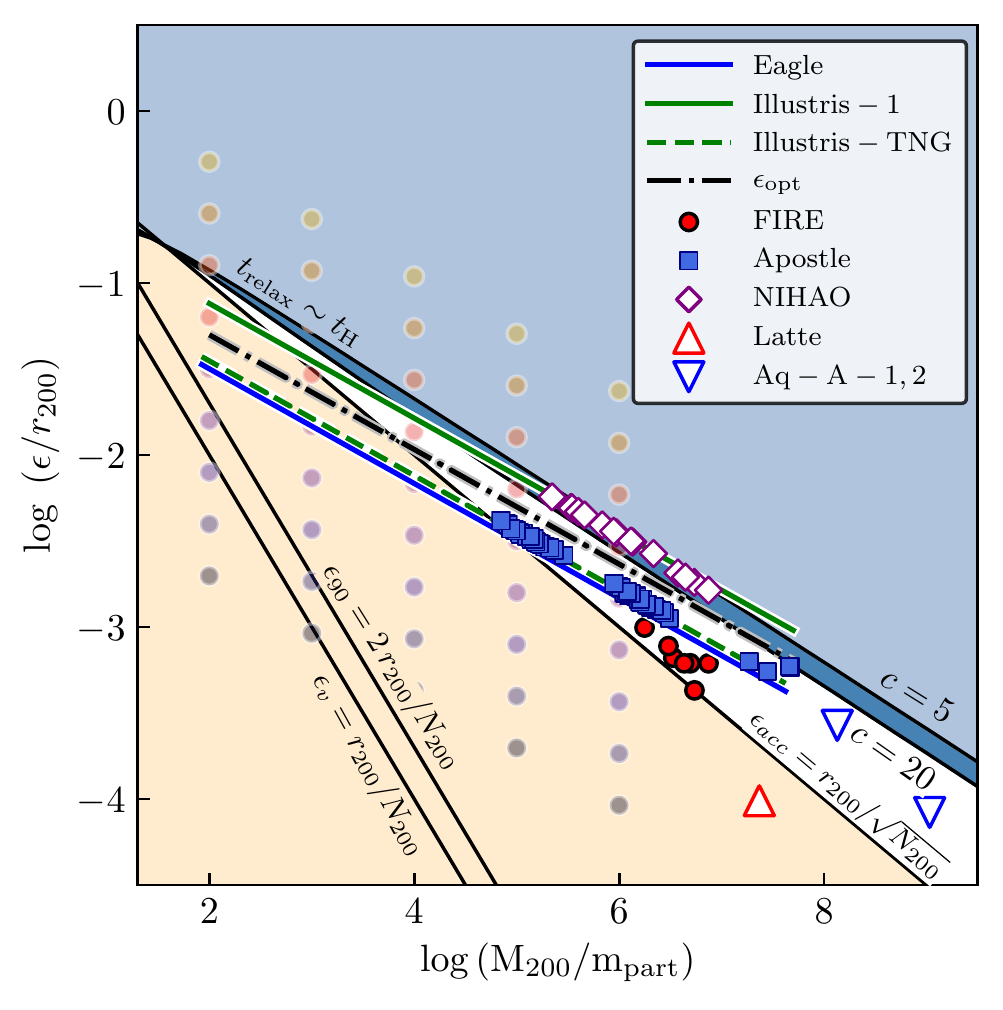} 
\caption{\label{fig:softening} 
    Diagonal black lines show three estimates for the minimum value of the gravitational softening $\epsilon_{90}$, $\epsilon_v$, $\epsilon_{\rm acc}$ required for the simulated dynamics to be collisionless. Thus, the beige region might be dominated by numerical errors. The dark blue region marks the minimum scale for which the 2-body collisional relaxation is equal to a Hubble time, for halos of different concentrations. Thus, an optimal value of the lie near the white region for halos resolved with at least $100$ particles. The optimal values of \cite{vandenBosch:2018b}, $\epsilon_{\rm opt}/r_{200} = 2.3\times N_{200}^{-1/3}$, which is approximately $\epsilon_{\rm opt} \sim 0.017 L/N^{1/3}$, is denoted by the dashed line, whereas the values adopted in various simulations are shown by coloured lines. Figure adapted courtesy \cite{Ludlow:2019}}
\end{centering}
\end{figure}

Another aspect affecting the numerical accuracy of the $N$-body solutions is related to the so-called softening length, where the gravitational interaction ceases to be Newtonian. As discussed also at length in Sect.~\ref{sec:ewald_p3m}, the Green's function of the Laplacian is $G(\vecb{r}) = -1/(4\pi\|\vecb{r}\|)$, which translates into the Green's function $\vecb{G}_a:= -\bnabla\nabla^{-2} \delta_D(\vecb{r})$ of the acceleration operator, i.e.,
\begin{equation}
\vecb{G}_a(\vecb{r}) = -\frac{\vecb{r}}{4\pi\| \vecb{r} \|^{3}} \label{eq:Green_accel}
\end{equation}
which are both divergent for $r\to0$. If the goal is to model dark matter in the continuum limit, then this is undesired behavior: it drives an effective collision term resulting in two-body relaxation, which leads to a deviation from collisionless Vlasov-Poisson dynamics. In practice, it also imposes arbitrarily small time steps if step criteria like Eq.~(\ref{eq:std_timestep_crit}) are used. The usual solution is to impose a small-scale cut-off. In the simplest case of `Plummer softening' (e.g., \citealt{Hernquist:1990}) one sets $G_\epsilon(\vecb{r}) = -1/(4\pi\sqrt{\|\vecb{r}\|^2+\epsilon^2})$ with a `softening length' $\epsilon$, so that the tamed acceleration becomes
\begin{equation}
\vecb{G}_{a,\epsilon}(\vecb{r}) := -\bnabla G_\epsilon = -\frac{\vecb{r}}{4\pi\left(\| \vecb{r} \|^2+\epsilon^2\right)^{3/2}},
\end{equation}
which in fact vanishes for $r\to0$, but is only asymptotically Newtonian (i.e. at $r \rightarrow \infty$). The maximum acceleration is reached for $r=\epsilon/\sqrt{2}$ with a value of $\max_r \|\vecb{G}_{a,\epsilon}\| = 2/(3\sqrt{3}\epsilon^2)$. Other popular ways of modifying Eq.~(\ref{eq:Green_accel}), such as Kernel softening, can guarantee a transition to Newtonian force at finite $r$, typically a few times the softening scale, e.g., \cite{Springel:2001b}. One should also remark that if grid-based methods are used, unless extremely aggressive adaptive mesh refinement strategies are employed (e.g. refinement on single particles), the grid scale directly provides a regularisation scale and suppresses two-particle interactions.

With softening, the collisionless $N$-body system becomes well behaved at the price of smoothing out structure on scales comparable to the size of $\epsilon$. Formally, the need for a softening length appears in cases where the $N$-body discretization is no longer valid -- separations smaller than the typical size of the volume chosen for the coarse-graining process. Under these arguments, the softening should be chosen such that $\epsilon \sim \ell := (V/N)^{1/3}$ \citep{Melott:1997,Splinter:1998,Romeo:2008}. Unfortunately, with current computational power, this is typically much larger than the regions of interest -- e.g. the virial radius of a halo resolved with $\sim 100$ particles is $\sim 0.3 \ell$ -- and large-scale simulations adopt values 30--100 times smaller. For instance, and in units of the mean inter-particle separation, the Bolshoi simulation employs 0.016 \citep{Klypin:2011}, the Multidark simulations 0.0150--0.026 \citep{Klypin:2016}, the set of Millennium simulations 0.022 \citep{Springel:2005b,Boylan-Kolchin:2009,Angulo:2012}, and the EUCLID flagship 0.02 \citep{Potter:2017}.

There are several possible criteria for setting the softening length in a simulation. The goal is to suppress discreteness effects and force errors while maximizing the range of scales and amount of resolved nonlinear structure \citep{Ludlow:2019}. One possible criterion, $\epsilon_v$, is to require that the binding energy of a halo is larger than that of two particles at a distance $\epsilon$, other criteria aim to suppress the effect of close two-body encounters by looking at the change in acceleration ($\epsilon_{\rm acc}$) or in large-angle deflections ($\epsilon_{90}$). However, these are qualitative estimates, and usually softening is chosen empirically from the observed convergence in numerical experiments, e.g. the behaviour of the profiles of halos or the small-scale power spectrum. Fig.~\ref{fig:softening} shows a diagram with various possible estimates of the softening length, including $\epsilon_v$, $\epsilon_{\rm acc}$, $\epsilon_{90}$ and the optimal values discussed below, along with the values adopted by some state-of-the-art simulations. 

\cite{Power:2003} carried out a systematic analysis of the impact of numerical parameters in the convergence of circular velocity profiles. They empirically showed that convergence to within $\sim10\%$ can be achieved on scales larger than a region enclosing a number of particles whose two-body relaxation time was larger than the Hubble time. This criterion has been validated and employed in multiple simulations afterwards \citep{Diemand:2004,Springel:2008,Navarro:2010,Gao:2012} and is usually adopted when setting the softening length of current simulations. In addition, the same authors \citep{Power:2003} proposed an empirical `optimal' softening length $\epsilon_{\rm opt}/r_{200} = 4 (N_{200})^{-1/2}$ -- this approximately translates to $\epsilon = 0.03 \ell$ -- which is widely used, especially in zoom simulations. Note, however, that \cite{ZhangSoftening:2019} argued that this criterion was overly conservative (see also \citealt{Ludlow:2019}) and advocated for a factor of 2 smaller softening, i.e. $\epsilon_{\rm opt}/r_{200} = 2 (N_{200})^{-1/2}$ -- based on the convergence of density and circular velocity profiles of $\sim10^{12}\Msun$ halos in cosmological boxes. An even smaller value of the softening length was advocated by \cite{Mansfield:2021}. By comparing a large number of publicly-available simulations, they found that the shape and peak of the circular velocity of halos appeared systematically biased for values larger than $\epsilon/\ell \sim 0.008$.

Using controlled idealised simulations, \cite{vandenBosch:2018a,vandenBosch:2018b} have also argued that current choices of the softening length are inadequate for simulating the evolution of dark matter subhalos. Specifically, they argue that subhalos resolved with less than 1000 particles suffer from instabilities and are artificially disrupted. This could be a serious "over-merging" problem (a classic $N$-body problem, see already \citealt{Moore:1996}) for simulations as their prediction for the abundance and spatial distribution of dark and luminous satellites could be unreliable. However, more recent results \citep{Green:2021} indicate that the effect is at most 10--20 per cent on the subhalo mass function. As a way to alleviate the overmerging problem, \cite{vandenBosch:2018a} argue for a softening length a few times smaller than usual choices. Specifically, they argue that the optimal softening is $\epsilon_{\rm opt}/r_{200} \simeq 2.3 (N_{200})^{-0.33}$ (which is very similar to that of \citealt{ZhangSoftening:2019} discussed above).

On cosmological scales, \cite{Joyce:2020} argued that self-similarity in statistics measured in scale-free simulations (i.e., simulations with a power-law initial power spectrum in an $\Omega_m=1$ cosmology) is a good indicator of the degree of convergence of a given simulation setup. With such, \cite{Garrison:2021} advocate for an optimal softening length of $\epsilon = 0.033 \ell$ but fixed in {\it physical} (as opposite to {\it comoving}) coordinates. The choice of physical softening is a usual practice in hydrodynamic and zoom simulations, but note that this formally yields a time-dependent Hamiltonian (cf.\ Sect.~\ref{sec:time_integration}) requiring some extra care in formulating the integrator since the gravitational force would receive an additional contribution \citep{Price:2007} (but note that in superconformal time, the potential is already the only time-dependent piece of the Hamiltonian). However, the magnitude of the error introduced by ignoring such details might be acceptable for large-scale simulations, with the advantage of requiring less timesteps at high redshifts.

In summary, there is consensus that force softening slightly smaller than what has been traditionally adopted is preferred to improve convergence at a fixed computational cost. On the other hand, it is important to keep in mind that there are a few examples of problems originating from too small softening lengths. One is the case of artificial fragmentation in warm dark matter and first halos, where filaments (expected to be completely smooth) break down in pieces in a mass-resolution dependent way. Another example is in the simulation of two cold fluids with different initial power spectra (as is the case of baryons and dark matter in the early Universe). One has to keep in mind however that whenever $\epsilon\ll\ell$, then the system evolves as a discrete system rather than in the continuous limit. This leads to significant deviations during the initial phase of a simulation (see our discussion in the next section on initial conditions), and contributes an additional source of numerical error in cosmological predictions.

There have been some attempts in the literature to fix the problems of the local evolution of the `mean' particle separation scale $\ell$ but with relatively limited success. One example is the case of a variable softening length which is determined by the local density \citep{Price:2007,Bagla:2009,Ianuzzi:2011}, or by the eigenvalues of the moment of the inertia tensor \citep{Hobbs:2016}. Unfortunately, anisotropic collapse is typical in cosmological simulations, and thus isotropic softening tends to under- or over-smooth forces. Isotropic adaptive softening is also already implicitly built into those codes that rely on adaptive mesh refinement and a particle-mesh based gravity solver, as e.g. {\sc RAMSES} or {\sc ART}, see also \cite{Knebe:2000}, since the refinement criterion is usually tied to the local particle density. A more conservative refinement criterion compared to the very small softening typically employed with tree codes is then also the reason why often such AMR codes display some suppression of the mass function for the smallest halos (see also \ref{sec:code_comparison} below).

It is clear from the above that a possible direction for progress could be an anisotropic softening determined by the local distortion tensor or by solving for the GDE. This, in fact, would be very similar to a low-order version of the cold sub-manifold method discussed earlier. 

\subsection{Accuracy of initial conditions}
\label{sec:accuracy_ic}

\begin{figure}
\begin{centering}
\includegraphics[width=\textwidth]{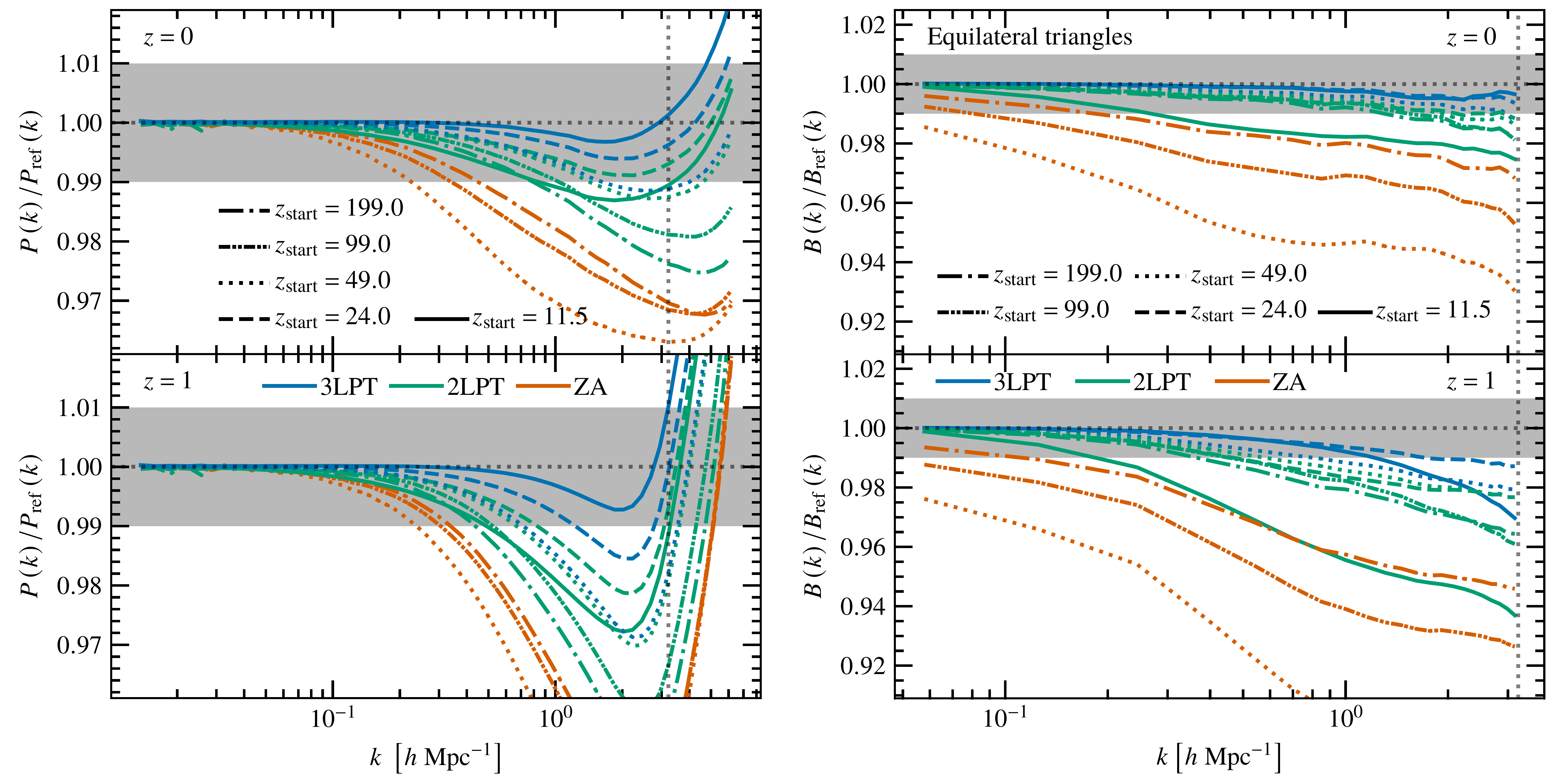} 
\caption{\label{fig:ics_convergence}
Dependence of the statistical properties of a simulated nonlinear density field on the details of how the initial conditions are constructed. Left and right panels show the results for the matter power spectrum and for the bispectrum with equilateral configuration ($\vecb{k}=\vecb{k_1}=\vecb{k_2}$); whereas top and bottom panels display $z=0$ and $z=1$, respectively. In each panel, results are shown for various choices of starting redshift $z_{\rm start} \in [11.5, 24, 49, 99, 199]$, and the order of the Lagrangian Perturbation Theory (LPT) used to compute the particle displacements at the respective redshift. The reference run is a simulation initialized with 3LPT at redshift 24 and using a face-centered-cubic lattice with four times as many particles as in the test simulations. Figure adapted courtesy \cite{Michaux:2020}}
\end{centering}
\end{figure}

Initial conditions that are set-up using Lagrangian perturbation theory (LPT), see Sect.~\ref{sec:ICs_from_LPT}, use a low-order truncation of LPT. The resulting truncation error appears as a `transient' error in the non-linear evolution of the cosmic density field. This truncation error is largest if LPT is truncated at low order and the simulation started at late times. As a consequence, second order (2LPT) ICs have long been advocated as a more accurate replacement of first order (Zel'dovich approximation) ICs for setting up initial conditions \citep{Scoccimarro:1998,Valageas:2002,Crocce:2006}. In principle, it is always possible to simply start the simulation at early times to reduce the truncation error and avoid having to use higher order LPT, but numerical errors during the early stages of the simulation  can be large for some schemes, since the density perturbations away from homogeneity become increasingly smaller and are easily overshadowed by numerical errors if inadequate force calculation is used (this is particularly a problem for the tree method (cf.\ Sect.~\ref{sec:tree}).

However, a further limitation of $N$-body simulations is due to discreteness effects. While LPT initialises the $N$-body simulation with the growing modes of the continuous fluid system (with perturbations truncated in the UV at some wave number usually not too far from the particle Nyquist wave number), the $N$-body system itself follows the dynamics of a discrete system. The resulting growing and decaying modes are different from the continuous fluid, and in fact anisotropic at the discretisation scale, as demonstrated by  \cite{Joyce:2005,Joyce:2007a,Joyce:2007b,Marcos:2006}. As a consequence, there is a secondary transient resulting from the transition of the $N$-body simulation from the continuous dynamics to the discrete dynamics, which is accompanied by a distinct suppression of the power spectrum on scales close to the particle Nyquist wave number during weakly non-linear phases of the simulation. To compensate for this effect, \cite{Garrison:2016} have proposed to explicitly correct for this error by projecting into the discrete eigenmodes and correcting the lack of growth with a boosted velocity that exactly compensates the leading error term at a specific target time. Arguably a downside of this approach is that, due to the boost, the evolution of the system prior to the target time is somewhat unphysical.

In order to circumvent both truncation and discreteness errors, \cite{Michaux:2020} have argued for the use of particularly late starting times employing higher order (3LPT) initial conditions. \cite{Michaux:2020} also showed that these errors are largest during the mildly non-linear evolution, but then decrease again once scales close to the particle Nyquist wave number are dominated by structures that are fully collapsed and virialised. The use of high order LPT and late starting times might therefore allow more economical simulations, i.e. fewer particles to reach a given accuracy up to a certain wave number. Even higher order initial conditions might not be necessary except in very special situations, since LPT has been shown to converge very quickly prior to shell-crossing \citep{Rampf:2020b}. The study by \cite{Rampf:2020b} also found that for higher order LPT it is preferable to exclude so-called `corner modes' (i.e., modes between 1 and $\sqrt{3}$ times the linear particle Nyquist wave number) in order to reduce artefacts in non-linear terms due to UV truncation of the perturbation spectrum and convolution integrals. The effect of these `corner modes' for low order ICs (i.e. up to 2LPT) on non-linear simulation results has been found to be small \citep{Falck:2017}.

All these aspects are summarised in Fig.~\ref{fig:ics_convergence} which compares the power spectrum and bispectrum as measured in simulations adopting various starting redshifts, $z_{\rm start}$, and LPT orders. We can see, for instance, that using 1LPT (i.e. the ``Zeldovich approximation'') leads to a systematic underestimation of  clustering statistics on both intermediate and small scales. This effect is somewhat reduced for higher starting redshifts, but an increase in the LPT order yields a much more significant improvement. Employing 3LPT initial conditions, even at $z_{\rm start} = 11.5$, agrees with the reference soluution at the subpercent level for $k\lesssim2\ihMpc$ at both redshifts and for the power spectrum and bispectrum in this case. 

%%%%%%%%%%%%%%%%%%%%%%%%%%%%%%%%%%%%%%%%%%%%%%%%%%%%%%%%%%%%

\subsection{Chaos and determinism in simulations}

So far we have discussed how differences in the setup, discretization, or numerical parameters of a simulation affect its predictions. However, even for a given set of choices and deterministic equations of motion, stochasticity can arise in the predictions for the nonlinear density field as a result of chaotic behavior. 

Chaos refers generally to a process in which exponentially divergent results appear from small differences in the initial state of a system. Chaos in a Hamiltonian system in a 3+3-dimensional phase space can be formally quantified in terms of three distinct Lyapunov exponents $\pm\lambda_{1,2,3}$ (due to the symplectic nature they come in pairs). A notion of predictability of a system is then given by the maximal Lyapunov exponent, defined as the most rapid characteristic separation rate of two trajectories separated by a vector $\delta\vecb{\xi}(t)$ in phase space, i.e. $\left| \delta \vecb{\xi}(t) \right| \sim \exp(\lambda t)\,\left| \delta \vecb{\xi}_0 \right|$ in a linearised sense so that

\begin{equation}
\lambda \sim \lim_{t\to\infty} \frac{1}{t} \log \frac{ \left| \delta \vecb{\xi}(t) \right| }{ \left| \delta \vecb{\xi}_0 \right| },
\end{equation}
where the initial separation $\delta\vecb{\xi}_0$ should be thought of as infinitesimal. 

In the case of numerical simulations, seeds for chaos and stochastic behavior could arise from round-off errors and/or from small variations in the initial conditions of the system. Additionally, errors associated with the force calculation and time integration could be exponentially amplified in chaotic systems. In fact, examples of chaotic behavior have been reported in the literature for $N$-body simulations of star clusters, satellite galaxies, halo stars, and planetary systems among others \citep{Heggie:1991,Goodman:1993,El-Zant:2019,Maffione:2015,Price-Whelan:2016}. In principle, Lyapunov exponents can be computed during the evolution of the $N$-body system to identify regions of chaos \citep{Habib:1995}, and the `GDE' approach can also give direct access to them \citep{Vogelsberger:2008}. Understanding chaos in a cosmological context is important to determine the accuracy and robustness of numerical simulations. As discussed earlier, the  evolution of a $N$-body system may not be guaranteed to converge to the continuum limit even for infinite particles. 

Chaos in a cosmological simulation context has been explored by several authors. By comparing an ensemble of runs with identical initial power spectrum but with small differences in their white-noise field, \cite{Thiebaut:2008} found that large scales and time-integrated halo properties such as position, mass and spin were robust predictions. On the other hand, the position of substructures, and the orientation of the spin and the velocity dispersion tensor showed larger variations. 

Using pairs of simulations with identical initial conditions up to a small perturbation, \cite{Genel:2019} quantified the role of chaos in the properties of well-resolved objects in numerical simulations. They found that differences in the mass and circular velocity of halos grows exponentially, but that it saturates at a level compatible with Poisson noise; $\sim 1/\sqrt{N}$ for a system resolved with $N$ particles. A similar conclusion was reached by \cite{El-Zant:2019} studying the time-reversibility of $N$-body systems: initial errors grow rapidly during an initial phase but they subsequently saturate. 

The results above suggest that numerical simulations of self-gravitating systems do indeed converge, albeit slowly, to the collisionless limit as $N$ tends to infinity. This is in contrast with cosmological hydrodynamical simulations, where initial differences, round-off errors and sometimes stochastic star formation prescriptions have been found to result in marked differences in the global properties of simulated galaxies, even contributing significantly to some scaling relations \citep{Genel:2019,Keller:2019}. 

%%%%%%%%%%%%%%%%%%%%%%%%%%%%%%%%%%%%%%%%%%%
%%%%%%%%%%%%%%%%%%%%%%%%%%%%%%%%%%%%%%%%%%%

\subsection{Convergence among codes}
\label{sec:code_comparison}

\begin{figure}
\includegraphics[width=\textwidth]{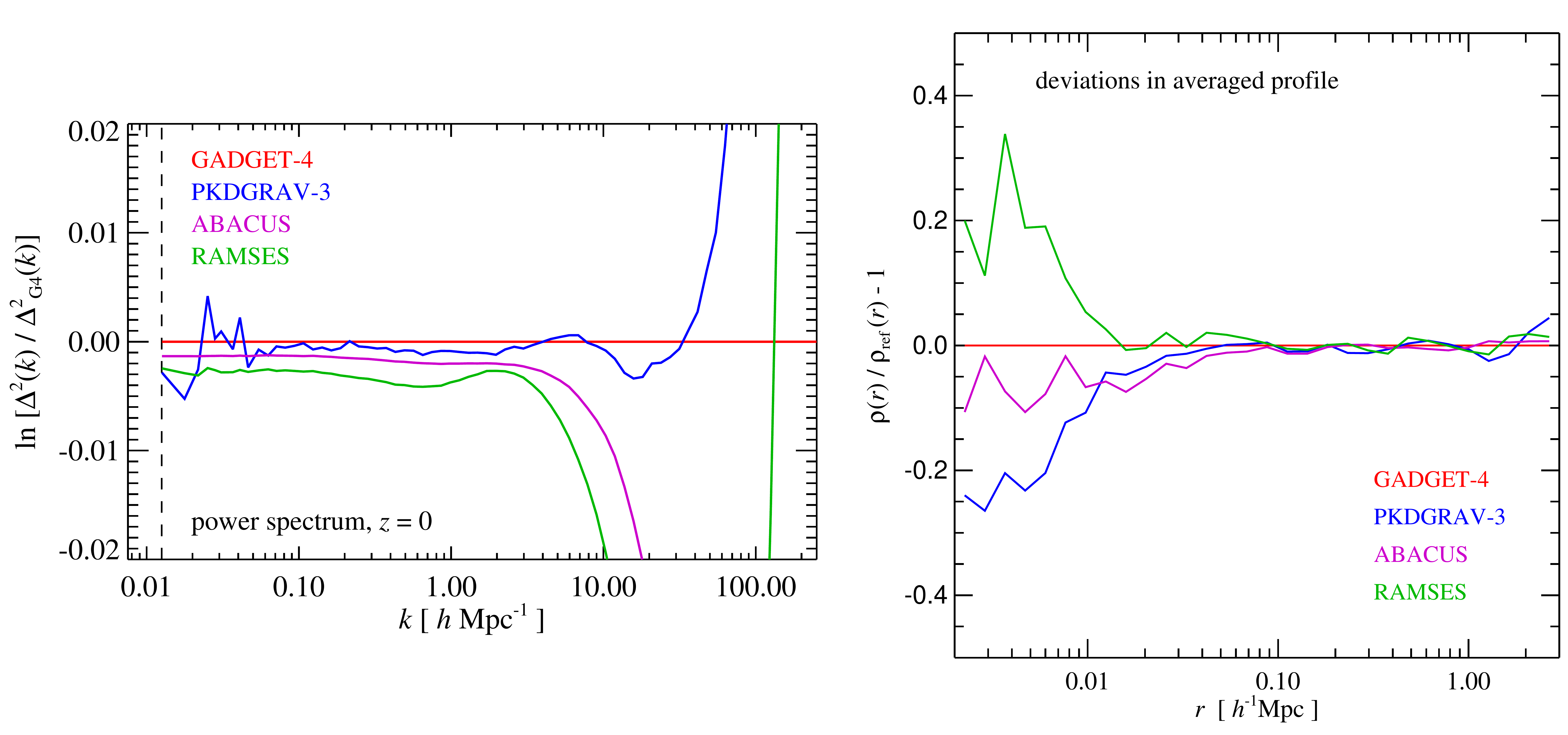}
\caption{Comparison between the nonlinear structure at $z=0$ as predicted by $4$ different $N$-body codes evolving the same initial conditions of the ``Euclid comparison project'' \citep{Schneider:2016}. This simulation corresponds to a cubical box of side $L=500\hMpc$ with $2048^3$ particles of mass $1.2\times10^9\Msun$. The left panel shows the nonlinear matter power spectrum whereas the right panel shows the stacked density profile for the 25 most massive halos in the simulation. Displayed results are obtained from {\sc GADGET-4} \citep{Springel:2020}, {\sc PKDGRAV-3} \citep{Potter:2016}, {\sc ABACUS} \citep{Garrison:2019}, and {\sc RAMSES} \citep{Teyssier:2002}. Note that all codes agree remarkably well even down to very small scales (<1\% up to $k \sim 10 \ihMpc$), and eventually disagree on smaller scales owing to different numerical parameters adopted in each run. Similarly, the density profiles agree to a percent level down to $10 h^{-1}$kpc. Figure adapted courtesy of \cite{Springel:2020} \label{fig:code_comparison}} 
\end{figure}

With few exceptions (see Sect.~\ref{sec:discretizations}), all modern simulation codes essentially adopt the same approximations to predict the nonlinear state of large-scale structure in the Universe: the $N$-body discretization of a single fluid (describing both baryons and dark matter) that only interacts gravitationally. However, different codes make different algorithmic choices regarding the force calculation, force softening, and time integration, and also could be subject to systematic errors (also programming bugs become more likely with the increasing algorithmic complexity of codes). Given the importance of the predictions and role of numerical simulations, it is important to compare the performance and predictions of different codes. 

An early comparison carried out by \cite{OShea:2005} between the codes {\sc ENZO} and {\sc GADGET} revealed systematic differences originating from the algorithm employed to compute gravitational interactions. The particle mesh simulation with adaptive mesh refinement yielded a lower power spectrum amplitude on small scales compared to the tree-PM simulation. It was found that this discrepancy came from the conservative refinement criteria missing the formation of small halos at high redshift. A more aggressive mesh refinement led to much better agreement, but this served as an example that, although codes are solving the same equations, numerical simulations are complex systems and numerical errors can easily propagate to large levels. In the same year, \cite{Heitmann:2005} compared several codes in the context of large-scale clustering predictions by carrying out both cosmological and idealised simulations.

A more systematic comparison among $N$-body codes was carried  out in 2008 by the \textit{Cosmic Code Comparison Project} \citep{Heitmann:2008}, in which ten codes evolved the same initial conditions of comoving boxes of $64\,\hMpc$ and $256\,\hMpc$, as well as the ``Santa Barbara cluster'' \citep{Frenk:1999}. The agreement of the codes had by then improved dramatically with all of them agreeing on the amplitude of the $z=0$ nonlinear power spectrum at the $10-20$\% level up to $k \sim 10\ihMpc$, and at $5-10$\% at $k \sim 5\ihMpc$. A comparison in 2014 of nine cosmological $N$-body codes simulating the same $\sim 10^{11}\Msun$ halo by \cite{Kim:2014} demonstrated small differences in the global density profile but pronounced differences in the subhalo mass functions between AMR and tree-based codes consistent with the earlier found suppression of small haloes when conservative refinement criteria are used.

\cite{Schneider:2013} carried out the ``Euclid comparison project'' -- a more demanding test simulating a box of $500\,\hMpc$ per side with $2048^3$ particles, which should have led to converged results with respect to boxsize and mass resolution. There were only three participant codes, {\sc GADGET-3}, {\sc PKDGRAV-3}, and {\sc RAMSES}, which showed a remarkable agreement of $< 5\%$ for $k \sim 5\ihMpc$. An important feature of this code comparison was that the initial conditions were released publicly \footnote{\url{https://www.ics.uzh.ch/~aurel/euclid.htm}}, so that any $N$-body code could carry out the same simulation and compare their predictions. In fact, subsequently, further codes have been added to this comparison as the initial conditions were run by several authors recently to include the codes {\sc ABACUS} \citep{Garrison:2019}, {\sc L-GADGET-3} \citep{Angulo:2020}, and {\sc GADGET-4} \citep{Springel:2020}. 

In Fig.~\ref{fig:code_comparison} we show a recent comparison between four state-of-the-art codes. Remarkably, the power spectrum on large scales agrees between all of them to better than $0.4\%$ which extends up to $k \sim 2\ihMpc$ and somewhat degrades to $2\%$ up to $k \sim 10\ihMpc$. Note that in the original comparison paper, {\sc GADGET-3} was an outlier, displaying systematic differences with respect to the other codes. However, the results of \cite{Angulo:2020} and \cite{Springel:2020} regarding the {\sc GADGET-3} code suggest that this run  was probably carried out with poorer time integration or force accuracy compared to the other codes.

Overall, the current agreement at the sub-percent level among codes is a remarkable achievement for the field of computational cosmology, which can now claim that the nonlinear evolution of collisionless matter can be predicted to better than one per cent, including sources of systematic error, over the full range of scales relevant for upcoming weak lensing surveys. This naturally has been the first necessary step to demonstrate the robustness of numerical predictions, in the future similar comparisons extended to other statistics (e.g. the bispectrum and higher-order correlations, velocity fields, halo mass functions, etc) and to a more realistic description of the universe (e.g. multiple fluids, including neutrinos and baryons), will be required to establish robustly the predictions of numerical simulations. In addition, a careful assessment of the impact of the ubiquitous $N$-body discretization itself will also be essential, which only now is becoming possible with the development of the alternatives that we have discussed previously. We discuss this in the next subsection.

\subsection{The \texorpdfstring{$N$}{N}-body approximation}

Throughout this review we have argued that simulations are essential in modern cosmology and are the only way to obtain accurate results in the nonlinear regime. Specifically, in this section we have discussed how the results of cosmological simulations converge as a function of numerical parameters; box size, mass resolution, and initial conditions, as well as conditions for choosing the softening length. Additionally, we discussed how different simulations codes do agree in their predictions for the nonlinear matter clustering down to small scales.

It is important to highlight that these statements regard convergence within the $N$-body discretisation. All codes employed in large-scale simulations do assume the same $N$-body discretisation of the underlying fluid equations, and they differ simply in algorithmic choices about how to compute gravitational forces and perform the time integration. Although they might differ in their computational efficiency and convergence rate, they are expected to show convergence. However, the $N$-body dynamics and final distribution function is not necessarily identical to that of the continuum limit. Actually there is no formal proof of the convergence, and any observed convergence through resolution studies could arise due to, for instance, the inherent noisiness of the problem or it could be that the convergence to the true solution is extremely slow. 

As we have discussed before, there are actually a few examples where the $N$-body discretization clearly introduces significant errors, even on large-scales. The spurious effects of discreteness noise and a finite softening length are very evident in warm dark matter cosmologies where filaments fragment and collapse into halos originated by discreteness effects. Another example is in two-fluid simulations, such as baryon and dark matter, where discretisation errors lead to the incorrect evolution, even on extremely large scales due to small-scale coupling. {Also, the differences between the continuous fluid solution and the discrete $N$-body system are noticeable as a pronounced starting redshift dependence of the small-scale power spectrum during the mildly non-linear evolution.

In fact, there are claims that $N$-body results might be strongly affected by discreteness noise, and overly strong phase-space diffusion, which could act as an attractor, thus showing convergence to the wrong solution \citep{Baushev:2015}. In this case, for instance, the actual density profile of collapsed objects might differ from the standard NFW profile, which could ultimately imply biases in, for instance, the comic parameters inferred from weak gravitational lensing.

Given the importance of simulations in cosmology, it is crucial to be aware of these limitations and explore whether the predictions relevant for large-scale structure and the interpretation of cosmological observations is actually correct. An important test  would be to compare the $N$-body results against those of simulations adopting other discretisation techniques. However, this has not yet been fully achieved in the highly nonlinear regime, since the alternative techniques are much more computationally expensive.

An important step in this direction was recently taken by \cite{Colombi:2020}. In this study, the formation of earth-mass microhalos and an idealised case of three collapsing sine waves were simulated using a $N$-body PM code and the Lagrangian phase-space tessellation code {\sc ColDICE} \citep{Sousbie:2016}. These authors showed that both approaches agree remarkably well, {\em as long as} there is more than one particle per force resolution element. This result is in agreement with the well known studies that argued that $N$-body simulations can be reliable only as long as the softening length is chosen of the order of the mean inter-particle separation \citep{Melott:1997,Splinter:1998}. This scale is also the one implicitly adopted by phase-space tessellations in Lagrangian space. 

Interestingly, \cite{Colombi:2020} showed that $N$-body and phase-space tessellations both predict a steep power law density profiles, in agreement with several previous studies (cf.\ Sect.~\ref{sec:wimp}). After an initial period of collapse, discreteness noise and perturbations of physical origin, such as mergers, drive the system towards an NFW-like profile \citep{SyerWhite:1998,Ishiyama:2014,Ogiya:2016,Angulo:2017,Ogiya:2018}. This is an interesting result since in some WDM models, the monolithic collapse of fluctuations occurs near the observational limits, thus it might be possible that this distinctive internal properties might leave observable signatures.

In summary, there is strong evidence that the $N$-body solution is correct as long as the softening is appropriately chosen and is of the order of the local mean inter-particle separation. Note, however, that as we have see in section \ref{sec:softening}, modern large-scale simulations typically use much smaller values. {This is mandated by the requirement that the gravitational force should be Newtonian down to the smallest possible scale in order to prevent a too slow growth of small-scale perturbation in CDM simulations. It will be interesting in the future to confirm that the overall predictions of large-scale simulations are quantitatively correct. This will be a crucial step towards robust inferences from cosmic observations that will progressively rely more on numerical simulations.

%% Analysis
%% \newpage 

\section{Analysis and postprocessing}
\label{sec:analysis}

In this section, we will review the most common statistics that can be computed from cosmological $N$-body simulations, and discuss various analysis and post-processing strategies to connect simulated universes to observables in the real Universe.

\subsection{The density field}
The most basic predictions from simulations are the statistics of the nonlinear matter density and velocity fields across cosmic time. These statistics are closely related to the interpretation of several observational measurements such as gravitational lensing, the abundance and spatial distribution of biased tracers, redshift space distortions, and the kinetic Sunyaev--Zel'dovich effect, and are important for an understanding of structure formation in general.

Using a mass deposit scheme (most commonly CIC interpolation, see Sect.~\ref{sec:mass-assignment-schemes}), the density field can be readily obtained on a three-dimensional grid from the $N$-body particle distribution. Such a procedure has however the problem that, if perturbations on scales $k>k_{\rm Ny}$ exist (the Nyquist wave number is defined as $k_{\rm Ny} := \pi/\Delta x$, where $\Delta x$ is the grid spacing) they will be misidentified as modes supported by the grid (i.e. $k<k_{\rm Ny}$), which leads to errors in Fourier space. This is referred to as aliasing (see \citealt{Hockney:1981} for a detailed discussion). This problem can be cured by interlacing, where the particles are additionally shifted (in the simplest case once by a vector $(\Delta x/2,\,\Delta x/2,\, \Delta x/2)$) before deposit and then shifted back using a Fourier space shift. If averaged with the original deposit, it can be shown that the leading order aliased contribution is cancelled out \citep{Hockney:1981,Sefusatti:2016}. 

Since the $N$-body scheme is Lagrangian, particles sample mass, and therefore the signal-to-noise ratio of the density field can be very low if a grid cells contain few (or no) particles. This shortcoming can in principle be circumvented by employing adaptive kernel estimators (such as an SPH)-like approach, a Voronoi-tessellation based on particle positions \citep{vandeweygaert:1994,Cautun:2011}, or by using the phase space sheet tessellation method (cf.\ Sect.~\ref{sec:dmsheetsims}, \citealt{Kaehler:2012}), which all can give a well defined density estimate everywhere in space. A disadvantage of adaptive softening is that the filter properties are not easily known in Fourier space, so that a de-convolution with the assignment kernel, which is a common step in power spectrum estimation, is not possible. Information from the full three-dimensional density field is then usually further compressed into various `summary statistics', traditionally those statistics that can be predicted from perturbation theory have been favoured (e.g., $n$-point spectra and correlation functions), but new statistics are being considered to maximise cosmological information content or discriminative power, e.g., \cite{Cheng:2020}.

Since primordial cosmological perturbations are (close to) Gaussian, non-Gaussianity arises in the late Universe through non-linear evolution, and the density field approaches a Gaussian field only on large scales. A Gaussian field is fully described by its two-point statistics so that power spectra and two-point correlation function take a central role in cosmological analysis, but need to be supplemented with higher order statistics to be sensitive to additional cosmological information in phase correlations, see Sect.~\ref{sec:spectra_npoint} below. By averaging over finite volumes (`counts in cells', \citealt{Peebles:1980}) and evaluating their one-point statistics (i.e., the variance, skewness, kurtosis of the density probability distribution, which is close to log-normal at late times, \citealt{Coles:1991}) one can probe similar information.  Also the statistics of peaks and troughs (or clusters and voids) is a sensitive probe of the underlying cosmology since these occupy the tails of the density distribution, see Section~\ref{sec:structure_finding}. This analysis has more recently been extended to include the statistics of all critical points (i.e. also saddle points) and critical lines to characterise the density field \citep{Sousbie:2011,Xu:2019}. Also excursion sets (i.e., isodensity surfaces) of the density field contain cosmological information, which is conveniently expressed in invariant form in the Minkowski functionals (scalars) \citep{Schmalzing:1997,Nakagami:2004,Aragon-Calvo:2010,Fang:2017,Lippich:2020}. These latter methods already quantify some aspects of the anisotropic cosmic web (see the separate discussion in Section \ref{sec:cosmic_web} below). 

\subsection{Power-spectra and correlation functions, \texorpdfstring{$n$}{n}-point and \texorpdfstring{$n$}{n}-spectra}
\label{sec:spectra_npoint}

The properties of the density field, or the spatial distribution of a (biased or unbiased) set of points drawn from it, e.g. $N$-body particles or halos, can be characterized by a hierarchy of $n$-point functions. For a non-Gaussian field, all the connected $n$-point functions are needed to fully describe the field. In Fourier space, the power spectrum corresponds to the two-point function; the bispectrum to the tree-point function; the tri-spectrum the four-point function, and so on.

The algorithmic complexity of computing these statistics increases as the $n$-th power of the number of points. For computational efficiency, it is common to estimate these correlations in Fourier space employing Fast Fourier Transforms (FFT) applied over the density field estimated on a uniform grid. This computation then scale with the number of grid points $N_g$ as $\mathcal{O}(N_g \log N_g)$, rather than with the number of points in the sample. FFTs are particularly suited to analyze cosmological simulations as they naturally incorporate the periodic boundary conditions assumed in simulations, and efficient numerical implementation exist. However, as mentioned earlier, some care is needed to avoid ``aliased'' results.

The impact of aliasing on the power spectrum has been studied by several authors. \cite{Jing:2005} analytically estimated the power spectrum including aliasing for generic mass assignments, and proposed an algorithm to iteratively correct for it. \cite{Colombi:2009} also proposed a correction based on a Taylor-Fourier expansion of the aliased spectrum. Aliasing can in principle be reduced if the mass assignment scheme is a low-pass filter, thus effectively zeroing all small scale Fourier modes not supported by the FFT (strictly all scales $k>2k_{\rm Ny}/3$ should be zeroed). \cite{Cui:2008,Yang:2009} argue that Daubechies wavelets and B-splines can fulfil this purpose. More recently, \cite{Sefusatti:2016} showed that {\it interlacing} (see above) is a very efficient way to reduce significantly the effect of aliasing for any mass assignment scheme.

Another issue related to uniform FFT grids is that the number of grid points increases as $\mathcal{O}((L/\Delta x)^{3})$, thus measuring small scales can quickly become prohibitively expensive. To circumvent this limitation, \cite{Jenkins:2001} proposed a \emph{folding} technique, where the particle distribution is periodically wrapped to a smaller box of size $L'\equiv L/\beta$ with $\beta\in\mathbb{N}$ replacing $x \rightarrow x \mod L'$. The new Nyquist wave number becomes a factor of $\beta$ higher for the same computational cost, at the price of an overall larger statistical error, which is however acceptable in many situations. More recently, the folding technique has been shown to also be applicable the calculation of the bispectrum \citep{Arico:2020a}. 

Note that in various situations, especially when estimating a density field from halos/galaxies, it can be preferable to compute correlation functions directly in configuration space, rather than through two DFTs (since the smallest resolved scale is not limited by the grid resolution, and the computational cost scales with the number of objects rather than the number of grid points). These direct algorithms can be very efficient if search trees are used for the range search. Therefore, another alternative for estimating small-scale power spectra and bispectra is to compute these Fourier statistics in configuration space, by counting pairs or triplets of objects \citep{Philcox:2020,Philcox:2021}. Currently, there exist several publicly available codes to efficiently compute correlation functions, power spectra and higher-order statistics from numerical simulations, some of these are {\sc POWMES} \citep{Colombi:2009}, {\sc NBODYKIT} \citep{Hand:2018}, {\sc BSKIT} \citep{foreman:2020}, {\sc PYLIANS} \citep{pylians:2018}, {\sc HIPSTER} \citep{Philcox:2021}, {\sc CORRFUNC} \citep{Sinha:2020}, {\sc CUTE} \citep{Alonso:2012}, and {\sc HALOTOOLS} \citep{Hearin:2017}.

An important property of a discrete sample (i.e. halos or galaxies) of the cosmic density field is its {\it bias}, $b(k)$, which quantifies the differences in the overdensity field $\delta_{\rm g}$ relative to that of the underlying (unbiased) dark matter $\delta_{\rm dm}$ as a (scale-dependent) multiplicative factor $b(k) := \delta_{\rm g}(k)/\delta_{\rm dm}$. The bias can be readily estimated from two point functions as $b(k) = \sqrt{P_{\rm g}(k)/P_{\rm dm}(k)}$, or from the cross-correlation between the sample and DM as $b(k) := P_{\rm g, dm}/P_{\rm dm}$, which has the advantage of being less affected by stochastic noise. The bias can also be defined for individual objects as the overdensity of DM around it relative to that around random locations (e.g., \citealt{Paranjape:2018}). Note that the average of the individual bias of objects in a sample is mathematically equivalent to the bias of the sample.

On large scales (where $|\delta| < 1$), it is possible to expand $b(k)$ perturbatively in terms of powers and derivatives of $\delta$ (see \citealt{Desjacques:2018} for a review). The coefficients of this expansion are usually referred to as {\it bias parameters}, and have been measured in $N$-body simulations using $n$-point functions, cumulants, and more recently also by employing the separate universe formalism (e.g., \citealt{Lazeyras:2016,Lazeyras:2018,Lazeyras:2019}). The perturbative expansion has the appeal that it provides a physically motivated basis for describing the clustering of a generic distribution of objects. In other words, the observed clustering of galaxies can be expressed as a weighted sum of the auto and cross-spectra of powers and derivatives of $\delta$. Traditionally, these density spectra have been computed using various flavours of perturbation theory (e.g., \citealt{Eggemeier:2021}), but recently there has been increased interest in measuring them directly from $N$-body simulations (e.g., \citealt{Modi:2020,Zennaro:2021}).

%%%%%%%%%%
%%%%%%%%%%

\subsection{Velocity fields}

\begin{figure}
\begin{centering}
\includegraphics[width=0.9\textwidth]{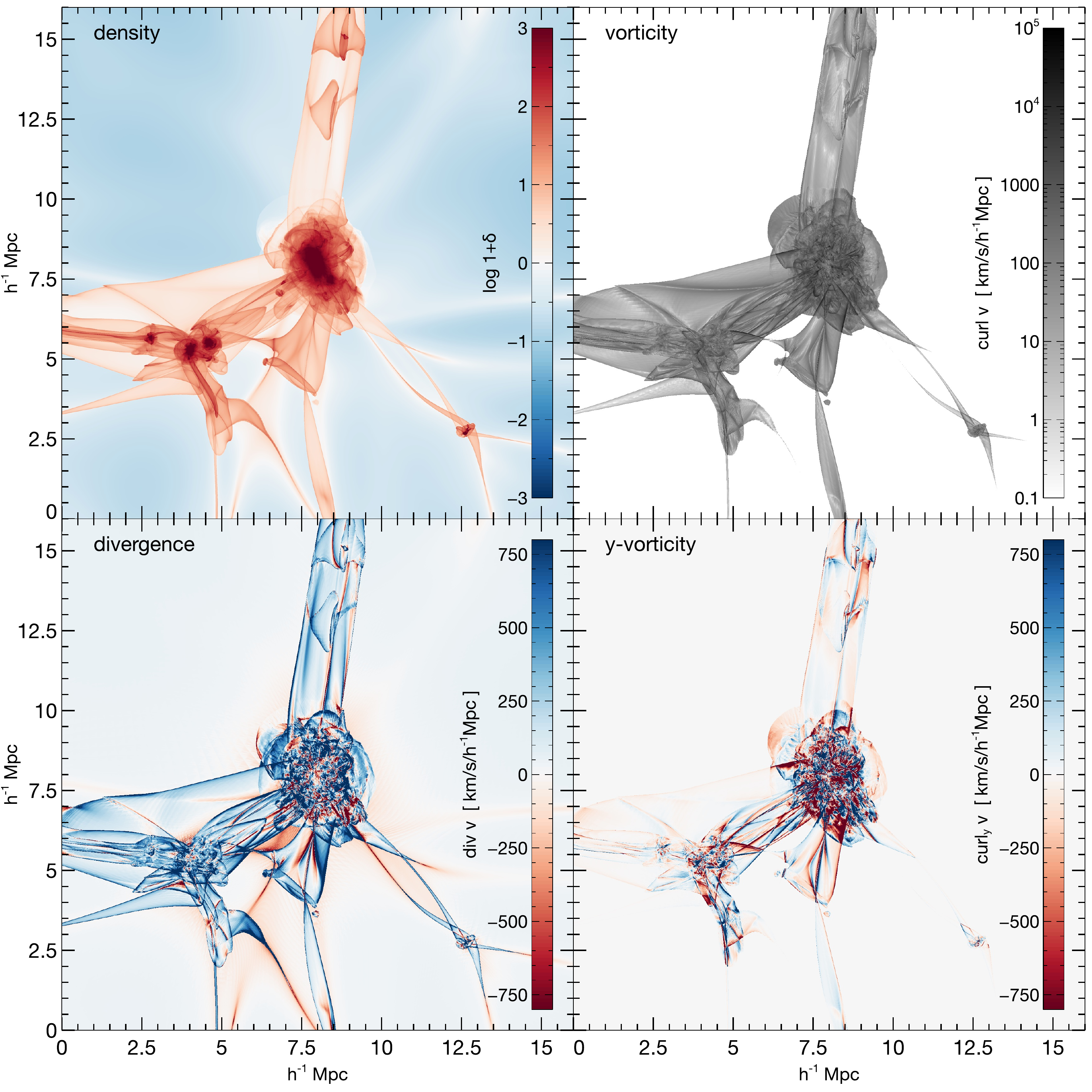} 
 \caption{ Projections of various fields in a region of $16\times16\hMpc$ around a massive halo of mass $1.4\times10^{14}\Msun$ in a Warm Dark Matter simulation. The top left panel shows the density field whereas the top right, bottom left, and bottom right show the voriticity, divergence, and the y-components of the vorticity field, respectively. Note the positive divergence in low density regions which have not yet shell-crossed. These regions can be clearly distinguished in the top-right panel as zero-vorticity. Figure adapted courtesy \cite{Hahn:2015} \label{fig:velocity_field}}
\end{centering}
\end{figure}

The estimation of the velocity field poses more challenges than that of density fields. The main problem is that the velocity field is only sampled at the position of $N$-body particles, which therefore implies an implicit mass weighing (i.e. one obtains the momentum field if particle velocities are co-added on a grid). In other words, although the velocity field should be defined everywhere in space, without further sophistication it is only possible to compute it where particles exist. Thus, for instance, in low density regions, signal-to-noise is low and there could be large empty regions where the velocity field is not estimated. This is the case even for high-resolution simulations, and the problem is even more serious when estimating velocity fields for biased tracers due to their higher sparsity compared to $N$-body particles \citep{Jennings:2015,Zhang:2015}.

This mass-weighted sampling of the velocity fields can potentially lead to large uncertainties when estimating their volume-weighted statistical properties (which is the ingredient entering in many frameworks for modelling redshift space distortions) \citep{Pueblas:2009,Zheng:2015,Zhang:2015,Jennings:2015}. As discussed, for instance, by \cite{Jennings:2011}, the standard cloud-in-cell estimation of the velocity field causes its power spectrum to strongly vary depending on the number of grid points defining the velocity field mesh. The power spectra reach convergence only if the vast majority of the cells contain at least one tracer particle. In practice this means that very large cells are needed, of the order of tens of Mpc, which prevents the investigation into the highly nonlinear regime. 

There have been several approaches in the literature to address this problem. One of them is to employ a large smoothing of the particle distribution. It is also possible to used adaptive smoothing based on the local density \citep{Colombi:2009}, or a Kriging interpolation \citep{Yu:2015,YuKriging:2017}. A different approach is to theoretically model these artefacts and then correct the measurements \citep{Zhang:2015,Zheng:2015}. In particular, Voronoi or Delaunay tessellations of the particle distribution have been shown to be a good estimator of the volume-weighted velocity field. In this approach, the space is partitioned into a set of disjoint regions with volumes tracing to local density without assuming isotropy. Specifically, several authors have shown that these tessellations allow to measure the statistical properties of velocity fields with significantly less noise and bias \citep{vandeweygaert:1998,Jennings:2018}. These approaches, however, define a coarse-grained velocity field, i.e., the average of the velocity of different streams in a given region of space. \cite{Hahn:2015} extended the phase-space tessellation method described in Section~\ref{sec:dmsheetsims} to define the velocity field of each dark matter stream everywhere in space, with which the average and higher-order moments and derivatives of the velocity field can be computed (cf. Fig.~\ref{fig:velocity_field}), this approach can also be used to estimate the anisotropic stress tensor \citep{Buehlmann:2019}

After these numerical artifacts are minimised or taken into account, it is possible to study the statistical properties of the volume weighted divergence and vorticity of the velocity fields, where analytic fits have been proposed \citep{Jennings:2012,Hahn:2015,Bel:2019}. In addition, it was shown that the halo velocity field is unbiased with respect to the matter velocity field on large scales ($k < 0.1 \ihMpc$). On smaller scales, there appear signs of velocity bias potentially originating from the fact that halos form at special locations of the density field \citep{ZhengVelBias:2015}. 

Furthermore, both vorticity and (generally anisotropic) stress in the velocity field are produced only through shell-crossing in multi-stream regions (both are decaying modes at linear order), so that they are thought to carry some information about the locally collapsed region \citep{Pichon:1999,Pueblas:2009,Laigle:2015,Hahn:2015,Jelic-Cizmek:2018,Buehlmann:2019}, and also potentially can give insight how perturbative approaches can be pushed beyond the shell-crossing frontier.

%%%%%%%%%%%%%%%%%%%%%%%%%%%%%%%%%%%%%%%%%%%%%%
%%%%%%%%%%%%%%%%%%%%%%%%%%%%%%%%%%%%%%%%%%%%%%

\subsection{Structure finding}
\label{sec:structure_finding}
\subsubsection{Halos}
\label{sec:halos}
`Dark matter halos' is the generic name given to regions of the universe which have undergone three-dimensional gravitational collapse. These halos are the fundamental nonlinear unit of the CDM Universe and, hence, their abundance, internal properties, and growth histories are at the core of several models for describing structure formation. This includes descriptions of the nonlinear power spectrum, modelling of biased tracers, and multiple models for the formation and spatial distribution of galaxies. Furthermore, large dark matter halos are employed to interpret observations of galaxy clusters, which can directly put constraints on cosmological parameters.

The somewhat arbitrary definition of a halo and its boundary has led to a multitude of operational definitions in $N$-body simulations and thus numerous algorithms exist \citep{Davis:1985,Eisenstein:1998,Stadel:2001,Bullock:2001,Springel:2001,Aubert:2004,Gill:2004,Weller:2005,Neyrinck:2005,KimPark:2006,Diemand:2006,Shaw:2007,Maciejewski:2009,KnollmannKnebe:2009,Planelles:2010,Habib:2009,Behroozi:2013a,Skory:2010,Falck2011,Roy:2014,Ivnik:2018,Elahi:2019a}; these methods can be roughly split into percolation algorithms in configuration or phase-space, and identification of peaks in the density field. \cite{Knebe:2011,Knebe:2013} performed a systematic comparison among 17 halo finders based on the same mock halos and $N$-body simulation. This study found a very good agreement in the recovered structure and their internal properties, although the halo mass returned by each algorithm varied greatly owing to the arbitrariness in the halo definition. Note that although the computational cost of performing group finding is small compared to carrying out a large simulation, storing all particle data imposes significant storage requirements, so that state-of-the-art $N$-body codes typically carry out on-the-fly halo finding during the simulation runtime thus avoiding to store the dark matter catalogues and allowing a finer time resolution in halo statistics, which is necessary for high quality merger trees (see below).

The three most common halo finders currently in use are the classical configuration space `friends-of-friends' (FoF) \citep{Press:1982,Davis:1985}, spherical overdensity ({\sc SO}) \citep{LaceyCole:1994}, and phase-space FoF (\citealt{Diemand:2006}; a widely used example being {\sc ROCKSTAR} \citealt{Behroozi:2013a}). FoF is a percolation algorithm that groups $N$-body particles whose maximum separation is below a threshold, usually referred to as the  ''linking length" ($\ell_l$) and given in terms of the mean inter-particle separation (formally a partitioning of all particles into equivalence classes whose members fullfill the distance criterion with at least one other member). FoF effectively identifies regions in terms of an isodensity contour with a value $\delta \sim 0.65\,\ell_l^{-3} - 1$ \citep{More:2011}. For typical values of $\ell_l = 0.2$, this corresponds to $\sim 80$ times the mean density. In contrast, SO identifies spherical regions whose mean enclosed density is above a threshold, usually matching the value expected from spherical collapse \citep{BryanNorman:1998}, although it also is common to use the threshold value for an Einstein-de-Sitter Universe: 200 times the background density and even 200 times the critical density of the Universe. More recently, phase-space FoF (and {\sc ROCKSTAR} in particular) has been proposed as a way to improve upon the FoF algorithm to include also sub-structure. The basic idea is to split a standard FoF group into a hierarchy of structures recursively identified with a FoF algorithm but with a linking length and distance metric defined in phase space, thus grouping together particles that are nearby both in position and velocity. Additionally, {\sc ROCKSTAR} discards those particles that are not gravitationally bound.

FoF and SO are popular methods but have shortcomings. In particular, FoF is known to suffer from numerical artefacts, where the mass of a given halo strongly depends on the mass resolution \citep{More:2011,Ludlow:2019,Leroy:2021}. Although corrections have been proposed \citep{Warren:2006}, they have not shown to be universal, but depend on the numerical resolution in addition to the redshift and cosmology of the simulation \citep{Ondaro:2021}. {\sc SO} on the other hand, suffers less from numerical artefacts, however it cannot readily adapt to the triaxial shape of DM halos (but note that this algorithm can be extended to identify ellipsoidal regions \citealt{Despali:2013}), there is an ambiguity in regions where the boundaries of two separate SO halos overlap \cite{Garcia:2019}, and there can be a time evolution of halo mass simply due to the time-dependence of the overdensity definition used \citep{Diemer:2013}. On the other hand, {\sc ROCKSTAR} has been shown to be relatively robust against numerical resolution \citep{Leroy:2021}, arguably thanks to the unbinding procedure. On the other hand, there is still arbitrariness in the definition of the phase-space metric, and the use of gravitational binding energy in the presence of accelerated frames and larger scale tidal fields \cite{Stucker:2021a}.

Another important issue regarding group finders concerns the so-called universality of the halo abundance as a function of mass, the ``halo mass function''. According to Press-Schechter theory and spherical collapse, the halo mass function should be ``universal'' in the sense that it is primarily given by statistics of peaks in Gaussian random fields -- thus a single set of $N$-body simulations could be used in principle to predict the halo mass function at any redshift and in any (reasonable) cosmology \citep{Jenkins:2001,Reed:2003,Bhattacharya:2011,Crocce:2010,Angulo:2012,Watson:2013,Bocquet:2016,Seppi:2020}. At the same time, $N$-body simulations, however, have detected departures from this universality at the $10-20\%$ level. While the precise amplitude of the effect varies with the halo mass definition and algorithm \citep{Tinker:2008,Despali:2016,Diemer:2020,Ondaro:2021}, it can be the leading systematic error in many theoretical models that rely on the halo mass function, and as a consequence in the parameter constraints from future galaxy cluster surveys (e.g., \citealt{Salvati:2020}).

The non-universality of the mass function and the ambiguity in the halo mass function have sparked recently a new wave of halo finders. These are relying on identifying caustic boundaries \citep{Shandarin:2021}, the splash-back radius where the slope of the density profile is the steepest \citep{DiemerKravtsov:2014,Diemer:2020}, and others seeking to define a halo via characteristic features in the halo density profiles \citep{Garcia:2020,Fong:2020}.

Even if these approaches succeed in defining more ``natural'' halo boundaries in the dark matter field and/or result in universal mass functions, it is not clear that they will necessarily lead to an advantage in the modelling of structure in the universe. Quite likely, the same halo (or mass) definition is not the best for all applications. For instance, one specific halo definition and boundary might be an excellent predictor for the properties of the galaxies it hosts, whereas another mass definition could be more tightly correlated with the $X$-ray luminosity associated to that halo. In any case, the standard halo definitions only approximately mark the boundary of a collapsed region, this new generation of halo finders might shed light into more fundamental relations between halos and observable counterparts.

\begin{figure}
\begin{centering}
\includegraphics[width=\textwidth]{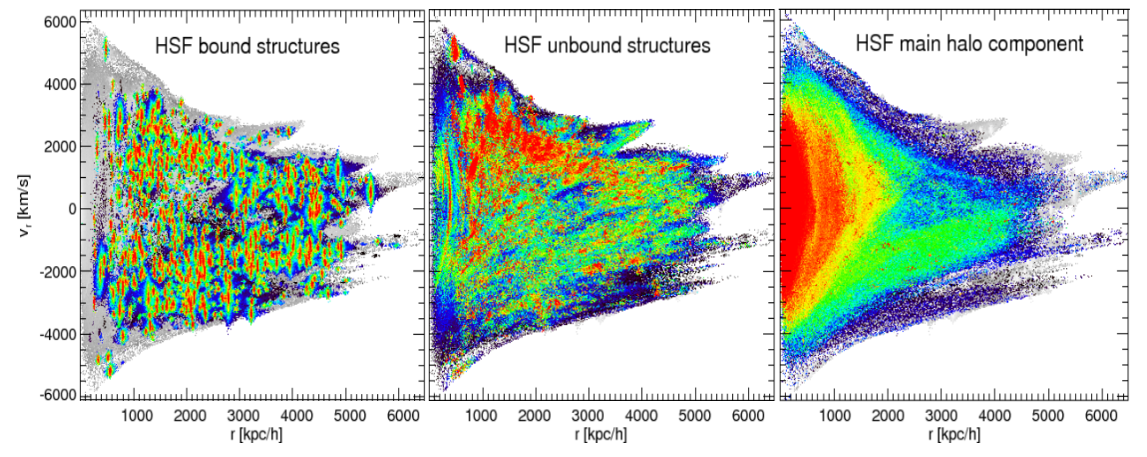} 
\caption{\label{fig:subhalo_phasespace}
Bound and unbound subtstructures of a CDM halo as identified by the {\sc HSF} phase-space subhalo finder \citep{Maciejewski:2009}. Shown is the density in radial phase space (i.e., radial distance from the halo center vs.\ radial velocity). The left panel shows the gravitationally bound subhalo structures identified, the central panel the coherent but not gravitationally bound structures (i.e. tidal streams) and the right panel the smooth main (host) halo component after removal of all identified substructure. A good substructure finder should leave no visible correlated small scale structure in the main halo. Figure adapted courtesy of \cite{Maciejewski:2009}.}
\end{centering}
\end{figure}

\subsubsection{Subhalos}
When a dark matter halo is accreted by a larger one, its outer parts are gradually tidally stripped but the remnant can still be identified as a separate structure, which is referred to as a ``subhalo''. These subhalos are expected to host satellite galaxies if they are massive enough, and to produce detectable perturbations in strong gravitational lensing, and potentially even emit electromagnetic radiation due to DM annihilation. An example of a subhalo population of a Milky-Way halo is shown in Fig.~\ref{fig:subhalo_phasespace} which displays structure in radial phase space coordinates, split into the bound subhaloes, still coherent tidally disrupted structures (tidal streams) and the smooth background halo, as identified by the {\sc HSF} subhalo finder  \citep{Maciejewski:2009}.

As for the case of halos, the definition of a subhalo is somewhat arbitrary which has led to multiple algorithms being proposed \citep{Onions:2012,Han:2012,Han:2018,Elahi:2019a}. They can be roughly grouped into phase-space percolation, peaks in the density field, and temporal algorithms which track the descendant particles of a previously-known dark matter halo. Examples of publicly-available subhalo finders are  {\sc ROCKSTAR} \citep{Behroozi:2015}, {\sc HBT+} \citep{Han:2018}, and {\sc VELOCIRAPTOR} \citep{Elahi:2019a}, and {\sc SUBFIND} \citep{Springel:2001} (updated along the ideas of {\sc HBT+}) which is now part of the public {\sc GADGET-4} release.

However, unlike for the case of halos, much larger differences exist between different group finders. This is particularly true in the case of major mergers \citep{Behroozi:2015}, where percolation algorithms in configuration space fail to distinguish two overlapping structures of similar mass. In the comparison performed by \cite{Knebe:2011}, it was found that subhalos are typically identified robustly by most algorithms down to $\sim40$ particles, which decreased to $\sim20$ when velocity information is included. 

Beyond identification completeness (which can be a function of radius in the host), also numerical effects lead to an `overmerging' of subhaloes (see the discussion in Section~\ref{sec:softening} on force softening). A robust characterisation of subhalo abundance in $N$-body simulations has to take into account such numerical selection effects. It is further common to replace completely disrupted subhalos with an `orphan' model that tracks them beyond the resolution limit (see Section~\ref{sec:SHAM}).

\subsubsection{Voids}
While most of the mass in the Universe is in halos and subhalos, most of the volume is found in low-density regions which have not undergone any kind of collapse. These are referred to as cosmic `voids', and can span regions of up to hundreds of Mpc in diameter.

Voids are capturing an increasing amount of attention in the community since they have been shown to provide useful cosmological tests in galaxy surveys (e.g., \citealt{Hamaus:2020}). In particular, they are expected to be sensitive to the non-Gaussian structure of the nonlinear universe, and thus are complementary to traditional cosmological probes based on two-point correlators. Specifically, they are particularly interesting in the context of screened versions of modified gravity which are identical to GR in high-density regions but display an additional fifth-force in low density regions such as voids (cf. Sect.~\ref{sec:modified_gravity}). Another motivation is that voids are earlier dominated by dark energy compared to average density regions, and a potential departure of gravity from general relativity can be more evident in low-density regions (see e.g., \citealt{Cai:2015}). They could also help in understanding dark energy trough their imprint in the CMB through the Integrated Sachs Wolfe (ISW) effect \citep{Cai:2016}. Another appeal of voids is that due to their comparatively simple dynamics and structure, they can be modelled accurately with simple analytic models (usually based on linear perturbation theory and excursion set theory) (see e.g., \citealt{Cai:2017}). This allows an optimal use of the data, nevertheless, the models employed need to be carefully tested and calibrated against numerical simulations. 

As for the other types of cosmic structures, the definition and boundary of a void is arbitrary and many algorithms have been proposed. Some algorithms rely on finding spherical underdense regions \citep{Padilla:2005}, others employ watershed approaches \citep{Platen:2007,Neyrinck:2008}, and others the eigenvalues of the tidal field \citep{Hahn:2007b}. In addition, some algorithms rely on an estimate of the three-dimensional cosmic density field, while others directly operate on halo/galaxy catalogues, which can facilitate a later comparison between results from observations and simulations. The differences among ``void finders'' were highlighted by \cite{Colberg:2008} who systematically compared 13 approaches applied to one large void from the Millennium simulation. A further specialised comparison of void finders focused on their discriminatory power to distinguish modified gravity models was carried out by \cite{Cautun:2018,Paillas:2019}.

Simulations are used to calibrate and develop theoretical models to use in data analysis: void size functions, the density and velocity profiles, and the geometric and dynamical distortions related to redshift space distortions, as well as their clustering. Due the low number density and sparsity of voids, very-large-volume simulations are required thus gravity-only simulations are commonly used (note no strong baryonic effects are expected \citep{Paillas:2017}), however, usually the dark matter field is used, which might show some dependence with the resolution of the simulation. The use of simulations has also helped in designing new probes, e.g., massive halos inside voids for constraining neutrino masses \citep{Zhang:2020} .

Ultimately, perhaps the biggest problem for cosmological inferences is that theoretical models might depend on the exact operational definition of voids, and it is not obvious how robust these are. Perhaps in the future, there will be emulators for voids (c.f. Section \ref{sec:emulators}) where the same definition can be used both in simulated and observed data. A more careful characterization of the covariances with other cosmological probes will become necessary, as the use of multi-probes and cross-correlation will be more common. New summary statistics could be able to extract non-Gaussian information more efficiently, for instance, the recently proposed k-Nearest-Neighbour (kNN) statistics \citep{Banerjee:2021a,Banerjee:2021b}, or statistics conditional on large-scale density \citep{Neyrinck:2018,Paillas:2021}.

%%%%%%%%%%%%%%%%%%%%%%%%%%%%%%%%%%%%%%%%%%%%%%%%%

%\subsection{Merger Trees}
%\TBD{incomplete section}
%In all the DM models discussed in this review, dark matter haloes form hierarchically. This means that haloes grow
%by accreting smaller haloes and uncollapsed mass. Merger trees are a data organization units that connects 

%Methods \cite{Springel:2005a,Tweed:2009,Han:2012,Lee:2013,Behroozi:2013b,Behroozi:2015,Ivkovic:2018,Elahi:2019b}.

%Comparison \cite{Srisawat:2013}

%Impact on SAM \cite{Lee:2014}, and the influence of halo finder \cite{Avila:2014}
%A minimum time resolution (i.e number of snapshots) is required for modelling galaxy properties that are converged \cite{Benson:2012}

%%%%%%%%%%%%%%%%%%%%%%%%%%%%%%%%%%%%%%%%%%%%%%%%%

\subsection{Cosmic web classification}
\label{sec:cosmic_web}

The filamentary cosmic web \citep{Bond:1996} has for decades been a focus of interest, since it might provide important insights into both environmental differences in the formation and evolution of dark matter haloes and galaxies, and also serve as a probe of cosmology. Its existence follows already from  Zel'dovich's \citeyearpar{Zeldovich:1970} famous formula for the Lagrangian density contrast
\begin{equation}
1+\delta(\vecb{q}) = \left| \left( 1-D_+ \lambda_1 \right)\left( 1-D_+ \lambda_2 \right)\left( 1-D_+ \lambda_3 \right) \right|^{-1}, \label{eq:density_zeldovich}
\end{equation}
where $\lambda_i$ are the eigenvalues of the Hessian $\phi^{(1)}_{,ij}$ of the initial scalar perturbation potential (cf.\ Sect.~\ref{sec:ICs_from_LPT}). Evidently, positive eigenvalues lead to singularities, while negative eigenvalues correspond to expansion, so that their signature gives rise to four dynamically distinct structures `nodes' (or `clusters'; with eigenvalue signature `$+++$'), `filaments' (`$++-$'), `sheets' (or `pancakes'; `$+--$') and `voids' (`$---$').

Due to the hierarchical nature of structure formation in CDM, one expects sheets to be made up of smaller scale filaments, and filaments to be made up of smaller scale halos, and the interesting question arises about how much mass is actually collapsed \citep{Stuecker:2018}. Since these signatures are defined w.r.t. the linear perturbation potential, they are in general non-trivially related to the late-time Eulerian density field, as measured in simulations or, more importantly, in the real Universe. For this reason, a large range of techniques have been developed that rely e.g. on a smoothed version of the Eulerian tidal field \citep{Hahn:2007}, or a multi-scale filtered version of the Hessian of the Eulerian density field \citep{Aragon-Calvo:2007,Cautun:2013}, the gradient tensor of the mean velocity field \citep{Hoffman:2012}, or the anisotropic velocity dispersion tensor \citep{Buehlmann:2019}, as well as graph based models \citep{Bonnaire:2020}, or an analysis of critical points and lines in the smoothed density field \citep{Sousbie:2011} based on Morse theory. 

Another class of techniques addresses the problem from a Lagrangian point of view, by noticing that the factors in Eq.~(\ref{eq:density_zeldovich}) undergo a sign flip when a particle crosses a caustic. Beyond the Zel'dovich approximation, on can either use the GDE tensor $\tnsr{D}_{{\rm xq},i}$ (see Eq.~\ref{eq:gde1}) or an estimate of it from a sheet tessellation (cf. Section~\ref{sec:dmsheetsims}) to track volume inversions. Algorithms based on tessellations where developed by \cite{Falck:2012,Ramachandra:2015,Shandarin:2017}, and based on a singular value decomposition of  $\tnsr{D}_{{\rm xq},i}$ by \cite{Stuecker:2019}. A detailed discussion of the relative performance of most of these methods can be found in the comparison paper of \citep{Libeskind:2018}.

While the formation and evolution of the cosmic web is reasonably well understood, the impact it leaves on galaxies and dark matter haloes is still a matter of ongoing research. Many studies have focused on differences of halo properties in the different components of the cosmic web (and its possibly intimate connection with assembly bias, see also Sect.~\ref{sec:SHAM}), e.g., \cite{Hahn:2007,Wang:2011,Goh:2019}, or the alignment of halo shapes or angular momenta with the web, e.g., \cite{Aragon-Calvo:2007b,Hahn:2007b,Codis:2012,Forero-Romero:2014,Chen:2016}, which could leave a signature in weak lensing measurements as an `intrinsic alignment' correlation, but note that the degree to which correlations found in dark matter carry over to galaxies is still unclear. An alternative approach has been followed by \cite{Paranjape:2018} who quantify only the relative anisotropy of tides as a measure of external influence on the formation of haloes and galaxies. 

While most of these web classification techniques have given interesting insights into environmental effects on the formation of haloes and galaxies, the dissected cosmic web, apart from voids and clusters, is not (yet) commonly used also as probe of fundamental physics. Other measures that are sensitive to the structure of the cosmic web, such as Minkowski functionals, are however quite commonly used also to constrain cosmological parameters, e.g., \cite{Shirasaki:2012}.

\subsection{Lightcones}

\begin{figure}
\begin{centering}
\includegraphics[width=0.6\textwidth]{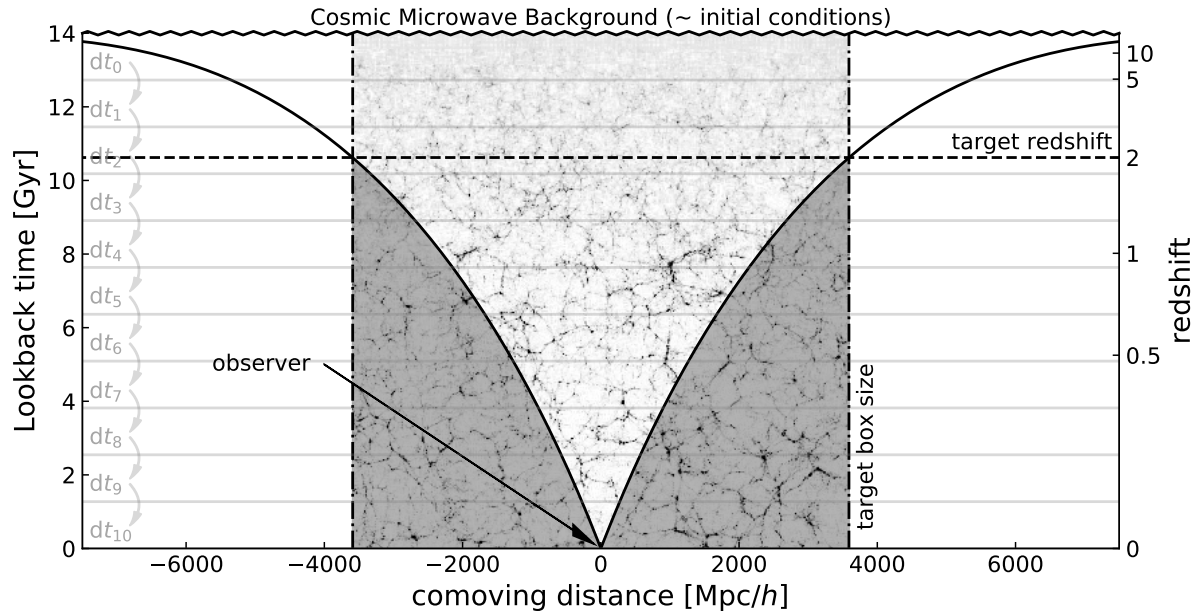} 
\includegraphics[width=0.35\textwidth]{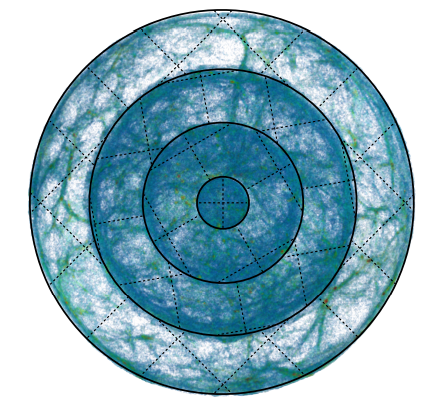} 
\caption{Schematic representation of a procedure to build lightcones from numerical simulations. In the left panel we show a 1+1 space-time diagram indicating the region inside the past lightcone of an observer at the origin. The vertical lines show the minimum simulation box size required to simulate structure in a full octant of the sky up to $z=2$, whereas the shaded area indicates the simulated regions which will never enter the lightcone. When building a lightcone, it is possible to identify and store the lightcone during runtime, or it is also possible to concatenate simulation outputs at discrete times, as those indicated by the horizontal grey lines, as a postprocessing step. Alternative to simulating large volumes, a small simulation box can be replicated and rotated such that it has a effectively a much volume, as exemplified in the figure on the right. Figure adapted courtesy of \cite{Garaldi:2020} and \cite{Sgier:2020}. \label{fig:lightcone}} 
\end{centering}
\end{figure}

Numerical simulations predict the phase-space coordinates of structure on constant-time slices and this information is commonly the output of a simulation, usually referred to as \emph{snapshots}. However, observations measure the universe along a null geodesic from our location at $z=0$, i.e. along the past lightcone of an observer. For a proper comparison, it is thus necessary to transform the simulation output to a lightcone in order to account for, among others, selection effects, line-of-sight contamination, etc., and thus increase the realism of the predictions from simulations (see e.g., \citealt{Izquierdo-Villalba:2019}).

There are essentially two approaches to build lightcones from simulations. In the first, one builds lightcones in a post-processing step by employing adjacent snapshots taken at $\left\{ a_n \middle| n=1\dots N_s \right\}$. In the simplest approach, structure in the lightcone at a distance $r$ is taken from the snapshot whose expansion factor is closest to $a_r$. This is implicitly defined for light-like distances from the FRW metric as
\begin{equation}
\label{eq:lightcone}
{\rm d}s^2=0\quad\Rightarrow\quad \int_{a_r}^1 {\rm d} a \frac{c}{a^2 H(a)} = r(a_r).
\end{equation}
Although simple, with a reasonable number of snapshots $N_s$, this approach recovers accurately the large-scale distribution, abundance of objects, etc. However, it creates small-scale discontinuities at the transition from one snapshot to another. To ameliorate this problem, it is possible to interpolate the phase-space coordinates of particles between snapshots to the precise intersection with the lightcone. This however, can lead to an underestimation of central densities in halos due to non-trivial motions in collapsed structures \citep{Merson:2013} for which a low-order interpolation in time is insufficient. An alternative is to interpolate directly halos and subhalos. Specifically, \cite{Merson:2013,Smith:2017,Izquierdo-Villalba:2019} employed merger trees to identify the descendant of each structure and in this way build its space-time trajectory. Note that despite these improvements, this approach creates pathological cases where the same object/particle appears twice in the lightcone or never appears at all. 

As can be easily seen from Eq.~(\ref{eq:lightcone}), a simulation of side-length $L$ can only represent structure up to a finite redshift, e.g. $z\sim1.4$ for $L=3000h$~Mpc. However, one can take advantage of the periodic boundary conditions and replicate the box to reach higher redshifts. Note that there can be optimal viewing angles and lightcone angular extensions that avoid line-of-sight structure repetition \citep{Kitzbichler:2007} (at least for surveys that cover a small angle). If the simulated box is large enough, structure can be considered uncorrelated, and it is common to replicate and rotate the simulation box, which has been adopted by several lightcone construction algorithms \citep{Blaizot:2005,Hollowed:2019,Rodriguez-Torres:2016,Sgier:2019,Sgier:2020}.  

An implicit requirement of the method above is that it requires a dense sampling of simulation outputs, $N_s\simeq 50$--$100$, to obtain accurate results, thus posing significant storage requirements for large state-of-the art simulations. A second option to build lightcones bypasses these requirements by constructing it directly during the runtime of the simulation. The idea is that particles are output only when they intersect the past lightcone, which is resolved with the time-resolution of the $N$-body time step. This approach was adopted, for instance, in the Hubble Volume simulation \citep{Evrard:2002} and more recently in the Euclid flagship simulation \citep{Potter:2017}, see also \cite{Fosalba:2008,Fosalba:2015}.

Since each $N$-body particle intersects the observer lightcone exactly once, in principle one is no longer interested in its evolution afterwards, since it is also now causally disconnected from the rest of the still `observable' particles. This idea was used by \cite{Llinares:2017} to propose a `shrinking domain' framework, in which particles outside the lightcone were simply discarded from the simulation, reducing the computational cost of a simulation. This assumption, however, is incompatible with the Newtonian approximation adopted in traditional simulations where information travels instantaneously. \cite{Garaldi:2020} thus proposed a `dynamic zoom simulation' in which particles outside the lightcone are not discarded but instead represented with progressively less resolution. Depending on the mass resolution and volume of the simulation, this approach could reduce the computational cost of a simulation by factors of a few, at the expense of not being able to store any full snapshot. Another short-coming of such optimisations is that only one lightcone can be produced per simulation, while with standard techniques in principle multiple distant observers can be used.

%%%%%%%%%%%%%%%%%%%%%%%%%%%%%%%%%%%%%%%%%%%%%%%%%

%%%%%%%%%%%%%%%%%%%%%%%%%%%%%%%%%%%%%%%%%%%%%%%%%

\subsection{Emulators and interpolators}
\label{sec:emulators}

Obtaining fast predictions for nonlinear structure for arbitrary cosmologies is extremely important in modern cosmology. This is an essential ingredient in the analysis of large extragalactic surveys, which could help to distinguish among competing explanations for the accelerated expansion of the Universe and play a crucial role in the hunt for new physics. 

In this review, we have argued that numerical simulations are the most precise, and often the only, way to predict nonlinear structure in the Universe. However, simulations are expensive computationally, and thus it is only possible to carry them out for a small number of specific cosmological parameters. The general strategy to address this problem is to carry out an ensemble of simulations covering a targeted space of cosmological parameters and then either interpolate the results, or calibrate a physically-motivated model with which fast predictions are obtained. This essentially comes down to non-parametric vs. parametric fitting based on simulation data.

A classical example of the parameterised approach is {\sc HALOFIT} \citep{Smith:2003} for the nonlinear power spectrum. This approach has proven to be very successful, and several revisions (recalibrations to improved modern simulations) exist \citep{Takahashi:2012,Mead:2015,Mead:2020} including extensions to massive neutrinos \citep{Bird:2012,Mead:2016}. Recently, the same idea has been applied to the matter bispectrum \citep{Takahashi:2020}. In the most recent parametric calibration carried out by \cite{Mead:2020}, 100 simulations of the Mira-Titan project \citep{Heitmann:2016} were used to calibrate the nonlinear matter power spectrum (based on the halo model and empirical corrections), reaching an accuracy of about $\sim 5-10$\% up to very small scales $k\lesssim5\hMpc$. The same authors have extended the approach to model the impact of baryons on these predictions.

The non-parametric approach is followed by `emulators' instead. These also correspond to interpolations of the simulation results but do not make any specific assumption about the underlying physical model or even its functional form. Thus, in general, they require a larger amount of data than parametric approaches, which however, is becoming less of a problem thanks to increasing computational resources. Emulators also have the advantage of being able to model complex trends in the data, so that they are being progressively constructed for many more kinds of large-scale structure statistics. For instance, emulators have been used in modelling the nonlinear matter power spectrum \citep{Heitmann:2009,Lawrence:2017,DeRose:2019,Knabenhans:2019,Angulo:2020} even beyond $\Lambda$CDM \citep{Winther:2019,Ramachandra:2020,Arnold:2021} and for baryonic effects \citep{Arico:2020b,Giri:2021}, the galaxy power spectrum and correlation function \citep{Kwan:2013,Zhai:2018}, weak lensing peak counts and power spectra \citep{Liu:2014,Petri:2015}, the 21-cm power spectrum \citep{Jennings:2018}, the halo mass function \citep{McClintock:2018,Bocquet:2020}, halo clustering statistics \citep{Nishimichi:2019,Kobayashi:2020}. Traditionally, emulators have been built with Gaussian processes (GP), but more recently feed forward neural networks are gaining popularity as they allow to deal with larger datasets and a high number of dimensions. 

A limitation of emulators is that they require a parameter space that is quite densely-sampled with simulations, and that they perform poorly outside the parameter ranges covered by this training set. As a consequence, most emulators cover a relatively small range of parameter values. Although the sampled area might be sufficient to cover the posterior distribution for parameters in upcoming large-scale structure surveys, they will definitely not be broad enough to cover the parameter prior distributions in data analysis. Therefore, a possible path in the future will be hybrid approaches where approximate and less accurate methods are employed for a broad parameter estimation, and emulators are used in a predefined region where high accuracy is required.

Another limitation of emulators is that they have a non-trivial uncertainty structure which can vary strongly across the parameter space (i.e. higher accuracy near the  parameters sampled by simulations and poorer in between them). This means that not only data has an uncertainty but so does the theoretical model, which propagates and could affect cosmological constraints. Although the uncertainty can be decreased in certain parts of the parameter space by iteratively adding new simulations \citep{Rogers:2019,Pellejero-Ibanez:2020} (or by combining simulations of different quality \citep{Ho:2021}), so that e.g. a high-likelihood region is better sampled than regions ruled out by data, this process depends on the summary statistic and scale in question. An alternative is to incorporate the emulator uncertainty in the data analysis, for which it will be very important to accurately quantify the emulator uncertainty in the first place. This will be a challenge per se since these are typically empirically measured with a small number of simulations.

A further challenge will be to construct accurate emulators for biased objects and in redshift space, which could aid in the interpretation of large-scale galaxy redshift surveys. This, however, poses several challenges. The first is related to the increase in the dimensionality of the problem, as an emulator would need to consider the properties of the object, e.g. the mass of the halo, and potentially other properties in addition to cosmological parameters. The second challenge is that numerical simulations are intrinsically noisier for discrete objects than for field quantities due to the shot noise associated with the latter. Nevertheless, recently, some progress towards emulators for dark matter haloes has been achieved \citep{Valcin:2019,Kobayashi:2020}. An interesting possibility will be to combine emulators with perturbative expansions of galaxy bias, such as those proposed by \cite{Modi:2020,Kokron:2021,Zennaro:2021}, which has been recently applied to data from the Dark Energy Survey \citep{Boryana:2021}.

%%%%%%%%%%%%%%%%%%%%%%%%%%%%%%%%%%%%%%%%%
%%%%%%%%%%%%%%%%%%%%%%%%%%%%%%%%%%%%%%%%%

\subsection{Machine learning}
\label{sec:machine_learning}

\begin{figure}
\begin{centering}
\includegraphics[width=0.95\textwidth]{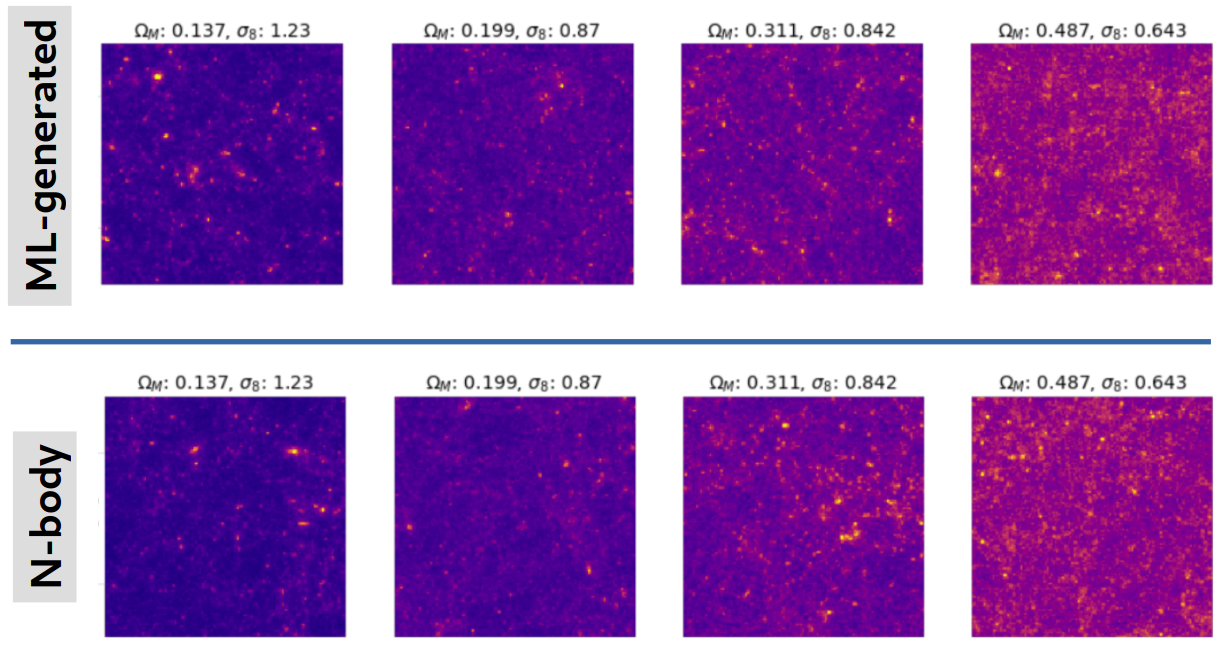} 
    \caption{Comparison of the projected two dimensional mass fields for different combinations of cosmological parameters, as computed from $N$-body simulations in the bottom row, and by a machine learning algorithm in the top row. Specifically, these maps were generated using a Generative Adversarial Network where two neural networks compete against each other; one seeking to generate ever more realistic results and the other attempting to distinguish the generated from real data. In this case, the GAN was trained with 46 different combinations of cosmological parameters ($\Omega_m$ and $\sigma_8$), and we can see it is able to generate data that correctly captures the cosmological dependence of such maps. \label{fig:cosmoGAN} Figure adapted courtesy of \cite{Perraudin:2020}}
\end{centering}
\end{figure}

%%%%%%%%%%%%%%%%%%%%%%%%%%%%%%%%%%%%%%%%%%%%%%%%%

Machine learning is a rapidly progressing field with an increasing impact on cosmology. In the specific context of cosmological numerical simulations, there are roughly three areas in which machine learning algorithms have been applied to date: artificial data generation, enhancement of $N$-body simulations, and cosmological parameter estimation.

The first area refers to the creation of artificial data, usually new realizations of the cosmic density field in two or three dimensions. Creating nonlinear fields with $N$-body simulations is expensive computationally, as we have discussed throughout this review. Machine learning can, instead, use a small set of simulations to learn its properties and then quickly create previously unseen realizations statistically consistent with this data. This is a common task in computer vision where, for instance, images of human faces can be automatically generated \citep{Radford:2015}. In a cosmological context, \cite{He:2019} employed U-net architectures -- in which several convolutional neural networks are used on different (downsampled) spatial scales, followed by an upsampling process and up-convolutions -- to create non-linear fields from a given initial linear field. Similar ideas can be applied to directly generate catalogues of dark matter halos, as proposed by \cite{BergerStein:2019,Bernardini:2020}, or of weak lensing convergence maps \citep{Mustafa:2019,Tamosiunas:2020}. 

Another pathway to create new realizations of the nonlinear matter field is based on Generative Adversarial Networks (GANs). The main idea of GANs is to have two competing neural networks; one of which creates fake data from a random process and another that seeks to distinguish them from those in the training sample. This architecture has shown to produce 2D and 3D fields quantitatively very similar to those obtained by direct numerical simulation \citep{Rodriguez:2018,Perraudin:2019,Tamosiunas:2020,Ullmo:2020,Feder:2020}, even in previously unseen cosmologies \citep{Perraudin:2020}. The architectures used to create fake data can also be combined with existing data from, for instance, low resolution simulations to artificially increase their resolution. This task, usually referred to as ``super resolution'' has been performed by e.g., \cite{KodiRamanah:2020,LiNi:2020} where small-scale high-resolution features were `in-painted' onto low-resolution $N$-body simulations, finding good agreement for a selection of density, velocity, and (sub)halo statistics \citep{Ni:2021}.

An alternative way to create new realisations was followed by \cite{He:2019}, who directly learned the displacement field from {\sc FastPM} simulations (relative to the Zel'dovich approximation) connecting Lagrangian and Eulerian coordinates of a given particle. Interestingly, these authors obtained density statistics that closely resemble {\sc FastPM} simulations, even adopting cosmological parameters different to those used in the training (see also \cite{Kaushal:2021} for a similar idea). \cite{DaiSeljak:2020} took this idea further by predicting the displacement field (including modifications due to baryonic effects), over a small number of layers (or time steps), but enforcing rotational and translational symmetries, as expected from the underlying physical laws.   

Neural networks can also be used to enhance the results of numerical simulations. For instance, \cite{Giusarma:2019} used a U-Net architecture to modify the outputs of standard $\Lambda$CDM $N$-body simulation in a way that mimics the presence of massive neutrinos. This approach appears accurate down to nonlinear scales, $k\sim 0.7 \ihMpc$, and for multiple statistics. Additionally, \cite{Troester:2019} used GANs and variational auto-encoders trained on the BAHAMAS hydrodynamical simulation to predict the distribution of gas and baryonic pressure in gravity-only simulations. These approaches can also be used to connect galaxies/halos to the initial density field. Specifically, \cite{Bernardini:2020,Zhang:2019,Icaza:2021} have used ``Convolutional Neural Networks'' or ``Sparse Regressions'' to predict either the dark matter halos or the galaxy distribution in a hydrodynamical simulation based on the properties of the dark matter field. Other examples are \cite{Nadler:2019} and \cite{Ogiya:2019} who used ``Random Forests'' to predict which subhalos in a gravity-only simulation would have been disrupted due to baryonic or tidal effects, respectively.

At the heart of many of the above algorithms is that there are nonlinear `features' that characterise the cosmic density field. Thus, it might be possible to use these features to also distinguish among cosmological models and constrain their parameters. The idea of using machine learning to obtain constraints on cosmological parameters has been explored in many recent works \citep{Ravanbakhsh:2017,Schmelzle:2017,Gupta:2018,Fluri:2018,Ribli:2019,Peel:2019,Ntampaka:2020,Pan:2020,Zorrilla-Mantilla:2020} and even applied to weak-lensing data \citep{Fluri:2019} of the KIDS-450 survey \citep{Hildebrandt:2017}. These works have used different flavours of deep convolutional neural networks trained on $N$-body simulations and were typically able to place stronger constraints than traditional clustering analyses based on two-point functions. Although they are usually carried out in idealistic scenarios and only for the matter field (thus bypassing the problem of galaxy bias), they highlight the potential of the methods and, more generally, the existence of valuable cosmological information in LSS data potentially currently unexploited. 

One of the drawbacks of this approach is that it is difficult to interpret what features a particular neural network is relying on when constraining cosmological parameters. In addition, as is known from the field of computer vision, it is possible to specifically design perturbations to images that can completely fool a well-trained neural network (adverse examples). This serves as a caution to the use of neural networks for robust parameter inference. An alternative is simply to define new summary statistics motivated by the typical machine learning filters; such as wavelet phase harmonics \citep{Allys:2020} or scattering transforms \citep{Cheng:2020}, or by analysing the feature maps in a trained network (such as the steepness of peaks in filtered convergence maps proposed by \cite{Ribli:2019}). These statistics appear to be able to capture most of the cosmological information in the fields, thus opening a promising analysis route for future large-scale structure surveys.

A key requirement for most machine learning algorithms is the existence of vast training data provided by numerical simulation. This has motivated the creation of large ensembles of simulations such as the {\sc CAMELS} suite \citep{Villaescusa-Navarro:2020}, or the {\sc MADLens} framework \citep{Bohm:2020}. Although they have so far been comprised of low-resolution or small simulated volumes, or even created with approximate methods such as {\sc COLA} and {\sc Fast-PM} (\S\ref{sec:cola_fastpm}), the continuous advances in numerical cosmology will certainly allow improvements in this direction in the future, which will enhance machine learning applications even further.

%%%%%%%%%%%%%%%%%%%%%%%%%%%%%%%%%%%%%%%%%%%%%%%%%

\subsection{Connecting dark matter to galaxies and baryons}
\label{sec:galaxies}

Understanding the galaxy-halo connection is at the very heart of using the large-scale structure of the Universe as a cosmological probe. On small scales, modelling the galaxy-halo connection can provide constraints on fundamental physics; for instance, on the free-streaming scale of dark matter or on possible dark matter-baryon interactions \citep{Nadler:2019}, on the allowed modifications of gravity \citep{He:2019} (cf. Sect.~\ref{sec:modified_gravity}), or on the dark matter mass \citep{Newton:2020}. We refer the reader to \cite{Baugh:2006,Wechsler:2018,Somerville:2015,Vogelsberger:2020} for specialised reviews of galaxy formation physics and here simply summarise those techniques most commonly used in large-scale dark matter simulations.

Currently, the arguably most realistic way to jointly model galaxies and the dark matter field is through cosmological hydro-dynamical simulations. These simulations attempt to explicitly simulate astrophysical processes related to baryonic physics such as gas cooling, star formation, and feedback energy injection from supernovae and supermassive black holes among others. State-of-the art examples are, in alphabetical order, the BAHAMAS \citep{McCarthy:2017}, Cosmo-OWLS \citep{LeBrun:2014}, EAGLE \citep{Schaye:2010,Crain:2015}, Horizon-AGN \citep{Dubois:2014}, Illustris \citep{Vogelsberger:2014}, Illustris-TNG \citep{Springel:2018}, Magneticum \citep{Hirschmann:2014}, MassiveBlack-II \citep{Khandai:2015}, and SIMBA \citep{Dave:2019} simulations.

Unfortunately, hydrodynamical simulations are notoriously expensive computationally, and thus they can typically only be carried out for relatively small cosmic volumes. Additionally, since galaxy formation cannot be predicted {\it ab initio}, many specific choices need to be adopted regarding  the physical processes to be modelled and their free parameters. In addition, the modelling of most of these processes is fine-tuned to the resolution of the simulation at hand, which raises the question how predictive these simulations are. This is perhaps the main limitation of hydrodynamical simulations in the context of large-scale structure. Given the current uncertainties in galaxy formation theory, it is therefore desirable to be able to explore a more general galaxy-halo connection, ideally incorporating the respective uncertainty when analysing and interpreting observational data. 

For these reasons, currently, there are various popular and useful models for the galaxy population in a given dark matter simulation which can be applied in post-processing. These models can be roughly split into two categories: empirical, where the relationship between dark matter halos and galaxies is simply given in parametric or non-parametric forms but that do not predict galaxy properties; and physically-motivated which attempt to directly model galaxy formation trough a set of simplified descriptions. In this subsection we review the most common approaches.

\subsubsection{HOD and CLF}

The simplest approach to connecting haloes with galaxies is the `halo occupation distribution' (HOD) model which assumes that the number of galaxies residing in a given halo is only a function of the host halo mass.
Note that a more sophisticated version of HOD is the conditional luminosity/mass function (CLF) which attempts to describe not only the total number of galaxies in a halo, but also the full distribution of a secondary galaxy property (e.g. stellar masses or luminosities) given by its distribution function conditional on halo mass \citep{vandenBosch:2003,vandenBosch:2007}

In the HOD formalism, the expected average number of satellite and central galaxies in a given halo (also referred to as ``occupation distribution") is given in a parametric form motivated by observations and galaxy formation models. To make a realisation of a galaxy population of a given halo in an $N$-body simulation, central galaxies are assumed to follow a Bernoulli distribution, whereas the number of satellite galaxies is assumed to be Poissonian.

Early HOD recipes contained only three free parameters and were able to describe the luminosity and color dependence of the correlation function of galaxies in SDSS \citep{Zehavi:2005}. Motivated by the measured occupation distribution in both hydrodynamical simulations and SAMs (cf.\ Sect.~\ref{sec:SAM}), the number of free parameters increased to five \citep{Zheng:2005}, whereas latest incarnations can include up to ten free parameters \citep{Hearin:2016}.

In the most basic HOD models, a central galaxy sits at rest in the center of the halo and the satellites are usually assumed to follow a NFW profile \citep{Zheng:2007} (or directly the phase-space distribution as dark matter particles). More sophisticated galaxy models have shown that many of these assumptions are not valid in general: the number and properties of galaxies depend also on halo properties other than mass, and satellite galaxies move and are located differently than DM particles \citep{DiemandVel:2004,WuHahn:2013,Orsi:2018}. All these impact the expected spatial distribution of galaxies, thus recent HOD implementations have relaxed some of the assumptions of this model to enable a more realistic description of galaxies. In particular, attempts are made to incorporate secondary correlations with other halo properties beyond mass or environmental properties \citep{Hadzhiyska:2020,Hadzhiyska:2021}, velocity bias \citep{Guo:2015}, or flexibility in the spatial distribution to incorporate assembly bias \citep{Hearin:2016,Zehavi:2019,Xu:2021}. 

Despite its limitations, HOD remains one of the most popular approaches to model galaxies in large-scale simulations, thanks to its flexibility to capture galaxies regardless of their type of selection function and without making strong assumptions about galaxy formation physics. A further advantage is that it can be applied to relatively low-resolution $N$-body simulations. In fact, the HOD and CLF have been used together with $N$-body simulations to provide the strongest cosmological constraints to date from galaxy clustering. For instance, \cite{Reid:2014} obtained 2.5\% measurements on the growth rate using the multipoles of the redshift-space correlation function as measured in SDSS-III BOSS CMASS and \cite{Lange:2021} 5\% with the LOWZ sample, whereas \cite{Cacciato:2013} obtained constraints on $\Omega_{\rm m}$ and $\sigma_8$ competitive with those derived from CMB analyses but using galaxy clustering and galaxy galaxy lensing in the SDSS survey. Although a careful assessment of the systematic sources of uncertainty in these constraints is required, they serve as an example for the statistical power obtained from the combination of $N$-body simulations and galaxy formation models.

%%%%%%%%%%%%%%%%%%%%%%%%%%%%%%%%%%%%%%%%%%%
%%%%%%%%%%%%%%%%%%%%%%%%%%%%%%%%%%%%%%%%%%%

\subsubsection{Subhalo abundance matching -- SHAM}
\label{sec:SHAM}
Several of the limitations of the HOD can be ameliorated by taking its assumption one step further.  In particular, one can assume that galaxies reside in dark matter {\it substructures} instead of simply residing in dark matter haloes. This seemingly small difference implies that there should be a one-to-one relation between galaxies and dark matter substructures, which makes DM simulations much more predictive about the phase-space coordinates of galaxies.

In the most general formulation, one could describe the galaxy population with the probability of residing in a subhalo with a given set of subhalo properties \citep{Yang:2012,Moster:2013,Behroozi:2013c}. Different authors have chosen different parametrisations, subhalo properties, and redshift evolution, but obtain similar results when inferring halo masses from the stellar mass of observed galaxies. Not only stellar mass can be modelled in this way, but also star formation rate, metallicity, dust content, etc.

Another, alternative avenue is followed in `subhalo abundance matching' (SHAM) by simply assuming a monotonic relation between subhalo and galaxy properties \citep{Kravtsov:2004,ValeOstriker:2004,Conroy:2006}. In this `rank-order' SHAM, the most massive subhalo hosts the galaxy with the largest corresponding property (e.g. the highest luminosity) and so on, thus providing a non-parametric galaxy-halo relation. In current implementations this relationship is not assumed to be perfect, but it is modelled with an additional parameter that allows for some scatter in the mapping, which appears to be demanded by the data. 

Exactly which global property of DM subhalo is the optimal parameter for a SHAM mapping has been a matter of debate. Modern implementations use parameters based on the circular velocity of the subhalo, $V_{\rm circ}(r) = \sqrt{G (M<r)/r}$, as it is less sensitive to a specific definition of halo mass and captures better the inner parts of a halo, which arguably, is better connected to the inner regions a galaxy is expected  to inhabit. In fact, $V_{\rm circ}$ appears to reproduce the small-scale clustering of galaxies much better than mass-based-SHAMs. Furthermore, since subhalo properties can evolve substantially once accreted due to the tidal heating and stripping, subhalo properties before accretion are preferred. Explicitly, several authors have argued that the highest value of the maximum of the circular velocity, $v_{\rm max} = max(V_{\rm circ})$ over the history of a halo, referred to as $v_{\rm peak}$, is the single property that correlates strongest with the stellar mass of galaxies in hydrodynamical simulations \citep{Chaves-Montero:2016,XuZheng:2020,He:2020}. There has also been attempts to combine more than one property for SHAM, e.g., including secondary dependences with $V_{\rm circ}(r=R_{200})$, the peak halo mass $M_{\rm peak}$, concentration, or large-scale density (e.g., \citealt{Mao:2015,Lehman:2017,Contreras:2020,Tonnesen:2021}), seeking to improve the accuracy of SHAM or capture better environmental dependences (e.g., as assembly bias).

Subhalo abundance matching has been shown to be in good agreement with multiple observations such as the two-point and three-point galaxy clustering and the Tully--Fisher relation \citep{ValeOstriker:2004,Conroy:2006,Marin:2008,Trujillo-Gomez:2011,Nuza:2013,Reddick:2013}, and to accurately describe the clustering of stellar-mass selected galaxies in hydrodynamical simulations \citep{Chaves-Montero:2016,Contreras:2020a,Favole:2021}, to capture the so-called assembly bias \citep{Contreras:2020a}, and to reduce the need of velocity bias \citep{Ye:2017} (but see \citealt{Hearin:2013} for difficulties regarding galaxy groups). SHAM, while accurate for describing stellar masses, is less so for star formation rates, or cold gas fraction, which are expected to display a non-monotonous relationships to dark matter halos. On the other hand, SHAM could be used as a starting point to describe these other properties by using the respective distribution functions conditional on stellar masses \citep{HearinWatson:2013,Favole:2017}, or account for e.g. observational incompleteness \citep{Favole:2016}. A similar idea has been implemented by \cite{Contreras:2020b,Favole:2021}, who extended SHAM with recipes for star formation rate based on the mass accretion rate, and showed a remarkably good agreement with the clustering of SFR-selected galaxies in the Illustris-TNG simulation \citep{Springel:2018}.

SHAM imposes stricter requirements than HOD on the numerical accuracy and mass resolution  of the parent numerical simulation. This is since the evolution of substructures inside halos needs to be followed for a considerable number of dynamical times, relatively high force and time integration accuracy, together with adequate mass resolution are necessary \citep{Klypin:2015,Guo:2014} (see also the discussion in Sect.~\ref{sec:softening}).

Roughly, in every pericentric passage, DM subhalos loose approximately 90\% of their mass. Thus, for any finite mass resolution, subhalo mass will eventually fall below the resolution limit of a simulation and be lost \citep{Moore:1996,vandenBosch:2017}, see also the discussion in Sect.~\ref{sec:softening}. Naturally, this does not necessarily mean that the respective galaxy has merged, thus one expects a population of subhalo-less galaxies, also known as {\it orphan galaxies} (see e.g., \citealt{Gao:2004,WuHahn:2013,Delfino:2021}). The relative abundance of these objects has been debated \citep{Klypin:2015,Guo:2014}, and might depend on the particular subhalo finder and merger tree algorithm used, it also depends on the target galaxy population that is modelled. Furthermore, even with an infinite resolution, baryons and stars modify the gravitational potential inside DM halos and of the infalling structures, thus SHAM and DM-only simulations cannot correctly capture the dynamics of satellite galaxies.  Thus, it is important to model orphan galaxies and their eventual disruption within SHAM to obtain precise predictions for the clustering of galaxies, specially on small scales \citep{Campbell:2018}.

Similarly to HODs, the SHAM technique has also been used together with $N$-body simulations to place strong constraints on cosmological parameters from the observed clustering of galaxies. For instance, \cite{Simha:2013} obtained constraints on $\Omega_{\rm m}$ and $\sigma_8$ from the projected correlation function in SDSS. Another example was provided by \cite{He:2018} who demonstrated that a basic HOD and $f(R)$ simulations was incompatible with the clustering in SDSS which allow them to place one of the strongest constraint on the amplitude of a hypotetical $f(R)$ scalar field. As for the HOD, further work is required to establish the robustness of these results, in particular with respect to degeneracies of galaxy formation physics, but these results suggest a very promising avenue for future cosmological data analysis.

%HOD vs SHAM \cite{Guo:2016} \TBD{missing}

%%%%%%%%%%%%%%%%%%%%%%%%%%%%%%%%%%%%%%%%%%%%%%%%%

\subsubsection{Semi-empirical models}

More recently, there have been attempts to build upon the success of SHAM and to extend it into a more physical model which is able to predict self-consistently galaxy properties across time. Such a path is followed in the models dubbed `Universe Machine' \citep{Behroozi:2019} and `EMERGE' \citep{Moster:2018}, and the `Surrogate-Baryonic-Universe' \citep{Chaves-Montero:2020}. The basic idea of these methods is to adopt an empirical model of the relationship between star formation rate (SFR) and the fraction of quenched galaxies, and subhalo properties (e.g. maximum circular velocity and halo formation time), together with a specific redshift evolution. Then, with the use of subhalo merger trees extracted from DM numerical simulations, it is possible to predict the star formation history of every modelled galaxy. As a result, this SFR history is then turned into the total stellar mass, from which other observable properties can be computed for a given stellar population synthesis model, initial mass function, and dust model. 

The free parameters ($15$ - $40$) of these relationships are constrained using Bayesian algorithms by requiring an agreement with observational data across multiple redshifts. In general, these models are more physical than purely empirical models in that galaxy predictions are self-consistent (the stellar mass of a galaxy is the integral of its star formation history), but they do not make any specific assumptions about the physics underlying the inferred relations.

These models are quite successful in reproducing and predicting multiple  properties of the observed galaxy population. For this reason, it would be very interesting to explore the kind of cosmological inferences that they would be able to place when employed with $N$-body simulations. For instance, the possible degeneracies between cosmological parameters and those physical parameters describing galaxy formation and evolution. This would also be interesting for informing which physical processes are the most important to understand and model for improved cosmological power. To our knowledge, this has not yet been attempted in the literature, but will certainly be an important step towards optimal exploitation of up upcoming large-scale structure data.

%\TBD{Maybe bathtub?}

%%%%%%%%%%%%%%%%%%%%%%%%%%%%%%%%%%%%%%%%%%%%%%%%%

%%%%%%%%%%%%%%%%%%%%%%%%%%%%%%%%%%%%%%%%%%%%%%%%%

\begin{figure}
\includegraphics[width=\textwidth]{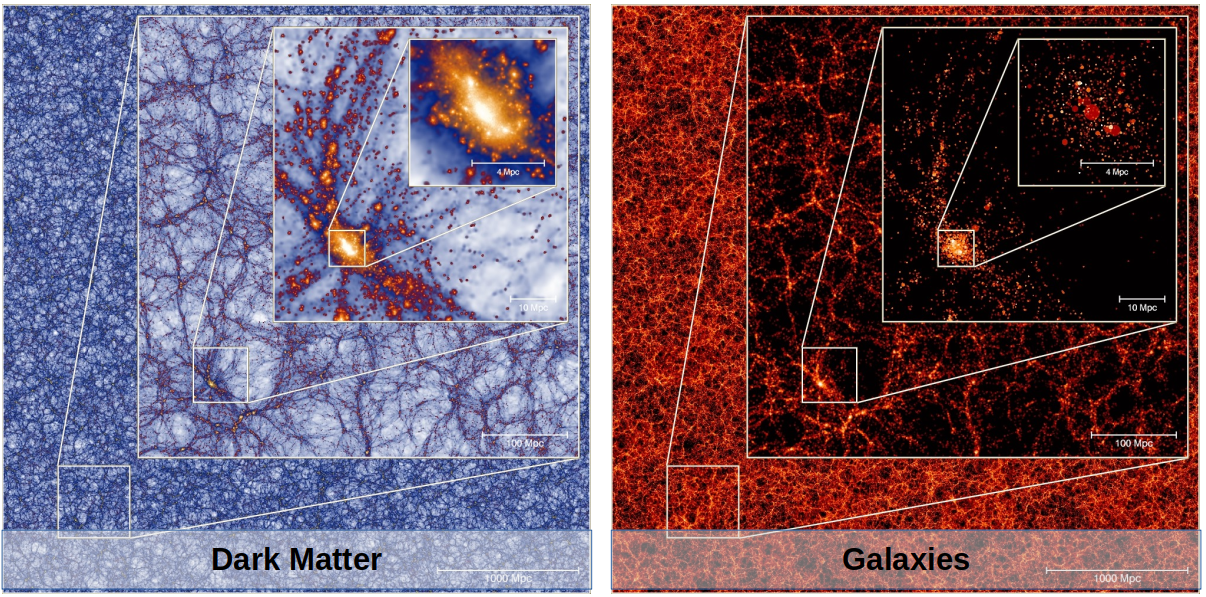} 
\caption{\label{fig:galaxies}
The spatial distribution of dark matter and stellar-mass selected galaxies with $M_* > 10^{10}\Msun$ at $z=0$ in the Millennium-XXL simulation \citep{Angulo:2012}. Galaxies were simulated using the {\sc L-Galaxies} semi-analitic galaxy formation model of \cite{Henriques:2013}, and are displayed as circles whose size is proportional to the expected half-mass radius. Figure adapted courtesy \cite{Angulo:2014}.}
\end{figure}

\subsubsection{Semi-analytic models -- SAMs}
\label{sec:SAM}

Arguably the most complete post-processing modelling of the galaxy population is provided by `semi-analytic' models (SAMs) of galaxy formation. This approach seeks to self-consistently evolve the properties of galaxies through a coupled system of ordinary differential (or difference) equations describing the amounts of stars, cold and hot gas, black holes, and star formation rate in galaxies. 

SAMs were in fact the first attempts to model galaxy formation and evolution, with the main ideas dating back to \cite{WhiteRees:1978,White:1991}. Early SAMs were built on top of analytic halo merger trees and captured only the main physics that was expected to be necessary to understand galaxy formation \citep{Kauffmann:1993,Cole:1994,Cole:2000}. Since then, SAMs have significantly evolved and have been improved in their assumptions, sophistication and scope. Modern examples are `L-Galaxies' \citep{Henriques:2015}, `Galform' \citep{Cole:2000}, `Galacticus' \citep{BensonGalacticus:2012}, `SAG' \citep{Cora:2018}, `SAGE' \citep{Croton:2016}, `SHARKS' \citep{Lagos:2018}, `Gaea' \citep{Hirschmann:2016}, as well as `$\nu^2$GC' \citep{Makiya:2016}; and typically include gas cooling, chemical evolution and enrichment for multiple elements, feedback from SNe type~I and II, stellar winds, binary evolution, environmental and RAM-pressure stripping, black hole formation, mergers, and feedback, dust formation, and radiative-transfer effects. For a comparison of models see \cite{deLucia:2010,Knebe:2015,Knebe:2018}. 

Schematically, the fundamental assumption underlying SAMs is that every time a mass element is accreted onto a DM halo, a corresponding amount of gas (set by the universal baryon fraction) will also be accreted. Baryons will be shock heated to $T_{\rm vir}$, the virial temperature of the host halo, and start to cool either via bremsstrahlung ($T_{\rm vir} > 10^7\,K$), electron recombination ($10^4\,K < T_{\rm vir} < 10^7\,K$), or metal-line cooling ($T_{\rm vir} < 10^4\,K$). If the cooling time of the halo is long compared to the age of the universe, then the accretion is said to be in a hot-mode, and a fraction of the gas will condense into stars in every dynamical time, otherwise the cooling is fast and system will develop cooling flows. 

The initial mass function (IMF) of newly formed stars is a degree of freedom of these models, and most modern implementations adopt a Chabrier IMF, but a Salpeter IMF, Kroupa IMF, and modifications of them have also been used in the past (e.g., \citealt{Baugh:2005}). Subsequently, stellar evolution models and stellar population synthesis models will then determine the typical lifetime and spectral energy distribution of stars of a given mass. During their life and death, stars return part of their mass to the intergalactic medium, which is typically assumed to be recycled instantaneously in the subsequent generation of starts. Energy injection could be thermal, kinetic, or radiative, and halts star formation.

In parallel to star formation, modern SAMs also follow the growth of supermassive black holes at the center of every galaxy. These black holes are assumed to grow from initial seeds by mergers and gas accretion. In most models, during most of their lifetime, black holes increase their mass through Bondi-Hoyle-Littleton accretion, and part of the accreted mass will then be re-injected as energy into the ISM reducing star formation, in what is known as a radio-mode feedback. During some moments of their history, a `quasar mode` black hole feedback cycle can be triggered by mergers and disk instabilities, which leads to high accretion rates and the emergence of quasars and AGNs \citep{Croton:2016,Bower:2006}. 

Finally, the halo and subhalo merger trees determine the mass accretion and mergers. As discussed above in the context of SHAM, subhalo mergers and disruption are not necessarily a good predictor for galaxy merger and disruptions. Thus, additional recipes are needed together with the modelling of orphan galaxies. 

A disadvantage of SAMs is the relatively large number of free parameters -- 20-50 typically. However, this arises mainly from our current uncertainty about many aspects of galaxy formation, and the large number of physical processes that are described. Typically, these free parameters are fixed by requiring agreement with a set of observations. Early SAMs were calibrated `by hand' where agreement with observations was subject to the criterion of a particular author.  Nowadays, SAMs are calibrated in a more objective manner using Bayesian statistics (Monte Carlo Markov chains, particle swarms; e.g. \citealt{Kampakoglou:2008,Henriques:2009,Bower:2010,Ruiz:2015}), however, since different codes and authors choose different observables for their calibration, large discrepancies still exist among different codes.

The strength of SAMs is that they provide self-consistent (physically motivated) predictions for galaxy properties over, in principle, the whole electromagnetic spectrum. Thus, SAMs have been employed to study and understand multiple aspects of galaxy formation: the mass-metallicity relation, mass-luminosity relations, galaxy clustering, galaxy colours, etc. SAMs have also been used to make predictions for upcoming large-scale galaxy surveys \citep{Merson:2018,Angulo:2014}, and to understand the physics of galaxy formation with dark energy beyond a cosmological constant \citep{Fontanot:2012,Fontanot:2015b}, $f(R)$ gravity \citep{Fontanot:2013}, and massive neutrinos \citep{Fontanot:2015}.

%%%%%%%%%%%%%%%%%%%%%%%%%%%%%%%%%%%%%%%%%%%%%%%%%

\subsubsection{Modelling of Baryonic and Gas physics}
\label{sec:baryonic_effects}

\begin{figure}
\includegraphics[width=\textwidth]{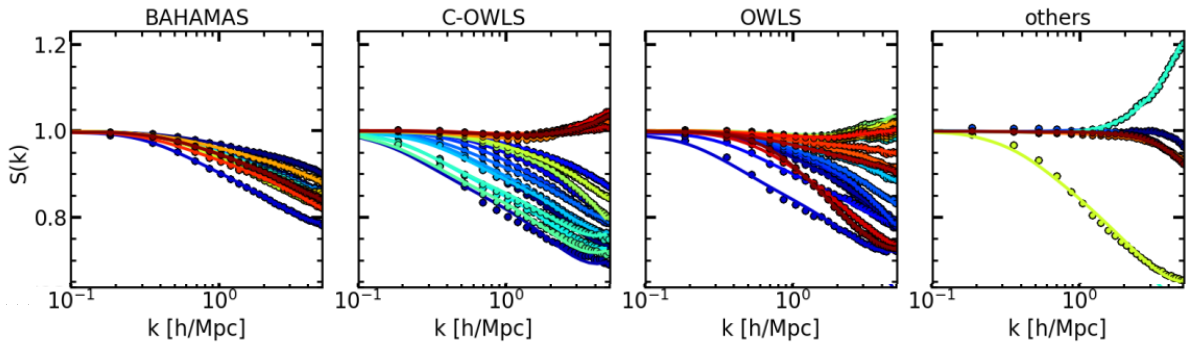} 
\caption{\label{fig:baryons.pdf}
The effect of baryonic physics quantified via $S(k) \equiv P(k)/P_{\rm GrO}(k)$: the ratio of the nonlinear power spectrum as measured in various hydrodynamical simulations over that in dark matter gravity-only $N$-body simulations. Each panel displays the measurements in various hydro-dynamical suites, BAHAMAS, C-OWLS, OWLS, whereas the rightmost panels shows the same measurement in the Illustris, EAGLE, Illustris-TNG, and Horizon-AGN. Additionally, in each case we display as solid lines the best-fitting description of a ``baryonification model'', in which the outputs of a gravity-only simulations are perturbed according to physically motivated recipes \citep{Schneider:2015,Schneider:2018,Arico:2019,Arico:2020a}. Figure adapted courtesy \cite{Arico:2020b}.}
\end{figure}

The main assumption of the galaxy formation modelling discussed in the previous section is that baryonic physics determines the properties of galaxies, but that the gravitational potential, and thus the dynamics, is fully determined by gravity without any back-reaction from galaxy astrophysics. Although gravitational interactions dominate the nonlinear evolution of cosmic structure, the accuracy required by future cosmological weak lensing observations is such that baryonic effects on the matter distribution cannot be neglected. 

Recently, the impact of effects such as feedback from supermassive black holes, star formation, gas cooling, etc, has been studied extensively by comparing the predictions of gravity-only simulations and full hydrodynamical calculations. The consensus is that baryons modify systematically the shape of the nonlinear density power spectrum on small scales. Specifically, baryons induce a suppression of the amplitude on intermediate scales ($k \sim 1 \ihMpc$) -- which is mostly resulting from the collisional nature of baryons and additional injection of energy through feedback which prevents the accretion of gas onto dark matter halos and modifies the baryonic density profiles -- and induce an enhancement on smaller scales ($k \sim 10 \ihMpc$) -- as a result of gas cooling and star formation \citep{vanDaalen:2019,Chisari:2019}. 

Although different hydrodynamical simulations agree qualitatively in their predictions, they differ in the magnitude and redshift dependence of these effects \citep{vanDaalen:2019}. Specifically, some simulations predict a relatively minor effect (less than 5\% at $k > 5\ihMpc$), e.g. Horizon-AGN \citep{Dubois:2014}, EAGLE \citep{Schaye:2015,Hellwing:2016}, and Illustris-TNG300 \citep{Springel:2018}. Other simulations, predict stronger effects  ($10$-$15\%$) on the same scales, e.g. BAHAMAS \citep{McCarthy:2017}, OWLS \citep{Schaye:2010,vanDaalen:2011}, and Cosmo-OWLS \citep{LeBrun:2014}. An extreme case is Illustris \citep{Vogelsberger:2014}, which predicts significant effects even at $k \sim 0.5\ihMpc$ but is known to suffer from unrealistically low gas fractions in massive haloes. Generally, it is striking that those simulations that are tuned to produce `realistic' galaxy properties, predict a weaker impact on the density power spectrum, whereas those tuned on cluster properties predict stronger baryonic effects. The discrepancy will have to be settled with the larger and more realistic next generation of hydrodynamic cosmological simulations. Perhaps, more interestingly will be the comparison with observational data. Although current constraint are able to only marginally rule out the most extreme feedback scenarios \citep{Huang:2021}, new prospects for estimating the magnitude of baryonic effects in the real Universe will open with the imminent improvement in weak lensing surveys and in the joint analysis of gas observables (kinetic and thermal Sunyaev-Zeldovich effects, and X-ray gas fractions; \citealt{Schneider:2021,Troster:2021}).

Due to this variety in predictions, the impact of baryonic physics currently should not be modelled based on a single simulation, and instead the focus is on creating flexible and general ways to model these effects. This motivation, along with the need to leverage the larger statistical power of gravity-only simulations (at lower computational cost), has triggered the development of several algorithms to modify the mass field in the presence of baryons. Such approaches are found in several extensions of the halo-model \citep{Semboloni:2011,Mohammed:2014,Fedeli:2014,Mead:2015,Debackere:2019}; in terms of response functions calibrated using Separate Universe simulations \citep{Barreira:2019}; displacing particles according to the expected gas pressure \citep{Dai:2018}; or even using machine learning \citep{Troester:2019,Villaescusa-Navarro:2020b}.

In the same spirit as galaxy formation models discussed above, another avenue is to model the impact of baryons directly from the dark matter field. Among such models is the baryonic correction model \citep{Schneider:2015,Schneider:2018,Arico:2019}. An advantage of this approach is that it does not depend on a specific set of simulations and/or observables, that its free parameters can be marginalised over to place cosmological constraints, and that they offer the opportunity to predict multiple observables that depend on the gas distribution, such as the Sunyaev-Zeldovich effect or the X-ray emission in galaxy clusters. Currently, these models appear flexible and accurate enough to be useful in the interpretation future datasets \citep{Arico:2020b}. Nevertheless, with the advent of more realistic and precise hydrodynamical simulations, these models are expected to be refined and improved, especially concerning the way that baryonic effects are modelled and parameterised, and improved physical priors on baryonic parameters. 

%%%%%%%%%%%%%%%%%%%%%%%%%%%%%%%%%%%%%%%%%%%%%%%%%

%% State-of-the-art Simulations and Results
%% \newpage

\section{Overview of state-of-the-art simulations and challenges}
\label{sec:state-of-the-art}

In this section, we will provide a brief overview of the largest gravity-only calculations to-date (i.e., by mid 2021), which unveil the most detailed views of nonlinear structure formation on large scales. We then discuss the challenges in carrying out these calculations, in obtaining predictions as a function of cosmology, and in the dissemination and sharing of their results and data products.

\begin{footnotesize}
  \begin{table}
    {\renewcommand{\arraystretch}{2.5}%
      \begin{tabular}{ cccccccc } 
      %  \hline
      Year & Simulation & \makecell{Code,\\  Algorithm} & \makecell{Supercomputer, \\ Location} & \makecell{Cores \\ $[10^{3}]$} & \makecell{$N_{\rm p}$ \\ $[10^{12}]$} & \makecell{Box \\ $[\hGpc]$} & \makecell{$\epsilon$ \\ $[\hkpc]$ }\\
          % &   &  &   & $[10^{3}]$  &  $[10^{12}]$ & $[\hGpc]$ & $[\hkpc]$ \\
        \hline\hline
        2014 & \makecell{Dark Sky \\ \citep{Skillman:2014}} & \makecell{{\sc 2HOT} \\ FMM}  &  \makecell{Titan \\ USA} & 20 & 1.1 & 8 & 36.8\\ 
        \hline
        2017 & \makecell{TianNu \\ \citep{Emberson:2017}}  & \makecell{{\sc ${\rm CUBEP^3M}$} \\ PM-PM-PP} &  \makecell{Tianhe-2 \\ China} & 331 & 2.97 &  1.2 & 13\\  
        \hline
          2017 & \makecell{Euclid Flagship \\ \citep{Potter:2017}} & \makecell{{\sc PKDGRAV3} \\ Tree-FMM} &  \makecell{PizDaint \\ Switzerland} & 4 & 2.0 &  3. & 4.8\\ 
          \hline
          2019 & \makecell{Outer Rim \\ \citep{Heitmann:2019}} & \makecell{{\sc HACC} \\ Tree-PM} &  \makecell{Mira \\ USA} & 524 & 1.07 &  3.0 & 2.84\\ 
          \hline
          2019 & \makecell{Cosmo-$\pi$ \\ \citep{Cheng:2020}} & \makecell{{\sc CUBE} \\ PM-PM} &  \makecell{$\pi$ 2.0 \\ China} & 20 & 4.39 & 3.2 & 195\\ 
          \hline
          2020 & \makecell{Uchuu \\ \citep{IshiyamaUshuu:2020}} & \makecell{{\sc GreeM} \\ Tree-PM} &  \makecell{ATERUI-II \\ Japan} & <40 & 2.0 &  2.0 & 4.3\\ 
          \hline
          2020 & \makecell{Last Journey \\ \citep{Heitmann:2021}} & \makecell{{\sc HACC} \\ Tree-PM} &  \makecell{Mira \\ USA} & 524 & 1.24 &  3.4 & 3.14\\ 
      \hline
          2021 & \makecell{Far Point \\ \citep{Frontiere:2021}} & \makecell{{\sc HACC} \\ Tree-PM} &  \makecell{Summit \\ USA} & ? & 1.86 &  1 & 0.8\\ 
      \hline
      \end{tabular}}
      \caption{List of cosmological simulations with a particle number in excess of 1 trillion ($10^{12}$)
      \label{tab:sims}
    }
\end{table}
\end{footnotesize}

\subsection{State-of-the-art simulations}

The rapid progress of large-scale structure surveys has dramatically pushed the need for on-par advances in the accuracy of theoretical modelling, and particularly so in cosmological simulations. Roughly speaking, it is necessary to simulate volumes similar to those observed with a mass resolution that allows to resolve, as a minimum, the halos expected to host galaxies targeted by observations but ideally to resolve the full merger history of those halos. For instance, the EUCLID survey will map a volume of 60~Gpc$^3$ and detect star-forming galaxies expected to reside in halos as small as $10^{10}$--$10^{11}\Msun$. This motivates ever larger and more accurate cosmological simulations.

The trend observed by \cite{Springel:2005a} in 2005 that the number of particles in $N$-body simulations doubled every $16.5$ months has somewhat flattened since but nevertheless cosmological simulations are at the forefront of what is possible with state-of-the-art infrastructure in national and international supercomputing centers. For instance, many cosmological codes have won or have been finalists to the Gordon Bell Performance Prize over the last 30 years, which is awarded for outstanding achievements in high-performance computing. Nowadays there are multiple simulations with more than a trillion particles -- several hundreds of times more than those used by the iconic \textit{Millennium} simulation. In Table~\ref{tab:sims} we compile a summary of recent efforts, all of which notably adopt different parallelisation and optimisation strategies as well as algorithms for the gravity calculation.

The first calculation to reach the 1-trillion ($10^{12}$) particle milestone was the \textit{Dark Sky} simulation \citep{Skillman:2014} in 2014. This simulation evolved a $10'240^{3}$ particle distribution from redshift 93 down to redshift zero in a $8\hGpc$ cubic volume. This simulation was carried out with 20,000~CPUs of the Titan Supercomputer at the Oak Ridge National Laboratory in USA. Gravitational forces were computed with the {\sc 2HOT} code \citep{Warren:2013} using a tree algorithm with multipole expansion, with a background subtraction to improve the performance at high redshifts, and a dual tree traversal  specifying cell-cell interactions in which multiple particles share the same interaction list. A key aspect enabling this simulation was the usage of Graphic Processing Units (GPUs) and many optimisations via SSE or AVX vector instructions. Additionally, catalogues of dark matter halos as well as a light-cone were constructed during the runtime of the simulation. 

The trillion particle mark was also reached by the \textit{TianNu} simulation using a different $N$-body code, {\sc ${\rm CUBEP^3M}$} \citep{HarnoisDeraps:2013}, which combines a 2-level PM with direct summation below the grid scale and including compression for the internal representation of coordinates \citep{YuCUBE:2018}. This simulation employed $86\%$ (331,776 CPUs) of the second largest supercomputer in the world at that time (4th place in early 2021): Tianhe-2 in China. The \textit{TianNu} simulated the cosmic density field with $2.97\times10^{12}$ particles in a $1.2\hGpc$ box which included the effects of massive neutrinos. The base simulation contained $6912^3$ CDM particles, and at $z=5$ a new set of $13824^3$ particles representing $M_\nu = 0.05eV$ neutrinos were added with which subtle differences between CDM and neutrinos were studied \citep{Inman:2015,Yu:2017}.

In the same year, 2017, the \textit{Euclid Flagship simulation} \citep{Potter:2017} was completed -- the first multi-trillion particle simulation with high force and mass resolution. The simulation comprised $10'000^3$ particles in a $3\hGpc$ box aimed at the preparation for the EUCLID mission. This simulation employed the {\sc PKDGRAV-3} code \citep{Potter:2016} which uses a binary tree with FMM expansion for the force calculation. As for other large simulations, it was possible thanks to the use of GPUs speeding up the gravity calculation.

The \textit{Outer Rim} was another simulation that reached the trillion particle count. It simulated a region of $3\hGpc$ across using the {\sc HACC} code \citep{Habib:2016} and a half million of cores of the Mira BG/Q supercomputer at the Argonne Leadership Computing Facility in the USA. The $10'240^3$ particles of $10^9\Msun$ were initialized at $z=200$ with 1LPT and later evolved generating approximately 5PB of data. The same group also used {\sc HACC} and the Mira supercomputer in 2020 to create another 1.24 trillion particle simulation, the \textit{Last Journey} \citep{Heitmann:2021}, which shares many specifics with Outer Rim but covers a slightly larger $3.4\hGpc$ cubic volume. Complementing these calculations, in 2021 the same group announced the \textit{FarPoint} simulation \citep{Frontiere:2021} with 1.86 trillion particles in a $1\hGpc$ box run on \textit{Summit} -- the 2nd largest supercomputer in the world at the time.

Similar to the \textit{Flagship} but with slightly better mass resolution is the \textit{Uchuu Simulation} \citep{IshiyamaUshuu:2020}, which is a $2000\hMpc$ simulation with 2.1 trillion particles ($12'800^3$) of mass $3.3\times10^8\,\Msun$ and employing $4.3h^{-1}$kpc softening length. The simulation was carried out on the ATERUI II supercomputer in Japan using the {\sc GreeM} tree-PM code \citep{Ishiyama:2012,IshiyamaUshuu:2020}, which makes extensive used of SIMD instructions and other low-level optimizations. 

The largest simulation to-date is Cosmo-$\pi$ which was performed in 2020 using the {\sc CUBE} code on $20'480$ cores of the Shanghai Jiao Tong University's $\pi$ 2.0 supercomputer in China. Cosmo-$\pi$ simulated a region of $3.2\hGpc$ with 4.39 trillion particles with a relatively modest force resolution, 195$\hkpc$. The force was computed with a 2-level particle mesh algorithm (PMPM) \citep{Merz:2005}, where a global PM mesh is used to compute long range forces which is complemented by independent local high-resolution PM meshes with isolated boundary condition to estimate short-range forces. The fine mesh can be stored completely locally in case of cubical domain decomposition, which has the advantage of reducing inter-node communication.

Note that the Cosmo-$\pi$ simulation was carried out on a comparatively modest supercomputer, this thanks to a heavily optimised memory consumption of the {\sc CUBE} code. This was realised mainly by using \textit{fixed-point precision} optimization where  phase-space coordinates are stored relative to the position and velocity of a coarse grid \citep{YuCUBE:2018}. Although this has additional computing demands (due to the compression/decompression), it reduced the required memory requirements from 24 to 6 bytes per particles, which is usually a limiting factor in these large-scale simulations.

\subsection{Computational trends and challenges}

It is interesting to note that the multi-trillion particle limit has been reached independently by several groups worldwide, employing different codes and different computational strategies. This is a remarkable achievement for the field of cosmological simulations. However, there are some trends and common features which we discuss next.

Firstly, these calculations are usually limited by the time spent in the force calculation. In this regard, it is finally becoming clear that hierarchical trees are sub-optimal compared to fast multipole methods, which display better scaling with particle number, by decreasing the algorithmic complexity of the (short-range) gravity interaction.  On the other hand, implementing FMM efficiently (including its parallelisation in distributed memory machines) is significantly more complex, and thus execution times can  vary significantly depending on the implementation. It also seems that highly optimized codes with ``hand-tuned'' parts using SIMD, AVX, and other low-level instructions are worth in this case, although this seems to depend on the target architecture.

Along these lines, the use of co-processors, and GPUs in particular, is proving extremely beneficial for these large calculations. We can see for instance that multi-trillion particle simulations are carried out with up to half-a-million CPUs (e.g. \textit{The Last Journey}) but with only 4 thousands CPUs when enhanced with GPUs (e.g. \textit{Euclid Flagship}). Specifically, {\sc PKDGRAV} claims that factors of 10 speed-ups can be achieved with the use of one GPU per node. The use of GPUs can also give the opportunity to researchers without access to the largest supercomputers in the world to carry out state-of-the-art simulations. This development is also in line with future supercomputer architectures, where exascale supercomputing is likely to be reached first with GPUs with the additional motivation provided by the training of machine learning algorithms.

Some degree of data compression seems to also be useful for these large calculations. This will be even more true in the future where the FLOP-to-byte ratio is expected to increase in future processors, more than the available RAM. Likewise, GPUs usually have much less memory than traditional CPUs, which provides further incentive for adopting some degree of internal compression. This is demonstrated by the Cosmo-$\pi$ simulation. However, it is unclear what degree of data compression can be afforded when increasing the spatial resolution, and whether this meets the accuracy requirement of upcoming LSS observations.

Another lesson learned from recent simulations is that I/O bandwidth and disk space is an important factor that could limit the usefulness of simulations. For instance, 200'000 CPU hours were needed only to read the initial conditions of the \textit{Outer Rim} simulation. Perhaps more importantly, the large raw datasets generated (easily in the petabyte scale), restrict the use of these simulations to a limited number of people, which unavoidably hinders their scientific exploitation. For this reason, more and more of the post-processing, including the construction of light-cones, structure identification, merger trees, and even the creation of maps for various observables (e.g. gravitational lensing or of the Sunyaev-Zeldovich effect) are carried out during the simulation runtime. 

Despite the extraordinary combination of volume and mass resolution of the simulations listed in the previous section, none of these multi-trillion particle calculations is actually sufficient to adequately match the characteristics e.g. of the EUCLID or Rubin-DESC surveys. However, the previous track record of continuous improvement and planned advances in supercomputer infrastructure means that it is likely that in the near future at least one calculation achieves the necessary combination of volume and resolution.  

Finally, a related challenge for carrying and storing simulation results will be in securing the necessary computational resources \textit{per se}. As we have argued throughout this review, gravity-only simulations are essential for modern cosmology and for large extragalactic surveys, where they can significantly enhance the scientific output. However, carrying out these simulations is left to individual groups or scientists who might or might not obtain competitively the required computational resources. This unavoidably creates a large degree of uncertainty and hinders a proper design and exploitation of simulations.

\subsection{Challenges for data dissemination}

State-of-the-art numerical simulations require huge computational efforts in terms of CPU/GPU power and storage facilities. For this reason, historically, simulations were only available to a small subset of the cosmological community. Over the last 10 years, this situation has changed thanks to the adoption of more open data policies together with the cost reduction for hardware and the advance of web technologies. 

It is now customary for simulation projects to put considerable effort into making their data publicly available and provide easy access to raw and secondary data products. In fact, the success of a simulation is many times judged not only by its direct scientific impact, but on the breadth of projects they enable and its use by the astronomical community as a whole. Sharing simulation data is also important in the context of cross-validation of simulation codes and results, since it readily allows a comparison and assessment of  the accuracy and potential sources of systematic errors. In principle, this extends also to the simulation codes themselves: a public code release strategy can leverage increased scrutiny of the methods employed in any single code, in terms of both correctness and economy. 

One of the pioneer efforts in the direction of data dissemination was that of the Virgo Consortium with the Millennium Database \citep{Lemson:2006}\footnote{\url{http://gavo.mpa-garching.mpg.de/Millennium/}}. In the context of the German Virtual Observatory, and following the example of the SkyServer in the Sloan Sky Digital Server, the Millennium Database served halo and semi-analytic galaxy catalogues via a relational database accessed trough the standard Structured Query Language (SQL). More recent examples are databases for the Bolshoi simulations \footnote{\url{https://www.cosmosim.org/}} which have additionally served full particle data, the Australian Theoretical Virtual Observatory\footnote{\url{https://tao.asvo.org.au/tao/}} which offers several post-processing and visualization tools \citep{Bernyk:2016}, and the INDRA database\footnote{\url{http://sciserver.org}} which provides access to a large ensemble of gravity-only simulations \citep{Falck:2021}. As an example of the significant value open data policies provide to simulation projects is the fact that even though the Millennium Simulation was carried out more than 15 years ago, it is still today being used in many research projects.

SQL databases are extremely efficient and useful for mining large datasets, however, for many applications it is desirable to serve large catalogues for which SQL is not optimal. In such cases, the increase of worldwide network bandwidths and the web developments have made practical to provide data in excess of tens of Terabytes. One example is http://skiesanduniverses.org/ where data from the \textit{Multidark} \citep{Klypin:2016}, \textit{Bolshoi} \citep{Klypin:2011}, \textit{GLAM} \citep{KlypinPrada:2018}, and \textit{Uchuu} \citep{IshiyamaUshuu:2020} simulations can be directly downloaded.  Another interesting alternative approach adopted to serve the data of the 1-trillion particles \textit{DarkSky} simulation \footnote{\url{https://darksky.slac.stanford.edu/}} is to directly map data in a given URL to a Python virtual memory buffer. In this way, local Python sessions can directly access data stored in a remote server. The \textit{Outer Rim} and \textit{Last Journey} simulations data, instead, was shared directly trough the \textit{Globus} technology \footnote{\url{http://www.globus.org}}. Globus is a non-profit research data management that allows triggering secure data transfers among different locations (e.g. personal computers, cloud storage, or supercomputer facilities) via GridFTP. In this way, a subset of the simulated data (halo catalogues and diluted snapshots, etc) can be easily shared with the research community.

The challenges faced by the current generation of simulations will become more acute in the future, with the increase in data volumes and complexity of the whole post-processing pipelines. Thus, it will be important that together with the data, there is a public release of data analysis software and, more generally, further emphasis in the reproducibility of the results. In this regard, the numerical simulation community has a good track record with open software and open comparison projects, all of which should enhance the robustness and quality of any potential cosmological inference derived from the analysis of numerical simulations. 

\subsection{Challenges for cosmological predictions}

Carrying out a single state-of-the-art simulation is challenging, yet if they are to be used for the exploitation of cosmological observations, we need a large number of them covering the full parameter space of plausible cosmological models. This means that the computational resources will be likely in excess of those of single grand-challenge simulations. 

Current campaigns aimed at providing predictions as a function of cosmology typical build emulators which are trained using $\mathcal{O}(100)$ simulations. This number can be decreased by more than an order of magnitude by using cosmology-rescaling algorithms \citep{Angulo:2010,Angulo:2015}, but still represent an important computational challenge. With this (relatively) low number of simulations, various groups have shown that it is already possible to achieve a $\lesssim1$\% accuracy for predicting the nonlinear power spectrum up to scales $k \sim 1\ihMpc$, for all the parameters of a minimal $\Lambda$CDM model \citep{Angulo:2020,Knabenhans:2020}. At smaller scales, they have an accuracy of 3\%, not too different from the accuracy of simulations themselves. The main limitation is the rather restricted cosmological parameter space they can cover, usually much smaller than the priors considered in data analysis. Covering large hyper-volumes quickly increases the number of required calculations and the computational cost associated, thus, probably in the future hybrid emulators will be built which display high accuracy in a given region of interest (e.g. near the best fitting $\Lambda$CDM parameters) but have less accuracy elsewhere. This could be achieved by using approximate simulations, physically-motivated extrapolations, or sampling less densely the parameter space. 

Although improving the gravity-only predictions on small scales will be an important task, baryons have a much more important impact on these scales and thus for the overall accuracy of the predictions. To address this, also emulators have been built based on physically-motivated models for baryonic effects, which claim that they can also obtain few percent accuracy \citep{Schneider:2020,Arico:2020}. Although they are close to the target accuracy, they are likely only sufficient for the first round of data analysis of upcoming surveys, but continued improvement in precision will be needed. This is probably within reach as better understanding from hydrodynamical simulations and observations is achieved. 

Note that hydrodynamical simulations are sometimes regarded as the ``ground truth'' and one might think that the best strategy for cosmological exploitation should be based on multiple hydrodynamic calculations with varying cosmology and baryonic parameters. However, it is important to be aware of the risks of this approach. First, most of the simulated physics rely on simplified ``sub-grid'' physics describing e.g. star formation, and feedback from black holes and supernovae. Thus the accuracy and realism of the cosmological predictions would be limited by these underlying recipes, which frequently are simply ad-hoc parametrisations with significant degrees of freedom that vary considerably among simulation codes and their physics implementations. These prescriptions are often valid only for the physical resolution of a given simulation and require re-tuning when the resolution changes. Second, many of these recipes have been calibrated or validated against observational data, thus there is the risk of a circular argument where observed data is used to calibrate simulations assuming a cosmology and then one uses those simulations to constrain cosmology. Finally, it is ultimately very difficult to estimate, let alone control, all possible sources of uncertainty and consistently propagate them to cosmological constraints. Due to all these reasons, a more conservative approach is to employ physically-motivated models which are informed by, and tested against, a broad range of cosmologies and galaxy formation models including those seemingly ruled out by data.

Reliably predicting galaxy clustering as a function of cosmology is another area where significant modelling uncertainty exists. One option to make such predictions would be to combine these simulations with one or more galaxy models described in Sect.~\ref{sec:galaxies}, so that the cosmological and astrophysical parameters could be jointly constrained. This however, poses several problems. The first is one of computational nature, as the target parameter space would be extremely large which would be a challenging task for emulators. Additionally, even at a fixed cosmology, the calibration of SAMs and empirical models can cost a few hundred thousands of CPU hours. An alternative could be to formulate these with automatic differentiation, for which several optimization algorithms can be employed speeding up parameter sampling considerably. Another option is to split the cosmological and galaxy formation parameters and emulate directly the ``Bayesian evidence" of a given cosmology \citep{Lange:2019,Lange:2021}.

As for our discussion regarding baryonic physics and hydrodynamical simulations, there is the risk of cosmological biases due to an inaccurate modelling of galaxy formation physics. Another potentially more robust option has been proposed recently in terms of combinations of parametric bias descriptions (motivated by perturbative expansion of galaxy bias) with $N$-body dynamics, which can be emulated efficiently and could be a robust way to extract information from the quasi-linear regime. In the future the main challenges will be the extension to include peculiar velocities and a self-consistent treatment of higher-order correlations \citep{Pellejero-Ibanez:2021,Banerjee:2021}. It is commonly argued that by using measurements of the late-time LSS, one can access a large number of modes in the Universe which can potentially outperform CMB analysis. It is, however, unclear how many of these modes actually have useful information about fundamental aspects of the universe. Likely a modelling based on simulations such as the one here described will ultimately answer this question and determine the maximum amount of cosmological information in the late-time Universe.

Another rich source of information is found in combining observations in different wavelengths and exploiting their cross-correlations, e.g. lensing-SZ, lensing-galaxy clustering, SZ-galaxy clustering, etc. For this, simulations would need to simultaneously predict multiple observables. In this regard, predictions for the halo mass function have also been possible either by emulators or by refined fitting functions. Future challenges will be about incorporating baryonic physics, and also observational errors (e.g. projection effects, off-centering, etc), since cluster cosmology depends sensitively on the observable-mass calibration (e.g. optical richness, X-ray, or Sunyaev-Zeldovich effects). 

In parallel, instead of building emulators, it will perhaps be possible to employ machine learning algorithms to design optimal summary statistics than go beyond classical $N$-point functions, or to train them to directly generate ensembles of fields with which a full forward modelling could be possible. In this way, the cosmological exploitation could be performed not only at the level of summary statistics, but also at the field level, e.g. incorporating the information in the phases of the cosmic field in Fourier space. However, to realise this possibility, more advances will be required demonstrating the robustness of the approach (against, e.g. observational systematic effects) and the quantification of modelling errors.

Common to all the above is that it will be very important to carefully quantify the emulators' accuracy and incorporate them into a Bayesian data analysis workflow. This assessment should entail all possible sources of systematic errors, errors arising from the $N$-body discretization, errors associated with the specific simulation, structure finding, interpolation method, and modelling of the relevant baryonic physics, and the projection to the observable space (redshift space and/or redshift errors). Since these errors are likely to have complex structures in parameter space (being more accurate in some regions and less so in others), they will need to be taken into account in Bayeasian parameter samplers and in the covariance matrices.

As the sophistication and accuracy of cosmological models improve, they will pose stricter requirements for the accuracy of the respective covariance matrices (see \citealt{Monaco:2016} for a review). Currently, there are indications that analytic or approximate methods for building covariances are enough for the precision of future LSS surveys and current data models \citep{Barreira:2018,Lippich:2019,Blot:2019,Colavincenzo:2019}. However, as these models improve in precision and become valid deep into the nonlinear regime (and for the cross-correlation of different observables), they will require equally accurate covariances and a re-assessment of the accuracy with which they are currently built will be very important.

Ultimately, simulations will be able to model simultaneously a broad range of observables over different wavelengths as a function of cosmology and over a wide set of scales an redshifts. This will also be important as more surveys scan the sky on different observables, the power of cross-correlations will become more evident by breaking degeneracies, identifying potential sources of systematic errors, and, perhaps more importantly, potentially detecting physics beyond $\Lambda$CDM from more than a single cosmic probe.

All this will likely be an iterative process between the numerical simulations and observational communities, where better predictions become available, data is interpreted which results into a better assessment of sources of noise which feeds back into more accurate and refined simulations.

%% Outlook and Conclusions
\newpage
\section{Conclusions and Outlook}

After decades of development, modern large-scale cosmological simulations have become a mature field that provides the most precise method for predicting and understanding nonlinear structure in the cosmos. In consequence, numerical simulations are essential in modern cosmology and in the ongoing efforts to understand the nature of our Universe.

Reaching this point required progress in various directions. The first concerns rigorous tests of the basic assumptions behind these calculations: from the impact of the $N$-body discretisation on results, to the validity of Newtonian limits of General Relativity. The second direction was a continuous improvement of the algorithms and the accuracy of the calculations: from a better understanding of the sources of error in the creation of initial conditions, in time integration, to the treatment of dark matter and baryons as a single fluid, and many others. The third direction concerned a continuous extension of physical models beyond the simplest $\Lambda$CDM model: from modifications of gravity to different dark matter models, this also had synergistic effects on the development of new cosmic probes and experiments. All these advances have been supported by a continuous increase in computational power, which allowed simulations of ever bigger cosmological volumes with better and better resolution. In parallel the development of algorithms continued to make best use of new generations of supercomputers, with cosmological codes continuously being among the few in science that can efficiently exploit the largest supercomputers worldwide.

As a consequence, predictions from gravity-only $N$-body simulations have become very reliable. For instance, different simulation codes, despite using very different numerical approaches for solving the underlying equations, now agree at the sub-percent level on the nonlinear matter distribution down to the internal structure of halos. Systematic comparison and convergence studies have also allowed the field to identify sources of noise and error, to improve the relevant approaches, and to validate simplifications and optimisations. In parallel, an important validation of the $N$-body approach itself is starting to be possible. Initial results from Lagrangian tessellation discretizations -- currently the only viable alternative to $N$-body -- are indicating the validity of $N$-body when following the gravitational collapse of CDM fluctuations. While there are still some problems that will require further elucidation, these results appear to provide an important support for the predictions of standard cosmological codes. 

In contrast, the connection of dark matter with visible objects such as galaxies and quasars, is comparatively still very much more uncertain. Nevertheless, in recent years tremendous progress has been made in the field of galaxy formation and increasing agreement on that front is being quickly achieved. This allowed the development of models for the impact of galaxy formation and baryonic physics on cosmological observables that can be applied to gravity-only simulations. However, some caution is necessary since hydrodynamical simulations usually rely on sub-resolution physics which have a number of parameters that are adjusted to reproduce observations: reproducing observations does not necessarily imply a modelling of the correct effective physics. Their predictive power can only be scrutinized by testing predictions far outside the calibrating space. Nevertheless, these galaxy formation simulations have aided in the development of physically-motivated models that could be applied to data in order to ultimately obtain constraints on cosmological parameters after a marginalization over baryonic physics. In addition, these baryonic models will keep improving in the future, which will naturally feed back into more accurate and deterministic models to be combined with gravity-only simulations.

All the development described throughout this review is preparing $N$-body simulations for their next big challenge: to be used directly in the data analysis pipeline of upcoming cosmological measurements. In this regard, we have seen fast progress on two fronts. The first is the development of approximate methods which, although not always formally correct, appear to reproduce large-scale statistics of the matter field at a fraction of the computational cost of a standard $N$-body simulation. The second front is the development of reliable interpolation techniques of $N$-body results in cosmological parameter space, either by constructing emulators over the outputs of hundreds of $N$-body results, or by training machine learning algorithms. With the steady growth in computer power, it is likely that we will see increasingly better predictions to the point that it will be possible to create predictions as a function of cosmology that will be indistinguishable from those of direct $N$-body simulations. This has the potential to significantly increase the scientific output of future extragalactic observations.

If the challenges described above can be successfully solved with high precision and convincing robustness over the next decade, large-scale $N$-body simulations could become a key piece in our effort to answer some of the most important questions in physics: the nature of dark matter, dark energy, and gravity.

%%%%%%%%%%%%%%%%%%%%%%%%%%%%%%%%%%%%%%%%%%%%%%%%%%%%
%%%%%%%%%%%%%%%%%%%%%%%%%%%%%%%%%%%%%%%%%%%%%%%%%%%%

\begin{acknowledgements}
We would like to thank the editor, Thorsten Naab, for his invitation to write this article and his enormous patience during the process. We would also like to thank Alexandre Barreira, Silvia Bonoli, Sergio Contreras, Baojiu Li, Go Ogiya, Aseem Paranjape, Cornelius Rampf, Jens St\"ucker, and Matteo Zennaro, for comments and suggestions. REA acknowledges the support of the ERC Starting-Grant 716151 (BACCO). O.H. acknowledges funding from the European Research Council (ERC) under the European Union's Horizon 2020 research and innovation programme, Grant agreement No. 679145 (COSMO-SIMS)
\end{acknowledgements}

%%%%%%%%%%%%%%%%%%%%%%%%%%%%%%%%%%%%%%%%%%%%%%%%%%%%
%%%%%%%%%%%%%%%%%%%%%%%%%%%%%%%%%%%%%%%%%%%%%%%%%%%%

\bibliographystyle{spbasic-FS}       % APS-like style for physics
\bibliography{DM_sim_review}   % name your BibTeX data base

\begin{thebibliography}{1018}
\expandafter\ifx\csname url\endcsname\relax
 \def\url#1{\burl{#1}}\fi
\expandafter\ifx\csname urlprefix\endcsname\relax\def\urlprefix{URL }\fi
\providecommand{\bibinfo}[2]{#2}
\providecommand{\eprint}[2][]{\url{#2}}
\providecommand{\doi}[1]{\urlstyle{rm}\url{https://doi.org/#1}}

\bibitem[{{Aarseth}(1963)}]{Aarseth:1963}
{Aarseth} SJ (1963) {Dynamical evolution of clusters of galaxies, I}. \mnras
  126:223. \doi{10.1093/mnras/126.3.223}

\bibitem[{Aarseth(2009)}]{Aarseth:2009}
Aarseth SJ (2009) Gravitational N-Body Simulations: Tools and Algorithms.
  Cambridge Monographs on Mathematical Physics, Cambridge University Press.
  \doi{10.1017/CBO9780511535246}

\bibitem[{{Abazajian} et~al.(2016){Abazajian}, {Adshead}, {Ahmed}, {Allen},
  {Alonso}, {Arnold}, {Baccigalupi}, {Bartlett}, {Battaglia}, {Benson},
  {Bischoff}, {Borrill}, {Buza}, {Calabrese}, {Caldwell}, {Carlstrom}, {Chang},
  {Crawford}, {Cyr-Racine}, {De Bernardis}, {de Haan}, {di Serego Alighieri},
  {Dunkley}, {Dvorkin}, {Errard}, {Fabbian}, {Feeney}, {Ferraro}, {Filippini},
  {Flauger}, {Fuller}, {Gluscevic}, {Green}, {Grin}, {Grohs}, {Henning},
  {Hill}, {Hlozek}, {Holder}, {Holzapfel}, {Hu}, {Huffenberger}, {Keskitalo},
  {Knox}, {Kosowsky}, {Kovac}, {Kovetz}, {Kuo}, {Kusaka}, {Le Jeune}, {Lee},
  {Lilley}, {Loverde}, {Madhavacheril}, {Mantz}, {Marsh}, {McMahon},
  {Meerburg}, {Meyers}, {Miller}, {Munoz}, {Nguyen}, {Niemack}, {Peloso},
  {Peloton}, {Pogosian}, {Pryke}, {Raveri}, {Reichardt}, {Rocha}, {Rotti},
  {Schaan}, {Schmittfull}, {Scott}, {Sehgal}, {Shandera}, {Sherwin}, {Smith},
  {Sorbo}, {Starkman}, {Story}, {van Engelen}, {Vieira}, {Watson}, {Whitehorn},
  and {Kimmy Wu}}]{CMBS4}
{Abazajian} KN, {Adshead} P, {Ahmed} Z, {Allen} SW, {Alonso} D, {Arnold} KS,
  {Baccigalupi} C, {Bartlett} JG, {Battaglia} N, {Benson} BA, {Bischoff} CA,
  {Borrill} J, {Buza} V, {Calabrese} E, {Caldwell} R, {Carlstrom} JE, {Chang}
  CL, {Crawford} TM, {Cyr-Racine} FY, {De Bernardis} F, {de Haan} T, {di Serego
  Alighieri} S, {Dunkley} J, {Dvorkin} C, {Errard} J, {Fabbian} G, {Feeney} S,
  {Ferraro} S, {Filippini} JP, {Flauger} R, {Fuller} GM, {Gluscevic} V, {Green}
  D, {Grin} D, {Grohs} E, {Henning} JW, {Hill} JC, {Hlozek} R, {Holder} G,
  {Holzapfel} W, {Hu} W, {Huffenberger} KM, {Keskitalo} R, {Knox} L, {Kosowsky}
  A, {Kovac} J, {Kovetz} ED, {Kuo} CL, {Kusaka} A, {Le Jeune} M, {Lee} AT,
  {Lilley} M, {Loverde} M, {Madhavacheril} MS, {Mantz} A, {Marsh} DJE,
  {McMahon} J, {Meerburg} PD, {Meyers} J, {Miller} AD, {Munoz} JB, {Nguyen} HN,
  {Niemack} MD, {Peloso} M, {Peloton} J, {Pogosian} L, {Pryke} C, {Raveri} M,
  {Reichardt} CL, {Rocha} G, {Rotti} A, {Schaan} E, {Schmittfull} MM, {Scott}
  D, {Sehgal} N, {Shandera} S, {Sherwin} BD, {Smith} TL, {Sorbo} L, {Starkman}
  GD, {Story} KT, {van Engelen} A, {Vieira} JD, {Watson} S, {Whitehorn} N,
  {Kimmy Wu} WL (2016) {CMB-S4 Science Book, First Edition}. arXiv e-prints
  arXiv:1610.02743.
  {\href{https://arxiv.org/abs/1610.02743}{{arXiv:1610.02743}}} {[astro-ph.CO]}

\bibitem[{{Abdallah} et~al.(2016){Abdallah}, {Abramowski}, {Aharonian}, {Ait
  Benkhali}, {Akhperjanian}, {Ang{\"u}ner}, {Arrieta}, {Aubert}, {Backes},
  {Balzer}, and et~al.}]{Abdallah:2016}
{Abdallah} H, {Abramowski} A, {Aharonian} F, {Ait Benkhali} F, {Akhperjanian}
  AG, {Ang{\"u}ner} E, {Arrieta} M, {Aubert} P, {Backes} M, {Balzer} A, et~al
  (2016) {Search for Dark Matter Annihilations towards the Inner Galactic Halo
  from 10 Years of Observations with H.E.S.S.} \prl 117(11):111301.
  \doi{10.1103/PhysRevLett.117.111301}.
  {\href{https://arxiv.org/abs/1607.08142}{{arXiv:1607.08142}}} {[astro-ph.HE]}

\bibitem[{{Abel} et~al.(2012){Abel}, {Hahn}, and {Kaehler}}]{Abel:2012}
{Abel} T, {Hahn} O, {Kaehler} R (2012) {Tracing the dark matter sheet in phase
  space}. \mnras 427:61--76. \doi{10.1111/j.1365-2966.2012.21754.x}.
  {\href{https://arxiv.org/abs/1111.3944}{{arXiv:1111.3944}}}

\bibitem[{{Ackermann} et~al.(2012){Ackermann}, {Ajello}, {Atwood}, {Baldini},
  {Barbiellini}, {Bastieri}, {Bechtol}, {Bellazzini}, {Blandford}, {Bloom},
  {Bonamente}, {Borgland}, {Bottacini}, {Brandt}, {Bregeon}, {Brigida},
  {Bruel}, {Buehler}, {Buson}, {Caliandro}, {Cameron}, {Caraveo}, {Casandjian},
  {Cecchi}, {Charles}, {Chekhtman}, {Chiang}, {Ciprini}, {Claus},
  {Cohen-Tanugi}, {Conrad}, {Cuoco}, {Cutini}, {D'Ammando}, {de Angelis}, {de
  Palma}, {Dermer}, {Silva}, {Drell}, {Drlica-Wagner}, {Falletti}, {Favuzzi},
  {Fegan}, {Focke}, {Fukazawa}, {Funk}, {Fusco}, {Gargano}, {Gasparrini},
  {Germani}, {Giglietto}, {Giordano}, {Giroletti}, {Glanzman}, {Godfrey},
  {Grenier}, {Guiriec}, {Gustafsson}, {Hadasch}, {Hayashida}, {Horan},
  {Hughes}, {Jackson}, {Jogler}, {J{\'o}hannesson}, {Johnson}, {Kamae},
  {Kn{\"o}dlseder}, {Kuss}, {Lande}, {Latronico}, {Lionetto}, {Llena Garde},
  {Longo}, {Loparco}, {Lott}, {Lovellette}, {Lubrano}, {Mazziotta}, {McEnery},
  {Mehault}, {Michelson}, {Mitthumsiri}, {Mizuno}, {Moiseev}, {Monte},
  {Monzani}, {Morselli}, {Moskalenko}, {Murgia}, {Naumann-Godo}, {Norris},
  {Nuss}, {Ohsugi}, {Orienti}, {Orlando}, {Ormes}, {Paneque}, {Panetta},
  {Pesce-Rollins}, {Pierbattista}, {Piron}, {Pivato}, {Poon}, {Rain{\`o}},
  {Rando}, {Razzano}, {Razzaque}, {Reimer}, {Reimer}, {Romoli}, {Sbarra},
  {Scargle}, {Sgr{\`o}}, {Siskind}, {Spandre}, {Spinelli}, {Stawarz}, {Strong},
  {Suson}, {Tajima}, {Takahashi}, {Tanaka}, {Thayer}, {Thayer}, {Tibaldo},
  {Tinivella}, {Tosti}, {Troja}, {Usher}, {Vandenbroucke}, {Vasileiou},
  {Vianello}, {Vitale}, {Waite}, {Wallace}, {Wood}, {Wood}, {Yang},
  {Zaharijas}, and {Zimmer}}]{Ackermann:2012}
{Ackermann} M, {Ajello} M, {Atwood} WB, {Baldini} L, {Barbiellini} G,
  {Bastieri} D, {Bechtol} K, {Bellazzini} R, {Blandford} RD, {Bloom} ED,
  {Bonamente} E, {Borgland} AW, {Bottacini} E, {Brandt} TJ, {Bregeon} J,
  {Brigida} M, {Bruel} P, {Buehler} R, {Buson} S, {Caliandro} GA, {Cameron} RA,
  {Caraveo} PA, {Casandjian} JM, {Cecchi} C, {Charles} E, {Chekhtman} A,
  {Chiang} J, {Ciprini} S, {Claus} R, {Cohen-Tanugi} J, {Conrad} J, {Cuoco} A,
  {Cutini} S, {D'Ammando} F, {de Angelis} A, {de Palma} F, {Dermer} CD, {Silva}
  EdCe, {Drell} PS, {Drlica-Wagner} A, {Falletti} L, {Favuzzi} C, {Fegan} SJ,
  {Focke} WB, {Fukazawa} Y, {Funk} S, {Fusco} P, {Gargano} F, {Gasparrini} D,
  {Germani} S, {Giglietto} N, {Giordano} F, {Giroletti} M, {Glanzman} T,
  {Godfrey} G, {Grenier} IA, {Guiriec} S, {Gustafsson} M, {Hadasch} D,
  {Hayashida} M, {Horan} D, {Hughes} RE, {Jackson} MS, {Jogler} T,
  {J{\'o}hannesson} G, {Johnson} AS, {Kamae} T, {Kn{\"o}dlseder} J, {Kuss} M,
  {Lande} J, {Latronico} L, {Lionetto} AM, {Llena Garde} M, {Longo} F,
  {Loparco} F, {Lott} B, {Lovellette} MN, {Lubrano} P, {Mazziotta} MN,
  {McEnery} JE, {Mehault} J, {Michelson} PF, {Mitthumsiri} W, {Mizuno} T,
  {Moiseev} AA, {Monte} C, {Monzani} ME, {Morselli} A, {Moskalenko} IV,
  {Murgia} S, {Naumann-Godo} M, {Norris} JP, {Nuss} E, {Ohsugi} T, {Orienti} M,
  {Orlando} E, {Ormes} JF, {Paneque} D, {Panetta} JH, {Pesce-Rollins} M,
  {Pierbattista} M, {Piron} F, {Pivato} G, {Poon} H, {Rain{\`o}} S, {Rando} R,
  {Razzano} M, {Razzaque} S, {Reimer} A, {Reimer} O, {Romoli} C, {Sbarra} C,
  {Scargle} JD, {Sgr{\`o}} C, {Siskind} EJ, {Spandre} G, {Spinelli} P,
  {Stawarz} {\L}, {Strong} AW, {Suson} DJ, {Tajima} H, {Takahashi} H, {Tanaka}
  T, {Thayer} JG, {Thayer} JB, {Tibaldo} L, {Tinivella} M, {Tosti} G, {Troja}
  E, {Usher} TL, {Vandenbroucke} J, {Vasileiou} V, {Vianello} G, {Vitale} V,
  {Waite} AP, {Wallace} E, {Wood} KS, {Wood} M, {Yang} Z, {Zaharijas} G,
  {Zimmer} S (2012) {Constraints on the Galactic Halo Dark Matter from
  Fermi-LAT Diffuse Measurements}. \apj 761(2):91.
  \doi{10.1088/0004-637X/761/2/91}.
  {\href{https://arxiv.org/abs/1205.6474}{{arXiv:1205.6474}}} {[astro-ph.CO]}

\bibitem[{{Acquaviva} et~al.(2003){Acquaviva}, {Bartolo}, {Matarrese}, and
  {Riotto}}]{Acquaviva:2003}
{Acquaviva} V, {Bartolo} N, {Matarrese} S, {Riotto} A (2003) {Gauge-invariant
  second-order perturbations and non-Gaussianity from inflation}. Nuclear
  Physics B 667(1-2):119--148. \doi{10.1016/S0550-3213(03)00550-9}.
  {\href{https://arxiv.org/abs/astro-ph/0209156}{{arXiv:astro-ph/0209156}}}
  {[astro-ph]}

\bibitem[{{Adamek} et~al.(2013){Adamek}, {Daverio}, {Durrer}, and
  {Kunz}}]{Adamek:2013}
{Adamek} J, {Daverio} D, {Durrer} R, {Kunz} M (2013) {General relativistic
  N-body simulations in the weak field limit}. \prd 88(10):103527.
  \doi{10.1103/PhysRevD.88.103527}.
  {\href{https://arxiv.org/abs/1308.6524}{{arXiv:1308.6524}}} {[astro-ph.CO]}

\bibitem[{{Adamek} et~al.(2016){Adamek}, {Daverio}, {Durrer}, and
  {Kunz}}]{Adamek:2016}
{Adamek} J, {Daverio} D, {Durrer} R, {Kunz} M (2016) {gevolution: a
  cosmological N-body code based on General Relativity}. J Cosmol Astro-Part
  Phys 2016(7):053. \doi{10.1088/1475-7516/2016/07/053}.
  {\href{https://arxiv.org/abs/1604.06065}{{arXiv:1604.06065}}} {[astro-ph.CO]}

\bibitem[{{Adamek} et~al.(2017){Adamek}, {Brandbyge}, {Fidler}, {Hannestad},
  {Rampf}, and {Tram}}]{Adamek:2017}
{Adamek} J, {Brandbyge} J, {Fidler} C, {Hannestad} S, {Rampf} C, {Tram} T
  (2017) {The effect of early radiation in N-body simulations of cosmic
  structure formation}. \mnras 470(1):303--313. \doi{10.1093/mnras/stx1157}.
  {\href{https://arxiv.org/abs/1703.08585}{{arXiv:1703.08585}}} {[astro-ph.CO]}

\bibitem[{Adamek et~al.(2017)Adamek, Durrer, and Kunz}]{Adamek2017}
Adamek J, Durrer R, Kunz M (2017) Relativistic n-body simulations with massive
  neutrinos. JCAP11(2017)004 \doi{10.1088/1475-7516/2017/11/004}.
  {\href{https://arxiv.org/abs/1707.06938}{{arXiv:1707.06938}}} {[astro-ph.CO]}

\bibitem[{{Adamek} et~al.(2019){Adamek}, {Byrnes}, {Gosenca}, and
  {Hotchkiss}}]{Adamek:2019b}
{Adamek} J, {Byrnes} CT, {Gosenca} M, {Hotchkiss} S (2019) {WIMPs and
  stellar-mass primordial black holes are incompatible}. \prd 100(2):023506.
  \doi{10.1103/PhysRevD.100.023506}.
  {\href{https://arxiv.org/abs/1901.08528}{{arXiv:1901.08528}}} {[astro-ph.CO]}

\bibitem[{{Ade} et~al.(2019){Ade}, {Aguirre}, {Ahmed}, {Aiola}, {Ali},
  {Alonso}, {Alvarez}, {Arnold}, {Ashton}, {Austermann}, and et~al.}]{SO}
{Ade} P, {Aguirre} J, {Ahmed} Z, {Aiola} S, {Ali} A, {Alonso} D, {Alvarez} MA,
  {Arnold} K, {Ashton} P, {Austermann} J, et~al (2019) {The Simons Observatory:
  science goals and forecasts}. \jcap 2019(2):056.
  \doi{10.1088/1475-7516/2019/02/056}.
  {\href{https://arxiv.org/abs/1808.07445}{{arXiv:1808.07445}}} {[astro-ph.CO]}

\bibitem[{{Agarwal} and {Feldman}(2011)}]{Agarwal:2011}
{Agarwal} S, {Feldman} HA (2011) {The effect of massive neutrinos on the matter
  power spectrum}. \mnras 410(3):1647--1654.
  \doi{10.1111/j.1365-2966.2010.17546.x}.
  {\href{https://arxiv.org/abs/1006.0689}{{arXiv:1006.0689}}} {[astro-ph.CO]}

\bibitem[{{Aker} et~al.(2021){Aker}, {Beglarian}, {Behrens}, {Berlev},
  {Besserer}, {Bieringer}, {Block}, {Bornschein}, {Bornschein}, {B{\"o}ttcher},
  {Brunst}, {Caldwell}, {Carney}, {La Cascio}, {Chilingaryan}, {Choi},
  {Debowski}, {Deffert}, {Descher}, {D{\'\i}az Barrero}, {Doe}, {Dragoun},
  {Drexlin}, {Eitel}, {Ellinger}, {Engel}, {Enomoto}, {Felden}, {Formaggio},
  {Fr{\"a}nkle}, {Franklin}, {Friedel}, {Fulst}, {Gauda}, {Gil}, {Gl{\"u}ck},
  {Gr{\"o}ssle}, {Gumbsheimer}, {Gupta}, {H{\"o}hn}, {Hannen}, {Hau{\ss}mann},
  {Helbing}, {Hickford}, {Hiller}, {Hillesheimer}, {Hinz}, {Houdy}, {Huber},
  {Jansen}, {Karl}, {Kellerer}, {Kellerer}, {Klein}, {K{\"o}hler},
  {K{\"o}llenberger}, {Kopmann}, {Korzeczek}, {Koval{\'\i}k}, {Krasch},
  {Krause}, {Kunka}, {Lasserre}, {Le}, {Lebeda}, {Lehnert}, {Lokhov},
  {Machatschek}, {Malcherek}, {Mark}, {Marsteller}, {Martin}, {Melzer},
  {Menshikov}, {Mertens}, {Mostafa}, {M{\"u}ller}, {Niemes}, {Oelpmann},
  {Parno}, {Poon}, {Poyato}, {Priester}, {R{\"o}llig}, {R{\"o}ttele},
  {Robertson}, {Rodejohann}, {Rodenbeck}, {Ry{\v{s}}av{\'y}}, {Sack}, {Saenz},
  {Sch{\"a}fer}, {Schaller}, {Schimpf}, {Schl{\"o}sser}, {Schl{\"o}sser},
  {Schl{\"u}ter}, {Schneidewind}, {Schrank}, {Schulz}, {Schwemmer},
  {{\v{S}}ef{\v{c}}{\'\i}k}, {Sibille}, {Siegmann}, {Slez{\'a}k}, {Steidl},
  {Sturm}, {Sun}, {Tcherniakhovski}, {Telle}, {Thorne}, {Th{\"u}mmler},
  {Titov}, {Tkachev}, {Urban}, {Valerius}, {V{\'e}nos}, {Vizcaya
  Hern{\'a}ndez}, {Weinheimer}, {Welte}, {Wendel}, {Wilkerson}, {Wolf},
  {W{\"u}stling}, {Xu}, {Yen}, {Zadoroghny}, and {Zeller}}]{Aker:2021}
{Aker} M, {Beglarian} A, {Behrens} J, {Berlev} A, {Besserer} U, {Bieringer} B,
  {Block} F, {Bornschein} B, {Bornschein} L, {B{\"o}ttcher} M, {Brunst} T,
  {Caldwell} TS, {Carney} RMD, {La Cascio} L, {Chilingaryan} S, {Choi} W,
  {Debowski} K, {Deffert} M, {Descher} M, {D{\'\i}az Barrero} D, {Doe} PJ,
  {Dragoun} O, {Drexlin} G, {Eitel} K, {Ellinger} E, {Engel} R, {Enomoto} S,
  {Felden} A, {Formaggio} JA, {Fr{\"a}nkle} FM, {Franklin} GB, {Friedel} F,
  {Fulst} A, {Gauda} K, {Gil} W, {Gl{\"u}ck} F, {Gr{\"o}ssle} R, {Gumbsheimer}
  R, {Gupta} V, {H{\"o}hn} T, {Hannen} V, {Hau{\ss}mann} N, {Helbing} K,
  {Hickford} S, {Hiller} R, {Hillesheimer} D, {Hinz} D, {Houdy} T, {Huber} A,
  {Jansen} A, {Karl} C, {Kellerer} F, {Kellerer} J, {Klein} M, {K{\"o}hler} C,
  {K{\"o}llenberger} L, {Kopmann} A, {Korzeczek} M, {Koval{\'\i}k} A, {Krasch}
  B, {Krause} H, {Kunka} N, {Lasserre} T, {Le} TL, {Lebeda} O, {Lehnert} B,
  {Lokhov} A, {Machatschek} M, {Malcherek} E, {Mark} M, {Marsteller} A,
  {Martin} EL, {Melzer} C, {Menshikov} A, {Mertens} S, {Mostafa} J,
  {M{\"u}ller} K, {Niemes} S, {Oelpmann} P, {Parno} DS, {Poon} AWP, {Poyato}
  JML, {Priester} F, {R{\"o}llig} M, {R{\"o}ttele} C, {Robertson} RGH,
  {Rodejohann} W, {Rodenbeck} C, {Ry{\v{s}}av{\'y}} M, {Sack} R, {Saenz} A,
  {Sch{\"a}fer} P, {Schaller} A, {Schimpf} L, {Schl{\"o}sser} K,
  {Schl{\"o}sser} M, {Schl{\"u}ter} L, {Schneidewind} S, {Schrank} M, {Schulz}
  B, {Schwemmer} A, {{\v{S}}ef{\v{c}}{\'\i}k} M, {Sibille} V, {Siegmann} D,
  {Slez{\'a}k} M, {Steidl} M, {Sturm} M, {Sun} M, {Tcherniakhovski} D, {Telle}
  HH, {Thorne} LA, {Th{\"u}mmler} T, {Titov} N, {Tkachev} I, {Urban} K,
  {Valerius} K, {V{\'e}nos} D, {Vizcaya Hern{\'a}ndez} AP, {Weinheimer} C,
  {Welte} S, {Wendel} J, {Wilkerson} JF, {Wolf} J, {W{\"u}stling} S, {Xu} W,
  {Yen} YR, {Zadoroghny} S, {Zeller} G (2021) {First direct neutrino-mass
  measurement with sub-eV sensitivity}. arXiv e-prints arXiv:2105.08533.
  {\href{https://arxiv.org/abs/2105.08533}{{arXiv:2105.08533}}} {[hep-ex]}

\bibitem[{{Akitsu} et~al.(2019){Akitsu}, {Sugiyama}, and
  {Shiraishi}}]{Akitsu:2019}
{Akitsu} K, {Sugiyama} NS, {Shiraishi} M (2019) {Super-sample tidal modes on
  the celestial sphere}. \prd 100(10):103515.
  \doi{10.1103/PhysRevD.100.103515}.
  {\href{https://arxiv.org/abs/1907.10591}{{arXiv:1907.10591}}} {[astro-ph.CO]}

\bibitem[{{Akitsu} et~al.(2020){Akitsu}, {Li}, and {Okumura}}]{Akitsu2020}
{Akitsu} K, {Li} Y, {Okumura} T (2020) {Cosmological simulation in tides: power
  spectrum and halo shape responses, and shape assembly bias}. arXiv e-prints
  arXiv:2011.06584.
  {\href{https://arxiv.org/abs/2011.06584}{{arXiv:2011.06584}}} {[astro-ph.CO]}

\bibitem[{{Alam} et~al.(2017){Alam}, {Ata}, {Bailey}, {Beutler}, {Bizyaev},
  {Blazek}, {Bolton}, {Brownstein}, {Burden}, {Chuang}, {Comparat}, {Cuesta},
  {Dawson}, {Eisenstein}, {Escoffier}, {Gil-Mar{\'\i}n}, {Grieb}, {Hand}, {Ho},
  {Kinemuchi}, {Kirkby}, {Kitaura}, {Malanushenko}, {Malanushenko}, {Maraston},
  {McBride}, {Nichol}, {Olmstead}, {Oravetz}, {Padmanabhan},
  {Palanque-Delabrouille}, {Pan}, {Pellejero-Ibanez}, {Percival}, {Petitjean},
  {Prada}, {Price-Whelan}, {Reid}, {Rodr{\'\i}guez-Torres}, {Roe}, {Ross},
  {Ross}, {Rossi}, {Rubi{\~n}o-Mart{\'\i}n}, {Saito}, {Salazar-Albornoz},
  {Samushia}, {S{\'a}nchez}, {Satpathy}, {Schlegel}, {Schneider},
  {Sc{\'o}ccola}, {Seo}, {Sheldon}, {Simmons}, {Slosar}, {Strauss}, {Swanson},
  {Thomas}, {Tinker}, {Tojeiro}, {Maga{\~n}a}, {Vazquez}, {Verde}, {Wake},
  {Wang}, {Weinberg}, {White}, {Wood-Vasey}, {Y{\`e}che}, {Zehavi}, {Zhai}, and
  {Zhao}}]{Alam:2017}
{Alam} S, {Ata} M, {Bailey} S, {Beutler} F, {Bizyaev} D, {Blazek} JA, {Bolton}
  AS, {Brownstein} JR, {Burden} A, {Chuang} CH, {Comparat} J, {Cuesta} AJ,
  {Dawson} KS, {Eisenstein} DJ, {Escoffier} S, {Gil-Mar{\'\i}n} H, {Grieb} JN,
  {Hand} N, {Ho} S, {Kinemuchi} K, {Kirkby} D, {Kitaura} F, {Malanushenko} E,
  {Malanushenko} V, {Maraston} C, {McBride} CK, {Nichol} RC, {Olmstead} MD,
  {Oravetz} D, {Padmanabhan} N, {Palanque-Delabrouille} N, {Pan} K,
  {Pellejero-Ibanez} M, {Percival} WJ, {Petitjean} P, {Prada} F, {Price-Whelan}
  AM, {Reid} BA, {Rodr{\'\i}guez-Torres} SA, {Roe} NA, {Ross} AJ, {Ross} NP,
  {Rossi} G, {Rubi{\~n}o-Mart{\'\i}n} JA, {Saito} S, {Salazar-Albornoz} S,
  {Samushia} L, {S{\'a}nchez} AG, {Satpathy} S, {Schlegel} DJ, {Schneider} DP,
  {Sc{\'o}ccola} CG, {Seo} HJ, {Sheldon} ES, {Simmons} A, {Slosar} A, {Strauss}
  MA, {Swanson} MEC, {Thomas} D, {Tinker} JL, {Tojeiro} R, {Maga{\~n}a} MV,
  {Vazquez} JA, {Verde} L, {Wake} DA, {Wang} Y, {Weinberg} DH, {White} M,
  {Wood-Vasey} WM, {Y{\`e}che} C, {Zehavi} I, {Zhai} Z, {Zhao} GB (2017) {The
  clustering of galaxies in the completed SDSS-III Baryon Oscillation
  Spectroscopic Survey: cosmological analysis of the DR12 galaxy sample}.
  \mnras 470(3):2617--2652. \doi{10.1093/mnras/stx721}.
  {\href{https://arxiv.org/abs/1607.03155}{{arXiv:1607.03155}}} {[astro-ph.CO]}

\bibitem[{{Alam} et~al.(2020){Alam}, {Aviles}, {Bean}, {Cai}, {Cautun},
  {Cervantes-Cota}, {Cuesta-Lazaro}, {Chandrachani Devi}, {Eggemeier},
  {Fromenteau}, {Gonzalez-Morales}, {Halenka}, {He}, {Hellwing},
  {Hernandez-Aguayo}, {Ishak}, {Koyama}, {Li}, {de la Macorra}, {Menesses
  Rizo}, {Miller}, {Mueller}, {Niz}, {Ntelis}, {Rodriguez Otero}, {Sabiu},
  {Slepian}, {Stark}, {Valenzuela}, {Valogiannis}, {Vargas-Magana}, {Winther},
  {Zarrouk}, {Zhao}, and {Zheng}}]{Alam:2020}
{Alam} S, {Aviles} A, {Bean} R, {Cai} YC, {Cautun} M, {Cervantes-Cota} JL,
  {Cuesta-Lazaro} C, {Chandrachani Devi} N, {Eggemeier} A, {Fromenteau} S,
  {Gonzalez-Morales} AX, {Halenka} V, {He} Jh, {Hellwing} WA,
  {Hernandez-Aguayo} C, {Ishak} M, {Koyama} K, {Li} B, {de la Macorra} A,
  {Menesses Rizo} J, {Miller} C, {Mueller} EM, {Niz} G, {Ntelis} P, {Rodriguez
  Otero} M, {Sabiu} CG, {Slepian} Z, {Stark} A, {Valenzuela} O, {Valogiannis}
  G, {Vargas-Magana} M, {Winther} HA, {Zarrouk} P, {Zhao} GB, {Zheng} Y (2020)
  {Testing the theory of gravity with DESI: estimators, predictions and
  simulation requirements}. arXiv e-prints arXiv:2011.05771.
  {\href{https://arxiv.org/abs/2011.05771}{{arXiv:2011.05771}}} {[astro-ph.CO]}

\bibitem[{{Ali-Ha{\"\i}moud} and {Bird}(2013)}]{AliHamoud:2013}
{Ali-Ha{\"\i}moud} Y, {Bird} S (2013) {An efficient implementation of massive
  neutrinos in non-linear structure formation simulations}. \mnras
  428(4):3375--3389. \doi{10.1093/mnras/sts286}.
  {\href{https://arxiv.org/abs/1209.0461}{{arXiv:1209.0461}}} {[astro-ph.CO]}

\bibitem[{{Allys} et~al.(2020){Allys}, {Marchand}, {Cardoso},
  {Villaescusa-Navarro}, {Ho}, and {Mallat}}]{Allys:2020}
{Allys} E, {Marchand} T, {Cardoso} JF, {Villaescusa-Navarro} F, {Ho} S,
  {Mallat} S (2020) {New interpretable statistics for large-scale structure
  analysis and generation}. \prd 102(10):103506.
  \doi{10.1103/PhysRevD.102.103506}.
  {\href{https://arxiv.org/abs/2006.06298}{{arXiv:2006.06298}}} {[astro-ph.CO]}

\bibitem[{{Alonso}(2012)}]{Alonso:2012}
{Alonso} D (2012) {CUTE solutions for two-point correlation functions from
  large cosmological datasets}. arXiv e-prints arXiv:1210.1833.
  {\href{https://arxiv.org/abs/1210.1833}{{arXiv:1210.1833}}} {[astro-ph.IM]}

\bibitem[{{Alvarez} et~al.(2014){Alvarez}, {Baldauf}, {Bond}, {Dalal}, {de
  Putter}, {Dor{\'e}}, {Green}, {Hirata}, {Huang}, {Huterer}, {Jeong},
  {Johnson}, {Krause}, {Loverde}, {Meyers}, {Meerburg}, {Senatore}, {Shandera},
  {Silverstein}, {Slosar}, {Smith}, {Zaldarriaga}, {Assassi}, {Braden},
  {Hajian}, {Kobayashi}, {Stein}, and {van Engelen}}]{Alvarez:2014}
{Alvarez} M, {Baldauf} T, {Bond} JR, {Dalal} N, {de Putter} R, {Dor{\'e}} O,
  {Green} D, {Hirata} C, {Huang} Z, {Huterer} D, {Jeong} D, {Johnson} MC,
  {Krause} E, {Loverde} M, {Meyers} J, {Meerburg} PD, {Senatore} L, {Shandera}
  S, {Silverstein} E, {Slosar} A, {Smith} K, {Zaldarriaga} M, {Assassi} V,
  {Braden} J, {Hajian} A, {Kobayashi} T, {Stein} G, {van Engelen} A (2014)
  {Testing Inflation with Large Scale Structure: Connecting Hopes with
  Reality}. arXiv e-prints arXiv:1412.4671.
  {\href{https://arxiv.org/abs/1412.4671}{{arXiv:1412.4671}}} {[astro-ph.CO]}

\bibitem[{{Anderhalden} and {Diemand}(2013)}]{Anderhalen:2013}
{Anderhalden} D, {Diemand} J (2013) {Density profiles of CDM microhalos and
  their implications for annihilation boost factors}. \jcap 2013(4):009.
  \doi{10.1088/1475-7516/2013/04/009}.
  {\href{https://arxiv.org/abs/1302.0003}{{arXiv:1302.0003}}} {[astro-ph.CO]}

\bibitem[{{Anderson} et~al.(2019){Anderson}, {Pontzen}, {Font-Ribera},
  {Villaescusa-Navarro}, {Rogers}, and {Genel}}]{Anderson:2019}
{Anderson} L, {Pontzen} A, {Font-Ribera} A, {Villaescusa-Navarro} F, {Rogers}
  KK, {Genel} S (2019) {Cosmological Hydrodynamic Simulations with Suppressed
  Variance in the Ly{\ensuremath{\alpha}} Forest Power Spectrum}. \apj
  871(2):144. \doi{10.3847/1538-4357/aaf576}

\bibitem[{{Angulo} and {Hilbert}(2015)}]{Angulo:2015}
{Angulo} RE, {Hilbert} S (2015) {Cosmological constraints from the CFHTLenS
  shear measurements using a new, accurate, and flexible way of predicting
  non-linear mass clustering}. \mnras 448(1):364--375.
  \doi{10.1093/mnras/stv050}.
  {\href{https://arxiv.org/abs/1405.5888}{{arXiv:1405.5888}}} {[astro-ph.CO]}

\bibitem[{{Angulo} and {Pontzen}(2016)}]{Angulo:2016}
{Angulo} RE, {Pontzen} A (2016) {Cosmological N-body simulations with
  suppressed variance}. \mnras 462(1):L1--L5. \doi{10.1093/mnrasl/slw098}.
  {\href{https://arxiv.org/abs/1603.05253}{{arXiv:1603.05253}}} {[astro-ph.CO]}

\bibitem[{{Angulo} and {White}(2010)}]{Angulo:2010}
{Angulo} RE, {White} SDM (2010) {One simulation to fit them all - changing the
  background parameters of a cosmological N-body simulation}. \mnras
  405(1):143--154. \doi{10.1111/j.1365-2966.2010.16459.x}.
  {\href{https://arxiv.org/abs/0912.4277}{{arXiv:0912.4277}}} {[astro-ph.CO]}

\bibitem[{{Angulo} et~al.(2012){Angulo}, {Springel}, {White}, {Jenkins},
  {Baugh}, and {Frenk}}]{Angulo:2012}
{Angulo} RE, {Springel} V, {White} SDM, {Jenkins} A, {Baugh} CM, {Frenk} CS
  (2012) {Scaling relations for galaxy clusters in the Millennium-XXL
  simulation}. \mnras 426(3):2046--2062.
  \doi{10.1111/j.1365-2966.2012.21830.x}.
  {\href{https://arxiv.org/abs/1203.3216}{{arXiv:1203.3216}}} {[astro-ph.CO]}

\bibitem[{{Angulo} et~al.(2013{\natexlab{a}}){Angulo}, {Hahn}, and
  {Abel}}]{Angulo:2013a}
{Angulo} RE, {Hahn} O, {Abel} T (2013{\natexlab{a}}) {How closely do baryons
  follow dark matter on large scales?} \mnras 434:1756--1764.
  \doi{10.1093/mnras/stt1135}.
  {\href{https://arxiv.org/abs/1301.7426}{{arXiv:1301.7426}}}

\bibitem[{{Angulo} et~al.(2013{\natexlab{b}}){Angulo}, {Hahn}, and
  {Abel}}]{Angulo:2013b}
{Angulo} RE, {Hahn} O, {Abel} T (2013{\natexlab{b}}) {The warm dark matter halo
  mass function below the cut-off scale}. \mnras 434(4):3337--3347.
  \doi{10.1093/mnras/stt1246}.
  {\href{https://arxiv.org/abs/1304.2406}{{arXiv:1304.2406}}} {[astro-ph.CO]}

\bibitem[{{Angulo} et~al.(2014){Angulo}, {White}, {Springel}, and
  {Henriques}}]{Angulo:2014}
{Angulo} RE, {White} SDM, {Springel} V, {Henriques} B (2014) {Galaxy formation
  on the largest scales: the impact of astrophysics on the baryonic acoustic
  oscillation peak}. \mnras 442(3):2131--2144. \doi{10.1093/mnras/stu905}.
  {\href{https://arxiv.org/abs/1311.7100}{{arXiv:1311.7100}}} {[astro-ph.CO]}

\bibitem[{{Angulo} et~al.(2017){Angulo}, {Hahn}, {Ludlow}, and
  {Bonoli}}]{Angulo:2017}
{Angulo} RE, {Hahn} O, {Ludlow} AD, {Bonoli} S (2017) {Earth-mass haloes and
  the emergence of NFW density profiles}. \mnras 471(4):4687--4701.
  \doi{10.1093/mnras/stx1658}.
  {\href{https://arxiv.org/abs/1604.03131}{{arXiv:1604.03131}}} {[astro-ph.CO]}

\bibitem[{{Angulo} et~al.(2020){Angulo}, {Zennaro}, {Contreras}, {Aric{\`o}},
  {Pellejero-Iba{\~n}ez}, and {St{\"u}cker}}]{Angulo:2020}
{Angulo} RE, {Zennaro} M, {Contreras} S, {Aric{\`o}} G, {Pellejero-Iba{\~n}ez}
  M, {St{\"u}cker} J (2020) {The BACCO Simulation Project: Exploiting the full
  power of large-scale structure for cosmology}. arXiv e-prints
  arXiv:2004.06245.
  {\href{https://arxiv.org/abs/2004.06245}{{arXiv:2004.06245}}} {[astro-ph.CO]}

\bibitem[{{Appel}(1985)}]{Appel:1985}
{Appel} AW (1985) {An Efficient Program for Many-Body Simulation}. SIAM J Sci
  Statist Computing 6(1):85--103

\bibitem[{{Arag{\'o}n-Calvo} et~al.(2007{\natexlab{a}}){Arag{\'o}n-Calvo},
  {Jones}, {van de Weygaert}, and {van der Hulst}}]{Aragon-Calvo:2007}
{Arag{\'o}n-Calvo} MA, {Jones} BJT, {van de Weygaert} R, {van der Hulst} JM
  (2007{\natexlab{a}}) {The multiscale morphology filter: identifying and
  extracting spatial patterns in the galaxy distribution}. \aap
  474(1):315--338. \doi{10.1051/0004-6361:20077880}.
  {\href{https://arxiv.org/abs/0705.2072}{{arXiv:0705.2072}}} {[astro-ph]}

\bibitem[{{Arag{\'o}n-Calvo} et~al.(2007{\natexlab{b}}){Arag{\'o}n-Calvo}, {van
  de Weygaert}, {Jones}, and {van der Hulst}}]{Aragon-Calvo:2007b}
{Arag{\'o}n-Calvo} MA, {van de Weygaert} R, {Jones} BJT, {van der Hulst} JM
  (2007{\natexlab{b}}) {Spin Alignment of Dark Matter Halos in Filaments and
  Walls}. \apjl 655(1):L5--L8. \doi{10.1086/511633}.
  {\href{https://arxiv.org/abs/astro-ph/0610249}{{arXiv:astro-ph/0610249}}}
  {[astro-ph]}

\bibitem[{{Arag{\'o}n-Calvo} et~al.(2010){Arag{\'o}n-Calvo}, {Shandarin}, and
  {Szalay}}]{Aragon-Calvo:2010}
{Arag{\'o}n-Calvo} MA, {Shandarin} SF, {Szalay} A (2010) Geometry of the cosmic
  web: Minkowski functionals from the delaunay tessellation. In: 2010
  International Symposium on Voronoi Diagrams in Science and Engineering. pp
  235--243. \doi{10.1109/ISVD.2010.33}

\bibitem[{{Archidiacono} et~al.(2020){Archidiacono}, {Hannestad}, and
  {Lesgourgues}}]{Archidiacono:2020}
{Archidiacono} M, {Hannestad} S, {Lesgourgues} J (2020) {What will it take to
  measure individual neutrino mass states using cosmology?} \jcap 2020(9):021.
  \doi{10.1088/1475-7516/2020/09/021}.
  {\href{https://arxiv.org/abs/2003.03354}{{arXiv:2003.03354}}} {[astro-ph.CO]}

\bibitem[{{Aric{\`o}} et~al.(2019){Aric{\`o}}, {Angulo},
  {Hern{\'a}ndez-Monteagudo}, {Contreras}, {Zennaro}, {Pellejero-Iba{\~n}ez},
  and {Rosas-Guevara}}]{Arico:2019}
{Aric{\`o}} G, {Angulo} RE, {Hern{\'a}ndez-Monteagudo} C, {Contreras} S,
  {Zennaro} M, {Pellejero-Iba{\~n}ez} M, {Rosas-Guevara} Y (2019) {Modelling
  the large scale structure of the Universe as a function of cosmology and
  baryonic physics}. arXiv e-prints arXiv:1911.08471.
  {\href{https://arxiv.org/abs/1911.08471}{{arXiv:1911.08471}}} {[astro-ph.CO]}

\bibitem[{{Aric{\`o}} et~al.(2020{\natexlab{a}}){Aric{\`o}}, {Angulo},
  {Contreras}, {Ondaro-Mallea}, {Pellejero-Iba{\~n}ez}, and
  {Zennaro}}]{Arico:2020b}
{Aric{\`o}} G, {Angulo} RE, {Contreras} S, {Ondaro-Mallea} L,
  {Pellejero-Iba{\~n}ez} M, {Zennaro} M (2020{\natexlab{a}}) {The BACCO
  Simulation Project: A baryonification emulator with Neural Networks}. arXiv
  e-prints arXiv:2011.15018.
  {\href{https://arxiv.org/abs/2011.15018}{{arXiv:2011.15018}}} {[astro-ph.CO]}

\bibitem[{{Aric{\`o}} et~al.(2020{\natexlab{b}}){Aric{\`o}}, {Angulo},
  {Contreras}, {Ondaro-Mallea}, {Pellejero-Iba{\~n}ez}, and
  {Zennaro}}]{Arico:2020}
{Aric{\`o}} G, {Angulo} RE, {Contreras} S, {Ondaro-Mallea} L,
  {Pellejero-Iba{\~n}ez} M, {Zennaro} M (2020{\natexlab{b}}) {The BACCO
  Simulation Project: A baryonification emulator with Neural Networks}. arXiv
  e-prints arXiv:2011.15018.
  {\href{https://arxiv.org/abs/2011.15018}{{arXiv:2011.15018}}} {[astro-ph.CO]}

\bibitem[{{Aric{\`o}} et~al.(2020{\natexlab{c}}){Aric{\`o}}, {Angulo},
  {Hern{\'a}ndez-Monteagudo}, {Contreras}, and {Zennaro}}]{Arico:2020a}
{Aric{\`o}} G, {Angulo} RE, {Hern{\'a}ndez-Monteagudo} C, {Contreras} S,
  {Zennaro} M (2020{\natexlab{c}}) {Simultaneous modelling of matter power
  spectrum and bispectrum in the presence of baryons}. arXiv e-prints
  arXiv:2009.14225.
  {\href{https://arxiv.org/abs/2009.14225}{{arXiv:2009.14225}}} {[astro-ph.CO]}

\bibitem[{{Armengaud} et~al.(2017){Armengaud}, {Palanque-Delabrouille},
  {Y{\`e}che}, {Marsh}, and {Baur}}]{Armengaud:2017}
{Armengaud} E, {Palanque-Delabrouille} N, {Y{\`e}che} C, {Marsh} DJE, {Baur} J
  (2017) {Constraining the mass of light bosonic dark matter using SDSS
  Lyman-{\ensuremath{\alpha}} forest}. \mnras 471(4):4606--4614.
  \doi{10.1093/mnras/stx1870}.
  {\href{https://arxiv.org/abs/1703.09126}{{arXiv:1703.09126}}} {[astro-ph.CO]}

\bibitem[{{Arnold} et~al.(2019){Arnold}, {Leo}, and {Li}}]{Arnold:2019}
{Arnold} C, {Leo} M, {Li} B (2019) {Realistic simulations of galaxy formation
  in f(R) modified gravity}. Nature Astronomy 3:945--954.
  \doi{10.1038/s41550-019-0823-y}.
  {\href{https://arxiv.org/abs/1907.02977}{{arXiv:1907.02977}}} {[astro-ph.CO]}

\bibitem[{{Arnold} et~al.(2021){Arnold}, {Li}, {Giblin}, {Harnois-D{\'e}raps},
  and {Cai}}]{Arnold:2021}
{Arnold} C, {Li} B, {Giblin} B, {Harnois-D{\'e}raps} J, {Cai} YC (2021) {FORGE
  -- the f(R) gravity cosmic emulator project I: Introduction and matter power
  spectrum emulator}. arXiv e-prints arXiv:2109.04984.
  {\href{https://arxiv.org/abs/2109.04984}{{arXiv:2109.04984}}} {[astro-ph.CO]}

\bibitem[{{Ascasibar} and {Gottl{\"o}ber}(2008)}]{Ascasibar:2008}
{Ascasibar} Y, {Gottl{\"o}ber} S (2008) {The dynamical structure of dark matter
  haloes}. \mnras 386(4):2022--2030. \doi{10.1111/j.1365-2966.2008.13160.x}.
  {\href{https://arxiv.org/abs/0802.4348}{{arXiv:0802.4348}}} {[astro-ph]}

\bibitem[{{Asgari} et~al.(2021){Asgari}, {Lin}, {Joachimi}, {Giblin},
  {Heymans}, {Hildebrandt}, {Kannawadi}, {St{\"o}lzner}, {Tr{\"o}ster}, {van
  den Busch}, {Wright}, {Bilicki}, {Blake}, {de Jong}, {Dvornik}, {Erben},
  {Getman}, {Hoekstra}, {K{\"o}hlinger}, {Kuijken}, {Miller}, {Radovich},
  {Schneider}, {Shan}, and {Valentijn}}]{Asgari:2021}
{Asgari} M, {Lin} CA, {Joachimi} B, {Giblin} B, {Heymans} C, {Hildebrandt} H,
  {Kannawadi} A, {St{\"o}lzner} B, {Tr{\"o}ster} T, {van den Busch} JL,
  {Wright} AH, {Bilicki} M, {Blake} C, {de Jong} J, {Dvornik} A, {Erben} T,
  {Getman} F, {Hoekstra} H, {K{\"o}hlinger} F, {Kuijken} K, {Miller} L,
  {Radovich} M, {Schneider} P, {Shan} H, {Valentijn} E (2021) {KiDS-1000
  cosmology: Cosmic shear constraints and comparison between two point
  statistics}. \aap 645:A104. \doi{10.1051/0004-6361/202039070}.
  {\href{https://arxiv.org/abs/2007.15633}{{arXiv:2007.15633}}} {[astro-ph.CO]}

\bibitem[{{Assassi} et~al.(2015){Assassi}, {Baumann}, and
  {Schmidt}}]{Assassi:2015}
{Assassi} V, {Baumann} D, {Schmidt} F (2015) {Galaxy bias and primordial
  non-Gaussianity}. \jcap 2015(12):043. \doi{10.1088/1475-7516/2015/12/043}.
  {\href{https://arxiv.org/abs/1510.03723}{{arXiv:1510.03723}}} {[astro-ph.CO]}

\bibitem[{{Ata} et~al.(2015){Ata}, {Kitaura}, and {M{\"u}ller}}]{Ata:2015}
{Ata} M, {Kitaura} FS, {M{\"u}ller} V (2015) {Bayesian inference of cosmic
  density fields from non-linear, scale-dependent, and stochastic biased
  tracers}. \mnras 446(4):4250--4259. \doi{10.1093/mnras/stu2347}.
  {\href{https://arxiv.org/abs/1408.2566}{{arXiv:1408.2566}}} {[astro-ph.CO]}

\bibitem[{{Aubert} et~al.(2004){Aubert}, {Pichon}, and {Colombi}}]{Aubert:2004}
{Aubert} D, {Pichon} C, {Colombi} S (2004) {The origin and implications of dark
  matter anisotropic cosmic infall on $\approx L_{\star}$ haloes}. \mnras
  352(2):376--398. \doi{10.1111/j.1365-2966.2004.07883.x}.
  {\href{https://arxiv.org/abs/astro-ph/0402405}{{arXiv:astro-ph/0402405}}}
  {[astro-ph]}

\bibitem[{{Audren} et~al.(2014){Audren}, {Lesgourgues}, {Mangano}, {Serpico},
  and {Tram}}]{Audren:2014}
{Audren} B, {Lesgourgues} J, {Mangano} G, {Serpico} PD, {Tram} T (2014)
  {Strongest model-independent bound on the lifetime of Dark Matter}. \jcap
  2014(12):028. \doi{10.1088/1475-7516/2014/12/028}.
  {\href{https://arxiv.org/abs/1407.2418}{{arXiv:1407.2418}}} {[astro-ph.CO]}

\bibitem[{{Avila} et~al.(2015){Avila}, {Murray}, {Knebe}, {Power}, {Robotham},
  and {Garcia-Bellido}}]{Avila:2015}
{Avila} S, {Murray} SG, {Knebe} A, {Power} C, {Robotham} ASG, {Garcia-Bellido}
  J (2015) {HALOGEN: a tool for fast generation of mock halo catalogues}.
  \mnras 450(2):1856--1867. \doi{10.1093/mnras/stv711}.
  {\href{https://arxiv.org/abs/1412.5228}{{arXiv:1412.5228}}} {[astro-ph.CO]}

\bibitem[{{Avila-Reese} et~al.(2001){Avila-Reese}, {Col{\'\i}n}, {Valenzuela},
  {D'Onghia}, and {Firmani}}]{Avila-Reese:2001}
{Avila-Reese} V, {Col{\'\i}n} P, {Valenzuela} O, {D'Onghia} E, {Firmani} C
  (2001) {Formation and Structure of Halos in a Warm Dark Matter Cosmology}.
  \apj 559(2):516--530. \doi{10.1086/322411}.
  {\href{https://arxiv.org/abs/astro-ph/0010525}{{arXiv:astro-ph/0010525}}}
  {[astro-ph]}

\bibitem[{{Babichev} et~al.(2009){Babichev}, {Deffayet}, and
  {Ziour}}]{Babichev:2009}
{Babichev} E, {Deffayet} C, {Ziour} R (2009) {k-MOUFLAGE Gravity}.
  International Journal of Modern Physics D 18(14):2147--2154.
  \doi{10.1142/S0218271809016107}.
  {\href{https://arxiv.org/abs/0905.2943}{{arXiv:0905.2943}}} {[hep-th]}

\bibitem[{{Bagla}(2002)}]{Bagla:2002}
{Bagla} JS (2002) {TreePM: A Code for Cosmological N-Body Simulations}. J
  Astrophys Astron 23:185--196. \doi{10.1007/BF02702282}.
  {\href{https://arxiv.org/abs/astro-ph/9911025}{{arXiv:astro-ph/9911025}}}
  {[astro-ph]}

\bibitem[{{Bagla} and {Khandai}(2009)}]{Bagla:2009}
{Bagla} JS, {Khandai} N (2009) {The Adaptive TreePM: an adaptive resolution
  code for cosmological N-body simulations}. \mnras 396(4):2211--2227.
  \doi{10.1111/j.1365-2966.2009.14880.x}.
  {\href{https://arxiv.org/abs/0811.4228}{{arXiv:0811.4228}}} {[astro-ph]}

\bibitem[{{Bagla} and {Ray}(2003)}]{Bagla:2003}
{Bagla} JS, {Ray} S (2003) {Performance characteristics of TreePM codes}. \na
  8(7):665--677. \doi{10.1016/S1384-1076(03)00056-3}.
  {\href{https://arxiv.org/abs/astro-ph/0212129}{{arXiv:astro-ph/0212129}}}
  {[astro-ph]}

\bibitem[{{Baldauf} et~al.(2011){Baldauf}, {Seljak}, {Senatore}, and
  {Zaldarriaga}}]{Baldauf:2011}
{Baldauf} T, {Seljak} U, {Senatore} L, {Zaldarriaga} M (2011) {Galaxy bias and
  non-linear structure formation in general relativity}. \jcap 2011(10):031.
  \doi{10.1088/1475-7516/2011/10/031}.
  {\href{https://arxiv.org/abs/1106.5507}{{arXiv:1106.5507}}} {[astro-ph.CO]}

\bibitem[{{Baldauf} et~al.(2016){Baldauf}, {Seljak}, {Senatore}, and
  {Zaldarriaga}}]{Baldauf:2016}
{Baldauf} T, {Seljak} U, {Senatore} L, {Zaldarriaga} M (2016) {Linear response
  to long wavelength fluctuations using curvature simulations}. \jcap
  2016(9):007. \doi{10.1088/1475-7516/2016/09/007}.
  {\href{https://arxiv.org/abs/1511.01465}{{arXiv:1511.01465}}} {[astro-ph.CO]}

\bibitem[{{Baldi}(2012)}]{Baldi:2012}
{Baldi} M (2012) {Dark Energy simulations}. Physics of the Dark Universe
  1(1-2):162--193. \doi{10.1016/j.dark.2012.10.004}.
  {\href{https://arxiv.org/abs/1210.6650}{{arXiv:1210.6650}}} {[astro-ph.CO]}

\bibitem[{{Baldi} and {Villaescusa-Navarro}(2018)}]{Baldi:2018}
{Baldi} M, {Villaescusa-Navarro} F (2018) {Cosmic degeneracies - II. Structure
  formation in joint simulations of warm dark matter and f(R) gravity}. \mnras
  473(3):3226--3240. \doi{10.1093/mnras/stx2594}

\bibitem[{{Baldi} et~al.(2014){Baldi}, {Villaescusa-Navarro}, {Viel},
  {Puchwein}, {Springel}, and {Moscardini}}]{Baldi:2014}
{Baldi} M, {Villaescusa-Navarro} F, {Viel} M, {Puchwein} E, {Springel} V,
  {Moscardini} L (2014) {Cosmic degeneracies - I. Joint N-body simulations of
  modified gravity and massive neutrinos}. \mnras 440(1):75--88.
  \doi{10.1093/mnras/stu259}.
  {\href{https://arxiv.org/abs/1311.2588}{{arXiv:1311.2588}}} {[astro-ph.CO]}

\bibitem[{Ballenegger(2014)}]{Ballenegger:2014}
Ballenegger V (2014) Communication: On the origin of the surface term in the
  ewald formula. J Chem Phys 140(16):161102. \doi{10.1063/1.4872019}

\bibitem[{{Banerjee} and {Abel}(2021{\natexlab{a}})}]{Banerjee:2021b}
{Banerjee} A, {Abel} T (2021{\natexlab{a}}) {Cosmological cross-correlations
  and nearest neighbor distributions}. arXiv e-prints arXiv:2102.01184.
  {\href{https://arxiv.org/abs/2102.01184}{{arXiv:2102.01184}}} {[astro-ph.CO]}

\bibitem[{{Banerjee} and {Abel}(2021{\natexlab{b}})}]{Banerjee:2021a}
{Banerjee} A, {Abel} T (2021{\natexlab{b}}) {Nearest neighbour distributions:
  New statistical measures for cosmological clustering}. \mnras
  500(4):5479--5499. \doi{10.1093/mnras/staa3604}.
  {\href{https://arxiv.org/abs/2007.13342}{{arXiv:2007.13342}}} {[astro-ph.CO]}

\bibitem[{{Banerjee} and {Dalal}(2016)}]{Banerjee:2016}
{Banerjee} A, {Dalal} N (2016) {Simulating nonlinear cosmological structure
  formation with massive neutrinos}. \jcap 2016(11):015.
  \doi{10.1088/1475-7516/2016/11/015}.
  {\href{https://arxiv.org/abs/1606.06167}{{arXiv:1606.06167}}} {[astro-ph.CO]}

\bibitem[{{Banerjee} et~al.(2018){Banerjee}, {Powell}, {Abel}, and
  {Villaescusa-Navarro}}]{Banerjee:2018}
{Banerjee} A, {Powell} D, {Abel} T, {Villaescusa-Navarro} F (2018) {Reducing
  noise in cosmological N-body simulations with neutrinos}. \jcap 9:028.
  \doi{10.1088/1475-7516/2018/09/028}.
  {\href{https://arxiv.org/abs/1801.03906}{{arXiv:1801.03906}}}

\bibitem[{{Banerjee} et~al.(2020){Banerjee}, {Adhikari}, {Dalal}, {More}, and
  {Kravtsov}}]{Banerjee:2020}
{Banerjee} A, {Adhikari} S, {Dalal} N, {More} S, {Kravtsov} A (2020)
  {Signatures of self-interacting dark matter on cluster density profile and
  subhalo distributions}. \jcap 2020(2):024.
  \doi{10.1088/1475-7516/2020/02/024}.
  {\href{https://arxiv.org/abs/1906.12026}{{arXiv:1906.12026}}} {[astro-ph.CO]}

\bibitem[{{Banerjee} et~al.(2021){Banerjee}, {Kokron}, and
  {Abel}}]{Banerjee:2021}
{Banerjee} A, {Kokron} N, {Abel} T (2021) {Modeling Nearest Neighbor
  distributions of biased tracers using Hybrid Effective Field Theory}. arXiv
  e-prints arXiv:2107.10287.
  {\href{https://arxiv.org/abs/2107.10287}{{arXiv:2107.10287}}} {[astro-ph.CO]}

\bibitem[{{Bar} et~al.(2018){Bar}, {Blas}, {Blum}, and {Sibiryakov}}]{Bar:2018}
{Bar} N, {Blas} D, {Blum} K, {Sibiryakov} S (2018) {Galactic rotation curves
  versus ultralight dark matter: Implications of the soliton-host halo
  relation}. \prd 98(8):083027. \doi{10.1103/PhysRevD.98.083027}.
  {\href{https://arxiv.org/abs/1805.00122}{{arXiv:1805.00122}}} {[astro-ph.CO]}

\bibitem[{{Barnes} and {Hut}(1986)}]{BarnesHut:1986}
{Barnes} J, {Hut} P (1986) {A hierarchical O(N log N) force-calculation
  algorithm}. \nat 324(6096):446--449. \doi{10.1038/324446a0}

\bibitem[{{Barreira}(2020)}]{BarreiraPNG:2020b}
{Barreira} A (2020) {On the impact of galaxy bias uncertainties on primordial
  non-Gaussianity constraints}. \jcap 2020(12):031.
  \doi{10.1088/1475-7516/2020/12/031}.
  {\href{https://arxiv.org/abs/2009.06622}{{arXiv:2009.06622}}} {[astro-ph.CO]}

\bibitem[{{Barreira}(2021)}]{Barreira:2021}
{Barreira} A (2021) {Predictions for local PNG bias in the galaxy power
  spectrum and bispectrum and the consequences for $f_{\rm NL}$ constraints}.
  arXiv e-prints arXiv:2107.06887.
  {\href{https://arxiv.org/abs/2107.06887}{{arXiv:2107.06887}}} {[astro-ph.CO]}

\bibitem[{{Barreira} et~al.(2015){Barreira}, {Bose}, and {Li}}]{Barreira:2015}
{Barreira} A, {Bose} S, {Li} B (2015) {Speeding up N-body simulations of
  modified gravity: Vainshtein screening models}. \jcap 2015(12):059.
  \doi{10.1088/1475-7516/2015/12/059}.
  {\href{https://arxiv.org/abs/1511.08200}{{arXiv:1511.08200}}} {[astro-ph.CO]}

\bibitem[{{Barreira} et~al.(2016){Barreira}, {S{\'a}nchez}, and
  {Schmidt}}]{Barreira:2016}
{Barreira} A, {S{\'a}nchez} AG, {Schmidt} F (2016) {Validating estimates of the
  growth rate of structure with modified gravity simulations}. \prd
  94(8):084022. \doi{10.1103/PhysRevD.94.084022}.
  {\href{https://arxiv.org/abs/1605.03965}{{arXiv:1605.03965}}} {[astro-ph.CO]}

\bibitem[{{Barreira} et~al.(2018){Barreira}, {Krause}, and
  {Schmidt}}]{Barreira:2018}
{Barreira} A, {Krause} E, {Schmidt} F (2018) {Accurate cosmic shear errors: do
  we need ensembles of simulations?} \jcap 2018(10):053.
  \doi{10.1088/1475-7516/2018/10/053}.
  {\href{https://arxiv.org/abs/1807.04266}{{arXiv:1807.04266}}} {[astro-ph.CO]}

\bibitem[{{Barreira} et~al.(2019){Barreira}, {Nelson}, {Pillepich}, {Springel},
  {Schmidt}, {Pakmor}, {Hernquist}, and {Vogelsberger}}]{Barreira:2019}
{Barreira} A, {Nelson} D, {Pillepich} A, {Springel} V, {Schmidt} F, {Pakmor} R,
  {Hernquist} L, {Vogelsberger} M (2019) {Separate Universe simulations with
  IllustrisTNG: baryonic effects on power spectrum responses and higher-order
  statistics}. \mnras 488:2079--2092. \doi{10.1093/mnras/stz1807}.
  {\href{https://arxiv.org/abs/1904.02070}{{arXiv:1904.02070}}}

\bibitem[{{Barreira} et~al.(2020{\natexlab{a}}){Barreira}, {Cabass}, {Lozanov},
  and {Schmidt}}]{BarreiraCabass:2020}
{Barreira} A, {Cabass} G, {Lozanov} KD, {Schmidt} F (2020{\natexlab{a}})
  {Compensated isocurvature perturbations in the galaxy power spectrum}. \jcap
  2020(7):049. \doi{10.1088/1475-7516/2020/07/049}.
  {\href{https://arxiv.org/abs/2002.12931}{{arXiv:2002.12931}}} {[astro-ph.CO]}

\bibitem[{{Barreira} et~al.(2020{\natexlab{b}}){Barreira}, {Cabass}, {Schmidt},
  {Pillepich}, and {Nelson}}]{BarreiraPNG:2020a}
{Barreira} A, {Cabass} G, {Schmidt} F, {Pillepich} A, {Nelson} D
  (2020{\natexlab{b}}) {Galaxy bias and primordial non-Gaussianity: insights
  from galaxy formation simulations with IllustrisTNG}. \jcap 2020(12):013.
  \doi{10.1088/1475-7516/2020/12/013}.
  {\href{https://arxiv.org/abs/2006.09368}{{arXiv:2006.09368}}} {[astro-ph.CO]}

\bibitem[{{Bartolo} et~al.(2004){Bartolo}, {Komatsu}, {Matarrese}, and
  {Riotto}}]{Bartolo:2004}
{Bartolo} N, {Komatsu} E, {Matarrese} S, {Riotto} A (2004) {Non-Gaussianity
  from inflation: theory and observations}. \physrep 402:103--266.
  \doi{10.1016/j.physrep.2004.08.022}.
  {\href{https://arxiv.org/abs/astro-ph/0406398}{{astro-ph/0406398}}}

\bibitem[{Bartolo et~al.(2007)Bartolo, Matarrese, and Riotto}]{Bartolo:2007}
Bartolo N, Matarrese S, Riotto A (2007) Course 5 - cosmic microwave background
  anisotropies up to second order. In: Bernardeau F, Grojean C, Dalibard J
  (eds) Particle Physics and Cosmology: The Fabric of Spacetime. Les Houches,
  vol~86. Elsevier, pp 233 -- 285.
  \doi{https://doi.org/10.1016/S0924-8099(07)80031-X}

\bibitem[{{Bastos de Senna Nascimento} and {Loverde}(2021)}]{Bastos:2021}
{Bastos de Senna Nascimento} C, {Loverde} M (2021) {Neutrinos in N-body
  simulations}. arXiv e-prints arXiv:2102.05690.
  {\href{https://arxiv.org/abs/2102.05690}{{arXiv:2102.05690}}} {[astro-ph.CO]}

\bibitem[{{Baugh}(2006)}]{Baugh:2006}
{Baugh} CM (2006) {A primer on hierarchical galaxy formation: the
  semi-analytical approach}. Reports on Progress in Physics 69(12):3101--3156.
  \doi{10.1088/0034-4885/69/12/R02}.
  {\href{https://arxiv.org/abs/astro-ph/0610031}{{arXiv:astro-ph/0610031}}}
  {[astro-ph]}

\bibitem[{{Baugh} et~al.(2005){Baugh}, {Lacey}, {Frenk}, {Granato}, {Silva},
  {Bressan}, {Benson}, and {Cole}}]{Baugh:2005}
{Baugh} CM, {Lacey} CG, {Frenk} CS, {Granato} GL, {Silva} L, {Bressan} A,
  {Benson} AJ, {Cole} S (2005) {Can the faint submillimetre galaxies be
  explained in the {\ensuremath{\Lambda}} cold dark matter model?} \mnras
  356(3):1191--1200. \doi{10.1111/j.1365-2966.2004.08553.x}.
  {\href{https://arxiv.org/abs/astro-ph/0406069}{{arXiv:astro-ph/0406069}}}
  {[astro-ph]}

\bibitem[{{Baumann} et~al.(2012){Baumann}, {Nicolis}, {Senatore}, and
  {Zaldarriaga}}]{Baumann:2012}
{Baumann} D, {Nicolis} A, {Senatore} L, {Zaldarriaga} M (2012) {Cosmological
  non-linearities as an effective fluid}. \jcap 2012(7):051.
  \doi{10.1088/1475-7516/2012/07/051}.
  {\href{https://arxiv.org/abs/1004.2488}{{arXiv:1004.2488}}} {[astro-ph.CO]}

\bibitem[{{Baumann} et~al.(2013){Baumann}, {Ferraro}, {Green}, and
  {Smith}}]{Baumann:2013}
{Baumann} D, {Ferraro} S, {Green} D, {Smith} KM (2013) {Stochastic bias from
  non-Gaussian initial conditions}. \jcap 2013(5):001.
  \doi{10.1088/1475-7516/2013/05/001}.
  {\href{https://arxiv.org/abs/1209.2173}{{arXiv:1209.2173}}} {[astro-ph.CO]}

\bibitem[{{Baushev}(2015)}]{Baushev:2015}
{Baushev} AN (2015) {The real and apparent convergence of N-body simulations of
  the dark matter structures: Is the Navarro-Frenk-White profile real?}
  Astroparticle Physics 62:47--53. \doi{10.1016/j.astropartphys.2014.07.012}.
  {\href{https://arxiv.org/abs/1312.0314}{{arXiv:1312.0314}}} {[astro-ph.CO]}

\bibitem[{{Bayer} et~al.(2020){Bayer}, {Banerjee}, and {Feng}}]{Bayer:2020}
{Bayer} AE, {Banerjee} A, {Feng} Y (2020) {A fast particle-mesh simulation of
  non-linear cosmological structure formation with massive neutrinos}. arXiv
  e-prints arXiv:2007.13394.
  {\href{https://arxiv.org/abs/2007.13394}{{arXiv:2007.13394}}} {[astro-ph.CO]}

\bibitem[{{Behroozi} et~al.(2015){Behroozi}, {Knebe}, {Pearce}, {Elahi}, {Han},
  {Lux}, {Mao}, {Muldrew}, {Potter}, and {Srisawat}}]{Behroozi:2015}
{Behroozi} P, {Knebe} A, {Pearce} FR, {Elahi} P, {Han} J, {Lux} H, {Mao} YY,
  {Muldrew} SI, {Potter} D, {Srisawat} C (2015) {Major mergers going Notts:
  challenges for modern halo finders}. \mnras 454(3):3020--3029.
  \doi{10.1093/mnras/stv2046}.
  {\href{https://arxiv.org/abs/1506.01405}{{arXiv:1506.01405}}} {[astro-ph.CO]}

\bibitem[{{Behroozi} et~al.(2019){Behroozi}, {Wechsler}, {Hearin}, and
  {Conroy}}]{Behroozi:2019}
{Behroozi} P, {Wechsler} RH, {Hearin} AP, {Conroy} C (2019) {UNIVERSEMACHINE:
  The correlation between galaxy growth and dark matter halo assembly from z =
  0-10}. \mnras 488(3):3143--3194. \doi{10.1093/mnras/stz1182}.
  {\href{https://arxiv.org/abs/1806.07893}{{arXiv:1806.07893}}} {[astro-ph.GA]}

\bibitem[{{Behroozi} et~al.(2013{\natexlab{a}}){Behroozi}, {Wechsler}, and
  {Conroy}}]{Behroozi:2013c}
{Behroozi} PS, {Wechsler} RH, {Conroy} C (2013{\natexlab{a}}) {The Average Star
  Formation Histories of Galaxies in Dark Matter Halos from z = 0-8}. \apj
  770(1):57. \doi{10.1088/0004-637X/770/1/57}.
  {\href{https://arxiv.org/abs/1207.6105}{{arXiv:1207.6105}}} {[astro-ph.CO]}

\bibitem[{{Behroozi} et~al.(2013{\natexlab{b}}){Behroozi}, {Wechsler}, and
  {Wu}}]{Behroozi:2013a}
{Behroozi} PS, {Wechsler} RH, {Wu} HY (2013{\natexlab{b}}) {The ROCKSTAR
  Phase-space Temporal Halo Finder and the Velocity Offsets of Cluster Cores}.
  \apj 762(2):109. \doi{10.1088/0004-637X/762/2/109}.
  {\href{https://arxiv.org/abs/1110.4372}{{arXiv:1110.4372}}} {[astro-ph.CO]}

\bibitem[{{Bel} et~al.(2019){Bel}, {Pezzotta}, {Carbone}, {Sefusatti}, and
  {Guzzo}}]{Bel:2019}
{Bel} J, {Pezzotta} A, {Carbone} C, {Sefusatti} E, {Guzzo} L (2019) {Accurate
  fitting functions for peculiar velocity spectra in standard and
  massive-neutrino cosmologies}. \aap 622:A109.
  \doi{10.1051/0004-6361/201834513}.
  {\href{https://arxiv.org/abs/1809.09338}{{arXiv:1809.09338}}} {[astro-ph.CO]}

\bibitem[{{Belgacem} et~al.(2019){Belgacem}, {Calcagni}, {Crisostomi},
  {Dalang}, {Dirian}, {Ezquiaga}, {Fasiello}, {Foffa}, {Ganz},
  {Garc{\'\i}a-Bellido}, {Lombriser}, {Maggiore}, {Tamanini}, {Tasinato},
  {Zumalac{\'a}rregui}, {Barausse}, {Bartolo}, {Bertacca}, {Klein},
  {Matarrese}, and {Sakellariadou}}]{Belgacem:2019}
{Belgacem} E, {Calcagni} G, {Crisostomi} M, {Dalang} C, {Dirian} Y, {Ezquiaga}
  JM, {Fasiello} M, {Foffa} S, {Ganz} A, {Garc{\'\i}a-Bellido} J, {Lombriser}
  L, {Maggiore} M, {Tamanini} N, {Tasinato} G, {Zumalac{\'a}rregui} M,
  {Barausse} E, {Bartolo} N, {Bertacca} D, {Klein} A, {Matarrese} S,
  {Sakellariadou} M (2019) {Testing modified gravity at cosmological distances
  with LISA standard sirens}. \jcap 2019(7):024.
  \doi{10.1088/1475-7516/2019/07/024}.
  {\href{https://arxiv.org/abs/1906.01593}{{arXiv:1906.01593}}} {[astro-ph.CO]}

\bibitem[{{Benson}(2012)}]{BensonGalacticus:2012}
{Benson} AJ (2012) {G ALACTICUS: A semi-analytic model of galaxy formation}.
  \na 17(2):175--197. \doi{10.1016/j.newast.2011.07.004}.
  {\href{https://arxiv.org/abs/1008.1786}{{arXiv:1008.1786}}} {[astro-ph.CO]}

\bibitem[{{Berezinsky} et~al.(2003){Berezinsky}, {Dokuchaev}, and
  {Eroshenko}}]{Berezinsky:2003}
{Berezinsky} V, {Dokuchaev} V, {Eroshenko} Y (2003) {Small-scale clumps in the
  galactic halo and dark matter annihilation}. \prd 68(10):103003.
  \doi{10.1103/PhysRevD.68.103003}.
  {\href{https://arxiv.org/abs/astro-ph/0301551}{{arXiv:astro-ph/0301551}}}
  {[astro-ph]}

\bibitem[{{Berger} and {Stein}(2019)}]{BergerStein:2019}
{Berger} P, {Stein} G (2019) {A volumetric deep Convolutional Neural Network
  for simulation of mock dark matter halo catalogues}. \mnras
  482(3):2861--2871. \doi{10.1093/mnras/sty2949}.
  {\href{https://arxiv.org/abs/1805.04537}{{arXiv:1805.04537}}} {[astro-ph.CO]}

\bibitem[{{Bernardeau}(1994)}]{Bernardeau:1994}
{Bernardeau} F (1994) {The Nonlinear Evolution of Rare Events}. \apj 427:51.
  \doi{10.1086/174121}.
  {\href{https://arxiv.org/abs/astro-ph/9311066}{{arXiv:astro-ph/9311066}}}
  {[astro-ph]}

\bibitem[{{Bernardeau} et~al.(2002){Bernardeau}, {Colombi}, {Gazta{\~n}aga},
  and {Scoccimarro}}]{Bernardeau:2002}
{Bernardeau} F, {Colombi} S, {Gazta{\~n}aga} E, {Scoccimarro} R (2002)
  {Large-scale structure of the Universe and cosmological perturbation theory}.
  \physrep 367(1-3):1--248. \doi{10.1016/S0370-1573(02)00135-7}.
  {\href{https://arxiv.org/abs/astro-ph/0112551}{{arXiv:astro-ph/0112551}}}
  {[astro-ph]}

\bibitem[{{Bernardini} et~al.(2020){Bernardini}, {Mayer}, {Reed}, and
  {Feldmann}}]{Bernardini:2020}
{Bernardini} M, {Mayer} L, {Reed} D, {Feldmann} R (2020) {Predicting dark
  matter halo formation in N-body simulations with deep regression networks}.
  \mnras 496(4):5116--5125. \doi{10.1093/mnras/staa1911}.
  {\href{https://arxiv.org/abs/1912.04299}{{arXiv:1912.04299}}} {[astro-ph.CO]}

\bibitem[{{Bernyk} et~al.(2016){Bernyk}, {Croton}, {Tonini}, {Hodkinson},
  {Hassan}, {Garel}, {Duffy}, {Mutch}, {Poole}, and {Hegarty}}]{Bernyk:2016}
{Bernyk} M, {Croton} DJ, {Tonini} C, {Hodkinson} L, {Hassan} AH, {Garel} T,
  {Duffy} AR, {Mutch} SJ, {Poole} GB, {Hegarty} S (2016) {The Theoretical
  Astrophysical Observatory: Cloud-based Mock Galaxy Catalogs}. \apjs 223(1):9.
  \doi{10.3847/0067-0049/223/1/9}.
  {\href{https://arxiv.org/abs/1403.5270}{{arXiv:1403.5270}}} {[astro-ph.GA]}

\bibitem[{{Bertone}(2010)}]{Bertone:2010}
{Bertone} G (2010) {The moment of truth for WIMP dark matter}. \nat
  468(7322):389--393. \doi{10.1038/nature09509}.
  {\href{https://arxiv.org/abs/1011.3532}{{arXiv:1011.3532}}} {[astro-ph.CO]}

\bibitem[{{Bertone} and {Tait}(2018)}]{Bertone:2018}
{Bertone} G, {Tait} TMP (2018) {A new era in the search for dark matter}. \nat
  562(7725):51--56. \doi{10.1038/s41586-018-0542-z}.
  {\href{https://arxiv.org/abs/1810.01668}{{arXiv:1810.01668}}} {[astro-ph.CO]}

\bibitem[{{Bertone} et~al.(2004){Bertone}, {Hooper}, and {Silk}}]{Bertone:2004}
{Bertone} G, {Hooper} D, {Silk} J (2004) {Particle dark matter: evidence,
  candidates and constraints.} \physrep 405:279--390

\bibitem[{{Bertschinger}(1985)}]{Bertschinger:1985}
{Bertschinger} E (1985) {Self-similar secondary infall and accretion in an
  Einstein-de Sitter universe}. \apjs 58:39--65. \doi{10.1086/191028}

\bibitem[{{Bertschinger}(2001)}]{Bertschinger:2001}
{Bertschinger} E (2001) {Multiscale Gaussian Random Fields and Their
  Application to Cosmological Simulations}. \apjs 137(1):1--20.
  \doi{10.1086/322526}.
  {\href{https://arxiv.org/abs/astro-ph/0103301}{{arXiv:astro-ph/0103301}}}
  {[astro-ph]}

\bibitem[{{Betoule} et~al.(2014){Betoule}, {Kessler}, {Guy}, {Mosher},
  {Hardin}, {Biswas}, {Astier}, {El-Hage}, {Konig}, {Kuhlmann}, {Marriner},
  {Pain}, {Regnault}, {Balland}, {Bassett}, {Brown}, {Campbell}, {Carlberg},
  {Cellier-Holzem}, {Cinabro}, {Conley}, {D'Andrea}, {DePoy}, {Doi}, {Ellis},
  {Fabbro}, {Filippenko}, {Foley}, {Frieman}, {Fouchez}, {Galbany}, {Goobar},
  {Gupta}, {Hill}, {Hlozek}, {Hogan}, {Hook}, {Howell}, {Jha}, {Le Guillou},
  {Leloudas}, {Lidman}, {Marshall}, {M{\"o}ller}, {Mour{\~a}o}, {Neveu},
  {Nichol}, {Olmstead}, {Palanque-Delabrouille}, {Perlmutter}, {Prieto},
  {Pritchet}, {Richmond}, {Riess}, {Ruhlmann-Kleider}, {Sako}, {Schahmaneche},
  {Schneider}, {Smith}, {Sollerman}, {Sullivan}, {Walton}, and
  {Wheeler}}]{Betoule:2014}
{Betoule} M, {Kessler} R, {Guy} J, {Mosher} J, {Hardin} D, {Biswas} R, {Astier}
  P, {El-Hage} P, {Konig} M, {Kuhlmann} S, {Marriner} J, {Pain} R, {Regnault}
  N, {Balland} C, {Bassett} BA, {Brown} PJ, {Campbell} H, {Carlberg} RG,
  {Cellier-Holzem} F, {Cinabro} D, {Conley} A, {D'Andrea} CB, {DePoy} DL, {Doi}
  M, {Ellis} RS, {Fabbro} S, {Filippenko} AV, {Foley} RJ, {Frieman} JA,
  {Fouchez} D, {Galbany} L, {Goobar} A, {Gupta} RR, {Hill} GJ, {Hlozek} R,
  {Hogan} CJ, {Hook} IM, {Howell} DA, {Jha} SW, {Le Guillou} L, {Leloudas} G,
  {Lidman} C, {Marshall} JL, {M{\"o}ller} A, {Mour{\~a}o} AM, {Neveu} J,
  {Nichol} R, {Olmstead} MD, {Palanque-Delabrouille} N, {Perlmutter} S,
  {Prieto} JL, {Pritchet} CJ, {Richmond} M, {Riess} AG, {Ruhlmann-Kleider} V,
  {Sako} M, {Schahmaneche} K, {Schneider} DP, {Smith} M, {Sollerman} J,
  {Sullivan} M, {Walton} NA, {Wheeler} CJ (2014) {Improved cosmological
  constraints from a joint analysis of the SDSS-II and SNLS supernova samples}.
  \aap 568:A22. \doi{10.1051/0004-6361/201423413}.
  {\href{https://arxiv.org/abs/1401.4064}{{arXiv:1401.4064}}} {[astro-ph.CO]}

\bibitem[{{Betti} et~al.(2019){Betti}, {Biasotti}, {Bosc{\'a}}, {Calle},
  {Canci}, {Cavoto}, {Chang}, {Cocco}, {Colijn}, {Conrad}, {D'Ambrosio}, {De
  Groot}, {de Salas}, {Faverzani}, {Ferella}, {Ferri}, {Garcia-Abia},
  {Garc{\'\i}a-Cort{\'e}s}, {Garcia Gomez-Tejedor}, {Gariazzo}, {Gatti},
  {Gentile}, {Giachero}, {Gudmundsson}, {Hochberg}, {Kahn}, {Kievsky},
  {Lisanti}, {Mancini-Terracciano}, {Mangano}, {Marcucci}, {Mariani},
  {Mart{\'\i}nez}, {Messina}, {Molinero-Vela}, {Monticone}, {Moro{\~n}o},
  {Nucciotti}, {Pandolfi}, {Parlati}, {Pastor}, {Pedr{\'o}s}, {P{\'e}rez de los
  Heros}, {Pisanti}, {Polosa}, {Puiu}, {Rago}, {Raitses}, {Rajteri}, {Rossi},
  {Rucandio}, {Santorelli}, {Schaeffner}, {Tully}, {Viviani}, {Zhao}, and
  {Zurek}}]{Betti:2019}
{Betti} MG, {Biasotti} M, {Bosc{\'a}} A, {Calle} F, {Canci} N, {Cavoto} G,
  {Chang} C, {Cocco} AG, {Colijn} AP, {Conrad} J, {D'Ambrosio} N, {De Groot} N,
  {de Salas} PF, {Faverzani} M, {Ferella} A, {Ferri} E, {Garcia-Abia} P,
  {Garc{\'\i}a-Cort{\'e}s} I, {Garcia Gomez-Tejedor} G, {Gariazzo} S, {Gatti}
  F, {Gentile} C, {Giachero} A, {Gudmundsson} JE, {Hochberg} Y, {Kahn} Y,
  {Kievsky} A, {Lisanti} M, {Mancini-Terracciano} C, {Mangano} G, {Marcucci}
  LE, {Mariani} C, {Mart{\'\i}nez} J, {Messina} M, {Molinero-Vela} A,
  {Monticone} E, {Moro{\~n}o} A, {Nucciotti} A, {Pandolfi} F, {Parlati} S,
  {Pastor} S, {Pedr{\'o}s} J, {P{\'e}rez de los Heros} C, {Pisanti} O, {Polosa}
  AD, {Puiu} A, {Rago} I, {Raitses} Y, {Rajteri} M, {Rossi} N, {Rucandio} I,
  {Santorelli} R, {Schaeffner} K, {Tully} CG, {Viviani} M, {Zhao} F, {Zurek} KM
  (2019) {Neutrino physics with the PTOLEMY project: active neutrino properties
  and the light sterile case}. \jcap 2019(7):047.
  \doi{10.1088/1475-7516/2019/07/047}.
  {\href{https://arxiv.org/abs/1902.05508}{{arXiv:1902.05508}}} {[astro-ph.CO]}

\bibitem[{{Bhattacharya} et~al.(2011){Bhattacharya}, {Heitmann}, {White},
  {Luki{\'c}}, {Wagner}, and {Habib}}]{Bhattacharya:2011}
{Bhattacharya} S, {Heitmann} K, {White} M, {Luki{\'c}} Z, {Wagner} C, {Habib} S
  (2011) {Mass Function Predictions Beyond {\ensuremath{\Lambda}}CDM}. \apj
  732(2):122. \doi{10.1088/0004-637X/732/2/122}.
  {\href{https://arxiv.org/abs/1005.2239}{{arXiv:1005.2239}}} {[astro-ph.CO]}

\bibitem[{{Bird} et~al.(2012){Bird}, {Viel}, and {Haehnelt}}]{Bird:2012}
{Bird} S, {Viel} M, {Haehnelt} MG (2012) {Massive neutrinos and the non-linear
  matter power spectrum}. \mnras 420(3):2551--2561.
  \doi{10.1111/j.1365-2966.2011.20222.x}.
  {\href{https://arxiv.org/abs/1109.4416}{{arXiv:1109.4416}}} {[astro-ph.CO]}

\bibitem[{{Bird} et~al.(2018){Bird}, {Ali-Ha{\"\i}moud}, {Feng}, and
  {Liu}}]{Bird2018}
{Bird} S, {Ali-Ha{\"\i}moud} Y, {Feng} Y, {Liu} J (2018) {An efficient and
  accurate hybrid method for simulating non-linear neutrino structure}. \mnras
  481(2):1486--1500. \doi{10.1093/mnras/sty2376}.
  {\href{https://arxiv.org/abs/1803.09854}{{arXiv:1803.09854}}} {[astro-ph.CO]}

\bibitem[{{Bird} et~al.(2020){Bird}, {Feng}, {Pedersen}, and
  {Font-Ribera}}]{Bird:2020}
{Bird} S, {Feng} Y, {Pedersen} C, {Font-Ribera} A (2020) {More accurate
  simulations with separate initial conditions for baryons and dark matter}.
  \jcap 2020(6):002. \doi{10.1088/1475-7516/2020/06/002}.
  {\href{https://arxiv.org/abs/2002.00015}{{arXiv:2002.00015}}} {[astro-ph.CO]}

\bibitem[{{Blaizot} et~al.(2005){Blaizot}, {Wadadekar}, {Guiderdoni},
  {Colombi}, {Bertin}, {Bouchet}, {Devriendt}, and {Hatton}}]{Blaizot:2005}
{Blaizot} J, {Wadadekar} Y, {Guiderdoni} B, {Colombi} ST, {Bertin} E, {Bouchet}
  FR, {Devriendt} JEG, {Hatton} S (2005) {MoMaF: the Mock Map Facility}. \mnras
  360(1):159--175. \doi{10.1111/j.1365-2966.2005.09019.x}.
  {\href{https://arxiv.org/abs/astro-ph/0309305}{{arXiv:astro-ph/0309305}}}
  {[astro-ph]}

\bibitem[{{Blake} et~al.(2020){Blake}, {Amon}, {Asgari}, {Bilicki}, {Dvornik},
  {Erben}, {Giblin}, {Glazebrook}, {Heymans}, {Hildebrandt}, {Joachimi},
  {Joudaki}, {Kannawadi}, {Kuijken}, {Lidman}, {Parkinson}, {Shan},
  {Tr{\"o}ster}, {van den Busch}, {Wolf}, and {Wright}}]{Blake:2020}
{Blake} C, {Amon} A, {Asgari} M, {Bilicki} M, {Dvornik} A, {Erben} T, {Giblin}
  B, {Glazebrook} K, {Heymans} C, {Hildebrandt} H, {Joachimi} B, {Joudaki} S,
  {Kannawadi} A, {Kuijken} K, {Lidman} C, {Parkinson} D, {Shan} H,
  {Tr{\"o}ster} T, {van den Busch} JL, {Wolf} C, {Wright} AH (2020) {Testing
  gravity using galaxy-galaxy lensing and clustering amplitudes in KiDS-1000,
  BOSS, and 2dFLenS}. \aap 642:A158. \doi{10.1051/0004-6361/202038505}.
  {\href{https://arxiv.org/abs/2005.14351}{{arXiv:2005.14351}}} {[astro-ph.CO]}

\bibitem[{{Blanes} and {Casas}(2016)}]{Blanes:2016}
{Blanes} S, {Casas} F (2016) A Concise Introduction to Geometric Numerical
  Integration, 1st edn. Chapman and Hall/CRC Press, New York

\bibitem[{{Blanes} et~al.(2009){Blanes}, {Casas}, {Oteo}, and
  {Ros}}]{Blanes:2009}
{Blanes} S, {Casas} F, {Oteo} JA, {Ros} J (2009) {The Magnus expansion and some
  of its applications}. \physrep 470:151--238.
  \doi{10.1016/j.physrep.2008.11.001}.
  {\href{https://arxiv.org/abs/0810.5488}{{arXiv:0810.5488}}} {[math-ph]}

\bibitem[{{Blas} et~al.(2011){Blas}, {Lesgourgues}, and {Tram}}]{Blas:2011}
{Blas} D, {Lesgourgues} J, {Tram} T (2011) {The Cosmic Linear Anisotropy
  Solving System (CLASS). Part II: Approximation schemes}. \jcap 2011(7):034.
  \doi{10.1088/1475-7516/2011/07/034}.
  {\href{https://arxiv.org/abs/1104.2933}{{arXiv:1104.2933}}} {[astro-ph.CO]}

\bibitem[{{Blas} et~al.(2017){Blas}, {Nacir}, and {Sibiryakov}}]{Blas:2017}
{Blas} D, {Nacir} DL, {Sibiryakov} S (2017) {Ultralight Dark Matter Resonates
  with Binary Pulsars}. \prl 118(26):261102.
  \doi{10.1103/PhysRevLett.118.261102}.
  {\href{https://arxiv.org/abs/1612.06789}{{arXiv:1612.06789}}} {[hep-ph]}

\bibitem[{{Blot} et~al.(2019){Blot}, {Crocce}, {Sefusatti}, {Lippich},
  {S{\'a}nchez}, {Colavincenzo}, {Monaco}, {Alvarez}, {Agrawal}, {Avila},
  {Balaguera-Antol{\'\i}nez}, {Bond}, {Codis}, {Dalla Vecchia}, {Dorta},
  {Fosalba}, {Izard}, {Kitaura}, {Pellejero-Ibanez}, {Stein}, {Vakili}, and
  {Yepes}}]{Blot:2019}
{Blot} L, {Crocce} M, {Sefusatti} E, {Lippich} M, {S{\'a}nchez} AG,
  {Colavincenzo} M, {Monaco} P, {Alvarez} MA, {Agrawal} A, {Avila} S,
  {Balaguera-Antol{\'\i}nez} A, {Bond} R, {Codis} S, {Dalla Vecchia} C, {Dorta}
  A, {Fosalba} P, {Izard} A, {Kitaura} FS, {Pellejero-Ibanez} M, {Stein} G,
  {Vakili} M, {Yepes} G (2019) {Comparing approximate methods for mock
  catalogues and covariance matrices II: power spectrum multipoles}. \mnras
  485(2):2806--2824. \doi{10.1093/mnras/stz507}.
  {\href{https://arxiv.org/abs/1806.09497}{{arXiv:1806.09497}}} {[astro-ph.CO]}

\bibitem[{{Bocquet} et~al.(2016){Bocquet}, {Saro}, {Dolag}, and
  {Mohr}}]{Bocquet:2016}
{Bocquet} S, {Saro} A, {Dolag} K, {Mohr} JJ (2016) {Halo mass function: baryon
  impact, fitting formulae, and implications for cluster cosmology}. \mnras
  456(3):2361--2373. \doi{10.1093/mnras/stv2657}.
  {\href{https://arxiv.org/abs/1502.07357}{{arXiv:1502.07357}}} {[astro-ph.CO]}

\bibitem[{{Bocquet} et~al.(2020){Bocquet}, {Heitmann}, {Habib}, {Lawrence},
  {Uram}, {Frontiere}, {Pope}, and {Finkel}}]{Bocquet:2020}
{Bocquet} S, {Heitmann} K, {Habib} S, {Lawrence} E, {Uram} T, {Frontiere} N,
  {Pope} A, {Finkel} H (2020) {The Mira-Titan Universe. III. Emulation of the
  Halo Mass Function}. \apj 901(1):5. \doi{10.3847/1538-4357/abac5c}.
  {\href{https://arxiv.org/abs/2003.12116}{{arXiv:2003.12116}}} {[astro-ph.CO]}

\bibitem[{{Bode} et~al.(2001){Bode}, {Ostriker}, and {Turok}}]{Bode:2001}
{Bode} P, {Ostriker} JP, {Turok} N (2001) {Halo Formation in Warm Dark Matter
  Models}. \apj 556(1):93--107. \doi{10.1086/321541}.
  {\href{https://arxiv.org/abs/astro-ph/0010389}{{arXiv:astro-ph/0010389}}}
  {[astro-ph]}

\bibitem[{{B{\"o}hm} et~al.(2020){B{\"o}hm}, {Feng}, {Lee}, and
  {Dai}}]{Bohm:2020}
{B{\"o}hm} V, {Feng} Y, {Lee} ME, {Dai} B (2020) {MADLens, a python package for
  fast and differentiable non-Gaussian lensing simulations}. arXiv e-prints
  arXiv:2012.07266.
  {\href{https://arxiv.org/abs/2012.07266}{{arXiv:2012.07266}}} {[astro-ph.CO]}

\bibitem[{{Bond} et~al.(1996){Bond}, {Kofman}, and {Pogosyan}}]{Bond:1996}
{Bond} JR, {Kofman} L, {Pogosyan} D (1996) {How filaments of galaxies are woven
  into the cosmic web}. \nat 380(6575):603--606. \doi{10.1038/380603a0}.
  {\href{https://arxiv.org/abs/astro-ph/9512141}{{arXiv:astro-ph/9512141}}}
  {[astro-ph]}

\bibitem[{{Bonnaire} et~al.(2020){Bonnaire}, {Aghanim}, {Decelle}, and
  {Douspis}}]{Bonnaire:2020}
{Bonnaire} T, {Aghanim} N, {Decelle} A, {Douspis} M (2020) {T-ReX: a
  graph-based filament detection method}. \aap 637:A18.
  \doi{10.1051/0004-6361/201936859}.
  {\href{https://arxiv.org/abs/1912.00732}{{arXiv:1912.00732}}} {[astro-ph.CO]}

\bibitem[{{Bonoli} et~al.(2021){Bonoli}, {Mar{\'\i}n-Franch}, {Varela},
  {V{\'a}zquez Rami{\'o}}, {Abramo}, {Cenarro}, {Dupke}, {V{\'\i}lchez},
  {Crist{\'o}bal-Hornillos}, {Gonz{\'a}lez Delgado},
  {Hern{\'a}ndez-Monteagudo}, {L{\'o}pez-Sanjuan}, {Muniesa}, {Civera},
  {Ederoclite}, {Hern{\'a}n-Caballero}, {Marra}, {Baqui}, {Cortesi},
  {Cypriano}, {Daflon}, {de Amorim}, {D{\'\i}az-Garc{\'\i}a}, {Diego},
  {Mart{\'\i}nez-Solaeche}, {P{\'e}rez}, {Placco}, {Prada}, {Queiroz},
  {Alcaniz}, {Alvarez-Candal}, {Cepa}, {Maroto}, {Roig}, {Siffert}, {Taylor},
  {Benitez}, {Moles}, {Sodr{\'e}}, {Carneiro}, {Mendes de Oliveira}, {Abdalla},
  {Angulo}, {Aparicio Resco}, {Balaguera-Antol{\'\i}nez}, {Ballesteros},
  {Brito-Silva}, {Broadhurst}, {Carrasco}, {Castro}, {Cid Fernandes}, {Coelho},
  {de Melo}, {Doubrawa}, {Fernandez-Soto}, {Ferrari}, {Finoguenov},
  {Garc{\'\i}a-Benito}, {Iglesias-P{\'a}ramo}, {Jim{\'e}nez-Teja}, {Kitaura},
  {Laur}, {Lopes}, {Lucatelli}, {Mart{\'\i}nez}, {Maturi}, {Overzier},
  {Pigozzo}, {Quartin}, {Rodr{\'\i}guez-Mart{\'\i}n}, {Salzano}, {Tamm},
  {Tempel}, {Umetsu}, {Valdivielso}, {von Marttens}, {Zitrin},
  {D{\'\i}az-Mart{\'\i}n}, {L{\'o}pez-Alegre}, {L{\'o}pez-Sainz},
  {Yanes-D{\'\i}az}, {Rueda-Teruel}, {Rueda-Teruel}, {Abril Iba{\~n}ez}, {L
  Ant{\'o}n Bravo}, {Bello Ferrer}, {Bielsa}, {Casino}, {Castillo}, {Chueca},
  {Cuesta}, {Garzar{\'a}n Calderaro}, {Iglesias-Marzoa}, {{\'I}niguez},
  {Lamadrid Gutierrez}, {Lopez-Martinez}, {Lozano-P{\'e}rez}, {Ma{\'\i}cas
  Sacrist{\'a}n}, {Molina-Ib{\'a}{\~n}ez}, {Moreno-Signes}, {Rodr{\'\i}guez
  Llano}, {Royo Navarro}, {Tilve Rua}, {Andrade}, {Alfaro}, {Akras},
  {Arnalte-Mur}, {Ascaso}, {Barbosa}, {Beltr{\'a}n Jim{\'e}nez}, {Benetti},
  {Bengaly}, {Bernui}, {Blanco-Pillado}, {Borges Fernandes}, {Bregman},
  {Bruzual}, {Calderone}, {Carvano}, {Casarini}, {Chaves-Montero},
  {Chies-Santos}, {Coutinho de Carvalho}, {Dimauro}, {Duarte Puertas},
  {Figueruelo}, {Gonz{\'a}lez-Serrano}, {Guerrero}, {Gurung-L{\'o}pez},
  {Herranz}, {Huertas-Company}, {Irwin}, {Izquierdo-Villalba}, {Kanaan},
  {Kehrig}, {Kirkpatrick}, {Lim}, {Lopes}, {Lopes de Oliveira},
  {Marcos-Caballero}, {Mart{\'\i}nez-Delgado}, {Mart{\'\i}nez-Gonz{\'a}lez},
  {Mart{\'\i}nez-Somonte}, {Oliveira}, {Orsi}, {Penna-Lima}, {Reis}, {Spinoso},
  {Tsujikawa}, {Vielva}, {Vitorelli}, {Xia}, {Yuan}, {Arroyo-Polonio},
  {Dantas}, {Galarza}, {Gon{\c{c}}alves}, {Gon{\c{c}}alves}, {Gonzalez},
  {Gonzalez}, {Greisel}, {Jim{\'e}nez-Esteban}, {Landim}, {Lazzaro}, {Magris},
  {Monteiro-Oliveira}, {Pereira}, {Rebou{\c{c}}as}, {Rodriguez-Espinosa},
  {Santos da Costa}, and {Telles}}]{JPAS}
{Bonoli} S, {Mar{\'\i}n-Franch} A, {Varela} J, {V{\'a}zquez Rami{\'o}} H,
  {Abramo} LR, {Cenarro} AJ, {Dupke} RA, {V{\'\i}lchez} JM,
  {Crist{\'o}bal-Hornillos} D, {Gonz{\'a}lez Delgado} RM,
  {Hern{\'a}ndez-Monteagudo} C, {L{\'o}pez-Sanjuan} C, {Muniesa} DJ, {Civera}
  T, {Ederoclite} A, {Hern{\'a}n-Caballero} A, {Marra} V, {Baqui} PO, {Cortesi}
  A, {Cypriano} ES, {Daflon} S, {de Amorim} AL, {D{\'\i}az-Garc{\'\i}a} LA,
  {Diego} JM, {Mart{\'\i}nez-Solaeche} G, {P{\'e}rez} E, {Placco} VM, {Prada}
  F, {Queiroz} C, {Alcaniz} J, {Alvarez-Candal} A, {Cepa} J, {Maroto} AL,
  {Roig} F, {Siffert} BB, {Taylor} K, {Benitez} N, {Moles} M, {Sodr{\'e}} L,
  {Carneiro} S, {Mendes de Oliveira} C, {Abdalla} E, {Angulo} RE, {Aparicio
  Resco} M, {Balaguera-Antol{\'\i}nez} A, {Ballesteros} FJ, {Brito-Silva} D,
  {Broadhurst} T, {Carrasco} ER, {Castro} T, {Cid Fernandes} R, {Coelho} P, {de
  Melo} RB, {Doubrawa} L, {Fernandez-Soto} A, {Ferrari} F, {Finoguenov} A,
  {Garc{\'\i}a-Benito} R, {Iglesias-P{\'a}ramo} J, {Jim{\'e}nez-Teja} Y,
  {Kitaura} FS, {Laur} J, {Lopes} PAA, {Lucatelli} G, {Mart{\'\i}nez} VJ,
  {Maturi} M, {Overzier} RA, {Pigozzo} C, {Quartin} M,
  {Rodr{\'\i}guez-Mart{\'\i}n} JE, {Salzano} V, {Tamm} A, {Tempel} E, {Umetsu}
  K, {Valdivielso} L, {von Marttens} R, {Zitrin} A, {D{\'\i}az-Mart{\'\i}n} MC,
  {L{\'o}pez-Alegre} G, {L{\'o}pez-Sainz} A, {Yanes-D{\'\i}az} A,
  {Rueda-Teruel} F, {Rueda-Teruel} S, {Abril Iba{\~n}ez} J, {L Ant{\'o}n Bravo}
  J, {Bello Ferrer} R, {Bielsa} S, {Casino} JM, {Castillo} J, {Chueca} S,
  {Cuesta} L, {Garzar{\'a}n Calderaro} J, {Iglesias-Marzoa} R, {{\'I}niguez} C,
  {Lamadrid Gutierrez} JL, {Lopez-Martinez} F, {Lozano-P{\'e}rez} D,
  {Ma{\'\i}cas Sacrist{\'a}n} N, {Molina-Ib{\'a}{\~n}ez} EL, {Moreno-Signes} A,
  {Rodr{\'\i}guez Llano} S, {Royo Navarro} M, {Tilve Rua} V, {Andrade} U,
  {Alfaro} EJ, {Akras} S, {Arnalte-Mur} P, {Ascaso} B, {Barbosa} CE,
  {Beltr{\'a}n Jim{\'e}nez} J, {Benetti} M, {Bengaly} CAP, {Bernui} A,
  {Blanco-Pillado} JJ, {Borges Fernandes} M, {Bregman} JN, {Bruzual} G,
  {Calderone} G, {Carvano} JM, {Casarini} L, {Chaves-Montero} J, {Chies-Santos}
  AL, {Coutinho de Carvalho} G, {Dimauro} P, {Duarte Puertas} S, {Figueruelo}
  D, {Gonz{\'a}lez-Serrano} JI, {Guerrero} MA, {Gurung-L{\'o}pez} S, {Herranz}
  D, {Huertas-Company} M, {Irwin} JA, {Izquierdo-Villalba} D, {Kanaan} A,
  {Kehrig} C, {Kirkpatrick} CC, {Lim} J, {Lopes} AR, {Lopes de Oliveira} R,
  {Marcos-Caballero} A, {Mart{\'\i}nez-Delgado} D, {Mart{\'\i}nez-Gonz{\'a}lez}
  E, {Mart{\'\i}nez-Somonte} G, {Oliveira} N, {Orsi} AA, {Penna-Lima} M, {Reis}
  RRR, {Spinoso} D, {Tsujikawa} S, {Vielva} P, {Vitorelli} AZ, {Xia} JQ, {Yuan}
  HB, {Arroyo-Polonio} A, {Dantas} MLL, {Galarza} CA, {Gon{\c{c}}alves} DR,
  {Gon{\c{c}}alves} RS, {Gonzalez} JE, {Gonzalez} AH, {Greisel} N,
  {Jim{\'e}nez-Esteban} F, {Landim} RG, {Lazzaro} D, {Magris} G,
  {Monteiro-Oliveira} R, {Pereira} CB, {Rebou{\c{c}}as} MJ,
  {Rodriguez-Espinosa} JM, {Santos da Costa} S, {Telles} E (2021) {The miniJPAS
  survey: A preview of the Universe in 56 colors}. \aap 653:A31.
  \doi{10.1051/0004-6361/202038841}.
  {\href{https://arxiv.org/abs/2007.01910}{{arXiv:2007.01910}}} {[astro-ph.CO]}

\bibitem[{{Borsanyi} et~al.(2016){Borsanyi}, {Fodor}, {Guenther}, {Kampert},
  {Katz}, {Kawanai}, {Kovacs}, {Mages}, {Pasztor}, {Pittler}, {Redondo},
  {Ringwald}, and {Szabo}}]{Borsanyi:2016}
{Borsanyi} S, {Fodor} Z, {Guenther} J, {Kampert} KH, {Katz} SD, {Kawanai} T,
  {Kovacs} TG, {Mages} SW, {Pasztor} A, {Pittler} F, {Redondo} J, {Ringwald} A,
  {Szabo} KK (2016) {Calculation of the axion mass based on high-temperature
  lattice quantum chromodynamics}. \nat 539(7627):69--71.
  \doi{10.1038/nature20115}

\bibitem[{{Bose} et~al.(2017){Bose}, {Li}, {Barreira}, {He}, {Hellwing},
  {Koyama}, {Llinares}, and {Zhao}}]{BoseLi:2017}
{Bose} S, {Li} B, {Barreira} A, {He} Jh, {Hellwing} WA, {Koyama} K, {Llinares}
  C, {Zhao} GB (2017) {Speeding up N-body simulations of modified gravity:
  chameleon screening models}. \jcap 2017(2):050.
  \doi{10.1088/1475-7516/2017/02/050}.
  {\href{https://arxiv.org/abs/1611.09375}{{arXiv:1611.09375}}} {[astro-ph.CO]}

\bibitem[{{Bose} et~al.(2019){Bose}, {Vogelsberger}, {Zavala}, {Pfrommer},
  {Cyr-Racine}, {Bohr}, and {Bringmann}}]{Bose:2019}
{Bose} S, {Vogelsberger} M, {Zavala} J, {Pfrommer} C, {Cyr-Racine} FY, {Bohr}
  S, {Bringmann} T (2019) {ETHOS - an Effective Theory of Structure Formation:
  detecting dark matter interactions through the Lyman-{\ensuremath{\alpha}}
  forest}. \mnras 487(1):522--536. \doi{10.1093/mnras/stz1276}.
  {\href{https://arxiv.org/abs/1811.10630}{{arXiv:1811.10630}}} {[astro-ph.CO]}

\bibitem[{{Bouchet} and {Hernquist}(1988)}]{Bouchet:1988}
{Bouchet} FR, {Hernquist} L (1988) {Cosmological Simulations Using the
  Hierarchical Tree Method}. \apjs 68:521. \doi{10.1086/191299}

\bibitem[{{Bouchet} et~al.(1995){Bouchet}, {Colombi}, {Hivon}, and
  {Juszkiewicz}}]{Bouchet:1995}
{Bouchet} FR, {Colombi} S, {Hivon} E, {Juszkiewicz} R (1995) {Perturbative
  Lagrangian approach to gravitational instability.} \aap 296:575.
  {\href{https://arxiv.org/abs/astro-ph/9406013}{{astro-ph/9406013}}}

\bibitem[{{Bower} et~al.(2006){Bower}, {Benson}, {Malbon}, {Helly}, {Frenk},
  {Baugh}, {Cole}, and {Lacey}}]{Bower:2006}
{Bower} RG, {Benson} AJ, {Malbon} R, {Helly} JC, {Frenk} CS, {Baugh} CM, {Cole}
  S, {Lacey} CG (2006) {Breaking the hierarchy of galaxy formation}. \mnras
  370(2):645--655. \doi{10.1111/j.1365-2966.2006.10519.x}.
  {\href{https://arxiv.org/abs/astro-ph/0511338}{{arXiv:astro-ph/0511338}}}
  {[astro-ph]}

\bibitem[{{Bower} et~al.(2010){Bower}, {Vernon}, {Goldstein}, {Benson},
  {Lacey}, {Baugh}, {Cole}, and {Frenk}}]{Bower:2010}
{Bower} RG, {Vernon} I, {Goldstein} M, {Benson} AJ, {Lacey} CG, {Baugh} CM,
  {Cole} S, {Frenk} CS (2010) {The parameter space of galaxy formation}. \mnras
  407(4):2017--2045. \doi{10.1111/j.1365-2966.2010.16991.x}.
  {\href{https://arxiv.org/abs/1004.0711}{{arXiv:1004.0711}}} {[astro-ph.CO]}

\bibitem[{{Boylan-Kolchin} et~al.(2009){Boylan-Kolchin}, {Springel}, {White},
  {Jenkins}, and {Lemson}}]{Boylan-Kolchin:2009}
{Boylan-Kolchin} M, {Springel} V, {White} SDM, {Jenkins} A, {Lemson} G (2009)
  {Resolving cosmic structure formation with the Millennium-II Simulation}.
  \mnras 398(3):1150--1164. \doi{10.1111/j.1365-2966.2009.15191.x}.
  {\href{https://arxiv.org/abs/0903.3041}{{arXiv:0903.3041}}} {[astro-ph.CO]}

\bibitem[{{Boylan-Kolchin} et~al.(2011){Boylan-Kolchin}, {Bullock}, and
  {Kaplinghat}}]{Boylan-Kolchin:2011}
{Boylan-Kolchin} M, {Bullock} JS, {Kaplinghat} M (2011) {Too big to fail? The
  puzzling darkness of massive Milky Way subhaloes}. \mnras 415(1):L40--L44.
  \doi{10.1111/j.1745-3933.2011.01074.x}.
  {\href{https://arxiv.org/abs/1103.0007}{{arXiv:1103.0007}}} {[astro-ph.CO]}

\bibitem[{{Boyle} and {Schmidt}(2020)}]{Boyle:2020}
{Boyle} A, {Schmidt} F (2020) {Neutrino mass constraints beyond linear order:
  cosmology dependence and systematic biases}. arXiv e-prints arXiv:2011.10594.
  {\href{https://arxiv.org/abs/2011.10594}{{arXiv:2011.10594}}} {[astro-ph.CO]}

\bibitem[{{Brandbyge} and {Hannestad}(2009)}]{Brandbyge:2009}
{Brandbyge} J, {Hannestad} S (2009) {Grid based linear neutrino perturbations
  in cosmological N-body simulations}. \jcap 2009(5):002.
  \doi{10.1088/1475-7516/2009/05/002}.
  {\href{https://arxiv.org/abs/0812.3149}{{arXiv:0812.3149}}} {[astro-ph]}

\bibitem[{{Brandbyge} et~al.(2008){Brandbyge}, {Hannestad}, {Haugb{\o}lle}, and
  {Thomsen}}]{Brandbyge:2008}
{Brandbyge} J, {Hannestad} S, {Haugb{\o}lle} T, {Thomsen} B (2008) {The effect
  of thermal neutrino motion on the non-linear cosmological matter power
  spectrum}. \jcap 2008(8):020. \doi{10.1088/1475-7516/2008/08/020}.
  {\href{https://arxiv.org/abs/0802.3700}{{arXiv:0802.3700}}} {[astro-ph]}

\bibitem[{{Brandbyge} et~al.(2017){Brandbyge}, {Rampf}, {Tram}, {Leclercq},
  {Fidler}, and {Hannestad}}]{Brandbyge:2017}
{Brandbyge} J, {Rampf} C, {Tram} T, {Leclercq} F, {Fidler} C, {Hannestad} S
  (2017) {Cosmological N -body simulations including radiation perturbations}.
  \mnras 466(1):L68--L72. \doi{10.1093/mnrasl/slw235}.
  {\href{https://arxiv.org/abs/1610.04236}{{arXiv:1610.04236}}} {[astro-ph.CO]}

\bibitem[{Brandt(1977)}]{Brandt:1977}
Brandt A (1977) Multi-level adaptive solutions to boundary-value problems.
  Mathematics of Computation 31(138):333--390

\bibitem[{{Brax} et~al.(2012){Brax}, {Davis}, {Li}, {Winther}, and
  {Zhao}}]{Brax:2012}
{Brax} P, {Davis} AC, {Li} B, {Winther} HA, {Zhao} GB (2012) {Systematic
  simulations of modified gravity: symmetron and dilaton models}. \jcap
  2012(10):002. \doi{10.1088/1475-7516/2012/10/002}.
  {\href{https://arxiv.org/abs/1206.3568}{{arXiv:1206.3568}}} {[astro-ph.CO]}

\bibitem[{{Breton} et~al.(2019){Breton}, {Rasera}, {Taruya}, {Lacombe}, and
  {Saga}}]{Breton:2019}
{Breton} MA, {Rasera} Y, {Taruya} A, {Lacombe} O, {Saga} S (2019) {Imprints of
  relativistic effects on the asymmetry of the halo cross-correlation function:
  from linear to non-linear scales}. \mnras 483(2):2671--2696.
  \doi{10.1093/mnras/sty3206}.
  {\href{https://arxiv.org/abs/1803.04294}{{arXiv:1803.04294}}} {[astro-ph.CO]}

\bibitem[{{Brinckmann} et~al.(2018){Brinckmann}, {Zavala}, {Rapetti}, {Hansen},
  and {Vogelsberger}}]{Brinckmann:2018}
{Brinckmann} T, {Zavala} J, {Rapetti} D, {Hansen} SH, {Vogelsberger} M (2018)
  {The structure and assembly history of cluster-sized haloes in
  self-interacting dark matter}. \mnras 474(1):746--759.
  \doi{10.1093/mnras/stx2782}.
  {\href{https://arxiv.org/abs/1705.00623}{{arXiv:1705.00623}}} {[astro-ph.CO]}

\bibitem[{{Bringmann} et~al.(2012){Bringmann}, {Scott}, and
  {Akrami}}]{Bringmann:2012}
{Bringmann} T, {Scott} P, {Akrami} Y (2012) {Improved constraints on the
  primordial power spectrum at small scales from ultracompact minihalos}. \prd
  85(12):125027. \doi{10.1103/PhysRevD.85.125027}.
  {\href{https://arxiv.org/abs/1110.2484}{{arXiv:1110.2484}}} {[astro-ph.CO]}

\bibitem[{{Brook} and {Di Cintio}(2015)}]{Brook:2015}
{Brook} CB, {Di Cintio} A (2015) {Expanded haloes, abundance matching and
  too-big-to-fail in the Local Group}. \mnras 450(4):3920--3934.
  \doi{10.1093/mnras/stv864}.
  {\href{https://arxiv.org/abs/1410.3825}{{arXiv:1410.3825}}} {[astro-ph.GA]}

\bibitem[{{Brooks} et~al.(2013){Brooks}, {Kuhlen}, {Zolotov}, and
  {Hooper}}]{Brooks:2013}
{Brooks} AM, {Kuhlen} M, {Zolotov} A, {Hooper} D (2013) {A Baryonic Solution to
  the Missing Satellites Problem}. \apj 765(1):22.
  \doi{10.1088/0004-637X/765/1/22}.
  {\href{https://arxiv.org/abs/1209.5394}{{arXiv:1209.5394}}} {[astro-ph.CO]}

\bibitem[{{Brown} et~al.(2020){Brown}, {McCarthy}, {Diemer}, {Font},
  {Stafford}, and {Pfeifer}}]{Brown:2020}
{Brown} ST, {McCarthy} IG, {Diemer} B, {Font} AS, {Stafford} SG, {Pfeifer} S
  (2020) {Connecting the structure of dark matter haloes to the primordial
  power spectrum}. \mnras 495(4):4994--5013. \doi{10.1093/mnras/staa1491}.
  {\href{https://arxiv.org/abs/2005.12933}{{arXiv:2005.12933}}} {[astro-ph.CO]}

\bibitem[{{Bryan} and {Norman}(1998)}]{BryanNorman:1998}
{Bryan} GL, {Norman} ML (1998) {Statistical Properties of X-Ray Clusters:
  Analytic and Numerical Comparisons}. \apj 495(1):80--99.
  \doi{10.1086/305262}.
  {\href{https://arxiv.org/abs/astro-ph/9710107}{{arXiv:astro-ph/9710107}}}
  {[astro-ph]}

\bibitem[{Buchert(1989)}]{Buchert:1989}
Buchert T (1989) {A class of solutions in Newtonian cosmology and the pancake
  theory}. \aap 223:9--24

\bibitem[{{Buchert}(1994)}]{Buchert:1994}
{Buchert} T (1994) {Lagrangian Theory of Gravitational Instability of
  Friedman-Lemaitre Cosmologies - a Generic Third-Order Model for Nonlinear
  Clustering}. \mnras 267:811. \doi{10.1093/mnras/267.4.811}.
  {\href{https://arxiv.org/abs/astro-ph/9309055}{{astro-ph/9309055}}}

\bibitem[{{Buehlmann} and {Hahn}(2019)}]{Buehlmann:2019}
{Buehlmann} M, {Hahn} O (2019) {Large-scale velocity dispersion and the cosmic
  web}. \mnras 487(1):228--245. \doi{10.1093/mnras/stz1243}.
  {\href{https://arxiv.org/abs/1812.07489}{{arXiv:1812.07489}}} {[astro-ph.CO]}

\bibitem[{{Bullock} et~al.(2001){Bullock}, {Kolatt}, {Sigad}, {Somerville},
  {Kravtsov}, {Klypin}, {Primack}, and {Dekel}}]{Bullock:2001}
{Bullock} JS, {Kolatt} TS, {Sigad} Y, {Somerville} RS, {Kravtsov} AV, {Klypin}
  AA, {Primack} JR, {Dekel} A (2001) {Profiles of dark haloes: evolution,
  scatter and environment}. \mnras 321(3):559--575.
  \doi{10.1046/j.1365-8711.2001.04068.x}.
  {\href{https://arxiv.org/abs/astro-ph/9908159}{{arXiv:astro-ph/9908159}}}
  {[astro-ph]}

\bibitem[{{Burkert}(2000)}]{Burkert:2000}
{Burkert} A (2000) {The Structure and Evolution of Weakly Self-interacting Cold
  Dark Matter Halos}. \apjl 534(2):L143--L146. \doi{10.1086/312674}.
  {\href{https://arxiv.org/abs/astro-ph/0002409}{{arXiv:astro-ph/0002409}}}
  {[astro-ph]}

\bibitem[{{Burkert}(2020)}]{Burkert:2020}
{Burkert} A (2020) {Fuzzy Dark Matter and Dark Matter Halo Cores}. \apj
  904(2):161. \doi{10.3847/1538-4357/abb242}.
  {\href{https://arxiv.org/abs/2006.11111}{{arXiv:2006.11111}}} {[astro-ph.GA]}

\bibitem[{{Buschmann} et~al.(2019){Buschmann}, {Foster}, and
  {Safdi}}]{Buschmann:2019}
{Buschmann} M, {Foster} JW, {Safdi} BR (2019) {Early-Universe Simulations of
  the Cosmological Axion}. arXiv e-prints arXiv:1906.00967.
  {\href{https://arxiv.org/abs/1906.00967}{{arXiv:1906.00967}}} {[astro-ph.CO]}

\bibitem[{{Byrnes} et~al.(2018){Byrnes}, {Hindmarsh}, {Young}, and
  {Hawkins}}]{Byrnes:2018}
{Byrnes} CT, {Hindmarsh} M, {Young} S, {Hawkins} MRS (2018) {Primordial black
  holes with an accurate QCD equation of state}. \jcap 2018(8):041.
  \doi{10.1088/1475-7516/2018/08/041}.
  {\href{https://arxiv.org/abs/1801.06138}{{arXiv:1801.06138}}} {[astro-ph.CO]}

\bibitem[{{Cacciato} et~al.(2013){Cacciato}, {van den Bosch}, {More}, {Mo}, and
  {Yang}}]{Cacciato:2013}
{Cacciato} M, {van den Bosch} FC, {More} S, {Mo} H, {Yang} X (2013)
  {Cosmological constraints from a combination of galaxy clustering and lensing
  - III. Application to SDSS data}. \mnras 430(2):767--786.
  \doi{10.1093/mnras/sts525}.
  {\href{https://arxiv.org/abs/1207.0503}{{arXiv:1207.0503}}} {[astro-ph.CO]}

\bibitem[{{Cai} et~al.(2015){Cai}, {Padilla}, and {Li}}]{Cai:2015}
{Cai} YC, {Padilla} N, {Li} B (2015) {Testing gravity using cosmic voids}.
  \mnras 451(1):1036--1055. \doi{10.1093/mnras/stv777}.
  {\href{https://arxiv.org/abs/1410.1510}{{arXiv:1410.1510}}} {[astro-ph.CO]}

\bibitem[{{Cai} et~al.(2016){Cai}, {Taylor}, {Peacock}, and
  {Padilla}}]{Cai:2016}
{Cai} YC, {Taylor} A, {Peacock} JA, {Padilla} N (2016) {Redshift-space
  distortions around voids}. \mnras 462(3):2465--2477.
  \doi{10.1093/mnras/stw1809}.
  {\href{https://arxiv.org/abs/1603.05184}{{arXiv:1603.05184}}} {[astro-ph.CO]}

\bibitem[{{Cai} et~al.(2017){Cai}, {Kaiser}, {Cole}, and {Frenk}}]{Cai:2017}
{Cai} YC, {Kaiser} N, {Cole} S, {Frenk} C (2017) {Gravitational redshift and
  asymmetric redshift-space distortions for stacked clusters}. \mnras
  468(2):1981--1993. \doi{10.1093/mnras/stx469}.
  {\href{https://arxiv.org/abs/1609.04864}{{arXiv:1609.04864}}} {[astro-ph.CO]}

\bibitem[{{Camera} et~al.(2015){Camera}, {Santos}, and
  {Maartens}}]{Camera:2015}
{Camera} S, {Santos} MG, {Maartens} R (2015) {Probing primordial
  non-Gaussianity with SKA galaxy redshift surveys: a fully relativistic
  analysis}. \mnras 448(2):1035--1043. \doi{10.1093/mnras/stv040}.
  {\href{https://arxiv.org/abs/1409.8286}{{arXiv:1409.8286}}} {[astro-ph.CO]}

\bibitem[{{Campbell} et~al.(2018){Campbell}, {van den Bosch}, {Padmanabhan},
  {Mao}, {Zentner}, {Lange}, {Jiang}, and {Villarreal}}]{Campbell:2018}
{Campbell} D, {van den Bosch} FC, {Padmanabhan} N, {Mao} YY, {Zentner} AR,
  {Lange} JU, {Jiang} F, {Villarreal} A (2018) {The galaxy clustering crisis in
  abundance matching}. \mnras 477(1):359--383. \doi{10.1093/mnras/sty495}.
  {\href{https://arxiv.org/abs/1705.06347}{{arXiv:1705.06347}}} {[astro-ph.GA]}

\bibitem[{{Carlesi} et~al.(2016){Carlesi}, {Sorce}, {Hoffman}, {Gottl{\"o}ber},
  {Yepes}, {Libeskind}, {Pilipenko}, {Knebe}, {Courtois}, {Tully}, and
  {Steinmetz}}]{Carlesi:2016}
{Carlesi} E, {Sorce} JG, {Hoffman} Y, {Gottl{\"o}ber} S, {Yepes} G, {Libeskind}
  NI, {Pilipenko} SV, {Knebe} A, {Courtois} H, {Tully} RB, {Steinmetz} M (2016)
  {Constrained Local UniversE Simulations: a Local Group factory}. \mnras
  458(1):900--911. \doi{10.1093/mnras/stw357}.
  {\href{https://arxiv.org/abs/1602.03919}{{arXiv:1602.03919}}} {[astro-ph.CO]}

\bibitem[{{Carr} and {K{\"u}hnel}(2020)}]{Carr:2020}
{Carr} B, {K{\"u}hnel} F (2020) {Primordial Black Holes as Dark Matter: Recent
  Developments}. Annual Review of Nuclear and Particle Science 70:355--394.
  \doi{10.1146/annurev-nucl-050520-125911}.
  {\href{https://arxiv.org/abs/2006.02838}{{arXiv:2006.02838}}} {[astro-ph.CO]}

\bibitem[{{Carr} et~al.(2016){Carr}, {K{\"u}hnel}, and {Sandstad}}]{Carr:2016}
{Carr} B, {K{\"u}hnel} F, {Sandstad} M (2016) {Primordial black holes as dark
  matter}. \prd 94(8):083504. \doi{10.1103/PhysRevD.94.083504}.
  {\href{https://arxiv.org/abs/1607.06077}{{arXiv:1607.06077}}}

\bibitem[{{Carr} et~al.(2020{\natexlab{a}}){Carr}, {Kohri}, {Sendouda}, and
  {Yokoyama}}]{Carr:2020c}
{Carr} B, {Kohri} K, {Sendouda} Y, {Yokoyama} J (2020{\natexlab{a}})
  {Constraints on Primordial Black Holes}. arXiv e-prints arXiv:2002.12778.
  {\href{https://arxiv.org/abs/2002.12778}{{arXiv:2002.12778}}} {[astro-ph.CO]}

\bibitem[{{Carr} et~al.(2020{\natexlab{b}}){Carr}, {Kuhnel}, and
  {Visinelli}}]{Carr:2020b}
{Carr} B, {Kuhnel} F, {Visinelli} L (2020{\natexlab{b}}) {Black Holes and
  WIMPs: All or Nothing or Something Else}. arXiv e-prints arXiv:2011.01930.
  {\href{https://arxiv.org/abs/2011.01930}{{arXiv:2011.01930}}} {[astro-ph.CO]}

\bibitem[{{Carr} et~al.(2021){Carr}, {Clesse}, {Garc{\'\i}a-Bellido}, and
  {K{\"u}hnel}}]{Carr:2021}
{Carr} B, {Clesse} S, {Garc{\'\i}a-Bellido} J, {K{\"u}hnel} F (2021) {Cosmic
  conundra explained by thermal history and primordial black holes}. Physics of
  the Dark Universe 31:100755. \doi{10.1016/j.dark.2020.100755}

\bibitem[{{Castorina} et~al.(2015){Castorina}, {Carbone}, {Bel}, {Sefusatti},
  and {Dolag}}]{Castorina:2015}
{Castorina} E, {Carbone} C, {Bel} J, {Sefusatti} E, {Dolag} K (2015) {DEMNUni:
  the clustering of large-scale structures in the presence of massive
  neutrinos}. \jcap 2015(7):043. \doi{10.1088/1475-7516/2015/07/043}.
  {\href{https://arxiv.org/abs/1505.07148}{{arXiv:1505.07148}}} {[astro-ph.CO]}

\bibitem[{{Castorina} et~al.(2019){Castorina}, {Hand}, {Seljak}, {Beutler},
  {Chuang}, {Zhao}, {Gil-Mar{\'\i}n}, {Percival}, {Ross}, {Choi}, {Dawson}, {de
  la Macorra}, {Rossi}, {Ruggeri}, {Schneider}, and {Zhao}}]{Castorina:2019}
{Castorina} E, {Hand} N, {Seljak} U, {Beutler} F, {Chuang} CH, {Zhao} C,
  {Gil-Mar{\'\i}n} H, {Percival} WJ, {Ross} AJ, {Choi} PD, {Dawson} K, {de la
  Macorra} A, {Rossi} G, {Ruggeri} R, {Schneider} D, {Zhao} GB (2019)
  {Redshift-weighted constraints on primordial non-Gaussianity from the
  clustering of the eBOSS DR14 quasars in Fourier space}. \jcap 2019(9):010.
  \doi{10.1088/1475-7516/2019/09/010}.
  {\href{https://arxiv.org/abs/1904.08859}{{arXiv:1904.08859}}} {[astro-ph.CO]}

\bibitem[{{Catelan}(1995)}]{Catelan:1995}
{Catelan} P (1995) {Lagrangian dynamics in non-flat universes and non-linear
  gravitational evolution}. \mnras 276:115--124. \doi{10.1093/mnras/276.1.115}.
  {\href{https://arxiv.org/abs/astro-ph/9406016}{{astro-ph/9406016}}}

\bibitem[{{Cautun} et~al.(2013){Cautun}, {van de Weygaert}, and
  {Jones}}]{Cautun:2013}
{Cautun} M, {van de Weygaert} R, {Jones} BJT (2013) {NEXUS: tracing the cosmic
  web connection}. \mnras 429(2):1286--1308. \doi{10.1093/mnras/sts416}.
  {\href{https://arxiv.org/abs/1209.2043}{{arXiv:1209.2043}}} {[astro-ph.CO]}

\bibitem[{{Cautun} et~al.(2018){Cautun}, {Paillas}, {Cai}, {Bose}, {Armijo},
  {Li}, and {Padilla}}]{Cautun:2018}
{Cautun} M, {Paillas} E, {Cai} YC, {Bose} S, {Armijo} J, {Li} B, {Padilla} N
  (2018) {The Santiago-Harvard-Edinburgh-Durham void comparison - I. SHEDding
  light on chameleon gravity tests}. \mnras 476(3):3195--3217.
  \doi{10.1093/mnras/sty463}.
  {\href{https://arxiv.org/abs/1710.01730}{{arXiv:1710.01730}}} {[astro-ph.CO]}

\bibitem[{{Cautun} and {van de Weygaert}(2011)}]{Cautun:2011}
{Cautun} MC, {van de Weygaert} R (2011) {The DTFE public software: The Delaunay
  Tessellation Field Estimator code}. arXiv e-prints arXiv:1105.0370.
  {\href{https://arxiv.org/abs/1105.0370}{{arXiv:1105.0370}}} {[astro-ph.IM]}

\bibitem[{{Centrella} and {Melott}(1983)}]{Centrella:1983}
{Centrella} J, {Melott} AL (1983) {Three-dimensional simulation of large-scale
  structure in the universe}. \nat 305:196--198. \doi{10.1038/305196a0}

\bibitem[{{Centrella} et~al.(1988){Centrella}, {Gallagher}, {Melott}, and
  {Bushouse}}]{Centrella:1988}
{Centrella} JM, {Gallagher} JS III, {Melott} AL, {Bushouse} HA (1988) {A case
  study of large-scale structure in a 'hot' model universe}. \apj 333:24--53.
  \doi{10.1086/166722}

\bibitem[{{Chan} et~al.(2020){Chan}, {Li}, {Biagetti}, and
  {Hamaus}}]{Chan:2020}
{Chan} KC, {Li} Y, {Biagetti} M, {Hamaus} N (2020) {Measurement of Void Bias
  Using Separate Universe Simulations}. \apj 889(2):89.
  \doi{10.3847/1538-4357/ab64ec}.
  {\href{https://arxiv.org/abs/1909.03736}{{arXiv:1909.03736}}} {[astro-ph.CO]}

\bibitem[{{Chan} et~al.(2015){Chan}, {Kere{\v{s}}}, {O{\~n}orbe}, {Hopkins},
  {Muratov}, {Faucher-Gigu{\`e}re}, and {Quataert}}]{Chan:2015}
{Chan} TK, {Kere{\v{s}}} D, {O{\~n}orbe} J, {Hopkins} PF, {Muratov} AL,
  {Faucher-Gigu{\`e}re} CA, {Quataert} E (2015) {The impact of baryonic physics
  on the structure of dark matter haloes: the view from the FIRE cosmological
  simulations}. \mnras 454(3):2981--3001. \doi{10.1093/mnras/stv2165}.
  {\href{https://arxiv.org/abs/1507.02282}{{arXiv:1507.02282}}} {[astro-ph.GA]}

\bibitem[{{Chavanis}(2011)}]{Chavanis:2011}
{Chavanis} PH (2011) {Mass-radius relation of Newtonian self-gravitating
  Bose-Einstein condensates with short-range interactions. I. Analytical
  results}. \prd 84(4):043531. \doi{10.1103/PhysRevD.84.043531}.
  {\href{https://arxiv.org/abs/1103.2050}{{arXiv:1103.2050}}} {[astro-ph.CO]}

\bibitem[{{Chavanis} and {Delfini}(2011)}]{Chavanis:2011b}
{Chavanis} PH, {Delfini} L (2011) {Mass-radius relation of Newtonian
  self-gravitating Bose-Einstein condensates with short-range interactions. II.
  Numerical results}. \prd 84(4):043532. \doi{10.1103/PhysRevD.84.043532}.
  {\href{https://arxiv.org/abs/1103.2054}{{arXiv:1103.2054}}} {[astro-ph.CO]}

\bibitem[{{Chaves-Montero} and {Hearin}(2020)}]{Chaves-Montero:2020}
{Chaves-Montero} J, {Hearin} A (2020) {Surrogate modelling the Baryonic
  Universe - I. The colour of star formation}. \mnras 495(2):2088--2104.
  \doi{10.1093/mnras/staa1230}.
  {\href{https://arxiv.org/abs/1910.11883}{{arXiv:1910.11883}}} {[astro-ph.GA]}

\bibitem[{{Chaves-Montero} et~al.(2016){Chaves-Montero}, {Angulo}, {Schaye},
  {Schaller}, {Crain}, {Furlong}, and {Theuns}}]{Chaves-Montero:2016}
{Chaves-Montero} J, {Angulo} RE, {Schaye} J, {Schaller} M, {Crain} RA,
  {Furlong} M, {Theuns} T (2016) {Subhalo abundance matching and assembly bias
  in the EAGLE simulation}. \mnras 460(3):3100--3118.
  \doi{10.1093/mnras/stw1225}.
  {\href{https://arxiv.org/abs/1507.01948}{{arXiv:1507.01948}}} {[astro-ph.GA]}

\bibitem[{{Chen} et~al.(1974){Chen}, {Bruce Langdon}, and
  {Birdsall}}]{Chen:1974}
{Chen} L, {Bruce Langdon} A, {Birdsall} CK (1974) {Reduction of the Grid
  Effects in Simulation Plasmas}. J Comput Phys 14(2):200--222.
  \doi{10.1016/0021-9991(74)90014-X}

\bibitem[{{Chen} et~al.(2016){Chen}, {Wang}, {Mo}, and {Shi}}]{Chen:2016}
{Chen} S, {Wang} H, {Mo} HJ, {Shi} J (2016) {Alignments of Dark Matter Halos
  with Large-scale Tidal Fields: Mass and Redshift Dependence}. \apj 825(1):49.
  \doi{10.3847/0004-637X/825/1/49}.
  {\href{https://arxiv.org/abs/1603.04152}{{arXiv:1603.04152}}} {[astro-ph.CO]}

\bibitem[{{Chen} et~al.(2019){Chen}, {Castorina}, and {White}}]{Chen:2019}
{Chen} SF, {Castorina} E, {White} M (2019) {Biased tracers of two fluids in the
  Lagrangian picture}. \jcap 2019(6):006. \doi{10.1088/1475-7516/2019/06/006}.
  {\href{https://arxiv.org/abs/1903.00437}{{arXiv:1903.00437}}} {[astro-ph.CO]}

\bibitem[{{Chen} et~al.(2021){Chen}, {Lee}, and {Dvorkin}}]{Chen:2021}
{Chen} SF, {Lee} H, {Dvorkin} C (2021) {Precise and Accurate Cosmology with
  CMBxLSS Power Spectra and Bispectra}. arXiv e-prints arXiv:2103.01229.
  {\href{https://arxiv.org/abs/2103.01229}{{arXiv:2103.01229}}} {[astro-ph.CO]}

\bibitem[{{Chen} et~al.(2017){Chen}, {Schive}, and {Chiueh}}]{Chen:2017}
{Chen} SR, {Schive} HY, {Chiueh} T (2017) {Jeans analysis for dwarf spheroidal
  galaxies in wave dark matter}. \mnras 468(2):1338--1348.
  \doi{10.1093/mnras/stx449}.
  {\href{https://arxiv.org/abs/1606.09030}{{arXiv:1606.09030}}} {[astro-ph.GA]}

\bibitem[{{Cheng} et~al.(2020){Cheng}, {Ting}, {M{\'e}nard}, and
  {Bruna}}]{Cheng:2020}
{Cheng} S, {Ting} YS, {M{\'e}nard} B, {Bruna} J (2020) {A new approach to
  observational cosmology using the scattering transform}. \mnras
  499(4):5902--5914. \doi{10.1093/mnras/staa3165}.
  {\href{https://arxiv.org/abs/2006.08561}{{arXiv:2006.08561}}} {[astro-ph.CO]}

\bibitem[{{Chernin} et~al.(2003){Chernin}, {Nagirner}, and
  {Starikova}}]{Chernin:2003}
{Chernin} AD, {Nagirner} DI, {Starikova} SV (2003) {Growth rate of cosmological
  perturbations in standard model: Explicit analytical solution}. \aap
  399:19--21. \doi{10.1051/0004-6361:20021763}.
  {\href{https://arxiv.org/abs/astro-ph/0110107}{{arXiv:astro-ph/0110107}}}
  {[astro-ph]}

\bibitem[{{Chiang} et~al.(2017){Chiang}, {Cieplak}, {Schmidt}, and
  {Slosar}}]{Chiang:2017}
{Chiang} CT, {Cieplak} AM, {Schmidt} F, {Slosar} A (2017) {Response approach to
  the squeezed-limit bispectrum: application to the correlation of quasar and
  Lyman-{\ensuremath{\alpha}} forest power spectrum}. \jcap 2017(6):022.
  \doi{10.1088/1475-7516/2017/06/022}.
  {\href{https://arxiv.org/abs/1701.03375}{{arXiv:1701.03375}}} {[astro-ph.CO]}

\bibitem[{{Chiang} et~al.(2018){Chiang}, {Hu}, {Li}, and
  {LoVerde}}]{Chiang:2018}
{Chiang} CT, {Hu} W, {Li} Y, {LoVerde} M (2018) {Scale-dependent bias and
  bispectrum in neutrino separate universe simulations}. \prd 97(12):123526.
  \doi{10.1103/PhysRevD.97.123526}.
  {\href{https://arxiv.org/abs/1710.01310}{{arXiv:1710.01310}}} {[astro-ph.CO]}

\bibitem[{{Chiang} et~al.(2019){Chiang}, {LoVerde}, and
  {Villaescusa-Navarro}}]{Chiang:2019}
{Chiang} CT, {LoVerde} M, {Villaescusa-Navarro} F (2019) {First Detection of
  Scale-Dependent Linear Halo Bias in N -Body Simulations with Massive
  Neutrinos}. \prl 122(4):041302. \doi{10.1103/PhysRevLett.122.041302}.
  {\href{https://arxiv.org/abs/1811.12412}{{arXiv:1811.12412}}} {[astro-ph.CO]}

\bibitem[{{Chin} and {Chen}(2001)}]{Chin:2001}
{Chin} SA, {Chen} CR (2001) {Fourth order gradient symplectic integrator
  methods for solving the time-dependent Schr{\"o}dinger equation}. \jcp
  114:7338--7341. \doi{10.1063/1.1362288}.
  {\href{https://arxiv.org/abs/physics/0012017}{{physics/0012017}}}

\bibitem[{{Chisari} and {Zaldarriaga}(2011)}]{Chisari:2011}
{Chisari} NE, {Zaldarriaga} M (2011) {Connection between Newtonian simulations
  and general relativity}. \prd 83(12):123505.
  \doi{10.1103/PhysRevD.83.123505}.
  {\href{https://arxiv.org/abs/1101.3555}{{arXiv:1101.3555}}} {[astro-ph.CO]}

\bibitem[{{Chisari} et~al.(2019){Chisari}, {Mead}, {Joudaki}, {Ferreira},
  {Schneider}, {Mohr}, {Tr{\"o}ster}, {Alonso}, {McCarthy}, {Martin-Alvarez},
  {Devriendt}, {Slyz}, and {van Daalen}}]{Chisari:2019}
{Chisari} NE, {Mead} AJ, {Joudaki} S, {Ferreira} P, {Schneider} A, {Mohr} J,
  {Tr{\"o}ster} T, {Alonso} D, {McCarthy} IG, {Martin-Alvarez} S, {Devriendt}
  J, {Slyz} A, {van Daalen} MP (2019) {Modelling baryonic feedback for survey
  cosmology}. arXiv e-prints
  {\href{https://arxiv.org/abs/1905.06082}{{arXiv:1905.06082}}}

\bibitem[{Choquet-Bruhat(1971)}]{Choquet-Bruhat:1971}
Choquet-Bruhat Y (1971) Probl\`eme de cauchy pour le syst\`eme
  int\'egro-diff\'erentiel d'einstein-liouville. Annales de l'Institut Fourier
  21(3):181--201. \doi{10.5802/aif.385},
  \urlprefix\url{http://www.numdam.org/articles/10.5802/aif.385/}

\bibitem[{Choquet-Bruhat(2015)}]{Choquet:2015}
Choquet-Bruhat Y (2015) Introduction to General Relativity, Black Holes, and
  Cosmology. Oxford University Press

\bibitem[{{Choquette} et~al.(2019){Choquette}, {Cline}, and
  {Cornell}}]{Choquette:2019}
{Choquette} J, {Cline} JM, {Cornell} JM (2019) {Early formation of supermassive
  black holes via dark matter self-interactions}. \jcap 2019(7):036.
  \doi{10.1088/1475-7516/2019/07/036}.
  {\href{https://arxiv.org/abs/1812.05088}{{arXiv:1812.05088}}} {[astro-ph.CO]}

\bibitem[{{Chuang} et~al.(2015{\natexlab{a}}){Chuang}, {Kitaura}, {Prada},
  {Zhao}, and {Yepes}}]{Chuang:2015}
{Chuang} CH, {Kitaura} FS, {Prada} F, {Zhao} C, {Yepes} G (2015{\natexlab{a}})
  {EZmocks: extending the Zel'dovich approximation to generate mock galaxy
  catalogues with accurate clustering statistics}. \mnras 446(3):2621--2628.
  \doi{10.1093/mnras/stu2301}.
  {\href{https://arxiv.org/abs/1409.1124}{{arXiv:1409.1124}}} {[astro-ph.CO]}

\bibitem[{{Chuang} et~al.(2015{\natexlab{b}}){Chuang}, {Zhao}, {Prada},
  {Munari}, {Avila}, {Izard}, {Kitaura}, {Manera}, {Monaco}, {Murray}, {Knebe},
  {Sc{\'o}ccola}, {Yepes}, {Garcia-Bellido}, {Mar{\'\i}n}, {M{\"u}ller},
  {Skibba}, {Crocce}, {Fosalba}, {Gottl{\"o}ber}, {Klypin}, {Power}, {Tao}, and
  {Turchaninov}}]{Chuang:2015b}
{Chuang} CH, {Zhao} C, {Prada} F, {Munari} E, {Avila} S, {Izard} A, {Kitaura}
  FS, {Manera} M, {Monaco} P, {Murray} S, {Knebe} A, {Sc{\'o}ccola} CG, {Yepes}
  G, {Garcia-Bellido} J, {Mar{\'\i}n} FA, {M{\"u}ller} V, {Skibba} R, {Crocce}
  M, {Fosalba} P, {Gottl{\"o}ber} S, {Klypin} AA, {Power} C, {Tao} C,
  {Turchaninov} V (2015{\natexlab{b}}) {nIFTy cosmology: Galaxy/halo mock
  catalogue comparison project on clustering statistics}. \mnras
  452(1):686--700. \doi{10.1093/mnras/stv1289}.
  {\href{https://arxiv.org/abs/1412.7729}{{arXiv:1412.7729}}} {[astro-ph.CO]}

\bibitem[{{Chuang} et~al.(2019){Chuang}, {Yepes}, {Kitaura},
  {Pellejero-Ibanez}, {Rodr{\'\i}guez-Torres}, {Feng}, {Metcalf}, {Wechsler},
  {Zhao}, {To}, {Alam}, {Banerjee}, {DeRose}, {Giocoli}, {Knebe}, and
  {Reyes}}]{Chuang:2019}
{Chuang} CH, {Yepes} G, {Kitaura} FS, {Pellejero-Ibanez} M,
  {Rodr{\'\i}guez-Torres} S, {Feng} Y, {Metcalf} RB, {Wechsler} RH, {Zhao} C,
  {To} CH, {Alam} S, {Banerjee} A, {DeRose} J, {Giocoli} C, {Knebe} A, {Reyes}
  G (2019) {UNIT project: Universe N-body simulations for the Investigation of
  Theoretical models from galaxy surveys}. \mnras 487(1):48--59.
  \doi{10.1093/mnras/stz1233}.
  {\href{https://arxiv.org/abs/1811.02111}{{arXiv:1811.02111}}} {[astro-ph.CO]}

\bibitem[{{Clark} et~al.(2016){Clark}, {Lewis}, and {Scott}}]{Clark:2016}
{Clark} HA, {Lewis} GF, {Scott} P (2016) {Investigating dark matter
  substructure with pulsar timing - I. Constraints on ultracompact minihaloes}.
  \mnras 456(2):1394--1401. \doi{10.1093/mnras/stv2743}.
  {\href{https://arxiv.org/abs/1509.02938}{{arXiv:1509.02938}}} {[astro-ph.CO]}

\bibitem[{{Clifton} et~al.(2012){Clifton}, {Ferreira}, {Padilla}, and
  {Skordis}}]{Clifton:2012}
{Clifton} T, {Ferreira} PG, {Padilla} A, {Skordis} C (2012) {Modified gravity
  and cosmology}. \physrep 513(1):1--189. \doi{10.1016/j.physrep.2012.01.001}.
  {\href{https://arxiv.org/abs/1106.2476}{{arXiv:1106.2476}}} {[astro-ph.CO]}

\bibitem[{{Coates} et~al.(2021){Coates}, {Adamek}, {Bull}, {Guandalin}, and
  {Clarkson}}]{Coates:2021}
{Coates} L, {Adamek} J, {Bull} P, {Guandalin} C, {Clarkson} C (2021) {Observing
  relativistic features in large-scale structure surveys - II. Doppler
  magnification in an ensemble of relativistic simulations}. \mnras
  504(3):3534--3543. \doi{10.1093/mnras/stab1076}.
  {\href{https://arxiv.org/abs/2011.12936}{{arXiv:2011.12936}}} {[astro-ph.CO]}

\bibitem[{{Codis} et~al.(2012){Codis}, {Pichon}, {Devriendt}, {Slyz},
  {Pogosyan}, {Dubois}, and {Sousbie}}]{Codis:2012}
{Codis} S, {Pichon} C, {Devriendt} J, {Slyz} A, {Pogosyan} D, {Dubois} Y,
  {Sousbie} T (2012) {Connecting the cosmic web to the spin of dark haloes:
  implications for galaxy formation}. \mnras 427(4):3320--3336.
  \doi{10.1111/j.1365-2966.2012.21636.x}.
  {\href{https://arxiv.org/abs/1201.5794}{{arXiv:1201.5794}}} {[astro-ph.CO]}

\bibitem[{{Colavincenzo} et~al.(2019){Colavincenzo}, {Sefusatti}, {Monaco},
  {Blot}, {Crocce}, {Lippich}, {S{\'a}nchez}, {Alvarez}, {Agrawal}, {Avila},
  {Balaguera-Antol{\'\i}nez}, {Bond}, {Codis}, {Dalla Vecchia}, {Dorta},
  {Fosalba}, {Izard}, {Kitaura}, {Pellejero-Ibanez}, {Stein}, {Vakili}, and
  {Yepes}}]{Colavincenzo:2019}
{Colavincenzo} M, {Sefusatti} E, {Monaco} P, {Blot} L, {Crocce} M, {Lippich} M,
  {S{\'a}nchez} AG, {Alvarez} MA, {Agrawal} A, {Avila} S,
  {Balaguera-Antol{\'\i}nez} A, {Bond} R, {Codis} S, {Dalla Vecchia} C, {Dorta}
  A, {Fosalba} P, {Izard} A, {Kitaura} FS, {Pellejero-Ibanez} M, {Stein} G,
  {Vakili} M, {Yepes} G (2019) {Comparing approximate methods for mock
  catalogues and covariance matrices - III: bispectrum}. \mnras
  482(4):4883--4905. \doi{10.1093/mnras/sty2964}.
  {\href{https://arxiv.org/abs/1806.09499}{{arXiv:1806.09499}}} {[astro-ph.CO]}

\bibitem[{{Colberg} et~al.(2008){Colberg}, {Pearce}, {Foster}, {Platen},
  {Brunino}, {Neyrinck}, {Basilakos}, {Fairall}, {Feldman}, {Gottl{\"o}ber},
  {Hahn}, {Hoyle}, {M{\"u}ller}, {Nelson}, {Plionis}, {Porciani}, {Shandarin},
  {Vogeley}, and {van de Weygaert}}]{Colberg:2008}
{Colberg} JM, {Pearce} F, {Foster} C, {Platen} E, {Brunino} R, {Neyrinck} M,
  {Basilakos} S, {Fairall} A, {Feldman} H, {Gottl{\"o}ber} S, {Hahn} O, {Hoyle}
  F, {M{\"u}ller} V, {Nelson} L, {Plionis} M, {Porciani} C, {Shandarin} S,
  {Vogeley} MS, {van de Weygaert} R (2008) {The Aspen-Amsterdam void finder
  comparison project}. \mnras 387(2):933--944.
  \doi{10.1111/j.1365-2966.2008.13307.x}.
  {\href{https://arxiv.org/abs/0803.0918}{{arXiv:0803.0918}}} {[astro-ph]}

\bibitem[{{Cole} et~al.(1994){Cole}, {Aragon-Salamanca}, {Frenk}, {Navarro},
  and {Zepf}}]{Cole:1994}
{Cole} S, {Aragon-Salamanca} A, {Frenk} CS, {Navarro} JF, {Zepf} SE (1994) {A
  recipe for galaxy formation.} \mnras 271:781--806.
  \doi{10.1093/mnras/271.4.781}.
  {\href{https://arxiv.org/abs/astro-ph/9402001}{{arXiv:astro-ph/9402001}}}
  {[astro-ph]}

\bibitem[{{Cole} et~al.(2000){Cole}, {Lacey}, {Baugh}, and {Frenk}}]{Cole:2000}
{Cole} S, {Lacey} CG, {Baugh} CM, {Frenk} CS (2000) {Hierarchical galaxy
  formation}. \mnras 319(1):168--204. \doi{10.1046/j.1365-8711.2000.03879.x}.
  {\href{https://arxiv.org/abs/astro-ph/0007281}{{arXiv:astro-ph/0007281}}}
  {[astro-ph]}

\bibitem[{{Coles} and {Bieri}(2020)}]{Coles:2020}
{Coles} JP, {Bieri} R (2020) {An optimizing symbolic algebra approach for
  generating fast multipole method operators}. Computer Physics Communications
  251:107081. \doi{10.1016/j.cpc.2019.107081}.
  {\href{https://arxiv.org/abs/1811.06332}{{arXiv:1811.06332}}}
  {[physics.comp-ph]}

\bibitem[{{Coles} and {Jones}(1991)}]{Coles:1991}
{Coles} P, {Jones} B (1991) {A lognormal model for the cosmological mass
  distribution.} \mnras 248:1--13. \doi{10.1093/mnras/248.1.1}

\bibitem[{{Col{\'\i}n} et~al.(2008){Col{\'\i}n}, {Valenzuela}, and
  {Avila-Reese}}]{Colin:2008}
{Col{\'\i}n} P, {Valenzuela} O, {Avila-Reese} V (2008) {On the Structure of
  Dark Matter Halos at the Damping Scale of the Power Spectrum with and without
  Relict Velocities}. \apj 673(1):203--214. \doi{10.1086/524030}.
  {\href{https://arxiv.org/abs/0709.4027}{{arXiv:0709.4027}}} {[astro-ph]}

\bibitem[{{Colombi}(2020)}]{Colombi:2020}
{Colombi} S (2020) {Phase-space structure of protohalos: Vlasov versus
  Particle-Mesh}. arXiv e-prints arXiv:2012.04409.
  {\href{https://arxiv.org/abs/2012.04409}{{arXiv:2012.04409}}} {[astro-ph.CO]}

\bibitem[{{Colombi} and {Alard}(2017)}]{Colombi:2017}
{Colombi} S, {Alard} C (2017) {A `metric' semi-Lagrangian Vlasov-Poisson
  solver}. J Plasma Phys 83(3):705830302. \doi{10.1017/S0022377817000393}.
  {\href{https://arxiv.org/abs/1705.03750}{{arXiv:1705.03750}}}

\bibitem[{{Colombi} and {Touma}(2008)}]{Colombi:2008}
{Colombi} S, {Touma} J (2008) {Vlasov Poisson: The waterbag method revisited}.
  Communications in Nonlinear Science and Numerical Simulations 13(1):46--52.
  \doi{10.1016/j.cnsns.2007.03.012}

\bibitem[{{Colombi} and {Touma}(2014)}]{Colombi:2014}
{Colombi} S, {Touma} J (2014) {Vlasov-Poisson in 1D: waterbags}. \mnras
  441:2414--2432. \doi{10.1093/mnras/stu739}

\bibitem[{{Colombi} et~al.(2009){Colombi}, {Jaffe}, {Novikov}, and
  {Pichon}}]{Colombi:2009}
{Colombi} S, {Jaffe} A, {Novikov} D, {Pichon} C (2009) {Accurate estimators of
  power spectra in N-body simulations}. \mnras 393(2):511--526.
  \doi{10.1111/j.1365-2966.2008.14176.x}.
  {\href{https://arxiv.org/abs/0811.0313}{{arXiv:0811.0313}}} {[astro-ph]}

\bibitem[{{Conroy} et~al.(2006){Conroy}, {Wechsler}, and
  {Kravtsov}}]{Conroy:2006}
{Conroy} C, {Wechsler} RH, {Kravtsov} AV (2006) {Modeling Luminosity-dependent
  Galaxy Clustering through Cosmic Time}. \apj 647(1):201--214.
  \doi{10.1086/503602}.
  {\href{https://arxiv.org/abs/astro-ph/0512234}{{arXiv:astro-ph/0512234}}}
  {[astro-ph]}

\bibitem[{{Contreras} et~al.(2020{\natexlab{a}}){Contreras}, {Angulo}, and
  {Zennaro}}]{Contreras:2020}
{Contreras} S, {Angulo} R, {Zennaro} M (2020{\natexlab{a}}) {A flexible
  modelling of galaxy assembly bias}. arXiv e-prints arXiv:2005.03672.
  {\href{https://arxiv.org/abs/2005.03672}{{arXiv:2005.03672}}} {[astro-ph.GA]}

\bibitem[{{Contreras} et~al.(2020{\natexlab{b}}){Contreras}, {Angulo}, and
  {Zennaro}}]{Contreras:2020b}
{Contreras} S, {Angulo} R, {Zennaro} M (2020{\natexlab{b}}) {A flexible subhalo
  abundance matching model for galaxy clustering in redshift space}. arXiv
  e-prints arXiv:2012.06596.
  {\href{https://arxiv.org/abs/2012.06596}{{arXiv:2012.06596}}} {[astro-ph.CO]}

\bibitem[{{Contreras} et~al.(2020{\natexlab{c}}){Contreras}, {Angulo},
  {Zennaro}, {Aric{\`o}}, and {Pellejero-Iba{\~n}ez}}]{Contreras:2020a}
{Contreras} S, {Angulo} RE, {Zennaro} M, {Aric{\`o}} G, {Pellejero-Iba{\~n}ez}
  M (2020{\natexlab{c}}) {3 per cent-accurate predictions for the clustering of
  dark matter, haloes, and subhaloes, over a wide range of cosmologies and
  scales}. \mnras 499(4):4905--4917. \doi{10.1093/mnras/staa3117}.
  {\href{https://arxiv.org/abs/2001.03176}{{arXiv:2001.03176}}} {[astro-ph.CO]}

\bibitem[{{Cora} et~al.(2018){Cora}, {Vega-Mart{\'\i}nez}, {Hough}, {Ruiz},
  {Orsi}, {Mu{\~n}oz Arancibia}, {Gargiulo}, {Collacchioni}, {Padilla},
  {Gottl{\"o}ber}, and {Yepes}}]{Cora:2018}
{Cora} SA, {Vega-Mart{\'\i}nez} CA, {Hough} T, {Ruiz} AN, {Orsi} {\'A}A,
  {Mu{\~n}oz Arancibia} AM, {Gargiulo} ID, {Collacchioni} F, {Padilla} ND,
  {Gottl{\"o}ber} S, {Yepes} G (2018) {Semi-analytic galaxies - I. Synthesis of
  environmental and star-forming regulation mechanisms}. \mnras 479(1):2--24.
  \doi{10.1093/mnras/sty1131}.
  {\href{https://arxiv.org/abs/1801.03883}{{arXiv:1801.03883}}} {[astro-ph.GA]}

\bibitem[{{Crain} et~al.(2015){Crain}, {Schaye}, {Bower}, {Furlong},
  {Schaller}, {Theuns}, {Dalla Vecchia}, {Frenk}, {McCarthy}, {Helly},
  {Jenkins}, {Rosas-Guevara}, {White}, and {Trayford}}]{Crain:2015}
{Crain} RA, {Schaye} J, {Bower} RG, {Furlong} M, {Schaller} M, {Theuns} T,
  {Dalla Vecchia} C, {Frenk} CS, {McCarthy} IG, {Helly} JC, {Jenkins} A,
  {Rosas-Guevara} YM, {White} SDM, {Trayford} JW (2015) {The EAGLE simulations
  of galaxy formation: calibration of subgrid physics and model variations}.
  \mnras 450(2):1937--1961. \doi{10.1093/mnras/stv725}.
  {\href{https://arxiv.org/abs/1501.01311}{{arXiv:1501.01311}}} {[astro-ph.GA]}

\bibitem[{{Creminelli}(2003)}]{Creminelli:2003}
{Creminelli} P (2003) {On non-Gaussianities in single-field inflation}. \jcap
  2003(10):003. \doi{10.1088/1475-7516/2003/10/003}.
  {\href{https://arxiv.org/abs/astro-ph/0306122}{{arXiv:astro-ph/0306122}}}
  {[astro-ph]}

\bibitem[{{Creminelli} and {Vernizzi}(2017)}]{Creminelli:2017}
{Creminelli} P, {Vernizzi} F (2017) {Dark Energy after GW170817 and
  GRB170817A}. \prl 119(25):251302. \doi{10.1103/PhysRevLett.119.251302}.
  {\href{https://arxiv.org/abs/1710.05877}{{arXiv:1710.05877}}} {[astro-ph.CO]}

\bibitem[{{Creminelli} et~al.(2007){Creminelli}, {Senatore}, and
  {Zaldarriaga}}]{Creminelli:2007}
{Creminelli} P, {Senatore} L, {Zaldarriaga} M (2007) {Estimators for local
  non-Gaussianities}. \jcap 3:019. \doi{10.1088/1475-7516/2007/03/019}.
  {\href{https://arxiv.org/abs/astro-ph/0606001}{{astro-ph/0606001}}}

\bibitem[{{Crocce} and {Scoccimarro}(2006)}]{Crocce:2006b}
{Crocce} M, {Scoccimarro} R (2006) {Memory of initial conditions in
  gravitational clustering}. \prd 73(6):063520.
  \doi{10.1103/PhysRevD.73.063520}.
  {\href{https://arxiv.org/abs/astro-ph/0509419}{{arXiv:astro-ph/0509419}}}
  {[astro-ph]}

\bibitem[{{Crocce} et~al.(2006){Crocce}, {Pueblas}, and
  {Scoccimarro}}]{Crocce:2006}
{Crocce} M, {Pueblas} S, {Scoccimarro} R (2006) {Transients from initial
  conditions in cosmological simulations}. \mnras 373:369--381.
  \doi{10.1111/j.1365-2966.2006.11040.x}.
  {\href{https://arxiv.org/abs/astro-ph/0606505}{{astro-ph/0606505}}}

\bibitem[{{Crocce} et~al.(2010){Crocce}, {Fosalba}, {Castander}, and
  {Gazta{\~n}aga}}]{Crocce:2010}
{Crocce} M, {Fosalba} P, {Castander} FJ, {Gazta{\~n}aga} E (2010) {Simulating
  the Universe with MICE: the abundance of massive clusters}. \mnras
  403(3):1353--1367. \doi{10.1111/j.1365-2966.2009.16194.x}.
  {\href{https://arxiv.org/abs/0907.0019}{{arXiv:0907.0019}}} {[astro-ph.CO]}

\bibitem[{{Croton} et~al.(2016){Croton}, {Stevens}, {Tonini}, {Garel},
  {Bernyk}, {Bibiano}, {Hodkinson}, {Mutch}, {Poole}, and
  {Shattow}}]{Croton:2016}
{Croton} DJ, {Stevens} ARH, {Tonini} C, {Garel} T, {Bernyk} M, {Bibiano} A,
  {Hodkinson} L, {Mutch} SJ, {Poole} GB, {Shattow} GM (2016) {Semi-Analytic
  Galaxy Evolution (SAGE): Model Calibration and Basic Results}. \apjs
  222(2):22. \doi{10.3847/0067-0049/222/2/22}.
  {\href{https://arxiv.org/abs/1601.04709}{{arXiv:1601.04709}}} {[astro-ph.GA]}

\bibitem[{{Cui} et~al.(2008){Cui}, {Liu}, {Yang}, {Wang}, {Feng}, and
  {Springel}}]{Cui:2008}
{Cui} W, {Liu} L, {Yang} X, {Wang} Y, {Feng} L, {Springel} V (2008) {An Ideal
  Mass Assignment Scheme for Measuring the Power Spectrum with Fast Fourier
  Transforms}. \apj 687(2):738--744. \doi{10.1086/592079}.
  {\href{https://arxiv.org/abs/0804.0070}{{arXiv:0804.0070}}} {[astro-ph]}

\bibitem[{{Cui} et~al.(2010){Cui}, {Zhang}, and {Yang}}]{Cui:2010}
{Cui} W, {Zhang} P, {Yang} X (2010) {Nonlinearities in modified gravity
  cosmology: Signatures of modified gravity in the nonlinear matter power
  spectrum}. \prd 81(10):103528. \doi{10.1103/PhysRevD.81.103528}.
  {\href{https://arxiv.org/abs/1001.5184}{{arXiv:1001.5184}}} {[astro-ph.CO]}

\bibitem[{{Cyr-Racine} et~al.(2014){Cyr-Racine}, {de Putter}, {Raccanelli}, and
  {Sigurdson}}]{Cyr-Racine:2014}
{Cyr-Racine} FY, {de Putter} R, {Raccanelli} A, {Sigurdson} K (2014)
  {Constraints on large-scale dark acoustic oscillations from cosmology}. \prd
  89(6):063517. \doi{10.1103/PhysRevD.89.063517}.
  {\href{https://arxiv.org/abs/1310.3278}{{arXiv:1310.3278}}} {[astro-ph.CO]}

\bibitem[{{Cyr-Racine} et~al.(2016){Cyr-Racine}, {Sigurdson}, {Zavala},
  {Bringmann}, {Vogelsberger}, and {Pfrommer}}]{Cyr-Racine:2016}
{Cyr-Racine} FY, {Sigurdson} K, {Zavala} J, {Bringmann} T, {Vogelsberger} M,
  {Pfrommer} C (2016) {ETHOS---an effective theory of structure formation: From
  dark particle physics to the matter distribution of the Universe}. \prd
  93(12):123527. \doi{10.1103/PhysRevD.93.123527}.
  {\href{https://arxiv.org/abs/1512.05344}{{arXiv:1512.05344}}} {[astro-ph.CO]}

\bibitem[{{Dai} and {Seljak}(2020)}]{DaiSeljak:2020}
{Dai} B, {Seljak} U (2020) {Learning effective physical laws for generating
  cosmological hydrodynamics with Lagrangian Deep Learning}. arXiv e-prints
  arXiv:2010.02926.
  {\href{https://arxiv.org/abs/2010.02926}{{arXiv:2010.02926}}} {[astro-ph.CO]}

\bibitem[{{Dai} et~al.(2018){Dai}, {Feng}, and {Seljak}}]{Dai:2018}
{Dai} B, {Feng} Y, {Seljak} U (2018) {A gradient based method for modeling
  baryons and matter in halos of fast simulations}. \jcap 11:009.
  \doi{10.1088/1475-7516/2018/11/009}.
  {\href{https://arxiv.org/abs/1804.00671}{{arXiv:1804.00671}}}

\bibitem[{{Dakin} et~al.(2019{\natexlab{a}}){Dakin}, {Brandbyge}, {Hannestad},
  {Haugb{\O}lle}, and {Tram}}]{Dakin2017}
{Dakin} J, {Brandbyge} J, {Hannestad} S, {Haugb{\O}lle} T, {Tram} T
  (2019{\natexlab{a}}) {{\ensuremath{\nu}}CONCEPT: cosmological neutrino
  simulations from the non-linear Boltzmann hierarchy}. \jcap 2019(2):052.
  \doi{10.1088/1475-7516/2019/02/052}.
  {\href{https://arxiv.org/abs/1712.03944}{{arXiv:1712.03944}}} {[astro-ph.CO]}

\bibitem[{{Dakin} et~al.(2019{\natexlab{b}}){Dakin}, {Hannestad}, and
  {Tram}}]{Dakin:2019}
{Dakin} J, {Hannestad} S, {Tram} T (2019{\natexlab{b}}) {Fully relativistic
  treatment of decaying cold dark matter in N-body simulations}. \jcap
  2019(6):032. \doi{10.1088/1475-7516/2019/06/032}.
  {\href{https://arxiv.org/abs/1904.11773}{{arXiv:1904.11773}}} {[astro-ph.CO]}

\bibitem[{{Dalal} et~al.(2008){Dalal}, {Dor{\'e}}, {Huterer}, and
  {Shirokov}}]{Dalal:2008}
{Dalal} N, {Dor{\'e}} O, {Huterer} D, {Shirokov} A (2008) {Imprints of
  primordial non-Gaussianities on large-scale structure: Scale-dependent bias
  and abundance of virialized objects}. \prd 77(12):123514.
  \doi{10.1103/PhysRevD.77.123514}.
  {\href{https://arxiv.org/abs/0710.4560}{{arXiv:0710.4560}}} {[astro-ph]}

\bibitem[{{Dalal} et~al.(2010{\natexlab{a}}){Dalal}, {Lithwick}, and
  {Kuhlen}}]{Dalal:2010}
{Dalal} N, {Lithwick} Y, {Kuhlen} M (2010{\natexlab{a}}) {The Origin of Dark
  Matter Halo Profiles}. arXiv e-prints arXiv:1010.2539.
  {\href{https://arxiv.org/abs/1010.2539}{{arXiv:1010.2539}}} {[astro-ph.CO]}

\bibitem[{{Dalal} et~al.(2010{\natexlab{b}}){Dalal}, {Pen}, and
  {Seljak}}]{Dalal:2010b}
{Dalal} N, {Pen} UL, {Seljak} U (2010{\natexlab{b}}) {Large-scale BAO
  signatures of the smallest galaxies}. \jcap 2010(11):007.
  \doi{10.1088/1475-7516/2010/11/007}.
  {\href{https://arxiv.org/abs/1009.4704}{{arXiv:1009.4704}}} {[astro-ph.CO]}

\bibitem[{{Dav{\'e}} et~al.(2019){Dav{\'e}}, {Angl{\'e}s-Alc{\'a}zar},
  {Narayanan}, {Li}, {Rafieferantsoa}, and {Appleby}}]{Dave:2019}
{Dav{\'e}} R, {Angl{\'e}s-Alc{\'a}zar} D, {Narayanan} D, {Li} Q,
  {Rafieferantsoa} MH, {Appleby} S (2019) {SIMBA: Cosmological simulations with
  black hole growth and feedback}. \mnras 486(2):2827--2849.
  \doi{10.1093/mnras/stz937}.
  {\href{https://arxiv.org/abs/1901.10203}{{arXiv:1901.10203}}} {[astro-ph.GA]}

\bibitem[{{Daverio} et~al.(2019){Daverio}, {Dirian}, and
  {Mitsou}}]{Daverio:2019}
{Daverio} D, {Dirian} Y, {Mitsou} E (2019) {General relativistic cosmological
  N-body simulations. Part I. Time integration}. \jcap 2019(10):065.
  \doi{10.1088/1475-7516/2019/10/065}.
  {\href{https://arxiv.org/abs/1904.07841}{{arXiv:1904.07841}}} {[astro-ph.CO]}

\bibitem[{{Davis} et~al.(1985){Davis}, {Efstathiou}, {Frenk}, and
  {White}}]{Davis:1985}
{Davis} M, {Efstathiou} G, {Frenk} CS, {White} SDM (1985) {The evolution of
  large-scale structure in a universe dominated by cold dark matter}. \apj
  292:371--394. \doi{10.1086/163168}

\bibitem[{{De Lucia} et~al.(2010){De Lucia}, {Boylan-Kolchin}, {Benson},
  {Fontanot}, and {Monaco}}]{deLucia:2010}
{De Lucia} G, {Boylan-Kolchin} M, {Benson} AJ, {Fontanot} F, {Monaco} P (2010)
  {A semi-analytic model comparison - gas cooling and galaxy mergers}. \mnras
  406(3):1533--1552. \doi{10.1111/j.1365-2966.2010.16806.x}.
  {\href{https://arxiv.org/abs/1003.3021}{{arXiv:1003.3021}}} {[astro-ph.CO]}

\bibitem[{{De Martino} et~al.(2017){De Martino}, {Broadhurst}, {Tye}, {Chiueh},
  {Schive}, and {Lazkoz}}]{deMartino:2017}
{De Martino} I, {Broadhurst} T, {Tye} SHH, {Chiueh} T, {Schive} HY, {Lazkoz} R
  (2017) {Recognising Axionic Dark Matter by Compton and de-Broglie Scale
  Modulation of Pulsar Timing}. \prl 119:221103.
  \doi{10.1103/PhysRevLett.119.221103}.
  {\href{https://arxiv.org/abs/1705.04367}{{arXiv:1705.04367}}} {[astro-ph.CO]}

\bibitem[{{De Martino} et~al.(2020){De Martino}, {Broadhurst}, {Henry Tye},
  {Chiueh}, and {Schive}}]{deMartino:2020}
{De Martino} I, {Broadhurst} T, {Henry Tye} SH, {Chiueh} T, {Schive} HY (2020)
  {Dynamical evidence of a dark solitonic core of $10^{9} M_{\odot}$ in the
  Milky Way}. Physics of the Dark Universe 28:100503.
  \doi{10.1016/j.dark.2020.100503}.
  {\href{https://arxiv.org/abs/1807.08153}{{arXiv:1807.08153}}} {[astro-ph.GA]}

\bibitem[{{Debackere} et~al.(2019){Debackere}, {Schaye}, and
  {Hoekstra}}]{Debackere:2019}
{Debackere} SNB, {Schaye} J, {Hoekstra} H (2019) {The impact of the observed
  baryon distribution in haloes on the total matter power spectrum}. arXiv
  e-prints {\href{https://arxiv.org/abs/1908.05765}{{arXiv:1908.05765}}}

\bibitem[{{Dehnen}(2000)}]{Dehnen:2000}
{Dehnen} W (2000) {A Very Fast and Momentum-conserving Tree Code}. \apjl
  536(1):L39--L42. \doi{10.1086/312724}.
  {\href{https://arxiv.org/abs/astro-ph/0003209}{{arXiv:astro-ph/0003209}}}
  {[astro-ph]}

\bibitem[{{Dehnen}(2002)}]{Dehnen:2002}
{Dehnen} W (2002) {A Hierarchical \&lt;E10\&gt;O\&lt;/E10\&gt;(N) Force
  Calculation Algorithm}. J Comput Phys 179(1):27--42.
  \doi{10.1006/jcph.2002.7026}.
  {\href{https://arxiv.org/abs/astro-ph/0202512}{{arXiv:astro-ph/0202512}}}
  {[astro-ph]}

\bibitem[{{Dehnen}(2014)}]{Dehnen:2014}
{Dehnen} W (2014) {A fast multipole method for stellar dynamics}. Computational
  Astrophysics and Cosmology 1:1. \doi{10.1186/s40668-014-0001-7}.
  {\href{https://arxiv.org/abs/1405.2255}{{arXiv:1405.2255}}} {[astro-ph.IM]}

\bibitem[{{Dehnen} and {Read}(2011)}]{Dehnen:2011}
{Dehnen} W, {Read} JI (2011) {N-body simulations of gravitational dynamics}.
  EPJ Plus 126:55. \doi{10.1140/epjp/i2011-11055-3}.
  {\href{https://arxiv.org/abs/1105.1082}{{arXiv:1105.1082}}} {[astro-ph.IM]}

\bibitem[{{Delfino} et~al.(2021){Delfino}, {Scoccola}, {Cora}, {Vega-Martinez},
  and {Gargiulo}}]{Delfino:2021}
{Delfino} FM, {Scoccola} CG, {Cora} SA, {Vega-Martinez} CA, {Gargiulo} ID
  (2021) {Orphan galaxies in semi-analytic models}. arXiv e-prints
  arXiv:2102.01837.
  {\href{https://arxiv.org/abs/2102.01837}{{arXiv:2102.01837}}} {[astro-ph.CO]}

\bibitem[{{Delos}(2019)}]{Delos:2019a}
{Delos} MS (2019) {Evolution of dark matter microhalos through stellar
  encounters}. \prd 100(8):083529. \doi{10.1103/PhysRevD.100.083529}.
  {\href{https://arxiv.org/abs/1907.13133}{{arXiv:1907.13133}}} {[astro-ph.CO]}

\bibitem[{{Delos} et~al.(2018{\natexlab{a}}){Delos}, {Erickcek}, {Bailey}, and
  {Alvarez}}]{Delos:2018a}
{Delos} MS, {Erickcek} AL, {Bailey} AP, {Alvarez} MA (2018{\natexlab{a}}) {Are
  ultracompact minihalos really ultracompact?} \prd 97(4):041303.
  \doi{10.1103/PhysRevD.97.041303}.
  {\href{https://arxiv.org/abs/1712.05421}{{arXiv:1712.05421}}} {[astro-ph.CO]}

\bibitem[{{Delos} et~al.(2018{\natexlab{b}}){Delos}, {Erickcek}, {Bailey}, and
  {Alvarez}}]{Delos:2018b}
{Delos} MS, {Erickcek} AL, {Bailey} AP, {Alvarez} MA (2018{\natexlab{b}})
  {Density profiles of ultracompact minihalos: Implications for constraining
  the primordial power spectrum}. \prd 98(6):063527.
  \doi{10.1103/PhysRevD.98.063527}.
  {\href{https://arxiv.org/abs/1806.07389}{{arXiv:1806.07389}}} {[astro-ph.CO]}

\bibitem[{{Delos} et~al.(2019){Delos}, {Bruff}, and {Erickcek}}]{Delos:2019b}
{Delos} MS, {Bruff} M, {Erickcek} AL (2019) {Predicting the density profiles of
  the first halos}. \prd 100(2):023523. \doi{10.1103/PhysRevD.100.023523}.
  {\href{https://arxiv.org/abs/1905.05766}{{arXiv:1905.05766}}} {[astro-ph.CO]}

\bibitem[{{Demianski} et~al.(2005){Demianski}, {Golda}, and
  {Woszczyna}}]{Demianski:2005}
{Demianski} M, {Golda} ZA, {Woszczyna} A (2005) {Evolution of density
  perturbations in a realistic universe}. General Relativity and Gravitation
  37(12):2063--2082. \doi{10.1007/s10714-005-0180-2}.
  {\href{https://arxiv.org/abs/gr-qc/0504089}{{arXiv:gr-qc/0504089}}} {[gr-qc]}

\bibitem[{{DeRose} et~al.(2019){DeRose}, {Wechsler}, {Tinker}, {Becker}, {Mao},
  {McClintock}, {McLaughlin}, {Rozo}, and {Zhai}}]{DeRose:2019}
{DeRose} J, {Wechsler} RH, {Tinker} JL, {Becker} MR, {Mao} YY, {McClintock} T,
  {McLaughlin} S, {Rozo} E, {Zhai} Z (2019) {The AEMULUS Project. I. Numerical
  Simulations for Precision Cosmology}. \apj 875(1):69.
  \doi{10.3847/1538-4357/ab1085}.
  {\href{https://arxiv.org/abs/1804.05865}{{arXiv:1804.05865}}} {[astro-ph.CO]}

\bibitem[{{DES Collaboration} et~al.(2021){DES Collaboration}, {Abbott},
  {Aguena}, {Alarcon}, {Allam}, {Alves}, {Amon}, {Andrade-Oliveira}, {Annis},
  {Avila}, {Bacon}, {Baxter}, {Bechtol}, {Becker}, {Bernstein}, {Bhargava},
  {Birrer}, {Blazek}, {Brandao-Souza}, {Bridle}, {Brooks}, {Buckley-Geer},
  {Burke}, {Camacho}, {Campos}, {Carnero Rosell}, {Carrasco Kind}, {Carretero},
  {Castander}, {Cawthon}, {Chang}, {Chen}, {Chen}, {Choi}, {Conselice},
  {Cordero}, {Costanzi}, {Crocce}, {da Costa}, {da Silva Pereira}, {Davis},
  {Davis}, {De Vicente}, {DeRose}, {Desai}, {Di Valentino}, {Diehl},
  {Dietrich}, {Dodelson}, {Doel}, {Doux}, {Drlica-Wagner}, {Eckert}, {Eifler},
  {Elsner}, {Elvin-Poole}, {Everett}, {Evrard}, {Fang}, {Farahi}, {Fernandez},
  {Ferrero}, {Fert{\'e}}, {Fosalba}, {Friedrich}, {Frieman},
  {Garc{\'\i}a-Bellido}, {Gatti}, {Gaztanaga}, {Gerdes}, {Giannantonio},
  {Giannini}, {Gruen}, {Gruendl}, {Gschwend}, {Gutierrez}, {Harrison},
  {Hartley}, {Herner}, {Hinton}, {Hollowood}, {Honscheid}, {Hoyle}, {Huff},
  {Huterer}, {Jain}, {James}, {Jarvis}, {Jeffrey}, {Jeltema}, {Kovacs},
  {Krause}, {Kron}, {Kuehn}, {Kuropatkin}, {Lahav}, {Leget}, {Lemos}, {Liddle},
  {Lidman}, {Lima}, {Lin}, {MacCrann}, {Maia}, {Marshall}, {Martini},
  {McCullough}, {Melchior}, {Mena-Fern{\'a}ndez}, {Menanteau}, {Miquel},
  {Mohr}, {Morgan}, {Muir}, {Myles}, {Nadathur}, {Navarro-Alsina}, {Nichol},
  {Ogando}, {Omori}, {Palmese}, {Pandey}, {Park}, {Paz-Chinch{\'o}n},
  {Petravick}, {Pieres}, {Plazas Malag{\'o}n}, {Porredon}, {Prat}, {Raveri},
  {Rodriguez-Monroy}, {Rollins}, {Romer}, {Roodman}, {Rosenfeld}, {Ross},
  {Rykoff}, {Samuroff}, {S{\'a}nchez}, {Sanchez}, {Sanchez}, {Sanchez Cid},
  {Scarpine}, {Schubnell}, {Scolnic}, {Secco}, {Serrano}, {Sevilla-Noarbe},
  {Sheldon}, {Shin}, {Smith}, {Soares-Santos}, {Suchyta}, {Swanson}, {Tabbutt},
  {Tarle}, {Thomas}, {To}, {Troja}, {Troxel}, {Tucker}, {Tutusaus}, {Varga},
  {Walker}, {Weaverdyck}, {Weller}, {Yanny}, {Yin}, {Zhang}, and
  {Zuntz}}]{Abbott:2021}
{DES Collaboration}, {Abbott} TMC, {Aguena} M, {Alarcon} A, {Allam} S, {Alves}
  O, {Amon} A, {Andrade-Oliveira} F, {Annis} J, {Avila} S, {Bacon} D, {Baxter}
  E, {Bechtol} K, {Becker} MR, {Bernstein} GM, {Bhargava} S, {Birrer} S,
  {Blazek} J, {Brandao-Souza} A, {Bridle} SL, {Brooks} D, {Buckley-Geer} E,
  {Burke} DL, {Camacho} H, {Campos} A, {Carnero Rosell} A, {Carrasco Kind} M,
  {Carretero} J, {Castander} FJ, {Cawthon} R, {Chang} C, {Chen} A, {Chen} R,
  {Choi} A, {Conselice} C, {Cordero} J, {Costanzi} M, {Crocce} M, {da Costa}
  LN, {da Silva Pereira} ME, {Davis} C, {Davis} TM, {De Vicente} J, {DeRose} J,
  {Desai} S, {Di Valentino} E, {Diehl} HT, {Dietrich} JP, {Dodelson} S, {Doel}
  P, {Doux} C, {Drlica-Wagner} A, {Eckert} K, {Eifler} TF, {Elsner} F,
  {Elvin-Poole} J, {Everett} S, {Evrard} AE, {Fang} X, {Farahi} A, {Fernandez}
  E, {Ferrero} I, {Fert{\'e}} A, {Fosalba} P, {Friedrich} O, {Frieman} J,
  {Garc{\'\i}a-Bellido} J, {Gatti} M, {Gaztanaga} E, {Gerdes} DW,
  {Giannantonio} T, {Giannini} G, {Gruen} D, {Gruendl} RA, {Gschwend} J,
  {Gutierrez} G, {Harrison} I, {Hartley} WG, {Herner} K, {Hinton} SR,
  {Hollowood} DL, {Honscheid} K, {Hoyle} B, {Huff} EM, {Huterer} D, {Jain} B,
  {James} DJ, {Jarvis} M, {Jeffrey} N, {Jeltema} T, {Kovacs} A, {Krause} E,
  {Kron} R, {Kuehn} K, {Kuropatkin} N, {Lahav} O, {Leget} PF, {Lemos} P,
  {Liddle} AR, {Lidman} C, {Lima} M, {Lin} H, {MacCrann} N, {Maia} MAG,
  {Marshall} JL, {Martini} P, {McCullough} J, {Melchior} P,
  {Mena-Fern{\'a}ndez} J, {Menanteau} F, {Miquel} R, {Mohr} JJ, {Morgan} R,
  {Muir} J, {Myles} J, {Nadathur} S, {Navarro-Alsina} A, {Nichol} RC, {Ogando}
  RLC, {Omori} Y, {Palmese} A, {Pandey} S, {Park} Y, {Paz-Chinch{\'o}n} F,
  {Petravick} D, {Pieres} A, {Plazas Malag{\'o}n} AA, {Porredon} A, {Prat} J,
  {Raveri} M, {Rodriguez-Monroy} M, {Rollins} RP, {Romer} AK, {Roodman} A,
  {Rosenfeld} R, {Ross} AJ, {Rykoff} ES, {Samuroff} S, {S{\'a}nchez} C,
  {Sanchez} E, {Sanchez} J, {Sanchez Cid} D, {Scarpine} V, {Schubnell} M,
  {Scolnic} D, {Secco} LF, {Serrano} S, {Sevilla-Noarbe} I, {Sheldon} E, {Shin}
  T, {Smith} M, {Soares-Santos} M, {Suchyta} E, {Swanson} MEC, {Tabbutt} M,
  {Tarle} G, {Thomas} D, {To} C, {Troja} A, {Troxel} MA, {Tucker} DL,
  {Tutusaus} I, {Varga} TN, {Walker} AR, {Weaverdyck} N, {Weller} J, {Yanny} B,
  {Yin} B, {Zhang} Y, {Zuntz} J (2021) {Dark Energy Survey Year 3 Results:
  Cosmological Constraints from Galaxy Clustering and Weak Lensing}. arXiv
  e-prints arXiv:2105.13549.
  {\href{https://arxiv.org/abs/2105.13549}{{arXiv:2105.13549}}} {[astro-ph.CO]}

\bibitem[{{DESI Collaboration} et~al.(2016){DESI Collaboration}, {Aghamousa},
  {Aguilar}, {Ahlen}, {Alam}, {Allen}, {Allende Prieto}, {Annis}, {Bailey},
  {Balland}, and et~al.}]{DESI}
{DESI Collaboration}, {Aghamousa} A, {Aguilar} J, {Ahlen} S, {Alam} S, {Allen}
  LE, {Allende Prieto} C, {Annis} J, {Bailey} S, {Balland} C, et~al (2016) {The
  DESI Experiment Part I: Science,Targeting, and Survey Design}. arXiv e-prints
  arXiv:1611.00036.
  {\href{https://arxiv.org/abs/1611.00036}{{arXiv:1611.00036}}} {[astro-ph.IM]}

\bibitem[{{Desjacques} and {Nusser}(2019)}]{Desjacques:2019}
{Desjacques} V, {Nusser} A (2019) {Axion core-halo mass and the black hole-halo
  mass relation: constraints on a few parsec scales}. \mnras 488(4):4497--4503.
  \doi{10.1093/mnras/stz1978}.
  {\href{https://arxiv.org/abs/1905.03450}{{arXiv:1905.03450}}} {[astro-ph.CO]}

\bibitem[{{Desjacques} and {Seljak}(2010)}]{Desjacques:2010}
{Desjacques} V, {Seljak} U (2010) {Signature of primordial non-Gaussianity of
  {\ensuremath{\phi}}$^{3}$ type in the mass function and bias of dark matter
  haloes}. \prd 81(2):023006. \doi{10.1103/PhysRevD.81.023006}.
  {\href{https://arxiv.org/abs/0907.2257}{{arXiv:0907.2257}}} {[astro-ph.CO]}

\bibitem[{{Desjacques} et~al.(2009){Desjacques}, {Seljak}, and
  {Iliev}}]{Desjacques:2009}
{Desjacques} V, {Seljak} U, {Iliev} IT (2009) {Scale-dependent bias induced by
  local non-Gaussianity: a comparison to N-body simulations}. \mnras
  396(1):85--96. \doi{10.1111/j.1365-2966.2009.14721.x}.
  {\href{https://arxiv.org/abs/0811.2748}{{arXiv:0811.2748}}} {[astro-ph]}

\bibitem[{{Desjacques} et~al.(2018){Desjacques}, {Jeong}, and
  {Schmidt}}]{Desjacques:2018}
{Desjacques} V, {Jeong} D, {Schmidt} F (2018) {Large-scale galaxy bias}.
  \physrep 733:1--193. \doi{10.1016/j.physrep.2017.12.002}.
  {\href{https://arxiv.org/abs/1611.09787}{{arXiv:1611.09787}}} {[astro-ph.CO]}

\bibitem[{{Despali} et~al.(2013){Despali}, {Tormen}, and
  {Sheth}}]{Despali:2013}
{Despali} G, {Tormen} G, {Sheth} RK (2013) {Ellipsoidal halo finders and
  implications for models of triaxial halo formation}. \mnras
  431(2):1143--1159. \doi{10.1093/mnras/stt235}.
  {\href{https://arxiv.org/abs/1212.4157}{{arXiv:1212.4157}}} {[astro-ph.CO]}

\bibitem[{{Despali} et~al.(2016){Despali}, {Giocoli}, {Angulo}, {Tormen},
  {Sheth}, {Baso}, and {Moscardini}}]{Despali:2016}
{Despali} G, {Giocoli} C, {Angulo} RE, {Tormen} G, {Sheth} RK, {Baso} G,
  {Moscardini} L (2016) {The universality of the virial halo mass function and
  models for non-universality of other halo definitions}. \mnras
  456(3):2486--2504. \doi{10.1093/mnras/stv2842}.
  {\href{https://arxiv.org/abs/1507.05627}{{arXiv:1507.05627}}} {[astro-ph.CO]}

\bibitem[{{D{\'\i}az Rivero} et~al.(2018){D{\'\i}az Rivero}, {Dvorkin},
  {Cyr-Racine}, {Zavala}, and {Vogelsberger}}]{DiazRivero:2018}
{D{\'\i}az Rivero} A, {Dvorkin} C, {Cyr-Racine} FY, {Zavala} J, {Vogelsberger}
  M (2018) {Gravitational lensing and the power spectrum of dark matter
  substructure: Insights from the ETHOS N -body simulations}. \prd
  98(10):103517. \doi{10.1103/PhysRevD.98.103517}.
  {\href{https://arxiv.org/abs/1809.00004}{{arXiv:1809.00004}}} {[astro-ph.CO]}

\bibitem[{{Diemand} et~al.(2004{\natexlab{a}}){Diemand}, {Moore}, and
  {Stadel}}]{DiemandVel:2004}
{Diemand} J, {Moore} B, {Stadel} J (2004{\natexlab{a}}) {Velocity and spatial
  biases in cold dark matter subhalo distributions}. \mnras 352(2):535--546.
  \doi{10.1111/j.1365-2966.2004.07940.x}.
  {\href{https://arxiv.org/abs/astro-ph/0402160}{{arXiv:astro-ph/0402160}}}
  {[astro-ph]}

\bibitem[{{Diemand} et~al.(2004{\natexlab{b}}){Diemand}, {Moore}, {Stadel}, and
  {Kazantzidis}}]{Diemand:2004}
{Diemand} J, {Moore} B, {Stadel} J, {Kazantzidis} S (2004{\natexlab{b}})
  {Two-body relaxation in cold dark matter simulations}. \mnras
  348(3):977--986. \doi{10.1111/j.1365-2966.2004.07424.x}.
  {\href{https://arxiv.org/abs/astro-ph/0304549}{{arXiv:astro-ph/0304549}}}
  {[astro-ph]}

\bibitem[{{Diemand} et~al.(2005){Diemand}, {Moore}, and
  {Stadel}}]{Diemand:2005}
{Diemand} J, {Moore} B, {Stadel} J (2005) {Earth-mass dark-matter haloes as the
  first structures in the early Universe}. \nat 433(7024):389--391.
  \doi{10.1038/nature03270}.
  {\href{https://arxiv.org/abs/astro-ph/0501589}{{arXiv:astro-ph/0501589}}}
  {[astro-ph]}

\bibitem[{{Diemand} et~al.(2006){Diemand}, {Kuhlen}, and
  {Madau}}]{Diemand:2006}
{Diemand} J, {Kuhlen} M, {Madau} P (2006) {Early Supersymmetric Cold Dark
  Matter Substructure}. \apj 649(1):1--13. \doi{10.1086/506377}.
  {\href{https://arxiv.org/abs/astro-ph/0603250}{{arXiv:astro-ph/0603250}}}
  {[astro-ph]}

\bibitem[{{Diemer}(2020)}]{Diemer:2020}
{Diemer} B (2020) {Universal at Last? The Splashback Mass Function of Dark
  Matter Halos}. \apj 903(2):87. \doi{10.3847/1538-4357/abbf52}.
  {\href{https://arxiv.org/abs/2007.10346}{{arXiv:2007.10346}}} {[astro-ph.CO]}

\bibitem[{{Diemer} and {Kravtsov}(2014)}]{DiemerKravtsov:2014}
{Diemer} B, {Kravtsov} AV (2014) {Dependence of the Outer Density Profiles of
  Halos on Their Mass Accretion Rate}. \apj 789(1):1.
  \doi{10.1088/0004-637X/789/1/1}.
  {\href{https://arxiv.org/abs/1401.1216}{{arXiv:1401.1216}}} {[astro-ph.CO]}

\bibitem[{{Diemer} et~al.(2013){Diemer}, {More}, and {Kravtsov}}]{Diemer:2013}
{Diemer} B, {More} S, {Kravtsov} AV (2013) {The Pseudo-evolution of Halo Mass}.
  \apj 766(1):25. \doi{10.1088/0004-637X/766/1/25}.
  {\href{https://arxiv.org/abs/1207.0816}{{arXiv:1207.0816}}} {[astro-ph.CO]}

\bibitem[{{Doro} et~al.(2013){Doro}, {Conrad}, {Emmanoulopoulos},
  {S{\`a}nchez-Conde}, {Barrio}, {Birsin}, {Bolmont}, {Brun}, {Colafrancesco},
  {Connell}, {Contreras}, {Daniel}, {Fornasa}, {Gaug}, {Glicenstein},
  {Gonz{\'a}lez-Mu{\~n}oz}, {Hassan}, {Horns}, {Jacholkowska}, {Jahn},
  {Mazini}, {Mirabal}, {Moralejo}, {Moulin}, {Nieto}, {Ripken}, {Sandaker},
  {Schwanke}, {Spengler}, {Stamerra}, {Viana}, {Zechlin}, {Zimmer}, and {CTA
  Consortium}}]{Doro:2013}
{Doro} M, {Conrad} J, {Emmanoulopoulos} D, {S{\`a}nchez-Conde} MA, {Barrio} JA,
  {Birsin} E, {Bolmont} J, {Brun} P, {Colafrancesco} S, {Connell} SH,
  {Contreras} JL, {Daniel} MK, {Fornasa} M, {Gaug} M, {Glicenstein} JF,
  {Gonz{\'a}lez-Mu{\~n}oz} A, {Hassan} T, {Horns} D, {Jacholkowska} A, {Jahn}
  C, {Mazini} R, {Mirabal} N, {Moralejo} A, {Moulin} E, {Nieto} D, {Ripken} J,
  {Sandaker} H, {Schwanke} U, {Spengler} G, {Stamerra} A, {Viana} A, {Zechlin}
  HS, {Zimmer} S, {CTA Consortium} (2013) {Dark matter and fundamental physics
  with the Cherenkov Telescope Array}. Astroparticle Physics 43:189--214.
  \doi{10.1016/j.astropartphys.2012.08.002}.
  {\href{https://arxiv.org/abs/1208.5356}{{arXiv:1208.5356}}} {[astro-ph.IM]}

\bibitem[{{Doroshkevich} et~al.(1973){Doroshkevich}, {Ryaben'kii}, and
  {Shandarin}}]{Doroshkevich:1973}
{Doroshkevich} AG, {Ryaben'kii} VS, {Shandarin} SF (1973) {Nonlinear theory of
  the development of potential perturbations}. Astrophysics 9:144--153.
  \doi{10.1007/BF01011421}

\bibitem[{{Doroshkevich} et~al.(1980){Doroshkevich}, {Kotok}, {Poliudov},
  {Shandarin}, {Sigov}, and {Novikov}}]{Doroshkevich:1980}
{Doroshkevich} AG, {Kotok} EV, {Poliudov} AN, {Shandarin} SF, {Sigov} IS,
  {Novikov} ID (1980) {Two-dimensional simulation of the gravitational system
  dynamics and formation of the large-scale structure of the universe}. \mnras
  192:321--337. \doi{10.1093/mnras/192.2.321}

\bibitem[{{Doumler} et~al.(2013{\natexlab{a}}){Doumler}, {Gottl{\"o}ber},
  {Hoffman}, and {Courtois}}]{Doumler:2013b}
{Doumler} T, {Gottl{\"o}ber} S, {Hoffman} Y, {Courtois} H (2013{\natexlab{a}})
  {Reconstructing cosmological initial conditions from galaxy peculiar
  velocities - III. Constrained simulations}. \mnras 430(2):912--923.
  \doi{10.1093/mnras/sts614}.
  {\href{https://arxiv.org/abs/1212.2810}{{arXiv:1212.2810}}} {[astro-ph.CO]}

\bibitem[{{Doumler} et~al.(2013{\natexlab{b}}){Doumler}, {Hoffman}, {Courtois},
  and {Gottl{\"o}ber}}]{Doumler:2013a}
{Doumler} T, {Hoffman} Y, {Courtois} H, {Gottl{\"o}ber} S (2013{\natexlab{b}})
  {Reconstructing cosmological initial conditions from galaxy peculiar
  velocities - I. Reverse Zeldovich Approximation}. \mnras 430(2):888--901.
  \doi{10.1093/mnras/sts613}.
  {\href{https://arxiv.org/abs/1212.2806}{{arXiv:1212.2806}}} {[astro-ph.CO]}

\bibitem[{{Dubois} et~al.(2014){Dubois}, {Pichon}, {Welker}, {Le Borgne},
  {Devriendt}, {Laigle}, {Codis}, {Pogosyan}, {Arnouts}, {Benabed}, {Bertin},
  {Blaizot}, {Bouchet}, {Cardoso}, {Colombi}, {de Lapparent}, {Desjacques},
  {Gavazzi}, {Kassin}, {Kimm}, {McCracken}, {Milliard}, {Peirani}, {Prunet},
  {Rouberol}, {Silk}, {Slyz}, {Sousbie}, {Teyssier}, {Tresse}, {Treyer},
  {Vibert}, and {Volonteri}}]{Dubois:2014}
{Dubois} Y, {Pichon} C, {Welker} C, {Le Borgne} D, {Devriendt} J, {Laigle} C,
  {Codis} S, {Pogosyan} D, {Arnouts} S, {Benabed} K, {Bertin} E, {Blaizot} J,
  {Bouchet} F, {Cardoso} JF, {Colombi} S, {de Lapparent} V, {Desjacques} V,
  {Gavazzi} R, {Kassin} S, {Kimm} T, {McCracken} H, {Milliard} B, {Peirani} S,
  {Prunet} S, {Rouberol} S, {Silk} J, {Slyz} A, {Sousbie} T, {Teyssier} R,
  {Tresse} L, {Treyer} M, {Vibert} D, {Volonteri} M (2014) {Dancing in the
  dark: galactic properties trace spin swings along the cosmic web}. \mnras
  444:1453--1468. \doi{10.1093/mnras/stu1227}.
  {\href{https://arxiv.org/abs/1402.1165}{{arXiv:1402.1165}}}

\bibitem[{{Duffy} and {van Bibber}(2009)}]{Duffy:2009}
{Duffy} LD, {van Bibber} K (2009) {Axions as dark matter particles}. New J Phys
  11(10):105008. \doi{10.1088/1367-2630/11/10/105008}.
  {\href{https://arxiv.org/abs/0904.3346}{{arXiv:0904.3346}}} {[hep-ph]}

\bibitem[{{Dupuy} and {Bernardeau}(2014)}]{Dupuy:2014}
{Dupuy} H, {Bernardeau} F (2014) {Describing massive neutrinos in cosmology as
  a collection of independent flows}. \jcap 1:030.
  \doi{10.1088/1475-7516/2014/01/030}.
  {\href{https://arxiv.org/abs/1311.5487}{{arXiv:1311.5487}}}

\bibitem[{{Dvali} et~al.(2000){Dvali}, {Gabadadze}, and {Porrati}}]{Dvali:2000}
{Dvali} G, {Gabadadze} G, {Porrati} M (2000) {4D gravity on a brane in 5D
  Minkowski space}. Phys Lett B 485(1-3):208--214.
  \doi{10.1016/S0370-2693(00)00669-9}.
  {\href{https://arxiv.org/abs/hep-th/0005016}{{arXiv:hep-th/0005016}}}
  {[hep-th]}

\bibitem[{{Earn} and {Tremaine}(1992)}]{Earn:1992}
{Earn} DJD, {Tremaine} S (1992) {Exact numerical studies of Hamiltonian maps:
  Iterating without roundoff error}. Physica D Nonlinear Phenomena 56(1):1--22.
  \doi{10.1016/0167-2789(92)90047-Q}

\bibitem[{{East} et~al.(2018){East}, {Wojtak}, and {Abel}}]{East:2018}
{East} WE, {Wojtak} R, {Abel} T (2018) {Comparing fully general relativistic
  and Newtonian calculations of structure formation}. \prd 97(4):043509.
  \doi{10.1103/PhysRevD.97.043509}.
  {\href{https://arxiv.org/abs/1711.06681}{{arXiv:1711.06681}}} {[astro-ph.CO]}

\bibitem[{{Eberhardt} et~al.(2020){Eberhardt}, {Banerjee}, {Kopp}, and
  {Abel}}]{Eberhardt:2020}
{Eberhardt} A, {Banerjee} A, {Kopp} M, {Abel} T (2020) {Investigating the use
  of field solvers for simulating classical systems}. \prd 101(4):043011.
  \doi{10.1103/PhysRevD.101.043011}.
  {\href{https://arxiv.org/abs/2001.05791}{{arXiv:2001.05791}}}
  {[physics.comp-ph]}

\bibitem[{{Edwards} et~al.(2018){Edwards}, {Kendall}, {Hotchkiss}, and
  {Easther}}]{Edwards:2018}
{Edwards} F, {Kendall} E, {Hotchkiss} S, {Easther} R (2018) {PyUltraLight: a
  pseudo-spectral solver for ultralight dark matter dynamics}. \jcap
  2018(10):027. \doi{10.1088/1475-7516/2018/10/027}.
  {\href{https://arxiv.org/abs/1807.04037}{{arXiv:1807.04037}}} {[astro-ph.CO]}

\bibitem[{{Eggemeier} et~al.(2021){Eggemeier}, {Scoccimarro}, {Smith},
  {Crocce}, {Pezzotta}, and {S{\'a}nchez}}]{Eggemeier:2021}
{Eggemeier} A, {Scoccimarro} R, {Smith} RE, {Crocce} M, {Pezzotta} A,
  {S{\'a}nchez} AG (2021) {Testing one-loop galaxy bias: joint analysis of
  power spectrum and bispectrum}. arXiv e-prints arXiv:2102.06902.
  {\href{https://arxiv.org/abs/2102.06902}{{arXiv:2102.06902}}} {[astro-ph.CO]}

\bibitem[{{Eggemeier} et~al.(2020){Eggemeier}, {Redondo}, {Dolag}, {Niemeyer},
  and {Vaquero}}]{Eggemeier:2020}
{Eggemeier} B, {Redondo} J, {Dolag} K, {Niemeyer} JC, {Vaquero} A (2020) {First
  Simulations of Axion Minicluster Halos}. \prl 125(4):041301.
  \doi{10.1103/PhysRevLett.125.041301}.
  {\href{https://arxiv.org/abs/1911.09417}{{arXiv:1911.09417}}} {[astro-ph.CO]}

\bibitem[{{Ehlers}(1971)}]{Ehlers:1971}
{Ehlers} J (1971) {Probl\`eme de Cauchy pour le syst\`eme
  int\'egro-diff\'erentiel d'Einstein-Liouville. (Cauchy problem for the
  Einstein-Liouville integro-differential system).} In: General Relativity and
  Cosmology. vol~21. Universit\'e Joseph Fourier, Grenoble; Association des
  Annales de l'Institut Fourier, Saint-Martin d'H\`eres

\bibitem[{{Eisenstein} and {Hut}(1998)}]{Eisenstein:1998}
{Eisenstein} DJ, {Hut} P (1998) {HOP: A New Group-Finding Algorithm for N-Body
  Simulations}. \apj 498(1):137--142. \doi{10.1086/305535}.
  {\href{https://arxiv.org/abs/astro-ph/9712200}{{arXiv:astro-ph/9712200}}}
  {[astro-ph]}

\bibitem[{{El Zant}(2013)}]{ElZant:2013}
{El Zant} AA (2013) {Dark Halos and Elliptical Galaxies as Marginally Stable
  Dynamical Systems}. \apj 779(1):64. \doi{10.1088/0004-637X/779/1/64}.
  {\href{https://arxiv.org/abs/1311.0129}{{arXiv:1311.0129}}} {[astro-ph.GA]}

\bibitem[{{El-Zant} et~al.(2019){El-Zant}, {Everitt}, and
  {Kassem}}]{El-Zant:2019}
{El-Zant} AA, {Everitt} MJ, {Kassem} SM (2019) {Errors, chaos, and the
  collisionless limit}. \mnras 484(2):1456--1474. \doi{10.1093/mnras/stz029}.
  {\href{https://arxiv.org/abs/1804.06920}{{arXiv:1804.06920}}} {[astro-ph.IM]}

\bibitem[{{Elahi} et~al.(2019){Elahi}, {Ca{\~n}as}, {Poulton}, {Tobar},
  {Willis}, {Lagos}, {Power}, and {Robotham}}]{Elahi:2019a}
{Elahi} PJ, {Ca{\~n}as} R, {Poulton} RJJ, {Tobar} RJ, {Willis} JS, {Lagos} CdP,
  {Power} C, {Robotham} ASG (2019) {Hunting for Galaxies and Halos in
  simulations with VELOCIraptor}. arXiv e-prints arXiv:1902.01010.
  {\href{https://arxiv.org/abs/1902.01010}{{arXiv:1902.01010}}} {[astro-ph.CO]}

\bibitem[{{Elbers} et~al.(2020){Elbers}, {Frenk}, {Jenkins}, {Li}, and
  {Pascoli}}]{Elbers:2020}
{Elbers} W, {Frenk} CS, {Jenkins} A, {Li} B, {Pascoli} S (2020) {An optimal
  nonlinear method for simulating relic neutrinos}. arXiv e-prints
  arXiv:2010.07321.
  {\href{https://arxiv.org/abs/2010.07321}{{arXiv:2010.07321}}} {[astro-ph.CO]}

\bibitem[{{Ellis} and {Buchert}(2005)}]{Buchert:2005}
{Ellis} GFR, {Buchert} T (2005) {The universe seen at different scales [rapid
  communication]}. Phys Lett A 347:38--46.
  \doi{10.1016/j.physleta.2005.06.087}.
  {\href{https://arxiv.org/abs/gr-qc/0506106}{{arXiv:gr-qc/0506106}}} {[gr-qc]}

\bibitem[{{Emberson} et~al.(2017){Emberson}, {Yu}, {Inman}, {Zhang}, {Pen},
  {Harnois-D{\'e}raps}, {Yuan}, {Teng}, {Zhu}, {Chen}, and
  {Xing}}]{Emberson:2017}
{Emberson} JD, {Yu} HR, {Inman} D, {Zhang} TJ, {Pen} UL, {Harnois-D{\'e}raps}
  J, {Yuan} S, {Teng} HY, {Zhu} HM, {Chen} X, {Xing} ZZ (2017) {Cosmological
  neutrino simulations at extreme scale}. Research in Astronomy and
  Astrophysics 17(8):085. \doi{10.1088/1674-4527/17/8/85}.
  {\href{https://arxiv.org/abs/1611.01545}{{arXiv:1611.01545}}} {[astro-ph.CO]}

\bibitem[{{Enzi} et~al.(2020){Enzi}, {Murgia}, {Newton}, {Vegetti}, {Frenk},
  {Viel}, {Cautun}, {Fassnacht}, {Auger}, {Despali}, {McKean}, {Koopmans}, and
  {Lovell}}]{Enzi:2020}
{Enzi} W, {Murgia} R, {Newton} O, {Vegetti} S, {Frenk} C, {Viel} M, {Cautun} M,
  {Fassnacht} CD, {Auger} M, {Despali} G, {McKean} J, {Koopmans} LVE, {Lovell}
  M (2020) {Joint constraints on thermal relic dark matter from a selection of
  astrophysical probes}. arXiv e-prints arXiv:2010.13802.
  {\href{https://arxiv.org/abs/2010.13802}{{arXiv:2010.13802}}} {[astro-ph.CO]}

\bibitem[{{Essig} et~al.(2019){Essig}, {McDermott}, {Yu}, and
  {Zhong}}]{Essig:2019}
{Essig} R, {McDermott} SD, {Yu} HB, {Zhong} YM (2019) {Constraining Dissipative
  Dark Matter Self-Interactions}. \prl 123(12):121102.
  \doi{10.1103/PhysRevLett.123.121102}.
  {\href{https://arxiv.org/abs/1809.01144}{{arXiv:1809.01144}}} {[hep-ph]}

\bibitem[{{Euclid Collaboration} et~al.(2019){Euclid Collaboration},
  {Knabenhans}, {Stadel}, {Marelli}, {Potter}, {Teyssier}, {Legrand},
  {Schneider}, {Sudret}, {Blot}, {Awan}, {Burigana}, {Carvalho},
  {Kurki-Suonio}, and {Sirri}}]{Knabenhans:2019}
{Euclid Collaboration}, {Knabenhans} M, {Stadel} J, {Marelli} S, {Potter} D,
  {Teyssier} R, {Legrand} L, {Schneider} A, {Sudret} B, {Blot} L, {Awan} S,
  {Burigana} C, {Carvalho} CS, {Kurki-Suonio} H, {Sirri} G (2019) {Euclid
  preparation: II. The EUCLIDEMULATOR - a tool to compute the cosmology
  dependence of the nonlinear matter power spectrum}. \mnras 484(4):5509--5529.
  \doi{10.1093/mnras/stz197}.
  {\href{https://arxiv.org/abs/1809.04695}{{arXiv:1809.04695}}} {[astro-ph.CO]}

\bibitem[{{Euclid Collaboration} et~al.(2020){Euclid Collaboration},
  {Knabenhans}, {Stadel}, {Potter}, {Dakin}, {Hannestad}, {Tram}, {Marelli},
  {Schneider}, {Teyssier}, {Andreon}, {Auricchio}, {Baccigalupi},
  {Balaguera-Antol{\'\i}nez}, {Baldi}, {Bardelli}, {Battaglia}, {Bender},
  {Biviano}, {Bodendorf}, {Bozzo}, {Branchini}, {Brescia}, {Burigana},
  {Cabanac}, {Camera}, {Capobianco}, {Cappi}, {Carbone}, {Carretero},
  {Carvalho}, {Casas}, {Casas}, {Castellano}, {Castignani}, {Cavuoti},
  {Cledassou}, {Colodro-Conde}, {Congedo}, {Conselice}, {Conversi}, {Copin},
  {Corcione}, {Coupon}, {Courtois}, {Da Silva}, {de la Torre}, {Di Ferdinando},
  {Duncan}, {Dupac}, {Fabbian}, {Farrens}, {Ferreira}, {Finelli}, {Frailis},
  {Franceschi}, {Galeotta}, {Garilli}, {Giocoli}, {Gozaliasl},
  {Graci{\'a}-Carpio}, {Grupp}, {Guzzo}, {Holmes}, {Hormuth}, {Israel},
  {Jahnke}, {Keihanen}, {Kermiche}, {Kirkpatrick}, {Kubik}, {Kunz},
  {Kurki-Suonio}, {Ligori}, {Lilje}, {Lloro}, {Maino}, {Marggraf}, {Markovic},
  {Martinet}, {Marulli}, {Massey}, {Mauri}, {Maurogordato}, {Medinaceli},
  {Meneghetti}, {Metcalf}, {Meylan}, {Moresco}, {Morin}, {Moscardini},
  {Munari}, {Neissner}, {Niemi}, {Padilla}, {Paltani}, {Pasian}, {Patrizii},
  {Pettorino}, {Pires}, {Polenta}, {Poncet}, {Raison}, {Renzi}, {Rhodes},
  {Riccio}, {Romelli}, {Roncarelli}, {Saglia}, {S{\'a}nchez}, {Sapone},
  {Schneider}, {Scottez}, {Secroun}, {Serrano}, {Sirignano}, {Sirri}, {Stanco},
  {Sureau}, {Tallada Cresp{\'\i}}, {Taylor}, {Tenti}, {Tereno}, {Toledo-Moreo},
  {Torradeflot}, {Valenziano}, {Valiviita}, {Vassallo}, {Viel}, {Wang},
  {Welikala}, {Whittaker}, {Zacchei}, and {Zucca}}]{Knabenhans:2020}
{Euclid Collaboration}, {Knabenhans} M, {Stadel} J, {Potter} D, {Dakin} J,
  {Hannestad} S, {Tram} T, {Marelli} S, {Schneider} A, {Teyssier} R, {Andreon}
  S, {Auricchio} N, {Baccigalupi} C, {Balaguera-Antol{\'\i}nez} A, {Baldi} M,
  {Bardelli} S, {Battaglia} P, {Bender} R, {Biviano} A, {Bodendorf} C, {Bozzo}
  E, {Branchini} E, {Brescia} M, {Burigana} C, {Cabanac} R, {Camera} S,
  {Capobianco} V, {Cappi} A, {Carbone} C, {Carretero} J, {Carvalho} CS, {Casas}
  R, {Casas} S, {Castellano} M, {Castignani} G, {Cavuoti} S, {Cledassou} R,
  {Colodro-Conde} C, {Congedo} G, {Conselice} CJ, {Conversi} L, {Copin} Y,
  {Corcione} L, {Coupon} J, {Courtois} HM, {Da Silva} A, {de la Torre} S, {Di
  Ferdinando} D, {Duncan} CAJ, {Dupac} X, {Fabbian} G, {Farrens} S, {Ferreira}
  PG, {Finelli} F, {Frailis} M, {Franceschi} E, {Galeotta} S, {Garilli} B,
  {Giocoli} C, {Gozaliasl} G, {Graci{\'a}-Carpio} J, {Grupp} F, {Guzzo} L,
  {Holmes} W, {Hormuth} F, {Israel} H, {Jahnke} K, {Keihanen} E, {Kermiche} S,
  {Kirkpatrick} CC, {Kubik} B, {Kunz} M, {Kurki-Suonio} H, {Ligori} S, {Lilje}
  PB, {Lloro} I, {Maino} D, {Marggraf} O, {Markovic} K, {Martinet} N, {Marulli}
  F, {Massey} R, {Mauri} N, {Maurogordato} S, {Medinaceli} E, {Meneghetti} M,
  {Metcalf} B, {Meylan} G, {Moresco} M, {Morin} B, {Moscardini} L, {Munari} E,
  {Neissner} C, {Niemi} SM, {Padilla} C, {Paltani} S, {Pasian} F, {Patrizii} L,
  {Pettorino} V, {Pires} S, {Polenta} G, {Poncet} M, {Raison} F, {Renzi} A,
  {Rhodes} J, {Riccio} G, {Romelli} E, {Roncarelli} M, {Saglia} R,
  {S{\'a}nchez} AG, {Sapone} D, {Schneider} P, {Scottez} V, {Secroun} A,
  {Serrano} S, {Sirignano} C, {Sirri} G, {Stanco} L, {Sureau} F, {Tallada
  Cresp{\'\i}} P, {Taylor} AN, {Tenti} M, {Tereno} I, {Toledo-Moreo} R,
  {Torradeflot} F, {Valenziano} L, {Valiviita} J, {Vassallo} T, {Viel} M,
  {Wang} Y, {Welikala} N, {Whittaker} L, {Zacchei} A, {Zucca} E (2020) {Euclid
  preparation: IX. EuclidEmulator2 -- Power spectrum emulation with massive
  neutrinos and self-consistent dark energy perturbations}. arXiv e-prints
  arXiv:2010.11288.
  {\href{https://arxiv.org/abs/2010.11288}{{arXiv:2010.11288}}} {[astro-ph.CO]}

\bibitem[{{Evrard} et~al.(2002){Evrard}, {MacFarland}, {Couchman}, {Colberg},
  {Yoshida}, {White}, {Jenkins}, {Frenk}, {Pearce}, {Peacock}, and
  {Thomas}}]{Evrard:2002}
{Evrard} AE, {MacFarland} TJ, {Couchman} HMP, {Colberg} JM, {Yoshida} N,
  {White} SDM, {Jenkins} A, {Frenk} CS, {Pearce} FR, {Peacock} JA, {Thomas} PA
  (2002) {Galaxy Clusters in Hubble Volume Simulations: Cosmological
  Constraints from Sky Survey Populations}. \apj 573(1):7--36.
  \doi{10.1086/340551}.
  {\href{https://arxiv.org/abs/astro-ph/0110246}{{arXiv:astro-ph/0110246}}}
  {[astro-ph]}

\bibitem[{{Ewald}(1921)}]{Ewald:1921}
{Ewald} PP (1921) {Die Berechnung optischer und elektrostatischer
  Gitterpotentiale}. Annalen der Physik 369(3):253--287.
  \doi{10.1002/andp.19213690304}

\bibitem[{{Ezquiaga} and {Zumalac{\'a}rregui}(2017)}]{Ezquiaga:2017}
{Ezquiaga} JM, {Zumalac{\'a}rregui} M (2017) {Dark Energy After GW170817: Dead
  Ends and the Road Ahead}. \prl 119(25):251304.
  \doi{10.1103/PhysRevLett.119.251304}.
  {\href{https://arxiv.org/abs/1710.05901}{{arXiv:1710.05901}}} {[astro-ph.CO]}

\bibitem[{{Fairbairn} et~al.(2018){Fairbairn}, {Marsh}, {Quevillon}, and
  {Rozier}}]{Fairbairn:2018}
{Fairbairn} M, {Marsh} DJE, {Quevillon} J, {Rozier} S (2018) {Structure
  formation and microlensing with axion miniclusters}. \prd 97(8):083502.
  \doi{10.1103/PhysRevD.97.083502}.
  {\href{https://arxiv.org/abs/1707.03310}{{arXiv:1707.03310}}} {[astro-ph.CO]}

\bibitem[{{Falck} et~al.(2017){Falck}, {McCullagh}, {Neyrinck}, {Wang}, and
  {Szalay}}]{Falck:2017}
{Falck} B, {McCullagh} N, {Neyrinck} MC, {Wang} J, {Szalay} AS (2017) {The
  Effect of Corner Modes in the Initial Conditions of Cosmological
  Simulations}. \apj 837(2):181. \doi{10.3847/1538-4357/aa60c7}.
  {\href{https://arxiv.org/abs/1610.04862}{{arXiv:1610.04862}}} {[astro-ph.CO]}

\bibitem[{{Falck} et~al.(2021){Falck}, {Wang}, {Jenkins}, {Lemson}, {Medvedev},
  {Neyrinck}, and {Szalay}}]{Falck:2021}
{Falck} B, {Wang} J, {Jenkins} A, {Lemson} G, {Medvedev} D, {Neyrinck} MC,
  {Szalay} AS (2021) {Indra: a Public Computationally-Accessible Suite of
  Cosmological $N$-body Simulations}. arXiv e-prints arXiv:2101.03631.
  {\href{https://arxiv.org/abs/2101.03631}{{arXiv:2101.03631}}} {[astro-ph.CO]}

\bibitem[{{Falck} et~al.(2012{\natexlab{a}}){Falck}, {Neyrinck}, and
  {Szalay}}]{Falck2011}
{Falck} BL, {Neyrinck} MC, {Szalay} AS (2012{\natexlab{a}}) {ORIGAMI:
  Delineating Halos Using Phase-space Folds}. \apj 754(2):126.
  \doi{10.1088/0004-637X/754/2/126}.
  {\href{https://arxiv.org/abs/1201.2353}{{arXiv:1201.2353}}} {[astro-ph.CO]}

\bibitem[{{Falck} et~al.(2012{\natexlab{b}}){Falck}, {Neyrinck}, and
  {Szalay}}]{Falck:2012}
{Falck} BL, {Neyrinck} MC, {Szalay} AS (2012{\natexlab{b}}) {ORIGAMI:
  Delineating Halos Using Phase-space Folds}. \apj 754(2):126.
  \doi{10.1088/0004-637X/754/2/126}.
  {\href{https://arxiv.org/abs/1201.2353}{{arXiv:1201.2353}}} {[astro-ph.CO]}

\bibitem[{{Fang} et~al.(2017){Fang}, {Li}, and {Zhao}}]{Fang:2017}
{Fang} W, {Li} B, {Zhao} GB (2017) {New Probe of Departures from General
  Relativity Using Minkowski Functionals}. \prl 118(18):181301.
  \doi{10.1103/PhysRevLett.118.181301}.
  {\href{https://arxiv.org/abs/1704.02325}{{arXiv:1704.02325}}} {[astro-ph.CO]}

\bibitem[{{Favole} et~al.(2016){Favole}, {Comparat}, {Prada}, {Yepes}, {Jullo},
  {Niemiec}, {Kneib}, {Rodr{\'\i}guez-Torres}, {Klypin}, {Skibba}, {McBride},
  {Eisenstein}, {Schlegel}, {Nuza}, {Chuang}, {Delubac}, {Y{\`e}che}, and
  {Schneider}}]{Favole:2016}
{Favole} G, {Comparat} J, {Prada} F, {Yepes} G, {Jullo} E, {Niemiec} A, {Kneib}
  JP, {Rodr{\'\i}guez-Torres} SA, {Klypin} A, {Skibba} RA, {McBride} CK,
  {Eisenstein} DJ, {Schlegel} DJ, {Nuza} SE, {Chuang} CH, {Delubac} T,
  {Y{\`e}che} C, {Schneider} DP (2016) {Clustering properties of $g$-selected
  galaxies at $z \sim 0.8$}. \mnras 461(4):3421--3431.
  \doi{10.1093/mnras/stw1483}.
  {\href{https://arxiv.org/abs/1507.04356}{{arXiv:1507.04356}}} {[astro-ph.CO]}

\bibitem[{{Favole} et~al.(2017){Favole}, {Rodr{\'\i}guez-Torres}, {Comparat},
  {Prada}, {Guo}, {Klypin}, and {Montero-Dorta}}]{Favole:2017}
{Favole} G, {Rodr{\'\i}guez-Torres} SA, {Comparat} J, {Prada} F, {Guo} H,
  {Klypin} A, {Montero-Dorta} AD (2017) {Galaxy clustering dependence on the [O
  II] emission line luminosity in the local Universe}. \mnras 472(1):550--558.
  \doi{10.1093/mnras/stx1980}.
  {\href{https://arxiv.org/abs/1611.05457}{{arXiv:1611.05457}}} {[astro-ph.GA]}

\bibitem[{{Favole} et~al.(2021){Favole}, {Montero-Dorta}, {Artale},
  {Contreras}, {Zehavi}, and {Xu}}]{Favole:2021}
{Favole} G, {Montero-Dorta} AD, {Artale} MC, {Contreras} S, {Zehavi} I, {Xu} X
  (2021) {SHAM through the lens of a hydrodynamical simulation}. arXiv e-prints
  arXiv:2101.10733.
  {\href{https://arxiv.org/abs/2101.10733}{{arXiv:2101.10733}}} {[astro-ph.GA]}

\bibitem[{{Fedeli}(2014)}]{Fedeli:2014}
{Fedeli} C (2014) {The clustering of baryonic matter. I: a halo-model
  approach}. \jcap 4:028. \doi{10.1088/1475-7516/2014/04/028}.
  {\href{https://arxiv.org/abs/1401.2997}{{arXiv:1401.2997}}}

\bibitem[{{Feder} et~al.(2020){Feder}, {Berger}, and {Stein}}]{Feder:2020}
{Feder} RM, {Berger} P, {Stein} G (2020) {Nonlinear 3D cosmic web simulation
  with heavy-tailed generative adversarial networks}. \prd 102(10):103504.
  \doi{10.1103/PhysRevD.102.103504}.
  {\href{https://arxiv.org/abs/2005.03050}{{arXiv:2005.03050}}} {[astro-ph.CO]}

\bibitem[{{Feng} et~al.(2016){Feng}, {Chu}, {Seljak}, and
  {McDonald}}]{Feng:2016}
{Feng} Y, {Chu} MY, {Seljak} U, {McDonald} P (2016) {FASTPM: a new scheme for
  fast simulations of dark matter and haloes}. \mnras 463(3):2273--2286.
  \doi{10.1093/mnras/stw2123}.
  {\href{https://arxiv.org/abs/1603.00476}{{arXiv:1603.00476}}} {[astro-ph.CO]}

\bibitem[{{Ferraro} and {Smith}(2015)}]{Ferraro:2015}
{Ferraro} S, {Smith} KM (2015) {Using large scale structure to measure f$_{NL}$
  , g$_{NL}$ and {\ensuremath{\tau}}$_{NL}$}. \prd 91(4):043506.
  \doi{10.1103/PhysRevD.91.043506}.
  {\href{https://arxiv.org/abs/1408.3126}{{arXiv:1408.3126}}} {[astro-ph.CO]}

\bibitem[{Ferreira(2021)}]{Ferreira:2021}
Ferreira EGM (2021) Ultra-light dark matter. The Astronomy and Astrophysics
  Review 29(1):7. \doi{10.1007/s00159-021-00135-6},
  \urlprefix\url{https://doi.org/10.1007/s00159-021-00135-6}

\bibitem[{{Ferrell} and {Bertschinger}(1994)}]{Ferrell:1994}
{Ferrell} R, {Bertschinger} E (1994) {Particle-Mesh Methods on the Connection
  Machine}. Int J Mod Phys C 5(6):933--956. \doi{10.1142/S0129183194001069}.
  {\href{https://arxiv.org/abs/comp-gas/9310002}{{arXiv:comp-gas/9310002}}}
  {[nlin.CG]}

\bibitem[{{Fidler} et~al.(2015){Fidler}, {Rampf}, {Tram}, {Crittenden},
  {Koyama}, and {Wands}}]{Fidler:2015}
{Fidler} C, {Rampf} C, {Tram} T, {Crittenden} R, {Koyama} K, {Wands} D (2015)
  {General relativistic corrections to N-body simulations and the Zel'dovich
  approximation}. \prd 92(12):123517. \doi{10.1103/PhysRevD.92.123517}.
  {\href{https://arxiv.org/abs/1505.04756}{{arXiv:1505.04756}}} {[astro-ph.CO]}

\bibitem[{{Fidler} et~al.(2016){Fidler}, {Tram}, {Rampf}, {Crittenden},
  {Koyama}, and {Wands}}]{Fidler:2016}
{Fidler} C, {Tram} T, {Rampf} C, {Crittenden} R, {Koyama} K, {Wands} D (2016)
  {Relativistic interpretation of Newtonian simulations for cosmic structure
  formation}. \jcap 2016(9):031. \doi{10.1088/1475-7516/2016/09/031}.
  {\href{https://arxiv.org/abs/1606.05588}{{arXiv:1606.05588}}} {[astro-ph.CO]}

\bibitem[{{Fidler} et~al.(2017{\natexlab{a}}){Fidler}, {Tram}, {Rampf},
  {Crittenden}, {Koyama}, and {Wands}}]{Fidler:2017b}
{Fidler} C, {Tram} T, {Rampf} C, {Crittenden} R, {Koyama} K, {Wands} D
  (2017{\natexlab{a}}) {General relativistic weak-field limit and Newtonian
  N-body simulations}. \jcap 2017(12):022. \doi{10.1088/1475-7516/2017/12/022}.
  {\href{https://arxiv.org/abs/1708.07769}{{arXiv:1708.07769}}} {[astro-ph.CO]}

\bibitem[{{Fidler} et~al.(2017{\natexlab{b}}){Fidler}, {Tram}, {Rampf},
  {Crittenden}, {Koyama}, and {Wands}}]{Fidler:2017}
{Fidler} C, {Tram} T, {Rampf} C, {Crittenden} R, {Koyama} K, {Wands} D
  (2017{\natexlab{b}}) {Relativistic initial conditions for N-body
  simulations}. \jcap 2017(6):043. \doi{10.1088/1475-7516/2017/06/043}.
  {\href{https://arxiv.org/abs/1702.03221}{{arXiv:1702.03221}}} {[astro-ph.CO]}

\bibitem[{{Fluri} et~al.(2018){Fluri}, {Kacprzak}, {Refregier}, {Amara},
  {Lucchi}, and {Hofmann}}]{Fluri:2018}
{Fluri} J, {Kacprzak} T, {Refregier} A, {Amara} A, {Lucchi} A, {Hofmann} T
  (2018) {Cosmological constraints from noisy convergence maps through deep
  learning}. \prd 98(12):123518. \doi{10.1103/PhysRevD.98.123518}.
  {\href{https://arxiv.org/abs/1807.08732}{{arXiv:1807.08732}}} {[astro-ph.CO]}

\bibitem[{{Fluri} et~al.(2019){Fluri}, {Kacprzak}, {Lucchi}, {Refregier},
  {Amara}, {Hofmann}, and {Schneider}}]{Fluri:2019}
{Fluri} J, {Kacprzak} T, {Lucchi} A, {Refregier} A, {Amara} A, {Hofmann} T,
  {Schneider} A (2019) {Cosmological constraints with deep learning from
  KiDS-450 weak lensing maps}. \prd 100(6):063514.
  \doi{10.1103/PhysRevD.100.063514}.
  {\href{https://arxiv.org/abs/1906.03156}{{arXiv:1906.03156}}} {[astro-ph.CO]}

\bibitem[{{Fong} and {Han}(2020)}]{Fong:2020}
{Fong} M, {Han} J (2020) {A natural boundary of dark matter haloes revealed
  around the minimum bias and maximum infall locations}. arXiv e-prints
  arXiv:2008.03477.
  {\href{https://arxiv.org/abs/2008.03477}{{arXiv:2008.03477}}} {[astro-ph.CO]}

\bibitem[{{Fontanot} et~al.(2012){Fontanot}, {Springel}, {Angulo}, and
  {Henriques}}]{Fontanot:2012}
{Fontanot} F, {Springel} V, {Angulo} RE, {Henriques} B (2012) {Semi-analytic
  galaxy formation in early dark energy cosmologies}. \mnras 426(3):2335--2341.
  \doi{10.1111/j.1365-2966.2012.21880.x}.
  {\href{https://arxiv.org/abs/1207.1723}{{arXiv:1207.1723}}} {[astro-ph.CO]}

\bibitem[{{Fontanot} et~al.(2013){Fontanot}, {Puchwein}, {Springel}, and
  {Bianchi}}]{Fontanot:2013}
{Fontanot} F, {Puchwein} E, {Springel} V, {Bianchi} D (2013) {Semi-analytic
  galaxy formation in f(R)-gravity cosmologies}. \mnras 436(3):2672--2679.
  \doi{10.1093/mnras/stt1763}.
  {\href{https://arxiv.org/abs/1307.5065}{{arXiv:1307.5065}}} {[astro-ph.CO]}

\bibitem[{{Fontanot} et~al.(2015{\natexlab{a}}){Fontanot}, {Baldi}, {Springel},
  and {Bianchi}}]{Fontanot:2015b}
{Fontanot} F, {Baldi} M, {Springel} V, {Bianchi} D (2015{\natexlab{a}})
  {Semi-analytic galaxy formation in coupled dark energy cosmologies}. \mnras
  452(1):978--985. \doi{10.1093/mnras/stv1345}.
  {\href{https://arxiv.org/abs/1505.02770}{{arXiv:1505.02770}}} {[astro-ph.CO]}

\bibitem[{{Fontanot} et~al.(2015{\natexlab{b}}){Fontanot},
  {Villaescusa-Navarro}, {Bianchi}, and {Viel}}]{Fontanot:2015}
{Fontanot} F, {Villaescusa-Navarro} F, {Bianchi} D, {Viel} M
  (2015{\natexlab{b}}) {Semi-analytic galaxy formation in massive neutrino
  cosmologies}. \mnras 447(4):3361--3367. \doi{10.1093/mnras/stu2705}.
  {\href{https://arxiv.org/abs/1409.6309}{{arXiv:1409.6309}}} {[astro-ph.CO]}

\bibitem[{{Foreman} et~al.(2020){Foreman}, {Coulton}, {Villaescusa-Navarro},
  and {Barreira}}]{foreman:2020}
{Foreman} S, {Coulton} W, {Villaescusa-Navarro} F, {Barreira} A (2020)
  {Baryonic effects on the matter bispectrum}. \mnras 498(2):2887--2911.
  \doi{10.1093/mnras/staa2523}.
  {\href{https://arxiv.org/abs/1910.03597}{{arXiv:1910.03597}}} {[astro-ph.CO]}

\bibitem[{{Forero-Romero} et~al.(2014){Forero-Romero}, {Contreras}, and
  {Padilla}}]{Forero-Romero:2014}
{Forero-Romero} JE, {Contreras} S, {Padilla} N (2014) {Cosmic web alignments
  with the shape, angular momentum and peculiar velocities of dark matter
  haloes}. \mnras 443(2):1090--1102. \doi{10.1093/mnras/stu1150}.
  {\href{https://arxiv.org/abs/1406.0508}{{arXiv:1406.0508}}} {[astro-ph.CO]}

\bibitem[{{Fornasa} and {S{\'a}nchez-Conde}(2015)}]{Fornasa:2015}
{Fornasa} M, {S{\'a}nchez-Conde} MA (2015) {The nature of the Diffuse Gamma-Ray
  Background}. \physrep 598:1--58. \doi{10.1016/j.physrep.2015.09.002}.
  {\href{https://arxiv.org/abs/1502.02866}{{arXiv:1502.02866}}} {[astro-ph.CO]}

\bibitem[{{Fosalba} et~al.(2008){Fosalba}, {Gazta{\~n}aga}, {Castander}, and
  {Manera}}]{Fosalba:2008}
{Fosalba} P, {Gazta{\~n}aga} E, {Castander} FJ, {Manera} M (2008) {The onion
  universe: all sky lightcone simulations in spherical shells}. \mnras
  391(1):435--446. \doi{10.1111/j.1365-2966.2008.13910.x}.
  {\href{https://arxiv.org/abs/0711.1540}{{arXiv:0711.1540}}} {[astro-ph]}

\bibitem[{{Fosalba} et~al.(2015){Fosalba}, {Crocce}, {Gazta{\~n}aga}, and
  {Castander}}]{Fosalba:2015}
{Fosalba} P, {Crocce} M, {Gazta{\~n}aga} E, {Castander} FJ (2015) {The MICE
  grand challenge lightcone simulation - I. Dark matter clustering}. \mnras
  448(4):2987--3000. \doi{10.1093/mnras/stv138}.
  {\href{https://arxiv.org/abs/1312.1707}{{arXiv:1312.1707}}} {[astro-ph.CO]}

\bibitem[{{Frenk} and {White}(2012)}]{Frenk:2012}
{Frenk} CS, {White} SDM (2012) {Dark matter and cosmic structure}. Annalen der
  Physik 524(9-10):507--534. \doi{10.1002/andp.201200212}.
  {\href{https://arxiv.org/abs/1210.0544}{{arXiv:1210.0544}}} {[astro-ph.CO]}

\bibitem[{{Frenk} et~al.(1999){Frenk}, {White}, {Bode}, {Bond}, {Bryan}, {Cen},
  {Couchman}, {Evrard}, {Gnedin}, {Jenkins}, {Khokhlov}, {Klypin}, {Navarro},
  {Norman}, {Ostriker}, {Owen}, {Pearce}, {Pen}, {Steinmetz}, {Thomas},
  {Villumsen}, {Wadsley}, {Warren}, {Xu}, and {Yepes}}]{Frenk:1999}
{Frenk} CS, {White} SDM, {Bode} P, {Bond} JR, {Bryan} GL, {Cen} R, {Couchman}
  HMP, {Evrard} AE, {Gnedin} N, {Jenkins} A, {Khokhlov} AM, {Klypin} A,
  {Navarro} JF, {Norman} ML, {Ostriker} JP, {Owen} JM, {Pearce} FR, {Pen} UL,
  {Steinmetz} M, {Thomas} PA, {Villumsen} JV, {Wadsley} JW, {Warren} MS, {Xu}
  G, {Yepes} G (1999) {The Santa Barbara Cluster Comparison Project: A
  Comparison of Cosmological Hydrodynamics Solutions}. \apj 525(2):554--582.
  \doi{10.1086/307908}.
  {\href{https://arxiv.org/abs/astro-ph/9906160}{{arXiv:astro-ph/9906160}}}
  {[astro-ph]}

\bibitem[{{Frieman} et~al.(1995){Frieman}, {Hill}, {Stebbins}, and
  {Waga}}]{Frieman:1995}
{Frieman} JA, {Hill} CT, {Stebbins} A, {Waga} I (1995) {Cosmology with
  Ultralight Pseudo Nambu-Goldstone Bosons}. Physical Review Letters
  75:2077--2080. \doi{10.1103/PhysRevLett.75.2077}.
  {\href{https://arxiv.org/abs/astro-ph/9505060}{{astro-ph/9505060}}}

\bibitem[{Frigo and Johnson(2005)}]{FFTW05}
Frigo M, Johnson SG (2005) The design and implementation of {FFTW3}.
  Proceedings of the IEEE 93(2):216--231. Special issue on ``Program
  Generation, Optimization, and Platform Adaptation''

\bibitem[{{Frontiere} et~al.(2021){Frontiere}, {Heitmann}, {Rangel}, {Larsen},
  {Pope}, {Sultan}, {Uram}, {Habib}, {Rizzi}, and {Insley}}]{Frontiere:2021}
{Frontiere} N, {Heitmann} K, {Rangel} E, {Larsen} P, {Pope} A, {Sultan} I,
  {Uram} T, {Habib} S, {Rizzi} S, {Insley} J (2021) {Farpoint: A
  High-Resolution Cosmology Simulation at the Gpc Scale}. arXiv e-prints
  arXiv:2109.01956.
  {\href{https://arxiv.org/abs/2109.01956}{{arXiv:2109.01956}}} {[astro-ph.CO]}

\bibitem[{{Fujiwara}(1981)}]{Fujiwara:1981}
{Fujiwara} T (1981) {Vlasov Simulations of Stellar Systems - Infinite
  Homogeneous Case}. \pasj 33:531

\bibitem[{{Gao} et~al.(2004){Gao}, {De Lucia}, {White}, and
  {Jenkins}}]{Gao:2004}
{Gao} L, {De Lucia} G, {White} SDM, {Jenkins} A (2004) {Galaxies and subhaloes
  in {\ensuremath{\Lambda}}CDM galaxy clusters}. \mnras 352(2):L1--L5.
  \doi{10.1111/j.1365-2966.2004.08098.x}.
  {\href{https://arxiv.org/abs/astro-ph/0405010}{{arXiv:astro-ph/0405010}}}
  {[astro-ph]}

\bibitem[{{Gao} et~al.(2005){Gao}, {White}, {Jenkins}, {Frenk}, and
  {Springel}}]{Gao:2005b}
{Gao} L, {White} SDM, {Jenkins} A, {Frenk} CS, {Springel} V (2005) {Early
  structure in {\ensuremath{\Lambda}}CDM}. \mnras 363(2):379--392.
  \doi{10.1111/j.1365-2966.2005.09509.x}.
  {\href{https://arxiv.org/abs/astro-ph/0503003}{{arXiv:astro-ph/0503003}}}
  {[astro-ph]}

\bibitem[{{Gao} et~al.(2012){Gao}, {Navarro}, {Frenk}, {Jenkins}, {Springel},
  and {White}}]{Gao:2012}
{Gao} L, {Navarro} JF, {Frenk} CS, {Jenkins} A, {Springel} V, {White} SDM
  (2012) {The Phoenix Project: the dark side of rich Galaxy clusters}. \mnras
  425(3):2169--2186. \doi{10.1111/j.1365-2966.2012.21564.x}.
  {\href{https://arxiv.org/abs/1201.1940}{{arXiv:1201.1940}}} {[astro-ph.CO]}

\bibitem[{{Garaldi} et~al.(2016){Garaldi}, {Baldi}, and
  {Moscardini}}]{Garaldi:2016}
{Garaldi} E, {Baldi} M, {Moscardini} L (2016) {Zoomed high-resolution
  simulations of Multi-coupled Dark Energy: cored galaxy density profiles at
  high redshift}. \jcap 2016(1):050. \doi{10.1088/1475-7516/2016/01/050}.
  {\href{https://arxiv.org/abs/1511.02239}{{arXiv:1511.02239}}} {[astro-ph.CO]}

\bibitem[{{Garaldi} et~al.(2020){Garaldi}, {Nori}, and {Baldi}}]{Garaldi:2020}
{Garaldi} E, {Nori} M, {Baldi} M (2020) {Dynamic zoom simulations: A fast,
  adaptive algorithm for simulating light-cones}. \mnras 499(2):2685--2700.
  \doi{10.1093/mnras/staa2064}.
  {\href{https://arxiv.org/abs/2005.05328}{{arXiv:2005.05328}}} {[astro-ph.CO]}

\bibitem[{{Garc{\'\i}a} and {Rozo}(2019)}]{Garcia:2019}
{Garc{\'\i}a} R, {Rozo} E (2019) {Halo exclusion criteria impacts halo
  statistics}. \mnras 489(3):4170--4175. \doi{10.1093/mnras/stz2458}.
  {\href{https://arxiv.org/abs/1903.01709}{{arXiv:1903.01709}}} {[astro-ph.CO]}

\bibitem[{{Garcia} et~al.(2020){Garcia}, {Rozo}, {Becker}, and
  {More}}]{Garcia:2020}
{Garcia} R, {Rozo} E, {Becker} MR, {More} S (2020) {A Redefinition of the Halo
  Boundary Leads to a Simple yet Accurate Halo Model of Large Scale Structure}.
  arXiv e-prints arXiv:2006.12751.
  {\href{https://arxiv.org/abs/2006.12751}{{arXiv:2006.12751}}} {[astro-ph.CO]}

\bibitem[{{Garny} et~al.(2020){Garny}, {Konstandin}, and {Rubira}}]{Garny:2020}
{Garny} M, {Konstandin} T, {Rubira} H (2020) {The Schr{\"o}dinger-Poisson
  method for Large-Scale Structure}. \jcap 2020(4):003.
  \doi{10.1088/1475-7516/2020/04/003}.
  {\href{https://arxiv.org/abs/1911.04505}{{arXiv:1911.04505}}} {[astro-ph.CO]}

\bibitem[{{Garrison} et~al.(2016){Garrison}, {Eisenstein}, {Ferrer},
  {Metchnik}, and {Pinto}}]{Garrison:2016}
{Garrison} LH, {Eisenstein} DJ, {Ferrer} D, {Metchnik} MV, {Pinto} PA (2016)
  {Improving initial conditions for cosmological N-body simulations}. \mnras
  461:4125--4145. \doi{10.1093/mnras/stw1594}.
  {\href{https://arxiv.org/abs/1605.02333}{{arXiv:1605.02333}}}

\bibitem[{{Garrison} et~al.(2018){Garrison}, {Eisenstein}, {Ferrer}, {Tinker},
  {Pinto}, and {Weinberg}}]{Garrison:2018}
{Garrison} LH, {Eisenstein} DJ, {Ferrer} D, {Tinker} JL, {Pinto} PA, {Weinberg}
  DH (2018) {The Abacus Cosmos: A Suite of Cosmological N-body Simulations}.
  \apjs 236(2):43. \doi{10.3847/1538-4365/aabfd3}.
  {\href{https://arxiv.org/abs/1712.05768}{{arXiv:1712.05768}}} {[astro-ph.CO]}

\bibitem[{{Garrison} et~al.(2019){Garrison}, {Eisenstein}, and
  {Pinto}}]{Garrison:2019}
{Garrison} LH, {Eisenstein} DJ, {Pinto} PA (2019) {A high-fidelity realization
  of the Euclid code comparison N-body simulation with ABACUS}. \mnras
  485(3):3370--3377. \doi{10.1093/mnras/stz634}.
  {\href{https://arxiv.org/abs/1810.02916}{{arXiv:1810.02916}}} {[astro-ph.CO]}

\bibitem[{{Garrison} et~al.(2021){Garrison}, {Joyce}, and
  {Eisenstein}}]{Garrison:2021}
{Garrison} LH, {Joyce} M, {Eisenstein} DJ (2021) {Good and Proper:
  Self-similarity of N-body Simulations with Proper Force Softening}. arXiv
  e-prints arXiv:2102.08972.
  {\href{https://arxiv.org/abs/2102.08972}{{arXiv:2102.08972}}} {[astro-ph.CO]}

\bibitem[{{Genel} et~al.(2019){Genel}, {Bryan}, {Springel}, {Hernquist},
  {Nelson}, {Pillepich}, {Weinberger}, {Pakmor}, {Marinacci}, and
  {Vogelsberger}}]{Genel:2019}
{Genel} S, {Bryan} GL, {Springel} V, {Hernquist} L, {Nelson} D, {Pillepich} A,
  {Weinberger} R, {Pakmor} R, {Marinacci} F, {Vogelsberger} M (2019) {A
  Quantification of the Butterfly Effect in Cosmological Simulations and
  Implications for Galaxy Scaling Relations}. \apj 871(1):21.
  \doi{10.3847/1538-4357/aaf4bb}.
  {\href{https://arxiv.org/abs/1807.07084}{{arXiv:1807.07084}}} {[astro-ph.GA]}

\bibitem[{{Giannantonio} and {Percival}(2014)}]{Giannantonio:2014}
{Giannantonio} T, {Percival} WJ (2014) {Using correlations between cosmic
  microwave background lensing and large-scale structure to measure primordial
  non-Gaussianity.} \mnras 441:L16--L20. \doi{10.1093/mnrasl/slu036}.
  {\href{https://arxiv.org/abs/1312.5154}{{arXiv:1312.5154}}} {[astro-ph.CO]}

\bibitem[{{Giannantonio} et~al.(2012){Giannantonio}, {Porciani}, {Carron},
  {Amara}, and {Pillepich}}]{Giannantonio:2012}
{Giannantonio} T, {Porciani} C, {Carron} J, {Amara} A, {Pillepich} A (2012)
  {Constraining primordial non-Gaussianity with future galaxy surveys}. \mnras
  422(4):2854--2877. \doi{10.1111/j.1365-2966.2012.20604.x}.
  {\href{https://arxiv.org/abs/1109.0958}{{arXiv:1109.0958}}} {[astro-ph.CO]}

\bibitem[{{Giblin} et~al.(2016){Giblin}, {Mertens}, and
  {Starkman}}]{Giblin:2016}
{Giblin} JT, {Mertens} JB, {Starkman} GD (2016) {Departures from the
  Friedmann-Lemaitre-Robertson-Walker Cosmological Model in an Inhomogeneous
  Universe: A Numerical Examination}. \prl 116(25):251301.
  \doi{10.1103/PhysRevLett.116.251301}.
  {\href{https://arxiv.org/abs/1511.01105}{{arXiv:1511.01105}}} {[gr-qc]}

\bibitem[{{Gill} et~al.(2004){Gill}, {Knebe}, and {Gibson}}]{Gill:2004}
{Gill} SPD, {Knebe} A, {Gibson} BK (2004) {The evolution of substructure - I. A
  new identification method}. \mnras 351(2):399--409.
  \doi{10.1111/j.1365-2966.2004.07786.x}.
  {\href{https://arxiv.org/abs/astro-ph/0404258}{{arXiv:astro-ph/0404258}}}
  {[astro-ph]}

\bibitem[{{Giri} and {Schneider}(2021)}]{Giri:2021}
{Giri} SK, {Schneider} A (2021) {Emulation of baryonic effects on the matter
  power spectrum and constraints from galaxy cluster data}. arXiv e-prints
  arXiv:2108.08863.
  {\href{https://arxiv.org/abs/2108.08863}{{arXiv:2108.08863}}} {[astro-ph.CO]}

\bibitem[{{Giusarma} et~al.(2019){Giusarma}, {Reyes Hurtado},
  {Villaescusa-Navarro}, {He}, {Ho}, and {Hahn}}]{Giusarma:2019}
{Giusarma} E, {Reyes Hurtado} M, {Villaescusa-Navarro} F, {He} S, {Ho} S,
  {Hahn} C (2019) {Learning neutrino effects in Cosmology with Convolutional
  Neural Networks}. arXiv e-prints arXiv:1910.04255.
  {\href{https://arxiv.org/abs/1910.04255}{{arXiv:1910.04255}}} {[astro-ph.CO]}

\bibitem[{{Gnedin}(2019)}]{Gnedin:2019}
{Gnedin} NY (2019) {Hierarchical Particle Mesh: An FFT-accelerated Fast
  Multipole Method}. \apjs 243(2):19. \doi{10.3847/1538-4365/ab2d24}.
  {\href{https://arxiv.org/abs/1906.10734}{{arXiv:1906.10734}}}
  {[physics.comp-ph]}

\bibitem[{{Gnedin} et~al.(2011){Gnedin}, {Kravtsov}, and {Rudd}}]{Gnedin:2011}
{Gnedin} NY, {Kravtsov} AV, {Rudd} DH (2011) {Implementing the DC Mode in
  Cosmological Simulations with Supercomoving Variables}. \apjs 194(2):46.
  \doi{10.1088/0067-0049/194/2/46}.
  {\href{https://arxiv.org/abs/1104.1428}{{arXiv:1104.1428}}} {[astro-ph.CO]}

\bibitem[{{Goh} et~al.(2019){Goh}, {Primack}, {Lee}, {Aragon-Calvo},
  {Hellinger}, {Behroozi}, {Rodriguez-Puebla}, {Eckholm}, and
  {Johnston}}]{Goh:2019}
{Goh} T, {Primack} J, {Lee} CT, {Aragon-Calvo} M, {Hellinger} D, {Behroozi} P,
  {Rodriguez-Puebla} A, {Eckholm} E, {Johnston} K (2019) {Dark matter halo
  properties versus local density and cosmic web location}. \mnras
  483(2):2101--2122. \doi{10.1093/mnras/sty3153}.
  {\href{https://arxiv.org/abs/1805.04943}{{arXiv:1805.04943}}} {[astro-ph.GA]}

\bibitem[{{Goldstein} et~al.(2002){Goldstein}, {Poole}, and
  {Safko}}]{Goldstein:2002}
{Goldstein} H, {Poole} C, {Safko} J (2002) {Classical mechanics}.
  Addison-Wesley, San Francisco

\bibitem[{{Goodman} et~al.(1993){Goodman}, {Heggie}, and {Hut}}]{Goodman:1993}
{Goodman} J, {Heggie} DC, {Hut} P (1993) {On the Exponential Instability of
  N-Body Systems}. \apj 415:715. \doi{10.1086/173196}

\bibitem[{{Goroff} et~al.(1986){Goroff}, {Grinstein}, {Rey}, and
  {Wise}}]{Goroff:1986}
{Goroff} MH, {Grinstein} B, {Rey} SJ, {Wise} MB (1986) {Coupling of modes of
  cosmological mass density fluctuations}. \apj 311:6--14. \doi{10.1086/164749}

\bibitem[{{Gosenca} et~al.(2017){Gosenca}, {Adamek}, {Byrnes}, and
  {Hotchkiss}}]{Gosenca:2017}
{Gosenca} M, {Adamek} J, {Byrnes} CT, {Hotchkiss} S (2017) {3D simulations with
  boosted primordial power spectra and ultracompact minihalos}. \prd
  96(12):123519. \doi{10.1103/PhysRevD.96.123519}.
  {\href{https://arxiv.org/abs/1710.02055}{{arXiv:1710.02055}}} {[astro-ph.CO]}

\bibitem[{{Gottloeber} et~al.(2010){Gottloeber}, {Hoffman}, and
  {Yepes}}]{Gottloeber:2010}
{Gottloeber} S, {Hoffman} Y, {Yepes} G (2010) {Constrained Local UniversE
  Simulations (CLUES)}. arXiv e-prints arXiv:1005.2687.
  {\href{https://arxiv.org/abs/1005.2687}{{arXiv:1005.2687}}} {[astro-ph.CO]}

\bibitem[{{Green} et~al.(2004){Green}, {Hofmann}, and {Schwarz}}]{Green:2004}
{Green} AM, {Hofmann} S, {Schwarz} DJ (2004) {The power spectrum of SUSY-CDM on
  subgalactic scales}. \mnras 353(3):L23--L27.
  \doi{10.1111/j.1365-2966.2004.08232.x}.
  {\href{https://arxiv.org/abs/astro-ph/0309621}{{arXiv:astro-ph/0309621}}}
  {[astro-ph]}

\bibitem[{{Green} et~al.(2021){Green}, {van den Bosch}, and
  {Jiang}}]{Green:2021}
{Green} SB, {van den Bosch} FC, {Jiang} F (2021) {The tidal evolution of dark
  matter substructure -- II. The impact of artificial disruption on subhalo
  mass functions and radial profiles}. arXiv e-prints arXiv:2103.01227.
  {\href{https://arxiv.org/abs/2103.01227}{{arXiv:2103.01227}}} {[astro-ph.GA]}

\bibitem[{Greengard and Rokhlin(1987)}]{Greengard:1987}
Greengard L, Rokhlin V (1987) A fast algorithm for particle simulations. J
  Comput Phys 73(2):325 -- 348. \doi{10.1016/0021-9991(87)90140-9}

\bibitem[{{Grossi} et~al.(2009){Grossi}, {Verde}, {Carbone}, {Dolag},
  {Branchini}, {Iannuzzi}, {Matarrese}, and {Moscardini}}]{Grossi:2009}
{Grossi} M, {Verde} L, {Carbone} C, {Dolag} K, {Branchini} E, {Iannuzzi} F,
  {Matarrese} S, {Moscardini} L (2009) {Large-scale non-Gaussian mass function
  and halo bias: tests on N-body simulations}. \mnras 398(1):321--332.
  \doi{10.1111/j.1365-2966.2009.15150.x}.
  {\href{https://arxiv.org/abs/0902.2013}{{arXiv:0902.2013}}} {[astro-ph.CO]}

\bibitem[{{Grudi{\'c}} and {Hopkins}(2019)}]{Grudic:2019}
{Grudi{\'c}} MY, {Hopkins} PF (2019) {A general-purpose timestep criterion for
  simulations with gravity}. arXiv e-prints arXiv:1910.06349.
  {\href{https://arxiv.org/abs/1910.06349}{{arXiv:1910.06349}}} {[astro-ph.IM]}

\bibitem[{{Guandalin} et~al.(2021){Guandalin}, {Adamek}, {Bull}, {Clarkson},
  {Abramo}, and {Coates}}]{Guandalin:2021}
{Guandalin} C, {Adamek} J, {Bull} P, {Clarkson} C, {Abramo} LR, {Coates} L
  (2021) {Observing relativistic features in large-scale structure surveys - I.
  Multipoles of the power spectrum}. \mnras 501(2):2547--2561.
  \doi{10.1093/mnras/staa3890}.
  {\href{https://arxiv.org/abs/2009.02284}{{arXiv:2009.02284}}} {[astro-ph.CO]}

\bibitem[{{Guillet} and {Teyssier}(2011)}]{Guillet:2011}
{Guillet} T, {Teyssier} R (2011) {A simple multigrid scheme for solving the
  Poisson equation with arbitrary domain boundaries}. J Comput Phys
  230(12):4756--4771. \doi{10.1016/j.jcp.2011.02.044}.
  {\href{https://arxiv.org/abs/1104.1703}{{arXiv:1104.1703}}}
  {[physics.comp-ph]}

\bibitem[{{Guo} et~al.(2015){Guo}, {Zheng}, {Zehavi}, {Dawson}, {Skibba},
  {Tinker}, {Weinberg}, {White}, and {Schneider}}]{Guo:2015}
{Guo} H, {Zheng} Z, {Zehavi} I, {Dawson} K, {Skibba} RA, {Tinker} JL,
  {Weinberg} DH, {White} M, {Schneider} DP (2015) {Velocity bias from the
  small-scale clustering of SDSS-III BOSS galaxies}. \mnras 446(1):578--594.
  \doi{10.1093/mnras/stu2120}.
  {\href{https://arxiv.org/abs/1407.4811}{{arXiv:1407.4811}}} {[astro-ph.CO]}

\bibitem[{{Guo} and {White}(2014)}]{Guo:2014}
{Guo} Q, {White} S (2014) {Numerical resolution limits on subhalo abundance
  matching}. \mnras 437(4):3228--3235. \doi{10.1093/mnras/stt2116}.
  {\href{https://arxiv.org/abs/1303.3586}{{arXiv:1303.3586}}} {[astro-ph.CO]}

\bibitem[{{Gupta} et~al.(2018){Gupta}, {Zorrilla Matilla}, {Hsu}, and
  {Haiman}}]{Gupta:2018}
{Gupta} A, {Zorrilla Matilla} JM, {Hsu} D, {Haiman} Z (2018) {Non-Gaussian
  information from weak lensing data via deep learning}. \prd 97(10):103515.
  \doi{10.1103/PhysRevD.97.103515}.
  {\href{https://arxiv.org/abs/1802.01212}{{arXiv:1802.01212}}} {[astro-ph.CO]}

\bibitem[{{Gurbatov} et~al.(1985){Gurbatov}, {Saichev}, and
  {Shandarin}}]{Gurbatov:1985}
{Gurbatov} SN, {Saichev} AI, {Shandarin} SF (1985) {A model for describing the
  development of the large-scale structure of the universe}. Soviet Physics
  Doklady 20:921

\bibitem[{{Habib} and {Ryne}(1995)}]{Habib:1995}
{Habib} S, {Ryne} RD (1995) {Symplectic Calculation of Lyapunov Exponents}.
  \prl 74(1):70--73. \doi{10.1103/PhysRevLett.74.70}.
  {\href{https://arxiv.org/abs/chao-dyn/9406010}{{arXiv:chao-dyn/9406010}}}
  {[nlin.CD]}

\bibitem[{Habib et~al.(2009)Habib, Pope, Luki{\'{c}}, Daniel, Fasel, Desai,
  Heitmann, Hsu, Ankeny, Mark, Bhattacharya, and Ahrens}]{Habib:2009}
Habib S, Pope A, Luki{\'{c}} Z, Daniel D, Fasel P, Desai N, Heitmann K, Hsu CH,
  Ankeny L, Mark G, Bhattacharya S, Ahrens J (2009) Hybrid petacomputing meets
  cosmology: The roadrunner universe project. J Phys Conf Ser 180:012019.
  \doi{10.1088/1742-6596/180/1/012019},
  \urlprefix\url{https://doi.org/10.1088/1742-6596/180/1/012019}

\bibitem[{{Habib} et~al.(2016){Habib}, {Pope}, {Finkel}, {Frontiere},
  {Heitmann}, {Daniel}, {Fasel}, {Morozov}, {Zagaris}, {Peterka}, {Vishwanath},
  {Luki{\'c}}, {Sehrish}, and {Liao}}]{Habib:2016}
{Habib} S, {Pope} A, {Finkel} H, {Frontiere} N, {Heitmann} K, {Daniel} D,
  {Fasel} P, {Morozov} V, {Zagaris} G, {Peterka} T, {Vishwanath} V, {Luki{\'c}}
  Z, {Sehrish} S, {Liao} Wk (2016) {HACC: Simulating sky surveys on
  state-of-the-art supercomputing architectures}. \na 42:49--65.
  \doi{10.1016/j.newast.2015.06.003}.
  {\href{https://arxiv.org/abs/1410.2805}{{arXiv:1410.2805}}} {[astro-ph.IM]}

\bibitem[{{Hadzhiyska} et~al.(2020){Hadzhiyska}, {Bose}, {Eisenstein},
  {Hernquist}, and {Spergel}}]{Hadzhiyska:2020}
{Hadzhiyska} B, {Bose} S, {Eisenstein} D, {Hernquist} L, {Spergel} DN (2020)
  {Limitations to the `basic' HOD model and beyond}. \mnras 493(4):5506--5519.
  \doi{10.1093/mnras/staa623}.
  {\href{https://arxiv.org/abs/1911.02610}{{arXiv:1911.02610}}} {[astro-ph.CO]}

\bibitem[{{Hadzhiyska} et~al.(2021{\natexlab{a}}){Hadzhiyska}, {Bose},
  {Eisenstein}, and {Hernquist}}]{Hadzhiyska:2021}
{Hadzhiyska} B, {Bose} S, {Eisenstein} D, {Hernquist} L (2021{\natexlab{a}})
  {Extensions to models of the galaxy-halo connection}. \mnras
  501(2):1603--1620. \doi{10.1093/mnras/staa3776}.
  {\href{https://arxiv.org/abs/2008.04913}{{arXiv:2008.04913}}} {[astro-ph.CO]}

\bibitem[{{Hadzhiyska} et~al.(2021{\natexlab{b}}){Hadzhiyska},
  {Garc{\'\i}a-Garc{\'\i}a}, {Alonso}, {Nicola}, and {Slosar}}]{Boryana:2021}
{Hadzhiyska} B, {Garc{\'\i}a-Garc{\'\i}a} C, {Alonso} D, {Nicola} A, {Slosar} A
  (2021{\natexlab{b}}) {Hefty enhancement of cosmological constraints from the
  DES Y1 data using a Hybrid Effective Field Theory approach to galaxy bias}.
  arXiv e-prints arXiv:2103.09820.
  {\href{https://arxiv.org/abs/2103.09820}{{arXiv:2103.09820}}} {[astro-ph.CO]}

\bibitem[{{Hahn} and {Abel}(2011)}]{Hahn:2011}
{Hahn} O, {Abel} T (2011) {Multi-scale initial conditions for cosmological
  simulations}. \mnras 415(3):2101--2121.
  \doi{10.1111/j.1365-2966.2011.18820.x}.
  {\href{https://arxiv.org/abs/1103.6031}{{arXiv:1103.6031}}} {[astro-ph.CO]}

\bibitem[{{Hahn} and {Angulo}(2016)}]{HahnAngulo:2016}
{Hahn} O, {Angulo} RE (2016) {An adaptively refined phase-space element method
  for cosmological simulations and collisionless dynamics}. \mnras
  455:1115--1133. \doi{10.1093/mnras/stv2304}.
  {\href{https://arxiv.org/abs/1501.01959}{{arXiv:1501.01959}}}

\bibitem[{{Hahn} and {Paranjape}(2016)}]{Hahn:2016}
{Hahn} O, {Paranjape} A (2016) {General relativistic screening in cosmological
  simulations}. \prd 94(8):083511. \doi{10.1103/PhysRevD.94.083511}.
  {\href{https://arxiv.org/abs/1602.07699}{{arXiv:1602.07699}}}

\bibitem[{{Hahn} et~al.(2007{\natexlab{a}}){Hahn}, {Carollo}, {Porciani}, and
  {Dekel}}]{Hahn:2007b}
{Hahn} O, {Carollo} CM, {Porciani} C, {Dekel} A (2007{\natexlab{a}}) {The
  evolution of dark matter halo properties in clusters, filaments, sheets and
  voids}. \mnras 381(1):41--51. \doi{10.1111/j.1365-2966.2007.12249.x}.
  {\href{https://arxiv.org/abs/0704.2595}{{arXiv:0704.2595}}} {[astro-ph]}

\bibitem[{{Hahn} et~al.(2007{\natexlab{b}}){Hahn}, {Porciani}, {Carollo}, and
  {Dekel}}]{Hahn:2007}
{Hahn} O, {Porciani} C, {Carollo} CM, {Dekel} A (2007{\natexlab{b}})
  {Properties of dark matter haloes in clusters, filaments, sheets and voids}.
  \mnras 375(2):489--499. \doi{10.1111/j.1365-2966.2006.11318.x}.
  {\href{https://arxiv.org/abs/astro-ph/0610280}{{arXiv:astro-ph/0610280}}}
  {[astro-ph]}

\bibitem[{{Hahn} et~al.(2013){Hahn}, {Abel}, and {Kaehler}}]{Hahn:2013}
{Hahn} O, {Abel} T, {Kaehler} R (2013) {A new approach to simulating
  collisionless dark matter fluids}. \mnras 434:1171--1191.
  \doi{10.1093/mnras/stt1061}.
  {\href{https://arxiv.org/abs/1210.6652}{{arXiv:1210.6652}}}

\bibitem[{{Hahn} et~al.(2015){Hahn}, {Angulo}, and {Abel}}]{Hahn:2015}
{Hahn} O, {Angulo} RE, {Abel} T (2015) {The properties of cosmic velocity
  fields}. \mnras 454(4):3920--3937. \doi{10.1093/mnras/stv2179}.
  {\href{https://arxiv.org/abs/1404.2280}{{arXiv:1404.2280}}} {[astro-ph.CO]}

\bibitem[{{Hahn} et~al.(2021){Hahn}, {Rampf}, and {Uhlemann}}]{Hahn:2020}
{Hahn} O, {Rampf} C, {Uhlemann} C (2021) {Higher order initial conditions for
  mixed baryon-CDM simulations}. \mnras 503(1):426--445.
  \doi{10.1093/mnras/staa3773}.
  {\href{https://arxiv.org/abs/2008.09124}{{arXiv:2008.09124}}} {[astro-ph.CO]}

\bibitem[{Hairer et~al.(2006)Hairer, Lubich, and Wanner}]{Hairer:2006}
Hairer E, Lubich C, Wanner G (2006) {Geometric Numerical Integration:
  Structure-Preserving Algorithms for Ordinary Differential Equations; 2nd ed.}
  Springer, Dordrecht

\bibitem[{{Halle} et~al.(2019){Halle}, {Colombi}, and {Peirani}}]{Halle:2019}
{Halle} A, {Colombi} S, {Peirani} S (2019) {Phase-space structure analysis of
  self-gravitating collisionless spherical systems}. \aap 621:A8.
  \doi{10.1051/0004-6361/201833460}.
  {\href{https://arxiv.org/abs/1701.01384}{{arXiv:1701.01384}}} {[astro-ph.GA]}

\bibitem[{{Hamaus} et~al.(2020){Hamaus}, {Pisani}, {Choi}, {Lavaux}, {Wandelt},
  and {Weller}}]{Hamaus:2020}
{Hamaus} N, {Pisani} A, {Choi} JA, {Lavaux} G, {Wandelt} BD, {Weller} J (2020)
  {Precision cosmology with voids in the final BOSS data}. \jcap 2020(12):023.
  \doi{10.1088/1475-7516/2020/12/023}.
  {\href{https://arxiv.org/abs/2007.07895}{{arXiv:2007.07895}}} {[astro-ph.CO]}

\bibitem[{{Han} et~al.(2012){Han}, {Jing}, {Wang}, and {Wang}}]{Han:2012}
{Han} J, {Jing} YP, {Wang} H, {Wang} W (2012) {Resolving subhaloes' lives with
  the Hierarchical Bound-Tracing algorithm}. \mnras 427(3):2437--2449.
  \doi{10.1111/j.1365-2966.2012.22111.x}.
  {\href{https://arxiv.org/abs/1103.2099}{{arXiv:1103.2099}}} {[astro-ph.CO]}

\bibitem[{{Han} et~al.(2018){Han}, {Cole}, {Frenk}, {Benitez-Llambay}, and
  {Helly}}]{Han:2018}
{Han} J, {Cole} S, {Frenk} CS, {Benitez-Llambay} A, {Helly} J (2018) {HBT+: an
  improved code for finding subhaloes and building merger trees in cosmological
  simulations}. \mnras 474(1):604--617. \doi{10.1093/mnras/stx2792}.
  {\href{https://arxiv.org/abs/1708.03646}{{arXiv:1708.03646}}} {[astro-ph.CO]}

\bibitem[{{Hand} et~al.(2018){Hand}, {Feng}, {Beutler}, {Li}, {Modi}, {Seljak},
  and {Slepian}}]{Hand:2018}
{Hand} N, {Feng} Y, {Beutler} F, {Li} Y, {Modi} C, {Seljak} U, {Slepian} Z
  (2018) {nbodykit: An Open-source, Massively Parallel Toolkit for Large-scale
  Structure}. \aj 156(4):160. \doi{10.3847/1538-3881/aadae0}.
  {\href{https://arxiv.org/abs/1712.05834}{{arXiv:1712.05834}}} {[astro-ph.IM]}

\bibitem[{{Hannestad} et~al.(2020){Hannestad}, {Upadhye}, and
  {Wong}}]{Hannestad:2020}
{Hannestad} S, {Upadhye} A, {Wong} YYY (2020) {Spoon or slide? The non-linear
  matter power spectrum in the presence of massive neutrinos}. \jcap
  2020(11):062. \doi{10.1088/1475-7516/2020/11/062}.
  {\href{https://arxiv.org/abs/2006.04995}{{arXiv:2006.04995}}} {[astro-ph.CO]}

\bibitem[{{Hansen} et~al.(2007){Hansen}, {Agertz}, {Joyce}, {Stadel}, {Moore},
  and {Potter}}]{Hansen:2007}
{Hansen} SH, {Agertz} O, {Joyce} M, {Stadel} J, {Moore} B, {Potter} D (2007)
  {An Alternative to Grids and Glasses: Quaquaversal Pre-Initial Conditions for
  N-Body Simulations}. \apj 656(2):631--635. \doi{10.1086/510477}.
  {\href{https://arxiv.org/abs/astro-ph/0606148}{{arXiv:astro-ph/0606148}}}
  {[astro-ph]}

\bibitem[{{Harnois-D{\'e}raps} et~al.(2013){Harnois-D{\'e}raps}, {Pen},
  {Iliev}, {Merz}, {Emberson}, and {Desjacques}}]{HarnoisDeraps:2013}
{Harnois-D{\'e}raps} J, {Pen} UL, {Iliev} IT, {Merz} H, {Emberson} JD,
  {Desjacques} V (2013) {High-performance P$^{3}$M N-body code: CUBEP$^{3}$M}.
  \mnras 436(1):540--559. \doi{10.1093/mnras/stt1591}.
  {\href{https://arxiv.org/abs/1208.5098}{{arXiv:1208.5098}}} {[astro-ph.CO]}

\bibitem[{{Harvey} et~al.(2015){Harvey}, {Massey}, {Kitching}, {Taylor}, and
  {Tittley}}]{Harvey:2015}
{Harvey} D, {Massey} R, {Kitching} T, {Taylor} A, {Tittley} E (2015) {The
  nongravitational interactions of dark matter in colliding galaxy clusters}.
  Science 347(6229):1462--1465. \doi{10.1126/science.1261381}.
  {\href{https://arxiv.org/abs/1503.07675}{{arXiv:1503.07675}}} {[astro-ph.CO]}

\bibitem[{{Hashim} et~al.(2018){Hashim}, {Giocoli}, {Baldi}, {Bertacca}, and
  {Maartens}}]{Hashim:2018}
{Hashim} M, {Giocoli} C, {Baldi} M, {Bertacca} D, {Maartens} R (2018) {Cosmic
  degeneracies III: N-body simulations of interacting dark energy with
  non-Gaussian initial conditions}. \mnras 481(3):2933--2945.
  \doi{10.1093/mnras/sty2450}.
  {\href{https://arxiv.org/abs/1806.02356}{{arXiv:1806.02356}}} {[astro-ph.CO]}

\bibitem[{{Hassani} and {Lombriser}(2020)}]{Hassani:2020}
{Hassani} F, {Lombriser} L (2020) {N-body simulations for parametrized modified
  gravity}. \mnras 497(2):1885--1894. \doi{10.1093/mnras/staa2083}.
  {\href{https://arxiv.org/abs/2003.05927}{{arXiv:2003.05927}}} {[astro-ph.CO]}

\bibitem[{{Hayli}(1967)}]{Hayli:1967}
{Hayli} A (1967) {Le probl{\`e}me des N corps dans un champ ext{\'e}rieur.
  Application {\`a} l'{\'e}volution dynamique des amas ouverts. I}. In: Les
  Nouvelles M\&eacute;thodes de la Dynamique Stellaire. p~67

\bibitem[{{Hayli}(1974)}]{Hayli:1974}
{Hayli} A (1974) {The method of the doubly individual step for N-body
  computations.} In: Numerical Solution of Ordinary Differential Equations. pp
  304--312

\bibitem[{{He}(2020)}]{He:2020}
{He} Jh (2020) {Modelling the tightest relation between galaxy properties and
  dark matter halo properties from hydrodynamical simulations of galaxy
  formation}. \mnras 493(3):4453--4462. \doi{10.1093/mnras/staa620}.
  {\href{https://arxiv.org/abs/1905.01612}{{arXiv:1905.01612}}} {[astro-ph.CO]}

\bibitem[{{He} et~al.(2016){He}, {Li}, and {Baugh}}]{He:2016}
{He} Jh, {Li} B, {Baugh} CM (2016) {Subhalo Abundance Matching in f (R )
  Gravity}. \prl 117(22):221101. \doi{10.1103/PhysRevLett.117.221101}.
  {\href{https://arxiv.org/abs/1605.04709}{{arXiv:1605.04709}}} {[astro-ph.CO]}

\bibitem[{{He} et~al.(2018){He}, {Guzzo}, {Li}, and {Baugh}}]{He:2018}
{He} Jh, {Guzzo} L, {Li} B, {Baugh} CM (2018) {No evidence for modifications of
  gravity from galaxy motions on cosmological scales}. Nature Astronomy
  2:967--972. \doi{10.1038/s41550-018-0573-2}.
  {\href{https://arxiv.org/abs/1809.09019}{{arXiv:1809.09019}}} {[astro-ph.CO]}

\bibitem[{{He} et~al.(2019){He}, {Li}, {Feng}, {Ho}, {Ravanbakhsh}, {Chen}, and
  {P{\'o}czos}}]{He:2019}
{He} S, {Li} Y, {Feng} Y, {Ho} S, {Ravanbakhsh} S, {Chen} W, {P{\'o}czos} B
  (2019) {Learning to predict the cosmological structure formation}.
  Proceedings of the National Academy of Science 116(28):13825--13832.
  \doi{10.1073/pnas.1821458116}.
  {\href{https://arxiv.org/abs/1811.06533}{{arXiv:1811.06533}}} {[astro-ph.CO]}

\bibitem[{{Hearin} and {Watson}(2013)}]{HearinWatson:2013}
{Hearin} AP, {Watson} DF (2013) {The dark side of galaxy colour}. \mnras
  435(2):1313--1324. \doi{10.1093/mnras/stt1374}.
  {\href{https://arxiv.org/abs/1304.5557}{{arXiv:1304.5557}}} {[astro-ph.CO]}

\bibitem[{{Hearin} et~al.(2013){Hearin}, {Zentner}, {Berlind}, and
  {Newman}}]{Hearin:2013}
{Hearin} AP, {Zentner} AR, {Berlind} AA, {Newman} JA (2013) {SHAM beyond
  clustering: new tests of galaxy-halo abundance matching with galaxy groups}.
  \mnras 433(1):659--680. \doi{10.1093/mnras/stt755}.
  {\href{https://arxiv.org/abs/1210.4927}{{arXiv:1210.4927}}} {[astro-ph.CO]}

\bibitem[{{Hearin} et~al.(2016){Hearin}, {Zentner}, {van den Bosch},
  {Campbell}, and {Tollerud}}]{Hearin:2016}
{Hearin} AP, {Zentner} AR, {van den Bosch} FC, {Campbell} D, {Tollerud} E
  (2016) {Introducing decorated HODs: modelling assembly bias in the
  galaxy-halo connection}. \mnras 460(3):2552--2570.
  \doi{10.1093/mnras/stw840}.
  {\href{https://arxiv.org/abs/1512.03050}{{arXiv:1512.03050}}} {[astro-ph.CO]}

\bibitem[{{Hearin} et~al.(2017){Hearin}, {Campbell}, {Tollerud}, {Behroozi},
  {Diemer}, {Goldbaum}, {Jennings}, {Leauthaud}, {Mao}, {More}, {Parejko},
  {Sinha}, {Sip{\"o}cz}, and {Zentner}}]{Hearin:2017}
{Hearin} AP, {Campbell} D, {Tollerud} E, {Behroozi} P, {Diemer} B, {Goldbaum}
  NJ, {Jennings} E, {Leauthaud} A, {Mao} YY, {More} S, {Parejko} J, {Sinha} M,
  {Sip{\"o}cz} B, {Zentner} A (2017) {Forward Modeling of Large-scale
  Structure: An Open-source Approach with Halotools}. \aj 154(5):190.
  \doi{10.3847/1538-3881/aa859f}.
  {\href{https://arxiv.org/abs/1606.04106}{{arXiv:1606.04106}}} {[astro-ph.IM]}

\bibitem[{{Heggie}(1991)}]{Heggie:1991}
{Heggie} DC (1991) {Chaos in the N-body problem of stellar dynamics.} In:
  Predictability, Stability, and Chaos in N-Body Dynamical Systems. NATO
  Advanced Study Institute (ASI) Series B, vol 272. pp 47--62

\bibitem[{{Heinesen} and {Buchert}(2020)}]{Heinesen:2020}
{Heinesen} A, {Buchert} T (2020) {Solving the curvature and Hubble parameter
  inconsistencies through structure formation-induced curvature}. Classical and
  Quantum Gravity 37(16):164001. \doi{10.1088/1361-6382/ab954b}.
  {\href{https://arxiv.org/abs/2002.10831}{{arXiv:2002.10831}}} {[gr-qc]}

\bibitem[{{Heitmann} et~al.(2005){Heitmann}, {Ricker}, {Warren}, and
  {Habib}}]{Heitmann:2005}
{Heitmann} K, {Ricker} PM, {Warren} MS, {Habib} S (2005) {Robustness of
  Cosmological Simulations. I. Large-Scale Structure}. \apjs 160(1):28--58.
  \doi{10.1086/432646}.
  {\href{https://arxiv.org/abs/astro-ph/0411795}{{arXiv:astro-ph/0411795}}}
  {[astro-ph]}

\bibitem[{{Heitmann} et~al.(2008){Heitmann}, {Luki{\'c}}, {Fasel}, {Habib},
  {Warren}, {White}, {Ahrens}, {Ankeny}, {Armstrong}, {O'Shea}, {Ricker},
  {Springel}, {Stadel}, and {Trac}}]{Heitmann:2008}
{Heitmann} K, {Luki{\'c}} Z, {Fasel} P, {Habib} S, {Warren} MS, {White} M,
  {Ahrens} J, {Ankeny} L, {Armstrong} R, {O'Shea} B, {Ricker} PM, {Springel} V,
  {Stadel} J, {Trac} H (2008) {The cosmic code comparison project}.
  Computational Science and Discovery 1(1):015003.
  \doi{10.1088/1749-4699/1/1/015003}.
  {\href{https://arxiv.org/abs/0706.1270}{{arXiv:0706.1270}}} {[astro-ph]}

\bibitem[{Heitmann et~al.(2009)Heitmann, Higdon, White, Habib, Williams, and
  Wagner}]{Heitmann:2009}
Heitmann K, Higdon D, White M, Habib S, Williams BJ, Wagner C (2009) {The
  Coyote Universe II: Cosmological Models and Precision Emulation of the
  Nonlinear Matter Power Spectrum}. Astrophys J 705:156--174.
  \doi{10.1088/0004-637X/705/1/156}.
  {\href{https://arxiv.org/abs/0902.0429}{{arXiv:0902.0429}}} {[astro-ph.CO]}

\bibitem[{{Heitmann} et~al.(2016){Heitmann}, {Bingham}, {Lawrence}, {Bergner},
  {Habib}, {Higdon}, {Pope}, {Biswas}, {Finkel}, {Frontiere}, and
  {Bhattacharya}}]{Heitmann:2016}
{Heitmann} K, {Bingham} D, {Lawrence} E, {Bergner} S, {Habib} S, {Higdon} D,
  {Pope} A, {Biswas} R, {Finkel} H, {Frontiere} N, {Bhattacharya} S (2016) {The
  Mira-Titan Universe: Precision Predictions for Dark Energy Surveys}. \apj
  820(2):108. \doi{10.3847/0004-637X/820/2/108}.
  {\href{https://arxiv.org/abs/1508.02654}{{arXiv:1508.02654}}} {[astro-ph.CO]}

\bibitem[{{Heitmann} et~al.(2019){Heitmann}, {Finkel}, {Pope}, {Morozov},
  {Frontiere}, {Habib}, {Rangel}, {Uram}, {Korytov}, {Child}, {Flender},
  {Insley}, and {Rizzi}}]{Heitmann:2019}
{Heitmann} K, {Finkel} H, {Pope} A, {Morozov} V, {Frontiere} N, {Habib} S,
  {Rangel} E, {Uram} T, {Korytov} D, {Child} H, {Flender} S, {Insley} J,
  {Rizzi} S (2019) {The Outer Rim Simulation: A Path to Many-core
  Supercomputers}. \apjs 245(1):16. \doi{10.3847/1538-4365/ab4da1}.
  {\href{https://arxiv.org/abs/1904.11970}{{arXiv:1904.11970}}} {[astro-ph.CO]}

\bibitem[{{Heitmann} et~al.(2021){Heitmann}, {Frontiere}, {Rangel}, {Larsen},
  {Pope}, {Sultan}, {Uram}, {Habib}, {Finkel}, {Korytov}, {Kovacs}, {Rizzi},
  {Insley}, and {Knowles}}]{Heitmann:2021}
{Heitmann} K, {Frontiere} N, {Rangel} E, {Larsen} P, {Pope} A, {Sultan} I,
  {Uram} T, {Habib} S, {Finkel} H, {Korytov} D, {Kovacs} E, {Rizzi} S, {Insley}
  J, {Knowles} JYK (2021) {The Last Journey. I. An Extreme-scale Simulation on
  the Mira Supercomputer}. \apjs 252(2):19. \doi{10.3847/1538-4365/abcc67}.
  {\href{https://arxiv.org/abs/2006.01697}{{arXiv:2006.01697}}} {[astro-ph.CO]}

\bibitem[{{Hellwing} et~al.(2016){Hellwing}, {Schaller}, {Frenk}, {Theuns},
  {Schaye}, {Bower}, and {Crain}}]{Hellwing:2016}
{Hellwing} WA, {Schaller} M, {Frenk} CS, {Theuns} T, {Schaye} J, {Bower} RG,
  {Crain} RA (2016) {The effect of baryons on redshift space distortions and
  cosmic density and velocity fields in the EAGLE simulation}. \mnras
  461:L11--L15. \doi{10.1093/mnrasl/slw081}.
  {\href{https://arxiv.org/abs/1603.03328}{{arXiv:1603.03328}}}

\bibitem[{{H{\'e}non}(1964)}]{Henon:1964}
{H{\'e}non} M (1964) {L'{\'e}volution initiale d'un amas sph{\'e}rique}.
  Annales d'Astrophysique 27:83

\bibitem[{{Henriques} et~al.(2009){Henriques}, {Thomas}, {Oliver}, and
  {Roseboom}}]{Henriques:2009}
{Henriques} BMB, {Thomas} PA, {Oliver} S, {Roseboom} I (2009) {Monte Carlo
  Markov Chain parameter estimation in semi-analytic models of galaxy
  formation}. \mnras 396(1):535--547. \doi{10.1111/j.1365-2966.2009.14730.x}.
  {\href{https://arxiv.org/abs/0810.2548}{{arXiv:0810.2548}}} {[astro-ph]}

\bibitem[{{Henriques} et~al.(2013){Henriques}, {White}, {Thomas}, {Angulo},
  {Guo}, {Lemson}, and {Springel}}]{Henriques:2013}
{Henriques} BMB, {White} SDM, {Thomas} PA, {Angulo} RE, {Guo} Q, {Lemson} G,
  {Springel} V (2013) {Simulations of the galaxy population constrained by
  observations from z = 3 to the present day: implications for galactic winds
  and the fate of their ejecta}. \mnras 431(4):3373--3395.
  \doi{10.1093/mnras/stt415}.
  {\href{https://arxiv.org/abs/1212.1717}{{arXiv:1212.1717}}} {[astro-ph.CO]}

\bibitem[{{Henriques} et~al.(2015){Henriques}, {White}, {Thomas}, {Angulo},
  {Guo}, {Lemson}, {Springel}, and {Overzier}}]{Henriques:2015}
{Henriques} BMB, {White} SDM, {Thomas} PA, {Angulo} R, {Guo} Q, {Lemson} G,
  {Springel} V, {Overzier} R (2015) {Galaxy formation in the Planck cosmology -
  I. Matching the observed evolution of star formation rates, colours and
  stellar masses}. \mnras 451(3):2663--2680. \doi{10.1093/mnras/stv705}.
  {\href{https://arxiv.org/abs/1410.0365}{{arXiv:1410.0365}}} {[astro-ph.GA]}

\bibitem[{{Hern{\'a}ndez-Aguayo} et~al.(2021){Hern{\'a}ndez-Aguayo}, {Ruan},
  {Li}, {Arnold}, {Baugh}, {Klypin}, and {Prada}}]{HAguayo:2021}
{Hern{\'a}ndez-Aguayo} C, {Ruan} CZ, {Li} B, {Arnold} C, {Baugh} CM, {Klypin}
  A, {Prada} F (2021) {Fast full $N$-body simulations of generic modified
  gravity: derivative coupling models}. arXiv e-prints arXiv:2110.00566.
  {\href{https://arxiv.org/abs/2110.00566}{{arXiv:2110.00566}}} {[astro-ph.CO]}

\bibitem[{{Hern{\'a}ndez-Monteagudo} et~al.(2020){Hern{\'a}ndez-Monteagudo},
  {Chaves-Montero}, {Angulo}, and {Aricc{\`o}}}]{HernandezMonteagudo:2020}
{Hern{\'a}ndez-Monteagudo} C, {Chaves-Montero} J, {Angulo} RE, {Aricc{\`o}} G
  (2020) {Tomographic Constraints on Gravity from Angular Redshift Fluctuations
  in the Late Universe}. arXiv e-prints arXiv:2005.06568.
  {\href{https://arxiv.org/abs/2005.06568}{{arXiv:2005.06568}}} {[astro-ph.CO]}

\bibitem[{{Hernquist} and {Barnes}(1990)}]{Hernquist:1990}
{Hernquist} L, {Barnes} JE (1990) {Are Some N-Body Algorithms Intrinsically
  Less Collisional than Others?} \apj 349:562. \doi{10.1086/168343}

\bibitem[{{He{\ss}} et~al.(2013){He{\ss}}, {Kitaura}, and
  {Gottl{\"o}ber}}]{Hess:2013}
{He{\ss}} S, {Kitaura} FS, {Gottl{\"o}ber} S (2013) {Simulating structure
  formation of the Local Universe}. \mnras 435(3):2065--2076.
  \doi{10.1093/mnras/stt1428}.
  {\href{https://arxiv.org/abs/1304.6565}{{arXiv:1304.6565}}} {[astro-ph.CO]}

\bibitem[{{Hildebrandt} et~al.(2017){Hildebrandt}, {Viola}, {Heymans},
  {Joudaki}, {Kuijken}, {Blake}, {Erben}, {Joachimi}, {Klaes}, {Miller},
  {Morrison}, {Nakajima}, {Verdoes Kleijn}, {Amon}, {Choi}, {Covone}, {de
  Jong}, {Dvornik}, {Fenech Conti}, {Grado}, {Harnois-D{\'e}raps}, {Herbonnet},
  {Hoekstra}, {K{\"o}hlinger}, {McFarland}, {Mead}, {Merten}, {Napolitano},
  {Peacock}, {Radovich}, {Schneider}, {Simon}, {Valentijn}, {van den Busch},
  {van Uitert}, and {Van Waerbeke}}]{Hildebrandt:2017}
{Hildebrandt} H, {Viola} M, {Heymans} C, {Joudaki} S, {Kuijken} K, {Blake} C,
  {Erben} T, {Joachimi} B, {Klaes} D, {Miller} L, {Morrison} CB, {Nakajima} R,
  {Verdoes Kleijn} G, {Amon} A, {Choi} A, {Covone} G, {de Jong} JTA, {Dvornik}
  A, {Fenech Conti} I, {Grado} A, {Harnois-D{\'e}raps} J, {Herbonnet} R,
  {Hoekstra} H, {K{\"o}hlinger} F, {McFarland} J, {Mead} A, {Merten} J,
  {Napolitano} N, {Peacock} JA, {Radovich} M, {Schneider} P, {Simon} P,
  {Valentijn} EA, {van den Busch} JL, {van Uitert} E, {Van Waerbeke} L (2017)
  {KiDS-450: cosmological parameter constraints from tomographic weak
  gravitational lensing}. \mnras 465(2):1454--1498.
  \doi{10.1093/mnras/stw2805}.
  {\href{https://arxiv.org/abs/1606.05338}{{arXiv:1606.05338}}} {[astro-ph.CO]}

\bibitem[{{Hirschmann} et~al.(2014){Hirschmann}, {Dolag}, {Saro}, {Bachmann},
  {Borgani}, and {Burkert}}]{Hirschmann:2014}
{Hirschmann} M, {Dolag} K, {Saro} A, {Bachmann} L, {Borgani} S, {Burkert} A
  (2014) {Cosmological simulations of black hole growth: AGN luminosities and
  downsizing}. \mnras 442(3):2304--2324. \doi{10.1093/mnras/stu1023}.
  {\href{https://arxiv.org/abs/1308.0333}{{arXiv:1308.0333}}} {[astro-ph.CO]}

\bibitem[{{Hirschmann} et~al.(2016){Hirschmann}, {De Lucia}, and
  {Fontanot}}]{Hirschmann:2016}
{Hirschmann} M, {De Lucia} G, {Fontanot} F (2016) {Galaxy assembly, stellar
  feedback and metal enrichment: the view from the GAEA model}. \mnras
  461(2):1760--1785. \doi{10.1093/mnras/stw1318}.
  {\href{https://arxiv.org/abs/1512.04531}{{arXiv:1512.04531}}} {[astro-ph.GA]}

\bibitem[{{Hlozek} et~al.(2015){Hlozek}, {Grin}, {Marsh}, and
  {Ferreira}}]{Hlozek:2015}
{Hlozek} R, {Grin} D, {Marsh} DJE, {Ferreira} PG (2015) {A search for
  ultralight axions using precision cosmological data}. \prd 91(10):103512.
  \doi{10.1103/PhysRevD.91.103512}.
  {\href{https://arxiv.org/abs/1410.2896}{{arXiv:1410.2896}}} {[astro-ph.CO]}

\bibitem[{{Ho} et~al.(2021){Ho}, {Bird}, and {Shelton}}]{Ho:2021}
{Ho} MF, {Bird} S, {Shelton} CR (2021) {Multi-Fidelity Emulation for the Matter
  Power Spectrum using Gaussian Processes}. arXiv e-prints arXiv:2105.01081.
  {\href{https://arxiv.org/abs/2105.01081}{{arXiv:2105.01081}}} {[astro-ph.CO]}

\bibitem[{{Ho} et~al.(2015){Ho}, {Agarwal}, {Myers}, {Lyons}, {Disbrow}, {Seo},
  {Ross}, {Hirata}, {Padmanabhan}, {O'Connell}, {Huff}, {Schlegel}, {Slosar},
  {Weinberg}, {Strauss}, {Ross}, {Schneider}, {Bahcall}, {Brinkmann},
  {Palanque-Delabrouille}, and {Y{\`e}che}}]{Ho:2015}
{Ho} S, {Agarwal} N, {Myers} AD, {Lyons} R, {Disbrow} A, {Seo} HJ, {Ross} A,
  {Hirata} C, {Padmanabhan} N, {O'Connell} R, {Huff} E, {Schlegel} D, {Slosar}
  A, {Weinberg} D, {Strauss} M, {Ross} NP, {Schneider} DP, {Bahcall} N,
  {Brinkmann} J, {Palanque-Delabrouille} N, {Y{\`e}che} C (2015) {Sloan Digital
  Sky Survey III photometric quasar clustering: probing the initial conditions
  of the Universe}. \jcap 2015(5):040. \doi{10.1088/1475-7516/2015/05/040}.
  {\href{https://arxiv.org/abs/1311.2597}{{arXiv:1311.2597}}} {[astro-ph.CO]}

\bibitem[{{Hobbs} et~al.(2016){Hobbs}, {Read}, {Agertz}, {Iannuzzi}, and
  {Power}}]{Hobbs:2016}
{Hobbs} A, {Read} JI, {Agertz} O, {Iannuzzi} F, {Power} C (2016) {NOVel
  Adaptive softening for collisionless N-body simulations: eliminating spurious
  haloes}. \mnras 458(1):468--479. \doi{10.1093/mnras/stw251}.
  {\href{https://arxiv.org/abs/1503.02689}{{arXiv:1503.02689}}} {[astro-ph.CO]}

\bibitem[{{Hockney} and {Eastwood}(1981)}]{Hockney:1981}
{Hockney} RW, {Eastwood} JW (1981) {Computer Simulation Using Particles}.
  Computer Simulation Using Particles, New York: McGraw-Hill, 1981

\bibitem[{{Hoffman} and {Ribak}(1991)}]{Hoffman:1991}
{Hoffman} Y, {Ribak} E (1991) {Constrained Realizations of Gaussian Fields: A
  Simple Algorithm}. \apjl 380:L5. \doi{10.1086/186160}

\bibitem[{{Hoffman} et~al.(2012){Hoffman}, {Metuki}, {Yepes}, {Gottl{\"o}ber},
  {Forero-Romero}, {Libeskind}, and {Knebe}}]{Hoffman:2012}
{Hoffman} Y, {Metuki} O, {Yepes} G, {Gottl{\"o}ber} S, {Forero-Romero} JE,
  {Libeskind} NI, {Knebe} A (2012) {A kinematic classification of the cosmic
  web}. \mnras 425(3):2049--2057. \doi{10.1111/j.1365-2966.2012.21553.x}.
  {\href{https://arxiv.org/abs/1201.3367}{{arXiv:1201.3367}}} {[astro-ph.CO]}

\bibitem[{{Hofmann} et~al.(2001){Hofmann}, {Schwarz}, and
  {St{\"o}cker}}]{Hofmann:2001}
{Hofmann} S, {Schwarz} DJ, {St{\"o}cker} H (2001) {Damping scales of neutralino
  cold dark matter}. \prd 64(8):083507. \doi{10.1103/PhysRevD.64.083507}.
  {\href{https://arxiv.org/abs/astro-ph/0104173}{{arXiv:astro-ph/0104173}}}
  {[astro-ph]}

\bibitem[{{Hojjati} et~al.(2011){Hojjati}, {Pogosian}, and
  {Zhao}}]{Hojjati:2011}
{Hojjati} A, {Pogosian} L, {Zhao} GB (2011) {Testing gravity with CAMB and
  CosmoMC}. \jcap 2011(8):005. \doi{10.1088/1475-7516/2011/08/005}.
  {\href{https://arxiv.org/abs/1106.4543}{{arXiv:1106.4543}}} {[astro-ph.CO]}

\bibitem[{{Hollowed}(2019)}]{Hollowed:2019}
{Hollowed} J (2019) {Lightcone Construction for HACC Cosmological Simulations
  with LANTERN}. arXiv e-prints arXiv:1906.08355.
  {\href{https://arxiv.org/abs/1906.08355}{{arXiv:1906.08355}}} {[astro-ph.CO]}

\bibitem[{{Hopkins}(2019)}]{Hopkins:2019}
{Hopkins} PF (2019) {A stable finite-volume method for scalar field dark
  matter}. \mnras 489(2):2367--2376. \doi{10.1093/mnras/stz1922}.
  {\href{https://arxiv.org/abs/1811.05583}{{arXiv:1811.05583}}} {[astro-ph.CO]}

\bibitem[{{Hotinli} et~al.(2019){Hotinli}, {Mertens}, {Johnson}, and
  {Kamionkowski}}]{Hotinli:2019}
{Hotinli} SC, {Mertens} JB, {Johnson} MC, {Kamionkowski} M (2019) {Probing
  correlated compensated isocurvature perturbations using scale-dependent
  galaxy bias}. \prd 100(10):103528. \doi{10.1103/PhysRevD.100.103528}.
  {\href{https://arxiv.org/abs/1908.08953}{{arXiv:1908.08953}}} {[astro-ph.CO]}

\bibitem[{{Howlett} et~al.(2015){Howlett}, {Manera}, and
  {Percival}}]{Howlett:2015}
{Howlett} C, {Manera} M, {Percival} WJ (2015) {L-PICOLA: A parallel code for
  fast dark matter simulation}. Astronomy and Computing 12:109--126.
  \doi{10.1016/j.ascom.2015.07.003}.
  {\href{https://arxiv.org/abs/1506.03737}{{arXiv:1506.03737}}} {[astro-ph.CO]}

\bibitem[{{Hu} et~al.(2000){Hu}, {Barkana}, and {Gruzinov}}]{Hu:2000}
{Hu} W, {Barkana} R, {Gruzinov} A (2000) {Fuzzy Cold Dark Matter: The Wave
  Properties of Ultralight Particles}. \prl 85(6):1158--1161.
  \doi{10.1103/PhysRevLett.85.1158}.
  {\href{https://arxiv.org/abs/astro-ph/0003365}{{arXiv:astro-ph/0003365}}}
  {[astro-ph]}

\bibitem[{{Hu} et~al.(2016){Hu}, {Chiang}, {Li}, and {LoVerde}}]{Hu:2016}
{Hu} W, {Chiang} CT, {Li} Y, {LoVerde} M (2016) {Separating the Universe into
  real and fake energy densities}. \prd 94(2):023002.
  \doi{10.1103/PhysRevD.94.023002}.
  {\href{https://arxiv.org/abs/1605.01412}{{arXiv:1605.01412}}} {[astro-ph.CO]}

\bibitem[{{Huang} et~al.(2021){Huang}, {Eifler}, {Mandelbaum}, {Bernstein},
  {Chen}, {Choi}, {Garc{\'\i}a-Bellido}, {Huterer}, {Krause}, {Rozo}, {Singh},
  {Bridle}, {DeRose}, {Elvin-Poole}, {Fang}, {Friedrich}, {Gatti}, {Gaztanaga},
  {Gruen}, {Hartley}, {Hoyle}, {Jarvis}, {MacCrann}, {Miranda}, {Rau}, {Prat},
  {S{\'a}nchez}, {Samuroff}, {Troxel}, {Zuntz}, {Abbott}, {Aguena}, {Annis},
  {Avila}, {Becker}, {Bertin}, {Brooks}, {Burke}, {Carnero Rosell}, {Carrasco
  Kind}, {Carretero}, {Castander}, {da Costa}, {De Vicente}, {Dietrich},
  {Doel}, {Everett}, {Flaugher}, {Fosalba}, {Frieman}, {Gruendl}, {Gutierrez},
  {Hinton}, {Honscheid}, {James}, {Kuehn}, {Lahav}, {Lima}, {Maia}, {Marshall},
  {Menanteau}, {Miquel}, {Paz-Chinch{\'o}n}, {Malag{\'o}n}, {Romer}, {Roodman},
  {Sanchez}, {Scarpine}, {Serrano}, {Sevilla}, {Smith}, {Soares-Santos},
  {Suchyta}, {Swanson}, {Tarle}, {Thomas}, {Weller}, and {DES
  Collaboration}}]{Huang:2021}
{Huang} HJ, {Eifler} T, {Mandelbaum} R, {Bernstein} GM, {Chen} A, {Choi} A,
  {Garc{\'\i}a-Bellido} J, {Huterer} D, {Krause} E, {Rozo} E, {Singh} S,
  {Bridle} S, {DeRose} J, {Elvin-Poole} J, {Fang} X, {Friedrich} O, {Gatti} M,
  {Gaztanaga} E, {Gruen} D, {Hartley} W, {Hoyle} B, {Jarvis} M, {MacCrann} N,
  {Miranda} V, {Rau} M, {Prat} J, {S{\'a}nchez} C, {Samuroff} S, {Troxel} M,
  {Zuntz} J, {Abbott} T, {Aguena} M, {Annis} J, {Avila} S, {Becker} M, {Bertin}
  E, {Brooks} D, {Burke} D, {Carnero Rosell} A, {Carrasco Kind} M, {Carretero}
  J, {Castander} FJ, {da Costa} L, {De Vicente} J, {Dietrich} J, {Doel} P,
  {Everett} S, {Flaugher} B, {Fosalba} P, {Frieman} J, {Gruendl} R, {Gutierrez}
  G, {Hinton} S, {Honscheid} K, {James} D, {Kuehn} K, {Lahav} O, {Lima} M,
  {Maia} M, {Marshall} J, {Menanteau} F, {Miquel} R, {Paz-Chinch{\'o}n} F,
  {Malag{\'o}n} AP, {Romer} K, {Roodman} A, {Sanchez} E, {Scarpine} V,
  {Serrano} S, {Sevilla} I, {Smith} M, {Soares-Santos} M, {Suchyta} E,
  {Swanson} M, {Tarle} G, {Thomas} DH, {Weller} J, {DES Collaboration} (2021)
  {Dark energy survey year 1 results: Constraining baryonic physics in the
  Universe}. \mnras 502(4):6010--6031. \doi{10.1093/mnras/stab357}.
  {\href{https://arxiv.org/abs/2007.15026}{{arXiv:2007.15026}}} {[astro-ph.CO]}

\bibitem[{{Hui}(2021)}]{Hui:2021}
{Hui} L (2021) {Wave Dark Matter}. arXiv e-prints arXiv:2101.11735.
  {\href{https://arxiv.org/abs/2101.11735}{{arXiv:2101.11735}}} {[astro-ph.CO]}

\bibitem[{{Hui} et~al.(2017){Hui}, {Ostriker}, {Tremaine}, and
  {Witten}}]{Hui:2017}
{Hui} L, {Ostriker} JP, {Tremaine} S, {Witten} E (2017) {Ultralight scalars as
  cosmological dark matter}. \prd 95(4):043541.
  \doi{10.1103/PhysRevD.95.043541}.
  {\href{https://arxiv.org/abs/1610.08297}{{arXiv:1610.08297}}}

\bibitem[{{Huo} et~al.(2020){Huo}, {Yu}, and {Zhong}}]{Huo:2020}
{Huo} R, {Yu} HB, {Zhong} YM (2020) {The structure of dissipative dark matter
  halos}. \jcap 2020(6):051. \doi{10.1088/1475-7516/2020/06/051}.
  {\href{https://arxiv.org/abs/1912.06757}{{arXiv:1912.06757}}} {[astro-ph.CO]}

\bibitem[{{Iannuzzi} and {Dolag}(2011)}]{Ianuzzi:2011}
{Iannuzzi} F, {Dolag} K (2011) {Adaptive gravitational softening in GADGET}.
  \mnras 417(4):2846--2859. \doi{10.1111/j.1365-2966.2011.19446.x}.
  {\href{https://arxiv.org/abs/1107.2942}{{arXiv:1107.2942}}} {[astro-ph.CO]}

\bibitem[{{Icaza-Lizaola} et~al.(2021){Icaza-Lizaola}, {Bower}, {Norberg},
  {Cole}, and {Egan}}]{Icaza:2021}
{Icaza-Lizaola} M, {Bower} RG, {Norberg} P, {Cole} S, {Egan} S (2021) {A sparse
  regression approach to modeling the relation between galaxy stellar masses
  and their host halos}. arXiv e-prints arXiv:2101.02986.
  {\href{https://arxiv.org/abs/2101.02986}{{arXiv:2101.02986}}} {[astro-ph.GA]}

\bibitem[{Igouchkine et~al.(2016)Igouchkine, Leaf, and Ma}]{Igouchkine:2016}
Igouchkine O, Leaf N, Ma KL (2016) Volume rendering dark matter simulations
  using cell projection and order-independent transparency. In: SIGGRAPH ASIA
  2016 Symposium on Visualization. SA '16. ACM, New York, NY, USA, pp 8:1--8:8.
  \doi{10.1145/3002151.3002163}

\bibitem[{{Inman} and {Ali-Ha{\"\i}moud}(2019)}]{Inman:2019}
{Inman} D, {Ali-Ha{\"\i}moud} Y (2019) {Early structunre formation in
  primordial black hole cosmologies}. \prd 100(8):083528.
  \doi{10.1103/PhysRevD.100.083528}.
  {\href{https://arxiv.org/abs/1907.08129}{{arXiv:1907.08129}}} {[astro-ph.CO]}

\bibitem[{{Inman} and {Yu}(2020)}]{Inman:2020}
{Inman} D, {Yu} HR (2020) {Simulating the Cosmic Neutrino Background Using
  Collisionless Hydrodynamics}. \apjs 250(1):21.
  \doi{10.3847/1538-4365/aba0b3}.
  {\href{https://arxiv.org/abs/2002.04601}{{arXiv:2002.04601}}} {[astro-ph.CO]}

\bibitem[{{Inman} et~al.(2015){Inman}, {Emberson}, {Pen}, {Farchi}, {Yu}, and
  {Harnois-D{\'e}raps}}]{Inman:2015}
{Inman} D, {Emberson} JD, {Pen} UL, {Farchi} A, {Yu} HR, {Harnois-D{\'e}raps} J
  (2015) {Precision reconstruction of the cold dark matter-neutrino relative
  velocity from N -body simulations}. \prd 92(2):023502.
  \doi{10.1103/PhysRevD.92.023502}.
  {\href{https://arxiv.org/abs/1503.07480}{{arXiv:1503.07480}}} {[astro-ph.CO]}

\bibitem[{{Ir{\v{s}}i{\v{c}}} et~al.(2017{\natexlab{a}}){Ir{\v{s}}i{\v{c}}},
  {Viel}, {Haehnelt}, {Bolton}, and {Becker}}]{Irsic:2017b}
{Ir{\v{s}}i{\v{c}}} V, {Viel} M, {Haehnelt} MG, {Bolton} JS, {Becker} GD
  (2017{\natexlab{a}}) {First Constraints on Fuzzy Dark Matter from
  Lyman-{\ensuremath{\alpha}} Forest Data and Hydrodynamical Simulations}. \prl
  119(3):031302. \doi{10.1103/PhysRevLett.119.031302}.
  {\href{https://arxiv.org/abs/1703.04683}{{arXiv:1703.04683}}} {[astro-ph.CO]}

\bibitem[{{Ir{\v{s}}i{\v{c}}} et~al.(2017{\natexlab{b}}){Ir{\v{s}}i{\v{c}}},
  {Viel}, {Haehnelt}, {Bolton}, {Cristiani}, {Becker}, {D'Odorico}, {Cupani},
  {Kim}, {Berg}, {L{\'o}pez}, {Ellison}, {Christensen}, {Denney}, and
  {Worseck}}]{Irsic:2017}
{Ir{\v{s}}i{\v{c}}} V, {Viel} M, {Haehnelt} MG, {Bolton} JS, {Cristiani} S,
  {Becker} GD, {D'Odorico} V, {Cupani} G, {Kim} TS, {Berg} TAM, {L{\'o}pez} S,
  {Ellison} S, {Christensen} L, {Denney} KD, {Worseck} G (2017{\natexlab{b}})
  {New constraints on the free-streaming of warm dark matter from intermediate
  and small scale Lyman-{\ensuremath{\alpha}} forest data}. \prd 96(2):023522.
  \doi{10.1103/PhysRevD.96.023522}.
  {\href{https://arxiv.org/abs/1702.01764}{{arXiv:1702.01764}}} {[astro-ph.CO]}

\bibitem[{{Ishiyama}(2014)}]{Ishiyama:2014}
{Ishiyama} T (2014) {Hierarchical Formation of Dark Matter Halos and the Free
  Streaming Scale}. \apj 788(1):27. \doi{10.1088/0004-637X/788/1/27}.
  {\href{https://arxiv.org/abs/1404.1650}{{arXiv:1404.1650}}} {[astro-ph.CO]}

\bibitem[{{Ishiyama} and {Ando}(2020)}]{Ishiyama:2020}
{Ishiyama} T, {Ando} S (2020) {The abundance and structure of subhaloes near
  the free streaming scale and their impact on indirect dark matter searches}.
  \mnras 492(3):3662--3671. \doi{10.1093/mnras/staa069}.
  {\href{https://arxiv.org/abs/1907.03642}{{arXiv:1907.03642}}} {[astro-ph.CO]}

\bibitem[{{Ishiyama} et~al.(2009){Ishiyama}, {Fukushige}, and
  {Makino}}]{Ishiyama:2009}
{Ishiyama} T, {Fukushige} T, {Makino} J (2009) {GreeM: Massively Parallel
  TreePM Code for Large Cosmological N -body Simulations}. \pasj 61:1319.
  \doi{10.1093/pasj/61.6.1319}.
  {\href{https://arxiv.org/abs/0910.0121}{{arXiv:0910.0121}}} {[astro-ph.IM]}

\bibitem[{{Ishiyama} et~al.(2012){Ishiyama}, {Nitadori}, and
  {Makino}}]{Ishiyama:2012}
{Ishiyama} T, {Nitadori} K, {Makino} J (2012) {4.45 Pflops Astrophysical N-Body
  Simulation on K computer -- The Gravitational Trillion-Body Problem}. arXiv
  e-prints arXiv:1211.4406.
  {\href{https://arxiv.org/abs/1211.4406}{{arXiv:1211.4406}}} {[astro-ph.CO]}

\bibitem[{{Ishiyama} et~al.(2021){Ishiyama}, {Prada}, {Klypin}, {Sinha},
  {Metcalf}, {Jullo}, {Altieri}, {Cora}, {Croton}, {de la Torre},
  {Mill{\'a}n-Calero}, {Oogi}, {Ruedas}, and
  {Vega-Mart{\'\i}nez}}]{IshiyamaUshuu:2020}
{Ishiyama} T, {Prada} F, {Klypin} AA, {Sinha} M, {Metcalf} RB, {Jullo} E,
  {Altieri} B, {Cora} SA, {Croton} D, {de la Torre} S, {Mill{\'a}n-Calero} DE,
  {Oogi} T, {Ruedas} J, {Vega-Mart{\'\i}nez} CA (2021) {The Uchuu simulations:
  Data Release 1 and dark matter halo concentrations}. \mnras
  506(3):4210--4231. \doi{10.1093/mnras/stab1755}.
  {\href{https://arxiv.org/abs/2007.14720}{{arXiv:2007.14720}}} {[astro-ph.CO]}

\bibitem[{{Ivezi{\'c}} et~al.(2019){Ivezi{\'c}}, {Kahn}, {Tyson}, {Abel},
  {Acosta}, {Allsman}, {Alonso}, {AlSayyad}, {Anderson}, {Andrew}, and
  et~al.}]{LSST}
{Ivezi{\'c}} {\v{Z}}, {Kahn} SM, {Tyson} JA, {Abel} B, {Acosta} E, {Allsman} R,
  {Alonso} D, {AlSayyad} Y, {Anderson} SF, {Andrew} J, et~al (2019) {LSST: From
  Science Drivers to Reference Design and Anticipated Data Products}. \apj
  873(2):111. \doi{10.3847/1538-4357/ab042c}.
  {\href{https://arxiv.org/abs/0805.2366}{{arXiv:0805.2366}}} {[astro-ph]}

\bibitem[{{Ivkin} et~al.(2018){Ivkin}, {Liu}, {Yang}, {Kumar}, {Lemson},
  {Neyrinck}, {Szalay}, {Braverman}, and {Budavari}}]{Ivnik:2018}
{Ivkin} N, {Liu} Z, {Yang} LF, {Kumar} SS, {Lemson} G, {Neyrinck} M, {Szalay}
  AS, {Braverman} V, {Budavari} T (2018) {Scalable streaming tools for
  analyzing N-body simulations: Finding halos and investigating excursion sets
  in one pass}. Astronomy and Computing 23:166.
  \doi{10.1016/j.ascom.2018.04.003}.
  {\href{https://arxiv.org/abs/1711.00975}{{arXiv:1711.00975}}} {[astro-ph.IM]}

\bibitem[{{Iwanus} et~al.(2017){Iwanus}, {Elahi}, and {Lewis}}]{Iwanus:2017}
{Iwanus} N, {Elahi} PJ, {Lewis} GF (2017) {Dark matter annihilation feedback in
  cosmological simulations - I: Code convergence and idealized haloes}. \mnras
  472(1):1214--1225. \doi{10.1093/mnras/stx1974}.
  {\href{https://arxiv.org/abs/1707.06770}{{arXiv:1707.06770}}} {[astro-ph.CO]}

\bibitem[{{Izard} et~al.(2016){Izard}, {Crocce}, and {Fosalba}}]{Izard:2016}
{Izard} A, {Crocce} M, {Fosalba} P (2016) {ICE-COLA: towards fast and accurate
  synthetic galaxy catalogues optimizing a quasi-N-body method}. \mnras
  459(3):2327--2341. \doi{10.1093/mnras/stw797}.
  {\href{https://arxiv.org/abs/1509.04685}{{arXiv:1509.04685}}} {[astro-ph.CO]}

\bibitem[{{Izquierdo-Villalba} et~al.(2019){Izquierdo-Villalba}, {Angulo},
  {Orsi}, {Hurier}, {Vilella-Rojo}, {Bonoli}, {L{\'o}pez-Sanjuan}, {Alcaniz},
  {Cenarro}, {Crist{\'o}bal-Hornillos}, {Dupke}, {Ederoclite},
  {Hern{\'a}ndez-Monteagudo}, {Mar{\'\i}n-Franch}, {Moles}, {Mendes de
  Oliveira}, {Sodr{\'e}}, {Varela}, and {V{\'a}zquez
  Rami{\'o}}}]{Izquierdo-Villalba:2019}
{Izquierdo-Villalba} D, {Angulo} RE, {Orsi} A, {Hurier} G, {Vilella-Rojo} G,
  {Bonoli} S, {L{\'o}pez-Sanjuan} C, {Alcaniz} J, {Cenarro} J,
  {Crist{\'o}bal-Hornillos} D, {Dupke} R, {Ederoclite} A,
  {Hern{\'a}ndez-Monteagudo} C, {Mar{\'\i}n-Franch} A, {Moles} M, {Mendes de
  Oliveira} C, {Sodr{\'e}} L, {Varela} J, {V{\'a}zquez Rami{\'o}} H (2019)
  {J-PLUS: Synthetic galaxy catalogues with emission lines for photometric
  surveys}. \aap 631:A82. \doi{10.1051/0004-6361/201936232}.
  {\href{https://arxiv.org/abs/1907.02111}{{arXiv:1907.02111}}} {[astro-ph.GA]}

\bibitem[{{Jamieson} and {Loverde}(2019{\natexlab{a}})}]{Jamieson:2019a}
{Jamieson} D, {Loverde} M (2019{\natexlab{a}}) {Quintessential isocurvature in
  separate universe simulations}. \prd 100(2):023516.
  \doi{10.1103/PhysRevD.100.023516}.
  {\href{https://arxiv.org/abs/1812.08765}{{arXiv:1812.08765}}} {[astro-ph.CO]}

\bibitem[{{Jamieson} and {Loverde}(2019{\natexlab{b}})}]{Jamieson:2019b}
{Jamieson} D, {Loverde} M (2019{\natexlab{b}}) {Separate universe void bias}.
  \prd 100(12):123528. \doi{10.1103/PhysRevD.100.123528}.
  {\href{https://arxiv.org/abs/1909.05313}{{arXiv:1909.05313}}} {[astro-ph.CO]}

\bibitem[{{Jasche} and {Lavaux}(2019)}]{Jasche:2019}
{Jasche} J, {Lavaux} G (2019) {Physical Bayesian modelling of the non-linear
  matter distribution: New insights into the nearby universe}. \aap 625:A64.
  \doi{10.1051/0004-6361/201833710}.
  {\href{https://arxiv.org/abs/1806.11117}{{arXiv:1806.11117}}} {[astro-ph.CO]}

\bibitem[{{Jasche} et~al.(2010){Jasche}, {Kitaura}, {Li}, and
  {En{\ss}lin}}]{Jasche:2010}
{Jasche} J, {Kitaura} FS, {Li} C, {En{\ss}lin} TA (2010) {Bayesian non-linear
  large-scale structure inference of the Sloan Digital Sky Survey Data Release
  7}. \mnras 409(1):355--370. \doi{10.1111/j.1365-2966.2010.17313.x}.
  {\href{https://arxiv.org/abs/0911.2498}{{arXiv:0911.2498}}} {[astro-ph.CO]}

\bibitem[{{Jelic-Cizmek} et~al.(2018){Jelic-Cizmek}, {Lepori}, {Adamek}, and
  {Durrer}}]{Jelic-Cizmek:2018}
{Jelic-Cizmek} G, {Lepori} F, {Adamek} J, {Durrer} R (2018) {The generation of
  vorticity in cosmological N-body simulations}. \jcap 2018(9):006.
  \doi{10.1088/1475-7516/2018/09/006}.
  {\href{https://arxiv.org/abs/1806.05146}{{arXiv:1806.05146}}} {[astro-ph.CO]}

\bibitem[{{Jenkins}(2010)}]{Jenkins:2010}
{Jenkins} A (2010) {Second-order Lagrangian perturbation theory initial
  conditions for resimulations}. \mnras 403(4):1859--1872.
  \doi{10.1111/j.1365-2966.2010.16259.x}.
  {\href{https://arxiv.org/abs/0910.0258}{{arXiv:0910.0258}}} {[astro-ph.CO]}

\bibitem[{{Jenkins}(2013)}]{Jenkins:2013}
{Jenkins} A (2013) {A new way of setting the phases for cosmological multiscale
  Gaussian initial conditions}. \mnras 434:2094--2120.
  \doi{10.1093/mnras/stt1154}.
  {\href{https://arxiv.org/abs/1306.5968}{{arXiv:1306.5968}}}

\bibitem[{{Jenkins} et~al.(2001){Jenkins}, {Frenk}, {White}, {Colberg}, {Cole},
  {Evrard}, {Couchman}, and {Yoshida}}]{Jenkins:2001}
{Jenkins} A, {Frenk} CS, {White} SDM, {Colberg} JM, {Cole} S, {Evrard} AE,
  {Couchman} HMP, {Yoshida} N (2001) {The mass function of dark matter haloes}.
  \mnras 321:372--384. \doi{10.1046/j.1365-8711.2001.04029.x}.
  {\href{https://arxiv.org/abs/astro-ph/0005260}{{astro-ph/0005260}}}

\bibitem[{{Jennings}(2012)}]{Jennings:2012}
{Jennings} E (2012) {An improved model for the non-linear velocity power
  spectrum}. \mnras 427(1):L25--L29. \doi{10.1111/j.1745-3933.2012.01338.x}.
  {\href{https://arxiv.org/abs/1207.1439}{{arXiv:1207.1439}}} {[astro-ph.CO]}

\bibitem[{{Jennings} et~al.(2011){Jennings}, {Baugh}, and
  {Pascoli}}]{Jennings:2011}
{Jennings} E, {Baugh} CM, {Pascoli} S (2011) {Modelling redshift space
  distortions in hierarchical cosmologies}. \mnras 410(3):2081--2094.
  \doi{10.1111/j.1365-2966.2010.17581.x}.
  {\href{https://arxiv.org/abs/1003.4282}{{arXiv:1003.4282}}} {[astro-ph.CO]}

\bibitem[{{Jennings} et~al.(2015){Jennings}, {Baugh}, and
  {Hatt}}]{Jennings:2015}
{Jennings} E, {Baugh} CM, {Hatt} D (2015) {Velocity and mass bias in the
  distribution of dark matter haloes}. \mnras 446(1):793--802.
  \doi{10.1093/mnras/stu2043}.
  {\href{https://arxiv.org/abs/1407.7296}{{arXiv:1407.7296}}} {[astro-ph.CO]}

\bibitem[{Jennings et~al.(2019)Jennings, Watkinson, Abdalla, and
  McEwen}]{Jennings:2018}
Jennings WD, Watkinson CA, Abdalla FB, McEwen JD (2019) {Evaluating machine
  learning techniques for predicting power spectra from reionization
  simulations}. Mon Not Roy Astron Soc 483(3):2907--2922.
  \doi{10.1093/mnras/sty3168}.
  {\href{https://arxiv.org/abs/1811.09141}{{arXiv:1811.09141}}} {[astro-ph.CO]}

\bibitem[{{Jessop} et~al.(1994){Jessop}, {Duncan}, and {Chau}}]{Jessop:1994}
{Jessop} C, {Duncan} M, {Chau} WY (1994) {Multigrid Methods for N-Body
  Gravitational Systems}. J Comput Phys 115(2):339--351.
  \doi{10.1006/jcph.1994.1200}

\bibitem[{{Jimenez} et~al.(2010){Jimenez}, {Kitching}, {Pe{\~n}a-Garay}, and
  {Verde}}]{Jimenez:2010}
{Jimenez} R, {Kitching} T, {Pe{\~n}a-Garay} C, {Verde} L (2010) {Can we measure
  the neutrino mass hierarchy in the sky?} \jcap 2010(5):035.
  \doi{10.1088/1475-7516/2010/05/035}.
  {\href{https://arxiv.org/abs/1003.5918}{{arXiv:1003.5918}}} {[astro-ph.CO]}

\bibitem[{{Jing}(2005)}]{Jing:2005}
{Jing} YP (2005) {Correcting for the Alias Effect When Measuring the Power
  Spectrum Using a Fast Fourier Transform}. \apj 620(2):559--563.
  \doi{10.1086/427087}.
  {\href{https://arxiv.org/abs/astro-ph/0409240}{{arXiv:astro-ph/0409240}}}
  {[astro-ph]}

\bibitem[{{Johansen} and {Colella}(1998)}]{Johansen:1998}
{Johansen} H, {Colella} P (1998) {A Cartesian Grid Embedded Boundary Method for
  Poisson's Equation on Irregular Domains}. J Comput Phys 147(1):60--85.
  \doi{10.1006/jcph.1998.5965}

\bibitem[{{Joyce} and {Marcos}(2007{\natexlab{a}})}]{Joyce:2007b}
{Joyce} M, {Marcos} B (2007{\natexlab{a}}) {Quantification of discreteness
  effects in cosmological N-body simulations. II. Evolution up to shell
  crossing}. \prd 76(10):103505. \doi{10.1103/PhysRevD.76.103505}.
  {\href{https://arxiv.org/abs/0704.3697}{{arXiv:0704.3697}}}

\bibitem[{{Joyce} and {Marcos}(2007{\natexlab{b}})}]{Joyce:2007a}
{Joyce} M, {Marcos} B (2007{\natexlab{b}}) {Quantification of discreteness
  effects in cosmological N-body simulations: Initial conditions}. \prd
  75(6):063516. \doi{10.1103/PhysRevD.75.063516}.
  {\href{https://arxiv.org/abs/astro-ph/0410451}{{astro-ph/0410451}}}

\bibitem[{{Joyce} et~al.(2005){Joyce}, {Marcos}, {Gabrielli}, {Baertschiger},
  and {Sylos Labini}}]{Joyce:2005}
{Joyce} M, {Marcos} B, {Gabrielli} A, {Baertschiger} T, {Sylos Labini} F (2005)
  {Gravitational Evolution of a Perturbed Lattice and its Fluid Limit}.
  Physical Review Letters 95(1):011304. \doi{10.1103/PhysRevLett.95.011304}.
  {\href{https://arxiv.org/abs/astro-ph/0504213}{{astro-ph/0504213}}}

\bibitem[{{Joyce} et~al.(2020){Joyce}, {Garrison}, and
  {Eisenstein}}]{Joyce:2020}
{Joyce} M, {Garrison} L, {Eisenstein} D (2020) {Quantifying resolution in
  cosmological N-body simulations using self-similarity}. \mnras
  \doi{10.1093/mnras/staa3434}.
  {\href{https://arxiv.org/abs/2004.07256}{{arXiv:2004.07256}}} {[astro-ph.CO]}

\bibitem[{{Juan} et~al.(2014){Juan}, {Salvador-Sol{\'e}}, {Dom{\`e}nech}, and
  {Manrique}}]{Juan:2014}
{Juan} E, {Salvador-Sol{\'e}} E, {Dom{\`e}nech} G, {Manrique} A (2014) {Fixing
  a rigorous formalism for the accurate analytic derivation of halo
  properties}. \mnras 439(1):719--724. \doi{10.1093/mnras/stt2493}.
  {\href{https://arxiv.org/abs/1401.7335}{{arXiv:1401.7335}}} {[astro-ph.CO]}

\bibitem[{{Kaehler}(2017)}]{Kaehler:2018}
{Kaehler} R (2017) {Massively parallel computation of accurate densities for
  N-body dark matter simulations using the phase-space-element method}.
  Astronomy and Computing 20:68--76. \doi{10.1016/j.ascom.2017.05.005}.
  {\href{https://arxiv.org/abs/1612.09491}{{arXiv:1612.09491}}}
  {[physics.comp-ph]}

\bibitem[{K{\"{a}}hler et~al.(2012)K{\"{a}}hler, Hahn, and Abel}]{Kaehler:2012}
K{\"{a}}hler R, Hahn O, Abel T (2012) A novel approach to visualizing dark
  matter simulations. {IEEE} Trans Vis Comput Graph 18(12):2078--2087.
  \doi{10.1109/TVCG.2012.187}

\bibitem[{{Kahlhoefer} et~al.(2015){Kahlhoefer}, {Schmidt-Hoberg}, {Kummer},
  and {Sarkar}}]{Kahlhoefer:2015}
{Kahlhoefer} F, {Schmidt-Hoberg} K, {Kummer} J, {Sarkar} S (2015) {On the
  interpretation of dark matter self-interactions in Abell 3827}. \mnras
  452(1):L54--L58. \doi{10.1093/mnrasl/slv088}.
  {\href{https://arxiv.org/abs/1504.06576}{{arXiv:1504.06576}}} {[astro-ph.CO]}

\bibitem[{{Kampakoglou} et~al.(2008){Kampakoglou}, {Trotta}, and
  {Silk}}]{Kampakoglou:2008}
{Kampakoglou} M, {Trotta} R, {Silk} J (2008) {Monolithic or hierarchical star
  formation? A new statistical analysis}. \mnras 384(4):1414--1426.
  \doi{10.1111/j.1365-2966.2007.12747.x}.
  {\href{https://arxiv.org/abs/0709.1104}{{arXiv:0709.1104}}} {[astro-ph]}

\bibitem[{{Kates-Harbeck} et~al.(2016){Kates-Harbeck}, {Totorica}, {Zrake}, and
  {Abel}}]{KatesHarbeck:2016}
{Kates-Harbeck} J, {Totorica} S, {Zrake} J, {Abel} T (2016) {Simplex-in-cell
  technique for collisionless plasma simulations}. J Comput Phys 304:231--251.
  \doi{10.1016/j.jcp.2015.10.017}.
  {\href{https://arxiv.org/abs/1506.07207}{{arXiv:1506.07207}}}
  {[physics.comp-ph]}

\bibitem[{{Kato} and {Soda}(2020)}]{Kato:2020}
{Kato} R, {Soda} J (2020) {Search for ultralight scalar dark matter with
  NANOGrav pulsar timing arrays}. \jcap 2020(9):036.
  \doi{10.1088/1475-7516/2020/09/036}.
  {\href{https://arxiv.org/abs/1904.09143}{{arXiv:1904.09143}}} {[astro-ph.HE]}

\bibitem[{{Katz} et~al.(1994){Katz}, {Quinn}, {Bertschinger}, and
  {Gelb}}]{Katz:1994}
{Katz} N, {Quinn} T, {Bertschinger} E, {Gelb} JM (1994) {Formation of Quasars
  at High Redshift}. \mnras 270:L71. \doi{10.1093/mnras/270.1.L71}

\bibitem[{{Kauffmann} et~al.(1993){Kauffmann}, {White}, and
  {Guiderdoni}}]{Kauffmann:1993}
{Kauffmann} G, {White} SDM, {Guiderdoni} B (1993) {The formation and evolution
  of galaxies within merging dark matter haloes.} \mnras 264:201--218.
  \doi{10.1093/mnras/264.1.201}

\bibitem[{{Kaushal} et~al.(2021){Kaushal}, {Villaescusa-Navarro}, {Giusarma},
  {Li}, {Hawry}, and {Reyes}}]{Kaushal:2021}
{Kaushal} N, {Villaescusa-Navarro} F, {Giusarma} E, {Li} Y, {Hawry} C, {Reyes}
  M (2021) {NECOLA: Towards a Universal Field-level Cosmological Emulator}.
  arXiv e-prints arXiv:2111.02441.
  {\href{https://arxiv.org/abs/2111.02441}{{arXiv:2111.02441}}} {[astro-ph.CO]}

\bibitem[{{Keller} et~al.(2019){Keller}, {Wadsley}, {Wang}, and
  {Kruijssen}}]{Keller:2019}
{Keller} BW, {Wadsley} JW, {Wang} L, {Kruijssen} JMD (2019) {Chaos and variance
  in galaxy formation}. \mnras 482(2):2244--2261. \doi{10.1093/mnras/sty2859}.
  {\href{https://arxiv.org/abs/1803.05445}{{arXiv:1803.05445}}} {[astro-ph.GA]}

\bibitem[{{Khandai} et~al.(2015){Khandai}, {Di Matteo}, {Croft}, {Wilkins},
  {Feng}, {Tucker}, {DeGraf}, and {Liu}}]{Khandai:2015}
{Khandai} N, {Di Matteo} T, {Croft} R, {Wilkins} S, {Feng} Y, {Tucker} E,
  {DeGraf} C, {Liu} MS (2015) {The MassiveBlack-II simulation: the evolution of
  haloes and galaxies to z {\ensuremath{\sim}} 0}. \mnras 450(2):1349--1374.
  \doi{10.1093/mnras/stv627}.
  {\href{https://arxiv.org/abs/1402.0888}{{arXiv:1402.0888}}} {[astro-ph.CO]}

\bibitem[{{Khmelnitsky} and {Rubakov}(2014)}]{Khmelnitsky:2014}
{Khmelnitsky} A, {Rubakov} V (2014) {Pulsar timing signal from ultralight
  scalar dark matter}. \jcap 2014(2):019. \doi{10.1088/1475-7516/2014/02/019}.
  {\href{https://arxiv.org/abs/1309.5888}{{arXiv:1309.5888}}} {[astro-ph.CO]}

\bibitem[{{Khoraminezhad} et~al.(2021){Khoraminezhad}, {Lazeyras}, {Angulo},
  {Hahn}, and {Viel}}]{Khoraminezhad:2020}
{Khoraminezhad} H, {Lazeyras} T, {Angulo} RE, {Hahn} O, {Viel} M (2021)
  {Quantifying the impact of baryon-CDM perturbations on halo clustering and
  baryon fraction}. \jcap 2021(3):023. \doi{10.1088/1475-7516/2021/03/023}.
  {\href{https://arxiv.org/abs/2011.01037}{{arXiv:2011.01037}}} {[astro-ph.CO]}

\bibitem[{{Khoury} and {Weltman}(2004{\natexlab{a}})}]{Khoury:2004b}
{Khoury} J, {Weltman} A (2004{\natexlab{a}}) {Chameleon cosmology}. \prd
  69(4):044026. \doi{10.1103/PhysRevD.69.044026}.
  {\href{https://arxiv.org/abs/astro-ph/0309411}{{arXiv:astro-ph/0309411}}}
  {[astro-ph]}

\bibitem[{{Khoury} and {Weltman}(2004{\natexlab{b}})}]{Khoury:2004a}
{Khoury} J, {Weltman} A (2004{\natexlab{b}}) {Chameleon Fields: Awaiting
  Surprises for Tests of Gravity in Space}. \prl 93(17):171104.
  \doi{10.1103/PhysRevLett.93.171104}.
  {\href{https://arxiv.org/abs/astro-ph/0309300}{{arXiv:astro-ph/0309300}}}
  {[astro-ph]}

\bibitem[{{Kim} and {Park}(2006)}]{KimPark:2006}
{Kim} J, {Park} C (2006) {A New Halo-finding Method for N-Body Simulations}.
  \apj 639(2):600--616. \doi{10.1086/499761}.
  {\href{https://arxiv.org/abs/astro-ph/0401386}{{arXiv:astro-ph/0401386}}}
  {[astro-ph]}

\bibitem[{{Kim} et~al.(2014){Kim}, {Abel}, {Agertz}, {Bryan}, {Ceverino},
  {Christensen}, {Conroy}, {Dekel}, {Gnedin}, {Goldbaum}, {Guedes}, {Hahn},
  {Hobbs}, {Hopkins}, {Hummels}, {Iannuzzi}, {Keres}, {Klypin}, {Kravtsov},
  {Krumholz}, {Kuhlen}, {Leitner}, {Madau}, {Mayer}, {Moody}, {Nagamine},
  {Norman}, {Onorbe}, {O'Shea}, {Pillepich}, {Primack}, {Quinn}, {Read},
  {Robertson}, {Rocha}, {Rudd}, {Shen}, {Smith}, {Szalay}, {Teyssier},
  {Thompson}, {Todoroki}, {Turk}, {Wadsley}, {Wise}, {Zolotov}, and {AGORA
  Collaboration29}}]{Kim:2014}
{Kim} Jh, {Abel} T, {Agertz} O, {Bryan} GL, {Ceverino} D, {Christensen} C,
  {Conroy} C, {Dekel} A, {Gnedin} NY, {Goldbaum} NJ, {Guedes} J, {Hahn} O,
  {Hobbs} A, {Hopkins} PF, {Hummels} CB, {Iannuzzi} F, {Keres} D, {Klypin} A,
  {Kravtsov} AV, {Krumholz} MR, {Kuhlen} M, {Leitner} SN, {Madau} P, {Mayer} L,
  {Moody} CE, {Nagamine} K, {Norman} ML, {Onorbe} J, {O'Shea} BW, {Pillepich}
  A, {Primack} JR, {Quinn} T, {Read} JI, {Robertson} BE, {Rocha} M, {Rudd} DH,
  {Shen} S, {Smith} BD, {Szalay} AS, {Teyssier} R, {Thompson} R, {Todoroki} K,
  {Turk} MJ, {Wadsley} JW, {Wise} JH, {Zolotov} A, {AGORA Collaboration29} t
  (2014) {The AGORA High-resolution Galaxy Simulations Comparison Project}.
  \apjs 210(1):14. \doi{10.1088/0067-0049/210/1/14}.
  {\href{https://arxiv.org/abs/1308.2669}{{arXiv:1308.2669}}} {[astro-ph.GA]}

\bibitem[{{Kitaura} and {En{\ss}lin}(2008)}]{Kitaura:2008}
{Kitaura} FS, {En{\ss}lin} TA (2008) {Bayesian reconstruction of the
  cosmological large-scale structure: methodology, inverse algorithms and
  numerical optimization}. \mnras 389(2):497--544.
  \doi{10.1111/j.1365-2966.2008.13341.x}.
  {\href{https://arxiv.org/abs/0705.0429}{{arXiv:0705.0429}}} {[astro-ph]}

\bibitem[{{Kitaura} and {Hess}(2013)}]{KitauraHess:2013}
{Kitaura} FS, {Hess} S (2013) {Cosmological structure formation with augmented
  lagrangian perturbation theory.} \mnras 435:L78--L82.
  \doi{10.1093/mnrasl/slt101}.
  {\href{https://arxiv.org/abs/1212.3514}{{arXiv:1212.3514}}} {[astro-ph.CO]}

\bibitem[{{Kitaura} et~al.(2014){Kitaura}, {Yepes}, and {Prada}}]{Kitaura:2014}
{Kitaura} FS, {Yepes} G, {Prada} F (2014) {Modelling baryon acoustic
  oscillations with perturbation theory and stochastic halo biasing.} \mnras
  439:L21--L25. \doi{10.1093/mnrasl/slt172}.
  {\href{https://arxiv.org/abs/1307.3285}{{arXiv:1307.3285}}} {[astro-ph.CO]}

\bibitem[{{Kitzbichler} and {White}(2007)}]{Kitzbichler:2007}
{Kitzbichler} MG, {White} SDM (2007) {The high-redshift galaxy population in
  hierarchical galaxy formation models}. \mnras 376(1):2--12.
  \doi{10.1111/j.1365-2966.2007.11458.x}.
  {\href{https://arxiv.org/abs/astro-ph/0609636}{{arXiv:astro-ph/0609636}}}
  {[astro-ph]}

\bibitem[{{Klypin} and {Prada}(2018{\natexlab{a}})}]{Klypin:2018}
{Klypin} A, {Prada} F (2018{\natexlab{a}}) {Dark matter statistics for large
  galaxy catalogues: power spectra and covariance matrices}. \mnras
  478(4):4602--4621. \doi{10.1093/mnras/sty1340}.
  {\href{https://arxiv.org/abs/1701.05690}{{arXiv:1701.05690}}} {[astro-ph.CO]}

\bibitem[{{Klypin} and {Prada}(2018{\natexlab{b}})}]{KlypinPrada:2018}
{Klypin} A, {Prada} F (2018{\natexlab{b}}) {Dark matter statistics for large
  galaxy catalogues: power spectra and covariance matrices}. \mnras
  478(4):4602--4621. \doi{10.1093/mnras/sty1340}.
  {\href{https://arxiv.org/abs/1701.05690}{{arXiv:1701.05690}}} {[astro-ph.CO]}

\bibitem[{{Klypin} and {Prada}(2019)}]{Klypin:2019}
{Klypin} A, {Prada} F (2019) {Effects of long-wavelength fluctuations in large
  galaxy surveys}. \mnras 489(2):1684--1696. \doi{10.1093/mnras/stz2194}.
  {\href{https://arxiv.org/abs/1809.03637}{{arXiv:1809.03637}}} {[astro-ph.CO]}

\bibitem[{{Klypin} et~al.(1993){Klypin}, {Holtzman}, {Primack}, and
  {Regos}}]{Klypin:1993}
{Klypin} A, {Holtzman} J, {Primack} J, {Regos} E (1993) {Structure Formation
  with Cold plus Hot Dark Matter}. \apj 416:1. \doi{10.1086/173210}.
  {\href{https://arxiv.org/abs/astro-ph/9305011}{{arXiv:astro-ph/9305011}}}
  {[astro-ph]}

\bibitem[{{Klypin} et~al.(1999{\natexlab{a}}){Klypin}, {Gottl{\"o}ber},
  {Kravtsov}, and {Khokhlov}}]{Klypin:1999b}
{Klypin} A, {Gottl{\"o}ber} S, {Kravtsov} AV, {Khokhlov} AM
  (1999{\natexlab{a}}) {Galaxies in N-Body Simulations: Overcoming the
  Overmerging Problem}. \apj 516(2):530--551. \doi{10.1086/307122}.
  {\href{https://arxiv.org/abs/astro-ph/9708191}{{arXiv:astro-ph/9708191}}}
  {[astro-ph]}

\bibitem[{{Klypin} et~al.(1999{\natexlab{b}}){Klypin}, {Kravtsov},
  {Valenzuela}, and {Prada}}]{Klypin:1999}
{Klypin} A, {Kravtsov} AV, {Valenzuela} O, {Prada} F (1999{\natexlab{b}})
  {Where Are the Missing Galactic Satellites?} \apj 522(1):82--92.
  \doi{10.1086/307643}.
  {\href{https://arxiv.org/abs/astro-ph/9901240}{{arXiv:astro-ph/9901240}}}
  {[astro-ph]}

\bibitem[{{Klypin} et~al.(2003){Klypin}, {Hoffman}, {Kravtsov}, and
  {Gottl{\"o}ber}}]{Klypin:2003}
{Klypin} A, {Hoffman} Y, {Kravtsov} AV, {Gottl{\"o}ber} S (2003) {Constrained
  Simulations of the Real Universe: The Local Supercluster}. \apj
  596(1):19--33. \doi{10.1086/377574}.
  {\href{https://arxiv.org/abs/astro-ph/0107104}{{arXiv:astro-ph/0107104}}}
  {[astro-ph]}

\bibitem[{{Klypin} et~al.(2015){Klypin}, {Prada}, {Yepes}, {He{\ss}}, and
  {Gottl{\"o}ber}}]{Klypin:2015}
{Klypin} A, {Prada} F, {Yepes} G, {He{\ss}} S, {Gottl{\"o}ber} S (2015) {Halo
  abundance matching: accuracy and conditions for numerical convergence}.
  \mnras 447(4):3693--3707. \doi{10.1093/mnras/stu2685}

\bibitem[{{Klypin} et~al.(2016){Klypin}, {Yepes}, {Gottl{\"o}ber}, {Prada}, and
  {He{\ss}}}]{Klypin:2016}
{Klypin} A, {Yepes} G, {Gottl{\"o}ber} S, {Prada} F, {He{\ss}} S (2016)
  {MultiDark simulations: the story of dark matter halo concentrations and
  density profiles}. \mnras 457(4):4340--4359. \doi{10.1093/mnras/stw248}.
  {\href{https://arxiv.org/abs/1411.4001}{{arXiv:1411.4001}}} {[astro-ph.CO]}

\bibitem[{{Klypin} et~al.(2020){Klypin}, {Prada}, and {Byun}}]{Klypin:2020}
{Klypin} A, {Prada} F, {Byun} J (2020) {Suppressing cosmic variance with
  paired-and-fixed cosmological simulations: average properties and covariances
  of dark matter clustering statistics}. \mnras 496(3):3862--3869.
  \doi{10.1093/mnras/staa734}.
  {\href{https://arxiv.org/abs/1903.08518}{{arXiv:1903.08518}}} {[astro-ph.CO]}

\bibitem[{{Klypin} and {Shandarin}(1983)}]{Klypin:1983}
{Klypin} AA, {Shandarin} SF (1983) {Three-dimensional numerical model of the
  formation of large-scale structure in the Universe}. \mnras 204:891--907.
  \doi{10.1093/mnras/204.3.891}

\bibitem[{{Klypin} et~al.(2011){Klypin}, {Trujillo-Gomez}, and
  {Primack}}]{Klypin:2011}
{Klypin} AA, {Trujillo-Gomez} S, {Primack} J (2011) {Dark Matter Halos in the
  Standard Cosmological Model: Results from the Bolshoi Simulation}. \apj
  740(2):102. \doi{10.1088/0004-637X/740/2/102}.
  {\href{https://arxiv.org/abs/1002.3660}{{arXiv:1002.3660}}} {[astro-ph.CO]}

\bibitem[{{Knebe} et~al.(2000){Knebe}, {Kravtsov}, {Gottl{\"o}ber}, and
  {Klypin}}]{Knebe:2000}
{Knebe} A, {Kravtsov} AV, {Gottl{\"o}ber} S, {Klypin} AA (2000) {On the effects
  of resolution in dissipationless cosmological simulations}. \mnras
  317(3):630--648. \doi{10.1046/j.1365-8711.2000.03673.x}.
  {\href{https://arxiv.org/abs/astro-ph/9912257}{{arXiv:astro-ph/9912257}}}
  {[astro-ph]}

\bibitem[{{Knebe} et~al.(2011){Knebe}, {Knollmann}, {Muldrew}, {Pearce},
  {Aragon-Calvo}, {Ascasibar}, {Behroozi}, {Ceverino}, {Colombi}, {Diemand},
  {Dolag}, {Falck}, {Fasel}, {Gardner}, {Gottl{\"o}ber}, {Hsu}, {Iannuzzi},
  {Klypin}, {Luki{\'c}}, {Maciejewski}, {McBride}, {Neyrinck}, {Planelles},
  {Potter}, {Quilis}, {Rasera}, {Read}, {Ricker}, {Roy}, {Springel}, {Stadel},
  {Stinson}, {Sutter}, {Turchaninov}, {Tweed}, {Yepes}, and
  {Zemp}}]{Knebe:2011}
{Knebe} A, {Knollmann} SR, {Muldrew} SI, {Pearce} FR, {Aragon-Calvo} MA,
  {Ascasibar} Y, {Behroozi} PS, {Ceverino} D, {Colombi} S, {Diemand} J, {Dolag}
  K, {Falck} BL, {Fasel} P, {Gardner} J, {Gottl{\"o}ber} S, {Hsu} CH,
  {Iannuzzi} F, {Klypin} A, {Luki{\'c}} Z, {Maciejewski} M, {McBride} C,
  {Neyrinck} MC, {Planelles} S, {Potter} D, {Quilis} V, {Rasera} Y, {Read} JI,
  {Ricker} PM, {Roy} F, {Springel} V, {Stadel} J, {Stinson} G, {Sutter} PM,
  {Turchaninov} V, {Tweed} D, {Yepes} G, {Zemp} M (2011) {Haloes gone MAD: The
  Halo-Finder Comparison Project}. \mnras 415(3):2293--2318.
  \doi{10.1111/j.1365-2966.2011.18858.x}.
  {\href{https://arxiv.org/abs/1104.0949}{{arXiv:1104.0949}}} {[astro-ph.CO]}

\bibitem[{{Knebe} et~al.(2013){Knebe}, {Pearce}, {Lux}, {Ascasibar},
  {Behroozi}, {Casado}, {Moran}, {Diemand}, {Dolag}, {Dominguez-Tenreiro},
  {Elahi}, {Falck}, {Gottl{\"o}ber}, {Han}, {Klypin}, {Luki{\'c}},
  {Maciejewski}, {McBride}, {Merch{\'a}n}, {Muldrew}, {Neyrinck}, {Onions},
  {Planelles}, {Potter}, {Quilis}, {Rasera}, {Ricker}, {Roy}, {Ruiz},
  {Sgr{\'o}}, {Springel}, {Stadel}, {Sutter}, {Tweed}, and {Zemp}}]{Knebe:2013}
{Knebe} A, {Pearce} FR, {Lux} H, {Ascasibar} Y, {Behroozi} P, {Casado} J,
  {Moran} CC, {Diemand} J, {Dolag} K, {Dominguez-Tenreiro} R, {Elahi} P,
  {Falck} B, {Gottl{\"o}ber} S, {Han} J, {Klypin} A, {Luki{\'c}} Z,
  {Maciejewski} M, {McBride} CK, {Merch{\'a}n} ME, {Muldrew} SI, {Neyrinck} M,
  {Onions} J, {Planelles} S, {Potter} D, {Quilis} V, {Rasera} Y, {Ricker} PM,
  {Roy} F, {Ruiz} AN, {Sgr{\'o}} MA, {Springel} V, {Stadel} J, {Sutter} PM,
  {Tweed} D, {Zemp} M (2013) {Structure finding in cosmological simulations:
  the state of affairs}. \mnras 435(2):1618--1658. \doi{10.1093/mnras/stt1403}.
  {\href{https://arxiv.org/abs/1304.0585}{{arXiv:1304.0585}}} {[astro-ph.CO]}

\bibitem[{{Knebe} et~al.(2015){Knebe}, {Pearce}, {Thomas}, {Benson}, {Blaizot},
  {Bower}, {Carretero}, {Castander}, {Cattaneo}, {Cora}, {Croton}, {Cui},
  {Cunnama}, {De Lucia}, {Devriendt}, {Elahi}, {Font}, {Fontanot},
  {Garcia-Bellido}, {Gargiulo}, {Gonzalez-Perez}, {Helly}, {Henriques},
  {Hirschmann}, {Lee}, {Mamon}, {Monaco}, {Onions}, {Padilla}, {Power},
  {Pujol}, {Skibba}, {Somerville}, {Srisawat}, {Vega-Mart{\'\i}nez}, and
  {Yi}}]{Knebe:2015}
{Knebe} A, {Pearce} FR, {Thomas} PA, {Benson} A, {Blaizot} J, {Bower} R,
  {Carretero} J, {Castander} FJ, {Cattaneo} A, {Cora} SA, {Croton} DJ, {Cui} W,
  {Cunnama} D, {De Lucia} G, {Devriendt} JE, {Elahi} PJ, {Font} A, {Fontanot}
  F, {Garcia-Bellido} J, {Gargiulo} ID, {Gonzalez-Perez} V, {Helly} J,
  {Henriques} B, {Hirschmann} M, {Lee} J, {Mamon} GA, {Monaco} P, {Onions} J,
  {Padilla} ND, {Power} C, {Pujol} A, {Skibba} RA, {Somerville} RS, {Srisawat}
  C, {Vega-Mart{\'\i}nez} CA, {Yi} SK (2015) {nIFTy cosmology: comparison of
  galaxy formation models}. \mnras 451(4):4029--4059.
  \doi{10.1093/mnras/stv1149}.
  {\href{https://arxiv.org/abs/1505.04607}{{arXiv:1505.04607}}} {[astro-ph.GA]}

\bibitem[{{Knebe} et~al.(2018){Knebe}, {Pearce}, {Gonzalez-Perez}, {Thomas},
  {Benson}, {Asquith}, {Blaizot}, {Bower}, {Carretero}, {Castander},
  {Cattaneo}, {Cora}, {Croton}, {Cui}, {Cunnama}, {Devriendt}, {Elahi}, {Font},
  {Fontanot}, {Gargiulo}, {Helly}, {Henriques}, {Lee}, {Mamon}, {Onions},
  {Padilla}, {Power}, {Pujol}, {Ruiz}, {Srisawat}, {Stevens}, {Tollet},
  {Vega-Mart{\'\i}nez}, and {Yi}}]{Knebe:2018}
{Knebe} A, {Pearce} FR, {Gonzalez-Perez} V, {Thomas} PA, {Benson} A, {Asquith}
  R, {Blaizot} J, {Bower} R, {Carretero} J, {Castander} FJ, {Cattaneo} A,
  {Cora} SA, {Croton} DJ, {Cui} W, {Cunnama} D, {Devriendt} JE, {Elahi} PJ,
  {Font} A, {Fontanot} F, {Gargiulo} ID, {Helly} J, {Henriques} B, {Lee} J,
  {Mamon} GA, {Onions} J, {Padilla} ND, {Power} C, {Pujol} A, {Ruiz} AN,
  {Srisawat} C, {Stevens} ARH, {Tollet} E, {Vega-Mart{\'\i}nez} CA, {Yi} SK
  (2018) {Cosmic CARNage I: on the calibration of galaxy formation models}.
  \mnras 475(3):2936--2954. \doi{10.1093/mnras/stx3274}.
  {\href{https://arxiv.org/abs/1712.06420}{{arXiv:1712.06420}}} {[astro-ph.GA]}

\bibitem[{{Knirck} et~al.(2018){Knirck}, {Millar}, {O'Hare}, {Redondo}, and
  {Steffen}}]{Knirck:2018}
{Knirck} S, {Millar} AJ, {O'Hare} CAJ, {Redondo} J, {Steffen} FD (2018)
  {Directional axion detection}. \jcap 2018(11):051.
  \doi{10.1088/1475-7516/2018/11/051}.
  {\href{https://arxiv.org/abs/1806.05927}{{arXiv:1806.05927}}} {[astro-ph.CO]}

\bibitem[{{Knollmann} and {Knebe}(2009)}]{KnollmannKnebe:2009}
{Knollmann} SR, {Knebe} A (2009) {AHF: Amiga's Halo Finder}. \apjs
  182(2):608--624. \doi{10.1088/0067-0049/182/2/608}.
  {\href{https://arxiv.org/abs/0904.3662}{{arXiv:0904.3662}}} {[astro-ph.CO]}

\bibitem[{{Kobayashi} et~al.(2017){Kobayashi}, {Murgia}, {De Simone},
  {Ir{\v{s}}i{\v{c}}}, and {Viel}}]{Kobayashi:2017}
{Kobayashi} T, {Murgia} R, {De Simone} A, {Ir{\v{s}}i{\v{c}}} V, {Viel} M
  (2017) {Lyman-{\ensuremath{\alpha}} constraints on ultralight scalar dark
  matter: Implications for the early and late universe}. \prd 96(12):123514.
  \doi{10.1103/PhysRevD.96.123514}.
  {\href{https://arxiv.org/abs/1708.00015}{{arXiv:1708.00015}}} {[astro-ph.CO]}

\bibitem[{{Kobayashi} et~al.(2020){Kobayashi}, {Nishimichi}, {Takada},
  {Takahashi}, and {Osato}}]{Kobayashi:2020}
{Kobayashi} Y, {Nishimichi} T, {Takada} M, {Takahashi} R, {Osato} K (2020)
  {Accurate emulator for the redshift-space power spectrum of dark matter halos
  and its application to galaxy power spectrum}. \prd 102(6):063504.
  \doi{10.1103/PhysRevD.102.063504}.
  {\href{https://arxiv.org/abs/2005.06122}{{arXiv:2005.06122}}} {[astro-ph.CO]}

\bibitem[{{Kochanek} and {White}(2000)}]{Kochanek:2000}
{Kochanek} CS, {White} M (2000) {A Quantitative Study of Interacting Dark
  Matter in Halos}. \apj 543:514--520. \doi{10.1086/317149}.
  {\href{https://arxiv.org/abs/astro-ph/0003483}{{astro-ph/0003483}}}

\bibitem[{{Koda} and {Shapiro}(2011)}]{Koda:2011}
{Koda} J, {Shapiro} PR (2011) {Gravothermal collapse of isolated
  self-interacting dark matter haloes: N-body simulation versus the fluid
  model}. \mnras 415(2):1125--1137. \doi{10.1111/j.1365-2966.2011.18684.x}.
  {\href{https://arxiv.org/abs/1101.3097}{{arXiv:1101.3097}}} {[astro-ph.CO]}

\bibitem[{{Koda} et~al.(2016){Koda}, {Blake}, {Beutler}, {Kazin}, and
  {Marin}}]{Koda:2016}
{Koda} J, {Blake} C, {Beutler} F, {Kazin} E, {Marin} F (2016) {Fast and
  accurate mock catalogue generation for low-mass galaxies}. \mnras
  459(2):2118--2129. \doi{10.1093/mnras/stw763}.
  {\href{https://arxiv.org/abs/1507.05329}{{arXiv:1507.05329}}} {[astro-ph.CO]}

\bibitem[{{KodiRamanah} et~al.(2020){KodiRamanah}, {Charnock},
  {Villaescusa-Navarro}, and {Wandelt}}]{KodiRamanah:2020}
{KodiRamanah} D, {Charnock} T, {Villaescusa-Navarro} F, {Wandelt} BD (2020)
  {Super-resolution emulator of cosmological simulations using deep physical
  models}. \mnras 495(4):4227--4236. \doi{10.1093/mnras/staa1428}.
  {\href{https://arxiv.org/abs/2001.05519}{{arXiv:2001.05519}}} {[astro-ph.CO]}

\bibitem[{{Kofman} et~al.(1992){Kofman}, {Pogosyan}, {Shandarin}, and
  {Melott}}]{Kofman:1992}
{Kofman} L, {Pogosyan} D, {Shandarin} SF, {Melott} AL (1992) {Coherent
  Structures in the Universe and the Adhesion Model}. \apj 393:437.
  \doi{10.1086/171517}

\bibitem[{{Kofman} and {Shandarin}(1988)}]{Kofman:1988}
{Kofman} LA, {Shandarin} SF (1988) {Theory of adhesion for the large-scale
  structure of the Universe}. \nat 334(6178):129--131. \doi{10.1038/334129a0}

\bibitem[{{Kokron} et~al.(2021){Kokron}, {DeRose}, {Chen}, {White}, and
  {Wechsler}}]{Kokron:2021}
{Kokron} N, {DeRose} J, {Chen} SF, {White} M, {Wechsler} RH (2021) {The
  cosmology dependence of galaxy clustering and lensing from a hybrid
  $N$-body-perturbation theory model}. arXiv e-prints arXiv:2101.11014.
  {\href{https://arxiv.org/abs/2101.11014}{{arXiv:2101.11014}}} {[astro-ph.CO]}

\bibitem[{{Kolb} and {Tkachev}(1996)}]{Kolb:1996}
{Kolb} EW, {Tkachev} II (1996) {Femtolensing and Picolensing by Axion
  Miniclusters}. \apjl 460:L25. \doi{10.1086/309962}.
  {\href{https://arxiv.org/abs/astro-ph/9510043}{{arXiv:astro-ph/9510043}}}
  {[astro-ph]}

\bibitem[{{Kopp} et~al.(2014){Kopp}, {Uhlemann}, and {Haugg}}]{Kopp:2014}
{Kopp} M, {Uhlemann} C, {Haugg} T (2014) {Newton to Einstein -- dust to dust}.
  \jcap 3:018. \doi{10.1088/1475-7516/2014/03/018}.
  {\href{https://arxiv.org/abs/1312.3638}{{arXiv:1312.3638}}} {[astro-ph.CO]}

\bibitem[{{Kopp} et~al.(2017){Kopp}, {Vattis}, and {Skordis}}]{Kopp:2017}
{Kopp} M, {Vattis} K, {Skordis} C (2017) {Solving the Vlasov equation in two
  spatial dimensions with the Schr{\"o}dinger method}. \prd 96(12):123532.
  \doi{10.1103/PhysRevD.96.123532}.
  {\href{https://arxiv.org/abs/1711.00140}{{arXiv:1711.00140}}}

\bibitem[{{Koyama}(2016)}]{Koyama:2016}
{Koyama} K (2016) {Cosmological tests of modified gravity}. Reports on Progress
  in Physics 79(4):046902. \doi{10.1088/0034-4885/79/4/046902}.
  {\href{https://arxiv.org/abs/1504.04623}{{arXiv:1504.04623}}} {[astro-ph.CO]}

\bibitem[{{Kravtsov} et~al.(1997){Kravtsov}, {Klypin}, and
  {Khokhlov}}]{Kravtsov:1997}
{Kravtsov} AV, {Klypin} AA, {Khokhlov} AM (1997) {Adaptive Refinement Tree: A
  New High-Resolution N-Body Code for Cosmological Simulations}. \apjs
  111(1):73--94. \doi{10.1086/313015}.
  {\href{https://arxiv.org/abs/astro-ph/9701195}{{arXiv:astro-ph/9701195}}}
  {[astro-ph]}

\bibitem[{{Kravtsov} et~al.(2002){Kravtsov}, {Klypin}, and
  {Hoffman}}]{Kravtsov:2002}
{Kravtsov} AV, {Klypin} A, {Hoffman} Y (2002) {Constrained Simulations of the
  Real Universe. II. Observational Signatures of Intergalactic Gas in the Local
  Supercluster Region}. \apj 571(2):563--575. \doi{10.1086/340046}.
  {\href{https://arxiv.org/abs/astro-ph/0109077}{{arXiv:astro-ph/0109077}}}
  {[astro-ph]}

\bibitem[{{Kravtsov} et~al.(2004){Kravtsov}, {Berlind}, {Wechsler}, {Klypin},
  {Gottl{\"o}ber}, {Allgood}, and {Primack}}]{Kravtsov:2004}
{Kravtsov} AV, {Berlind} AA, {Wechsler} RH, {Klypin} AA, {Gottl{\"o}ber} S,
  {Allgood} Bo, {Primack} JR (2004) {The Dark Side of the Halo Occupation
  Distribution}. \apj 609(1):35--49. \doi{10.1086/420959}.
  {\href{https://arxiv.org/abs/astro-ph/0308519}{{arXiv:astro-ph/0308519}}}
  {[astro-ph]}

\bibitem[{{Kuhlen} et~al.(2012){Kuhlen}, {Vogelsberger}, and
  {Angulo}}]{Kuhlen:2012}
{Kuhlen} M, {Vogelsberger} M, {Angulo} R (2012) {Numerical simulations of the
  dark universe: State of the art and the next decade}. Physics of the Dark
  Universe 1:50--93. \doi{10.1016/j.dark.2012.10.002}.
  {\href{https://arxiv.org/abs/1209.5745}{{arXiv:1209.5745}}} {[astro-ph.CO]}

\bibitem[{{Kuo} et~al.(2018){Kuo}, {Lattanzi}, {Cheung}, and
  {Valle}}]{Kuo:2018}
{Kuo} JL, {Lattanzi} M, {Cheung} K, {Valle} JWF (2018) {Decaying warm dark
  matter and structure formation}. \jcap 2018(12):026.
  \doi{10.1088/1475-7516/2018/12/026}.
  {\href{https://arxiv.org/abs/1803.05650}{{arXiv:1803.05650}}} {[astro-ph.CO]}

\bibitem[{{{Kurzak}, J.} and {{Pettitt}, B.~M.}(2006)}]{Kurzak:2006}
{{Kurzak}, J}, {{Pettitt}, B~M} (2006) Fast multipole methods for particle
  dynamics. Molecular simulation 32(10-11):775--790

\bibitem[{Kwan et~al.(2015)Kwan, Heitmann, Habib, Padmanabhan, Finkel,
  Lawrence, Frontiere, and Pope}]{Kwan:2013}
Kwan J, Heitmann K, Habib S, Padmanabhan N, Finkel H, Lawrence E, Frontiere N,
  Pope A (2015) {Cosmic Emulation: Fast Predictions for the Galaxy Power
  Spectrum}. Astrophys J 810(1):35. \doi{10.1088/0004-637X/810/1/35}.
  {\href{https://arxiv.org/abs/1311.6444}{{arXiv:1311.6444}}} {[astro-ph.CO]}

\bibitem[{{Lacey} and {Cole}(1994)}]{LaceyCole:1994}
{Lacey} C, {Cole} S (1994) {Merger Rates in Hierarchical Models of Galaxy
  Formation - Part Two - Comparison with N-Body Simulations}. \mnras 271:676.
  \doi{10.1093/mnras/271.3.676}.
  {\href{https://arxiv.org/abs/astro-ph/9402069}{{arXiv:astro-ph/9402069}}}
  {[astro-ph]}

\bibitem[{{Lagos} et~al.(2018){Lagos}, {Tobar}, {Robotham}, {Obreschkow},
  {Mitchell}, {Power}, and {Elahi}}]{Lagos:2018}
{Lagos} CdP, {Tobar} RJ, {Robotham} ASG, {Obreschkow} D, {Mitchell} PD, {Power}
  C, {Elahi} PJ (2018) {Shark: introducing an open source, free, and flexible
  semi-analytic model of galaxy formation}. \mnras 481(3):3573--3603.
  \doi{10.1093/mnras/sty2440}.
  {\href{https://arxiv.org/abs/1807.11180}{{arXiv:1807.11180}}} {[astro-ph.GA]}

\bibitem[{{Laigle} et~al.(2015){Laigle}, {Pichon}, {Codis}, {Dubois}, {Le
  Borgne}, {Pogosyan}, {Devriendt}, {Peirani}, {Prunet}, {Rouberol}, {Slyz},
  and {Sousbie}}]{Laigle:2015}
{Laigle} C, {Pichon} C, {Codis} S, {Dubois} Y, {Le Borgne} D, {Pogosyan} D,
  {Devriendt} J, {Peirani} S, {Prunet} S, {Rouberol} S, {Slyz} A, {Sousbie} T
  (2015) {Swirling around filaments: are large-scale structure vortices
  spinning up dark haloes?} \mnras 446(3):2744--2759.
  \doi{10.1093/mnras/stu2289}.
  {\href{https://arxiv.org/abs/1310.3801}{{arXiv:1310.3801}}} {[astro-ph.CO]}

\bibitem[{Lanczos(1986)}]{Lanczos:1986}
Lanczos C (1986) The Variational Principles of Mechanics. Dover Books On
  Physics, Dover Publications

\bibitem[{{Lange} et~al.(2019){Lange}, {van den Bosch}, {Zentner}, {Wang},
  {Hearin}, and {Guo}}]{Lange:2019}
{Lange} JU, {van den Bosch} FC, {Zentner} AR, {Wang} K, {Hearin} AP, {Guo} H
  (2019) {Cosmological Evidence Modelling: a new simulation-based approach to
  constrain cosmology on non-linear scales}. \mnras 490(2):1870--1878.
  \doi{10.1093/mnras/stz2664}.
  {\href{https://arxiv.org/abs/1909.03107}{{arXiv:1909.03107}}} {[astro-ph.CO]}

\bibitem[{{Lange} et~al.(2021){Lange}, {Hearin}, {Leauthaud}, {van den Bosch},
  {Guo}, and {DeRose}}]{Lange:2021}
{Lange} JU, {Hearin} AP, {Leauthaud} A, {van den Bosch} FC, {Guo} H, {DeRose} J
  (2021) {Five-percent measurements of the growth rate from simulation-based
  modelling of redshift-space clustering in BOSS LOWZ}. arXiv e-prints
  arXiv:2101.12261.
  {\href{https://arxiv.org/abs/2101.12261}{{arXiv:2101.12261}}} {[astro-ph.CO]}

\bibitem[{{Laskar}(1993)}]{Laskar:1993}
{Laskar} J (1993) {Frequency analysis for multi-dimensional systems. Global
  dynamics and diffusion}. Physica D Nonlinear Phenomena 67:257--281.
  \doi{10.1016/0167-2789(93)90210-R}

\bibitem[{{Laureijs} et~al.(2011){Laureijs}, {Amiaux}, {Arduini},
  {Augu{\`e}res}, {Brinchmann}, {Cole}, {Cropper}, {Dabin}, {Duvet}, {Ealet},
  and et~al.}]{Euclid}
{Laureijs} R, {Amiaux} J, {Arduini} S, {Augu{\`e}res} JL, {Brinchmann} J,
  {Cole} R, {Cropper} M, {Dabin} C, {Duvet} L, {Ealet} A, et~al (2011) {Euclid
  Definition Study Report}. arXiv e-prints arXiv:1110.3193.
  {\href{https://arxiv.org/abs/1110.3193}{{arXiv:1110.3193}}} {[astro-ph.CO]}

\bibitem[{{Lavaux} and {Hudson}(2011)}]{Lavaux:2011}
{Lavaux} G, {Hudson} MJ (2011) {The 2M++ galaxy redshift catalogue}. \mnras
  416(4):2840--2856. \doi{10.1111/j.1365-2966.2011.19233.x}.
  {\href{https://arxiv.org/abs/1105.6107}{{arXiv:1105.6107}}} {[astro-ph.CO]}

\bibitem[{{Lavaux} and {Jasche}(2016)}]{Lavaux:2016}
{Lavaux} G, {Jasche} J (2016) {Unmasking the masked Universe: the 2M++
  catalogue through Bayesian eyes}. \mnras 455(3):3169--3179.
  \doi{10.1093/mnras/stv2499}.
  {\href{https://arxiv.org/abs/1509.05040}{{arXiv:1509.05040}}} {[astro-ph.CO]}

\bibitem[{{Lawrence} et~al.(2017){Lawrence}, {Heitmann}, {Kwan}, {Upadhye},
  {Bingham}, {Habib}, {Higdon}, {Pope}, {Finkel}, and
  {Frontiere}}]{Lawrence:2017}
{Lawrence} E, {Heitmann} K, {Kwan} J, {Upadhye} A, {Bingham} D, {Habib} S,
  {Higdon} D, {Pope} A, {Finkel} H, {Frontiere} N (2017) {The Mira-Titan
  Universe. II. Matter Power Spectrum Emulation}. \apj 847(1):50.
  \doi{10.3847/1538-4357/aa86a9}.
  {\href{https://arxiv.org/abs/1705.03388}{{arXiv:1705.03388}}} {[astro-ph.CO]}

\bibitem[{{Lazeyras} and {Schmidt}(2018)}]{Lazeyras:2018}
{Lazeyras} T, {Schmidt} F (2018) {Beyond LIMD bias: a measurement of the
  complete set of third-order halo bias parameters}. \jcap 2018(9):008.
  \doi{10.1088/1475-7516/2018/09/008}.
  {\href{https://arxiv.org/abs/1712.07531}{{arXiv:1712.07531}}} {[astro-ph.CO]}

\bibitem[{{Lazeyras} and {Schmidt}(2019)}]{Lazeyras:2019}
{Lazeyras} T, {Schmidt} F (2019) {A robust measurement of the first
  higher-derivative bias of dark matter halos}. \jcap 2019(11):041.
  \doi{10.1088/1475-7516/2019/11/041}.
  {\href{https://arxiv.org/abs/1904.11294}{{arXiv:1904.11294}}} {[astro-ph.CO]}

\bibitem[{{Lazeyras} et~al.(2016){Lazeyras}, {Wagner}, {Baldauf}, and
  {Schmidt}}]{Lazeyras:2016}
{Lazeyras} T, {Wagner} C, {Baldauf} T, {Schmidt} F (2016) {Precision
  measurement of the local bias of dark matter halos}. \jcap 2016(2):018.
  \doi{10.1088/1475-7516/2016/02/018}.
  {\href{https://arxiv.org/abs/1511.01096}{{arXiv:1511.01096}}} {[astro-ph.CO]}

\bibitem[{{Lazeyras} et~al.(2017){Lazeyras}, {Musso}, and
  {Schmidt}}]{Lazeyras:2017}
{Lazeyras} T, {Musso} M, {Schmidt} F (2017) {Large-scale assembly bias of dark
  matter halos}. \jcap 2017(3):059. \doi{10.1088/1475-7516/2017/03/059}.
  {\href{https://arxiv.org/abs/1612.04360}{{arXiv:1612.04360}}} {[astro-ph.CO]}

\bibitem[{{Le Brun} et~al.(2014){Le Brun}, {McCarthy}, {Schaye}, and
  {Ponman}}]{LeBrun:2014}
{Le Brun} AMC, {McCarthy} IG, {Schaye} J, {Ponman} TJ (2014) {Towards a
  realistic population of simulated galaxy groups and clusters}. \mnras
  441:1270--1290. \doi{10.1093/mnras/stu608}.
  {\href{https://arxiv.org/abs/1312.5462}{{arXiv:1312.5462}}}

\bibitem[{{Leane}(2020)}]{Leane:2020}
{Leane} RK (2020) {Indirect Detection of Dark Matter in the Galaxy}. arXiv
  e-prints arXiv:2006.00513.
  {\href{https://arxiv.org/abs/2006.00513}{{arXiv:2006.00513}}} {[hep-ph]}

\bibitem[{{Leclercq} et~al.(2020){Leclercq}, {Faure}, {Lavaux}, {Wandelt},
  {Jaffe}, {Heavens}, and {Percival}}]{Leclercq:2020}
{Leclercq} F, {Faure} B, {Lavaux} G, {Wandelt} BD, {Jaffe} AH, {Heavens} AF,
  {Percival} WJ (2020) {Perfectly parallel cosmological simulations using
  spatial comoving Lagrangian acceleration}. \aap 639:A91.
  \doi{10.1051/0004-6361/202037995}.
  {\href{https://arxiv.org/abs/2003.04925}{{arXiv:2003.04925}}} {[astro-ph.CO]}

\bibitem[{{Lehmann} et~al.(2017){Lehmann}, {Mao}, {Becker}, {Skillman}, and
  {Wechsler}}]{Lehman:2017}
{Lehmann} BV, {Mao} YY, {Becker} MR, {Skillman} SW, {Wechsler} RH (2017) {The
  Concentration Dependence of the Galaxy-Halo Connection: Modeling Assembly
  Bias with Abundance Matching}. \apj 834(1):37.
  \doi{10.3847/1538-4357/834/1/37}.
  {\href{https://arxiv.org/abs/1510.05651}{{arXiv:1510.05651}}} {[astro-ph.CO]}

\bibitem[{{Leistedt} et~al.(2014){Leistedt}, {Peiris}, and
  {Roth}}]{Leistedt:2014}
{Leistedt} B, {Peiris} HV, {Roth} N (2014) {Constraints on Primordial
  Non-Gaussianity from 800 000 Photometric Quasars}. \prl 113(22):221301.
  \doi{10.1103/PhysRevLett.113.221301}.
  {\href{https://arxiv.org/abs/1405.4315}{{arXiv:1405.4315}}} {[astro-ph.CO]}

\bibitem[{{Lemson} and {Virgo Consortium}(2006)}]{Lemson:2006}
{Lemson} G, {Virgo Consortium} t (2006) {Halo and Galaxy Formation Histories
  from the Millennium Simulation: Public release of a VO-oriented and
  SQL-queryable database for studying the evolution of galaxies in the
  LambdaCDM cosmogony}. arXiv e-prints astro-ph/0608019.
  {\href{https://arxiv.org/abs/astro-ph/0608019}{{arXiv:astro-ph/0608019}}}
  {[astro-ph]}

\bibitem[{{Leo} et~al.(2017){Leo}, {Baugh}, {Li}, and {Pascoli}}]{Leo:2017}
{Leo} M, {Baugh} CM, {Li} B, {Pascoli} S (2017) {The effect of thermal
  velocities on structure formation in N-body simulations of warm dark matter}.
  \jcap 2017(11):017. \doi{10.1088/1475-7516/2017/11/017}.
  {\href{https://arxiv.org/abs/1706.07837}{{arXiv:1706.07837}}} {[astro-ph.CO]}

\bibitem[{{Lepori} et~al.(2021){Lepori}, {Adamek}, and {Durrer}}]{Lepori:2021}
{Lepori} F, {Adamek} J, {Durrer} R (2021) {Cosmological Simulations of Number
  Counts}. arXiv e-prints arXiv:2106.01347.
  {\href{https://arxiv.org/abs/2106.01347}{{arXiv:2106.01347}}} {[astro-ph.CO]}

\bibitem[{{Leroy} et~al.(2021){Leroy}, {Garrison}, {Eisenstein}, {Joyce}, and
  {Maleubre}}]{Leroy:2021}
{Leroy} M, {Garrison} L, {Eisenstein} D, {Joyce} M, {Maleubre} S (2021)
  {Testing dark matter halo properties using self-similarity}. \mnras
  501(4):5064--5072. \doi{10.1093/mnras/staa3435}.
  {\href{https://arxiv.org/abs/2004.08406}{{arXiv:2004.08406}}} {[astro-ph.CO]}

\bibitem[{{Lesgourgues}(2011)}]{Lesgourgues:2011}
{Lesgourgues} J (2011) {The Cosmic Linear Anisotropy Solving System (CLASS) I:
  Overview}. arXiv e-prints arXiv:1104.2932.
  {\href{https://arxiv.org/abs/1104.2932}{{arXiv:1104.2932}}} {[astro-ph.IM]}

\bibitem[{{Lesgourgues} et~al.(2004){Lesgourgues}, {Pastor}, and
  {Perotto}}]{Lesgourgues:2004}
{Lesgourgues} J, {Pastor} S, {Perotto} L (2004) {Probing neutrino masses with
  future galaxy redshift surveys}. \prd 70(4):045016.
  \doi{10.1103/PhysRevD.70.045016}.
  {\href{https://arxiv.org/abs/hep-ph/0403296}{{arXiv:hep-ph/0403296}}}
  {[hep-ph]}

\bibitem[{Lewis et~al.(2000)Lewis, Challinor, and Lasenby}]{Lewis:1999bs}
Lewis A, Challinor A, Lasenby A (2000) {Efficient computation of CMB
  anisotropies in closed FRW models}. \apj 538:473--476. \doi{10.1086/309179}.
  {\href{https://arxiv.org/abs/astro-ph/9911177}{{arXiv:astro-ph/9911177}}}
  {[astro-ph]}

\bibitem[{{Li}(2018)}]{BaojiuLi:2018}
{Li} B (2018) {Simulating Large-Scale Structure for Models of Cosmic
  Acceleration}. \doi{10.1088/978-0-7503-1587-6}

\bibitem[{{Li} et~al.(2012{\natexlab{a}}){Li}, {Zhao}, {Teyssier}, and
  {Koyama}}]{Li:2012}
{Li} B, {Zhao} GB, {Teyssier} R, {Koyama} K (2012{\natexlab{a}}) {ECOSMOG: an
  Efficient COde for Simulating MOdified Gravity}. \jcap 2012(1):051.
  \doi{10.1088/1475-7516/2012/01/051}.
  {\href{https://arxiv.org/abs/1110.1379}{{arXiv:1110.1379}}} {[astro-ph.CO]}

\bibitem[{{Li} et~al.(2013{\natexlab{a}}){Li}, {Barreira}, {Baugh}, {Hellwing},
  {Koyama}, {Pascoli}, and {Zhao}}]{BLi:2013a}
{Li} B, {Barreira} A, {Baugh} CM, {Hellwing} WA, {Koyama} K, {Pascoli} S,
  {Zhao} GB (2013{\natexlab{a}}) {Simulating the quartic Galileon gravity model
  on adaptively refined meshes}. \jcap 2013(11):012.
  \doi{10.1088/1475-7516/2013/11/012}.
  {\href{https://arxiv.org/abs/1308.3491}{{arXiv:1308.3491}}} {[astro-ph.CO]}

\bibitem[{{Li} et~al.(2013{\natexlab{b}}){Li}, {Zhao}, and
  {Koyama}}]{BLi:2013b}
{Li} B, {Zhao} GB, {Koyama} K (2013{\natexlab{b}}) {Exploring Vainshtein
  mechanism on adaptively refined meshes}. \jcap 2013(5):023.
  \doi{10.1088/1475-7516/2013/05/023}.
  {\href{https://arxiv.org/abs/1303.0008}{{arXiv:1303.0008}}} {[astro-ph.CO]}

\bibitem[{{Li} et~al.(2012{\natexlab{b}}){Li}, {Erickcek}, and
  {Law}}]{LiUCMH:2012}
{Li} F, {Erickcek} AL, {Law} NM (2012{\natexlab{b}}) {A new probe of the
  small-scale primordial power spectrum: Astrometric microlensing by
  ultracompact minihalos}. \prd 86(4):043519. \doi{10.1103/PhysRevD.86.043519}.
  {\href{https://arxiv.org/abs/1202.1284}{{arXiv:1202.1284}}} {[astro-ph.CO]}

\bibitem[{{Li} et~al.(2014){Li}, {Hu}, and {Takada}}]{Li:2014}
{Li} Y, {Hu} W, {Takada} M (2014) {Super-sample covariance in simulations}.
  \prd 89(8):083519. \doi{10.1103/PhysRevD.89.083519}.
  {\href{https://arxiv.org/abs/1401.0385}{{arXiv:1401.0385}}} {[astro-ph.CO]}

\bibitem[{{Li} et~al.(2018){Li}, {Schmittfull}, and {Seljak}}]{Li:2018}
{Li} Y, {Schmittfull} M, {Seljak} U (2018) {Galaxy power-spectrum responses and
  redshift-space super-sample effect}. \jcap 2018(2):022.
  \doi{10.1088/1475-7516/2018/02/022}.
  {\href{https://arxiv.org/abs/1711.00018}{{arXiv:1711.00018}}} {[astro-ph.CO]}

\bibitem[{{Li} et~al.(2020){Li}, {Ni}, {Croft}, {Di Matteo}, {Bird}, and
  {Feng}}]{LiNi:2020}
{Li} Y, {Ni} Y, {Croft} RAC, {Di Matteo} T, {Bird} S, {Feng} Y (2020)
  {AI-assisted super-resolution cosmological simulations}. arXiv e-prints
  arXiv:2010.06608.
  {\href{https://arxiv.org/abs/2010.06608}{{arXiv:2010.06608}}} {[astro-ph.CO]}

\bibitem[{{Liao}(2018)}]{Liao:2018}
{Liao} S (2018) {An alternative method to generate pre-initial conditions for
  cosmological N-body simulations}. \mnras 481(3):3750--3760.
  \doi{10.1093/mnras/sty2523}.
  {\href{https://arxiv.org/abs/1807.03574}{{arXiv:1807.03574}}} {[astro-ph.GA]}

\bibitem[{{Libeskind} et~al.(2018){Libeskind}, {van de Weygaert}, {Cautun},
  {Falck}, {Tempel}, {Abel}, {Alpaslan}, {Arag{\'o}n-Calvo}, {Forero-Romero},
  {Gonzalez}, {Gottl{\"o}ber}, {Hahn}, {Hellwing}, {Hoffman}, {Jones},
  {Kitaura}, {Knebe}, {Manti}, {Neyrinck}, {Nuza}, {Padilla}, {Platen},
  {Ramachandra}, {Robotham}, {Saar}, {Shand arin}, {Steinmetz}, {Stoica},
  {Sousbie}, and {Yepes}}]{Libeskind:2018}
{Libeskind} NI, {van de Weygaert} R, {Cautun} M, {Falck} B, {Tempel} E, {Abel}
  T, {Alpaslan} M, {Arag{\'o}n-Calvo} MA, {Forero-Romero} JE, {Gonzalez} R,
  {Gottl{\"o}ber} S, {Hahn} O, {Hellwing} WA, {Hoffman} Y, {Jones} BJT,
  {Kitaura} F, {Knebe} A, {Manti} S, {Neyrinck} M, {Nuza} SE, {Padilla} N,
  {Platen} E, {Ramachandra} N, {Robotham} A, {Saar} E, {Shand arin} S,
  {Steinmetz} M, {Stoica} RS, {Sousbie} T, {Yepes} G (2018) {Tracing the cosmic
  web}. \mnras 473(1):1195--1217. \doi{10.1093/mnras/stx1976}.
  {\href{https://arxiv.org/abs/1705.03021}{{arXiv:1705.03021}}} {[astro-ph.CO]}

\bibitem[{{Lippich} and {S{\'a}nchez}(2020)}]{Lippich:2020}
{Lippich} M, {S{\'a}nchez} AG (2020) {MEDUSA: Minkowski functionals estimated
  from Delaunay tessellations of the three-dimensional large-scale structure}.
  arXiv e-prints arXiv:2012.08529.
  {\href{https://arxiv.org/abs/2012.08529}{{arXiv:2012.08529}}} {[astro-ph.CO]}

\bibitem[{{Lippich} et~al.(2019){Lippich}, {S{\'a}nchez}, {Colavincenzo},
  {Sefusatti}, {Monaco}, {Blot}, {Crocce}, {Alvarez}, {Agrawal}, {Avila},
  {Balaguera-Antol{\'\i}nez}, {Bond}, {Codis}, {Dalla Vecchia}, {Dorta},
  {Fosalba}, {Izard}, {Kitaura}, {Pellejero-Ibanez}, {Stein}, {Vakili}, and
  {Yepes}}]{Lippich:2019}
{Lippich} M, {S{\'a}nchez} AG, {Colavincenzo} M, {Sefusatti} E, {Monaco} P,
  {Blot} L, {Crocce} M, {Alvarez} MA, {Agrawal} A, {Avila} S,
  {Balaguera-Antol{\'\i}nez} A, {Bond} R, {Codis} S, {Dalla Vecchia} C, {Dorta}
  A, {Fosalba} P, {Izard} A, {Kitaura} FS, {Pellejero-Ibanez} M, {Stein} G,
  {Vakili} M, {Yepes} G (2019) {Comparing approximate methods for mock
  catalogues and covariance matrices - I. Correlation function}. \mnras
  482(2):1786--1806. \doi{10.1093/mnras/sty2757}.
  {\href{https://arxiv.org/abs/1806.09477}{{arXiv:1806.09477}}} {[astro-ph.CO]}

\bibitem[{{List} et~al.(2019{\natexlab{a}}){List}, {Bhat}, and
  {Lewis}}]{List:2019}
{List} F, {Bhat} I, {Lewis} GF (2019{\natexlab{a}}) {A black box for dark
  sector physics: predicting dark matter annihilation feedback with conditional
  GANs}. \mnras 490(3):3134--3143. \doi{10.1093/mnras/stz2759}.
  {\href{https://arxiv.org/abs/1910.00291}{{arXiv:1910.00291}}} {[astro-ph.CO]}

\bibitem[{{List} et~al.(2019{\natexlab{b}}){List}, {Iwanus}, {Elahi}, and
  {Lewis}}]{List:2019a}
{List} F, {Iwanus} N, {Elahi} PJ, {Lewis} GF (2019{\natexlab{b}}) {A novel
  scheme for Dark Matter Annihilation Feedback in cosmological simulations}.
  \mnras 489(3):4217--4232. \doi{10.1093/mnras/stz2287}.
  {\href{https://arxiv.org/abs/1908.05812}{{arXiv:1908.05812}}} {[astro-ph.CO]}

\bibitem[{Liu et~al.(2015)Liu, Petri, Haiman, Hui, Kratochvil, and
  May}]{Liu:2014}
Liu J, Petri A, Haiman Z, Hui L, Kratochvil JM, May M (2015) {Cosmology
  constraints from the weak lensing peak counts and the power spectrum in
  CFHTLenS data}. Phys Rev D91(6):063507. \doi{10.1103/PhysRevD.91.063507}.
  {\href{https://arxiv.org/abs/1412.0757}{{arXiv:1412.0757}}} {[astro-ph.CO]}

\bibitem[{{Llinares}(2017)}]{Llinares:2017}
{Llinares} C (2017) {The shrinking domain framework I: a new, faster, more
  efficient approach to cosmological simulations}. arXiv e-prints
  arXiv:1709.04703.
  {\href{https://arxiv.org/abs/1709.04703}{{arXiv:1709.04703}}} {[astro-ph.CO]}

\bibitem[{{Llinares}(2021)}]{Llinares:2021}
{Llinares} C (2021) {Simulation techniques for modified gravity}. arXiv
  e-prints arXiv:2103.10890.
  {\href{https://arxiv.org/abs/2103.10890}{{arXiv:2103.10890}}} {[astro-ph.CO]}

\bibitem[{{Llinares} and {Mota}(2014)}]{LlinaresMota:2014}
{Llinares} C, {Mota} DF (2014) {Cosmological simulations of screened modified
  gravity out of the static approximation: Effects on matter distribution}.
  \prd 89(8):084023. \doi{10.1103/PhysRevD.89.084023}.
  {\href{https://arxiv.org/abs/1312.6016}{{arXiv:1312.6016}}} {[astro-ph.CO]}

\bibitem[{{Llinares} et~al.(2014){Llinares}, {Mota}, and
  {Winther}}]{Llinares:2014}
{Llinares} C, {Mota} DF, {Winther} HA (2014) {ISIS: a new N-body cosmological
  code with scalar fields based on RAMSES. Code presentation and application to
  the shapes of clusters}. \aap 562:A78. \doi{10.1051/0004-6361/201322412}.
  {\href{https://arxiv.org/abs/1307.6748}{{arXiv:1307.6748}}} {[astro-ph.CO]}

\bibitem[{{Loeb} and {Zaldarriaga}(2005)}]{Loeb:2005}
{Loeb} A, {Zaldarriaga} M (2005) {Small-scale power spectrum of cold dark
  matter}. \prd 71(10):103520. \doi{10.1103/PhysRevD.71.103520}.
  {\href{https://arxiv.org/abs/astro-ph/0504112}{{arXiv:astro-ph/0504112}}}
  {[astro-ph]}

\bibitem[{{Lombriser}(2016)}]{Lombriser:2016}
{Lombriser} L (2016) {A parametrisation of modified gravity on nonlinear
  cosmological scales}. \jcap 2016(11):039.
  \doi{10.1088/1475-7516/2016/11/039}.
  {\href{https://arxiv.org/abs/1608.00522}{{arXiv:1608.00522}}} {[astro-ph.CO]}

\bibitem[{{Lovell} et~al.(2012){Lovell}, {Eke}, {Frenk}, {Gao}, {Jenkins},
  {Theuns}, {Wang}, {White}, {Boyarsky}, and {Ruchayskiy}}]{Lovell:2012}
{Lovell} MR, {Eke} V, {Frenk} CS, {Gao} L, {Jenkins} A, {Theuns} T, {Wang} J,
  {White} SDM, {Boyarsky} A, {Ruchayskiy} O (2012) {The haloes of bright
  satellite galaxies in a warm dark matter universe}. \mnras 420(3):2318--2324.
  \doi{10.1111/j.1365-2966.2011.20200.x}.
  {\href{https://arxiv.org/abs/1104.2929}{{arXiv:1104.2929}}} {[astro-ph.CO]}

\bibitem[{{Lovell} et~al.(2014){Lovell}, {Frenk}, {Eke}, {Jenkins}, {Gao}, and
  {Theuns}}]{Lovell:2014}
{Lovell} MR, {Frenk} CS, {Eke} VR, {Jenkins} A, {Gao} L, {Theuns} T (2014) {The
  properties of warm dark matter haloes}. \mnras 439(1):300--317.
  \doi{10.1093/mnras/stt2431}.
  {\href{https://arxiv.org/abs/1308.1399}{{arXiv:1308.1399}}} {[astro-ph.CO]}

\bibitem[{{Lovell} et~al.(2018){Lovell}, {Zavala}, {Vogelsberger}, {Shen},
  {Cyr-Racine}, {Pfrommer}, {Sigurdson}, {Boylan-Kolchin}, and
  {Pillepich}}]{Lovell:2018}
{Lovell} MR, {Zavala} J, {Vogelsberger} M, {Shen} X, {Cyr-Racine} FY,
  {Pfrommer} C, {Sigurdson} K, {Boylan-Kolchin} M, {Pillepich} A (2018) {ETHOS
  - an effective theory of structure formation: predictions for the
  high-redshift Universe - abundance of galaxies and reionization}. \mnras
  477(3):2886--2899. \doi{10.1093/mnras/sty818}.
  {\href{https://arxiv.org/abs/1711.10497}{{arXiv:1711.10497}}} {[astro-ph.CO]}

\bibitem[{{Lovell} et~al.(2019){Lovell}, {Zavala}, and
  {Vogelsberger}}]{Lovell:2019}
{Lovell} MR, {Zavala} J, {Vogelsberger} M (2019) {ETHOS - an effective theory
  of structure formation: formation of the first haloes and their stars}.
  \mnras 485(4):5474--5489. \doi{10.1093/mnras/stz766}.
  {\href{https://arxiv.org/abs/1812.04627}{{arXiv:1812.04627}}} {[astro-ph.GA]}

\bibitem[{{LoVerde} and {Smith}(2011)}]{LoVerde:2011}
{LoVerde} M, {Smith} KM (2011) {The non-Gaussian halo mass function with
  f$_{NL}$, g$_{NL}$ and {\ensuremath{\tau}}$_{NL}$}. \jcap 2011(8):003.
  \doi{10.1088/1475-7516/2011/08/003}.
  {\href{https://arxiv.org/abs/1102.1439}{{arXiv:1102.1439}}} {[astro-ph.CO]}

\bibitem[{{Ludlow} et~al.(2014){Ludlow}, {Navarro}, {Angulo}, {Boylan-Kolchin},
  {Springel}, {Frenk}, and {White}}]{Ludlow:2014}
{Ludlow} AD, {Navarro} JF, {Angulo} RE, {Boylan-Kolchin} M, {Springel} V,
  {Frenk} C, {White} SDM (2014) {The mass-concentration-redshift relation of
  cold dark matter haloes}. \mnras 441(1):378--388. \doi{10.1093/mnras/stu483}.
  {\href{https://arxiv.org/abs/1312.0945}{{arXiv:1312.0945}}} {[astro-ph.CO]}

\bibitem[{{Ludlow} et~al.(2016){Ludlow}, {Bose}, {Angulo}, {Wang}, {Hellwing},
  {Navarro}, {Cole}, and {Frenk}}]{Ludlow:2016}
{Ludlow} AD, {Bose} S, {Angulo} RE, {Wang} L, {Hellwing} WA, {Navarro} JF,
  {Cole} S, {Frenk} CS (2016) {The mass-concentration-redshift relation of cold
  and warm dark matter haloes}. \mnras 460(2):1214--1232.
  \doi{10.1093/mnras/stw1046}.
  {\href{https://arxiv.org/abs/1601.02624}{{arXiv:1601.02624}}} {[astro-ph.CO]}

\bibitem[{{Ludlow} et~al.(2019){Ludlow}, {Schaye}, and {Bower}}]{Ludlow:2019}
{Ludlow} AD, {Schaye} J, {Bower} R (2019) {Numerical convergence of simulations
  of galaxy formation: the abundance and internal structure of cold dark matter
  haloes}. \mnras 488(3):3663--3684. \doi{10.1093/mnras/stz1821}.
  {\href{https://arxiv.org/abs/1812.05777}{{arXiv:1812.05777}}} {[astro-ph.CO]}

\bibitem[{{Ma} and {Bertschinger}(1994)}]{Ma:1994}
{Ma} CP, {Bertschinger} E (1994) {A Calculation of the Full Neutrino Phase
  Space in Cold + Hot Dark Matter Models}. \apj 429:22. \doi{10.1086/174298}.
  {\href{https://arxiv.org/abs/astro-ph/9308006}{{arXiv:astro-ph/9308006}}}
  {[astro-ph]}

\bibitem[{{Macci{\`o}} et~al.(2012){Macci{\`o}}, {Paduroiu}, {Anderhalden},
  {Schneider}, and {Moore}}]{Maccio:2012}
{Macci{\`o}} AV, {Paduroiu} S, {Anderhalden} D, {Schneider} A, {Moore} B (2012)
  {Cores in warm dark matter haloes: a Catch 22 problem}. \mnras
  424(2):1105--1112. \doi{10.1111/j.1365-2966.2012.21284.x}.
  {\href{https://arxiv.org/abs/1202.1282}{{arXiv:1202.1282}}} {[astro-ph.CO]}

\bibitem[{{Maciejewski} et~al.(2009){Maciejewski}, {Colombi}, {Springel},
  {Alard}, and {Bouchet}}]{Maciejewski:2009}
{Maciejewski} M, {Colombi} S, {Springel} V, {Alard} C, {Bouchet} FR (2009)
  {Phase-space structures - II. Hierarchical Structure Finder}. \mnras
  396(3):1329--1348. \doi{10.1111/j.1365-2966.2009.14825.x}.
  {\href{https://arxiv.org/abs/0812.0288}{{arXiv:0812.0288}}} {[astro-ph]}

\bibitem[{{Macpherson} et~al.(2019){Macpherson}, {Price}, and
  {Lasky}}]{Macpherson:2019}
{Macpherson} HJ, {Price} DJ, {Lasky} PD (2019) {Einstein's Universe:
  Cosmological structure formation in numerical relativity}. \prd 99(6):063522.
  \doi{10.1103/PhysRevD.99.063522}.
  {\href{https://arxiv.org/abs/1807.01711}{{arXiv:1807.01711}}} {[astro-ph.CO]}

\bibitem[{{Madelung}(1927)}]{Madelung:1927}
{Madelung} E (1927) {Quantentheorie in hydrodynamischer Form}. Zeitschrift fur
  Physik 40:322--326. \doi{10.1007/BF01400372}

\bibitem[{{Maffione} et~al.(2015){Maffione}, {G{\'o}mez}, {Cincotta},
  {Giordano}, {Cooper}, and {O'Shea}}]{Maffione:2015}
{Maffione} NP, {G{\'o}mez} FA, {Cincotta} PM, {Giordano} CM, {Cooper} AP,
  {O'Shea} BW (2015) {On the relevance of chaos for halo stars in the solar
  neighbourhood}. \mnras 453(3):2830--2847. \doi{10.1093/mnras/stv1778}.
  {\href{https://arxiv.org/abs/1508.00579}{{arXiv:1508.00579}}} {[astro-ph.GA]}

\bibitem[{Magnus(1954)}]{Magnus:1954}
Magnus W (1954) On the exponential solution of differential equations for a
  linear operator. Communications on Pure and Applied Mathematics
  7(4):649--673. \doi{10.1002/cpa.3160070404}

\bibitem[{{Makiya} et~al.(2016){Makiya}, {Enoki}, {Ishiyama}, {Kobayashi},
  {Nagashima}, {Okamoto}, {Okoshi}, {Oogi}, and {Shirakata}}]{Makiya:2016}
{Makiya} R, {Enoki} M, {Ishiyama} T, {Kobayashi} MAR, {Nagashima} M, {Okamoto}
  T, {Okoshi} K, {Oogi} T, {Shirakata} H (2016) {The New Numerical Galaxy
  Catalog ({\ensuremath{\nu}}$^{2}$GC): An updated semi-analytic model of
  galaxy and active galactic nucleus formation with large cosmological N-body
  simulations}. \pasj 68(2):25. \doi{10.1093/pasj/psw005}.
  {\href{https://arxiv.org/abs/1508.07215}{{arXiv:1508.07215}}} {[astro-ph.GA]}

\bibitem[{{Maldacena}(2003)}]{Maldacena:2003}
{Maldacena} J (2003) {Non-gaussian features of primordial fluctuations in
  single field inflationary models}. J High Energy Phys 2003(5):013.
  \doi{10.1088/1126-6708/2003/05/013}.
  {\href{https://arxiv.org/abs/astro-ph/0210603}{{arXiv:astro-ph/0210603}}}
  {[astro-ph]}

\bibitem[{{Manera} et~al.(2013){Manera}, {Scoccimarro}, {Percival}, {Samushia},
  {McBride}, {Ross}, {Sheth}, {White}, {Reid}, {S{\'a}nchez}, {de Putter},
  {Xu}, {Berlind}, {Brinkmann}, {Maraston}, {Nichol}, {Montesano},
  {Padmanabhan}, {Skibba}, {Tojeiro}, and {Weaver}}]{Manera:2013}
{Manera} M, {Scoccimarro} R, {Percival} WJ, {Samushia} L, {McBride} CK, {Ross}
  AJ, {Sheth} RK, {White} M, {Reid} BA, {S{\'a}nchez} AG, {de Putter} R, {Xu}
  X, {Berlind} AA, {Brinkmann} J, {Maraston} C, {Nichol} B, {Montesano} F,
  {Padmanabhan} N, {Skibba} RA, {Tojeiro} R, {Weaver} BA (2013) {The clustering
  of galaxies in the SDSS-III Baryon Oscillation Spectroscopic Survey: a large
  sample of mock galaxy catalogues}. \mnras 428(2):1036--1054.
  \doi{10.1093/mnras/sts084}.
  {\href{https://arxiv.org/abs/1203.6609}{{arXiv:1203.6609}}} {[astro-ph.CO]}

\bibitem[{{Mansfield} and {Avestruz}(2021)}]{Mansfield:2021}
{Mansfield} P, {Avestruz} C (2021) {How biased are halo properties in
  cosmological simulations?} \mnras 500(3):3309--3328.
  \doi{10.1093/mnras/staa3388}.
  {\href{https://arxiv.org/abs/2008.08591}{{arXiv:2008.08591}}} {[astro-ph.CO]}

\bibitem[{{Mao} et~al.(2015){Mao}, {Williamson}, and {Wechsler}}]{Mao:2015}
{Mao} YY, {Williamson} M, {Wechsler} RH (2015) {The Dependence of Subhalo
  Abundance on Halo Concentration}. \apj 810(1):21.
  \doi{10.1088/0004-637X/810/1/21}.
  {\href{https://arxiv.org/abs/1503.02637}{{arXiv:1503.02637}}} {[astro-ph.CO]}

\bibitem[{{Marcos}(2008)}]{Marcos:2008}
{Marcos} B (2008) {Particle linear theory on a self-gravitating perturbed cubic
  Bravais lattice}. \prd 78(4):043536. \doi{10.1103/PhysRevD.78.043536}.
  {\href{https://arxiv.org/abs/0804.2570}{{arXiv:0804.2570}}} {[astro-ph]}

\bibitem[{{Marcos} et~al.(2006){Marcos}, {Baertschiger}, {Joyce}, {Gabrielli},
  and {Sylos Labini}}]{Marcos:2006}
{Marcos} B, {Baertschiger} T, {Joyce} M, {Gabrielli} A, {Sylos Labini} F (2006)
  {Linear perturbative theory of the discrete cosmological N-body problem}.
  \prd 73(10):103507. \doi{10.1103/PhysRevD.73.103507}.
  {\href{https://arxiv.org/abs/astro-ph/0601479}{{astro-ph/0601479}}}

\bibitem[{{Mar{\'\i}n} et~al.(2008){Mar{\'\i}n}, {Wechsler}, {Frieman}, and
  {Nichol}}]{Marin:2008}
{Mar{\'\i}n} FA, {Wechsler} RH, {Frieman} JA, {Nichol} RC (2008) {Modeling the
  Galaxy Three-Point Correlation Function}. \apj 672(2):849--860.
  \doi{10.1086/523628}.
  {\href{https://arxiv.org/abs/0704.0255}{{arXiv:0704.0255}}} {[astro-ph]}

\bibitem[{{Marsh}(2016)}]{Marsh:2016}
{Marsh} DJE (2016) {Axion cosmology}. \physrep 643:1--79.
  \doi{10.1016/j.physrep.2016.06.005}.
  {\href{https://arxiv.org/abs/1510.07633}{{arXiv:1510.07633}}}

\bibitem[{{Martel} and {Shapiro}(1998)}]{Martel:1998}
{Martel} H, {Shapiro} PR (1998) {A convenient set of comoving cosmological
  variables and their application}. \mnras 297:467--485.
  \doi{10.1046/j.1365-8711.1998.01497.x}.
  {\href{https://arxiv.org/abs/astro-ph/9710119}{{astro-ph/9710119}}}

\bibitem[{{Masaki} et~al.(2020){Masaki}, {Nishimichi}, and
  {Takada}}]{Masaki:2020}
{Masaki} S, {Nishimichi} T, {Takada} M (2020) {Anisotropic separate universe
  simulations}. \mnras 496(1):483--496. \doi{10.1093/mnras/staa1579}.
  {\href{https://arxiv.org/abs/2003.10052}{{arXiv:2003.10052}}} {[astro-ph.CO]}

\bibitem[{{Matarrese} and {Verde}(2008)}]{Matarrese:2008}
{Matarrese} S, {Verde} L (2008) {The Effect of Primordial Non-Gaussianity on
  Halo Bias}. \apjl 677(2):L77. \doi{10.1086/587840}.
  {\href{https://arxiv.org/abs/0801.4826}{{arXiv:0801.4826}}} {[astro-ph]}

\bibitem[{Matsubara(2015)}]{Matsubara:2015}
Matsubara T (2015) {Recursive Solutions of Lagrangian Perturbation Theory}.
  \prd 92(2):023534. \doi{10.1103/PhysRevD.92.023534}.
  {\href{https://arxiv.org/abs/1505.01481}{{arXiv:1505.01481}}} {[astro-ph.CO]}

\bibitem[{{May} and {Springel}(2021)}]{May:2021}
{May} S, {Springel} V (2021) {Structure formation in large-volume cosmological
  simulations of fuzzy dark matter: Impact of the non-linear dynamics}. arXiv
  e-prints arXiv:2101.01828.
  {\href{https://arxiv.org/abs/2101.01828}{{arXiv:2101.01828}}} {[astro-ph.CO]}

\bibitem[{{McCarthy} et~al.(2017){McCarthy}, {Schaye}, {Bird}, and {Le
  Brun}}]{McCarthy:2017}
{McCarthy} IG, {Schaye} J, {Bird} S, {Le Brun} AMC (2017) {The BAHAMAS project:
  calibrated hydrodynamical simulations for large-scale structure cosmology}.
  \mnras 465:2936--2965. \doi{10.1093/mnras/stw2792}.
  {\href{https://arxiv.org/abs/1603.02702}{{arXiv:1603.02702}}}

\bibitem[{McClintock et~al.(2019)McClintock, Rozo, Becker, DeRose, Mao,
  McLaughlin, Tinker, Wechsler, and Zhai}]{McClintock:2018}
McClintock T, Rozo E, Becker MR, DeRose J, Mao YY, McLaughlin S, Tinker JL,
  Wechsler RH, Zhai Z (2019) {The Aemulus Project II: Emulating the Halo Mass
  Function}. Astrophys J 872(1):53. \doi{10.3847/1538-4357/aaf568}.
  {\href{https://arxiv.org/abs/1804.05866}{{arXiv:1804.05866}}} {[astro-ph.CO]}

\bibitem[{{McDonald}(2008)}]{McDonald:2008}
{McDonald} P (2008) {Primordial non-Gaussianity: Large-scale structure
  signature in the perturbative bias model}. \prd 78(12):123519.
  \doi{10.1103/PhysRevD.78.123519}.
  {\href{https://arxiv.org/abs/0806.1061}{{arXiv:0806.1061}}} {[astro-ph]}

\bibitem[{{McMillan}(1986)}]{McMillan:1986}
{McMillan} SLW (1986) {The Vectorization of Small-N Integrators}. In: {Hut} P,
  {McMillan} SLW (eds) The Use of Supercomputers in Stellar Dynamics. Lecture
  Notes in Physics, vol 267. Springer, Berlin, Heidelberg, p 156.
  \doi{10.1007/BFb0116406}

\bibitem[{{Mead} et~al.(2020){Mead}, {Brieden}, {Tr{\"o}ster}, and
  {Heymans}}]{Mead:2020}
{Mead} A, {Brieden} S, {Tr{\"o}ster} T, {Heymans} C (2020) {HMcode-2020:
  Improved modelling of non-linear cosmological power spectra with baryonic
  feedback}. arXiv e-prints arXiv:2009.01858.
  {\href{https://arxiv.org/abs/2009.01858}{{arXiv:2009.01858}}} {[astro-ph.CO]}

\bibitem[{{Mead} et~al.(2015{\natexlab{a}}){Mead}, {Peacock}, {Heymans},
  {Joudaki}, and {Heavens}}]{Mead:2015}
{Mead} AJ, {Peacock} JA, {Heymans} C, {Joudaki} S, {Heavens} AF
  (2015{\natexlab{a}}) {An accurate halo model for fitting non-linear
  cosmological power spectra and baryonic feedback models}. \mnras
  454(2):1958--1975. \doi{10.1093/mnras/stv2036}.
  {\href{https://arxiv.org/abs/1505.07833}{{arXiv:1505.07833}}} {[astro-ph.CO]}

\bibitem[{{Mead} et~al.(2015{\natexlab{b}}){Mead}, {Peacock}, {Lombriser}, and
  {Li}}]{MeadFR:2015}
{Mead} AJ, {Peacock} JA, {Lombriser} L, {Li} B (2015{\natexlab{b}}) {Rapid
  simulation rescaling from standard to modified gravity models}. \mnras
  452(4):4203--4221. \doi{10.1093/mnras/stv1484}.
  {\href{https://arxiv.org/abs/1412.5195}{{arXiv:1412.5195}}} {[astro-ph.CO]}

\bibitem[{{Mead} et~al.(2016){Mead}, {Heymans}, {Lombriser}, {Peacock},
  {Steele}, and {Winther}}]{Mead:2016}
{Mead} AJ, {Heymans} C, {Lombriser} L, {Peacock} JA, {Steele} OI, {Winther} HA
  (2016) {Accurate halo-model matter power spectra with dark energy, massive
  neutrinos and modified gravitational forces}. \mnras 459(2):1468--1488.
  \doi{10.1093/mnras/stw681}.
  {\href{https://arxiv.org/abs/1602.02154}{{arXiv:1602.02154}}} {[astro-ph.CO]}

\bibitem[{{Melott}(2007)}]{Melott:2007}
{Melott} AL (2007) {Comment on 'Discreteness Effects in Simulations of Hot/Warm
  Dark Matter' by J. Wang {\&} S.D.M. White}. arXiv e-prints 07090745
  {\href{https://arxiv.org/abs/0709.0745}{{arXiv:0709.0745}}}

\bibitem[{{Melott} and {Shandarin}(1989{\natexlab{a}})}]{Melott:1989}
{Melott} AL, {Shandarin} SF (1989{\natexlab{a}}) {Gravitational instability
  with high resolution}. \apj 343:26--30. \doi{10.1086/167681}

\bibitem[{{Melott} and
  {Shandarin}(1989{\natexlab{b}})}]{MelottFragmentation:2007}
{Melott} AL, {Shandarin} SF (1989{\natexlab{b}}) {Gravitational instability
  with high resolution}. \apj 343:26--30. \doi{10.1086/167681}

\bibitem[{{Melott} et~al.(1997){Melott}, {Shandarin}, {Splinter}, and
  {Suto}}]{Melott:1997}
{Melott} AL, {Shandarin} SF, {Splinter} RaJ, {Suto} Y (1997) {Demonstrating
  Discreteness and Collision Error in Cosmological N-Body Simulations of Dark
  Matter Gravitational Clustering}. \apjl 479(2):L79--L83.
  \doi{10.1086/310590}.
  {\href{https://arxiv.org/abs/astro-ph/9609152}{{arXiv:astro-ph/9609152}}}
  {[astro-ph]}

\bibitem[{{Merloni} et~al.(2012){Merloni}, {Predehl}, {Becker},
  {B{\"o}hringer}, {Boller}, {Brunner}, {Brusa}, {Dennerl}, {Freyberg},
  {Friedrich}, {Georgakakis}, {Haberl}, {Hasinger}, {Meidinger}, {Mohr},
  {Nandra}, {Rau}, {Reiprich}, {Robrade}, {Salvato}, {Santangelo}, {Sasaki},
  {Schwope}, {Wilms}, and {German eROSITA Consortium}}]{eROSITA}
{Merloni} A, {Predehl} P, {Becker} W, {B{\"o}hringer} H, {Boller} T, {Brunner}
  H, {Brusa} M, {Dennerl} K, {Freyberg} M, {Friedrich} P, {Georgakakis} A,
  {Haberl} F, {Hasinger} G, {Meidinger} N, {Mohr} J, {Nandra} K, {Rau} A,
  {Reiprich} TH, {Robrade} J, {Salvato} M, {Santangelo} A, {Sasaki} M,
  {Schwope} A, {Wilms} J, {German eROSITA Consortium} t (2012) {eROSITA Science
  Book: Mapping the Structure of the Energetic Universe}. arXiv e-prints
  arXiv:1209.3114. {\href{https://arxiv.org/abs/1209.3114}{{arXiv:1209.3114}}}
  {[astro-ph.HE]}

\bibitem[{{Merson} et~al.(2018){Merson}, {Wang}, {Benson}, {Faisst}, {Masters},
  {Kiessling}, and {Rhodes}}]{Merson:2018}
{Merson} A, {Wang} Y, {Benson} A, {Faisst} A, {Masters} D, {Kiessling} A,
  {Rhodes} J (2018) {Predicting H{\ensuremath{\alpha}} emission-line galaxy
  counts for future galaxy redshift surveys}. \mnras 474(1):177--196.
  \doi{10.1093/mnras/stx2649}.
  {\href{https://arxiv.org/abs/1710.00833}{{arXiv:1710.00833}}} {[astro-ph.GA]}

\bibitem[{{Merson} et~al.(2013){Merson}, {Baugh}, {Helly}, {Gonzalez-Perez},
  {Cole}, {Bielby}, {Norberg}, {Frenk}, {Benson}, {Bower}, {Lacey}, and
  {Lagos}}]{Merson:2013}
{Merson} AI, {Baugh} CM, {Helly} JC, {Gonzalez-Perez} V, {Cole} S, {Bielby} R,
  {Norberg} P, {Frenk} CS, {Benson} AJ, {Bower} RG, {Lacey} CG, {Lagos} CdP
  (2013) {Lightcone mock catalogues from semi-analytic models of galaxy
  formation - I. Construction and application to the BzK colour selection}.
  \mnras 429(1):556--578. \doi{10.1093/mnras/sts355}.
  {\href{https://arxiv.org/abs/1206.4049}{{arXiv:1206.4049}}} {[astro-ph.CO]}

\bibitem[{{Merz} et~al.(2005){Merz}, {Pen}, and {Trac}}]{Merz:2005}
{Merz} H, {Pen} UL, {Trac} H (2005) {Towards optimal parallel PM N-body codes:
  PMFAST}. \na 10(5):393--407. \doi{10.1016/j.newast.2005.02.001}.
  {\href{https://arxiv.org/abs/astro-ph/0402443}{{arXiv:astro-ph/0402443}}}
  {[astro-ph]}

\bibitem[{{Michaux} et~al.(2021){Michaux}, {Hahn}, {Rampf}, and
  {Angulo}}]{Michaux:2020}
{Michaux} M, {Hahn} O, {Rampf} C, {Angulo} RE (2021) {Accurate initial
  conditions for cosmological N-body simulations: minimizing truncation and
  discreteness errors}. \mnras 500(1):663--683. \doi{10.1093/mnras/staa3149}.
  {\href{https://arxiv.org/abs/2008.09588}{{arXiv:2008.09588}}} {[astro-ph.CO]}

\bibitem[{{Milillo} et~al.(2015){Milillo}, {Bertacca}, {Bruni}, and
  {Maselli}}]{Milillo:2015}
{Milillo} I, {Bertacca} D, {Bruni} M, {Maselli} A (2015) {Missing link: A
  nonlinear post-Friedmann framework for small and large scales}. \prd
  92(2):023519. \doi{10.1103/PhysRevD.92.023519}.
  {\href{https://arxiv.org/abs/1502.02985}{{arXiv:1502.02985}}} {[gr-qc]}

\bibitem[{{Miller}(1983)}]{Miller:1983}
{Miller} RH (1983) {Numerical experiments on the clustering of galaxies}. \apj
  270:390--409. \doi{10.1086/161133}

\bibitem[{{Miller} and {Prendergast}(1968)}]{Miller:1968}
{Miller} RH, {Prendergast} KH (1968) {Stellar Dynamics in a Discrete Phase
  Space}. \apj 151:699. \doi{10.1086/149469}

\bibitem[{{Mina} et~al.(2020){Mina}, {Mota}, and {Winther}}]{Mina:2020}
{Mina} M, {Mota} DF, {Winther} HA (2020) {SCALAR: an AMR code to simulate
  axion-like dark matter models}. \aap 641:A107.
  \doi{10.1051/0004-6361/201936272}.
  {\href{https://arxiv.org/abs/1906.12160}{{arXiv:1906.12160}}}
  {[physics.comp-ph]}

\bibitem[{{Miniati} and {Colella}(2007)}]{Miniati:2007}
{Miniati} F, {Colella} P (2007) {Block structured adaptive mesh and time
  refinement for hybrid, hyperbolic + N-body systems}. J Comput Phys
  227(1):400--430. \doi{10.1016/j.jcp.2007.07.035}.
  {\href{https://arxiv.org/abs/astro-ph/0608156}{{arXiv:astro-ph/0608156}}}
  {[astro-ph]}

\bibitem[{Misner et~al.(1973)Misner, Thorne, and Wheeler}]{Misner:1973}
Misner C, Thorne K, Wheeler J (1973) Gravitation. W. H. Freeman

\bibitem[{{Miyoshi} and {Kihara}(1975)}]{Miyoshi:1975}
{Miyoshi} K, {Kihara} T (1975) {Development of the correlation of galaxies in
  an expanding universe}. \pasj 27:333--346

\bibitem[{{Mocz} and {Succi}(2015)}]{Mocz:2015}
{Mocz} P, {Succi} S (2015) {Numerical solution of the nonlinear Schr{\"o}dinger
  equation using smoothed-particle hydrodynamics}. \pre 91(5):053304.
  \doi{10.1103/PhysRevE.91.053304}.
  {\href{https://arxiv.org/abs/1503.03869}{{arXiv:1503.03869}}}
  {[physics.comp-ph]}

\bibitem[{{Mocz} and {Succi}(2017)}]{MoczLattice:2017}
{Mocz} P, {Succi} S (2017) {Integer lattice dynamics for Vlasov-Poisson}.
  \mnras 465:3154--3162. \doi{10.1093/mnras/stw2928}.
  {\href{https://arxiv.org/abs/1611.02757}{{arXiv:1611.02757}}}

\bibitem[{{Mocz} et~al.(2017){Mocz}, {Vogelsberger}, {Robles}, {Zavala},
  {Boylan-Kolchin}, {Fialkov}, and {Hernquist}}]{Mocz:2017}
{Mocz} P, {Vogelsberger} M, {Robles} VH, {Zavala} J, {Boylan-Kolchin} M,
  {Fialkov} A, {Hernquist} L (2017) {Galaxy formation with BECDM - I.
  Turbulence and relaxation of idealized haloes}. \mnras 471(4):4559--4570.
  \doi{10.1093/mnras/stx1887}.
  {\href{https://arxiv.org/abs/1705.05845}{{arXiv:1705.05845}}} {[astro-ph.CO]}

\bibitem[{{Mocz} et~al.(2018){Mocz}, {Lancaster}, {Fialkov}, {Becerra}, and
  {Chavanis}}]{Mocz:2018}
{Mocz} P, {Lancaster} L, {Fialkov} A, {Becerra} F, {Chavanis} PH (2018)
  {Schr{\"o}dinger-Poisson-Vlasov-Poisson correspondence}. \prd 97(8):083519.
  \doi{10.1103/PhysRevD.97.083519}.
  {\href{https://arxiv.org/abs/1801.03507}{{arXiv:1801.03507}}}

\bibitem[{{Mocz} et~al.(2020){Mocz}, {Fialkov}, {Vogelsberger}, {Becerra},
  {Shen}, {Robles}, {Amin}, {Zavala}, {Boylan-Kolchin}, {Bose}, {Marinacci},
  {Chavanis}, {Lancaster}, and {Hernquist}}]{Mocz:2020}
{Mocz} P, {Fialkov} A, {Vogelsberger} M, {Becerra} F, {Shen} X, {Robles} VH,
  {Amin} MA, {Zavala} J, {Boylan-Kolchin} M, {Bose} S, {Marinacci} F,
  {Chavanis} PH, {Lancaster} L, {Hernquist} L (2020) {Galaxy formation with
  BECDM - II. Cosmic filaments and first galaxies}. \mnras 494(2):2027--2044.
  \doi{10.1093/mnras/staa738}.
  {\href{https://arxiv.org/abs/1911.05746}{{arXiv:1911.05746}}} {[astro-ph.CO]}

\bibitem[{{Modi} et~al.(2020{\natexlab{a}}){Modi}, {Chen}, and
  {White}}]{Modi:2020}
{Modi} C, {Chen} SF, {White} M (2020{\natexlab{a}}) {Simulations and
  symmetries}. \mnras 492(4):5754--5763. \doi{10.1093/mnras/staa251}.
  {\href{https://arxiv.org/abs/1910.07097}{{arXiv:1910.07097}}} {[astro-ph.CO]}

\bibitem[{{Modi} et~al.(2020{\natexlab{b}}){Modi}, {Lanusse}, and
  {Seljak}}]{ModiTensorFlow:2020}
{Modi} C, {Lanusse} F, {Seljak} U (2020{\natexlab{b}}) {FlowPM: Distributed
  TensorFlow Implementation of the FastPM Cosmological N-body Solver}. arXiv
  e-prints arXiv:2010.11847.
  {\href{https://arxiv.org/abs/2010.11847}{{arXiv:2010.11847}}} {[astro-ph.CO]}

\bibitem[{{Mohammed} et~al.(2014){Mohammed}, {Martizzi}, {Teyssier}, and
  {Amara}}]{Mohammed:2014}
{Mohammed} I, {Martizzi} D, {Teyssier} R, {Amara} A (2014) {Baryonic effects on
  weak-lensing two-point statistics and its cosmological implications}. arXiv
  e-prints {\href{https://arxiv.org/abs/1410.6826}{{arXiv:1410.6826}}}

\bibitem[{{Mohayaee} et~al.(2006){Mohayaee}, {Mathis}, {Colombi}, and
  {Silk}}]{Mohayaee:2006}
{Mohayaee} R, {Mathis} H, {Colombi} S, {Silk} J (2006) {Reconstruction of
  primordial density fields}. \mnras 365(3):939--959.
  \doi{10.1111/j.1365-2966.2005.09774.x}.
  {\href{https://arxiv.org/abs/astro-ph/0501217}{{arXiv:astro-ph/0501217}}}
  {[astro-ph]}

\bibitem[{{Monaco}(2016)}]{Monaco:2016}
{Monaco} P (2016) {Approximate Methods for the Generation of Dark Matter Halo
  Catalogs in the Age of Precision Cosmology}. Galaxies 4(4):53.
  \doi{10.3390/galaxies4040053}.
  {\href{https://arxiv.org/abs/1605.07752}{{arXiv:1605.07752}}} {[astro-ph.CO]}

\bibitem[{{Monaco} et~al.(2002){Monaco}, {Theuns}, and {Taffoni}}]{Monaco:2002}
{Monaco} P, {Theuns} T, {Taffoni} G (2002) {The pinocchio algorithm:
  pinpointing orbit-crossing collapsed hierarchical objects in a linear density
  field}. \mnras 331(3):587--608. \doi{10.1046/j.1365-8711.2002.05162.x}.
  {\href{https://arxiv.org/abs/astro-ph/0109323}{{arXiv:astro-ph/0109323}}}
  {[astro-ph]}

\bibitem[{{Monaco} et~al.(2013){Monaco}, {Sefusatti}, {Borgani}, {Crocce},
  {Fosalba}, {Sheth}, and {Theuns}}]{Monaco:2013}
{Monaco} P, {Sefusatti} E, {Borgani} S, {Crocce} M, {Fosalba} P, {Sheth} RK,
  {Theuns} T (2013) {An accurate tool for the fast generation of dark matter
  halo catalogues}. \mnras 433(3):2389--2402. \doi{10.1093/mnras/stt907}.
  {\href{https://arxiv.org/abs/1305.1505}{{arXiv:1305.1505}}} {[astro-ph.CO]}

\bibitem[{{Moore} et~al.(1996){Moore}, {Katz}, and {Lake}}]{Moore:1996}
{Moore} B, {Katz} N, {Lake} G (1996) {On the Destruction and Overmerging of
  Dark Halos in Dissipationless N-Body Simulations}. \apj 457:455.
  \doi{10.1086/176745}.
  {\href{https://arxiv.org/abs/astro-ph/9503088}{{arXiv:astro-ph/9503088}}}
  {[astro-ph]}

\bibitem[{{Moore} et~al.(1999){Moore}, {Ghigna}, {Governato}, {Lake}, {Quinn},
  {Stadel}, and {Tozzi}}]{Moore:1999}
{Moore} B, {Ghigna} S, {Governato} F, {Lake} G, {Quinn} T, {Stadel} J, {Tozzi}
  P (1999) {Dark Matter Substructure within Galactic Halos}. \apjl
  524(1):L19--L22. \doi{10.1086/312287}.
  {\href{https://arxiv.org/abs/astro-ph/9907411}{{arXiv:astro-ph/9907411}}}
  {[astro-ph]}

\bibitem[{{More} et~al.(2011){More}, {Kravtsov}, {Dalal}, and
  {Gottl{\"o}ber}}]{More:2011}
{More} S, {Kravtsov} AV, {Dalal} N, {Gottl{\"o}ber} S (2011) {The Overdensity
  and Masses of the Friends-of-friends Halos and Universality of Halo Mass
  Function}. \apjs 195(1):4. \doi{10.1088/0067-0049/195/1/4}.
  {\href{https://arxiv.org/abs/1103.0005}{{arXiv:1103.0005}}} {[astro-ph.CO]}

\bibitem[{{Moster} et~al.(2013){Moster}, {Naab}, and {White}}]{Moster:2013}
{Moster} BP, {Naab} T, {White} SDM (2013) {Galactic star formation and
  accretion histories from matching galaxies to dark matter haloes}. \mnras
  428(4):3121--3138. \doi{10.1093/mnras/sts261}.
  {\href{https://arxiv.org/abs/1205.5807}{{arXiv:1205.5807}}} {[astro-ph.CO]}

\bibitem[{{Moster} et~al.(2018){Moster}, {Naab}, and {White}}]{Moster:2018}
{Moster} BP, {Naab} T, {White} SDM (2018) {{EMERGE} -- an empirical model for
  the formation of galaxies since $z \sim 10$}. \mnras 477(2):1822--1852.
  \doi{10.1093/mnras/sty655}.
  {\href{https://arxiv.org/abs/1705.05373}{{arXiv:1705.05373}}} {[astro-ph.GA]}

\bibitem[{{Mueller} et~al.(2018){Mueller}, {Percival}, {Linder}, {Alam},
  {Zhao}, {S{\'a}nchez}, {Beutler}, and {Brinkmann}}]{Mueller:2018}
{Mueller} EM, {Percival} W, {Linder} E, {Alam} S, {Zhao} GB, {S{\'a}nchez} AG,
  {Beutler} F, {Brinkmann} J (2018) {The clustering of galaxies in the
  completed SDSS-III Baryon Oscillation Spectroscopic Survey: constraining
  modified gravity}. \mnras 475(2):2122--2131. \doi{10.1093/mnras/stx3232}.
  {\href{https://arxiv.org/abs/1612.00812}{{arXiv:1612.00812}}} {[astro-ph.CO]}

\bibitem[{{Mueller} et~al.(2021){Mueller}, {Rezaie}, {Percival}, {Ross},
  {Ruggeri}, {Seo}, {Gil-Mar{\i}n}, {Bautista}, {Brownstein}, {Dawson}, {de la
  Macorra}, {Palanque-Delabrouille}, {Rossi}, {Schneider}, and
  {Yeche}}]{Mueller:2021}
{Mueller} EM, {Rezaie} M, {Percival} WJ, {Ross} AJ, {Ruggeri} R, {Seo} HJ,
  {Gil-Mar{\i}n} H, {Bautista} J, {Brownstein} JR, {Dawson} K, {de la Macorra}
  A, {Palanque-Delabrouille} N, {Rossi} G, {Schneider} DP, {Yeche} C (2021)
  {The clustering of galaxies in the completed SDSS-IV extended Baryon
  Oscillation Spectroscopic Survey: Primordial non-Gaussianity in Fourier
  Space}. arXiv e-prints arXiv:2106.13725.
  {\href{https://arxiv.org/abs/2106.13725}{{arXiv:2106.13725}}} {[astro-ph.CO]}

\bibitem[{{Munshi} et~al.(1994){Munshi}, {Sahni}, and
  {Starobinsky}}]{Munshi:1994}
{Munshi} D, {Sahni} V, {Starobinsky} AA (1994) {Nonlinear approximations to
  gravitational instability: A comparison in the quasi-linear regime}. \apj
  436:517--527. \doi{10.1086/174925}.
  {\href{https://arxiv.org/abs/astro-ph/9402065}{{astro-ph/9402065}}}

\bibitem[{{Murgia} et~al.(2017){Murgia}, {Merle}, {Viel}, {Totzauer}, and
  {Schneider}}]{Murgia:2017}
{Murgia} R, {Merle} A, {Viel} M, {Totzauer} M, {Schneider} A (2017)
  {``Non-cold'' dark matter at small scales: a general approach}. \jcap
  2017(11):046. \doi{10.1088/1475-7516/2017/11/046}.
  {\href{https://arxiv.org/abs/1704.07838}{{arXiv:1704.07838}}} {[astro-ph.CO]}

\bibitem[{{Mustafa} et~al.(2019){Mustafa}, {Bard}, {Bhimji}, {Luki{\'c}},
  {Al-Rfou}, and {Kratochvil}}]{Mustafa:2019}
{Mustafa} M, {Bard} D, {Bhimji} W, {Luki{\'c}} Z, {Al-Rfou} R, {Kratochvil} JM
  (2019) {CosmoGAN: creating high-fidelity weak lensing convergence maps using
  Generative Adversarial Networks}. Computational Astrophysics and Cosmology
  6(1):1. \doi{10.1186/s40668-019-0029-9}.
  {\href{https://arxiv.org/abs/1706.02390}{{arXiv:1706.02390}}} {[astro-ph.IM]}

\bibitem[{{Myers} et~al.(2016){Myers}, {Colella}, and {Van
  Straalen}}]{Myers:2016}
{Myers} A, {Colella} P, {Van Straalen} B (2016) {The Convergence of
  Particle-in-Cell Schemes for Cosmological Dark Matter Simulations}. \apj
  816(2):56. \doi{10.3847/0004-637X/816/2/56}.
  {\href{https://arxiv.org/abs/1503.05969}{{arXiv:1503.05969}}} {[astro-ph.CO]}

\bibitem[{{Nadler} et~al.(2019){Nadler}, {Gluscevic}, {Boddy}, and
  {Wechsler}}]{Nadler:2019}
{Nadler} EO, {Gluscevic} V, {Boddy} KK, {Wechsler} RH (2019) {Constraints on
  Dark Matter Microphysics from the Milky Way Satellite Population}. arXiv
  e-prints arXiv:1904.10000.
  {\href{https://arxiv.org/abs/1904.10000}{{arXiv:1904.10000}}} {[astro-ph.CO]}

\bibitem[{{Nakagami} et~al.(2004){Nakagami}, {Matsubara}, {Schmalzing}, and
  {Jing}}]{Nakagami:2004}
{Nakagami} T, {Matsubara} T, {Schmalzing} J, {Jing} Y (2004) {An Analysis of
  the Large Scale N-body Simulation using the Minkowski Functionals}. arXiv
  e-prints astro-ph/0408428.
  {\href{https://arxiv.org/abs/astro-ph/0408428}{{arXiv:astro-ph/0408428}}}
  {[astro-ph]}

\bibitem[{{Navarro} et~al.(1996){Navarro}, {Eke}, and {Frenk}}]{Navarro:1996}
{Navarro} JF, {Eke} VR, {Frenk} CS (1996) {The cores of dwarf galaxy haloes}.
  \mnras 283(3):L72--L78. \doi{10.1093/mnras/283.3.L72}.
  {\href{https://arxiv.org/abs/astro-ph/9610187}{{arXiv:astro-ph/9610187}}}
  {[astro-ph]}

\bibitem[{{Navarro} et~al.(1997){Navarro}, {Frenk}, and {White}}]{Navarro:1997}
{Navarro} JF, {Frenk} CS, {White} SDM (1997) {A Universal Density Profile from
  Hierarchical Clustering}. \apj 490:493--508. \doi{10.1086/304888}.
  {\href{https://arxiv.org/abs/astro-ph/9611107}{{astro-ph/9611107}}}

\bibitem[{{Navarro} et~al.(2010){Navarro}, {Ludlow}, {Springel}, {Wang},
  {Vogelsberger}, {White}, {Jenkins}, {Frenk}, and {Helmi}}]{Navarro:2010}
{Navarro} JF, {Ludlow} A, {Springel} V, {Wang} J, {Vogelsberger} M, {White}
  SDM, {Jenkins} A, {Frenk} CS, {Helmi} A (2010) {The diversity and similarity
  of simulated cold dark matter haloes}. \mnras 402(1):21--34.
  \doi{10.1111/j.1365-2966.2009.15878.x}.
  {\href{https://arxiv.org/abs/0810.1522}{{arXiv:0810.1522}}} {[astro-ph]}

\bibitem[{{Neto} et~al.(2007){Neto}, {Gao}, {Bett}, {Cole}, {Navarro}, {Frenk},
  {White}, {Springel}, and {Jenkins}}]{Neto:2007}
{Neto} AF, {Gao} L, {Bett} P, {Cole} S, {Navarro} JF, {Frenk} CS, {White} SDM,
  {Springel} V, {Jenkins} A (2007) {The statistics of {\ensuremath{\Lambda}}
  CDM halo concentrations}. \mnras 381(4):1450--1462.
  \doi{10.1111/j.1365-2966.2007.12381.x}.
  {\href{https://arxiv.org/abs/0706.2919}{{arXiv:0706.2919}}} {[astro-ph]}

\bibitem[{{Newton} et~al.(2020){Newton}, {Leo}, {Cautun}, {Jenkins}, {Frenk},
  {Lovell}, {Helly}, and {Benson}}]{Newton:2020}
{Newton} O, {Leo} M, {Cautun} M, {Jenkins} A, {Frenk} CS, {Lovell} MR, {Helly}
  JC, {Benson} AJ (2020) {Constraints on the properties of warm dark matter
  using the satellite galaxies of the Milky Way}. arXiv e-prints
  arXiv:2011.08865.
  {\href{https://arxiv.org/abs/2011.08865}{{arXiv:2011.08865}}} {[astro-ph.CO]}

\bibitem[{{Neyrinck}(2008)}]{Neyrinck:2008}
{Neyrinck} MC (2008) {ZOBOV: a parameter-free void-finding algorithm}. \mnras
  386(4):2101--2109. \doi{10.1111/j.1365-2966.2008.13180.x}.
  {\href{https://arxiv.org/abs/0712.3049}{{arXiv:0712.3049}}} {[astro-ph]}

\bibitem[{{Neyrinck}(2016)}]{Neyrinck:2016}
{Neyrinck} MC (2016) {Truthing the stretch: non-perturbative cosmological
  realizations with multiscale spherical collapse}. \mnras 455(1):L11--L15.
  \doi{10.1093/mnrasl/slv141}.
  {\href{https://arxiv.org/abs/1503.07534}{{arXiv:1503.07534}}} {[astro-ph.CO]}

\bibitem[{{Neyrinck} et~al.(2005){Neyrinck}, {Gnedin}, and
  {Hamilton}}]{Neyrinck:2005}
{Neyrinck} MC, {Gnedin} NY, {Hamilton} AJS (2005) {VOBOZ: an
  almost-parameter-free halo-finding algorithm}. \mnras 356(4):1222--1232.
  \doi{10.1111/j.1365-2966.2004.08505.x}.
  {\href{https://arxiv.org/abs/astro-ph/0402346}{{arXiv:astro-ph/0402346}}}
  {[astro-ph]}

\bibitem[{{Neyrinck} et~al.(2018){Neyrinck}, {Szapudi}, {McCullagh}, {Szalay},
  {Falck}, and {Wang}}]{Neyrinck:2018}
{Neyrinck} MC, {Szapudi} I, {McCullagh} N, {Szalay} AS, {Falck} B, {Wang} J
  (2018) {Density-dependent clustering - I. Pullingback the curtains on motions
  of the BAO peak}. \mnras 478(2):2495--2504. \doi{10.1093/mnras/sty1074}.
  {\href{https://arxiv.org/abs/1610.06215}{{arXiv:1610.06215}}} {[astro-ph.CO]}

\bibitem[{{Ni} et~al.(2021){Ni}, {Li}, {Lachance}, {Croft}, {Di Matteo},
  {Bird}, and {Feng}}]{Ni:2021}
{Ni} Y, {Li} Y, {Lachance} P, {Croft} RAC, {Di Matteo} T, {Bird} S, {Feng} Y
  (2021) {AI-assisted superresolution cosmological simulations - II. Halo
  substructures, velocities, and higher order statistics}. \mnras
  507(1):1021--1033. \doi{10.1093/mnras/stab2113}.
  {\href{https://arxiv.org/abs/2105.01016}{{arXiv:2105.01016}}} {[astro-ph.CO]}

\bibitem[{{Nishimichi} et~al.(2019){Nishimichi}, {Takada}, {Takahashi},
  {Osato}, {Shirasaki}, {Oogi}, {Miyatake}, {Oguri}, {Murata}, {Kobayashi}, and
  {Yoshida}}]{Nishimichi:2019}
{Nishimichi} T, {Takada} M, {Takahashi} R, {Osato} K, {Shirasaki} M, {Oogi} T,
  {Miyatake} H, {Oguri} M, {Murata} R, {Kobayashi} Y, {Yoshida} N (2019) {Dark
  Quest. I. Fast and Accurate Emulation of Halo Clustering Statistics and Its
  Application to Galaxy Clustering}. \apj 884(1):29.
  \doi{10.3847/1538-4357/ab3719}.
  {\href{https://arxiv.org/abs/1811.09504}{{arXiv:1811.09504}}} {[astro-ph.CO]}

\bibitem[{{Nori} and {Baldi}(2018)}]{Nori:2018}
{Nori} M, {Baldi} M (2018) {AX-GADGET: a new code for cosmological simulations
  of Fuzzy Dark Matter and Axion models}. \mnras 478(3):3935--3951.
  \doi{10.1093/mnras/sty1224}.
  {\href{https://arxiv.org/abs/1801.08144}{{arXiv:1801.08144}}} {[astro-ph.CO]}

\bibitem[{{Nori} et~al.(2019){Nori}, {Murgia}, {Ir{\v{s}}i{\v{c}}}, {Baldi},
  and {Viel}}]{Nori:2019}
{Nori} M, {Murgia} R, {Ir{\v{s}}i{\v{c}}} V, {Baldi} M, {Viel} M (2019) {Lyman
  {\ensuremath{\alpha}} forest and non-linear structure characterization in
  Fuzzy Dark Matter cosmologies}. \mnras 482(3):3227--3243.
  \doi{10.1093/mnras/sty2888}.
  {\href{https://arxiv.org/abs/1809.09619}{{arXiv:1809.09619}}} {[astro-ph.CO]}

\bibitem[{{Ntampaka} et~al.(2020){Ntampaka}, {Eisenstein}, {Yuan}, and
  {Garrison}}]{Ntampaka:2020}
{Ntampaka} M, {Eisenstein} DJ, {Yuan} S, {Garrison} LH (2020) {A Hybrid Deep
  Learning Approach to Cosmological Constraints from Galaxy Redshift Surveys}.
  \apj 889(2):151. \doi{10.3847/1538-4357/ab5f5e}.
  {\href{https://arxiv.org/abs/1909.10527}{{arXiv:1909.10527}}} {[astro-ph.CO]}

\bibitem[{{Nuza} et~al.(2013){Nuza}, {S{\'a}nchez}, {Prada}, {Klypin},
  {Schlegel}, {Gottl{\"o}ber}, {Montero-Dorta}, {Manera}, {McBride}, {Ross},
  {Angulo}, {Blanton}, {Bolton}, {Favole}, {Samushia}, {Montesano}, {Percival},
  {Padmanabhan}, {Steinmetz}, {Tinker}, {Skibba}, {Schneider}, {Guo}, {Zehavi},
  {Zheng}, {Bizyaev}, {Malanushenko}, {Malanushenko}, {Oravetz}, {Oravetz}, and
  {Shelden}}]{Nuza:2013}
{Nuza} SE, {S{\'a}nchez} AG, {Prada} F, {Klypin} A, {Schlegel} DJ,
  {Gottl{\"o}ber} S, {Montero-Dorta} AD, {Manera} M, {McBride} CK, {Ross} AJ,
  {Angulo} R, {Blanton} M, {Bolton} A, {Favole} G, {Samushia} L, {Montesano} F,
  {Percival} WJ, {Padmanabhan} N, {Steinmetz} M, {Tinker} J, {Skibba} R,
  {Schneider} DP, {Guo} H, {Zehavi} I, {Zheng} Z, {Bizyaev} D, {Malanushenko}
  O, {Malanushenko} V, {Oravetz} AE, {Oravetz} DJ, {Shelden} AC (2013) {The
  clustering of galaxies at z {\ensuremath{\approx}} 0.5 in the SDSS-III Data
  Release 9 BOSS-CMASS sample: a test for the {\ensuremath{\Lambda}}CDM
  cosmology}. \mnras 432(1):743--760. \doi{10.1093/mnras/stt513}.
  {\href{https://arxiv.org/abs/1202.6057}{{arXiv:1202.6057}}} {[astro-ph.CO]}

\bibitem[{{Ogiya} and {Hahn}(2018)}]{Ogiya:2018}
{Ogiya} G, {Hahn} O (2018) {What sets the central structure of dark matter
  haloes?} \mnras 473(4):4339--4359. \doi{10.1093/mnras/stx2639}.
  {\href{https://arxiv.org/abs/1707.07693}{{arXiv:1707.07693}}} {[astro-ph.CO]}

\bibitem[{{Ogiya} et~al.(2016){Ogiya}, {Nagai}, and {Ishiyama}}]{Ogiya:2016}
{Ogiya} G, {Nagai} D, {Ishiyama} T (2016) {Dynamical evolution of primordial
  dark matter haloes through mergers}. \mnras 461(3):3385--3396.
  \doi{10.1093/mnras/stw1551}.
  {\href{https://arxiv.org/abs/1604.02866}{{arXiv:1604.02866}}} {[astro-ph.CO]}

\bibitem[{{Ogiya} et~al.(2019){Ogiya}, {van den Bosch}, {Hahn}, {Green},
  {Miller}, and {Burkert}}]{Ogiya:2019}
{Ogiya} G, {van den Bosch} FC, {Hahn} O, {Green} SB, {Miller} TB, {Burkert} A
  (2019) {{DASH}: a library of dynamical subhalo evolution}. \mnras
  485(1):189--202. \doi{10.1093/mnras/stz375}.
  {\href{https://arxiv.org/abs/1901.08601}{{arXiv:1901.08601}}} {[astro-ph.GA]}

\bibitem[{{O'Hare} and {Green}(2017)}]{OHare:2017}
{O'Hare} CAJ, {Green} AM (2017) {Axion astronomy with microwave cavity
  experiments}. \prd 95(6):063017. \doi{10.1103/PhysRevD.95.063017}.
  {\href{https://arxiv.org/abs/1701.03118}{{arXiv:1701.03118}}} {[astro-ph.CO]}

\bibitem[{{O'Leary} and {McQuinn}(2012)}]{OLeary:2012}
{O'Leary} RM, {McQuinn} M (2012) {The Formation of the First Cosmic Structures
  and the Physics of the $z \sim 20$ Universe}. \apj 760(1):4.
  \doi{10.1088/0004-637X/760/1/4}.
  {\href{https://arxiv.org/abs/1204.1344}{{arXiv:1204.1344}}} {[astro-ph.CO]}

\bibitem[{{Oman} et~al.(2015){Oman}, {Navarro}, {Fattahi}, {Frenk}, {Sawala},
  {White}, {Bower}, {Crain}, {Furlong}, {Schaller}, {Schaye}, and
  {Theuns}}]{Oman:2015}
{Oman} KA, {Navarro} JF, {Fattahi} A, {Frenk} CS, {Sawala} T, {White} SDM,
  {Bower} R, {Crain} RA, {Furlong} M, {Schaller} M, {Schaye} J, {Theuns} T
  (2015) {The unexpected diversity of dwarf galaxy rotation curves}. \mnras
  452(4):3650--3665. \doi{10.1093/mnras/stv1504}.
  {\href{https://arxiv.org/abs/1504.01437}{{arXiv:1504.01437}}} {[astro-ph.GA]}

\bibitem[{{Ondaro-Mallea} et~al.(2021){Ondaro-Mallea}, {Angulo}, {Zennaro},
  {Contreras}, and {Aric{\`o}}}]{Ondaro:2021}
{Ondaro-Mallea} L, {Angulo} RE, {Zennaro} M, {Contreras} S, {Aric{\`o}} G
  (2021) {Non-universality of the mass function: dependence on the growth rate
  and power spectrum shape}. arXiv e-prints arXiv:2102.08958.
  {\href{https://arxiv.org/abs/2102.08958}{{arXiv:2102.08958}}} {[astro-ph.CO]}

\bibitem[{{Onions} et~al.(2012){Onions}, {Knebe}, {Pearce}, {Muldrew}, {Lux},
  {Knollmann}, {Ascasibar}, {Behroozi}, {Elahi}, {Han}, {Maciejewski},
  {Merch{\'a}n}, {Neyrinck}, {Ruiz}, {Sgr{\'o}}, {Springel}, and
  {Tweed}}]{Onions:2012}
{Onions} J, {Knebe} A, {Pearce} FR, {Muldrew} SI, {Lux} H, {Knollmann} SR,
  {Ascasibar} Y, {Behroozi} P, {Elahi} P, {Han} J, {Maciejewski} M,
  {Merch{\'a}n} ME, {Neyrinck} M, {Ruiz} AN, {Sgr{\'o}} MA, {Springel} V,
  {Tweed} D (2012) {Subhaloes going Notts: the subhalo-finder comparison
  project}. \mnras 423(2):1200--1214. \doi{10.1111/j.1365-2966.2012.20947.x}.
  {\href{https://arxiv.org/abs/1203.3695}{{arXiv:1203.3695}}} {[astro-ph.CO]}

\bibitem[{{Orban}(2013)}]{Orban:2013}
{Orban} C (2013) {Keeping it real: revisiting a real-space approach to running
  ensembles of cosmological N-body simulations}. \jcap 2013(5):032.
  \doi{10.1088/1475-7516/2013/05/032}.
  {\href{https://arxiv.org/abs/1201.2082}{{arXiv:1201.2082}}} {[astro-ph.CO]}

\bibitem[{{Orsi} and {Angulo}(2018)}]{Orsi:2018}
{Orsi} {\'A}A, {Angulo} RE (2018) {The impact of galaxy formation on satellite
  kinematics and redshift-space distortions}. \mnras 475(2):2530--2544.
  \doi{10.1093/mnras/stx3349}.
  {\href{https://arxiv.org/abs/1708.00956}{{arXiv:1708.00956}}} {[astro-ph.CO]}

\bibitem[{{Orszag}(1971)}]{Orszag:1971}
{Orszag} SA (1971) {On the Elimination of Aliasing in Finite-Difference Schemes
  by Filtering High-Wavenumber Components.} J Atmosph Sci 28:1074--1074.
  \doi{10.1175/1520-0469(1971)028<1074:OTEOAI>2.0.CO;2}

\bibitem[{{O'Shea} et~al.(2005){O'Shea}, {Nagamine}, {Springel}, {Hernquist},
  and {Norman}}]{OShea:2005}
{O'Shea} BW, {Nagamine} K, {Springel} V, {Hernquist} L, {Norman} ML (2005)
  {Comparing AMR and SPH Cosmological Simulations. I. Dark Matter and Adiabatic
  Simulations}. \apjs 160(1):1--27. \doi{10.1086/432645}.
  {\href{https://arxiv.org/abs/astro-ph/0312651}{{arXiv:astro-ph/0312651}}}
  {[astro-ph]}

\bibitem[{{Oteo} and {Ros}(1991)}]{Oteo:1991}
{Oteo} JA, {Ros} J (1991) {The Magnus expansion for classical Hamiltonian
  systems}. J Phys A 24:5751--5762. \doi{10.1088/0305-4470/24/24/011}

\bibitem[{{Oyaizu}(2008)}]{Oyaizu:2008}
{Oyaizu} H (2008) {Nonlinear evolution of f(R) cosmologies. I. Methodology}.
  \prd 78(12):123523. \doi{10.1103/PhysRevD.78.123523}.
  {\href{https://arxiv.org/abs/0807.2449}{{arXiv:0807.2449}}} {[astro-ph]}

\bibitem[{{Padilla} et~al.(2005){Padilla}, {Ceccarelli}, and
  {Lambas}}]{Padilla:2005}
{Padilla} ND, {Ceccarelli} L, {Lambas} DG (2005) {Spatial and dynamical
  properties of voids in a {\ensuremath{\Lambda}} cold dark matter universe}.
  \mnras 363(3):977--990. \doi{10.1111/j.1365-2966.2005.09500.x}.
  {\href{https://arxiv.org/abs/astro-ph/0508297}{{arXiv:astro-ph/0508297}}}
  {[astro-ph]}

\bibitem[{{Paillas} et~al.(2017){Paillas}, {Lagos}, {Padilla}, {Tissera},
  {Helly}, and {Schaller}}]{Paillas:2017}
{Paillas} E, {Lagos} CDP, {Padilla} N, {Tissera} P, {Helly} J, {Schaller} M
  (2017) {Baryon effects on void statistics in the EAGLE simulation}. \mnras
  470(4):4434--4452. \doi{10.1093/mnras/stx1514}.
  {\href{https://arxiv.org/abs/1609.00101}{{arXiv:1609.00101}}} {[astro-ph.CO]}

\bibitem[{{Paillas} et~al.(2019){Paillas}, {Cautun}, {Li}, {Cai}, {Padilla},
  {Armijo}, and {Bose}}]{Paillas:2019}
{Paillas} E, {Cautun} M, {Li} B, {Cai} YC, {Padilla} N, {Armijo} J, {Bose} S
  (2019) {The Santiago-Harvard-Edinburgh-Durham void comparison II: unveiling
  the Vainshtein screening using weak lensing}. \mnras 484(1):1149--1165.
  \doi{10.1093/mnras/stz022}.
  {\href{https://arxiv.org/abs/1810.02864}{{arXiv:1810.02864}}} {[astro-ph.CO]}

\bibitem[{{Paillas} et~al.(2021){Paillas}, {Cai}, {Padilla}, and
  {S{\'a}nchez}}]{Paillas:2021}
{Paillas} E, {Cai} YC, {Padilla} N, {S{\'a}nchez} A (2021) {Redshift-space
  distortions with split densities}. arXiv e-prints arXiv:2101.09854.
  {\href{https://arxiv.org/abs/2101.09854}{{arXiv:2101.09854}}} {[astro-ph.CO]}

\bibitem[{{Palanque-Delabrouille} et~al.(2019){Palanque-Delabrouille},
  {Y{\`e}che}, {Sch{\"o}neberg}, {Lesgourgues}, {Walther}, {Chabanier}, and
  {Armengaud}}]{Palanque:2019}
{Palanque-Delabrouille} N, {Y{\`e}che} C, {Sch{\"o}neberg} N, {Lesgourgues} J,
  {Walther} M, {Chabanier} S, {Armengaud} E (2019) {Hints, neutrino bounds and
  WDM constraints from SDSS DR14 Lyman-$\alpha$ and Planck full-survey data}.
  arXiv e-prints arXiv:1911.09073.
  {\href{https://arxiv.org/abs/1911.09073}{{arXiv:1911.09073}}} {[astro-ph.CO]}

\bibitem[{{Pan} et~al.(2020){Pan}, {Liu}, {Forero-Romero}, {Sabiu}, {Li},
  {Miao}, and {Li}}]{Pan:2020}
{Pan} S, {Liu} M, {Forero-Romero} J, {Sabiu} CG, {Li} Z, {Miao} H, {Li} XD
  (2020) {Cosmological parameter estimation from large-scale structure deep
  learning}. Science China Physics, Mechanics, and Astronomy 63(11):110412.
  \doi{10.1007/s11433-020-1586-3}.
  {\href{https://arxiv.org/abs/1908.10590}{{arXiv:1908.10590}}} {[astro-ph.CO]}

\bibitem[{{Papastergis} et~al.(2011){Papastergis}, {Martin}, {Giovanelli}, and
  {Haynes}}]{Papastergis:2011}
{Papastergis} E, {Martin} AM, {Giovanelli} R, {Haynes} MP (2011) {The Velocity
  Width Function of Galaxies from the 40\% ALFALFA Survey: Shedding Light on
  the Cold Dark Matter Overabundance Problem}. \apj 739(1):38.
  \doi{10.1088/0004-637X/739/1/38}.
  {\href{https://arxiv.org/abs/1106.0710}{{arXiv:1106.0710}}} {[astro-ph.CO]}

\bibitem[{{Papastergis} et~al.(2015){Papastergis}, {Giovanelli}, {Haynes}, and
  {Shankar}}]{Papastergis:2015}
{Papastergis} E, {Giovanelli} R, {Haynes} MP, {Shankar} F (2015) {Is there a
  ``too big to fail'' problem in the field?} \aap 574:A113.
  \doi{10.1051/0004-6361/201424909}.
  {\href{https://arxiv.org/abs/1407.4665}{{arXiv:1407.4665}}} {[astro-ph.GA]}

\bibitem[{{Paranjape}(2009)}]{Paranjape:2009}
{Paranjape} A (2009) {The Averaging Problem in Cosmology}. PhD thesis, Tata
  Institute of Fundamental Research, Mumbai, India

\bibitem[{{Paranjape} and {Padmanabhan}(2017)}]{Paranjape:2017}
{Paranjape} A, {Padmanabhan} N (2017) {Halo assembly bias from Separate
  Universe simulations}. \mnras 468(3):2984--2999. \doi{10.1093/mnras/stx659}.
  {\href{https://arxiv.org/abs/1612.02833}{{arXiv:1612.02833}}} {[astro-ph.CO]}

\bibitem[{{Paranjape} et~al.(2018){Paranjape}, {Hahn}, and
  {Sheth}}]{Paranjape:2018}
{Paranjape} A, {Hahn} O, {Sheth} RK (2018) {Halo assembly bias and the tidal
  anisotropy of the local halo environment}. \mnras 476(3):3631--3647.
  \doi{10.1093/mnras/sty496}.
  {\href{https://arxiv.org/abs/1706.09906}{{arXiv:1706.09906}}} {[astro-ph.CO]}

\bibitem[{{Park} et~al.(2020){Park}, {Ahn}, {Yoshida}, and
  {Hirano}}]{Park:2020}
{Park} H, {Ahn} K, {Yoshida} N, {Hirano} S (2020) {First Structure Formation
  under the Influence of Gas-Dark Matter Streaming Velocity and Density: Impact
  of the ``Baryons Trace Dark Matter'' Approximation}. \apj 900(1):30.
  \doi{10.3847/1538-4357/aba26e}.
  {\href{https://arxiv.org/abs/2004.00863}{{arXiv:2004.00863}}} {[astro-ph.CO]}

\bibitem[{{Partmann} et~al.(2020){Partmann}, {Fidler}, {Rampf}, and
  {Hahn}}]{Partmann:2020}
{Partmann} C, {Fidler} C, {Rampf} C, {Hahn} O (2020) {Fast simulations of
  cosmic large-scale structure with massive neutrinos}. \jcap 2020(9):018.
  \doi{10.1088/1475-7516/2020/09/018}.
  {\href{https://arxiv.org/abs/2003.07387}{{arXiv:2003.07387}}} {[astro-ph.CO]}

\bibitem[{{Pearce} and {Couchman}(1997)}]{Pearce:1997}
{Pearce} FR, {Couchman} HMP (1997) {Hydra: a parallel adaptive grid code}. \na
  2(5):411--427. \doi{10.1016/S1384-1076(97)00025-0}.
  {\href{https://arxiv.org/abs/astro-ph/9703183}{{arXiv:astro-ph/9703183}}}
  {[astro-ph]}

\bibitem[{{Peebles}(1971)}]{Peebles:1971}
{Peebles} PJE (1971) {Rotation of Galaxies and the Gravitational Instability
  Picture}. \aap 11:377

\bibitem[{{Peebles}(1980)}]{Peebles:1980}
{Peebles} PJE (1980) {The large-scale structure of the universe}. Princeton
  University Press, Princeton, N.J.

\bibitem[{{Peebles} et~al.(1989){Peebles}, {Melott}, {Holmes}, and
  {Jiang}}]{Peebles:1989}
{Peebles} PJE, {Melott} AL, {Holmes} MR, {Jiang} LR (1989) {A model for the
  formation of the Local Group}. \apj 345:108--121. \doi{10.1086/167885}

\bibitem[{{Peel} et~al.(2019){Peel}, {Lalande}, {Starck}, {Pettorino},
  {Merten}, {Giocoli}, {Meneghetti}, and {Baldi}}]{Peel:2019}
{Peel} A, {Lalande} F, {Starck} JL, {Pettorino} V, {Merten} J, {Giocoli} C,
  {Meneghetti} M, {Baldi} M (2019) {Distinguishing standard and modified
  gravity cosmologies with machine learning}. \prd 100(2):023508.
  \doi{10.1103/PhysRevD.100.023508}.
  {\href{https://arxiv.org/abs/1810.11030}{{arXiv:1810.11030}}} {[astro-ph.CO]}

\bibitem[{{Pellejero-Iba{\~n}ez} et~al.(2020){Pellejero-Iba{\~n}ez}, {Angulo},
  {Aric{\'o}}, {Zennaro}, {Contreras}, and
  {St{\"u}cker}}]{Pellejero-Ibanez:2020}
{Pellejero-Iba{\~n}ez} M, {Angulo} RE, {Aric{\'o}} G, {Zennaro} M, {Contreras}
  S, {St{\"u}cker} J (2020) {Cosmological parameter estimation via iterative
  emulation of likelihoods}. \mnras 499(4):5257--5268.
  \doi{10.1093/mnras/staa3075}.
  {\href{https://arxiv.org/abs/1912.08806}{{arXiv:1912.08806}}} {[astro-ph.CO]}

\bibitem[{{Pellejero-Ibanez} et~al.(2021){Pellejero-Ibanez}, {Stuecker},
  {Angulo}, {Zennaro}, {Contreras}, and {Arico}}]{Pellejero-Ibanez:2021}
{Pellejero-Ibanez} M, {Stuecker} J, {Angulo} RE, {Zennaro} M, {Contreras} S,
  {Arico} G (2021) {Modelling galaxy clustering in redshift space with a
  Lagrangian bias formalism and $N$-body simulations}. arXiv e-prints
  arXiv:2109.08699.
  {\href{https://arxiv.org/abs/2109.08699}{{arXiv:2109.08699}}} {[astro-ph.CO]}

\bibitem[{{Pen}(1997)}]{Pen:1997}
{Pen} UL (1997) {Generating Cosmological Gaussian Random Fields}. \apjl
  490(2):L127--L130. \doi{10.1086/311042}.
  {\href{https://arxiv.org/abs/astro-ph/9709261}{{arXiv:astro-ph/9709261}}}
  {[astro-ph]}

\bibitem[{{Perraudin} et~al.(2019){Perraudin}, {Srivastava}, {Lucchi},
  {Kacprzak}, {Hofmann}, and {R{\'e}fr{\'e}gier}}]{Perraudin:2019}
{Perraudin} N, {Srivastava} A, {Lucchi} A, {Kacprzak} T, {Hofmann} T,
  {R{\'e}fr{\'e}gier} A (2019) {Cosmological N-body simulations: a challenge
  for scalable generative models}. Computational Astrophysics and Cosmology
  6(1):5. \doi{10.1186/s40668-019-0032-1}.
  {\href{https://arxiv.org/abs/1908.05519}{{arXiv:1908.05519}}}
  {[physics.comp-ph]}

\bibitem[{{Perraudin} et~al.(2020){Perraudin}, {Marcon}, {Lucchi}, and
  {Kacprzak}}]{Perraudin:2020}
{Perraudin} N, {Marcon} S, {Lucchi} A, {Kacprzak} T (2020) {Emulation of
  cosmological mass maps with conditional generative adversarial networks}.
  arXiv e-prints arXiv:2004.08139.
  {\href{https://arxiv.org/abs/2004.08139}{{arXiv:2004.08139}}} {[astro-ph.CO]}

\bibitem[{Petri et~al.(2015)Petri, Liu, Haiman, May, Hui, and
  Kratochvil}]{Petri:2015}
Petri A, Liu J, Haiman Z, May M, Hui L, Kratochvil JM (2015) {Emulating the
  CFHTLenS Weak Lensing data: Cosmological Constraints from moments and
  Minkowski functionals}. Phys Rev D91(10):103511.
  \doi{10.1103/PhysRevD.91.103511}.
  {\href{https://arxiv.org/abs/1503.06214}{{arXiv:1503.06214}}} {[astro-ph.CO]}

\bibitem[{{Philcox}(2021)}]{Philcox:2021}
{Philcox} OHE (2021) {A faster Fourier transform? Computing small-scale power
  spectra and bispectra for cosmological simulations in $\mathcal{O}(N^2)$
  time}. \mnras 501(3):4004--4034. \doi{10.1093/mnras/staa3882}.
  {\href{https://arxiv.org/abs/2005.01739}{{arXiv:2005.01739}}} {[astro-ph.CO]}

\bibitem[{{Philcox} and {Eisenstein}(2020)}]{Philcox:2020}
{Philcox} OHE, {Eisenstein} DJ (2020) {Computing the small-scale galaxy power
  spectrum and bispectrum in configuration space}. \mnras 492(1):1214--1242.
  \doi{10.1093/mnras/stz3335}.
  {\href{https://arxiv.org/abs/1912.01010}{{arXiv:1912.01010}}} {[astro-ph.CO]}

\bibitem[{{Pichon} and {Bernardeau}(1999)}]{Pichon:1999}
{Pichon} C, {Bernardeau} F (1999) {Vorticity generation in large-scale
  structure caustics}. \aap 343:663--681.
  {\href{https://arxiv.org/abs/astro-ph/9902142}{{arXiv:astro-ph/9902142}}}
  {[astro-ph]}

\bibitem[{{Pietroni}(2018)}]{Pietroni:2018}
{Pietroni} M (2018) {Structure formation beyond shell-crossing: nonperturbative
  expansions and late-time attractors}. \jcap 2018(6):028.
  \doi{10.1088/1475-7516/2018/06/028}.
  {\href{https://arxiv.org/abs/1804.09140}{{arXiv:1804.09140}}} {[astro-ph.CO]}

\bibitem[{{Pillepich} et~al.(2010){Pillepich}, {Porciani}, and
  {Hahn}}]{Pillepich:2010}
{Pillepich} A, {Porciani} C, {Hahn} O (2010) {Halo mass function and
  scale-dependent bias from N-body simulations with non-Gaussian initial
  conditions}. \mnras 402:191--206. \doi{10.1111/j.1365-2966.2009.15914.x}.
  {\href{https://arxiv.org/abs/0811.4176}{{arXiv:0811.4176}}}

\bibitem[{{Planck Collaboration}(2020)}]{PlanckfNL:2020}
{Planck Collaboration} (2020) {Planck 2018 results. IX. Constraints on
  primordial non-Gaussianity}. \aap 641:A9. \doi{10.1051/0004-6361/201935891}.
  {\href{https://arxiv.org/abs/1905.05697}{{arXiv:1905.05697}}} {[astro-ph.CO]}

\bibitem[{{Planck Collaboration} et~al.(2016){Planck Collaboration}, {Ade},
  {Aghanim}, {Arnaud}, {Arroja}, {Ashdown}, {Aumont}, {Baccigalupi},
  {Ballardini}, {Banday}, and et~al.}]{PlanckfNL:2016}
{Planck Collaboration}, {Ade} PAR, {Aghanim} N, {Arnaud} M, {Arroja} F,
  {Ashdown} M, {Aumont} J, {Baccigalupi} C, {Ballardini} M, {Banday} AJ, et~al
  (2016) {Planck 2015 results. XVII. Constraints on primordial
  non-Gaussianity}. \aap 594:A17. \doi{10.1051/0004-6361/201525836}.
  {\href{https://arxiv.org/abs/1502.01592}{{arXiv:1502.01592}}} {[astro-ph.CO]}

\bibitem[{{Planck Collaboration} et~al.(2020){Planck Collaboration}, {Aghanim},
  {Akrami}, {Ashdown}, {Aumont}, {Baccigalupi}, {Ballardini}, {Banday},
  {Barreiro}, {Bartolo}, {Basak}, {Battye}, {Benabed}, {Bernard}, {Bersanelli},
  {Bielewicz}, {Bock}, {Bond}, {Borrill}, {Bouchet}, {Boulanger}, {Bucher},
  {Burigana}, {Butler}, {Calabrese}, {Cardoso}, {Carron}, {Challinor},
  {Chiang}, {Chluba}, {Colombo}, {Combet}, {Contreras}, {Crill}, {Cuttaia}, {de
  Bernardis}, {de Zotti}, {Delabrouille}, {Delouis}, {Di Valentino}, {Diego},
  {Dor{\'e}}, {Douspis}, {Ducout}, {Dupac}, {Dusini}, {Efstathiou}, {Elsner},
  {En{\ss}lin}, {Eriksen}, {Fantaye}, {Farhang}, {Fergusson},
  {Fernandez-Cobos}, {Finelli}, {Forastieri}, {Frailis}, {Fraisse},
  {Franceschi}, {Frolov}, {Galeotta}, {Galli}, {Ganga}, {G{\'e}nova-Santos},
  {Gerbino}, {Ghosh}, {Gonz{\'a}lez-Nuevo}, {G{\'o}rski}, {Gratton},
  {Gruppuso}, {Gudmundsson}, {Hamann}, {Handley}, {Hansen}, {Herranz},
  {Hildebrandt}, {Hivon}, {Huang}, {Jaffe}, {Jones}, {Karakci}, {Keih{\"a}nen},
  {Keskitalo}, {Kiiveri}, {Kim}, {Kisner}, {Knox}, {Krachmalnicoff}, {Kunz},
  {Kurki-Suonio}, {Lagache}, {Lamarre}, {Lasenby}, {Lattanzi}, {Lawrence}, {Le
  Jeune}, {Lemos}, {Lesgourgues}, {Levrier}, {Lewis}, {Liguori}, {Lilje},
  {Lilley}, {Lindholm}, {L{\'o}pez-Caniego}, {Lubin}, {Ma},
  {Mac{\'\i}as-P{\'e}rez}, {Maggio}, {Maino}, {Mandolesi}, {Mangilli},
  {Marcos-Caballero}, {Maris}, {Martin}, {Martinelli},
  {Mart{\'\i}nez-Gonz{\'a}lez}, {Matarrese}, {Mauri}, {McEwen}, {Meinhold},
  {Melchiorri}, {Mennella}, {Migliaccio}, {Millea}, {Mitra},
  {Miville-Desch{\^e}nes}, {Molinari}, {Montier}, {Morgante}, {Moss}, {Natoli},
  {N{\o}rgaard-Nielsen}, {Pagano}, {Paoletti}, {Partridge}, {Patanchon},
  {Peiris}, {Perrotta}, {Pettorino}, {Piacentini}, {Polastri}, {Polenta},
  {Puget}, {Rachen}, {Reinecke}, {Remazeilles}, {Renzi}, {Rocha}, {Rosset},
  {Roudier}, {Rubi{\~n}o-Mart{\'\i}n}, {Ruiz-Granados}, {Salvati}, {Sandri},
  {Savelainen}, {Scott}, {Shellard}, {Sirignano}, {Sirri}, {Spencer},
  {Sunyaev}, {Suur-Uski}, {Tauber}, {Tavagnacco}, {Tenti}, {Toffolatti},
  {Tomasi}, {Trombetti}, {Valenziano}, {Valiviita}, {Van Tent}, {Vibert},
  {Vielva}, {Villa}, {Vittorio}, {Wandelt}, {Wehus}, {White}, {White},
  {Zacchei}, and {Zonca}}]{Planck:2020}
{Planck Collaboration}, {Aghanim} N, {Akrami} Y, {Ashdown} M, {Aumont} J,
  {Baccigalupi} C, {Ballardini} M, {Banday} AJ, {Barreiro} RB, {Bartolo} N,
  {Basak} S, {Battye} R, {Benabed} K, {Bernard} JP, {Bersanelli} M, {Bielewicz}
  P, {Bock} JJ, {Bond} JR, {Borrill} J, {Bouchet} FR, {Boulanger} F, {Bucher}
  M, {Burigana} C, {Butler} RC, {Calabrese} E, {Cardoso} JF, {Carron} J,
  {Challinor} A, {Chiang} HC, {Chluba} J, {Colombo} LPL, {Combet} C,
  {Contreras} D, {Crill} BP, {Cuttaia} F, {de Bernardis} P, {de Zotti} G,
  {Delabrouille} J, {Delouis} JM, {Di Valentino} E, {Diego} JM, {Dor{\'e}} O,
  {Douspis} M, {Ducout} A, {Dupac} X, {Dusini} S, {Efstathiou} G, {Elsner} F,
  {En{\ss}lin} TA, {Eriksen} HK, {Fantaye} Y, {Farhang} M, {Fergusson} J,
  {Fernandez-Cobos} R, {Finelli} F, {Forastieri} F, {Frailis} M, {Fraisse} AA,
  {Franceschi} E, {Frolov} A, {Galeotta} S, {Galli} S, {Ganga} K,
  {G{\'e}nova-Santos} RT, {Gerbino} M, {Ghosh} T, {Gonz{\'a}lez-Nuevo} J,
  {G{\'o}rski} KM, {Gratton} S, {Gruppuso} A, {Gudmundsson} JE, {Hamann} J,
  {Handley} W, {Hansen} FK, {Herranz} D, {Hildebrandt} SR, {Hivon} E, {Huang}
  Z, {Jaffe} AH, {Jones} WC, {Karakci} A, {Keih{\"a}nen} E, {Keskitalo} R,
  {Kiiveri} K, {Kim} J, {Kisner} TS, {Knox} L, {Krachmalnicoff} N, {Kunz} M,
  {Kurki-Suonio} H, {Lagache} G, {Lamarre} JM, {Lasenby} A, {Lattanzi} M,
  {Lawrence} CR, {Le Jeune} M, {Lemos} P, {Lesgourgues} J, {Levrier} F, {Lewis}
  A, {Liguori} M, {Lilje} PB, {Lilley} M, {Lindholm} V, {L{\'o}pez-Caniego} M,
  {Lubin} PM, {Ma} YZ, {Mac{\'\i}as-P{\'e}rez} JF, {Maggio} G, {Maino} D,
  {Mandolesi} N, {Mangilli} A, {Marcos-Caballero} A, {Maris} M, {Martin} PG,
  {Martinelli} M, {Mart{\'\i}nez-Gonz{\'a}lez} E, {Matarrese} S, {Mauri} N,
  {McEwen} JD, {Meinhold} PR, {Melchiorri} A, {Mennella} A, {Migliaccio} M,
  {Millea} M, {Mitra} S, {Miville-Desch{\^e}nes} MA, {Molinari} D, {Montier} L,
  {Morgante} G, {Moss} A, {Natoli} P, {N{\o}rgaard-Nielsen} HU, {Pagano} L,
  {Paoletti} D, {Partridge} B, {Patanchon} G, {Peiris} HV, {Perrotta} F,
  {Pettorino} V, {Piacentini} F, {Polastri} L, {Polenta} G, {Puget} JL,
  {Rachen} JP, {Reinecke} M, {Remazeilles} M, {Renzi} A, {Rocha} G, {Rosset} C,
  {Roudier} G, {Rubi{\~n}o-Mart{\'\i}n} JA, {Ruiz-Granados} B, {Salvati} L,
  {Sandri} M, {Savelainen} M, {Scott} D, {Shellard} EPS, {Sirignano} C, {Sirri}
  G, {Spencer} LD, {Sunyaev} R, {Suur-Uski} AS, {Tauber} JA, {Tavagnacco} D,
  {Tenti} M, {Toffolatti} L, {Tomasi} M, {Trombetti} T, {Valenziano} L,
  {Valiviita} J, {Van Tent} B, {Vibert} L, {Vielva} P, {Villa} F, {Vittorio} N,
  {Wandelt} BD, {Wehus} IK, {White} M, {White} SDM, {Zacchei} A, {Zonca} A
  (2020) {Planck 2018 results. VI. Cosmological parameters}. \aap 641:A6.
  \doi{10.1051/0004-6361/201833910}.
  {\href{https://arxiv.org/abs/1807.06209}{{arXiv:1807.06209}}} {[astro-ph.CO]}

\bibitem[{{Planelles} and {Quilis}(2010)}]{Planelles:2010}
{Planelles} S, {Quilis} V (2010) {ASOHF: a new adaptive spherical overdensity
  halo finder}. \aap 519:A94. \doi{10.1051/0004-6361/201014214}.
  {\href{https://arxiv.org/abs/1006.3205}{{arXiv:1006.3205}}} {[astro-ph.CO]}

\bibitem[{{Platen} et~al.(2007){Platen}, {van de Weygaert}, and
  {Jones}}]{Platen:2007}
{Platen} E, {van de Weygaert} R, {Jones} BJT (2007) {A cosmic watershed: the
  WVF void detection technique}. \mnras 380(2):551--570.
  \doi{10.1111/j.1365-2966.2007.12125.x}.
  {\href{https://arxiv.org/abs/0706.2788}{{arXiv:0706.2788}}} {[astro-ph]}

\bibitem[{{Polisensky} and {Ricotti}(2015)}]{Polisensky:2015}
{Polisensky} E, {Ricotti} M (2015) {Fingerprints of the initial conditions on
  the density profiles of cold and warm dark matter haloes}. \mnras
  450(2):2172--2184. \doi{10.1093/mnras/stv736}.
  {\href{https://arxiv.org/abs/1504.02126}{{arXiv:1504.02126}}} {[astro-ph.GA]}

\bibitem[{{Pontzen} and {Governato}(2012)}]{Pontzen:2012}
{Pontzen} A, {Governato} F (2012) {How supernova feedback turns dark matter
  cusps into cores}. \mnras 421(4):3464--3471.
  \doi{10.1111/j.1365-2966.2012.20571.x}.
  {\href{https://arxiv.org/abs/1106.0499}{{arXiv:1106.0499}}} {[astro-ph.CO]}

\bibitem[{{Pontzen} and {Governato}(2013)}]{Pontzen:2013}
{Pontzen} A, {Governato} F (2013) {Conserved actions, maximum entropy and dark
  matter haloes}. \mnras 430(1):121--133. \doi{10.1093/mnras/sts529}.
  {\href{https://arxiv.org/abs/1210.1849}{{arXiv:1210.1849}}} {[astro-ph.CO]}

\bibitem[{{Pontzen} et~al.(2016){Pontzen}, {Slosar}, {Roth}, and
  {Peiris}}]{Pontzen:2016}
{Pontzen} A, {Slosar} A, {Roth} N, {Peiris} HV (2016) {Inverted initial
  conditions: Exploring the growth of cosmic structure and voids}. \prd
  93(10):103519. \doi{10.1103/PhysRevD.93.103519}.
  {\href{https://arxiv.org/abs/1511.04090}{{arXiv:1511.04090}}} {[astro-ph.CO]}

\bibitem[{{Porayko} and {Postnov}(2014)}]{Porayko:2014}
{Porayko} NK, {Postnov} KA (2014) {Constraints on ultralight scalar dark matter
  from pulsar timing}. \prd 90(6):062008. \doi{10.1103/PhysRevD.90.062008}.
  {\href{https://arxiv.org/abs/1408.4670}{{arXiv:1408.4670}}} {[astro-ph.CO]}

\bibitem[{{Porayko} et~al.(2018){Porayko}, {Zhu}, {Levin}, {Hui}, {Hobbs},
  {Grudskaya}, {Postnov}, {Bailes}, {Bhat}, {Coles}, {Dai}, {Dempsey}, {Keith},
  {Kerr}, {Kramer}, {Lasky}, {Manchester}, {Os{\l}owski}, {Parthasarathy},
  {Ravi}, {Reardon}, {Rosado}, {Russell}, {Shannon}, {Spiewak}, {van Straten},
  {Toomey}, {Wang}, {Wen}, {You}, and {PPTA Collaboration}}]{Porayko:2018}
{Porayko} NK, {Zhu} X, {Levin} Y, {Hui} L, {Hobbs} G, {Grudskaya} A, {Postnov}
  K, {Bailes} M, {Bhat} NDR, {Coles} W, {Dai} S, {Dempsey} J, {Keith} MJ,
  {Kerr} M, {Kramer} M, {Lasky} PD, {Manchester} RN, {Os{\l}owski} S,
  {Parthasarathy} A, {Ravi} V, {Reardon} DJ, {Rosado} PA, {Russell} CJ,
  {Shannon} RM, {Spiewak} R, {van Straten} W, {Toomey} L, {Wang} J, {Wen} L,
  {You} X, {PPTA Collaboration} (2018) {Parkes Pulsar Timing Array constraints
  on ultralight scalar-field dark matter}. \prd 98(10):102002.
  \doi{10.1103/PhysRevD.98.102002}.
  {\href{https://arxiv.org/abs/1810.03227}{{arXiv:1810.03227}}} {[astro-ph.CO]}

\bibitem[{{Potter} and {Stadel}(2016)}]{Potter:2016}
{Potter} D, {Stadel} J (2016) {PKDGRAV3: Parallel gravity code}.
  {\href{https://arxiv.org/abs/1609.016}{{ascl:1609.016}}}

\bibitem[{{Potter} et~al.(2017){Potter}, {Stadel}, and
  {Teyssier}}]{Potter:2017}
{Potter} D, {Stadel} J, {Teyssier} R (2017) {PKDGRAV3: beyond trillion particle
  cosmological simulations for the next era of galaxy surveys}. Computational
  Astrophysics and Cosmology 4(1):2. \doi{10.1186/s40668-017-0021-1}.
  {\href{https://arxiv.org/abs/1609.08621}{{arXiv:1609.08621}}} {[astro-ph.IM]}

\bibitem[{{Powell} and {Abel}(2015)}]{Powell:2015}
{Powell} D, {Abel} T (2015) {An exact general remeshing scheme applied to
  physically conservative voxelization}. J Comput Phys 297:340--356.
  \doi{10.1016/j.jcp.2015.05.022}.
  {\href{https://arxiv.org/abs/1412.4941}{{arXiv:1412.4941}}}
  {[physics.comp-ph]}

\bibitem[{{Power}(2013)}]{Power:2013}
{Power} C (2013) {Seeking Observable Imprints of Small-Scale Structure on the
  Properties of Dark Matter Haloes}. \pasa 30:e053. \doi{10.1017/pasa.2013.32}.
  {\href{https://arxiv.org/abs/1309.1591}{{arXiv:1309.1591}}} {[astro-ph.CO]}

\bibitem[{{Power} and {Knebe}(2006)}]{Power:2006}
{Power} C, {Knebe} A (2006) {The impact of box size on the properties of dark
  matter haloes in cosmological simulations}. \mnras 370(2):691--701.
  \doi{10.1111/j.1365-2966.2006.10562.x}.
  {\href{https://arxiv.org/abs/astro-ph/0512281}{{arXiv:astro-ph/0512281}}}
  {[astro-ph]}

\bibitem[{{Power} et~al.(2003){Power}, {Navarro}, {Jenkins}, {Frenk}, {White},
  {Springel}, {Stadel}, and {Quinn}}]{Power:2003}
{Power} C, {Navarro} JF, {Jenkins} A, {Frenk} CS, {White} SDM, {Springel} V,
  {Stadel} J, {Quinn} T (2003) {The inner structure of
  {\ensuremath{\Lambda}}CDM haloes - I. A numerical convergence study}. \mnras
  338(1):14--34. \doi{10.1046/j.1365-8711.2003.05925.x}.
  {\href{https://arxiv.org/abs/astro-ph/0201544}{{arXiv:astro-ph/0201544}}}
  {[astro-ph]}

\bibitem[{{Power} et~al.(2016){Power}, {Robotham}, {Obreschkow}, {Hobbs}, and
  {Lewis}}]{Power:2016}
{Power} C, {Robotham} ASG, {Obreschkow} D, {Hobbs} A, {Lewis} GF (2016)
  {Spurious haloes and discreteness-driven relaxation in cosmological
  simulations}. \mnras 462(1):474--489. \doi{10.1093/mnras/stw1644}.
  {\href{https://arxiv.org/abs/1606.02038}{{arXiv:1606.02038}}} {[astro-ph.CO]}

\bibitem[{{Pozo} et~al.(2020){Pozo}, {Broadhurst}, {de Martino}, {Chiueh},
  {Smoot}, {Bonoli}, and {Angulo}}]{Pozo:2020}
{Pozo} A, {Broadhurst} T, {de Martino} I, {Chiueh} T, {Smoot} GF, {Bonoli} S,
  {Angulo} R (2020) {Detection of a universal core-halo transition in dwarf
  galaxies as predicted by Bose-Einstein dark matter}. arXiv e-prints
  arXiv:2010.10337.
  {\href{https://arxiv.org/abs/2010.10337}{{arXiv:2010.10337}}} {[astro-ph.GA]}

\bibitem[{{Press} and {Davis}(1982)}]{Press:1982}
{Press} WH, {Davis} M (1982) {How to identify and weigh virialized clusters of
  galaxies in a complete redshift catalog}. \apj 259:449--473.
  \doi{10.1086/160183}

\bibitem[{{Press} and {Schechter}(1974)}]{Press:1974}
{Press} WH, {Schechter} P (1974) {Formation of Galaxies and Clusters of
  Galaxies by Self-Similar Gravitational Condensation}. \apj 187:425--438.
  \doi{10.1086/152650}

\bibitem[{{Press} et~al.(1990){Press}, {Ryden}, and {Spergel}}]{Press:1990}
{Press} WH, {Ryden} BS, {Spergel} DN (1990) {Single mechanism for generating
  large-scale structure and providing dark missing matter}. Physical Review
  Letters 64:1084--1087. \doi{10.1103/PhysRevLett.64.1084}

\bibitem[{{Price} and {Monaghan}(2007)}]{Price:2007}
{Price} DJ, {Monaghan} JJ (2007) {An energy-conserving formalism for adaptive
  gravitational force softening in smoothed particle hydrodynamics and N-body
  codes}. \mnras 374(4):1347--1358. \doi{10.1111/j.1365-2966.2006.11241.x}.
  {\href{https://arxiv.org/abs/astro-ph/0610872}{{arXiv:astro-ph/0610872}}}
  {[astro-ph]}

\bibitem[{{Price-Whelan} et~al.(2016){Price-Whelan}, {Johnston}, {Valluri},
  {Pearson}, {K{\"u}pper}, and {Hogg}}]{Price-Whelan:2016}
{Price-Whelan} AM, {Johnston} KV, {Valluri} M, {Pearson} S, {K{\"u}pper} AHW,
  {Hogg} DW (2016) {Chaotic dispersal of tidal debris}. \mnras
  455(1):1079--1098. \doi{10.1093/mnras/stv2383}.
  {\href{https://arxiv.org/abs/1507.08662}{{arXiv:1507.08662}}} {[astro-ph.GA]}

\bibitem[{{Puchwein} et~al.(2013){Puchwein}, {Baldi}, and
  {Springel}}]{Puchwein:2013}
{Puchwein} E, {Baldi} M, {Springel} V (2013) {Modified-Gravity-GADGET: a new
  code for cosmological hydrodynamical simulations of modified gravity models}.
  \mnras 436(1):348--360. \doi{10.1093/mnras/stt1575}.
  {\href{https://arxiv.org/abs/1305.2418}{{arXiv:1305.2418}}} {[astro-ph.CO]}

\bibitem[{{Pueblas} and {Scoccimarro}(2009)}]{Pueblas:2009}
{Pueblas} S, {Scoccimarro} R (2009) {Generation of vorticity and velocity
  dispersion by orbit crossing}. \prd 80(4):043504.
  \doi{10.1103/PhysRevD.80.043504}.
  {\href{https://arxiv.org/abs/0809.4606}{{arXiv:0809.4606}}} {[astro-ph]}

\bibitem[{{Quinn} et~al.(1997){Quinn}, {Katz}, {Stadel}, and
  {Lake}}]{Quinn:1997}
{Quinn} T, {Katz} N, {Stadel} J, {Lake} G (1997) {Time stepping N-body
  simulations}. arXiv astro-ph/9710043
  {\href{https://arxiv.org/abs/astro-ph/9710043}{{astro-ph/9710043}}}

\bibitem[{{R{\'a}cz} et~al.(2018){R{\'a}cz}, {Szapudi}, {Csabai}, and
  {Dobos}}]{Racz:2018}
{R{\'a}cz} G, {Szapudi} I, {Csabai} I, {Dobos} L (2018) {Compactified
  cosmological simulations of the infinite universe}. \mnras 477(2):1949--1957.
  \doi{10.1093/mnras/sty695}.
  {\href{https://arxiv.org/abs/1711.04959}{{arXiv:1711.04959}}} {[astro-ph.CO]}

\bibitem[{{R{\'a}cz} et~al.(2019){R{\'a}cz}, {Szapudi}, {Dobos}, {Csabai}, and
  {Szalay}}]{Racz:2019}
{R{\'a}cz} G, {Szapudi} I, {Dobos} L, {Csabai} I, {Szalay} AS (2019) {StePS: A
  multi-GPU cosmological N-body Code for compactified simulations}. Astronomy
  and Computing 28:100303. \doi{10.1016/j.ascom.2019.100303}.
  {\href{https://arxiv.org/abs/1811.05903}{{arXiv:1811.05903}}} {[astro-ph.CO]}

\bibitem[{{Radford} et~al.(2015){Radford}, {Metz}, and
  {Chintala}}]{Radford:2015}
{Radford} A, {Metz} L, {Chintala} S (2015) {Unsupervised Representation
  Learning with Deep Convolutional Generative Adversarial Networks}. arXiv
  e-prints arXiv:1511.06434.
  {\href{https://arxiv.org/abs/1511.06434}{{arXiv:1511.06434}}} {[cs.LG]}

\bibitem[{{Ramachandra} et~al.(2020){Ramachandra}, {Valogiannis}, {Ishak}, and
  {Heitmann}}]{Ramachandra:2020}
{Ramachandra} N, {Valogiannis} G, {Ishak} M, {Heitmann} K (2020) {Matter Power
  Spectrum Emulator for f(R) Modified Gravity Cosmologies}. arXiv e-prints
  arXiv:2010.00596.
  {\href{https://arxiv.org/abs/2010.00596}{{arXiv:2010.00596}}} {[astro-ph.CO]}

\bibitem[{{Ramachandra} and {Shandarin}(2015)}]{Ramachandra:2015}
{Ramachandra} NS, {Shandarin} SF (2015) {Multi-stream portrait of the cosmic
  web}. \mnras 452(2):1643--1653. \doi{10.1093/mnras/stv1389}.
  {\href{https://arxiv.org/abs/1412.7768}{{arXiv:1412.7768}}} {[astro-ph.CO]}

\bibitem[{{Ramakrishnan} and {Paranjape}(2020)}]{Ramakrishnan:2020}
{Ramakrishnan} S, {Paranjape} A (2020) {Separate Universe calibration of the
  dependence of halo bias on cosmic web anisotropy}. \mnras 499(3):4418--4431.
  \doi{10.1093/mnras/staa2999}.
  {\href{https://arxiv.org/abs/2007.03711}{{arXiv:2007.03711}}} {[astro-ph.CO]}

\bibitem[{Rampf(2012)}]{Rampf:2012a}
Rampf C (2012) {The recursion relation in Lagrangian perturbation theory}.
  \jcap 1212:004. \doi{10.1088/1475-7516/2012/12/004}.
  {\href{https://arxiv.org/abs/1205.5274}{{arXiv:1205.5274}}} {[astro-ph.CO]}

\bibitem[{{Rampf} and {Buchert}(2012)}]{Rampf:2012b}
{Rampf} C, {Buchert} T (2012) {Lagrangian perturbations and the matter
  bispectrum I: fourth-order model for non-linear clustering}. \jcap 6:021.
  \doi{10.1088/1475-7516/2012/06/021}.
  {\href{https://arxiv.org/abs/1203.4260}{{arXiv:1203.4260}}} {[astro-ph.CO]}

\bibitem[{{Rampf} and {Hahn}(2021)}]{Rampf:2020b}
{Rampf} C, {Hahn} O (2021) {Shell-crossing in a {\ensuremath{\Lambda}}CDM
  Universe}. \mnras 501(1):L71--L75. \doi{10.1093/mnrasl/slaa198}.
  {\href{https://arxiv.org/abs/2010.12584}{{arXiv:2010.12584}}} {[astro-ph.CO]}

\bibitem[{{Rampf} et~al.(2021{\natexlab{a}}){Rampf}, {Frisch}, and
  {Hahn}}]{Rampf:2019}
{Rampf} C, {Frisch} U, {Hahn} O (2021{\natexlab{a}}) {Unveiling the singular
  dynamics in the cosmic large-scale structure}. \mnras 505(1):L90--L94.
  \doi{10.1093/mnrasl/slab053}

\bibitem[{{Rampf} et~al.(2021{\natexlab{b}}){Rampf}, {Uhlemann}, and
  {Hahn}}]{Rampf:2020}
{Rampf} C, {Uhlemann} C, {Hahn} O (2021{\natexlab{b}}) {Cosmological
  perturbations for two cold fluids in {\ensuremath{\Lambda}}CDM}. \mnras
  503(1):406--425. \doi{10.1093/mnras/staa3605}.
  {\href{https://arxiv.org/abs/2008.09123}{{arXiv:2008.09123}}} {[astro-ph.CO]}

\bibitem[{{Rantala} et~al.(2021){Rantala}, {Naab}, and
  {Springel}}]{Rantala:2021}
{Rantala} A, {Naab} T, {Springel} V (2021) {FROST: a momentum-conserving CUDA
  implementation of a hierarchical fourth-order forward symplectic integrator}.
  \mnras \doi{10.1093/mnras/stab057}.
  {\href{https://arxiv.org/abs/2011.14984}{{arXiv:2011.14984}}} {[astro-ph.IM]}

\bibitem[{{Ravanbakhsh} et~al.(2017){Ravanbakhsh}, {Oliva}, {Fromenteau},
  {Price}, {Ho}, {Schneider}, and {Poczos}}]{Ravanbakhsh:2017}
{Ravanbakhsh} S, {Oliva} J, {Fromenteau} S, {Price} LC, {Ho} S, {Schneider} J,
  {Poczos} B (2017) {Estimating Cosmological Parameters from the Dark Matter
  Distribution}. arXiv e-prints arXiv:1711.02033.
  {\href{https://arxiv.org/abs/1711.02033}{{arXiv:1711.02033}}} {[astro-ph.CO]}

\bibitem[{{Reddick} et~al.(2013){Reddick}, {Wechsler}, {Tinker}, and
  {Behroozi}}]{Reddick:2013}
{Reddick} RM, {Wechsler} RH, {Tinker} JL, {Behroozi} PS (2013) {The Connection
  between Galaxies and Dark Matter Structures in the Local Universe}. \apj
  771(1):30. \doi{10.1088/0004-637X/771/1/30}.
  {\href{https://arxiv.org/abs/1207.2160}{{arXiv:1207.2160}}} {[astro-ph.CO]}

\bibitem[{{Reed} et~al.(2003){Reed}, {Gardner}, {Quinn}, {Stadel}, {Fardal},
  {Lake}, and {Governato}}]{Reed:2003}
{Reed} D, {Gardner} J, {Quinn} T, {Stadel} J, {Fardal} M, {Lake} G, {Governato}
  F (2003) {Evolution of the mass function of dark matter haloes}. \mnras
  346(2):565--572. \doi{10.1046/j.1365-2966.2003.07113.x}.
  {\href{https://arxiv.org/abs/astro-ph/0301270}{{arXiv:astro-ph/0301270}}}
  {[astro-ph]}

\bibitem[{{Regan} et~al.(2012){Regan}, {Schmittfull}, {Shellard}, and
  {Fergusson}}]{Regan:2012}
{Regan} DM, {Schmittfull} MM, {Shellard} EPS, {Fergusson} JR (2012) {Universal
  non-Gaussian initial conditions for N-body simulations}. \prd 86(12):123524.
  \doi{10.1103/PhysRevD.86.123524}.
  {\href{https://arxiv.org/abs/1108.3813}{{arXiv:1108.3813}}} {[astro-ph.CO]}

\bibitem[{{Reid} et~al.(2010){Reid}, {Verde}, {Dolag}, {Matarrese}, and
  {Moscardini}}]{Reid:2010}
{Reid} BA, {Verde} L, {Dolag} K, {Matarrese} S, {Moscardini} L (2010)
  {Non-Gaussian halo assembly bias}. \jcap 2010(7):013.
  \doi{10.1088/1475-7516/2010/07/013}.
  {\href{https://arxiv.org/abs/1004.1637}{{arXiv:1004.1637}}} {[astro-ph.CO]}

\bibitem[{{Reid} et~al.(2014){Reid}, {Seo}, {Leauthaud}, {Tinker}, and
  {White}}]{Reid:2014}
{Reid} BA, {Seo} HJ, {Leauthaud} A, {Tinker} JL, {White} M (2014) {A 2.5 per
  cent measurement of the growth rate from small-scale redshift space
  clustering of SDSS-III CMASS galaxies}. \mnras 444(1):476--502.
  \doi{10.1093/mnras/stu1391}.
  {\href{https://arxiv.org/abs/1404.3742}{{arXiv:1404.3742}}} {[astro-ph.CO]}

\bibitem[{{Ribli} et~al.(2019){Ribli}, {Pataki}, and {Csabai}}]{Ribli:2019}
{Ribli} D, {Pataki} B{\'A}, {Csabai} I (2019) {An improved cosmological
  parameter inference scheme motivated by deep learning}. Nature Astronomy
  3:93--98. \doi{10.1038/s41550-018-0596-8}.
  {\href{https://arxiv.org/abs/1806.05995}{{arXiv:1806.05995}}} {[astro-ph.CO]}

\bibitem[{{Richardson} et~al.(2021){Richardson}, {St{\"u}cker}, {Angulo}, and
  {Hahn}}]{Richardson:2021}
{Richardson} TRG, {St{\"u}cker} J, {Angulo} RE, {Hahn} O (2021) {Non-Halo
  Structures and their Effects on Gravitational Lensing}. arXiv e-prints
  arXiv:2101.07806.
  {\href{https://arxiv.org/abs/2101.07806}{{arXiv:2101.07806}}} {[astro-ph.CO]}

\bibitem[{{Ricotti} and {Gould}(2009)}]{Ricotti:2009}
{Ricotti} M, {Gould} A (2009) {A New Probe of Dark Matter and High-Energy
  Universe Using Microlensing}. \apj 707(2):979--987.
  \doi{10.1088/0004-637X/707/2/979}.
  {\href{https://arxiv.org/abs/0908.0735}{{arXiv:0908.0735}}} {[astro-ph.CO]}

\bibitem[{{Riess} et~al.(2019){Riess}, {Casertano}, {Yuan}, {Macri}, and
  {Scolnic}}]{Riess:2019}
{Riess} AG, {Casertano} S, {Yuan} W, {Macri} LM, {Scolnic} D (2019) {Large
  Magellanic Cloud Cepheid Standards Provide a 1\% Foundation for the
  Determination of the Hubble Constant and Stronger Evidence for Physics beyond
  {\ensuremath{\Lambda}}CDM}. \apj 876(1):85. \doi{10.3847/1538-4357/ab1422}.
  {\href{https://arxiv.org/abs/1903.07603}{{arXiv:1903.07603}}} {[astro-ph.CO]}

\bibitem[{{Robertson} et~al.(2017){Robertson}, {Massey}, and
  {Eke}}]{Robertson:2017}
{Robertson} A, {Massey} R, {Eke} V (2017) {What does the Bullet Cluster tell us
  about self-interacting dark matter?} \mnras 465(1):569--587.
  \doi{10.1093/mnras/stw2670}.
  {\href{https://arxiv.org/abs/1605.04307}{{arXiv:1605.04307}}} {[astro-ph.CO]}

\bibitem[{{Rocha} et~al.(2013){Rocha}, {Peter}, {Bullock}, {Kaplinghat},
  {Garrison-Kimmel}, {O{\~n}orbe}, and {Moustakas}}]{Rocha:2013}
{Rocha} M, {Peter} AHG, {Bullock} JS, {Kaplinghat} M, {Garrison-Kimmel} S,
  {O{\~n}orbe} J, {Moustakas} LA (2013) {Cosmological simulations with
  self-interacting dark matter - I. Constant-density cores and substructure}.
  \mnras 430:81--104. \doi{10.1093/mnras/sts514}.
  {\href{https://arxiv.org/abs/1208.3025}{{arXiv:1208.3025}}}

\bibitem[{{Rodr{\'\i}guez} et~al.(2018){Rodr{\'\i}guez}, {Kacprzak}, {Lucchi},
  {Amara}, {Sgier}, {Fluri}, {Hofmann}, and
  {R{\'e}fr{\'e}gier}}]{Rodriguez:2018}
{Rodr{\'\i}guez} AC, {Kacprzak} T, {Lucchi} A, {Amara} A, {Sgier} R, {Fluri} J,
  {Hofmann} T, {R{\'e}fr{\'e}gier} A (2018) {Fast cosmic web simulations with
  generative adversarial networks}. Computational Astrophysics and Cosmology
  5(1):4. \doi{10.1186/s40668-018-0026-4}.
  {\href{https://arxiv.org/abs/1801.09070}{{arXiv:1801.09070}}} {[astro-ph.CO]}

\bibitem[{{Rodr{\'\i}guez-Torres} et~al.(2016){Rodr{\'\i}guez-Torres},
  {Chuang}, {Prada}, {Guo}, {Klypin}, {Behroozi}, {Hahn}, {Comparat}, {Yepes},
  {Montero-Dorta}, {Brownstein}, {Maraston}, {McBride}, {Tinker},
  {Gottl{\"o}ber}, {Favole}, {Shu}, {Kitaura}, {Bolton}, {Scoccimarro},
  {Samushia}, {Schlegel}, {Schneider}, and {Thomas}}]{Rodriguez-Torres:2016}
{Rodr{\'\i}guez-Torres} SA, {Chuang} CH, {Prada} F, {Guo} H, {Klypin} A,
  {Behroozi} P, {Hahn} CH, {Comparat} J, {Yepes} G, {Montero-Dorta} AD,
  {Brownstein} JR, {Maraston} C, {McBride} CK, {Tinker} J, {Gottl{\"o}ber} S,
  {Favole} G, {Shu} Y, {Kitaura} FS, {Bolton} A, {Scoccimarro} R, {Samushia} L,
  {Schlegel} D, {Schneider} DP, {Thomas} D (2016) {The clustering of galaxies
  in the SDSS-III Baryon Oscillation Spectroscopic Survey: modelling the
  clustering and halo occupation distribution of BOSS CMASS galaxies in the
  Final Data Release}. \mnras 460(2):1173--1187. \doi{10.1093/mnras/stw1014}.
  {\href{https://arxiv.org/abs/1509.06404}{{arXiv:1509.06404}}} {[astro-ph.CO]}

\bibitem[{{Rogers} et~al.(2019){Rogers}, {Peiris}, {Pontzen}, {Bird}, {Verde},
  and {Font-Ribera}}]{Rogers:2019}
{Rogers} KK, {Peiris} HV, {Pontzen} A, {Bird} S, {Verde} L, {Font-Ribera} A
  (2019) {Bayesian emulator optimisation for cosmology: application to the
  Lyman-alpha forest}. J Cosmol Astro-Part Phys 2019(2):031.
  \doi{10.1088/1475-7516/2019/02/031}.
  {\href{https://arxiv.org/abs/1812.04631}{{arXiv:1812.04631}}} {[astro-ph.CO]}

\bibitem[{{Romano-D{\'\i}az} et~al.(2008){Romano-D{\'\i}az}, {Shlosman},
  {Hoffman}, and {Heller}}]{RomanoDiaz:2008}
{Romano-D{\'\i}az} E, {Shlosman} I, {Hoffman} Y, {Heller} C (2008) {Erasing
  Dark Matter Cusps in Cosmological Galactic Halos with Baryons}. \apjl
  685(2):L105. \doi{10.1086/592687}.
  {\href{https://arxiv.org/abs/0808.0195}{{arXiv:0808.0195}}} {[astro-ph]}

\bibitem[{{Romeo} et~al.(2008){Romeo}, {Agertz}, {Moore}, and
  {Stadel}}]{Romeo:2008}
{Romeo} AB, {Agertz} O, {Moore} B, {Stadel} J (2008) {Discreteness Effects in
  {\ensuremath{\Lambda}}CDM Simulations: A Wavelet-Statistical View}. \apj
  686(1):1--12. \doi{10.1086/591236}.
  {\href{https://arxiv.org/abs/0804.0294}{{arXiv:0804.0294}}} {[astro-ph]}

\bibitem[{{Roszkowski} et~al.(2018){Roszkowski}, {Sessolo}, and
  {Trojanowski}}]{Roszkowski:2018}
{Roszkowski} L, {Sessolo} EM, {Trojanowski} S (2018) {WIMP dark matter
  candidates and searches---current status and future prospects}. Reports on
  Progress in Physics 81(6):066201. \doi{10.1088/1361-6633/aab913}.
  {\href{https://arxiv.org/abs/1707.06277}{{arXiv:1707.06277}}} {[hep-ph]}

\bibitem[{{Roth} et~al.(2016){Roth}, {Pontzen}, and {Peiris}}]{Roth:2016}
{Roth} N, {Pontzen} A, {Peiris} HV (2016) {Genetically modified haloes: towards
  controlled experiments in {\ensuremath{\Lambda}}CDM galaxy formation}. \mnras
  455(1):974--986. \doi{10.1093/mnras/stv2375}.
  {\href{https://arxiv.org/abs/1504.07250}{{arXiv:1504.07250}}} {[astro-ph.GA]}

\bibitem[{{Roy} et~al.(2014){Roy}, {Bouillot}, and {Rasera}}]{Roy:2014}
{Roy} F, {Bouillot} VR, {Rasera} Y (2014) {pFoF: a highly scalable halo-finder
  for large cosmological data sets}. \aap 564:A13.
  \doi{10.1051/0004-6361/201322555}

\bibitem[{{Ruan} et~al.(2021){Ruan}, {Hern{\'a}ndez-Aguayo}, {Li}, {Arnold},
  {Baugh}, {Klypin}, and {Prada}}]{Ruan:2021}
{Ruan} CZ, {Hern{\'a}ndez-Aguayo} C, {Li} B, {Arnold} C, {Baugh} CM, {Klypin}
  A, {Prada} F (2021) {Fast full N-body simulations of generic modified
  gravity: conformal coupling models}. arXiv e-prints arXiv:2110.00328.
  {\href{https://arxiv.org/abs/2110.00328}{{arXiv:2110.00328}}} {[astro-ph.CO]}

\bibitem[{{Ruiz} et~al.(2015){Ruiz}, {Cora}, {Padilla}, {Dom{\'\i}nguez},
  {Vega-Mart{\'\i}nez}, {Tecce}, {Orsi}, {Yaryura}, {Garc{\'\i}a Lambas},
  {Gargiulo}, and {Mu{\~n}oz Arancibia}}]{Ruiz:2015}
{Ruiz} AN, {Cora} SA, {Padilla} ND, {Dom{\'\i}nguez} MJ, {Vega-Mart{\'\i}nez}
  CA, {Tecce} TE, {Orsi} {\'A}, {Yaryura} Y, {Garc{\'\i}a Lambas} D, {Gargiulo}
  ID, {Mu{\~n}oz Arancibia} AM (2015) {Calibration of Semi-analytic Models of
  Galaxy Formation Using Particle Swarm Optimization}. \apj 801(2):139.
  \doi{10.1088/0004-637X/801/2/139}.
  {\href{https://arxiv.org/abs/1310.7034}{{arXiv:1310.7034}}} {[astro-ph.CO]}

\bibitem[{{Salvati} et~al.(2020){Salvati}, {Douspis}, and
  {Aghanim}}]{Salvati:2020}
{Salvati} L, {Douspis} M, {Aghanim} N (2020) {Impact of systematics on
  cosmological parameters from future galaxy cluster surveys}. \aap 643:A20.
  \doi{10.1051/0004-6361/202038465}.
  {\href{https://arxiv.org/abs/2005.10204}{{arXiv:2005.10204}}} {[astro-ph.CO]}

\bibitem[{{Sawala} et~al.(2016){Sawala}, {Frenk}, {Fattahi}, {Navarro},
  {Bower}, {Crain}, {Dalla Vecchia}, {Furlong}, {Helly}, {Jenkins}, {Oman},
  {Schaller}, {Schaye}, {Theuns}, {Trayford}, and {White}}]{Sawala:2016}
{Sawala} T, {Frenk} CS, {Fattahi} A, {Navarro} JF, {Bower} RG, {Crain} RA,
  {Dalla Vecchia} C, {Furlong} M, {Helly} JC, {Jenkins} A, {Oman} KA,
  {Schaller} M, {Schaye} J, {Theuns} T, {Trayford} J, {White} SDM (2016) {The
  APOSTLE simulations: solutions to the Local Group's cosmic puzzles}. \mnras
  457(2):1931--1943. \doi{10.1093/mnras/stw145}.
  {\href{https://arxiv.org/abs/1511.01098}{{arXiv:1511.01098}}} {[astro-ph.GA]}

\bibitem[{{Sawala} et~al.(2021{\natexlab{a}}){Sawala}, {Jenkins}, {McAlpine},
  {Jasche}, {Lavaux}, {Johansson}, and {Frenk}}]{Sawala:2021}
{Sawala} T, {Jenkins} A, {McAlpine} S, {Jasche} J, {Lavaux} G, {Johansson} PH,
  {Frenk} CS (2021{\natexlab{a}}) {Setting the stage: structures from Gaussian
  random fields}. \mnras 501(4):4759--4776. \doi{10.1093/mnras/staa3568}.
  {\href{https://arxiv.org/abs/2003.04321}{{arXiv:2003.04321}}} {[astro-ph.CO]}

\bibitem[{{Sawala} et~al.(2021{\natexlab{b}}){Sawala}, {McAlpine}, {Jasche},
  {Lavaux}, {Jenkins}, {Johansson}, and {Frenk}}]{Sawala:2021b}
{Sawala} T, {McAlpine} S, {Jasche} J, {Lavaux} G, {Jenkins} A, {Johansson} PH,
  {Frenk} CS (2021{\natexlab{b}}) {The SIBELIUS Project: E Pluribus Unum}.
  arXiv e-prints arXiv:2103.12073.
  {\href{https://arxiv.org/abs/2103.12073}{{arXiv:2103.12073}}} {[astro-ph.CO]}

\bibitem[{{Schaller} et~al.(2014){Schaller}, {Becker}, {Ruchayskiy},
  {Boyarsky}, and {Shaposhnikov}}]{Schaller:2014}
{Schaller} M, {Becker} C, {Ruchayskiy} O, {Boyarsky} A, {Shaposhnikov} M (2014)
  {A new framework for numerical simulations of structure formation}. \mnras
  442(4):3073--3095. \doi{10.1093/mnras/stu1069}.
  {\href{https://arxiv.org/abs/1310.5102}{{arXiv:1310.5102}}} {[astro-ph.CO]}

\bibitem[{{Schaye} et~al.(2010){Schaye}, {Dalla Vecchia}, {Booth}, {Wiersma},
  {Theuns}, {Haas}, {Bertone}, {Duffy}, {McCarthy}, and {van de
  Voort}}]{Schaye:2010}
{Schaye} J, {Dalla Vecchia} C, {Booth} CM, {Wiersma} RPC, {Theuns} T, {Haas}
  MR, {Bertone} S, {Duffy} AR, {McCarthy} IG, {van de Voort} F (2010) {The
  physics driving the cosmic star formation history}. \mnras 402(3):1536--1560.
  \doi{10.1111/j.1365-2966.2009.16029.x}.
  {\href{https://arxiv.org/abs/0909.5196}{{arXiv:0909.5196}}} {[astro-ph.CO]}

\bibitem[{{Schaye} et~al.(2015){Schaye}, {Crain}, {Bower}, {Furlong},
  {Schaller}, {Theuns}, {Dalla Vecchia}, {Frenk}, {McCarthy}, {Helly},
  {Jenkins}, {Rosas-Guevara}, {White}, {Baes}, {Booth}, {Camps}, {Navarro},
  {Qu}, {Rahmati}, {Sawala}, {Thomas}, and {Trayford}}]{Schaye:2015}
{Schaye} J, {Crain} RA, {Bower} RG, {Furlong} M, {Schaller} M, {Theuns} T,
  {Dalla Vecchia} C, {Frenk} CS, {McCarthy} IG, {Helly} JC, {Jenkins} A,
  {Rosas-Guevara} YM, {White} SDM, {Baes} M, {Booth} CM, {Camps} P, {Navarro}
  JF, {Qu} Y, {Rahmati} A, {Sawala} T, {Thomas} PA, {Trayford} J (2015) {The
  EAGLE project: simulating the evolution and assembly of galaxies and their
  environments}. \mnras 446:521--554. \doi{10.1093/mnras/stu2058}.
  {\href{https://arxiv.org/abs/1407.7040}{{arXiv:1407.7040}}}

\bibitem[{{Schive} et~al.(2014){Schive}, {Chiueh}, and
  {Broadhurst}}]{Schive:2014}
{Schive} HY, {Chiueh} T, {Broadhurst} T (2014) {Cosmic structure as the quantum
  interference of a coherent dark wave}. Nature Physics 10:496--499.
  \doi{10.1038/nphys2996}.
  {\href{https://arxiv.org/abs/1406.6586}{{arXiv:1406.6586}}}

\bibitem[{{Schmalzing} and {Buchert}(1997)}]{Schmalzing:1997}
{Schmalzing} J, {Buchert} T (1997) {Beyond Genus Statistics: A Unifying
  Approach to the Morphology of Cosmic Structure}. \apjl 482(1):L1--L4.
  \doi{10.1086/310680}.
  {\href{https://arxiv.org/abs/astro-ph/9702130}{{arXiv:astro-ph/9702130}}}
  {[astro-ph]}

\bibitem[{{Schmelzle} et~al.(2017){Schmelzle}, {Lucchi}, {Kacprzak}, {Amara},
  {Sgier}, {R{\'e}fr{\'e}gier}, and {Hofmann}}]{Schmelzle:2017}
{Schmelzle} J, {Lucchi} A, {Kacprzak} T, {Amara} A, {Sgier} R,
  {R{\'e}fr{\'e}gier} A, {Hofmann} T (2017) {Cosmological model discrimination
  with Deep Learning}. arXiv e-prints arXiv:1707.05167.
  {\href{https://arxiv.org/abs/1707.05167}{{arXiv:1707.05167}}} {[astro-ph.CO]}

\bibitem[{{Schmidt} et~al.(2018){Schmidt}, {White}, {Schmidt}, and
  {St{\"u}cker}}]{Schmidt:2018}
{Schmidt} AS, {White} SDM, {Schmidt} F, {St{\"u}cker} J (2018) {Cosmological
  N-body simulations with a large-scale tidal field}. \mnras 479:162--170.
  \doi{10.1093/mnras/sty1430}.
  {\href{https://arxiv.org/abs/1803.03274}{{arXiv:1803.03274}}}

\bibitem[{{Schmidt}(2009)}]{Schmidt:2009}
{Schmidt} F (2009) {Self-consistent cosmological simulations of DGP braneworld
  gravity}. \prd 80(4):043001. \doi{10.1103/PhysRevD.80.043001}.
  {\href{https://arxiv.org/abs/0905.0858}{{arXiv:0905.0858}}} {[astro-ph.CO]}

\bibitem[{{Schmidt}(2016)}]{Schmidt:2016}
{Schmidt} F (2016) {Effect of relative velocity and density perturbations
  between baryons and dark matter on the clustering of galaxies}. \prd
  94(6):063508. \doi{10.1103/PhysRevD.94.063508}.
  {\href{https://arxiv.org/abs/1602.09059}{{arXiv:1602.09059}}} {[astro-ph.CO]}

\bibitem[{{Schneider} and {Teyssier}(2015)}]{Schneider:2015}
{Schneider} A, {Teyssier} R (2015) {A new method to quantify the effects of
  baryons on the matter power spectrum}. \jcap 12:049.
  \doi{10.1088/1475-7516/2015/12/049}.
  {\href{https://arxiv.org/abs/1510.06034}{{arXiv:1510.06034}}}

\bibitem[{{Schneider} et~al.(2013){Schneider}, {Smith}, and
  {Reed}}]{Schneider:2013}
{Schneider} A, {Smith} RE, {Reed} D (2013) {Halo mass function and the free
  streaming scale}. \mnras 433(2):1573--1587. \doi{10.1093/mnras/stt829}.
  {\href{https://arxiv.org/abs/1303.0839}{{arXiv:1303.0839}}} {[astro-ph.CO]}

\bibitem[{{Schneider} et~al.(2016){Schneider}, {Teyssier}, {Potter}, {Stadel},
  {Onions}, {Reed}, {Smith}, {Springel}, {Pearce}, and
  {Scoccimarro}}]{Schneider:2016}
{Schneider} A, {Teyssier} R, {Potter} D, {Stadel} J, {Onions} J, {Reed} DS,
  {Smith} RE, {Springel} V, {Pearce} FR, {Scoccimarro} R (2016) {Matter power
  spectrum and the challenge of percent accuracy}. \jcap 2016(4):047.
  \doi{10.1088/1475-7516/2016/04/047}.
  {\href{https://arxiv.org/abs/1503.05920}{{arXiv:1503.05920}}} {[astro-ph.CO]}

\bibitem[{{Schneider} et~al.(2018){Schneider}, {Teyssier}, {Stadel}, {Chisari},
  {Le Brun}, {Amara}, and {Refregier}}]{Schneider:2018}
{Schneider} A, {Teyssier} R, {Stadel} J, {Chisari} NE, {Le Brun} AMC, {Amara}
  A, {Refregier} A (2018) {Quantifying baryon effects on the matter power
  spectrum and the weak lensing shear correlation}. arXiv e-prints
  {\href{https://arxiv.org/abs/1810.08629}{{arXiv:1810.08629}}}

\bibitem[{{Schneider} et~al.(2020){Schneider}, {Stoira}, {Refregier}, {Weiss},
  {Knabenhans}, {Stadel}, and {Teyssier}}]{Schneider:2020}
{Schneider} A, {Stoira} N, {Refregier} A, {Weiss} AJ, {Knabenhans} M, {Stadel}
  J, {Teyssier} R (2020) {Baryonic effects for weak lensing. Part I. Power
  spectrum and covariance matrix}. \jcap 2020(4):019.
  \doi{10.1088/1475-7516/2020/04/019}.
  {\href{https://arxiv.org/abs/1910.11357}{{arXiv:1910.11357}}} {[astro-ph.CO]}

\bibitem[{{Schneider} et~al.(2021){Schneider}, {Giri}, {Amodeo}, and
  {Refregier}}]{Schneider:2021}
{Schneider} A, {Giri} SK, {Amodeo} S, {Refregier} A (2021) {Constraining
  baryonic feedback and cosmology with weak-lensing, X-ray, and kinematic
  Sunyaev-Zeldovich observations}. arXiv e-prints arXiv:2110.02228.
  {\href{https://arxiv.org/abs/2110.02228}{{arXiv:2110.02228}}} {[astro-ph.CO]}

\bibitem[{{Sch{\"o}n} et~al.(2015){Sch{\"o}n}, {Mack}, {Avram}, {Wyithe}, and
  {Barberio}}]{Schoen:2015}
{Sch{\"o}n} S, {Mack} KJ, {Avram} CA, {Wyithe} JSB, {Barberio} E (2015) {Dark
  matter annihilation in the first galaxy haloes}. \mnras 451(3):2840--2850.
  \doi{10.1093/mnras/stv1056}.
  {\href{https://arxiv.org/abs/1411.3783}{{arXiv:1411.3783}}} {[astro-ph.CO]}

\bibitem[{{Schutz} et~al.(2018){Schutz}, {Lin}, {Safdi}, and
  {Wu}}]{Schutz:2018}
{Schutz} K, {Lin} T, {Safdi} BR, {Wu} CL (2018) {Constraining a Thin Dark
  Matter Disk with G a i a}. \prl 121(8):081101.
  \doi{10.1103/PhysRevLett.121.081101}.
  {\href{https://arxiv.org/abs/1711.03103}{{arXiv:1711.03103}}} {[astro-ph.GA]}

\bibitem[{{Schwabe} et~al.(2016){Schwabe}, {Niemeyer}, and
  {Engels}}]{Schwabe:2016}
{Schwabe} B, {Niemeyer} JC, {Engels} JF (2016) {Simulations of solitonic core
  mergers in ultralight axion dark matter cosmologies}. \prd 94(4):043513.
  \doi{10.1103/PhysRevD.94.043513}.
  {\href{https://arxiv.org/abs/1606.05151}{{arXiv:1606.05151}}} {[astro-ph.CO]}

\bibitem[{{Schwabe} et~al.(2020){Schwabe}, {Gosenca}, {Behrens}, {Niemeyer},
  and {Easther}}]{Schwabe:2020}
{Schwabe} B, {Gosenca} M, {Behrens} C, {Niemeyer} JC, {Easther} R (2020)
  {Simulating mixed fuzzy and cold dark matter}. \prd 102(8):083518.
  \doi{10.1103/PhysRevD.102.083518}.
  {\href{https://arxiv.org/abs/2007.08256}{{arXiv:2007.08256}}} {[astro-ph.CO]}

\bibitem[{{Scoccimarro}(1998)}]{Scoccimarro:1998}
{Scoccimarro} R (1998) {Transients from initial conditions: a perturbative
  analysis}. \mnras 299:1097--1118. \doi{10.1046/j.1365-8711.1998.01845.x}.
  {\href{https://arxiv.org/abs/astro-ph/9711187}{{astro-ph/9711187}}}

\bibitem[{{Scoccimarro} and {Sheth}(2002)}]{Scoccimarro:2002}
{Scoccimarro} R, {Sheth} RK (2002) {PTHALOS: a fast method for generating mock
  galaxy distributions}. \mnras 329(3):629--640.
  \doi{10.1046/j.1365-8711.2002.04999.x}.
  {\href{https://arxiv.org/abs/astro-ph/0106120}{{arXiv:astro-ph/0106120}}}
  {[astro-ph]}

\bibitem[{{Scoccimarro} et~al.(2012){Scoccimarro}, {Hui}, {Manera}, and
  {Chan}}]{Scoccimarro:2012}
{Scoccimarro} R, {Hui} L, {Manera} M, {Chan} KC (2012) {Large-scale bias and
  efficient generation of initial conditions for nonlocal primordial
  non-Gaussianity}. \prd 85(8):083002. \doi{10.1103/PhysRevD.85.083002}.
  {\href{https://arxiv.org/abs/1108.5512}{{arXiv:1108.5512}}} {[astro-ph.CO]}

\bibitem[{{Sefusatti} et~al.(2016){Sefusatti}, {Crocce}, {Scoccimarro}, and
  {Couchman}}]{Sefusatti:2016}
{Sefusatti} E, {Crocce} M, {Scoccimarro} R, {Couchman} HMP (2016) {Accurate
  estimators of correlation functions in Fourier space}. \mnras
  460(4):3624--3636. \doi{10.1093/mnras/stw1229}.
  {\href{https://arxiv.org/abs/1512.07295}{{arXiv:1512.07295}}} {[astro-ph.CO]}

\bibitem[{{Semboloni} et~al.(2011){Semboloni}, {Hoekstra}, {Schaye}, {van
  Daalen}, and {McCarthy}}]{Semboloni:2011}
{Semboloni} E, {Hoekstra} H, {Schaye} J, {van Daalen} MP, {McCarthy} IG (2011)
  {Quantifying the effect of baryon physics on weak lensing tomography}. \mnras
  417:2020--2035. \doi{10.1111/j.1365-2966.2011.19385.x}.
  {\href{https://arxiv.org/abs/1105.1075}{{arXiv:1105.1075}}}

\bibitem[{{Seppi} et~al.(2020){Seppi}, {Comparat}, {Nandra}, {Bulbul}, {Prada},
  {Klypin}, {Merloni}, {Predehl}, and {Ider Chitham}}]{Seppi:2020}
{Seppi} R, {Comparat} J, {Nandra} K, {Bulbul} E, {Prada} F, {Klypin} A,
  {Merloni} A, {Predehl} P, {Ider Chitham} J (2020) {The mass function
  dependence on dark matter haloes dynamical state}. arXiv e-prints
  arXiv:2008.03179.
  {\href{https://arxiv.org/abs/2008.03179}{{arXiv:2008.03179}}} {[astro-ph.CO]}

\bibitem[{{Sgier} et~al.(2020){Sgier}, {Fluri}, {Herbel}, {R{\'e}fr{\'e}gier},
  {Amara}, {Kacprzak}, and {Nicola}}]{Sgier:2020}
{Sgier} R, {Fluri} J, {Herbel} J, {R{\'e}fr{\'e}gier} A, {Amara} A, {Kacprzak}
  T, {Nicola} A (2020) {Fast Lightcones for Combined Cosmological Probes}.
  arXiv e-prints arXiv:2007.05735.
  {\href{https://arxiv.org/abs/2007.05735}{{arXiv:2007.05735}}} {[astro-ph.CO]}

\bibitem[{{Sgier} et~al.(2019){Sgier}, {R{\'e}fr{\'e}gier}, {Amara}, and
  {Nicola}}]{Sgier:2019}
{Sgier} RJ, {R{\'e}fr{\'e}gier} A, {Amara} A, {Nicola} A (2019) {Fast
  generation of covariance matrices for weak lensing}. \jcap 2019(1):044.
  \doi{10.1088/1475-7516/2019/01/044}.
  {\href{https://arxiv.org/abs/1801.05745}{{arXiv:1801.05745}}} {[astro-ph.CO]}

\bibitem[{{Shandarin} et~al.(2012){Shandarin}, {Habib}, and
  {Heitmann}}]{Shandarin:2012}
{Shandarin} S, {Habib} S, {Heitmann} K (2012) {Cosmic web, multistream flows,
  and tessellations}. \prd 85(8):083005. \doi{10.1103/PhysRevD.85.083005}.
  {\href{https://arxiv.org/abs/1111.2366}{{arXiv:1111.2366}}}

\bibitem[{{Shandarin}(2021)}]{Shandarin:2021}
{Shandarin} SF (2021) {Identifying dark matter haloes by the caustic boundary}.
  \jcap 2021(1):044. \doi{10.1088/1475-7516/2021/01/044}.
  {\href{https://arxiv.org/abs/2005.14548}{{arXiv:2005.14548}}} {[astro-ph.CO]}

\bibitem[{{Shandarin} and {Medvedev}(2017)}]{Shandarin:2017}
{Shandarin} SF, {Medvedev} MV (2017) {The features of the Cosmic Web unveiled
  by the flip-flop field}. \mnras 468(4):4056--4076.
  \doi{10.1093/mnras/stx699}.
  {\href{https://arxiv.org/abs/1609.08554}{{arXiv:1609.08554}}} {[astro-ph.CO]}

\bibitem[{{Shandera} et~al.(2011){Shandera}, {Dalal}, and
  {Huterer}}]{Shandera:2011}
{Shandera} S, {Dalal} N, {Huterer} D (2011) {A generalized local ansatz and its
  effect on halo bias}. \jcap 2011(3):017. \doi{10.1088/1475-7516/2011/03/017}.
  {\href{https://arxiv.org/abs/1010.3722}{{arXiv:1010.3722}}} {[astro-ph.CO]}

\bibitem[{{Shapiro} et~al.(1983){Shapiro}, {Struck-Marcell}, and
  {Melott}}]{Shapiro:1983}
{Shapiro} PR, {Struck-Marcell} C, {Melott} AL (1983) {Pancakes and the
  formation of galaxies in a neutrino-dominated universe}. \apj 275:413--429.
  \doi{10.1086/161543}

\bibitem[{{Shaw} et~al.(2007){Shaw}, {Weller}, {Ostriker}, and
  {Bode}}]{Shaw:2007}
{Shaw} LD, {Weller} J, {Ostriker} JP, {Bode} P (2007) {The Bound Mass of
  Substructures in Dark Matter Halos}. \apj 659(2):1082--1095.
  \doi{10.1086/511849}.
  {\href{https://arxiv.org/abs/astro-ph/0603150}{{arXiv:astro-ph/0603150}}}
  {[astro-ph]}

\bibitem[{{Sherwin} and {Zaldarriaga}(2012)}]{Sherwin:2012}
{Sherwin} BD, {Zaldarriaga} M (2012) {Shift of the baryon acoustic oscillation
  scale: A simple physical picture}. \prd 85(10):103523.
  \doi{10.1103/PhysRevD.85.103523}.
  {\href{https://arxiv.org/abs/1202.3998}{{arXiv:1202.3998}}} {[astro-ph.CO]}

\bibitem[{{Shirasaki} et~al.(2012){Shirasaki}, {Yoshida}, {Hamana}, and
  {Nishimichi}}]{Shirasaki:2012}
{Shirasaki} M, {Yoshida} N, {Hamana} T, {Nishimichi} T (2012) {Probing
  Primordial Non-Gaussianity with Weak-lensing Minkowski Functionals}. \apj
  760(1):45. \doi{10.1088/0004-637X/760/1/45}.
  {\href{https://arxiv.org/abs/1204.4981}{{arXiv:1204.4981}}} {[astro-ph.CO]}

\bibitem[{{Simha} and {Cole}(2013)}]{Simha:2013}
{Simha} V, {Cole} S (2013) {Cosmological constraints from applying SHAM to
  rescaled cosmological simulations}. \mnras 436(2):1142--1151.
  \doi{10.1093/mnras/stt1643}.
  {\href{https://arxiv.org/abs/1302.0852}{{arXiv:1302.0852}}} {[astro-ph.CO]}

\bibitem[{{Sinha} and {Garrison}(2020)}]{Sinha:2020}
{Sinha} M, {Garrison} LH (2020) {CORRFUNC - a suite of blazing fast correlation
  functions on the CPU}. \mnras 491(2):3022--3041. \doi{10.1093/mnras/stz3157}.
  {\href{https://arxiv.org/abs/1911.03545}{{arXiv:1911.03545}}} {[astro-ph.CO]}

\bibitem[{{Sirko}(2005)}]{Sirko:2005}
{Sirko} E (2005) {Initial Conditions to Cosmological N-Body Simulations, or,
  How to Run an Ensemble of Simulations}. \apj 634:728--743.
  \doi{10.1086/497090}.
  {\href{https://arxiv.org/abs/astro-ph/0503106}{{astro-ph/0503106}}}

\bibitem[{{Skillman} et~al.(2014){Skillman}, {Warren}, {Turk}, {Wechsler},
  {Holz}, and {Sutter}}]{Skillman:2014}
{Skillman} SW, {Warren} MS, {Turk} MJ, {Wechsler} RH, {Holz} DE, {Sutter} PM
  (2014) {Dark Sky Simulations: Early Data Release}. arXiv e-prints
  arXiv:1407.2600. {\href{https://arxiv.org/abs/1407.2600}{{arXiv:1407.2600}}}
  {[astro-ph.CO]}

\bibitem[{{Skory} et~al.(2010){Skory}, {Turk}, {Norman}, and
  {Coil}}]{Skory:2010}
{Skory} S, {Turk} MJ, {Norman} ML, {Coil} AL (2010) {Parallel HOP: A Scalable
  Halo Finder for Massive Cosmological Data Sets}. \apjs 191(1):43--57.
  \doi{10.1088/0067-0049/191/1/43}.
  {\href{https://arxiv.org/abs/1001.3411}{{arXiv:1001.3411}}} {[astro-ph.CO]}

\bibitem[{{Slosar} et~al.(2008){Slosar}, {Hirata}, {Seljak}, {Ho}, and
  {Padmanabhan}}]{Slosar:2008}
{Slosar} A, {Hirata} C, {Seljak} U, {Ho} S, {Padmanabhan} N (2008) {Constraints
  on local primordial non-Gaussianity from large scale structure}. \jcap
  2008(8):031. \doi{10.1088/1475-7516/2008/08/031}.
  {\href{https://arxiv.org/abs/0805.3580}{{arXiv:0805.3580}}} {[astro-ph]}

\bibitem[{{Smith} et~al.(2017){Smith}, {Cole}, {Baugh}, {Zheng}, {Angulo},
  {Norberg}, and {Zehavi}}]{Smith:2017}
{Smith} A, {Cole} S, {Baugh} C, {Zheng} Z, {Angulo} R, {Norberg} P, {Zehavi} I
  (2017) {A lightcone catalogue from the Millennium-XXL simulation}. \mnras
  470(4):4646--4661. \doi{10.1093/mnras/stx1432}.
  {\href{https://arxiv.org/abs/1701.06581}{{arXiv:1701.06581}}} {[astro-ph.CO]}

\bibitem[{{Smith} et~al.(2003){Smith}, {Peacock}, {Jenkins}, {White}, {Frenk},
  {Pearce}, {Thomas}, {Efstathiou}, and {Couchman}}]{Smith:2003}
{Smith} RE, {Peacock} JA, {Jenkins} A, {White} SDM, {Frenk} CS, {Pearce} FR,
  {Thomas} PA, {Efstathiou} G, {Couchman} HMP (2003) {Stable clustering, the
  halo model and non-linear cosmological power spectra}. \mnras
  341(4):1311--1332. \doi{10.1046/j.1365-8711.2003.06503.x}.
  {\href{https://arxiv.org/abs/astro-ph/0207664}{{arXiv:astro-ph/0207664}}}
  {[astro-ph]}

\bibitem[{{Somerville} and {Dav{\'e}}(2015)}]{Somerville:2015}
{Somerville} RS, {Dav{\'e}} R (2015) {Physical Models of Galaxy Formation in a
  Cosmological Framework}. \araa 53:51--113.
  \doi{10.1146/annurev-astro-082812-140951}.
  {\href{https://arxiv.org/abs/1412.2712}{{arXiv:1412.2712}}} {[astro-ph.GA]}

\bibitem[{{Sorce}(2020)}]{Sorce:2020}
{Sorce} JG (2020) {Efficiently estimating mean, uncertainty, and unconstrained
  large-scale fraction of local Universe simulations with paired fixed fields}.
  \mnras 495(4):4463--4474. \doi{10.1093/mnras/staa1432}.
  {\href{https://arxiv.org/abs/2006.01838}{{arXiv:2006.01838}}} {[astro-ph.CO]}

\bibitem[{{Sorce} et~al.(2014){Sorce}, {Courtois}, {Gottl{\"o}ber}, {Hoffman},
  and {Tully}}]{Sorce:2014}
{Sorce} JG, {Courtois} HM, {Gottl{\"o}ber} S, {Hoffman} Y, {Tully} RB (2014)
  {Simulations of the Local Universe constrained by observational peculiar
  velocities}. \mnras 437(4):3586--3595. \doi{10.1093/mnras/stt2153}.
  {\href{https://arxiv.org/abs/1311.2253}{{arXiv:1311.2253}}} {[astro-ph.CO]}

\bibitem[{{Sousbie}(2011)}]{Sousbie:2011}
{Sousbie} T (2011) {The persistent cosmic web and its filamentary structure -
  I. Theory and implementation}. \mnras 414(1):350--383.
  \doi{10.1111/j.1365-2966.2011.18394.x}.
  {\href{https://arxiv.org/abs/1009.4015}{{arXiv:1009.4015}}} {[astro-ph.CO]}

\bibitem[{{Sousbie} and {Colombi}(2016)}]{Sousbie:2016}
{Sousbie} T, {Colombi} S (2016) {ColDICE: A parallel Vlasov-Poisson solver
  using moving adaptive simplicial tessellation}. J Comput Phys 321:644--697.
  \doi{10.1016/j.jcp.2016.05.048}.
  {\href{https://arxiv.org/abs/1509.07720}{{arXiv:1509.07720}}}
  {[physics.comp-ph]}

\bibitem[{{Splinter} et~al.(1998){Splinter}, {Melott}, {Shand arin}, and
  {Suto}}]{Splinter:1998}
{Splinter} RJ, {Melott} AL, {Shand arin} SF, {Suto} Y (1998) {Fundamental
  Discreteness Limitations of Cosmological N-Body Clustering Simulations}. \apj
  497(1):38--61. \doi{10.1086/305450}.
  {\href{https://arxiv.org/abs/astro-ph/9706099}{{arXiv:astro-ph/9706099}}}
  {[astro-ph]}

\bibitem[{{Springel}(2005)}]{Springel:2005b}
{Springel} V (2005) {The cosmological simulation code GADGET-2}. \mnras
  364:1105--1134. \doi{10.1111/j.1365-2966.2005.09655.x}.
  {\href{https://arxiv.org/abs/astro-ph/0505010}{{astro-ph/0505010}}}

\bibitem[{{Springel} et~al.(2001{\natexlab{a}}){Springel}, {White}, {Tormen},
  and {Kauffmann}}]{Springel:2001}
{Springel} V, {White} SDM, {Tormen} G, {Kauffmann} G (2001{\natexlab{a}})
  {Populating a cluster of galaxies - I. Results at [formmu2]z=0}. \mnras
  328(3):726--750. \doi{10.1046/j.1365-8711.2001.04912.x}.
  {\href{https://arxiv.org/abs/astro-ph/0012055}{{arXiv:astro-ph/0012055}}}
  {[astro-ph]}

\bibitem[{{Springel} et~al.(2001{\natexlab{b}}){Springel}, {Yoshida}, and
  {White}}]{Springel:2001b}
{Springel} V, {Yoshida} N, {White} SDM (2001{\natexlab{b}}) {GADGET: a code for
  collisionless and gasdynamical cosmological simulations}. \na 6(2):79--117.
  \doi{10.1016/S1384-1076(01)00042-2}.
  {\href{https://arxiv.org/abs/astro-ph/0003162}{{arXiv:astro-ph/0003162}}}
  {[astro-ph]}

\bibitem[{{Springel} et~al.(2005){Springel}, {White}, {Jenkins}, {Frenk},
  {Yoshida}, {Gao}, {Navarro}, {Thacker}, {Croton}, {Helly}, {Peacock}, {Cole},
  {Thomas}, {Couchman}, {Evrard}, {Colberg}, and {Pearce}}]{Springel:2005a}
{Springel} V, {White} SDM, {Jenkins} A, {Frenk} CS, {Yoshida} N, {Gao} L,
  {Navarro} J, {Thacker} R, {Croton} D, {Helly} J, {Peacock} JA, {Cole} S,
  {Thomas} P, {Couchman} H, {Evrard} A, {Colberg} J, {Pearce} F (2005)
  {Simulations of the formation, evolution and clustering of galaxies and
  quasars}. \nat 435(7042):629--636. \doi{10.1038/nature03597}.
  {\href{https://arxiv.org/abs/astro-ph/0504097}{{arXiv:astro-ph/0504097}}}
  {[astro-ph]}

\bibitem[{{Springel} et~al.(2006){Springel}, {Frenk}, and
  {White}}]{Springel:2006}
{Springel} V, {Frenk} CS, {White} SDM (2006) {The large-scale structure of the
  Universe}. \nat 440(7088):1137--1144. \doi{10.1038/nature04805}.
  {\href{https://arxiv.org/abs/astro-ph/0604561}{{arXiv:astro-ph/0604561}}}
  {[astro-ph]}

\bibitem[{{Springel} et~al.(2008{\natexlab{a}}){Springel}, {Wang},
  {Vogelsberger}, {Ludlow}, {Jenkins}, {Helmi}, {Navarro}, {Frenk}, and
  {White}}]{Springel:2008}
{Springel} V, {Wang} J, {Vogelsberger} M, {Ludlow} A, {Jenkins} A, {Helmi} A,
  {Navarro} JF, {Frenk} CS, {White} SDM (2008{\natexlab{a}}) {The Aquarius
  Project: the subhaloes of galactic haloes}. \mnras 391(4):1685--1711.
  \doi{10.1111/j.1365-2966.2008.14066.x}.
  {\href{https://arxiv.org/abs/0809.0898}{{arXiv:0809.0898}}} {[astro-ph]}

\bibitem[{{Springel} et~al.(2008{\natexlab{b}}){Springel}, {White}, {Frenk},
  {Navarro}, {Jenkins}, {Vogelsberger}, {Wang}, {Ludlow}, and
  {Helmi}}]{Springel:2008b}
{Springel} V, {White} SDM, {Frenk} CS, {Navarro} JF, {Jenkins} A,
  {Vogelsberger} M, {Wang} J, {Ludlow} A, {Helmi} A (2008{\natexlab{b}})
  {Prospects for detecting supersymmetric dark matter in the Galactic halo}.
  \nat 456(7218):73--76. \doi{10.1038/nature07411}.
  {\href{https://arxiv.org/abs/0809.0894}{{arXiv:0809.0894}}} {[astro-ph]}

\bibitem[{{Springel} et~al.(2018){Springel}, {Pakmor}, {Pillepich},
  {Weinberger}, {Nelson}, {Hernquist}, {Vogelsberger}, {Genel}, {Torrey},
  {Marinacci}, and {Naiman}}]{Springel:2018}
{Springel} V, {Pakmor} R, {Pillepich} A, {Weinberger} R, {Nelson} D,
  {Hernquist} L, {Vogelsberger} M, {Genel} S, {Torrey} P, {Marinacci} F,
  {Naiman} J (2018) {First results from the IllustrisTNG simulations: matter
  and galaxy clustering}. \mnras 475(1):676--698. \doi{10.1093/mnras/stx3304}.
  {\href{https://arxiv.org/abs/1707.03397}{{arXiv:1707.03397}}} {[astro-ph.GA]}

\bibitem[{{Springel} et~al.(2020){Springel}, {Pakmor}, {Zier}, and
  {Reinecke}}]{Springel:2020}
{Springel} V, {Pakmor} R, {Zier} O, {Reinecke} M (2020) {Simulating cosmic
  structure formation with the GADGET-4 code}. arXiv e-prints arXiv:2010.03567.
  {\href{https://arxiv.org/abs/2010.03567}{{arXiv:2010.03567}}} {[astro-ph.IM]}

\bibitem[{{Srinivasan} et~al.(2021){Srinivasan}, {Thomas}, {Pace}, and
  {Battye}}]{Srinivasan:2021}
{Srinivasan} S, {Thomas} DB, {Pace} F, {Battye} R (2021) {Cosmological gravity
  on all scales II: Model independent modified gravity $N$-body simulations}.
  arXiv e-prints arXiv:2103.05051.
  {\href{https://arxiv.org/abs/2103.05051}{{arXiv:2103.05051}}} {[astro-ph.CO]}

\bibitem[{{Stadel}(2001)}]{Stadel:2001}
{Stadel} JG (2001) {Cosmological N-body simulations and their analysis}. PhD
  thesis, UNIVERSITY OF WASHINGTON

\bibitem[{{Stein} et~al.(2020){Stein}, {Alvarez}, {Bond}, {van Engelen}, and
  {Battaglia}}]{Stein:2020}
{Stein} G, {Alvarez} MA, {Bond} JR, {van Engelen} A, {Battaglia} N (2020) {The
  Websky extragalactic CMB simulations}. \jcap 2020(10):012.
  \doi{10.1088/1475-7516/2020/10/012}.
  {\href{https://arxiv.org/abs/2001.08787}{{arXiv:2001.08787}}} {[astro-ph.CO]}

\bibitem[{{Stopyra} et~al.(2021){Stopyra}, {Pontzen}, {Peiris}, {Roth}, and
  {Rey}}]{Stopyra:2020}
{Stopyra} S, {Pontzen} A, {Peiris} H, {Roth} N, {Rey} MP (2021) {GenetIC---A
  New Initial Conditions Generator to Support Genetically Modified Zoom
  Simulations}. \apjs 252(2):28. \doi{10.3847/1538-4365/abcd94}.
  {\href{https://arxiv.org/abs/2006.01841}{{arXiv:2006.01841}}} {[astro-ph.IM]}

\bibitem[{{Straumann}(2013)}]{Straumann:2013}
{Straumann} N (2013) {General Relativity}. Springer.
  \doi{10.1007/978-94-007-5410-2}

\bibitem[{{St{\"u}cker} et~al.(2018{\natexlab{a}}){St{\"u}cker}, {Busch}, and
  {White}}]{Stucker:2018}
{St{\"u}cker} J, {Busch} P, {White} SDM (2018{\natexlab{a}}) {The median
  density of the Universe}. \mnras 477(3):3230--3246.
  \doi{10.1093/mnras/sty815}.
  {\href{https://arxiv.org/abs/1710.09881}{{arXiv:1710.09881}}} {[astro-ph.CO]}

\bibitem[{{St{\"u}cker} et~al.(2018{\natexlab{b}}){St{\"u}cker}, {Busch}, and
  {White}}]{Stuecker:2018}
{St{\"u}cker} J, {Busch} P, {White} SDM (2018{\natexlab{b}}) {The median
  density of the Universe}. \mnras 477(3):3230--3246.
  \doi{10.1093/mnras/sty815}.
  {\href{https://arxiv.org/abs/1710.09881}{{arXiv:1710.09881}}} {[astro-ph.CO]}

\bibitem[{{St{\"u}cker} et~al.(2020{\natexlab{a}}){St{\"u}cker}, {Hahn},
  {Angulo}, and {White}}]{Stuecker:2019}
{St{\"u}cker} J, {Hahn} O, {Angulo} RE, {White} SDM (2020{\natexlab{a}})
  {Simulating the complexity of the dark matter sheet I: numerical algorithms}.
  \mnras 495(4):4943--4964. \doi{10.1093/mnras/staa1468}.
  {\href{https://arxiv.org/abs/1909.00008}{{arXiv:1909.00008}}} {[astro-ph.CO]}

\bibitem[{{St{\"u}cker} et~al.(2020{\natexlab{b}}){St{\"u}cker}, {Schmidt},
  {White}, {Schmidt}, and {Hahn}}]{Stuecker:2020}
{St{\"u}cker} J, {Schmidt} AS, {White} SDM, {Schmidt} F, {Hahn} O
  (2020{\natexlab{b}}) {Measuring the Tidal Response of Structure Formation:
  Anisotropic Separate Universe Simulations using TreePM}. arXiv e-prints
  arXiv:2003.06427.
  {\href{https://arxiv.org/abs/2003.06427}{{arXiv:2003.06427}}} {[astro-ph.CO]}

\bibitem[{{St{\"u}cker} et~al.(2021{\natexlab{a}}){St{\"u}cker}, {Angulo}, and
  {Busch}}]{Stucker:2021a}
{St{\"u}cker} J, {Angulo} RE, {Busch} P (2021{\natexlab{a}}) {The boosted
  potential}. \mnras \doi{10.1093/mnras/stab2913}.
  {\href{https://arxiv.org/abs/2107.13008}{{arXiv:2107.13008}}} {[astro-ph.CO]}

\bibitem[{{St{\"u}cker} et~al.(2021{\natexlab{b}}){St{\"u}cker}, {Angulo},
  {Hahn}, and {White}}]{Stucker:2021}
{St{\"u}cker} J, {Angulo} RE, {Hahn} O, {White} SDM (2021{\natexlab{b}})
  {Simulating the complexity of the dark matter sheet II: halo and subhalo mass
  functions for non-cold dark matter models}. \mnras
  \doi{10.1093/mnras/stab3078}.
  {\href{https://arxiv.org/abs/2109.09760}{{arXiv:2109.09760}}} {[astro-ph.CO]}

\bibitem[{{St{\"u}cker} et~al.(2021{\natexlab{c}}){St{\"u}cker}, {Angulo},
  {Hahn}, and {White}}]{Stucker:2021b}
{St{\"u}cker} J, {Angulo} RE, {Hahn} O, {White} SDM (2021{\natexlab{c}})
  {Simulating the complexity of the dark matter sheet II: halo and subhalo mass
  functions for non-cold dark matter models}. \mnras
  \doi{10.1093/mnras/stab3078}.
  {\href{https://arxiv.org/abs/2109.09760}{{arXiv:2109.09760}}} {[astro-ph.CO]}

\bibitem[{{Suisalu} and {Saar}(1995)}]{Suisalu:1995}
{Suisalu} I, {Saar} E (1995) {An adaptive multigrid solver for high-resolution
  cosmological simulations}. \mnras 274(1):287--299.
  \doi{10.1093/mnras/274.1.287}.
  {\href{https://arxiv.org/abs/astro-ph/9412043}{{arXiv:astro-ph/9412043}}}
  {[astro-ph]}

\bibitem[{{Sunayama} et~al.(2016){Sunayama}, {Padmanabhan}, {Heitmann},
  {Habib}, and {Rangel}}]{Sunayama:2016}
{Sunayama} T, {Padmanabhan} N, {Heitmann} K, {Habib} S, {Rangel} E (2016)
  {Efficient construction of mock catalogs for baryon acoustic oscillation
  surveys}. \jcap 2016(5):051. \doi{10.1088/1475-7516/2016/05/051}.
  {\href{https://arxiv.org/abs/1510.06665}{{arXiv:1510.06665}}} {[astro-ph.CO]}

\bibitem[{{Svrcek} and {Witten}(2006)}]{Svrcek:2006}
{Svrcek} P, {Witten} E (2006) {Axions in string theory}. J High Energy Phys
  6:051. \doi{10.1088/1126-6708/2006/06/051}.
  {\href{https://arxiv.org/abs/hep-th/0605206}{{hep-th/0605206}}}

\bibitem[{{Syer} and {White}(1998)}]{SyerWhite:1998}
{Syer} D, {White} SDM (1998) {Dark halo mergers and the formation of a
  universal profile}. \mnras 293(4):337--342.
  \doi{10.1046/j.1365-8711.1998.01285.x}

\bibitem[{Taha and Ablowitz(1984)}]{Taha:1984}
Taha TR, Ablowitz MI (1984) Analytical and numerical aspects of certain
  nonlinear evolution equations. ii. numerical, nonlinear schr{\"o}dinger
  equation. Journal of Computational Physics 55(2):203--230.
  \doi{https://doi.org/10.1016/0021-9991(84)90003-2},
  \urlprefix\url{https://www.sciencedirect.com/science/article/pii/0021999184900032}

\bibitem[{{Takahashi} et~al.(2012){Takahashi}, {Sato}, {Nishimichi}, {Taruya},
  and {Oguri}}]{Takahashi:2012}
{Takahashi} R, {Sato} M, {Nishimichi} T, {Taruya} A, {Oguri} M (2012) {Revising
  the Halofit Model for the Nonlinear Matter Power Spectrum}. \apj 761(2):152.
  \doi{10.1088/0004-637X/761/2/152}.
  {\href{https://arxiv.org/abs/1208.2701}{{arXiv:1208.2701}}} {[astro-ph.CO]}

\bibitem[{{Takahashi} et~al.(2020){Takahashi}, {Nishimichi}, {Namikawa},
  {Taruya}, {Kayo}, {Osato}, {Kobayashi}, and {Shirasaki}}]{Takahashi:2020}
{Takahashi} R, {Nishimichi} T, {Namikawa} T, {Taruya} A, {Kayo} I, {Osato} K,
  {Kobayashi} Y, {Shirasaki} M (2020) {Fitting the Nonlinear Matter Bispectrum
  by the Halofit Approach}. \apj 895(2):113. \doi{10.3847/1538-4357/ab908d}.
  {\href{https://arxiv.org/abs/1911.07886}{{arXiv:1911.07886}}} {[astro-ph.CO]}

\bibitem[{{Tamosiunas} et~al.(2020){Tamosiunas}, {Winther}, {Koyama}, {Bacon},
  {Nichol}, and {Mawdsley}}]{Tamosiunas:2020}
{Tamosiunas} A, {Winther} HA, {Koyama} K, {Bacon} DJ, {Nichol} RC, {Mawdsley} B
  (2020) {Investigating Cosmological GAN Emulators Using Latent Space
  Interpolation}. arXiv e-prints arXiv:2004.10223.
  {\href{https://arxiv.org/abs/2004.10223}{{arXiv:2004.10223}}} {[astro-ph.CO]}

\bibitem[{{Tanaka} et~al.(2017){Tanaka}, {Yoshikawa}, {Minoshima}, and
  {Yoshida}}]{Tanaka:2017}
{Tanaka} S, {Yoshikawa} K, {Minoshima} T, {Yoshida} N (2017) {Multidimensional
  Vlasov-Poisson Simulations with High-order Monotonicity- and
  Positivity-preserving Schemes}. \apj 849:76. \doi{10.3847/1538-4357/aa901f}.
  {\href{https://arxiv.org/abs/1702.08521}{{arXiv:1702.08521}}}
  {[physics.comp-ph]}

\bibitem[{{Taruya} and {Colombi}(2017)}]{Taruya:2017}
{Taruya} A, {Colombi} S (2017) {Post-collapse perturbation theory in 1D
  cosmology - beyond shell-crossing}. \mnras 470(4):4858--4884.
  \doi{10.1093/mnras/stx1501}.
  {\href{https://arxiv.org/abs/1701.09088}{{arXiv:1701.09088}}} {[astro-ph.CO]}

\bibitem[{{Tassev} et~al.(2013){Tassev}, {Zaldarriaga}, and
  {Eisenstein}}]{Tassev:2013}
{Tassev} S, {Zaldarriaga} M, {Eisenstein} DJ (2013) {Solving large scale
  structure in ten easy steps with COLA}. \jcap 2013(6):036.
  \doi{10.1088/1475-7516/2013/06/036}.
  {\href{https://arxiv.org/abs/1301.0322}{{arXiv:1301.0322}}} {[astro-ph.CO]}

\bibitem[{{Tassev} et~al.(2015){Tassev}, {Eisenstein}, {Wandelt}, and
  {Zaldarriaga}}]{Tassev:2015}
{Tassev} S, {Eisenstein} DJ, {Wandelt} BD, {Zaldarriaga} M (2015) {sCOLA: The
  N-body COLA Method Extended to the Spatial Domain}. arXiv e-prints
  arXiv:1502.07751.
  {\href{https://arxiv.org/abs/1502.07751}{{arXiv:1502.07751}}} {[astro-ph.CO]}

\bibitem[{{Taylor} and {Navarro}(2001)}]{Taylor:2001}
{Taylor} JE, {Navarro} JF (2001) {The Phase-Space Density Profiles of Cold Dark
  Matter Halos}. \apj 563(2):483--488. \doi{10.1086/324031}.
  {\href{https://arxiv.org/abs/astro-ph/0104002}{{arXiv:astro-ph/0104002}}}
  {[astro-ph]}

\bibitem[{{Teyssier}(2002)}]{Teyssier:2002}
{Teyssier} R (2002) {Cosmological hydrodynamics with adaptive mesh refinement.
  A new high resolution code called RAMSES}. \aap 385:337--364.
  \doi{10.1051/0004-6361:20011817}.
  {\href{https://arxiv.org/abs/astro-ph/0111367}{{astro-ph/0111367}}}

\bibitem[{{Thi{\'e}baut} et~al.(2008){Thi{\'e}baut}, {Pichon}, {Sousbie},
  {Prunet}, and {Pogosyan}}]{Thiebaut:2008}
{Thi{\'e}baut} J, {Pichon} C, {Sousbie} T, {Prunet} S, {Pogosyan} D (2008) {On
  the onset of stochasticity in {\ensuremath{\Lambda}} cold dark matter
  cosmological simulations}. \mnras 387(1):397--406.
  \doi{10.1111/j.1365-2966.2008.13250.x}.
  {\href{https://arxiv.org/abs/0803.3120}{{arXiv:0803.3120}}} {[astro-ph]}

\bibitem[{{Thomas}(2020)}]{Thomas:2020}
{Thomas} DB (2020) {Cosmological gravity on all scales: Simple equations,
  required conditions, and a framework for modified gravity}. \prd
  101(12):123517. \doi{10.1103/PhysRevD.101.123517}.
  {\href{https://arxiv.org/abs/2004.13051}{{arXiv:2004.13051}}} {[gr-qc]}

\bibitem[{{Tinker} et~al.(2008){Tinker}, {Kravtsov}, {Klypin}, {Abazajian},
  {Warren}, {Yepes}, {Gottl{\"o}ber}, and {Holz}}]{Tinker:2008}
{Tinker} J, {Kravtsov} AV, {Klypin} A, {Abazajian} K, {Warren} M, {Yepes} G,
  {Gottl{\"o}ber} S, {Holz} DE (2008) {Toward a Halo Mass Function for
  Precision Cosmology: The Limits of Universality}. \apj 688(2):709--728.
  \doi{10.1086/591439}.
  {\href{https://arxiv.org/abs/0803.2706}{{arXiv:0803.2706}}} {[astro-ph]}

\bibitem[{{Tinyakov} et~al.(2016){Tinyakov}, {Tkachev}, and
  {Zioutas}}]{Tinyakov:2016}
{Tinyakov} P, {Tkachev} I, {Zioutas} K (2016) {Tidal streams from axion
  miniclusters and direct axion searches}. \jcap 2016(1):035.
  \doi{10.1088/1475-7516/2016/01/035}.
  {\href{https://arxiv.org/abs/1512.02884}{{arXiv:1512.02884}}} {[astro-ph.CO]}

\bibitem[{{Tkachev} et~al.(2020){Tkachev}, {Pilipenko}, and
  {Yepes}}]{Tkachev:2020}
{Tkachev} MV, {Pilipenko} SV, {Yepes} G (2020) {Dark matter simulations with
  primordial black holes in the early Universe}. \mnras 499(4):4854--4862.
  \doi{10.1093/mnras/staa3103}.
  {\href{https://arxiv.org/abs/2009.07813}{{arXiv:2009.07813}}} {[astro-ph.CO]}

\bibitem[{{Tonnesen} and {Ostriker}(2021)}]{Tonnesen:2021}
{Tonnesen} S, {Ostriker} JP (2021) {An Improved and Physically-Motivated Scheme
  for Matching Galaxies with Dark Matter Halos}. arXiv e-prints
  arXiv:2102.13122.
  {\href{https://arxiv.org/abs/2102.13122}{{arXiv:2102.13122}}} {[astro-ph.GA]}

\bibitem[{{Tosone} et~al.(2021){Tosone}, {Neyrinck}, {Granett}, {Guzzo}, and
  {Vittorio}}]{Tosone:2021}
{Tosone} F, {Neyrinck} MC, {Granett} BR, {Guzzo} L, {Vittorio} N (2021)
  {MUSCLE-UPS: improved approximations of the matter field with the extended
  Press-Schechter formalism and Lagrangian perturbation theory}. \mnras
  505(2):2999--3015. \doi{10.1093/mnras/stab1517}.
  {\href{https://arxiv.org/abs/2012.14446}{{arXiv:2012.14446}}} {[astro-ph.CO]}

\bibitem[{{Tram} et~al.(2019){Tram}, {Brandbyge}, {Dakin}, and
  {Hannestad}}]{Tram2018}
{Tram} T, {Brandbyge} J, {Dakin} J, {Hannestad} S (2019) {Fully relativistic
  treatment of light neutrinos in N-body simulations}. \jcap 2019(3):022.
  \doi{10.1088/1475-7516/2019/03/022}.
  {\href{https://arxiv.org/abs/1811.00904}{{arXiv:1811.00904}}} {[astro-ph.CO]}

\bibitem[{{Tremaine} and {Gunn}(1979)}]{Tremaine:1979}
{Tremaine} S, {Gunn} JE (1979) {Dynamical role of light neutral leptons in
  cosmology}. \prl 42(6):407--410. \doi{10.1103/PhysRevLett.42.407}

\bibitem[{{Tr{\"o}ster} et~al.(2019){Tr{\"o}ster}, {Ferguson},
  {Harnois-D{\'e}raps}, and {McCarthy}}]{Troester:2019}
{Tr{\"o}ster} T, {Ferguson} C, {Harnois-D{\'e}raps} J, {McCarthy} IG (2019)
  {Painting with baryons: augmenting N-body simulations with gas using deep
  generative models}. \mnras 487(1):L24--L29. \doi{10.1093/mnrasl/slz075}.
  {\href{https://arxiv.org/abs/1903.12173}{{arXiv:1903.12173}}} {[astro-ph.CO]}

\bibitem[{{Tr{\"o}ster} et~al.(2021){Tr{\"o}ster}, {Mead}, {Heymans}, {Yan},
  {Alonso}, {Asgari}, {Bilicki}, {Dvornik}, {Hildebrandt}, {Joachimi},
  {Kannawadi}, {Kuijken}, {Schneider}, {Shan}, {van Waerbeke}, and
  {Wright}}]{Troster:2021}
{Tr{\"o}ster} T, {Mead} AJ, {Heymans} C, {Yan} Z, {Alonso} D, {Asgari} M,
  {Bilicki} M, {Dvornik} A, {Hildebrandt} H, {Joachimi} B, {Kannawadi} A,
  {Kuijken} K, {Schneider} P, {Shan} H, {van Waerbeke} L, {Wright} AH (2021)
  {Joint constraints on cosmology and the impact of baryon feedback: combining
  KiDS-1000 lensing with the thermal Sunyaev-Zeldovich effect from Planck and
  ACT}. arXiv e-prints arXiv:2109.04458.
  {\href{https://arxiv.org/abs/2109.04458}{{arXiv:2109.04458}}} {[astro-ph.CO]}

\bibitem[{Trottenberg et~al.(2001)Trottenberg, Oosterlee, and
  Sch{\"u}ller}]{Trottenberg:2001}
Trottenberg U, Oosterlee CW, Sch{\"u}ller A (2001) Multigrid, Texts in Applied
  Mathematics. Bd., vol~33. Academic Press, San Diego. With contributions by A.
  Brandt, P. Oswald and K. St{\"u}ben

\bibitem[{{Trujillo-Gomez} et~al.(2011){Trujillo-Gomez}, {Klypin}, {Primack},
  and {Romanowsky}}]{Trujillo-Gomez:2011}
{Trujillo-Gomez} S, {Klypin} A, {Primack} J, {Romanowsky} AJ (2011) {Galaxies
  in {\ensuremath{\Lambda}}CDM with Halo Abundance Matching:
  Luminosity-Velocity Relation, Baryonic Mass-Velocity Relation, Velocity
  Function, and Clustering}. \apj 742(1):16. \doi{10.1088/0004-637X/742/1/16}.
  {\href{https://arxiv.org/abs/1005.1289}{{arXiv:1005.1289}}} {[astro-ph.CO]}

\bibitem[{{Tseliakhovich} and {Hirata}(2010)}]{Tseliakhovich:2010}
{Tseliakhovich} D, {Hirata} C (2010) {Relative velocity of dark matter and
  baryonic fluids and the formation of the first structures}. \prd
  82(8):083520. \doi{10.1103/PhysRevD.82.083520}.
  {\href{https://arxiv.org/abs/1005.2416}{{arXiv:1005.2416}}}

\bibitem[{{Tulin} and {Yu}(2018)}]{Tulin:2018}
{Tulin} S, {Yu} HB (2018) {Dark matter self-interactions and small scale
  structure}. \physrep 730:1--57. \doi{10.1016/j.physrep.2017.11.004}.
  {\href{https://arxiv.org/abs/1705.02358}{{arXiv:1705.02358}}} {[hep-ph]}

\bibitem[{{Tully} et~al.(2016){Tully}, {Courtois}, and {Sorce}}]{Tully:2016}
{Tully} RB, {Courtois} HM, {Sorce} JG (2016) {Cosmicflows-3}. \aj 152(2):50.
  \doi{10.3847/0004-6256/152/2/50}.
  {\href{https://arxiv.org/abs/1605.01765}{{arXiv:1605.01765}}} {[astro-ph.CO]}

\bibitem[{{Tweed} et~al.(2017){Tweed}, {Yang}, {Wang}, {Cui}, {Zhang}, {Li},
  {Jing}, and {Mo}}]{ElucidII:2017}
{Tweed} D, {Yang} X, {Wang} H, {Cui} W, {Zhang} Y, {Li} S, {Jing} YP, {Mo} HJ
  (2017) {ELUCID---Exploring the Local Universe with the reConstructed Initial
  Density Field. II. Reconstruction Diagnostics, Applied to Numerical Halo
  Catalogs}. \apj 841(1):55. \doi{10.3847/1538-4357/aa6bf8}.
  {\href{https://arxiv.org/abs/1704.03675}{{arXiv:1704.03675}}} {[astro-ph.CO]}

\bibitem[{{Uhlemann} et~al.(2014){Uhlemann}, {Kopp}, and
  {Haugg}}]{Uhlemann:2014}
{Uhlemann} C, {Kopp} M, {Haugg} T (2014) {Schr{\"o}dinger method as N-body
  double and UV completion of dust}. \prd 90(2):023517.
  \doi{10.1103/PhysRevD.90.023517}.
  {\href{https://arxiv.org/abs/1403.5567}{{arXiv:1403.5567}}}

\bibitem[{{Uhlemann} et~al.(2019){Uhlemann}, {Rampf}, {Gosenca}, and
  {Hahn}}]{Uhlemann:2019}
{Uhlemann} C, {Rampf} C, {Gosenca} M, {Hahn} O (2019) {Semiclassical path to
  cosmic large-scale structure}. \prd 99(8):083524.
  \doi{10.1103/PhysRevD.99.083524}.
  {\href{https://arxiv.org/abs/1812.05633}{{arXiv:1812.05633}}}

\bibitem[{{Ullmo} et~al.(2020){Ullmo}, {Decelle}, and {Aghanim}}]{Ullmo:2020}
{Ullmo} M, {Decelle} A, {Aghanim} N (2020) {Encoding large scale cosmological
  structure with Generative Adversarial Networks}. arXiv e-prints
  arXiv:2011.05244.
  {\href{https://arxiv.org/abs/2011.05244}{{arXiv:2011.05244}}} {[astro-ph.CO]}

\bibitem[{{Upadhye} et~al.(2014){Upadhye}, {Biswas}, {Pope}, {Heitmann},
  {Habib}, {Finkel}, and {Frontiere}}]{Upadhye:2014}
{Upadhye} A, {Biswas} R, {Pope} A, {Heitmann} K, {Habib} S, {Finkel} H,
  {Frontiere} N (2014) {Large-scale structure formation with massive neutrinos
  and dynamical dark energy}. \prd 89(10):103515.
  \doi{10.1103/PhysRevD.89.103515}.
  {\href{https://arxiv.org/abs/1309.5872}{{arXiv:1309.5872}}} {[astro-ph.CO]}

\bibitem[{Vainshtein(1972)}]{Vainshtein:1972}
Vainshtein AI (1972) {To the problem of nonvanishing gravitation mass}. Phys
  Lett B 39:393--394. \doi{10.1016/0370-2693(72)90147-5}

\bibitem[{{Valageas}(2002)}]{Valageas:2002}
{Valageas} P (2002) {Transients from Zel'dovich initial conditions}. \aap
  385:761--767. \doi{10.1051/0004-6361:20020187}.
  {\href{https://arxiv.org/abs/astro-ph/0112102}{{astro-ph/0112102}}}

\bibitem[{{Valcin} et~al.(2019){Valcin}, {Villaescusa-Navarro}, {Verde}, and
  {Raccanelli}}]{Valcin:2019}
{Valcin} D, {Villaescusa-Navarro} F, {Verde} L, {Raccanelli} A (2019)
  {BE-HaPPY: bias emulator for halo power spectrum including massive
  neutrinos}. \jcap 2019(12):057. \doi{10.1088/1475-7516/2019/12/057}.
  {\href{https://arxiv.org/abs/1901.06045}{{arXiv:1901.06045}}} {[astro-ph.CO]}

\bibitem[{{Vale} and {Ostriker}(2004)}]{ValeOstriker:2004}
{Vale} A, {Ostriker} JP (2004) {Linking halo mass to galaxy luminosity}. \mnras
  353(1):189--200. \doi{10.1111/j.1365-2966.2004.08059.x}.
  {\href{https://arxiv.org/abs/astro-ph/0402500}{{arXiv:astro-ph/0402500}}}
  {[astro-ph]}

\bibitem[{{Valogiannis} and {Bean}(2017)}]{Valogiannis:2017}
{Valogiannis} G, {Bean} R (2017) {Efficient simulations of large-scale
  structure in modified gravity cosmologies with comoving Lagrangian
  acceleration}. \prd 95(10):103515. \doi{10.1103/PhysRevD.95.103515}.
  {\href{https://arxiv.org/abs/1612.06469}{{arXiv:1612.06469}}} {[astro-ph.CO]}

\bibitem[{{van Daalen} et~al.(2011){van Daalen}, {Schaye}, {Booth}, and {Dalla
  Vecchia}}]{vanDaalen:2011}
{van Daalen} MP, {Schaye} J, {Booth} CM, {Dalla Vecchia} C (2011) {The effects
  of galaxy formation on the matter power spectrum: a challenge for precision
  cosmology}. \mnras 415:3649--3665. \doi{10.1111/j.1365-2966.2011.18981.x}.
  {\href{https://arxiv.org/abs/1104.1174}{{arXiv:1104.1174}}} {[astro-ph.CO]}

\bibitem[{{van Daalen} et~al.(2019){van Daalen}, {McCarthy}, and
  {Schaye}}]{vanDaalen:2019}
{van Daalen} MP, {McCarthy} IG, {Schaye} J (2019) {Exploring the effects of
  galaxy formation on matter clustering through a library of simulation power
  spectra}. arXiv e-prints
  {\href{https://arxiv.org/abs/1906.00968}{{arXiv:1906.00968}}}

\bibitem[{{van de Weygaert}(1994)}]{vandeweygaert:1994}
{van de Weygaert} R (1994) {Fragmenting the Universe. 3: The constructions and
  statistics of 3-D Voronoi tessellations}. \aap 283(2):361--406

\bibitem[{{van de Weygaert} and {Bernardeau}(1998)}]{vandeweygaert:1998}
{van de Weygaert} R, {Bernardeau} F (1998) {Velocity Fields and Tessellation
  Techniques: Unbiased Estimators of Omega}. In: {Mueller} V, {Gottloeber} S,
  {Muecket} JP, {Wambsganss} J (eds) Large Scale Structure: Tracks and Traces.
  pp 207--216

\bibitem[{{van de Weygaert} and {Bertschinger}(1996)}]{VanDeWeygaert:1996}
{van de Weygaert} R, {Bertschinger} E (1996) {Peak and gravity constraints in
  Gaussian primordial density fields: An application of the Hoffman-Ribak
  method}. \mnras 281:84. \doi{10.1093/mnras/281.1.84}.
  {\href{https://arxiv.org/abs/astro-ph/9507024}{{arXiv:astro-ph/9507024}}}
  {[astro-ph]}

\bibitem[{{van den Bosch}(2017)}]{vandenBosch:2017}
{van den Bosch} FC (2017) {Dissecting the evolution of dark matter subhaloes in
  the Bolshoi simulation}. \mnras 468(1):885--909. \doi{10.1093/mnras/stx520}.
  {\href{https://arxiv.org/abs/1611.02657}{{arXiv:1611.02657}}} {[astro-ph.GA]}

\bibitem[{{van den Bosch} and {Ogiya}(2018)}]{vandenBosch:2018b}
{van den Bosch} FC, {Ogiya} G (2018) {Dark matter substructure in numerical
  simulations: a tale of discreteness noise, runaway instabilities, and
  artificial disruption}. \mnras 475(3):4066--4087. \doi{10.1093/mnras/sty084}.
  {\href{https://arxiv.org/abs/1801.05427}{{arXiv:1801.05427}}} {[astro-ph.GA]}

\bibitem[{{van den Bosch} et~al.(2003){van den Bosch}, {Yang}, and
  {Mo}}]{vandenBosch:2003}
{van den Bosch} FC, {Yang} X, {Mo} HJ (2003) {Linking early- and late-type
  galaxies to their dark matter haloes}. \mnras 340(3):771--792.
  \doi{10.1046/j.1365-8711.2003.06335.x}.
  {\href{https://arxiv.org/abs/astro-ph/0210495}{{arXiv:astro-ph/0210495}}}
  {[astro-ph]}

\bibitem[{{van den Bosch} et~al.(2007){van den Bosch}, {Yang}, {Mo},
  {Weinmann}, {Macci{\`o}}, {More}, {Cacciato}, {Skibba}, and
  {Kang}}]{vandenBosch:2007}
{van den Bosch} FC, {Yang} X, {Mo} HJ, {Weinmann} SM, {Macci{\`o}} AV, {More}
  S, {Cacciato} M, {Skibba} R, {Kang} X (2007) {Towards a concordant model of
  halo occupation statistics}. \mnras 376(2):841--860.
  \doi{10.1111/j.1365-2966.2007.11493.x}.
  {\href{https://arxiv.org/abs/astro-ph/0610686}{{arXiv:astro-ph/0610686}}}
  {[astro-ph]}

\bibitem[{{van den Bosch} et~al.(2018){van den Bosch}, {Ogiya}, {Hahn}, and
  {Burkert}}]{vandenBosch:2018a}
{van den Bosch} FC, {Ogiya} G, {Hahn} O, {Burkert} A (2018) {Disruption of dark
  matter substructure: fact or fiction?} \mnras 474(3):3043--3066.
  \doi{10.1093/mnras/stx2956}.
  {\href{https://arxiv.org/abs/1711.05276}{{arXiv:1711.05276}}} {[astro-ph.GA]}

\bibitem[{{Vaquero} et~al.(2019){Vaquero}, {Redondo}, and
  {Stadler}}]{Vaquero:2018}
{Vaquero} A, {Redondo} J, {Stadler} J (2019) {Early seeds of axion
  miniclusters}. \jcap 2019(4):012. \doi{10.1088/1475-7516/2019/04/012}.
  {\href{https://arxiv.org/abs/1809.09241}{{arXiv:1809.09241}}} {[astro-ph.CO]}

\bibitem[{{Veltmaat} and {Niemeyer}(2016)}]{Veltmaat:2016}
{Veltmaat} J, {Niemeyer} JC (2016) {Cosmological particle-in-cell simulations
  with ultralight axion dark matter}. \prd 94(12):123523.
  \doi{10.1103/PhysRevD.94.123523}.
  {\href{https://arxiv.org/abs/1608.00802}{{arXiv:1608.00802}}} {[astro-ph.CO]}

\bibitem[{{Veltmaat} et~al.(2018){Veltmaat}, {Niemeyer}, and
  {Schwabe}}]{Veltmaat:2018}
{Veltmaat} J, {Niemeyer} JC, {Schwabe} B (2018) {Formation and structure of
  ultralight bosonic dark matter halos}. \prd 98(4):043509.
  \doi{10.1103/PhysRevD.98.043509}.
  {\href{https://arxiv.org/abs/1804.09647}{{arXiv:1804.09647}}} {[astro-ph.CO]}

\bibitem[{{Viel} et~al.(2005){Viel}, {Lesgourgues}, {Haehnelt}, {Matarrese},
  and {Riotto}}]{Viel:2005}
{Viel} M, {Lesgourgues} J, {Haehnelt} MG, {Matarrese} S, {Riotto} A (2005)
  {Constraining warm dark matter candidates including sterile neutrinos and
  light gravitinos with WMAP and the Lyman-{\ensuremath{\alpha}} forest}. \prd
  71(6):063534. \doi{10.1103/PhysRevD.71.063534}.
  {\href{https://arxiv.org/abs/astro-ph/0501562}{{arXiv:astro-ph/0501562}}}
  {[astro-ph]}

\bibitem[{{Viel} et~al.(2010){Viel}, {Haehnelt}, and {Springel}}]{Viel:2010}
{Viel} M, {Haehnelt} MG, {Springel} V (2010) {The effect of neutrinos on the
  matter distribution as probed by the intergalactic medium}. \jcap
  2010(6):015. \doi{10.1088/1475-7516/2010/06/015}.
  {\href{https://arxiv.org/abs/1003.2422}{{arXiv:1003.2422}}} {[astro-ph.CO]}

\bibitem[{{Villaescusa-Navarro}(2018)}]{pylians:2018}
{Villaescusa-Navarro} F (2018) {Pylians: Python libraries for the analysis of
  numerical simulations}.
  {\href{https://arxiv.org/abs/1811.008}{{ascl:1811.008}}}

\bibitem[{{Villaescusa-Navarro} et~al.(2018){Villaescusa-Navarro}, {Naess},
  {Genel}, {Pontzen}, {Wandelt}, {Anderson}, {Font-Ribera}, {Battaglia}, and
  {Spergel}}]{Villaescusa-Navarro:2018}
{Villaescusa-Navarro} F, {Naess} S, {Genel} S, {Pontzen} A, {Wandelt} B,
  {Anderson} L, {Font-Ribera} A, {Battaglia} N, {Spergel} DN (2018)
  {Statistical Properties of Paired Fixed Fields}. \apj 867(2):137.
  \doi{10.3847/1538-4357/aae52b}.
  {\href{https://arxiv.org/abs/1806.01871}{{arXiv:1806.01871}}} {[astro-ph.CO]}

\bibitem[{{Villaescusa-Navarro}
  et~al.(2020{\natexlab{a}}){Villaescusa-Navarro}, {Angl{\'e}s-Alc{\'a}zar},
  {Genel}, {Spergel}, {Somerville}, {Dave}, {Pillepich}, {Hernquist}, {Nelson},
  {Torrey}, {Narayanan}, {Li}, {Philcox}, {La Torre}, {Delgado}, {Ho},
  {Hassan}, {Burkhart}, {Wadekar}, {Battaglia}, and
  {Contardo}}]{Villaescusa-Navarro:2020}
{Villaescusa-Navarro} F, {Angl{\'e}s-Alc{\'a}zar} D, {Genel} S, {Spergel} DN,
  {Somerville} RS, {Dave} R, {Pillepich} A, {Hernquist} L, {Nelson} D, {Torrey}
  P, {Narayanan} D, {Li} Y, {Philcox} O, {La Torre} V, {Delgado} AM, {Ho} S,
  {Hassan} S, {Burkhart} B, {Wadekar} D, {Battaglia} N, {Contardo} G
  (2020{\natexlab{a}}) {The CAMELS project: Cosmology and Astrophysics with
  MachinE Learning Simulations}. arXiv e-prints arXiv:2010.00619.
  {\href{https://arxiv.org/abs/2010.00619}{{arXiv:2010.00619}}} {[astro-ph.CO]}

\bibitem[{{Villaescusa-Navarro}
  et~al.(2020{\natexlab{b}}){Villaescusa-Navarro}, {Wandelt},
  {Angl{\'e}s-Alc{\'a}zar}, {Genel}, {Zorrilla Mantilla}, {Ho}, and
  {Spergel}}]{Villaescusa-Navarro:2020b}
{Villaescusa-Navarro} F, {Wandelt} BD, {Angl{\'e}s-Alc{\'a}zar} D, {Genel} S,
  {Zorrilla Mantilla} JM, {Ho} S, {Spergel} DN (2020{\natexlab{b}}) {Neural
  networks as optimal estimators to marginalize over baryonic effects}. arXiv
  e-prints arXiv:2011.05992.
  {\href{https://arxiv.org/abs/2011.05992}{{arXiv:2011.05992}}} {[astro-ph.CO]}

\bibitem[{{Visscher} and {Apalkov}(2010)}]{Visscher:2010}
{Visscher} PB, {Apalkov} DM (2010) {Simple recursive implementation of fast
  multipole method}. J Magnetism Magnetic Materials 322(2):275--281.
  \doi{10.1016/j.jmmm.2009.09.033}

\bibitem[{{Vogelsberger} and {White}(2011)}]{Vogelsberger:2011}
{Vogelsberger} M, {White} SDM (2011) {Streams and caustics: the fine-grained
  structure of {\ensuremath{\Lambda}} cold dark matter haloes}. \mnras
  413(2):1419--1438. \doi{10.1111/j.1365-2966.2011.18224.x}.
  {\href{https://arxiv.org/abs/1002.3162}{{arXiv:1002.3162}}} {[astro-ph.CO]}

\bibitem[{{Vogelsberger} et~al.(2008){Vogelsberger}, {White}, {Helmi}, and
  {Springel}}]{Vogelsberger:2008}
{Vogelsberger} M, {White} SDM, {Helmi} A, {Springel} V (2008) {The fine-grained
  phase-space structure of cold dark matter haloes}. \mnras 385(1):236--254.
  \doi{10.1111/j.1365-2966.2007.12746.x}.
  {\href{https://arxiv.org/abs/0711.1105}{{arXiv:0711.1105}}} {[astro-ph]}

\bibitem[{{Vogelsberger} et~al.(2012){Vogelsberger}, {Zavala}, and
  {Loeb}}]{Vogelsberger:2012}
{Vogelsberger} M, {Zavala} J, {Loeb} A (2012) {Subhaloes in self-interacting
  galactic dark matter haloes}. \mnras 423:3740--3752.
  \doi{10.1111/j.1365-2966.2012.21182.x}.
  {\href{https://arxiv.org/abs/1201.5892}{{arXiv:1201.5892}}}

\bibitem[{{Vogelsberger} et~al.(2014){Vogelsberger}, {Genel}, {Springel},
  {Torrey}, {Sijacki}, {Xu}, {Snyder}, {Nelson}, and
  {Hernquist}}]{Vogelsberger:2014}
{Vogelsberger} M, {Genel} S, {Springel} V, {Torrey} P, {Sijacki} D, {Xu} D,
  {Snyder} G, {Nelson} D, {Hernquist} L (2014) {Introducing the Illustris
  Project: simulating the coevolution of dark and visible matter in the
  Universe}. \mnras 444(2):1518--1547. \doi{10.1093/mnras/stu1536}.
  {\href{https://arxiv.org/abs/1405.2921}{{arXiv:1405.2921}}} {[astro-ph.CO]}

\bibitem[{{Vogelsberger} et~al.(2016){Vogelsberger}, {Zavala}, {Cyr-Racine},
  {Pfrommer}, {Bringmann}, and {Sigurdson}}]{Vogelsberger:2016}
{Vogelsberger} M, {Zavala} J, {Cyr-Racine} FY, {Pfrommer} C, {Bringmann} T,
  {Sigurdson} K (2016) {ETHOS - an effective theory of structure formation:
  dark matter physics as a possible explanation of the small-scale CDM
  problems}. \mnras 460(2):1399--1416. \doi{10.1093/mnras/stw1076}.
  {\href{https://arxiv.org/abs/1512.05349}{{arXiv:1512.05349}}} {[astro-ph.CO]}

\bibitem[{{Vogelsberger} et~al.(2020){Vogelsberger}, {Marinacci}, {Torrey}, and
  {Puchwein}}]{Vogelsberger:2020}
{Vogelsberger} M, {Marinacci} F, {Torrey} P, {Puchwein} E (2020) {Cosmological
  simulations of galaxy formation}. Nature Reviews Physics 2(1):42--66.
  \doi{10.1038/s42254-019-0127-2}.
  {\href{https://arxiv.org/abs/1909.07976}{{arXiv:1909.07976}}} {[astro-ph.GA]}

\bibitem[{{Voivodic} and {Barreira}(2020)}]{Voivodic:2020}
{Voivodic} R, {Barreira} A (2020) {Responses of Halo Occupation Distributions:
  a new ingredient in the halo model \& the impact on galaxy bias}. arXiv
  e-prints arXiv:2012.04637.
  {\href{https://arxiv.org/abs/2012.04637}{{arXiv:2012.04637}}} {[astro-ph.CO]}

\bibitem[{{von Hoerner}(1960)}]{Hoerner:1960}
{von Hoerner} S (1960) {Die numerische {I}ntegration des
  n-{K}{\"o}rper-{P}roblemes f{\"u}r {S}ternhaufen. I}. \zap 50

\bibitem[{{Wagner} and {Verde}(2012)}]{Wagner:2012b}
{Wagner} C, {Verde} L (2012) {N-body simulations with generic non-Gaussian
  initial conditions II: halo bias}. \jcap 2012(3):002.
  \doi{10.1088/1475-7516/2012/03/002}.
  {\href{https://arxiv.org/abs/1102.3229}{{arXiv:1102.3229}}} {[astro-ph.CO]}

\bibitem[{{Wagner} et~al.(2010){Wagner}, {Verde}, and
  {Boubekeur}}]{Wagner:2010}
{Wagner} C, {Verde} L, {Boubekeur} L (2010) {N-body simulations with generic
  non-Gaussian initial conditions I: power spectrum and halo mass function}.
  \jcap 2010(10):022. \doi{10.1088/1475-7516/2010/10/022}.
  {\href{https://arxiv.org/abs/1006.5793}{{arXiv:1006.5793}}} {[astro-ph.CO]}

\bibitem[{{Wagner} et~al.(2012){Wagner}, {Verde}, and {Jimenez}}]{Wagner:2012}
{Wagner} C, {Verde} L, {Jimenez} R (2012) {Effects of the Neutrino Mass
  Splitting on the Nonlinear Matter Power Spectrum}. \apjl 752(2):L31.
  \doi{10.1088/2041-8205/752/2/L31}.
  {\href{https://arxiv.org/abs/1203.5342}{{arXiv:1203.5342}}} {[astro-ph.CO]}

\bibitem[{{Wagner} et~al.(2015){Wagner}, {Schmidt}, {Chiang}, and
  {Komatsu}}]{Wagner:2015}
{Wagner} C, {Schmidt} F, {Chiang} CT, {Komatsu} E (2015) {Separate universe
  simulations.} \mnras 448:L11--L15. \doi{10.1093/mnrasl/slu187}.
  {\href{https://arxiv.org/abs/1409.6294}{{arXiv:1409.6294}}} {[astro-ph.CO]}

\bibitem[{{Wang} et~al.(2011){Wang}, {Mo}, {Jing}, {Yang}, and
  {Wang}}]{Wang:2011}
{Wang} H, {Mo} HJ, {Jing} YP, {Yang} X, {Wang} Y (2011) {Internal properties
  and environments of dark matter haloes}. \mnras 413(3):1973--1990.
  \doi{10.1111/j.1365-2966.2011.18301.x}.
  {\href{https://arxiv.org/abs/1007.0612}{{arXiv:1007.0612}}} {[astro-ph.CO]}

\bibitem[{{Wang} et~al.(2014){Wang}, {Mo}, {Yang}, {Jing}, and
  {Lin}}]{ElucidI:2014}
{Wang} H, {Mo} HJ, {Yang} X, {Jing} YP, {Lin} WP (2014) {ELUCID---Exploring the
  Local Universe with the Reconstructed Initial Density Field. I. Hamiltonian
  Markov Chain Monte Carlo Method with Particle Mesh Dynamics}. \apj 794(1):94.
  \doi{10.1088/0004-637X/794/1/94}.
  {\href{https://arxiv.org/abs/1407.3451}{{arXiv:1407.3451}}} {[astro-ph.CO]}

\bibitem[{{Wang} et~al.(2016){Wang}, {Mo}, {Yang}, {Zhang}, {Shi}, {Jing},
  {Liu}, {Li}, {Kang}, and {Gao}}]{ElucidIII:2016}
{Wang} H, {Mo} HJ, {Yang} X, {Zhang} Y, {Shi} J, {Jing} YP, {Liu} C, {Li} S,
  {Kang} X, {Gao} Y (2016) {ELUCID - Exploring the Local Universe with
  ReConstructed Initial Density Field III: Constrained Simulation in the SDSS
  Volume}. \apj 831(2):164. \doi{10.3847/0004-637X/831/2/164}.
  {\href{https://arxiv.org/abs/1608.01763}{{arXiv:1608.01763}}} {[astro-ph.CO]}

\bibitem[{{Wang} and {White}(2007)}]{WangWhite:2007}
{Wang} J, {White} SDM (2007) {Discreteness effects in simulations of hot/warm
  dark matter}. \mnras 380(1):93--103. \doi{10.1111/j.1365-2966.2007.12053.x}.
  {\href{https://arxiv.org/abs/astro-ph/0702575}{{arXiv:astro-ph/0702575}}}
  {[astro-ph]}

\bibitem[{{Wang} et~al.(2020){Wang}, {Bose}, {Frenk}, {Gao}, {Jenkins},
  {Springel}, and {White}}]{Wang:2020}
{Wang} J, {Bose} S, {Frenk} CS, {Gao} L, {Jenkins} A, {Springel} V, {White} SDM
  (2020) {Universal structure of dark matter haloes over a mass range of 20
  orders of magnitude}. \nat 585(7823):39--42. \doi{10.1038/s41586-020-2642-9}.
  {\href{https://arxiv.org/abs/1911.09720}{{arXiv:1911.09720}}} {[astro-ph.CO]}

\bibitem[{{Wang}(2021)}]{WangFMM:2020}
{Wang} Q (2021) {A hybrid Fast Multipole Method for cosmological N-body
  simulations}. Research in Astronomy and Astrophysics 21(1):003.
  \doi{10.1088/1674-4527/21/1/3}.
  {\href{https://arxiv.org/abs/2006.14952}{{arXiv:2006.14952}}}
  {[physics.comp-ph]}

\bibitem[{{Warren}(2013)}]{Warren:2013}
{Warren} MS (2013) {2HOT: An Improved Parallel Hashed Oct-Tree N-Body Algorithm
  for Cosmological Simulation}. arXiv e-prints arXiv:1310.4502.
  {\href{https://arxiv.org/abs/1310.4502}{{arXiv:1310.4502}}} {[astro-ph.IM]}

\bibitem[{{Warren} et~al.(2006){Warren}, {Abazajian}, {Holz}, and
  {Teodoro}}]{Warren:2006}
{Warren} MS, {Abazajian} K, {Holz} DE, {Teodoro} L (2006) {Precision
  Determination of the Mass Function of Dark Matter Halos}. \apj
  646(2):881--885. \doi{10.1086/504962}.
  {\href{https://arxiv.org/abs/astro-ph/0506395}{{arXiv:astro-ph/0506395}}}
  {[astro-ph]}

\bibitem[{{Watson} et~al.(2013){Watson}, {Iliev}, {D'Aloisio}, {Knebe},
  {Shapiro}, and {Yepes}}]{Watson:2013}
{Watson} WA, {Iliev} IT, {D'Aloisio} A, {Knebe} A, {Shapiro} PR, {Yepes} G
  (2013) {The halo mass function through the cosmic ages}. \mnras
  433(2):1230--1245. \doi{10.1093/mnras/stt791}.
  {\href{https://arxiv.org/abs/1212.0095}{{arXiv:1212.0095}}} {[astro-ph.CO]}

\bibitem[{{Wechsler} and {Tinker}(2018)}]{Wechsler:2018}
{Wechsler} RH, {Tinker} JL (2018) {The Connection Between Galaxies and Their
  Dark Matter Halos}. \araa 56:435--487.
  \doi{10.1146/annurev-astro-081817-051756}.
  {\href{https://arxiv.org/abs/1804.03097}{{arXiv:1804.03097}}} {[astro-ph.GA]}

\bibitem[{{Weller} et~al.(2005){Weller}, {Ostriker}, {Bode}, and
  {Shaw}}]{Weller:2005}
{Weller} J, {Ostriker} JP, {Bode} P, {Shaw} L (2005) {Fast identification of
  bound structures in large N-body simulations}. \mnras 364(3):823--832.
  \doi{10.1111/j.1365-2966.2005.09602.x}.
  {\href{https://arxiv.org/abs/astro-ph/0405445}{{arXiv:astro-ph/0405445}}}
  {[astro-ph]}

\bibitem[{{White} et~al.(2014){White}, {Tinker}, and {McBride}}]{White:2014}
{White} M, {Tinker} JL, {McBride} CK (2014) {Mock galaxy catalogues using the
  quick particle mesh method}. \mnras 437(3):2594--2606.
  \doi{10.1093/mnras/stt2071}.
  {\href{https://arxiv.org/abs/1309.5532}{{arXiv:1309.5532}}} {[astro-ph.CO]}

\bibitem[{{White}(1994)}]{White:1994}
{White} SDM (1994) {Formation and Evolution of Galaxies: Les Houches Lectures}.
  arXiv e-prints astro-ph/9410043.
  {\href{https://arxiv.org/abs/astro-ph/9410043}{{arXiv:astro-ph/9410043}}}
  {[astro-ph]}

\bibitem[{{White} and {Frenk}(1991)}]{White:1991}
{White} SDM, {Frenk} CS (1991) {Galaxy Formation through Hierarchical
  Clustering}. \apj 379:52. \doi{10.1086/170483}

\bibitem[{{White} and {Rees}(1978)}]{WhiteRees:1978}
{White} SDM, {Rees} MJ (1978) {Core condensation in heavy halos: a two-stage
  theory for galaxy formation and clustering.} \mnras 183:341--358.
  \doi{10.1093/mnras/183.3.341}

\bibitem[{{White} and {Vogelsberger}(2009)}]{White:2009}
{White} SDM, {Vogelsberger} M (2009) {Dark matter caustics}. \mnras
  392(1):281--286. \doi{10.1111/j.1365-2966.2008.14038.x}.
  {\href{https://arxiv.org/abs/0809.0497}{{arXiv:0809.0497}}} {[astro-ph]}

\bibitem[{{White} et~al.(1983){White}, {Frenk}, and {Davis}}]{White:1983}
{White} SDM, {Frenk} CS, {Davis} M (1983) {Clustering in a neutrino-dominated
  universe}. \apjl 274:L1--L5. \doi{10.1086/184139}

\bibitem[{{Widrow} and {Kaiser}(1993)}]{Widrow:1993}
{Widrow} LM, {Kaiser} N (1993) {Using the Schr\"odinger Equation to Simulate
  Collisionless Matter}. \apjl 416:L71. \doi{10.1086/187073}

\bibitem[{{Winther} et~al.(2019){Winther}, {Casas}, {Baldi}, {Koyama}, {Li},
  {Lombriser}, and {Zhao}}]{Winther:2019}
{Winther} H, {Casas} S, {Baldi} M, {Koyama} K, {Li} B, {Lombriser} L, {Zhao} GB
  (2019) {Emulators for the non-linear matter power spectrum beyond
  $\Lambda$CDM}. arXiv e-prints arXiv:1903.08798.
  {\href{https://arxiv.org/abs/1903.08798}{{arXiv:1903.08798}}} {[astro-ph.CO]}

\bibitem[{{Winther} and {Ferreira}(2015)}]{Winther:2014}
{Winther} HA, {Ferreira} PG (2015) {Fast route to nonlinear clustering
  statistics in modified gravity theories}. \prd 91(12):123507.
  \doi{10.1103/PhysRevD.91.123507}.
  {\href{https://arxiv.org/abs/1403.6492}{{arXiv:1403.6492}}} {[astro-ph.CO]}

\bibitem[{{Winther} et~al.(2015){Winther}, {Schmidt}, {Barreira}, {Arnold},
  {Bose}, {Llinares}, {Baldi}, {Falck}, {Hellwing}, {Koyama}, {Li}, {Mota},
  {Puchwein}, {Smith}, and {Zhao}}]{Winther:2015}
{Winther} HA, {Schmidt} F, {Barreira} A, {Arnold} C, {Bose} S, {Llinares} C,
  {Baldi} M, {Falck} B, {Hellwing} WA, {Koyama} K, {Li} B, {Mota} DF,
  {Puchwein} E, {Smith} RE, {Zhao} GB (2015) {Modified gravity N-body code
  comparison project}. \mnras 454(4):4208--4234. \doi{10.1093/mnras/stv2253}.
  {\href{https://arxiv.org/abs/1506.06384}{{arXiv:1506.06384}}} {[astro-ph.CO]}

\bibitem[{{Winther} et~al.(2017){Winther}, {Koyama}, {Manera}, {Wright}, and
  {Zhao}}]{Winther:2017}
{Winther} HA, {Koyama} K, {Manera} M, {Wright} BS, {Zhao} GB (2017) {COLA with
  scale-dependent growth: applications to screened modified gravity models}.
  \jcap 2017(8):006. \doi{10.1088/1475-7516/2017/08/006}.
  {\href{https://arxiv.org/abs/1703.00879}{{arXiv:1703.00879}}} {[astro-ph.CO]}

\bibitem[{{Woo} and {Chiueh}(2009)}]{Woo:2009}
{Woo} TP, {Chiueh} T (2009) {High-Resolution Simulation on Structure Formation
  with Extremely Light Bosonic Dark Matter}. \apj 697:850--861.
  \doi{10.1088/0004-637X/697/1/850}.
  {\href{https://arxiv.org/abs/0806.0232}{{arXiv:0806.0232}}}

\bibitem[{{Wright} et~al.(2017){Wright}, {Winther}, and {Koyama}}]{Wright:2017}
{Wright} BS, {Winther} HA, {Koyama} K (2017) {COLA with massive neutrinos}.
  \jcap 2017(10):054. \doi{10.1088/1475-7516/2017/10/054}.
  {\href{https://arxiv.org/abs/1705.08165}{{arXiv:1705.08165}}} {[astro-ph.CO]}

\bibitem[{{Wu} et~al.(2013){Wu}, {Hahn}, {Evrard}, {Wechsler}, and
  {Dolag}}]{WuHahn:2013}
{Wu} HY, {Hahn} O, {Evrard} AE, {Wechsler} RH, {Dolag} K (2013) {Virial scaling
  of galaxies in clusters: bright to faint is cool to hot}. \mnras
  436(1):460--469. \doi{10.1093/mnras/stt1582}.
  {\href{https://arxiv.org/abs/1307.0011}{{arXiv:1307.0011}}} {[astro-ph.CO]}

\bibitem[{{Xiao} et~al.(2021){Xiao}, {Williams}, and {McQuinn}}]{Xiao:2021}
{Xiao} H, {Williams} I, {McQuinn} M (2021) {Simulations of Axion Minihalos}.
  arXiv e-prints arXiv:2101.04177.
  {\href{https://arxiv.org/abs/2101.04177}{{arXiv:2101.04177}}} {[astro-ph.CO]}

\bibitem[{{Xu} and {Zheng}(2020)}]{XuZheng:2020}
{Xu} X, {Zheng} Z (2020) {Galaxy assembly bias of central galaxies in the
  Illustris simulation}. \mnras 492(2):2739--2754. \doi{10.1093/mnras/staa009}.
  {\href{https://arxiv.org/abs/1812.11210}{{arXiv:1812.11210}}} {[astro-ph.GA]}

\bibitem[{{Xu} et~al.(2019){Xu}, {Cisewski-Kehe}, {Green}, and
  {Nagai}}]{Xu:2019}
{Xu} X, {Cisewski-Kehe} J, {Green} SB, {Nagai} D (2019) {Finding cosmic voids
  and filament loops using topological data analysis}. Astronomy and Computing
  27:34. \doi{10.1016/j.ascom.2019.02.003}.
  {\href{https://arxiv.org/abs/1811.08450}{{arXiv:1811.08450}}} {[astro-ph.CO]}

\bibitem[{{Xu} et~al.(2021){Xu}, {Zehavi}, and {Contreras}}]{Xu:2021}
{Xu} X, {Zehavi} I, {Contreras} S (2021) {Dissecting and modelling galaxy
  assembly bias}. \mnras 502(3):3242--3263. \doi{10.1093/mnras/stab100}.
  {\href{https://arxiv.org/abs/2007.05545}{{arXiv:2007.05545}}} {[astro-ph.GA]}

\bibitem[{{Yamauchi} et~al.(2014){Yamauchi}, {Takahashi}, and
  {Oguri}}]{Yamauchi:2014}
{Yamauchi} D, {Takahashi} K, {Oguri} M (2014) {Constraining primordial
  non-Gaussianity via a multitracer technique with surveys by Euclid and the
  Square Kilometre Array}. \prd 90(8):083520. \doi{10.1103/PhysRevD.90.083520}.
  {\href{https://arxiv.org/abs/1407.5453}{{arXiv:1407.5453}}} {[astro-ph.CO]}

\bibitem[{{Yang} et~al.(2012){Yang}, {Mo}, {van den Bosch}, {Zhang}, and
  {Han}}]{Yang:2012}
{Yang} X, {Mo} HJ, {van den Bosch} FC, {Zhang} Y, {Han} J (2012) {Evolution of
  the Galaxy-Dark Matter Connection and the Assembly of Galaxies in Dark Matter
  Halos}. \apj 752(1):41. \doi{10.1088/0004-637X/752/1/41}.
  {\href{https://arxiv.org/abs/1110.1420}{{arXiv:1110.1420}}} {[astro-ph.CO]}

\bibitem[{{Yang} et~al.(2009){Yang}, {Feng}, {Pan}, and {Yang}}]{Yang:2009}
{Yang} YB, {Feng} LL, {Pan} J, {Yang} XH (2009) {An optimal method for the
  power spectrum measurement}. Research in Astronomy and Astrophysics
  9(2):227--236. \doi{10.1088/1674-4527/9/2/012}

\bibitem[{{Ye} et~al.(2017){Ye}, {Guo}, {Zheng}, and {Zehavi}}]{Ye:2017}
{Ye} JN, {Guo} H, {Zheng} Z, {Zehavi} I (2017) {Properties and Origin of Galaxy
  Velocity Bias in the Illustris Simulation}. \apj 841(1):45.
  \doi{10.3847/1538-4357/aa70e7}.
  {\href{https://arxiv.org/abs/1705.02071}{{arXiv:1705.02071}}} {[astro-ph.CO]}

\bibitem[{{Yepes} et~al.(2014){Yepes}, {Gottl{\"o}ber}, and
  {Hoffman}}]{Yepes:2014}
{Yepes} G, {Gottl{\"o}ber} S, {Hoffman} Y (2014) {Dark matter in the Local
  Universe}. \nar 58:1--18. \doi{10.1016/j.newar.2013.11.001}.
  {\href{https://arxiv.org/abs/1312.0105}{{arXiv:1312.0105}}} {[astro-ph.CO]}

\bibitem[{Yokota and Barba(2012)}]{Yokota:2012}
Yokota R, Barba LA (2012) A tuned and scalable fast multipole method as a
  preeminent algorithm for exascale systems. Int J High Perform Comput Appl
  26(4):337--346. \doi{10.1177/1094342011429952}.
  {\href{https://arxiv.org/abs/https://doi.org/10.1177/1094342011429952}{{https://doi.org/10.1177/1094342011429952}}}

\bibitem[{{Yoshida}(1990)}]{Yoshida:1990}
{Yoshida} H (1990) {Construction of higher order symplectic integrators}. Phys
  Lett A 150:262--268. \doi{10.1016/0375-9601(90)90092-3}

\bibitem[{{Yoshida} et~al.(2000){Yoshida}, {Springel}, {White}, and
  {Tormen}}]{Yoshida:2000}
{Yoshida} N, {Springel} V, {White} SDM, {Tormen} G (2000) {Weakly
  Self-interacting Dark Matter and the Structure of Dark Halos}. \apjl
  544:L87--L90. \doi{10.1086/317306}.
  {\href{https://arxiv.org/abs/astro-ph/0006134}{{astro-ph/0006134}}}

\bibitem[{{Yoshida} et~al.(2003){Yoshida}, {Sugiyama}, and
  {Hernquist}}]{Yoshida:2003}
{Yoshida} N, {Sugiyama} N, {Hernquist} L (2003) {The evolution of baryon
  density fluctuations in multicomponent cosmological simulations}. \mnras
  344(2):481--491. \doi{10.1046/j.1365-8711.2003.06829.x}.
  {\href{https://arxiv.org/abs/astro-ph/0305210}{{arXiv:astro-ph/0305210}}}
  {[astro-ph]}

\bibitem[{{Yoshikawa} et~al.(2013){Yoshikawa}, {Yoshida}, and
  {Umemura}}]{Yoshikawa:2013}
{Yoshikawa} K, {Yoshida} N, {Umemura} M (2013) {Direct Integration of the
  Collisionless Boltzmann Equation in Six-dimensional Phase Space:
  Self-gravitating Systems}. \apj 762:116. \doi{10.1088/0004-637X/762/2/116}.
  {\href{https://arxiv.org/abs/1206.6152}{{arXiv:1206.6152}}} {[astro-ph.IM]}

\bibitem[{{Yoshikawa} et~al.(2020){Yoshikawa}, {Tanaka}, {Yoshida}, and
  {Saito}}]{Yoshikawa:2020}
{Yoshikawa} K, {Tanaka} S, {Yoshida} N, {Saito} S (2020) {Cosmological
  Vlasov--Poisson Simulations of Structure Formation with Relic Neutrinos:
  Nonlinear Clustering and the Neutrino Mass}. \apj 904(2):159.
  \doi{10.3847/1538-4357/abbd46}.
  {\href{https://arxiv.org/abs/2010.00248}{{arXiv:2010.00248}}} {[astro-ph.CO]}

\bibitem[{{Yoshikawa} et~al.(2021){Yoshikawa}, {Tanaka}, and
  {Yoshida}}]{Yoshikawa:2021}
{Yoshikawa} K, {Tanaka} S, {Yoshida} N (2021) {A 400 Trillion-Grid Vlasov
  Simulation on Fugaku Supercomputer: Large-Scale Distribution of Cosmic Relic
  Neutrinos in a Six-dimensional Phase Space}. arXiv e-prints arXiv:2110.15867.
  {\href{https://arxiv.org/abs/2110.15867}{{arXiv:2110.15867}}} {[astro-ph.CO]}

\bibitem[{{Yu} et~al.(2017{\natexlab{a}}){Yu}, {Emberson}, {Inman}, {Zhang},
  {Pen}, {Harnois-D{\'e}raps}, {Yuan}, {Teng}, {Zhu}, {Chen}, {Xing}, {Du},
  {Zhang}, {Lu}, and {Liao}}]{Yu:2017}
{Yu} HR, {Emberson} JD, {Inman} D, {Zhang} TJ, {Pen} UL, {Harnois-D{\'e}raps}
  J, {Yuan} S, {Teng} HY, {Zhu} HM, {Chen} X, {Xing} ZZ, {Du} Y, {Zhang} L,
  {Lu} Y, {Liao} X (2017{\natexlab{a}}) {Differential neutrino condensation
  onto cosmic structure}. Nature Astronomy 1:0143.
  \doi{10.1038/s41550-017-0143}.
  {\href{https://arxiv.org/abs/1609.08968}{{arXiv:1609.08968}}} {[astro-ph.CO]}

\bibitem[{{Yu} et~al.(2018){Yu}, {Pen}, and {Wang}}]{YuCUBE:2018}
{Yu} HR, {Pen} UL, {Wang} X (2018) {CUBE: An Information-optimized Parallel
  Cosmological N-body Algorithm}. \apjs 237(2):24.
  \doi{10.3847/1538-4365/aac830}.
  {\href{https://arxiv.org/abs/1712.06121}{{arXiv:1712.06121}}} {[astro-ph.CO]}

\bibitem[{{Yu} et~al.(2015){Yu}, {Zhang}, {Jing}, and {Zhang}}]{Yu:2015}
{Yu} Y, {Zhang} J, {Jing} Y, {Zhang} P (2015) {Kriging interpolating cosmic
  velocity field}. \prd 92(8):083527. \doi{10.1103/PhysRevD.92.083527}.
  {\href{https://arxiv.org/abs/1505.06827}{{arXiv:1505.06827}}} {[astro-ph.CO]}

\bibitem[{{Yu} et~al.(2017{\natexlab{b}}){Yu}, {Zhang}, {Jing}, and
  {Zhang}}]{YuKriging:2017}
{Yu} Y, {Zhang} J, {Jing} Y, {Zhang} P (2017{\natexlab{b}}) {Kriging
  interpolating cosmic velocity field. II. Taking anistropies and
  multistreaming into account}. \prd 95(4):043536.
  \doi{10.1103/PhysRevD.95.043536}.
  {\href{https://arxiv.org/abs/1603.05363}{{arXiv:1603.05363}}} {[astro-ph.CO]}

\bibitem[{{Zavala} and {Frenk}(2019)}]{Zavala:2019}
{Zavala} J, {Frenk} CS (2019) {Dark Matter Haloes and Subhaloes}. Galaxies
  7(4):81. \doi{10.3390/galaxies7040081}.
  {\href{https://arxiv.org/abs/1907.11775}{{arXiv:1907.11775}}} {[astro-ph.CO]}

\bibitem[{{Zavala} et~al.(2009){Zavala}, {Jing}, {Faltenbacher}, {Yepes},
  {Hoffman}, {Gottl{\"o}ber}, and {Catinella}}]{Zavala:2009}
{Zavala} J, {Jing} YP, {Faltenbacher} A, {Yepes} G, {Hoffman} Y,
  {Gottl{\"o}ber} S, {Catinella} B (2009) {The Velocity Function in the Local
  Environment from {\ensuremath{\Lambda}}CDM and {\ensuremath{\Lambda}}WDM
  Constrained Simulations}. \apj 700(2):1779--1793.
  \doi{10.1088/0004-637X/700/2/1779}.
  {\href{https://arxiv.org/abs/0906.0585}{{arXiv:0906.0585}}} {[astro-ph.CO]}

\bibitem[{{Zavala} et~al.(2010){Zavala}, {Springel}, and
  {Boylan-Kolchin}}]{Zavala:2010}
{Zavala} J, {Springel} V, {Boylan-Kolchin} M (2010) {Extragalactic gamma-ray
  background radiation from dark matter annihilation}. \mnras 405(1):593--612.
  \doi{10.1111/j.1365-2966.2010.16482.x}.
  {\href{https://arxiv.org/abs/0908.2428}{{arXiv:0908.2428}}} {[astro-ph.CO]}

\bibitem[{{Zavala} et~al.(2013){Zavala}, {Vogelsberger}, and
  {Walker}}]{Zavala:2013}
{Zavala} J, {Vogelsberger} M, {Walker} MG (2013) {Constraining self-interacting
  dark matter with the Milky way's dwarf spheroidals.} \mnras 431:L20--L24.
  \doi{10.1093/mnrasl/sls053}.
  {\href{https://arxiv.org/abs/1211.6426}{{arXiv:1211.6426}}} {[astro-ph.CO]}

\bibitem[{{Zehavi} et~al.(2005){Zehavi}, {Zheng}, {Weinberg}, {Frieman},
  {Berlind}, {Blanton}, {Scoccimarro}, {Sheth}, {Strauss}, {Kayo}, {Suto},
  {Fukugita}, {Nakamura}, {Bahcall}, {Brinkmann}, {Gunn}, {Hennessy},
  {Ivezi{\'c}}, {Knapp}, {Loveday}, {Meiksin}, {Schlegel}, {Schneider},
  {Szapudi}, {Tegmark}, {Vogeley}, {York}, and {SDSS
  Collaboration}}]{Zehavi:2005}
{Zehavi} I, {Zheng} Z, {Weinberg} DH, {Frieman} JA, {Berlind} AA, {Blanton} MR,
  {Scoccimarro} R, {Sheth} RK, {Strauss} MA, {Kayo} I, {Suto} Y, {Fukugita} M,
  {Nakamura} O, {Bahcall} NA, {Brinkmann} J, {Gunn} JE, {Hennessy} GS,
  {Ivezi{\'c}} {\v{Z}}, {Knapp} GR, {Loveday} J, {Meiksin} A, {Schlegel} DJ,
  {Schneider} DP, {Szapudi} I, {Tegmark} M, {Vogeley} MS, {York} DG, {SDSS
  Collaboration} (2005) {The Luminosity and Color Dependence of the Galaxy
  Correlation Function}. \apj 630(1):1--27. \doi{10.1086/431891}.
  {\href{https://arxiv.org/abs/astro-ph/0408569}{{arXiv:astro-ph/0408569}}}
  {[astro-ph]}

\bibitem[{{Zehavi} et~al.(2019){Zehavi}, {Kerby}, {Contreras}, {Jim{\'e}nez},
  {Padilla}, and {Baugh}}]{Zehavi:2019}
{Zehavi} I, {Kerby} SE, {Contreras} S, {Jim{\'e}nez} E, {Padilla} N, {Baugh} CM
  (2019) {On the Prospect of Using the Maximum Circular Velocity of Halos to
  Encapsulate Assembly Bias in the Galaxy-Halo Connection}. \apj 887(1):17.
  \doi{10.3847/1538-4357/ab4d4d}.
  {\href{https://arxiv.org/abs/1907.05424}{{arXiv:1907.05424}}} {[astro-ph.GA]}

\bibitem[{{Zel'dovich}(1970)}]{Zeldovich:1970}
{Zel'dovich} YB (1970) {Gravitational instability: An approximate theory for
  large density perturbations.} \aap 5:84--89

\bibitem[{{Zennaro} et~al.(2017){Zennaro}, {Bel}, {Villaescusa-Navarro},
  {Carbone}, {Sefusatti}, and {Guzzo}}]{Zennaro:2017}
{Zennaro} M, {Bel} J, {Villaescusa-Navarro} F, {Carbone} C, {Sefusatti} E,
  {Guzzo} L (2017) {Initial conditions for accurate N-body simulations of
  massive neutrino cosmologies}. \mnras 466:3244--3258.
  \doi{10.1093/mnras/stw3340}.
  {\href{https://arxiv.org/abs/1605.05283}{{arXiv:1605.05283}}}

\bibitem[{{Zennaro} et~al.(2019){Zennaro}, {Angulo}, {Aric{\`o}}, {Contreras},
  and {Pellejero-Ib{\'a}{\~n}ez}}]{Zennaro:2019}
{Zennaro} M, {Angulo} RE, {Aric{\`o}} G, {Contreras} S,
  {Pellejero-Ib{\'a}{\~n}ez} M (2019) {How to add massive neutrinos to your
  {\ensuremath{\Lambda}}CDM simulation - extending cosmology rescaling
  algorithms}. \mnras 489(4):5938--5951. \doi{10.1093/mnras/stz2612}.
  {\href{https://arxiv.org/abs/1905.08696}{{arXiv:1905.08696}}} {[astro-ph.CO]}

\bibitem[{{Zennaro} et~al.(2021){Zennaro}, {Angulo},
  {Pellejero-Ib{\'a}{\~n}ez}, {St{\"u}cker}, {Contreras}, and
  {Aric{\`o}}}]{Zennaro:2021}
{Zennaro} M, {Angulo} RE, {Pellejero-Ib{\'a}{\~n}ez} M, {St{\"u}cker} J,
  {Contreras} S, {Aric{\`o}} G (2021) {The BACCO simulation project: biased
  tracers in real space}. arXiv e-prints arXiv:2101.12187.
  {\href{https://arxiv.org/abs/2101.12187}{{arXiv:2101.12187}}} {[astro-ph.CO]}

\bibitem[{Zhai et~al.(2019)Zhai, Tinker, Becker, DeRose, Mao, McClintock,
  McLaughlin, Rozo, and Wechsler}]{Zhai:2018}
Zhai Z, Tinker JL, Becker MR, DeRose J, Mao YY, McClintock T, McLaughlin S,
  Rozo E, Wechsler RH (2019) {The Aemulus Project III: Emulation of the Galaxy
  Correlation Function}. Astrophys J 874(1):95. \doi{10.3847/1538-4357/ab0d7b}.
  {\href{https://arxiv.org/abs/1804.05867}{{arXiv:1804.05867}}} {[astro-ph.CO]}

\bibitem[{{Zhang} et~al.(2020){Zhang}, {Li}, {Liu}, {Spergel}, {Kreisch},
  {Pisani}, and {Wandelt}}]{Zhang:2020}
{Zhang} G, {Li} Z, {Liu} J, {Spergel} DN, {Kreisch} CD, {Pisani} A, {Wandelt}
  BD (2020) {Void halo mass function: A promising probe of neutrino mass}. \prd
  102(8):083537. \doi{10.1103/PhysRevD.102.083537}.
  {\href{https://arxiv.org/abs/1910.07553}{{arXiv:1910.07553}}} {[astro-ph.CO]}

\bibitem[{{Zhang} et~al.(2018{\natexlab{a}}){Zhang}, {Kuo}, {Liu}, {Sming
  Tsai}, {Cheung}, and {Chu}}]{ZhangQP:2018}
{Zhang} J, {Kuo} JL, {Liu} H, {Sming Tsai} YL, {Cheung} K, {Chu} MC
  (2018{\natexlab{a}}) {The Importance of Quantum Pressure of Fuzzy Dark Matter
  on Ly{\ensuremath{\alpha}} Forest}. \apj 863(1):73.
  \doi{10.3847/1538-4357/aacf3f}.
  {\href{https://arxiv.org/abs/1708.04389}{{arXiv:1708.04389}}} {[astro-ph.CO]}

\bibitem[{{Zhang} et~al.(2018{\natexlab{b}}){Zhang}, {Sming Tsai}, {Kuo},
  {Cheung}, and {Chu}}]{Zhang:2018}
{Zhang} J, {Sming Tsai} YL, {Kuo} JL, {Cheung} K, {Chu} MC (2018{\natexlab{b}})
  {Ultralight Axion Dark Matter and Its Impact on Dark Halo Structure in N-body
  Simulations}. \apj 853(1):51. \doi{10.3847/1538-4357/aaa485}.
  {\href{https://arxiv.org/abs/1611.00892}{{arXiv:1611.00892}}} {[astro-ph.CO]}

\bibitem[{Zhang et~al.(2002)Zhang, Zheng, and Mauser}]{Zhang:2002}
Zhang P, Zheng Y, Mauser NJ (2002) The limit from the schr{\"o}dinger-poisson
  to the vlasov-poisson equations with general data in one dimension.
  Communications on Pure and Applied Mathematics 55(5):582--632.
  \doi{https://doi.org/10.1002/cpa.3017}

\bibitem[{{Zhang} et~al.(2015){Zhang}, {Zheng}, and {Jing}}]{Zhang:2015}
{Zhang} P, {Zheng} Y, {Jing} Y (2015) {Sampling artifact in volume weighted
  velocity measurement. I. Theoretical modeling}. \prd 91(4):043522.
  \doi{10.1103/PhysRevD.91.043522}.
  {\href{https://arxiv.org/abs/1405.7125}{{arXiv:1405.7125}}} {[astro-ph.CO]}

\bibitem[{{Zhang} et~al.(2019{\natexlab{a}}){Zhang}, {Liao}, {Li}, and
  {Gao}}]{ZhangSoftening:2019}
{Zhang} T, {Liao} S, {Li} M, {Gao} L (2019{\natexlab{a}}) {The optimal
  gravitational softening length for cosmological N-body simulations}. \mnras
  487(1):1227--1232. \doi{10.1093/mnras/stz1370}.
  {\href{https://arxiv.org/abs/1810.07055}{{arXiv:1810.07055}}} {[astro-ph.CO]}

\bibitem[{{Zhang} et~al.(2019{\natexlab{b}}){Zhang}, {Wang}, {Zhang}, {Sun},
  {He}, {Contardo}, {Villaescusa-Navarro}, and {Ho}}]{Zhang:2019}
{Zhang} X, {Wang} Y, {Zhang} W, {Sun} Y, {He} S, {Contardo} G,
  {Villaescusa-Navarro} F, {Ho} S (2019{\natexlab{b}}) {From Dark Matter to
  Galaxies with Convolutional Networks}. arXiv e-prints arXiv:1902.05965.
  {\href{https://arxiv.org/abs/1902.05965}{{arXiv:1902.05965}}} {[astro-ph.CO]}

\bibitem[{{Zhao} et~al.(2011){Zhao}, {Li}, and {Koyama}}]{Zhao:2011}
{Zhao} GB, {Li} B, {Koyama} K (2011) {N-body simulations for f(R) gravity using
  a self-adaptive particle-mesh code}. \prd 83(4):044007.
  \doi{10.1103/PhysRevD.83.044007}.
  {\href{https://arxiv.org/abs/1011.1257}{{arXiv:1011.1257}}} {[astro-ph.CO]}

\bibitem[{{Zheligovsky} and {Frisch}(2014)}]{Zheligovsky:2014}
{Zheligovsky} V, {Frisch} U (2014) {Time-analyticity of Lagrangian particle
  trajectories in ideal fluid flow}. J Fluid Mech 749:404--430.
  \doi{10.1017/jfm.2014.221}.
  {\href{https://arxiv.org/abs/1312.6320}{{arXiv:1312.6320}}} {[math.AP]}

\bibitem[{{Zheng} et~al.(2015{\natexlab{a}}){Zheng}, {Zhang}, and
  {Jing}}]{ZhengVelBias:2015}
{Zheng} Y, {Zhang} P, {Jing} Y (2015{\natexlab{a}}) {Determination of the large
  scale volume weighted halo velocity bias in simulations}. \prd 91(12):123512.
  \doi{10.1103/PhysRevD.91.123512}.
  {\href{https://arxiv.org/abs/1410.1256}{{arXiv:1410.1256}}} {[astro-ph.CO]}

\bibitem[{{Zheng} et~al.(2015{\natexlab{b}}){Zheng}, {Zhang}, and
  {Jing}}]{Zheng:2015}
{Zheng} Y, {Zhang} P, {Jing} Y (2015{\natexlab{b}}) {Sampling artifact in
  volume weighted velocity measurement. II. Detection in simulations and
  comparison with theoretical modeling}. \prd 91(4):043523.
  \doi{10.1103/PhysRevD.91.043523}.
  {\href{https://arxiv.org/abs/1409.6809}{{arXiv:1409.6809}}} {[astro-ph.CO]}

\bibitem[{{Zheng} et~al.(2005){Zheng}, {Berlind}, {Weinberg}, {Benson},
  {Baugh}, {Cole}, {Dav{\'e}}, {Frenk}, {Katz}, and {Lacey}}]{Zheng:2005}
{Zheng} Z, {Berlind} AA, {Weinberg} DH, {Benson} AJ, {Baugh} CM, {Cole} S,
  {Dav{\'e}} R, {Frenk} CS, {Katz} N, {Lacey} CG (2005) {Theoretical Models of
  the Halo Occupation Distribution: Separating Central and Satellite Galaxies}.
  \apj 633(2):791--809. \doi{10.1086/466510}.
  {\href{https://arxiv.org/abs/astro-ph/0408564}{{arXiv:astro-ph/0408564}}}
  {[astro-ph]}

\bibitem[{{Zheng} et~al.(2007){Zheng}, {Coil}, and {Zehavi}}]{Zheng:2007}
{Zheng} Z, {Coil} AL, {Zehavi} I (2007) {Galaxy Evolution from Halo Occupation
  Distribution Modeling of DEEP2 and SDSS Galaxy Clustering}. \apj
  667(2):760--779. \doi{10.1086/521074}.
  {\href{https://arxiv.org/abs/astro-ph/0703457}{{arXiv:astro-ph/0703457}}}
  {[astro-ph]}

\bibitem[{{Zhiyu Chen} et~al.(2020){Zhiyu Chen}, {Upadhye}, and
  {Wong}}]{Chen:2020}
{Zhiyu Chen} J, {Upadhye} A, {Wong} YYY (2020) {One line to run them all:
  SuperEasy massive neutrino linear response in $N$-body simulations}. arXiv
  e-prints arXiv:2011.12504.
  {\href{https://arxiv.org/abs/2011.12504}{{arXiv:2011.12504}}} {[astro-ph.CO]}

\bibitem[{{Zorrilla-Matilla} et~al.(2020){Zorrilla-Matilla}, {Sharma}, {Hsu},
  and {Haiman}}]{Zorrilla-Mantilla:2020}
{Zorrilla-Matilla} JM, {Sharma} M, {Hsu} D, {Haiman} Z (2020) {Interpreting
  deep learning models for weak lensing}. arXiv e-prints arXiv:2007.06529.
  {\href{https://arxiv.org/abs/2007.06529}{{arXiv:2007.06529}}} {[astro-ph.CO]}

\bibitem[{{Zumalac{\'a}rregui} et~al.(2017){Zumalac{\'a}rregui}, {Bellini},
  {Sawicki}, {Lesgourgues}, and {Ferreira}}]{Zumalacarregui:2017}
{Zumalac{\'a}rregui} M, {Bellini} E, {Sawicki} I, {Lesgourgues} J, {Ferreira}
  PG (2017) {hi\_class: Horndeski in the Cosmic Linear Anisotropy Solving
  System}. \jcap 2017(8):019. \doi{10.1088/1475-7516/2017/08/019}.
  {\href{https://arxiv.org/abs/1605.06102}{{arXiv:1605.06102}}} {[astro-ph.CO]}

\end{thebibliography}

\end{document}